\documentclass[draft,final]{vutinfth} 

\usepackage{lmodern}        
\usepackage[T1]{fontenc}    
\usepackage[utf8]{inputenc} 

\usepackage{amsmath}    
\usepackage{amssymb}    
\usepackage{mathtools}  
\usepackage{microtype}  
\usepackage[inline]{enumitem} 
\usepackage{multirow}   
\usepackage{booktabs}   
\usepackage{subcaption} 
\usepackage[ruled,linesnumbered,algochapter]{algorithm2e} 
\usepackage[usenames,dvipsnames,table]{xcolor} 
\usepackage{nag}       
\usepackage{todonotes} 
\usepackage{hyperref}  
\usepackage[acronym,toc]{glossaries} 

\usepackage{bussproofs, multicol, upgreek, amsthm, bm, stmaryrd} 
\usepackage[all]{xy}

\newtheorem{definition}{Definition}
\newtheorem{example}{Example}
\newtheorem{lemma}{Lemma}
\newtheorem{theorem}{Theorem}
\newtheorem{remark}{Remark}
\newtheorem{proposition}{Proposition}
\newtheorem{corollary}{Corollary}

\newtheorem{innercustomlem}{Lemma}
\newenvironment{customlem}[1]
  {\renewcommand\theinnercustomlem{#1}\innercustomlem}
  {\endinnercustomlem}
  
  \newtheorem{innercustomthm}{Theorem}
\newenvironment{customthm}[1]
  {\renewcommand\theinnercustomthm{#1}\innercustomthm}
  {\endinnercustomthm}

\newcommand{\gtk}{\mathsf{G3K}}

\newcommand{\h}{\mathsf{H}}

\newcommand{\fcomp}[1]{|#1|} 
\newcommand{\negnnf}[1]{\neg{#1}} 
\newcommand{\seqcomp}{\circ}

\newcommand{\prop}{\mathrm{Prop}} 
\newcommand{\pred}{\mathrm{Pred}} 
\newcommand{\lab}{\mathrm{Lab}} 
\newcommand{\premises}{\mathrm{X}} 
\newcommand{\var}{\mathrm{Var}} 
\newcommand{\para}{\mathrm{Par}} 
\newcommand{\ag}{\mathrm{Ag}} 

\newcommand{\unda}{\underline{a}}
\newcommand{\undb}{\underline{b}}
\newcommand{\undc}{\underline{c}}

\newcommand{\uc}{\forall \vec{x}}
\newcommand{\ec}{\exists \vec{y}}
\newcommand{\lb}{\langle}
\newcommand{\rb}{\rangle}
\newcommand{\binop}{\circ}

\newcommand{\sar}{\Rightarrow} 
\newcommand{\rel}{\mathcal{R}} 

\newcommand{\seqempstr}{\epsilon} 

\def\wk{(wk)} 
\def\ctrr{(ctr_{r})} 
\def\ctrl{(ctr_{l})} 
\def\ctrrel{(ctr_{\mathcal{R}})} 
\newcommand{\lsub}{(lsb)} 
\newcommand{\ru}{(r)} 
\newcommand{\ruone}{(r_{1})} 
\newcommand{\rutwo}{(r_{2})} 
\newcommand{\ruthree}{(r_{3})} 
\newcommand{\rufour}{(r_{4})} 

\newcommand{\etc}{$\ldots$ } 
\newcommand{\sect}{Sect.} 
\newcommand{\dfn}{Def.} 
\newcommand{\lem}{Lem.} 
\newcommand{\thm}{Thm.} 
\newcommand{\cor}{Cor.} 
\newcommand{\prp}{Prop.} 
\newcommand{\fig}{Fig.} 
\newcommand{\ifandonlyif}{iff } 
\newcommand{\suchthat}{ such that } 
\newcommand{\ih}{IH } 
\newcommand{\rem}{Rmk.} 
\newcommand{\resp}{resp.}
\newcommand{\cptr}{Ch.}
\newcommand{\exmpl}{Ex.}
\newcommand{\alg}{Alg.}

\newcommand{\val}{\mathbb{V}}
\newcommand{\sub}{\sigma}

\newcommand{\lnt}{\mathfrak{N}}
\newcommand{\nlt}{\mathfrak{L}}

\newcommand{\nbr}{]}
\newcommand{\nbl}{[}
\newcommand{\nbbr}{\}}
\newcommand{\nbbl}{\{}
\newcommand{\na}{X}
\newcommand{\nb}{Y}
\newcommand{\nc}{Z}

\newcommand{\skms}{\text{SKm(S)}}
\newcommand{\dkms}{\mathsf{DKm}(S)}

\newcommand{\wancondi}{\mathbf{(W_{1})}}
\newcommand{\wancondii}{\mathbf{(W_{2})}}
\newcommand{\wancondiii}{\mathbf{(W_{3})}}
\newcommand{\wancondiv}{\mathbf{(W_{4})}}
\newcommand{\wancondv}{\mathbf{(W_{5})}}
\newcommand{\wancondvi}{\mathbf{(W_{6})}}

\newcommand{\li}{\blacktriangleright}
\newcommand{\cfcst}{CFCST } 
\newcommand{\stit}{STIT } 
\newcommand{\lang}{\mathcal{L}}

\newcommand{\seqgraph}{G}
\newcommand{\simul}{\mathfrak{S}}
\newcommand{\simulates}{ \rightrightarrows_{\simul} }
\newcommand{\numseq}{\sigma} 
\newcommand{\iso}{\cong}

\newcommand{\concat}{\cdot} 
\newcommand{\prgr}[1]{PG(#1)}
\newcommand{\nlkms}{\mathfrak{L}}
\newcommand{\lnkms}{\mathfrak{N}}
\newcommand{\provecpl}{\mathtt{ProveCPL}}

\newcommand{\gtkms}{\mathsf{G3Km}(S)} 
\newcommand{\kmsl}{\mathsf{Km}(S)\mathsf{L}} 

\newcommand{\dis}{(\lor)} 
\newcommand{\charaboxr}{(\charabox)} 
\newcommand{\charadiar}{(\charadia)}
\newcommand{\adiar}{(\adia)} 
\newcommand{\psr}{(p^{\chara}_{\stra})} 
\newcommand{\psrp}[2]{(p^{#1}_{#2})} 

\newcommand{\convr}{(C_{1}^{\chara})}
\newcommand{\psrcc}{(p^{\chara}_{\stra})_{\ddag}}

\newcommand{\pradiar}{(Pr_{\adia})} 

\newcommand{\prcharadiar}{(Pr_{\charadia})}
\newcommand{\prcharbdiar}{(Pr_{\charbdia})}

\newcommand{\adia}{\langle a \rangle} 
\newcommand{\abox}{[a]} 
\newcommand{\bdia}{\langle b \rangle} 
\newcommand{\adiac}{\langle \overline{a} \rangle} 
\newcommand{\aboxc}{[\overline{a}]} 
\newcommand{\cimp}{\rightarrow} 
\newcommand{\cbiimp}{\leftrightarrow} 
\newcommand{\charabox}{[\chara]}
\newcommand{\charadia}{\langle \chara \rangle}
\newcommand{\charbdia}{\langle \charb \rangle}

\newcommand{\charadiac}{\langle \conv{\chara} \rangle}
\newcommand{\strabox}{[\stra]}
\newcommand{\stradia}{\langle \stra \rangle}

\newcommand{\km}{\mathsf{Km}} 
\newcommand{\kms}{\mathsf{Km}(S)} 

\newcommand{\langkm}[1]{\mathcal{L}_{\km}(#1)} 

\newcommand{\albet}{\sum} 
\newcommand{\albetstr}{\albet^{\ast}} 
\newcommand{\strucsetkms}{\mathrm{Str}({\kms})} 
\newcommand{\prgrdom}{\mathrm{V}} 
\newcommand{\prgredges}{\mathrm{E}} 

\newcommand{\pto}{\longrightarrow} 
\newcommand{\dto}{\twoheadrightarrow} 
\newcommand{\dtoann}{\dto_{\thuesys}^{*}} 
\newcommand{\stra}{s} 
\newcommand{\strb}{t} 
\newcommand{\strc}{r} 
\newcommand{\empstr}{\varepsilon} 
\newcommand{\thuesys}{S} 
\newcommand{\thuesyslang}[1]{L_{\thuesys}(#1)} 

\newcommand{\conv}[1]{\overline{#1}} 
\newcommand{\chara}{x} 
\newcommand{\charb}{y} 
\newcommand{\charc}{z} 
\newcommand{\lenstr}[1]{|#1|}
\newcommand{\cate}{} 

\newcommand{\convcond}{\textbf{(C\textsubscript{1})} } 
\newcommand{\thueclcond}{\textbf{(C\textsubscript{2})} } 
\newcommand{\convcondns}{\textbf{(C\textsubscript{1})}} 
\newcommand{\thueclcondns}{\textbf{(C\textsubscript{2})}} 

\newcommand{\ppath}{\pi}
\newcommand{\gtkmsder}{\vdash_{\gtkms}}
\newcommand{\emppath}{\lambda}

\def\exc{\ensuremath{- \! \! \! <}} 
\def\imp{\supset} 

\def\R{\mathcal{R}} 

\newcommand{\gtintfond}{\mathsf{G3IntQ}} 
\newcommand{\gtintfocd}{\mathsf{G3IntQC}} 
\newcommand{\intfondl}{\mathsf{IntQL}} 
\newcommand{\intfocdl}{\mathsf{IntQCL}} 
\newcommand{\intfondlq}{\mathsf{IntQL}^{*}}
\newcommand{\intfocdlq}{\mathsf{IntQCL}^{*}}
\newcommand{\nintfond}{\mathsf{DIntQ}}
\newcommand{\nintfocd}{\mathsf{DIntQC}}

\newcommand{\psub}{(psb)} 
\newcommand{\idfo}{(id)} 
\def\trans{(tra)}

\def\nd{(nd)}
\newcommand{\ned}{(ned)}

\def\cd{(cd)}

\def\cut{(cut)}
\def\id{(id)}
\def\botl{(\bot_{l})}

\newcommand{\disr}{(\vee_{r})}
\newcommand{\conr}{(\wedge_{r})}
\newcommand{\disl}{(\vee_{l})}
\newcommand{\conl}{(\wedge_{l})}
\newcommand{\impr}{(\supset_{r})}
\newcommand{\impl}{(\supset_{l})}

\newcommand{\refl}{(ref)}

\def\lift{(lift)}

\def\allr{(\forall_{r})}
\def\alll{(\forall_{l})}
\def\existsl{(\exists_{l})}
\def\existsr{(\exists_{r})}

\newcommand{\idfonc}{(id_{*})}
\newcommand{\existsrnc}{(\exists_{r}^{\norc})}

\newcommand{\alllnc}{(\forall_{l}^{\norc})}
\newcommand{\primp}{(Pr_{\imp})}
\newcommand{\allrnc}{(\forall_{r}^{\norc})}
\newcommand{\existslnc}{(\exists_{l}^{\norc})}
\newcommand{\allrn}{(\forall_{r}^{\nnn})}
\newcommand{\allln}{(\forall_{l}^{\nnn})}
\newcommand{\existsln}{(\exists_{l}^{\nnn})}
\newcommand{\existsrn}{(\exists_{r}^{\nnn})}
\newcommand{\allrc}{(\forall_{r}^{\ccc})}
\newcommand{\alllc}{(\forall_{l}^{\ccc})}
\newcommand{\existslc}{(\exists_{l}^{\ccc})}
\newcommand{\existsrc}{(\exists_{r}^{\ccc})}

\newcommand{\idfonca}{(id^{\norc})}
\newcommand{\existsrnca}{(\exists_{r1}^{\norc})}
\newcommand{\existsrncia}{(\exists_{r2}^{\norc})}
\newcommand{\alllnca}{(\forall_{l1}^{\norc})}
\newcommand{\alllncia}{(\forall_{l2}^{\norc})}

\newcommand{\idfona}{(id^{\nnn})}
\newcommand{\existsrna}{(\exists_{r1}^{\nnn})}
\newcommand{\existsrnia}{(\exists_{r2}^{\nnn})}
\newcommand{\alllna}{(\forall_{l1}^{\nnn})}
\newcommand{\alllnia}{(\forall_{l2}^{\nnn})}
    
\newcommand{\idfoca}{(id^{\ccc})}
\newcommand{\existsrca}{(\exists_{r1}^{\ccc})}
\newcommand{\existsrcia}{(\exists_{r2}^{\ccc})}
\newcommand{\alllca}{(\forall_{l1}^{\ccc})}
\newcommand{\alllcia}{(\forall_{l2}^{\ccc})}

\newcommand{\negr}{(\neg_{r})}
\newcommand{\negl}{(\neg_{l})}

\renewcommand{\int}{\mathsf{Int}}
\newcommand{\intfond}{\mathsf{IntQ}}
\newcommand{\intfocd}{\mathsf{IntQC}}

\newcommand{\langintfo}{\mathcal{L}_{\mathsf{Int}}}

\newcommand{\ndcond}{\textbf{(ND)} } 
\newcommand{\cdcond}{\textbf{(CD)} } 
\newcommand{\moncond}{\textbf{(M)} } 
\newcommand{\ndcondns}{\textbf{(ND)}} 
\newcommand{\cdcondns}{\textbf{(CD)}} 
\newcommand{\moncondns}{\textbf{(M)}} 

\newcommand{\strucsetint}{\mathrm{Str}({\int})}

\newcommand{\overalldom}{\mathbf{D}}

\newcommand{\rulesna}{\mathrm{R}(\nnn)}
\newcommand{\rulesca}{\mathrm{R}(\ccc)}

\newcommand{\lnint}{\mathfrak{N}}
\newcommand{\nlint}{\mathfrak{L}}
\newcommand{\domcl}{\mathtt{DomCL}}

\newcommand{\thuesysi}{\thuesys4}
\newcommand{\thuesysii}{\thuesys5}
\newcommand{\thuesysilang}[1]{L_{\thuesysi}(#1)}
\newcommand{\thuesysiilang}[1]{L_{\thuesysii}(#1)}

\newcommand{\norc}{\mathsf{X}} 
\newcommand{\nnn}{\mathsf{Q}}
\newcommand{\ccc}{\mathsf{QC}}

\newcommand{\dsn}{\mathsf{DS}_{n}^{k}}
\newcommand{\dsnnz}{\mathsf{DS}_{0}^{k}}

\newcommand{\langdsn}{\mathcal{L}_{\dsn}}
\newcommand{\langdsnz}{\mathcal{L}_{\dsnnz}}

\newcommand{\donecond}{\textbf{(D\textsubscript{1})} } 
\newcommand{\dtwocond}{\textbf{(D\textsubscript{2})} } 
\newcommand{\dthreecond}{\textbf{(D\textsubscript{3})} } 
\newcommand{\partcond}{\textbf{(S\textsubscript{1})} } 
\newcommand{\ioacond}{\textbf{(S\textsubscript{2})} } 
\newcommand{\choicecond}{\textbf{(S\textsubscript{3})} } 

\newcommand{\donecondns}{\textbf{(D\textsubscript{1})}} 
\newcommand{\dtwocondns}{\textbf{(D\textsubscript{2})}} 
\newcommand{\dthreecondns}{\textbf{(D\textsubscript{3})}} 
\newcommand{\partcondns}{\textbf{(S\textsubscript{1})}} 
\newcommand{\ioacondns}{\textbf{(S\textsubscript{2})}} 
\newcommand{\choicecondns}{\textbf{(S\textsubscript{3})}} 

\newcommand{\agbox}{[i]}
\newcommand{\agdia}{\langle i \rangle}
\newcommand{\Oi}{\otimes_ {i}}
\newcommand{\ODi}{\ominus_{i}}
\newcommand{\agboxz}{[0]}
\newcommand{\agdiaz}{\langle 0 \rangle}
\newcommand{\Oiz}{\otimes_ {0}}
\newcommand{\ODiz}{\ominus_{0}}

\newcommand{\mcan}{M^{\premises}}
\newcommand{\wcan}{W^{\premises}}
\newcommand{\rcan}{R_{[i]}^{\premises}}
\newcommand{\ican}{\opt_{\Oi}^{\premises}}
\newcommand{\vcan}{V^{\premises}}

\newcommand{\gtdsn}{\mathsf{G3DS}_{n}^{k}}
\newcommand{\dsnl}{\mathsf{DS}_{n}^{k}\mathsf{L}}
\newcommand{\dsnlkz}{\mathsf{DS}_{n}^{0}\mathsf{L}}
\newcommand{\dsnlnz}{\mathsf{DS}_{0}^{k}\mathsf{L}}

\newcommand{\ioa}{(IOA)}
\newcommand{\ioar}{(IOA)}
\newcommand{\boxr}{(\Box)}
\newcommand{\diar}{(\Diamond)}
\newcommand{\agboxr}{(\agbox)}
\newcommand{\agdiar}{(\agdia)}
\newcommand{\agboxrz}{(\agboxz)}
\newcommand{\agdiarz}{(\agdiaz)}
\newcommand{\Oir}{(\Oi)}
\newcommand{\ODir}{(\ODi)}
\newcommand{\Oirz}{(\Oiz)}
\newcommand{\ODirz}{(\ODiz)}

\newcommand{\choicer}{(APC_{i}^{k})}
\newcommand{\refli}{(ref_{i})}
\newcommand{\transi}{(tra_{i})}
\newcommand{\symi}{(sym_{i})}
\newcommand{\eucli}{(euc_{i})}
\newcommand{\dtwoir}{(D_{2}^{i})}
\newcommand{\dthreeir}{(D_{3}^{i})}
    
\newcommand{\pragdiar}{(Pr_{\agdia})}
\newcommand{\prODirone}{(Pr_{\ODi}^{1})}
\newcommand{\prODirtwo}{(Pr_{\ODi}^{2})}
\newcommand{\pragdiarz}{(Pr_{\agdiaz})}
\newcommand{\prODironez}{(Pr_{\ODiz}^{1})}
\newcommand{\prODirtwoz}{(Pr_{\ODiz}^{2})}

\newcommand{\strucsetdsn}{\mathrm{Str}({\dsn})}
\newcommand{\prset}{\mathrm{Pr}({\dsn})}

\newcommand{\opt}{I}
\newcommand{\ideal}{I_{\Oi}}
\newcommand{\idealz}{I_{\Oiz}}
\newcommand{\uipath}{\sim_{i}}
\newcommand{\uipathrel}{\uipath^{\rel}}
\newcommand{\uipathz}{\sim_{0}}
\newcommand{\uipathrelz}{\uipathz^{\rel}}

\newcommand{\dsnsa}{\mathsf{DS}_{0}^{k}}

\newcommand{\dsnsal}{\mathsf{DS}_{0}^{k}\mathsf{L}}

\newcommand{\choicerz}{(APC_{0}^{k})}

\newcommand{\ct}{\mathrm{CT}}
\newcommand{\provedsk}{$\mathtt{ProveDS^{k}}$}
\newcommand{\provedsks}{$\mathtt{ProveDS^{k}}$ }
\newcommand{\sa}{0}
\newcommand{\countmod}{\mathbf{M}}
\newcommand{\countw}{\mathbf{W}}
\newcommand{\countrel}{\mathbf{R}_{\agboxz}}
\newcommand{\countideal}{\mathbf{I}_{\Oiz}}
\newcommand{\countval}{\mathbf{V}}

\newcommand{\kmsli}{\mathsf{Km}(\thuesys)\mathsf{LI}}

\newcommand{\iseq}[2]{#1 \ \| \ \{#2\}} 
\newcommand{\iiseq}[2]{#1 \ \| \ #2} 

\newcommand{\orth}[1]{(#1)^{\bot}} 
\newcommand{\sep}{ \ | \ } 
\newcommand{\sepgi}{ \ \| \ } 
\newcommand{\charaboxgi}{\charabox \gi^{w}_{u}}

\newcommand{\orthru}{(orth)} 
\newcommand{\idi}{(id_{1})}
\newcommand{\idii}{(id_{2})}

\newcommand{\lit}{\mathrm{Lit}}

\newcommand{\gi}{\mathcal{I}} 

\newcommand{\authorname}{Timothy Stephen Lyon} 
\newcommand{\thesistitle}{Refining Labelled Systems for Modal and Constructive Logics with Applications} 

\hypersetup{
    pdfpagelayout   = TwoPageRight,           
    linkbordercolor = {Melon},                
    pdfauthor       = {\authorname},          
    pdftitle        = {\thesistitle},         
    pdfsubject      = {Subject},              
    pdfkeywords     = {a, list, of, keywords} 
}

\setpnumwidth{2.5em}        
\setsecnumdepth{subsection} 

\nonzeroparskip             
\makeindex      
\makeglossaries 

\setauthor{}{\authorname}{}{male}
\setadvisor{Prof. Dr.}{Agata Ciabattoni}{}{female}

\setfirstassistant{Pretitle}{Forename Surname}{Posttitle}{male}
\setsecondassistant{Pretitle}{Forename Surname}{Posttitle}{male}
\setthirdassistant{Pretitle}{Forename Surname}{Posttitle}{male}

\setfirstreviewer{Dr.}{Alwen Tiu}{}{male}
\setsecondreviewer{Dr.}{Lutz Stra{\ss}burger}{}{male}

\setsecondadvisor{Dr.}{Revantha Ramanayake}{}{male} 

\setregnumber{11701257}
\setdate{30}{06}{2021} 
\settitle{\thesistitle}{Refining Labelled Systems for Modal and Constructive Logics with Applications} 

\setthesis{doctor}
\setdoctordegree{techn.}
\setfirstreviewerdata{Australian National University, Australia}
\setsecondreviewerdata{Laboratoire d'informatique de l'École polytechnique, France}

\begin{document}

\frontmatter 

\addtitlepage{naustrian} 
\addtitlepage{english} 
\addstatementpage

\begin{danksagung*}

Ich möchte meinen Eltern Lisa C. Lyon und Stephen W. Lyon, sowie meinen Großeltern Charlotte Chesnutt und Lee Chesnutt und dem Rest meiner Familie für ihre kontinuierliche Unterstützung und Beratung danken, ohne die diese Arbeit niemals möglich gewesen wäre. Des Weiteren bin ich Mr. William Ashby und Mrs. Vaughnene Ashby für ihre Großzügigkeit über all die Jahre sehr dankbar.

Meiner Betreuerin, Frau Prof. Agata Ciabattoni möchte ich von ganzem Herzen für ihre Beratung und Freundlichkeit danken, ohne die ich niemals so viel hätte erreichen können. Ich bin sehr dankbar für die Begleitung und das Wissen von Dr. Revantha Ramanayake. Ich danke Ihnen beiden, dass Sie sich die Zeit genommen haben, meine Arbeit zu überprüfen und mir wertvolles Feedback zu geben.

Ich bin dankbar für die sorgfältige Durchsicht meiner Arbeit und die hilfreichen Kommentare von Dr. Alwen Tiu und Dr. Lutz Stra{\ss}burger. Des Weiteren möchte ich allen Teilnehmern des TICAMORE-Projekts meinen aufrichtigen Dank aussprechen.

Danke an Dr. Anna Prianichnikova und die gesamte Fakultät der DK LogiCS für die Leitung eines Doktorandenprogramms, das für uns Studenten von großem Nutzen ist. Ein besonderer Dank geht auch an Doris Hotz, die mir bei bürokratischen Angelegenheiten immer zur Seite stand.

Außerdem hatte ich das große Vergnügen, mit Prof. Nick Galatos an der University of Denver sowie mit Dr. Alwen Tiu und Prof. Rajeev Gor\'e an der Australian National University zusammenzuarbeiten. Ich danke Ihnen für Ihre Gastfreundschaft und dafür, dass Sie Ihr Fachwissen mit mir geteilt haben.

\end{danksagung*}

\begin{acknowledgements*}

I would like to thank my parents Lisa C. Lyon and Stephen W. Lyon, my grandparents Charlotte Chesnutt and Lee Chesnutt, and the rest of my family, for their continuous support and guidance, without whom this thesis would never be possible. Also, I will forever be indebted to Mr. William Ashby and Mrs. Vaughnene Ashby for their generosity throughout the years.

I would also like to whole-heatedly thank my supervisor Prof. Agata Ciabattoni for her mentorship and kindness, without whom I could have never achieved so much. I am also very thankful for the guidance and knowledge of Dr. Revantha Ramanayake. Thank you both for taking the time to review my thesis and provide me with valuable feedback. 

I am grateful for the careful reviews of my thesis and useful comments given by Dr. Alwen Tiu and Dr. Lutz Stra{\ss}burger. Furthermore, I would like to express my sincere thanks to all of the participants of the TICAMORE project.

Thank you to Dr. Anna Prianichnikova and all of the faculty of DK LogiCS for managing a PhD program that greatly benefits the lives of us students. Also, special thanks to Doris Hotz for always assisting me with bureaucratic matters.

I also had the great pleasure of working with Prof. Nick Galatos at the University of Denver, and Dr. Alwen Tiu and Prof. Rajeev Gor\'e at the Australian National University. Thank you for your hospitality and for sharing your expertise with me.

\end{acknowledgements*}

\begin{kurzfassung}

In dieser Arbeit wird die \emph{Methode der strukturellen Verfeinerung} (im Folgenden \emph{Verfeinerung} genannt) vorgestellt, die dazu dient, die relationale (Kripke-)Semantik einer modalen und/oder konstruktiven Logik in ein ``sparsames'' Beweissystem zu transformieren, indem zwei beweistheoretische Paradigmen verbunden werden: gelabelte Sequenzen und verschachtelte Sequenzkalküle. Der Formalismus der gelabelten Sequenzen hat sich insofern bewährt, als dass schnittfreie Kalküle im Besitz wünschenswerter beweistheoretischer Eigenschaften (z. B. Zulässigkeit von Strukturregeln, Invertierbarkeit von Regeln, etc.) für große Klassen von Logiken automatisch generiert werden können. Trotz dieser Eigenschaften verwenden gelabelte Systeme eine komplizierte Syntax, die die Semantik der zugehörigen Logik explizit einbezieht, und solche Systeme verletzen normalerweise die Subformel-Eigenschaft in hohem Maße. Im Gegensatz dazu verwenden verschachtelte Sequenzkalküle eine einfachere Syntax und halten sich an eine strenge Lesart der Subformel-Eigenschaft, welche solche Systeme für den Entwurf von automatischen Schlussfolgerungsalgorithmen sinnvoll macht. Der Nachteil des Paradigmas der verschachtelten Sequenzen ist jedoch, dass eine allgemeine Theorie zur automatischen Konstruktion solcher Kalküle (wie im gelabelten Formalismus) im Wesentlichen fehlt, was bedeutet, dass die Konstruktion von verschachtelten Systemen und die Bestätigung ihrer Eigenschaften in der Regel auf einer Fall-zu-Fall-Basis erfolgt. Die Verfeinerungsmethode verbindet beide Paradigmen erfolgreich, indem sie gelabelte Systeme in geschachtelte (oder verfeinerte gelabelte) Systeme transformiert, wobei die Eigenschaften der Ersteren während des Transformationsprozesses erhalten bleiben. Die Eigenschaften von verschachtelten und verfeinerten gelabelten Systeme erleichtern die Arbeit mit ihnen, können zu einer Platzersparnis führen und eine gesteigerte Effizienz beim Automatisieren und Lösen von Argumentationsaufgaben bewirken (z. B. Beweissuche, effektive Interpolation, etc.).

Um die Methode der Verfeinerung und einige ihrer Anwendungen zu demonstrieren, betrachten wir eine vielfältige Gruppe von modalen und konstruktiven Logiken: kontextfreie Grammatiklogiken mit Konversen, intuitionistische Logiken erster Ordnung und deontische \stit-Logiken. Die vorgestellten verfeinerten gelabelten Kalküle werden verwendet, um die ersten Algorithmen zur Beweissuche und zur automatischen Extraktion von Gegenmodellen für deontische \stit-Logiken bereitzustellen und damit Entscheidungsprozeduren für die Logiken zu erhalten. Darüber hinaus verwenden wir unsere verfeinerten gelabelten Kalküle für kontextfreie Grammatiklogiken mit Konversen, um zu zeigen, dass jede Logik in der Klasse die effektive Lyndon-Interpolationseigenschaft besitzt. Um dieses Ergebnis zu realisieren, verwenden wir ein syntaktisches, beweistheoretisches Verfahren der Lyndon-Interpolation.

\end{kurzfassung}

\begin{abstract}

This thesis introduces the \emph{method of structural refinement} (that will be referred to more simply as \emph{refinement}), which serves as a means of transforming the relational (Kripke) semantics of a modal and/or constructive logic into an `economical' proof system by connecting two proof-theoretic paradigms: labelled sequent and nested sequent calculi. The formalism of labelled sequents has been successful in that cut-free calculi in possession of desirable proof-theoretic properties (e.g. admissibility of structural rules, invertibility of rules, etc.) can be automatically generated for large classes of logics. Despite these qualities, labelled systems make use of a complicated syntax that explicitly incorporates the semantics of the associated logic, and such systems typically violate the subformula property to a high degree. By contrast, nested sequent calculi employ a simpler syntax and adhere to a strict reading of the subformula property, making such systems useful in the design of automated reasoning algorithms. However, the downside of the nested sequent paradigm is that a general theory concerning the automated construction of such calculi (as in the labelled setting) is essentially absent, meaning that the construction of nested systems and the confirmation of their properties is usually done on a case-by-case basis. The refinement method connects both paradigms in a fruitful way, by transforming labelled systems into nested (or, refined labelled) systems with the properties of the former preserved throughout the transformation process. The qualities of nested and refined labelled systems makes them easier to work with, can lead to a savings in space, and can bring about an increased efficiency in automating and solving reasoning tasks (e.g. proof-search, effective interpolation, etc.).

To demonstrate the method of refinement and some of its applications, we consider a varied group of modal and constructive logics: context-free grammar logics with converse, first-order intuitionistic logics, and deontic \stit logics. The introduced refined labelled calculi will be used to provide the first proof-search and automated counter-model extraction algorithms for deontic \stit logics, thus yielding decision procedures for the logics. Furthermore, we employ our refined labelled calculi for context-free grammar logics with converse to show that every logic in the class possesses the effective Lyndon interpolation property. In order to carry out this result, we make use of a syntactic, proof-theoretic method of Lyndon interpolation.

\end{abstract}

\selectlanguage{english}

\tableofcontents 

\newpage
\section{List of the Logics and Proof Systems Discussed}\label{app:logics-proof-systems}

\begin{tabular}{|c|c|c|}
\hline
\textbf{Logic} & \textbf{Description} & \textbf{Page} \\
\hline
$\km$ & Base grammar logic & p.~\pageref{def:axiomatization-km} \\
\hline
$\kms$ & The grammar logic of the \cfcst system $\thuesys$ & p.~\pageref{def:axiomatization-km} \\
\hline
$\intfond$ & First-order intuitionistic logic with non-constant domains & p.~\pageref{def:axiomatization-IntFO} \\
\hline
$\intfocd$ & First-order intuitionistic logic with constant domains & p.~\pageref{def:axiomatization-IntFO} \\
\hline
$\dsn$ & Deontic \stit logic for $n$ agents with a maximum of $k$ choices & p.~\pageref{def:axiomatization-dsn} \\
\hline
\end{tabular}

\begin{tabular}{|c|c|c|}
\hline
\textbf{Calculus} & \textbf{Description} & \textbf{Page} \\
\hline
$\gtkms$ & Labelled calculus for $\kms$ & p.~\pageref{fig:G3Km(S)} \\
\hline
$\gtintfond$ & Labelled calculus for $\intfond$ & p.~\pageref{fig:labelled-calculi-FO-Int} \\
\hline
$\gtintfocd$ & Labelled calculus for $\intfocd$ & p.~\pageref{fig:labelled-calculi-FO-Int} \\
\hline
$\gtdsn$ & Labelled calculus for $\dsn$ & p.~\pageref{fig:base-Gcalculus} \\
\hline
$\kmsl$ & Refined labelled calculus for $\kms$ & p.~\pageref{fig:Refined-Calculus-i-kms} \\
\hline
$\intfondl$ & Refined labelled calculus for $\intfond$ & p.~\pageref{fig:refined-calculi-FO-Int} \\
\hline
$\intfocdl$ & Refined labelled calculus for $\intfocd$ & p.~\pageref{fig:refined-calculi-FO-Int} \\
\hline
$\dsnl$ & Refined labelled calculus for $\dsn$ & p.~\pageref{fig:refined-calculus-dsn} \\
\hline
$\dkms$ & Nested calculus for $\kms$ & p.~\pageref{fig:DKm(S)} \\
\hline
$\nintfond$ & Nested calculus for $\intfond$ & p.~\pageref{fig:nested-calculi-FOInt} \\
\hline
$\nintfocd$ & Nested calculus for $\intfocd$ & p.~\pageref{fig:nested-calculi-FOInt} \\
\hline
\end{tabular}

\mainmatter

\chapter{Introduction}\label{CPTR:Intro}

Between 1932 and 1935, Gerhard Gentzen introduced the \emph{natural deduction} and \emph{sequent calculus} frameworks for classical and intuitionistic logic, which broke with the proof-theoretic paradigm of the time~\cite{Gen35a,Gen35b,Pla18}. Up until that point, proof systems---based on the work of Frege, Peano, and Russell~\cite{Fre79,Pea89,Rus06}---consisted primarily of \emph{axioms} and few inference rules (e.g. modus ponens and universal generalization). By contrast, Gentzen's formalisms consisted primarily of \emph{rules}, and explicitly defined proofs as trees with assumptions, or trivial logical truths, as the leaves, and inference rules as the edges synthesizing the information via repeated applications into a concluding formula (i.e. the root of the tree)---taken to be the theorem derived~\cite{NegPla11}. 

Both of Gentzen's formalisms possessed a significant advantage over the proof systems of the antecedent paradigm, namely, proofs constructed within Gentzen's natural deduction and sequent calculus frameworks 
 were found to enjoy the so-called \emph{subformula property}~\cite{Gen35a,Gen35b,Pra65,Rag65} (for a historical discussion, see~\cite{Pla18}). The subformula property states that any theorem derivable within the deductive system is derivable with a proof consisting solely of subformulae of the derived theorem, that is to say, every formula used to reach the conclusion, occurs as a subformula of the conclusion. Typically, proofs and proof systems that possess this property (or a variation thereof) are qualified as \emph{analytic}. 
 This property is not only useful in establishing the consistency of a system's associated logic, but can be harnessed for other applications (e.g. decidability). 


The subformula property, however, does not come for free, as the natural deduction systems contain elimination rules 
 and the sequent calculi contain the \emph{cut} rule, which delete formulae when going from the premises to the conclusion; thus, both frameworks \emph{prima facie} break the subformula property. To overcome this obstacle in the sequent calculus framework, Gentzen proved the celebrated \emph{Hauptsatz} (now called the \emph{cut elimination theorem}) showing that any formula derivable in the sequent calculus (for classical and intuitionistic logic) is derivable \emph{without the cut rule}~\cite{Gen35a,Gen35b}. The subformula property of the calculi followed as a corollary. Later work by Prawitz~\cite{Pra65} and Raggio~\cite{Rag65} showed that natural deduction proofs could be \emph{normalized}, i.e. put into a form where no formula occurs as the major premise of an elimination rule and as the conclusion of an introduction rule.

Due to its utility and simplicity, Gentzen's sequent formalism has continued to be implemented, leading to the creation (or, discovery) of analytic calculi for a wide variety of logics (e.g.~\cite{Cor89,Cur52,Gir87,Rau80,Ten87}), and leading even to the discovery of new logics (e.g.~\cite{Gir87}). Such calculi have found a range of applications: from establishing logical properties (e.g. consistency, decidability, interpolation), to automating reasoning with the associated logic (e.g.~\cite{Dyc92,Sla97}). In spite of these advantages, Gentzen's sequent formalism does have its drawbacks; namely, the formalism appears to be \emph{too simple} to provide proof systems for the vast classes of non-classical logics studied by contemporary logicians. For example, despite effort to the contrary, a perspicuous and modular framework---in the style of Gentzen---that uniformly covers normal modal and tense logics (e.g.~$\mathsf{Kt}$, $\mathsf{KB}$, and $\mathsf{S5}$) has proven to be elusive (see~\cite{Wan02}). 
The inability of the Gentzen sequent formalism to uniformly provide analytic calculi for large classes of non-classical logics is a serious limitation, as the study and application of such logics has grown substantially over the last few decades (see~\cite{Avr96}), implying the need for a suitable proof theory.

In this thesis, we will focus on refining labelled systems within the context of a particular group of non-classical logics, that is, \emph{modal and constructive logics}. We will consider grammar logics~\cite{CerPen88,DemNiv05}, first-order intuitionistic logics~\cite{Kle52,Grz64}, and deontic \stit logics~\cite{BerLyo19b,BerLyo21,Hor01,Mur04}; we now briefly describe these logics and their utility. Grammar logics are multi-modal logics that were introduced in~\cite{CerPen88}, and were used---in that paper---to establish an equivalence between the problem of checking if a grammar generates a word and the problem of checking if a formula is a logical theorem. The strong relationship between grammar logics and formal language theory has allowed for (un)decidability results to be transferred between the two settings~\cite{Bal00,CerPen88,DemNiv05}. Moreover, the class of grammar logics is interesting in that it contains a large assortment of modal logics which can be viewed as epistemic logics~\cite{Bal00,Hal92}, temporal logics~\cite{CerHer95}, and even description logics (used in knowledge representation) with inverse roles and complex role inclusions axioms~\cite{HorSat04,TiuIanGor12}. First-order (and propositional) intuitionistic logics have been employed in the study of constructive reasoning and mathematics~\cite{TroDal88}, and are highly relevant in computer science; e.g. the Curry-Howard correspondence~\cite{How80}, and answer-set programming~\cite{OsoNavArr04}. A set of logics that will be of particular interest in this work is the logic of STIT (an acronym for `Seeing To It That') introduced by Belnap and Perloff in~\cite{BelPer90} to study and clarify agentive sentences. Since then, their formalism has been extended and applied to model agent-based choice making and interaction in the domain of epistemic reasoning~\cite{Bro11}, deontic reasoning~\cite{BerLyo19b,BerLyo21,Hor01,Mur04}, the formal analysis of legal reasoning~\cite{Bro11,LorSar15}, and has been used to model and verify autonomous systems~\cite{SheAbb20}. Thus, the logics we consider are not only diverse in their nature, but in their applications.

In search of a suitable and uniform proof-theoretic framework for modal and constructive logics, a 
 diverse number of formalisms have been assembled and proposed---examples include (prefixed) tableaux~\cite{Fit72,Fit14}, display calculi~\cite{Bel82,Wan94}, hypersequents~\cite{Pot83,Avr96}, labelled sequents~\cite{Gab96,Neg05,Vig00}, 2-sequents~\cite{Mas92,Mas93}, nested sequents~\cite{Bul92,Bru09,Kas94,Pog09}, and linear nested sequents~\cite{Lel15,LelPim15}. Each formalism extends or reworks the structure of the Gentzen-style sequent in a distinct way, incorporating additional bureaucracy that allows for proof-calculi to be constructed for certain classes of logics within the formalism, and which tend to uniformly possess fundamental proof-theoretic properties such as cut-admissibility, invertibility of rules, etc. (see, e.g.~\cite{GorPosTiu11,Neg05}). (NB. We will discuss the proof-theoretic formalisms relevant for, and used within, this thesis in more detail below.) As the number and diversity of proof-theoretic formalisms and systems increased, logicians began to ponder which criteria should be used to distinguish `nice' proof systems from less desirable ones, leading to some notable proposals~\cite{Avr96,Dos88,Wan94}. 

One significant proposal, which we will consider, comes from Wansing~\cite{Wan94}, who argues on philosophical and functional grounds, that proof systems ought to satisfy the following desiderata:

\begin{itemize}

\item[$\wancondi$] \emph{Separation}: Each logical rule exhibits no other logical connectives than the one to be introduced.

\item[$\wancondii$] \emph{(Weak) Symmetry}: 
 Each logical rule should be a left or right introduction rule, and symmetry stipulates that that the calculus is weakly symmetric and each logical connective has a left and right introduction rule.

\item[$\wancondiii$] \emph{(Weak) Explicitness}: 
 Each introduced logical connective appears only in the conclusion of its corresponding logical rule, and explicitness holds if the calculus is weakly explicit and there is only one occurrence of the logical connective in the conclusion of each corresponding logical rule.

\item[$\wancondiv$] \emph{Unique Characterization}: Each logical connective should be uniquely characterized by its corresponding logical rules.

\item[$\wancondv$] \emph{Do\v{s}en's Principle}: ``[T]he rules for the logical operations are never changed: all changes are made in the structural rules''~\cite[p.~352]{Dos88}.

\item[$\wancondvi$] \emph{Subformula property and cut-freedom}: The cut rule should be admissible and the system should possess the subformula property.

\end{itemize}

Desiderata $\wancondi$--$\wancondiii$ fix the meaning of a logical connective independent of the other logical connectives present in the language, and $\wancondiv$ ensures that two logical connectives collapse to the same connective if their logical rules are essentially identical (with the only distinguishing feature being that each set of logical rules introduces their own version of the connective). In addition, Wansing remarks (\cite[p.~129]{Wan94}) that if $\wancondi$--$\wancondiii$ hold of a system and the cut rule is \emph{admissible} (meaning that anything provable with cut, is provable without cut), then the subformula property holds for the system as a consequence. Do\v{s}en's principle $\wancondv$ stipulates that the logical rules should fix a base system for a base logic (e.g. the modal logic $\mathsf{K}$) with structural rule extensions yielding new systems for extensions of the base logic (e.g. the modal logics $\mathsf{S4}$ and $\mathsf{S5}$). If a formalism satisfies such a principle, then a high degree of modularity is obtained, allowing for a uniform proof-theoretic presentation for the corresponding set of logics. The last desideratum $\wancondvi$ is desirable as the subformula property is useful in automated reasoning, and cut-freedom/elimination has relevance to the Curry-Howard correspondence~\cite{How80}, in proving completeness, and in proving interpolation~\cite{Mae60}, among other things.

Wansing's desiderata are mentioned not only due to their historical relevance in characterizing desirable proof systems, but due to their relevance in judging the systems obtained via the \emph{method of structural refinement} (the central topic of this thesis), which we refer to more simply as \emph{refinement} and which begets calculi that satisfy Wansing's proof-theoretic design principles to a high degree (though, we will argue below that Do\v{s}en's principle is too strong and ought to be weakened). 
 Refinement is intimately connected with the formalism of labelled sequents, and is a strategy for transforming such systems into systems of a `simpler' formalism. Before making this strategy precise, we discuss the two main proof-theoretic formalisms that refinement concerns, namely, the labelled sequent and nested sequent formalisms.

The formalism of labelled sequents is rooted in the work of Kanger~\cite{Kan57}, who introduced sequent calculi for the modal logics $\mathsf{T}$, $\mathsf{S4}$, and $\mathsf{S5}$ that made use of \emph{spotted formulae} (i.e. formulae annotated with natural numbers). This method of annotating, or prefixing, formulae with \emph{labels} became the characteristic feature of the labelled paradigm, and is commonly used to incorporate semantic information into the syntax of the associated proof systems. There are numerous examples of calculi that fall within the labelled formalism, such as tableaux~\cite{BalGioMar98,BecGor97,Gab96,Ner91}, natural deduction systems~\cite{BasMatVig97,Sim94,Vig00}, and sequent-style systems~\cite{DycNeg12,Min97,Neg05,NegPla11,Sim94,Vig00}. We will be chiefly concerned with the latter---\emph{labelled sequent systems}.

The labelled sequent formalism offers many advantages: first, for sizable classes of modal and constructive logics, it has been shown that the relational semantics of each logic can be straightforwardly transformed into an associated labelled sequent calculus~\cite{DycNeg12,Neg05,NegPla11,Sim94,Vig00}; in fact, it was shown that the process of constructing labelled sequent calculi can be automated~\cite{CiaMafSpe13}. Second, calculi built within the labelled paradigm tend to be exceptionally modular, meaning that such systems allow for the addition or deletion of rules to obtain calculi for logics with desired properties. Third, general results have been given, which show that for a wide array of modal and constructive logics, their associated labelled calculi uniformly possess proof-theoretic properties such as height-preserving admissibility of structural rules, height-preserving invertibility of rules, and syntactic cut-elimination~\cite{DycNeg12,Neg05,NegPla11,Vig00}. Although such characteristics are certainly desirable, the labelled formalism does have its shortcomings. For example, labelled calculi commonly involve a complicated syntax (due to the explicit incorporation of semantic information), labelled sequents encode general graphs (which contrasts with other formalisms that employ simpler data structures such as linear graphs~\cite{Lel15}, or trees~\cite{Fit14,GorRam12}), and labelled systems usually contain larger sets of rules compared to other calculi (within different formalisms) for the same logics. The more complex data structures encoded in labelled sequents and the larger number of inference rules (which increases the size of proofs) can cause proof-search algorithms based on such systems to consume more spatial and temporal resources than is optimally required. Not only this, their complex nature can make labelled proof systems unwieldy both in use and in applications. Last, labelled calculi typically contain inference rules that delete formulae from premise to conclusion, thus violating the subformula property to a high degree---in contradiction with Wansing's desideratum $\wancondvi$.

While labelled calculi customarily employ sequents that encode a general graph structure, nested sequents encode a \emph{tree} of formulae, and therefore, employ a simpler data structure. The origin of the nested sequent formalism is often attributed to Bull~\cite{Bul92} and Kashima~\cite{Kas94}, though it should be noted that the prefixed tableaux of Fitting~\cite{Fit72} can be viewed as `upside-down' versions of nested sequent calculi. (NB. For a discussion on the relationship between prefixed tableaux and nested sequent calculi, see~\cite{Fit14}). In recent times, the use of nested sequents has become more widespread, with nested calculi constructed for normal modal logics~\cite{Bru09,Pog09}, tense logics~\cite{GorPosTiu11}, grammar logics~\cite{TiuIanGor12}, intuitionistic modal logics~\cite{Str13}, and (first-order) intuitionistic logics~\cite{Fit14}. Moreover, such calculi have been used in applications such as providing proof-search and counter-model construction algorithms for logics (e.g.~\cite{GorPosTiu11,TiuIanGor12}), and confirming interpolation as well as automating the extraction of interpolants (e.g.~\cite{FitKuz15,LyoTiuGorClo20}). A strength of such calculi is that they employ a relatively simple data structure, minimizing the bureaucracy occurring in proofs and making the calculi better suited for applications (relative to labelled calculi). Despite these advantages, a major drawback of the nested formalism is that the  construction of such calculi, and the confirmation of their proof-theoretic properties (e.g. cut-admissibility), is often done on a case-by-case basis. That is to say, the nested paradigm does not currently possess the same generality of results as those enjoyed in the labelled setting concerning the automatic construction of calculi in possession of fundamental properties.



Since the labelled formalism is well-suited for constructing calculi in possession of favorable proof-theoretic properties, and the nested formalism is simpler and better-suited for applications, a method of transforming calculi of the former formalism into the latter formalism---with proof-theoretic properties sufficiently preserved---is highly desirable. By connecting the two formalisms, we effectively obtain the best of both worlds: we may invoke the general results of the labelled paradigm to construct large classes of calculi for modal and constructive logics, and then transform these calculi into more \emph{refined} versions of themselves (e.g. nested calculi) that are suitable for applications. We will see that such calculi utilize simpler structures, leading to a compression in proof size and allowing for uniform presentations of terminating proof-search. Similar relationships between labelled and `more refined' systems have been discussed in the literature~\cite{GorRam12,Lyo20a,Lyo21,LyoBer19,Pim18}, where nested, tree-hypersequent, and `forestlike' labelled calculi were derived from proper labelled calculi for modal and constructive logics.


The \textbf{first} component of the \emph{method of refinement} consists of extracting labelled sequent calculi from the semantics of a class of logics for which one is interested. Such methods of extraction have been known for some time, with large classes of labelled sequent calculi being produced for large classes of modal and constructive logics~\cite{BerLyo19a,CasSma02,DycNeg12,KusOka03,Neg05,Sim94,Vig00}. If the labelled calculi are of a certain form (as will be discussed in \cptr~\ref{CPTR:Refinment-Modal} and~\ref{CPTR:Refinment-Constructive}), then through the elimination of structural rules (encoding properties such as reflexivity, transitivity, etc.), we obtain classes of refined labelled calculi complete relative to their respective classes of logics. (NB. It will be seen that refined labelled calculi are ordinarily notational variants of nested sequent systems, though not necessarily so, as discussed in \sect~\ref{SECT:Refine-STIT}.) 

The \textbf{second} step in the refinement procedure---\emph{structural rule elimination}---proceeds by considering how logical rules in the labelled calculi ought to be strengthened to ensure the elimination of structural rules. The strengthened logical rules required to complete the second step are frequently found to be \emph{propagation rules}. Such rules have been employed in prefixed tableaux~\cite{CasCerGasHer97,Fit72}, nested~\cite{TiuIanGor12}, and labelled sequent systems~\cite{CiaLyoRamTiu20,LyoBer19} in the literature, and acquire their name on the basis of their functionality---when applied bottom-up, the rules propagate formulae to the end of `paths' occurring within sequents or tableaux. What is interesting however, is that the investigations offered in this thesis will also identify a new class of rules---\emph{reachability rules}. These rules generalize the behavior of propagation rules by not only propagating formulae along paths, but additionally checking if data occurs along (potentially alternative) paths within a sequent. Reachability rules will be discussed in \cptr~\ref{CPTR:Refinment-Constructive} on refining labelled calculi for first-order intuitionistic logics. 

It should be noted that in the first-order setting structural rule elimination is composed with a further step in the refinement process---\emph{domain atom removal}. Although structural rule elimination does yield `simpler' proof calculi in the first-order setting, the syntax of sequents can be simplified further by showing the superfluity of certain syntactic structures that encode information about domains (hence, this step is unnecessary in the propositional setting where quantification over domains is absent). 

The \textbf{last} component of refinement regards the establishment of proof-theoretic properties for the refined labelled calculi. This step can be carried out by either showing that the refined labelled calculi inherit the properties of their `parent' labelled calculi, or by directly showing that each refined calculus possesses desirable properties without making reference to the original labelled systems that begat them. The former approach proceeds by reversing the structural rule elimination process to show that not only can the derivations in the parental labelled systems be transformed into derivations in the refined labelled systems, but the reverse transformation is possible as well. We will make use of both approaches in this thesis. 

Furthermore, the calculi obtained via the method of refinement satisfy Wansing's desiderata completely (or to a high degree) with the only exception being Do\v{s}en's principle $\wancondv$. Do\v{s}en's principle implicitly equates modularity and uniform coverage of logics with the addition and subtraction of structural rules. Nevertheless, our refined labelled calculi will be highly modular and provide uniform coverage over large classes of logics by making use of \emph{grammar theoretic machinery} within propagation and reachability rules, as opposed to using structural rules. This idea is motivated by and based upon the work in~\cite{CiaLyoRamTiu20,GorPosTiu11,LyoBer19,TiuIanGor12}. As explained above, propagation and reachability rules operate by considering `paths' within sequents, and then propagate formulae accordingly. Since we will encode these paths as strings generated by formal grammars, we can change the operation of our propagation and reachability rules by simply changing the formal grammar considered by the rule. This offers an alternative approach to the type of modularity expressed in Do\v{s}en's principle, which has become widespread (for good reason) in the construction of proof systems. Still, the benefits of this alternative approach to modularity and uniform coverage are many: first, propagation and reachability rules are \emph{formula driven}, meaning that bottom-up applications of rules only rely on logical formulae occurring within sequents (as opposed to other structures), typically making proof-search procedures easier to write. Second, systems built with grammar theoretic machinery tend to have fewer rules since certain structural rules are rendered superfluous---the omission of such structural rules brings about a compression in proof size and minimization of sequential structure. Third, exploiting formal grammars provides information about the complexity and (un)decidability of the associated logics by recognizing which formal grammars are required in a logic's propagation and reachability rules. Therefore, although the refined calculi we will obtain are in violation of Do\v{s}en's principle, we argue that our calculi capture the essence of the principle, namely, modularity and uniform coverage of logics. 



Beyond the construction of `nice' proof systems, refining labelled systems also brings about an intriguing observation: many of the refined labelled calculi happen to be notational variants of existing nested systems. For example, in~\cite{CiaLyoRamTiu20} it was found that refining labelled sequent systems for tense logics produced the nested sequent systems of~\cite{GorPosTiu11}; in~\cite{Lyo20a,Lyo21} refining labelled systems for propositional and first-order intuitionistic logics yielded the nested systems of Fitting~\cite{Fit14}; and, as will be shown in \sect~\ref{SECT:Refine-Grammar}, refining labelled calculi for grammar logics gives the nested calculi of~\cite{TiuIanGor12}. 
 These observations suggest a naturalness to the method of refinement, which can be seen as an underlying procedure unifying various nested sequent systems within a single theoretical framework. 





Although the thesis will primarily focus on the method of refinement, we will also put our refined systems to work, showing that the class of grammar logics considered admit effective Lyndon interpolation, and providing proof-search and counter-model extraction algorithms for a class of deontic \stit logics. The interpolation method is an additional, novel feature of this thesis and is based on the author's joint work in~\cite{LyoTiuGorClo20}. The interpolation method provides a purely syntactic and uniform approach to proving that logics possess the Lyndon (and Craig) interpolation property by harnessing their nested sequent systems, i.e. such systems can be used to show that for any valid implication $\phi \cimp \psi$, there exists a formula $\chi$ built with propositional atoms from $\phi$ and $\psi$ (in Craig interpolation) and propositional atoms from $\phi$ and $\psi$ with the same polarity (in Lyndon interpolation) such that $\phi \cimp \chi$ and $\chi \cimp \psi$ are valid; such properties have been used in verification~\cite{McM18}, to establish Beth definability~\cite{KihOno10}, and to conceal or forget information in ontology querying~\cite{LutWol11}. This method is both a generalization of Maehara's method~\cite{Mae60} and a variant of the successful semantic method of interpolation, initially provided by Fitting and Kuznets in~\cite{FitKuz15}, and expanded upon by the latter author in a sequence of papers~\cite{Kuz16,Kuz16b,Kuz18,KuzLel18}. 



The main contributions of this thesis are as follows: first, the refinement method is put forth, which transforms relational semantics into a nested or refined labelled proof system for a diverse class of logics. To illustrate this method, labelled calculi are constructed for grammar logics, first-order intuitionistic logics, and deontic \stit logics. The thesis dedicates many pages to explaining \emph{how} propagation and reachability rules are discovered through the process of structural rule elimination, giving the thesis added explanatory value. Furthermore, comparisons are provided for labelled and refined labelled calculi, showing and justifying why the latter require less structure in their sequents compared to the former. Also, although the nested calculi for grammar logics (obtained via refinement) already exist in Tiu \textit{et al.}'s paper~\cite{TiuIanGor12}, the nested and refined labelled calculi for first-order intuitionistic logics and deontic \stit logics are new. Even though the nested calculi for first-order intuitionistic logics resemble Fitting's nested calculi to a degree (cf.~\cite{Fit14}), it will be argued that the use of propagation and reachability rules permit the calculi to be easily transformed into calculi for alternative logics. Last, the first proof-search and counter-model extraction algorithms for (single-agent) deontic \stit logics are provided, along with an application of a syntactic method of interpolation to uniformly prove that context-free grammar logics with converse have the effective Lyndon interpolation property, thus generalizing the results of~\cite{GorNgu05}, which proved effective Craig interpolation for regular grammar logics.




\section{Outline of Dissertation}

The dissertation is structured as follows:

In \cptr~\ref{CPTR:Logics}, we introduce the semantics and axiomatizations of context-free grammar logics with converse, first-order intuitionistic logics, and deontic \stit logics. \cptr~\ref{CPTR:Labelled} presents labelled calculi for the three classes of logics and argues that the calculi possess proof-theoretic properties such as hp-admissibility of structural rules, hp-invertibility of rules, and syntactic cut-elimination. In \cptr~\ref{CPTR:Refinment-Modal}, the method of refinement is introduced and applied to the labelled calculi for grammar logics and deontic \stit logics. The first section of \cptr~\ref{CPTR:Refinment-Modal} will also show that all refined labelled calculi for grammar logics are notational variants of (slight reformulations of) the existing nested systems provided in~\cite{TiuIanGor12}, and will touch on the relationship between such systems and display calculi. The second section of \cptr~\ref{CPTR:Refinment-Modal} will show how to refine the labelled calculi for deontic \stit logics. In \cptr~\ref{CPTR:Refinment-Constructive}, we will cover refinement in the first-order setting and will show that the refined labelled calculi obtained are labelled versions of nested systems. \cptr~\ref{CPTR:Applications} discusses applications of refined labelled calculi, giving proof-search and counter-model extraction procedures for deontic \stit logics in the first section, followed by a uniform proof of effective Lyndon interpolation for grammar logics in the second section; also, we will briefly compare the syntactic method of interpolation from~\cite{LyoTiuGorClo20} and the semantic method from~\cite{FitKuz15}. The last chapter (\cptr~\ref{CPTR:Conclusion}) concludes and discusses future work. Also, we note that all logics and proof systems discussed within the thesis are presented in tables prior to this introduction (p.~\pageref{app:logics-proof-systems}) along with short descriptions of each and the number of the page where the logic or proof system is introduced.

\section{Publications}

This dissertation is based on work from the following papers:

\begin{enumerate}

\item Ciabattoni, A., Lyon, T., \& Ramanayake, R. (2018). From Display to Labelled Proofs for Tense Logics. In International Symposium on Logical Foundations of Computer Science (pp. 120-139). Springer, Cham.

\item Berkel, K., \& Lyon, T. (2019). Cut-Free Calculi and Relational Semantics for Temporal STIT Logics. In European Conference on Logics in Artificial Intelligence (pp. 803-819). Springer, Cham.

\item Lyon, T., \& Berkel, K. (2019). Automating Agential Reasoning: Proof-Calculi and Syntactic Decidability for STIT Logics. In International Conference on Principles and Practice of Multi-Agent Systems (pp. 202-218). Springer, Cham.

\item Lyon, T., Tiu, A., Goré, R., \& Clouston, R. (2020). Syntactic Interpolation for Tense Logics and Bi-Intuitionistic Logic via Nested Sequents. In 28th EACSL Annual Conference on Computer Science Logic (CSL 2020). Schloss Dagstuhl-Leibniz-Zentrum f\"ur Informatik.

\item Lyon, T. (2020). On Deriving Nested Calculi for Intuitionistic Logics from Semantic Systems. In International Symposium on Logical Foundations of Computer Science (pp. 177-194). Springer, Cham.

\item Lyon, T. (2021). On the Correspondence between Nested Calculi and Semantic Systems for Intuitionistic Logics. Journal of Logic and Computation. Oxford University Press.

\item Berkel, K. \& Lyon, T. (2021). The Varieties of Ought-Implies-Can and Deontic STIT Logic. In: Fenrong Liu, Alessandra Marra, Paul Portner, and Frederik Van De Putte (eds.). Deontic Logic and Normative Systems: 15th International Conference (DEON2020/2021, Munich). London: College Publications.

\item Ciabattoni, A., Lyon, T., Ramanayake, R., \& Tiu, A. (2021). Display to Labelled Proofs and Back Again for Tense Logics. ACM Transactions on Computational Logic (TOCL).

\end{enumerate}

\chapter{Preliminaries for Modal and Constructive Logics}
\label{CPTR:Logics} 



Contemporary modal logic is often traced back to the work of C.I. Lewis who attempted to resolve paradoxes of material implication via the formulation of strict implication~\cite{Lew18}. Languages of modal logics are characterized by their incorporation of \emph{modalities}---expressions that qualify the truth of a proposition. Common examples of modalities (see~\cite{Gar13}) include \emph{alethic} modalities such as ``It is necessary that'' (often denoted with $\Box$) and ``It is possible that'' (often denoted with $\Diamond$), \emph{temporal} modalities such as ``It will always be the case that'' (often denoted by $\mathsf{G}$) and ``It has been the case that'' (often denoted by $\mathsf{P}$), and \emph{deontic} modalities such as ``It is obligatory that'' (often denoted by $\mathcal{O}$) and ``It is permissible that'' (often denoted by $\mathcal{P}$). The extension of a propositional language with modalities makes for a more expressive 
 language for modeling, though, a benefit of modal logics is that despite their increased expressivity over classical propositional logic, decidability commonly holds~\cite{Var97}. Moreover, such logics are usually equipped with a relational semantics, which evaluates formulae at points in a relational structure. It can be seen that modal logics possess many advantages then: such logics allow for the qualification of truth, permitting one to model phenomena whose logical consequences depend on such qualifications, such formalisms increase the expressivity of a language while retaining decidability, and their evaluation over relational structures (ubiquitous in mathematics and computer science) is of practical consequence~\cite{BlaRijVen01}. 

Related to modal logics, intuitionistic logics employ a version of implication that is stronger than its classical counterpart, as well as a stronger version of universal quantification (in the first-order setting), both of which can be viewed as modalities. The advent of intuitionistic reasoning came in 1907/08 with the work of L.E.J. Brouwer, who put forth a philosophy of mathematics arguing that the truth of a mathematical statement rests upon a mental construction demonstrating its truth~\cite{Bro75}. Motivated by the work of Brouwer, axiomatic systems for propositional intuitionistic logic (based on Brouwer's intuitionism) were provided by Kolmogorov~\cite{Kol25}, Orlov~\cite{Orl28}, and Glivenko~\cite{Gli29}, and a first-order system was given by Heyting~\cite{Hey30}. Such logics are essential in the field of constructive mathematics~\cite{TroDal88}, and have important applications in computer science~\cite{How80,OsoNavArr04}.


In this thesis, we study the the refinement of labelled proof calculi for grammar logics~\cite{CerPen88}, first-order intuitionistic logics~\cite{Grz64,Hey30}, and deontic \stit logics~\cite{Hor01,Mur04,BerLyo21}. These logics are similar enough to permit a parallel and uniform investigation of their proof theory while also possessing enough distinguishing features from each other to make the investigation interesting and to justify the generality of the refinement method. In addition, as mentioned in the introduction (\cptr~\ref{CPTR:Intro}) such logics have a wide variety of applications, thus allowing for the work in this thesis to have potential practical effect.

Each class of logics will be introduced accordingly in the three subsequent sections: we introduce context-free grammar logics with converse in \sect~\ref{SEC:Grammar-Logics}, first-order intuitionistic logics in \sect~\ref{SEC:FO-Int-Logics}, and deontic \stit logics in \sect~\ref{SEC:STIT-Logics}. In each section, we define the language of each logic, supply a semantics, introduce fundamental concepts, and confirm soundness and completeness of each logic's axiomatization. We also adapt the method of canonical models (see~\cite{BlaRijVen01}) in \sect~\ref{SEC:STIT-Logics} to prove a new strong completeness result for our deontic \stit axiom systems.



\section{Grammar Logics}\label{SEC:Grammar-Logics}

Grammar logics\index{Grammar logics} are multi-modal logics that were introduced in 1988 by Fari\~nas del Cerro and Penttonen~\cite{CerPen88}. In that paper, the authors established equivalences between the validity problem for certain classes of grammar logics and the problem of checking if a word (or, string) is generated from a formal grammar. This relationship between modal logic and formal language theory has allowed for (un)decidability results to be transferred between the two settings~\cite{BalGioMar98,CerPen88,DemNiv05}. Grammar logics have been widely studied~\cite{BalGioMar98,CerPen88,Dem01,DemNiv05,GorNgu05,HorSat04,NguSza11} and are significant as they cover many well-known logics such as: description logics with complex role inclusion axioms and inverse roles~\cite{HorSat04}, epistemic logics~\cite{FagMosHalVar95}, information logics~\cite{Vak86}, temporal logics~\cite{CerHer95}, and standard modal logics (e.g. $\mathsf{K}$, $\mathsf{K4}$, $\mathsf{T}$, $ \mathsf{B}$, $\mathsf{S4}$, and $\mathsf{S5}$~\cite{DemNiv05}).

The language of grammar logics is defined relative to an \emph{alphabet}\index{Alphabet} $\albet$ consisting of a (non-empty) countable set of characters. Following~\cite{DemNiv05}, we assume that $\albet$ can be partitioned into a \emph{forward} part $\albet^{+} = \{a, b, c, \ldots\}$ and \emph{backward} part $\albet^{-} = \{\conv{a}, \conv{b}, \conv{c}, \ldots\}$, of the same cardinality, satisfying the following:
\begin{center}
$
\albet := \albet^{+} \cup \albet^{-} \text{ where } \albet^{+} \cap \albet^{-} = \emptyset \text{ , and } a \in \albet^{+} \text{ \ifandonlyif } \conv{a} \in \albet^{-}.
$
\end{center}
We use $a$, $b$, $c$, \etc (occasionally with subscripts) to denote the \emph{forward characters}\index{Forward character} of $\albet^{+}$, $\conv{a}$, $\conv{b}$, $\conv{c}$, \etc. (occasionally with subscripts) to denote the \emph{backward characters}\index{Backward character} of $\albet^{-}$, and refer to both forward and backward characters as \emph{characters} more generally, using $\chara$, $\charb$, $\charc$, \etc to range over such characters in $\albet^{+} \cup \albet^{-} = \albet$. In other words, our alphabet $\albet$ consists of the concrete characters $a$, $\conv{a}$, $b$, $\conv{b}$, \etc but we use $\chara$, $\charb$, $\charc$, \etc as `meta-characters' ranging over such concrete characters. This notation will be convenient in the following definitions and results, as it lets us reference characters in $\albet$ without specifying if they are forward or backward characters.

We define an involutory \emph{converse} operation\index{Converse!Characters} $\conv{\cdot}$ on characters that maps each forward character $a \in \albet^{+}$ to its \emph{converse} $\conv{a} \in \albet^{-}$, and maps each backward character $\conv{a} \in \albet^{-}$ to its \emph{converse} $a \in \albet^{+}$ (cf.~\cite{DemNiv05}). Notice that this operation is in fact an involution, since it satisfies the equation $\conv{\conv{\chara}} = \chara$. We stipulate that the symbol $\albet$ will be reserved to denote an alphabet for the remainder of the document. Using such an alphabet, we may define our language, which employs two modalities: $\charadia$ and $\charabox$.\footnote{We note that the language of common modal logics (e.g. $\mathsf{K}$, $\mathsf{S4}$, and $\mathsf{S5}$) may be obtained by restricting our alphabet to a single character without its corresponding converse.} Similar to the interpretation of standard modalities (e.g. $\Diamond$ and $\Box$~\cite{BlaRijVen01}), $\charadia \phi$ is interpreted as saying that there exists an $\chara$ successor state where $\phi$ holds, and $\charabox \phi$ is interpreted as saying that $\phi$ holds in all $\chara$ successor states; these interpretations are made explicit in \dfn~\ref{def:semantics-kms}.

\begin{definition}[The Language $\langkm{\albet}$] The language $\langkm{\albet}$\index{Language!for grammar logics} for our grammar logics is defined via the following grammar in BNF:
$$
\phi ::= p \ | \ \neg p \ | \ (\phi \lor \phi) \ | \ (\phi \land \phi) \ | \ \charadia \phi \ | \ \charabox \phi
$$
where $p$ is among a denumerable set of propositional variables $\prop = \{p,q,r, \ldots\}$, and $\chara \in \albet$. We use $\phi$, $\psi$, $\chi$, \etc (occasionally with subscripts) to denote formulae in $\langkm{\albet}$, and refer to formulae of the form $p$ and $\negnnf{p}$ (with $p \in \prop$) as \emph{literals}\index{Literal}.
\end{definition}


The formulae in our language $\langkm{\albet}$ are given in negation normal form\index{Negation normal form!for grammar logics}, meaning that applications of negations are restricted to propositional variables. This will simplify the structure of the sequents employed in our calculi as well as reduce the number of cases we need to consider when proving certain proof-theoretic results (see \sect~\ref{sec:lab-calc-kms}). 
Furthermore, it will be helpful to define the \emph{complexity}\index{Formula complexity!for grammar logics} of a formula $\phi$ from $\langkm{\albet}$, which corresponds to the number of binary connectives and modalities present in $\phi$. This measure will occasionally be employed 
 as a parameter in proofs by induction.

\begin{definition}[Complexity of an $\langkm{\albet}$ Formula] We define the \emph{complexity} $\fcomp{\phi}$ of a formula $\phi \in \langkm{\albet}$ inductively as follows:
\begin{itemize}

\item[$\li$] $\fcomp{p} = \fcomp{\neg p} = 0$ 

\item[$\li$] $\fcomp{\psi \land \chi} = \fcomp{\psi \lor \chi}  = max\{\fcomp{\psi},\fcomp{\chi}\} + 1$

\item[$\li$] $\fcomp{\charabox \psi} = \fcomp{\charadia \psi}  = \fcomp{\psi} + 1$

\end{itemize}
\end{definition}

We take the literals $p$ and $\negnnf{p}$ (for each $p \in \prop$), the binary operators $\land$ and $\lor$, and the modalities $\charabox$ and $\charadia$ (for each $\chara \in \albet$) to be \emph{duals} of each other. Using the notion of duality\index{Duality!for grammar logics}, we may define the negation $\neg \phi$ of a formula $\phi$ as the replacement of each literal, binary connective, and modality with its corresponding dual. The formal definition of negation is given below:

\begin{definition}[Negation of a $\langkm{\albet}$ Formula]\label{def:negation-kms} We define the \emph{negation} $\negnnf{\phi}$ of a formula $\phi \in \langkm{\albet}$ in the usual way, inductively as follows:
\begin{multicols}{2}
\begin{itemize}

\item[$\li$] If $\phi = p$, then $\negnnf{\phi} := \neg p$

\item[$\li$] If $\phi = \neg p$, then $\negnnf{\phi} := p$

\item[$\li$] If $\phi = \psi \land \chi$, then $\negnnf{\phi} := \negnnf{\psi} \lor \negnnf{\chi}$

\item[$\li$] If $\phi = \psi \lor \chi$, then $\negnnf{\phi} := \negnnf{\psi} \land \negnnf{\chi}$

\item[$\li$] If $\phi = \charabox \psi$, then $\negnnf{\phi} := \charadia \negnnf{\psi}$

\item[$\li$] If $\phi = \charadia \psi$, then $\negnnf{\phi} := \charabox \negnnf{\psi}$

\end{itemize}
\end{multicols}
\end{definition}

For example, if $\phi = \abox \negnnf{p} \land q$, then $\negnnf{\phi} = \adia p \lor \negnnf{q}$. Negation lets us further define the logical constants ($\top$ and $\bot$) as well as classical implication and bi-implication ($\cimp$ and $\leftrightarrow$, \resp), which we will use later on (e.g. \dfn~\ref{def:axiomatization-km}). For the definition of $\top$ and $\bot$, we fix an arbitrary propositional variable $p \in \prop$; the definitions are as follows:
$$
\top := p \lor \negnnf{p}, \quad \bot := p \land \negnnf{p}, \quad \phi \cimp \psi := \negnnf{\phi} \lor \psi, \quad \text{and} \quad \phi \cbiimp \psi := (\phi \cimp \psi) \land (\psi \cimp \phi).
$$

Due to our partitioning of $\albet$, our language consists of \emph{forward modalities}\index{Forward modality} $\abox$ and $\adia$ where $a \in \albet^{+}$, and \emph{backward modalities}\index{Backward modality} $\aboxc$ and $\adiac$ where $\conv{a} \in \albet^{-}$. We say that these modalities are \emph{converse} to one another since, as will be seen via the semantics of our language (\dfn~\ref{def:semantics-kms}), each class of modalities is interpreted relative to accessibility relations that are converse to one another. 
 We formally define, and make use of, the relational semantics for grammar logics provided in~\cite{DemNiv05} below. Note that these semantics are an extension and slight reformulation of the semantics for grammar logics that were given in Fari\~nas del Cerro and Penttonen's seminal paper~\cite{CerPen88}.


\begin{definition}[Frames and Models for Grammar Logics~\cite{DemNiv05}]\label{def:frames-models-kms} A \emph{$\albet$-frame}\index{$\albet$-frame} is an ordered pair $F = (W,\{R_{\chara} \ | \ \chara \in \albet\})$ such that (i) $W$ is a non-empty set of worlds $w$, $u$, $v$, \ldots \ and (ii) for each $\chara \in \albet$, $R_{\chara} \subseteq W \times W$ is a binary relation on $W$ satisfying the \emph{converse condition}\index{Converse condition}:
\begin{flushleft}
\emph{\convcond} \quad $(w,u) \in R_{\chara}$ \ifandonlyif $(u,w) \in R_{\conv{\chara}}$. 
\end{flushleft}
A \emph{$\albet$-model}\index{$\albet$-model} is a tuple $M = (F,V)$ such that $F$ is a $\albet$-frame and $V : \prop \mapsto 2^{W}$ is a \emph{valuation function}\index{Valuation function!$\albet$-model} mapping propositional variables from $\prop$ to subsets of $W$.
\end{definition}

The converse condition \convcond ensures that the backward modalities behave as expected and are truly converse to the forward modalities. That is to say, if we think of the `forward' relations $R_{a}$ (indexed with a forward character $a \in \albet^{+}$) as relating states to successor states, then the converse condition \convcond ensures that the `backward' relations $R_{\conv{a}}$ (indexed with a backward character $\conv{a} \in \albet^{-}$) relate states to predecessor states. This has the consequence that forward modalities such as $\abox$ and $\adia$ reference truth in successor states, whereas the backward modalities $\aboxc$ and $\adiac$ reference truth in predecessor states. This semantic relationship between the forward and backward modalities is made explicit via the following definition: 

\begin{definition}[Satisfaction, Global Truth~\cite{DemNiv05}]\label{def:semantics-kms} Let $M = (W,\{R_{\chara} \ | \ \chara \in \albet \}, V)$ be a $\albet$-model with $w \in W$, and define $R_{\chara}(w) := \{u \ | \ (w,u) \in R_{\chara}\}$ for $\chara \in \albet$. We define the \emph{satisfaction}\index{Formula satisfaction!for grammar logics} of a formula $\phi \in \langkm{\albet}$ on $M$ at $w$ (written $M,w \Vdash \phi$) inductively as follows:
\begin{itemize}

\item[$\li$] $M,w \Vdash p$ \ifandonlyif $w \in V(p)$;

\item[$\li$] $M,w \Vdash \negnnf{p}$ \ifandonlyif $w \not\in V(p)$;

\item[$\li$] $M,w \Vdash \phi \land \psi$ \ifandonlyif $M,w \Vdash \phi$ and $M,w \Vdash \psi$;

\item[$\li$] $M,w \Vdash \phi \lor \psi$ \ifandonlyif $M,w \Vdash \phi$ or $M,w \Vdash \psi$;

\item[$\li$] $M,w \Vdash \charabox \phi$ \ifandonlyif for all $u \in R_{\chara}(w)$, $M,u \Vdash \phi$;

\item[$\li$] $M,w \Vdash \charadia \phi$ \ifandonlyif for some $u \in R_{\chara}(w)$, $M,u \Vdash \phi$.

\end{itemize}
We say that a formula $\phi \in \langkm{\albet}$ is \emph{globally true}\index{Global truth!for grammar logics} on $M$ (written $M \Vdash \phi$) \ifandonlyif $M,w \Vdash \phi$ for all $w \in W$. 
\end{definition}




As mentioned previously (and as is implied by the name), grammar logics connect concepts concerning formal grammars to logical concepts. 
 Therefore, due to the intimate connection between formal grammars and grammar logics, we will introduce additional formal language theoretic concepts below. Such concepts will allow us to establish a correspondence between formal grammars---in particular, specific types of \emph{Semi-Thue Systems} (\dfn~\ref{def:CFCST-kms})---with properties imposed on $\albet$-frames (\dfn~\ref{fig:frame-conditions-production-rules}), and to define new classes of grammar logics (\dfn~\ref{def:axiomatization-km}). Most definitions are taken from~\cite{DemNiv05}.

\begin{definition}[Strings over $\albet$] We let $\concat$ represent the usual \emph{concatenation} operation\index{Concatenation} and let $\varepsilon$ be the \emph{empty string}\index{Empty string!in $\albet^{*}$}. The set $\albet^{*}$ of \emph{strings over $\albet$}\index{Strings} is defined to be the smallest set of strings such that:
\begin{itemize}

\item[$\li$] $\albet \cup \{\varepsilon\} \subseteq \albet^{*}$

\item[$\li$] $\text{If } \stra, \strb \in \albet^{*} \text{, then } \stra \concat \strb \in \albet^{*}$

\end{itemize}
As usual, we define $\stra \concat \empstr = \empstr \concat \stra = \stra$ for $\stra \in \albet^{*}$, showing that the empty string $\empstr$ is an identity element for the concatenation $\concat$ operation. Furthermore, for strings $\stra, \strb \in \albet^{*}$, we let $\stra \concat \strb := \stra \cate \strb$, that is, we will omit explicit mention of the concatenation operation, and simply stick strings together when performing concatenation, as is typically done.

We use $\stra$, $\strb$, $\strc$, \etc (possibly annotated) to denote strings in $\albetstr$. Moreover, we extend the converse operation\index{Converse!Strings} to strings as follows:
\begin{itemize}

\item[$\li$] $\conv{\varepsilon} := \varepsilon$

\item[$\li$] $\text{If } \stra = \chara_{0} \cate \chara_{1} \cdots \chara_{n-1} \cate \chara_{n} \text{, then } \conv{\stra} := \conv{\chara}_{n} \cate \conv{\chara}_{n-1} \cdots \conv{\chara}_{1} \cate \conv{\chara}_{0}$

\end{itemize}
Last, we define the \emph{length}\index{Length!String} of a string in $\stra \in \albet^{*}$ inductively as follows: 
\begin{itemize}

\item[$\li$] $\lenstr{\stra} := 0$, if $\stra = \empstr$

\item[$\li$] $\lenstr{\stra} = \lenstr{\strb \cate \chara} := \lenstr{\strb} + 1$, if $\stra = \strb \cate \chara$ with $\strb \in \albet^{*}$ and $\chara \in \albet$

\end{itemize}
The context will easily determine if the notation $| \cdot |$ is being used to denote the complexity of a formula or the length of a string.
\end{definition}

Since characters from $\albet$ are used to index the accessibility relations in a $\albet$-frame or $\albet$-model, strings from $\albetstr$ denote paths. We extend our relations $R_{\chara}$ to relations $R_{\stra}$ defined relative to strings below. Correspondingly, we define strings of modalities as follows: if $\stra = \chara_{0} \cate \chara_{1} \cdots \chara_{n-1} \cate \chara_{n}$, then $[ \stra ] =  [ \chara_{0} ] [ \chara_{1} ] \cdots [ \chara_{n-1} ] [ \chara_{n} ]$ and $\langle \stra \rangle =  \langle \chara_{0} \rangle \langle \chara_{1} \rangle \cdots \langle \chara_{n-1} \rangle \langle \chara_{n} \rangle$, and if $\stra = \empstr$, then $\strabox \phi = \stradia \phi = \phi$.

\begin{definition}[Generalized Relations]\label{def:generalized-relations-kms} Let $F = (W,\{R_{\chara} \ | \ \chara \in \albet\})$ be a $\albet$-frame, $\chara \in \albet$, and $\stra \in \albetstr$. We extend the definition of an accessibility relation\index{Generalized accessibility relation} (defined in \dfn~\ref{def:semantics-kms}) to strings and define $R_{\stra}$ inductively as follows:
\begin{itemize}

\item[$\li$] $R_{\empstr} := \{(w,w) \ | \ w \in W\}$;

\item[$\li$] $R_{\stra \cate \chara} := \{(w,u) \ | \ \exists v \in W, (w,v) \in R_{\stra} \text{ and } (v,u) \in R_{\chara}\}$.

\end{itemize}
\end{definition}

\begin{remark}\label{rmk:R-Converses-kms}
The above definition immediately implies that $(w,u) \in R_{\stra}$ \ifandonlyif $(u,w) \in R_{\conv{\stra}}$.
\end{remark}


As in~\cite{DemNiv05}, we utilize a language-theoretic framework to define the supplementary frame conditions we will impose on $\albet$-frames. For example, we can encode the well-known symmetry frame condition $\forall w, u (R_{a}wu \cimp R_{\conv{a}}wu)$ by the production rule\index{Production rule} $\conv{a} \pto a$, or a version of three-to-one transitivity $\forall w, u, v, z (R_{a}wu \land R_{b}uv \land R_{c}vz \cimp R_{d}wz)$ (equivalently, represented as $\forall w, u (R_{a \cate b \cate c}wu \cimp R_{d}wu)$) by the production rule $d \pto a \cate b \cate c$. (NB. See \fig~\ref{fig:frame-conditions-production-rules} for a collection of common frame conditions and their associated production rules.) In general, a production rule of the form $\chara \pto \stra$ corresponds to a frame condition of the form $\forall w,u (R_{\stra}wu \cimp R_{\chara}wu)$.

To make such frame conditions and their consequences precise, we define a restricted version of a \emph{Semi-Thue System}\index{Semi-Thue system} (cf.~\cite{Pos47}), i.e. we define a string rewriting system that is \emph{context-free}\index{Context-free!Semi-Thue system}---meaning the head of each production rule is a single character---and which is \emph{closed}\index{Closed!Semi-Thue system}---meaning that the production rules are closed under converses (cf.~\cite{DemNiv05,TiuIanGor12}).

\begin{definition}[\cfcst System]\label{def:CFCST-kms} We let $\chara \pto \stra$ represent a \emph{production rule} that rewrites the character $\chara \in \albet$ to the string $\stra \in \albet^{*}$, and refer to $\chara$ as the \emph{head} and $\stra$ as the \emph{tail} of the production rule. A \emph{context-free, closed, Semi-Thue (CFCST) System}\index{CFCST system} is a set $S$ of production rules satisfying the following \emph{closure condition}\index{Closure condition}:
\begin{flushleft}
\emph{\thueclcond} \quad $\chara \pto \stra \in S$ \ifandonlyif $\conv{\chara} \pto \conv{\stra} \in S$.
\end{flushleft}
\end{definition}


A \cfcst system $\thuesys$ re-writes strings in the following manner: if $\chara \pto \stra \in \thuesys$, then the string $\strb \cate \chara \cate \strc$ may be re-written as $\strb \cate \stra \cate \strc$. For example, if our \cfcst system is $\thuesys = \{a \pto b \cate b, \conv{a} \pto \conv{b} \cate \conv{b}, b \pto \conv{b}, \conv{b} \pto b\}$, then $a \cate b$ can be re-written as $b \cate b\cate  b$ or $a \cate \conv{b}$ by applying the first and third production rule in $\thuesys$, respectively; that is, the strings $b \cate b\cate  b$ and $a \cate \conv{b}$ can be derived from $a \cate b$ in one-step. As usual, through successive applications of the production rules, one can derive all possible strings produced by taking a string $\stra \in \albetstr$ as initial, thus generating a set of strings (i.e. language) determined on the basis of the \cfcst system and $\stra$. Continuing the above example, we could derive the strings $b \cate b \cate \conv{b}$, $a \cate b \cate b$, or $a \cate b \cate \conv{b} \cate b$ by taking $a \cate b$ as initial. We make the above notions formally precise with the following definition:

\begin{definition}[Derivation Relation, Language]\label{def:derivation-relation-language-kms} Let $\thuesys$ be a \cfcst system. We define the \emph{one-step derivation relation}\index{One-step derivation relation} $\dto_{\thuesys}$ relative to $\thuesys$ as follows: $\stra \dto_{\thuesys} \strb$ \ifandonlyif there exist $\stra_{0}, \stra_{1} \in \albet^{*}$, and $\chara \pto \strc \in \thuesys$, such that $\stra = \stra_{0} \cate \chara \cate \stra_{1}$ and $\strb = \stra_{0} \cate \strc \cate \stra_{1}$. 

We define the \emph{derivation relation}\index{Derivation relation} $\dto_{\thuesys}^{*}$ to be the reflexive and transitive closure of the one-step derivation relation $\dto_{\thuesys}$. 

The \emph{length of a derivation $\stra \dto_{\thuesys}^{*} \strb$}\index{Length!Derivation} is defined to be equal to the minimal number of one-step derivations needed to derive $\strb$ from $\stra$ in $\thuesys$.

Last, for each $\stra \in \albet^{*}$, we define the \emph{language of $\stra$ relative to $\thuesys$}\index{Language!of a string} to be $\thuesyslang{\stra} := \{\strb \ | \ \stra \dtoann \strb \}$.
\end{definition}

\begin{figure}[t]
\begin{center}
\begin{tabular}{| c | c | c |}
\hline
Name & Frame Property & Production Rule\\
\hline
Reflexivity & $\forall w R_{a}ww$ & $a \pto \empstr$ \\
Symmetry & $\forall w, u (R_{a}wu \cimp R_{\conv{a}}wu)$ & $\conv{a} \pto a$ \\
Transitivity & $\forall w, v, u (R_{a}wv \land R_{a}vu \cimp R_{a}wu)$ & $a \pto a \cate a$ \\
Euclideanity & $\forall w, v, u (R_{\conv{a}}vw \land R_{a}wu \cimp R_{a}vu)$ & $a \pto \conv{a} \cate a$ \\
\hline
\end{tabular}
\end{center}
\caption{Common frame properties and their associated production rules.}
\label{fig:frame-conditions-production-rules}
\end{figure}

The definition below explains how \cfcst systems are exploited to impose certain conditions on $\albet$-frames and models. This leads to a notion of $\thuesys$-validity for formulae (with $\thuesys$ a \cfcst system) that is provided in the ensuing definition.

\begin{definition}[Production Rule Satisfaction]\label{def:production-rule-sat-kms} Let the following ordered pair $F = (W,\{R_{\chara} \ | \ \chara \in \albet\})$ be a $\albet$-frame. A $\albet$-frame \emph{satisfies a production rule $a \pto \stra$}\index{Satisfaction!Production rule} \ifandonlyif $R_{\stra} \subseteq R_{a}$. 

A $\albet$-frame \emph{satisfies a \cfcst system $\thuesys$} \ifandonlyif it satisfies all production rules in $\thuesys$. 

Last, a $\albet$-model satisfies a production rule or \cfcst system \ifandonlyif its underlying $\albet$-frame does.
\end{definition}

\begin{definition}[$\thuesys$-validity]\label{def:S-validity-kms} Let $\thuesys$ be a \cfcst system. We say that a formula $\phi \in \langkm{\albet}$ is \emph{$\thuesys$-valid}\index{$\thuesys$-validity} \ifandonlyif for every $\albet$-model $M$, if $M$ satisfies $\thuesys$, then $M \Vdash \phi$.
\end{definition}


A favorable feature of the grammar logics we are considering is that a set of $\thuesys$-validities (with $\thuesys$ a \cfcst system) is always axiomatizable. Below, we present an axiomatization and definition for the grammar logic $\kms$, defined relative to a \cfcst system $\thuesys$. Similar axiomatizations for grammar logics have been provided in the literature~\cite{CerPen88,DemNiv05,TiuIanGor12}, albeit, such axiom systems were provided for languages based on a different signature, whereas our axiomatization is suited for formulae in negation normal form.

\begin{definition}[Axiomatization $\h\kms$]\label{def:axiomatization-km} Let $\thuesys$ be a \cfcst system with alphabet $\albet$. The axiomatization $\h\kms$\index{$\h\kms$} for the logic $\kms$ is as follows:
\begin{itemize}





\item[A0] All instances of (classical) propositional tautologies

\item[A1] For all $\chara \in \albet$, $\charabox (\phi \cimp \psi) \cimp (\charabox \phi \cimp \charabox \psi)$


\item[A2] For all $\chara \in \albet$, $\phi \cimp \charabox \charadiac \phi$

\item[A3] For all $\chara \pto \stra \in S$, $\langle \stra \rangle \phi \cimp \charadia \phi$

\item[R0] \AxiomC{$\phi$}
\AxiomC{$\phi \cimp \psi$}
\BinaryInfC{$\psi$}
\DisplayProof

\item[R1] For all $\chara \in \albet$, \AxiomC{$\phi$}
\UnaryInfC{$\charabox \phi$}
\DisplayProof
\end{itemize}
We define the \emph{logic $\kms$}\index{$\kms$} to be the smallest set of formulae from $\langkm{\albet}$ that is closed under substitutions of the axioms and applications of the inference rules. Also, we refer to a logic $\km(\emptyset)$ (relative to an alphabet $\albet$) as a \emph{minimal grammar logic} and let $\km := \km(\emptyset)$\index{$\km$}. Last, we say that a formula $\phi \in \langkm{\albet}$ is a \emph{$\kms$-theorem}\index{$\kms$-theorem} (written $\vdash_{\kms} \phi$) \ifandonlyif $\phi \in \kms$.
\end{definition}

The axioms A0 ensure that the logic $\kms$ is an extension of classical propositional logic, and R0 is the well-known rule \emph{modus ponens}. The A1 axioms are the typical \emph{K axioms}, which, together with the R1 \emph{necessitation} rules, ensure that all modalities are \emph{normal}.\footnote{See~\cite{BlaRijVen01} for a discussion and definition of \emph{normal modal logics}\index{Normal modal logic}.} 
 The A2 axioms encode the fact that the $\abox$ and $\adia$ modalities are converses of the $\aboxc$ and $\adiac$ modalities, respectively. Last, the A3 axioms are determined by the \cfcst system $\thuesys$ and add an axiom---called a \emph{path axiom}\index{Path axiom} (cf.~\cite{CiaLyoRamTiu20,TiuIanGor12})---for each production rule present in $\thuesys$. By standard methods for normal modal logics~\cite{BlaRijVen01}, the following soundness and completeness theorem for $\kms$ can be shown:



\begin{theorem}[Soundness and Completeness]
For all $\phi \in \langkm{\albet}$, $\phi$ is $\thuesys$-valid \ifandonlyif $\vdash_{\kms} \phi$.
\end{theorem}




\section{First-Order Intuitionistic Logics}\label{SEC:FO-Int-Logics}

First-order intuitionistic logic \emph{proper}\index{First-order intuitionistic logic!Proper}\index{First-order intuitionistic logic!Non-constant domains} (also called \emph{first-order intuitionistic logic with non-constant domains}) was axiomatized by Heyting early in the development of intuitionistic logic~\cite{Hey30}. Over three decades later, a variation of the logic, called \emph{first-order intuitionistic logic with constant domains}\index{First-order intuitionistic logic!Constant domains} was introduced by Grzgorczyk~\cite{Grz64}, and axiomatized independently by Klemke~\cite{Kle69} and G\"orneman~\cite{Goe71}. Both logics admit a relational semantics, with Kripke providing the first relational semantics for first-order (and propositional) intuitionistic logic in 1965~\cite{Kri65}. In this section, we will introduce the semantics and axiomatizations for both first-order intuitionistic logic with constant ($\intfond$) and non-constant domains ($\intfocd$), and when reference is made to \emph{first-order intuitionistic logics}\index{First-order intuitionistic logics}, these two logics are taken to be the referents. Such logics will be of interest when refining labelled calculi since they will demonstrate how the method is applied in the first-order setting.

\begin{definition}[The Language $\langintfo$]\label{def:langintfo} The language $\langintfo$ for first-order intuitionistic logics\index{Language!for first-order intuitionistic logics} is defined via the following grammar in BNF:
$$
\phi ::= p(x_{1}, \ldots, x_{n}) \ | \ \bot \ | \ (\phi \vee \phi) \ | \ (\phi \land \phi) \ | \ (\phi \imp \phi) \ | \ (\exists x) \phi \ | \ (\forall x) \phi
$$
where $p$ is among a denumerable set of $n$-ary \emph{predicate symbols}\index{Predicate symbol} $\pred = \{p, q, r, \ldots\}$ and $x_{1}, \ldots, x_{n}, x$ (with $n \in \mathbb{N}$) are among a denumerable set of \emph{variables} $\var = \{x, y, z, \ldots\}$.

We refer to formulae of the form $p(x_{1}, \ldots, x_{n})$ as \emph{atomic formulae}\index{Atomic formula}, and also refer to formulae of the form $p$ as \emph{propositional variables} when $n=0$, that is to say, a $0$-ary predicate $p$ is a propositional variable. We will use $\phi$, $\psi$, $\chi$, \etc to denote formulae from $\langintfo$, and define intuitionistic negation\index{Intuitionistic negation} in the usual way as $\neg \phi := \phi \imp \bot$.
\end{definition}

As usual, we impose an operator precedence\index{Operator precedence} on our logical operators occurring in formulae of $\langintfo$ as follows: $\exists$ and $\forall$ bind tighter than $\lor$ and $\land$, which bind tighter than $\imp$. For example, the formula $\forall x p(x) \lor q \imp r(x) \land q$ would be disambiguated as $((\forall x p(x)) \lor q) \imp (r(x) \land q)$. Imposing a precedence lets us omit parentheses to improve the readability of formulae.

Another concept that will come in handy when studying and leveraging the proof theory of first-order intuitionistic logics is the \emph{complexity}\index{Formula complexity!for first-order intuitionistic logics} of a formula in $\langintfo$. The complexity of a formula in $\langintfo$ is equal to the number of unary and binary logical operators occurring in a formula, and is useful as it provides a well-founded measure that may be invoked to prove certain results by induction (e.g. the cut-elimination theorem for first-order intuitionistic logics, \thm~\ref{thm:cut-admiss-FO-Int}). The formal definition is given below:

\begin{definition}[Complexity of an $\langintfo$ Formula] We define the \emph{complexity} $\fcomp{\phi}$ of a formula $\phi \in \langintfo$ inductively as follows:
\begin{itemize}

\item[$\li$] $\fcomp{p(x_{1}, \ldots, x_{n})} = \fcomp{\bot} = 0$ 

\item[$\li$] $\fcomp{\psi \lor \chi} = \fcomp{\psi \land \chi} = \fcomp{\psi \imp \chi}  = max\{\fcomp{\psi},\fcomp{\chi}\} + 1$

\item[$\li$] $\fcomp{\exists x \psi} = \fcomp{\forall x \psi}  = \fcomp{\psi} + 1$

\end{itemize}
\end{definition}

A useful concept will be the notion of \emph{free} and \emph{bound variables} occurring in a formula $\phi$. Intuitively, a variable $x$ in $\phi$ is \emph{free} \ifandonlyif it is not within the scope of a quantifier, and it is \emph{bound} \ifandonlyif it is within the scope of a quantifier (cf.~\cite{GabSheSkv09}). The formal definition of free and bound variables is given below:

\begin{definition}[Free and Bound Variables]\label{def:free-bound-variables} We define the set $FV(\phi)$ of free variables\index{Free variable} in a formula $\phi \in \langintfo$ inductively as follows:
\begin{itemize}

\item[$\li$] $FV(p(x_{1}, \ldots, x_{n})) = \{x_{1}, \ldots, x_{n}\}$

\item[$\li$] $FV(\bot) = \emptyset$

\item[$\li$] $FV(\psi \lor \chi) = FV(\psi \land \chi) = FV(\psi \imp \chi) = FV(\psi) \cup FV(\chi)$

\item[$\li$] $FV(\forall x \psi) = FV(\exists x \psi) = FV(\psi) - \{x\}$

\end{itemize}
We say that a variable $x$ is \emph{free} in $\phi$ \ifandonlyif $x \in FV(\phi)$. Similarly, we define the set $BV(\phi)$ of bound variables\index{Bound variable} in a formula $\phi \in \langintfo$ inductively as follows:
\begin{itemize}

\item[$\li$] $BV(p(x_{1}, \ldots, x_{n})) = \emptyset$

\item[$\li$] $BV(\bot) = \emptyset$

\item[$\li$] $BV(\psi \lor \chi) = BV(\psi \land \chi) = BV(\psi \imp \chi) = BV(\psi) \cup BV(\chi)$

\item[$\li$] $BV(\forall x \psi) = BV(\exists x \psi) = BV(\psi) \cup \{x\}$

\end{itemize}
We say that a variable $x$ is \emph{bound} in $\phi$ \ifandonlyif $x \in BV(\phi)$. 
\end{definition}

Note that the above definition allows for $FV(\phi) \cap BV(\phi) \neq \emptyset$, that is, variables are allowed to be both free and bound. For instance, the variable $x$ is both free and bound in the formula $\forall x p(x) \land q(x)$. Occasionally, we will refer to the \emph{occurrence} of a variable $x$ as being \emph{free} or \emph{bound} in a formula $\phi$, in which case, the occurrence\index{Variable occurrence} of the variable will be exclusively free or bound.

As explained previously, Kripke provided a relational semantics for first-order intuitionistic logic with non-constant domains~\cite{Kri65}. Here, we put forth a variant of Kripke's semantics provided by Gabbay \textit{et al.} in~\cite{GabSheSkv09} for both first-order intuitionistic logic with non-constant and constant domains (see~\dfn~\ref{def:axiomatization-IntFO} for a definition of these logics).

\begin{definition}[First-Order Intuitionistic Frames and Models~\cite{GabSheSkv09}]\label{def:IntFO-frame-model} We define an \emph{$\intfond$-frame}\index{$\intfond$-frame} to be a tuple $F = (W,\leq,D)$ such that:
\begin{itemize}

\item[$\blacktriangleright$] $W$ is a non-empty set of worlds $w$, $u
$, $v$, \etc 

\item[$\blacktriangleright$] $\leq \ \subseteq W \times W$ is a reflexive and transitive binary relation on $W$.\footnote{The properties imposed on $\leq$ are defined as follows: (reflexivity) for all $w \in W$, $w \leq w$, and (transitivity) for all $w, u, v \in W$, if $w \leq v$ and $v \leq u$, then $w \leq u$.}

\item[$\blacktriangleright$] $D$ is a \emph{domain function} mapping a world $w \in W$ to a non-empty set of objects $D_{w} = \{a, b, c, \dots\}$ satisfying the \emph{nested domain condition}\index{Nested domain condition} shown below:
\end{itemize}
\begin{flushleft}
\emph{\ndcond} \quad If $a \in D_{w}$ and $w \leq v$, then $a \in D_{v}$.
\end{flushleft}

A \emph{$\intfocd$-frame}\index{$\intfocd$-frame} is an \emph{$\intfond$-frame} that additionally satisfies the following \emph{constant domain condition}\index{Constant domain condition} shown below:
\begin{flushleft}
\emph{\cdcond} \quad If $a \in D_{v}$ and $w \leq v$, then $a \in D_{w}$.
\end{flushleft}

An \emph{$\intfond$-model}\index{$\intfond$-model} (\emph{$\intfocd$-model})\index{$\intfocd$-model} $M$ is an ordered pair $(F,V)$ where $F$ is an $\intfond$-frame ($\intfocd$-frame) and $V$ is a \emph{valuation function}\index{Valuation function!$\intfond$-model}\index{Valuation function!$\intfocd$-model} such that $V(p,w) \subseteq (D_{w})^{n}$ (with $n \in \mathbb{N}$) satisfying the following \emph{monotonicity condition}\index{Monotonicity condition}:
\begin{flushleft}
\emph{\moncond} \quad 
 If $w \leq u$, then $V(p,w) \subseteq V(p,u)$. 
\end{flushleft}
We uphold the convention in~\cite{GabSheSkv09} and assume that for each world $w \in W$, $(D_{w})^{0} = \{w\}$, so $V(p,w) = \{w\}$ or $V(p,w) = \emptyset$, for a propositional variable $p$.
\end{definition}

As in~\cite{GabSheSkv09}, we forgo the direct interpretation of formulae from $\langintfo$ on $\intfond$- and $\intfocd$-models, and instead, introduce $D_{w}$-sentences\index{$D_{w}$-sentence} (see Def.~\ref{def:d-sentence} below). Defining satisfaction relative to $D_{w}$-sentences gives rise to notions of validity for formulae in $\langintfo$ (Def.~\ref{def:semantics-intfo}). Additionally, this notion depends on the \emph{universal closure}\index{Universal closure} of a formula, which is defined as follows: For $\phi \in \langintfo$ such that $FV(\phi) = \{x_{1}, \ldots, x_{n}\}$, the universal closure $\uc \phi$ is taken to be the formula $\forall x_{1} \ldots \forall x_{n} \phi$.

\begin{definition}[$D_{w}$-Sentence]\label{def:d-sentence} Let $M = (W, \leq,D,V)$ be an $\intfond$- or $\intfocd$-model with $w \in W$. We define the set of \emph{parameters} $\para_{w} := \{\unda \ | \ a \in D_{w}\}$. The language $\langintfo({D_{w}})$ is defined the same as $\langintfo$, but with atomic formulae of the form $p(t_{1}, \ldots, t_{n})$, 
 where $p \in \pred$ and $t_{1}, \ldots, t_{n} \in \var \cup \para_{w}$. 
 
A \emph{$D_{w}$-formula}\index{$D_{w}$-formula} is simply a formula in $\langintfo(D_{w})$, and a \emph{$D_{w}$-sentence} is a $D_{w}$-formula that does not contain any free variables. (NB. Free variables are defined for formulae in $\langintfo({D_{w}})$ in the same way as for formulae in $\langintfo$, but with the exception that parameters from $\para_{w}$ are ignored.) 

Last, it should be made explicit that we use $\unda, \undb, \undc, \ldots$ to denote parameters corresponding to objects in $D_{w} = \{a, b, c, \ldots\}$.
\end{definition}

\begin{definition}[Satisfaction, Global Truth]\label{def:semantics-intfo} Let $M = (W, \leq,D,V)$ be an $\intfond$- or $\intfocd$-model with $w \in W$. The \emph{satisfaction relation}\index{Formula satisfaction!for first-order intuitionistic logics} $M,w \Vdash \phi$ between $w$ and a $D_{w}$-sentence $\phi$ is inductively defined as follows:
\begin{itemize}


\item[$\blacktriangleright$] 
 $M,w \Vdash p(\unda_{1}, \cdots, \unda_{n})$ iff $(a_{1}, \cdots, a_{n}) \in V(p,w)$;

\item[$\blacktriangleright$] $M, w \not\Vdash \bot$;

\item[$\blacktriangleright$] $M,w \Vdash \phi \vee \psi$ iff $M,w \Vdash \phi$ or $M,w \Vdash \psi$;

\item[$\blacktriangleright$] $M,w \Vdash \phi \land \psi$ iff $M,w \Vdash \phi$ and $M,w \Vdash \psi$;

\item[$\blacktriangleright$] $M,w \Vdash \phi \imp \psi$ iff for all $u \in W$, if $w \leq u$ and $M,u \Vdash \phi$, then $M,u \Vdash \psi$;

\item[$\blacktriangleright$] $M,w \Vdash \forall x \phi$ iff for all $u \in W$ and all $a \in D_{u}$, if $w \leq u$, then $M,u \Vdash \phi(\unda / x)$;

\item[$\blacktriangleright$] $M,w \Vdash \exists x \phi$ iff there exists an $a \in D_{w}$ such that $M,w \Vdash \phi(\unda / x)$.

\end{itemize}
We say that a formula $\phi$ is \emph{globally true on $M$},\index{Global truth!for first-order intuitionistic logics} written $M \Vdash \phi$, iff $M,u \Vdash \uc \phi$ for all worlds $u \in W$. A formula $\phi$ is \emph{$\intfond$-valid} (\emph{$\intfocd$-valid}), written $ \Vdash_{\intfond} \phi$ ($ \Vdash_{\intfocd} \phi$, resp.), iff it is globally true on all $\intfond$-models ($\intfocd$-models).
\end{definition}

We note that in the intuitionistic setting the universal quantifier is stronger than the existential quantifier, that is, unlike in the classical setting the two operators fail to be interdefinable (see~\cite{Kri65}). The monotonicity condition \moncondns, together with the semantic clauses of \dfn~\ref{def:semantics-intfo}, necessitates a general form of monotonicity, detailed below:

\begin{lemma}[General Monotonicity] 
\label{lm:fo-gen-monotonicity} Let $M$ be an $\intfond$- or $\intfocd$-model with $w,v \in W$ of $M$. For any $D_{w}$-sentence $\phi$, if $M,w \Vdash \phi$ and $w \leq v$, then $M,v \Vdash \phi$.
\end{lemma}

\begin{proof} See~{\cite[Lem.~3.2.16]{GabSheSkv09}} for details.
\end{proof}


Below (\dfn~\ref{def:axiomatization-IntFO}), we provide axiomatizations for first-order intuitionistic logic with non-constant domains~\cite[p.~119]{GabSheSkv09} and first-order intuitionistic logic with constant domains~\cite[p.~136]{GabSheSkv09}. Both axiomatizations are extensions of the axiomatization for propositional intuitionistic logic which can be found in~\cite[p.~6]{GabSheSkv09}. Axioms A9 and A10 make use of a \emph{substitution} $(y/x)$ of the variable $y$ for the free variable $x$ on a formula $\phi$. We define $\phi(y/x)$ in the standard way below as the replacement of all free occurrences of $x$ in $\phi$ with $y$, and define additional substitutions that will be used in the sequel. Also, the side condition \emph{$y$ is free for $x$ in $\phi$} is imposed on both axioms, and is taken to mean that $y$ does not become bound by a quantifier if substituted for $x$ in $\phi$.\footnote{See \cite[pp. 64--66]{Dal04} for a formal definition 
one variable being \emph{free for}\index{Free for} another variable in a formula.}

\begin{definition}[Substitutions]\label{def:substitutions-FO-Int} Let $\phi \in \langintfo (D_{w})$ and $t, t', s, t_{1}, \ldots,t_{n} \in \var \cup \para_{w}$. We define the \emph{substitution} $(t/s)$\index{Substitution!for first-order intuitionistic formulae} inductively as follows:
\begin{itemize}

\item[$\li$] $s(t/t') := t$ if $t' = s$ and $s(t/t') := s$ otherwise.

\item[$\li$] $p(t_{1}, \ldots,t_{n})(t/s) := p(t_{1}(t/s), \ldots,t_{n}(t/s))$.

\item[$\li$] $\bot(t/s) := \bot$.

\item[$\li$] $(\psi \binop \chi)(t/s) := \psi(t/s) \binop \chi(t/s)$ for $\binop \in \{\lor, \land, \imp\}$.

\item[$\li$] $(\forall x \psi)(t/s) := \forall x \psi(t/s)$ and $(\exists x \psi)(t/s) := \exists x \psi(t/s)$.

\end{itemize}
\end{definition}


\begin{definition}[Axiomatizations $\h\intfond$ and $\h\intfocd$]\label{def:axiomatization-IntFO} The axiomatization $\h\intfond$\index{$\h\intfond$} for the logic $\intfond$ is obtained by taking axioms A0--A12 and rules R0 and R1, whereas the axiomatization $\h\intfocd$\index{$\h\intfocd$} for the logic $\intfocd$ is obtained by taking all axioms and inference rules.
\begin{center}
\begin{multicols}{2}
\begin{itemize}

\item[A0] $\phi \supset (\psi \supset \phi)$

\item[A1] $(\phi \imp (\psi \imp \chi)){ \imp }((\phi \imp \psi) \imp (\phi \imp \chi))$

\item[A2] $\phi \supset (\psi \supset (\phi \land \psi))$

\item[A3] $(\phi \land \psi) \supset \phi$

\item[A4] $(\phi \land \psi) \supset \psi$

\item[A5] $\phi \supset (\phi \lor \psi)$

\item[A6] $\psi \supset (\phi \lor \psi)$

\item[A7] $\bot \imp \phi$

\item[A8] $(\phi \supset \chi) \supset ((\psi \supset \chi) \supset ((\phi \lor \psi) \supset \chi))$

\item[A9] $\forall x \phi \supset \phi(y/x)~\textit{y free for x in $\phi$}$

\item[A10] $\phi(y/x) \supset \exists x \phi~\textit{y free for x in $\phi$}$

\item[A11] $\forall x (\psi \imp \phi) \imp (\psi \imp \forall x \phi)$~$x \not\in FV(\psi)$

\item[A12] $\forall x (\phi \imp \psi) \imp (\exists x \phi \imp \psi)$~$x \not\in FV(\psi)$

\item[A13] $\forall x (\phi \vee \psi) \imp \forall x \phi \vee \psi$~$x \not\in FV(\psi)$

\item[R0] \AxiomC{$\phi$}
\AxiomC{$\phi \imp \psi$}
\RightLabel{$(mp)$}
\BinaryInfC{$\psi$}
\DisplayProof

\item[R1] \AxiomC{$\phi$}
\UnaryInfC{$\forall x \phi$}
\DisplayProof
\end{itemize}
\end{multicols}
\end{center}
We define the \emph{logic $\intfond$}\index{$\intfond$} (and $\intfocd$)\index{$\intfocd$} to be the smallest set of formulae from $\langintfo$ that is closed under substitutions of the axioms and applications of the inference rules in $\h\intfond$ ($\h\intfocd$, resp.). We say that a formula $\phi \in \langintfo$ is an \emph{$\intfond$-theorem}\index{$\intfond$-theorem} (\emph{$\intfocd$-theorem})\index{$\intfocd$-theorem}, written $\vdash_{\intfond} \phi$ ($\vdash_{\intfocd} \phi$, resp.) \ifandonlyif $\phi \in \intfond$ ($\phi \in \intfocd$, resp.).
\end{definition}

The axioms A0--A8, together with modus ponens (rule R0), is an axiomatization for propositional intuitionistic logic $\int$\index{$\int$} (cf.~\cite[p.~6]{GabSheSkv09}). (NB. We define \emph{propositional intuitionistic logic $\int$} to be the smallest set of formulae derivable from the axioms A0--A8 with rule R0.) The sole difference between the axiomatizations for $\intfond$ and $\intfocd$ is that the former omits the \emph{constant domain axiom}\index{Constant domain axiom} $\forall x (\phi \vee \psi) \imp \forall x \phi \vee \psi$ with $x \not\in FV(\psi)$ (first introduced by A. Grzegorczyk; cf.~\cite[p. 136]{GabSheSkv09}) whereas the latter includes it. The inclusion of this axiom in $\h\intfocd$ is what causes the logic $\intfocd$ to be sound and complete relative to frames with constant domains. Soundness and completeness of each of the above systems can be found, for example, in~\cite{GabSheSkv09}.

\begin{theorem}[Soundness and Completeness] For any $\phi \in \langintfo$, $\Vdash_{\intfond} \phi$ ($\Vdash_{\intfocd} \phi$) iff $\vdash_{\intfond} \phi$ ($\vdash_{\intfocd} \phi$, resp.).
\end{theorem}

\begin{proof} The forward direction follows from \cite[Cor.~6.2.21]{GabSheSkv09}, \cite[Prop.~6.2.22]{GabSheSkv09}, \cite[Prop.~7.2.9]{GabSheSkv09}, and \cite[Prop.~7.3.6]{GabSheSkv09}. The backward direction is obtained from~{\cite[Lem.~3.2.31]{GabSheSkv09}}.
\end{proof}

\section{Deontic STIT Logics}\label{SEC:STIT-Logics}


Traditional multi-agent \stit logics (with \stit an acronym for `Seeing To It That') have been used to model multi-agent choice making~\cite{BelPer90,BelPerXu01,Hor01}. Such logics employ an atemporal \emph{choice}\index{Choice operator} operator $\agbox$ expressing that `agent $i$ sees to it that' (some proposition is realized). For example, the formula $\agbox \mathtt{door\_Closed}$ might be interpreted as saying that `agent $i$ sees to it that the apartment door is closed', thus expressing that the proposition `the door is closed' is realized via a choice available to (and selected by) the agent $i$.

Since their inception, \stit logics have been augmented with deontic notions to additionally allow for normative reasoning in multi-agent scenarios~\cite{BelPer90}. Numerous proposals have been put forth regarding the extension of \stit logics with deontic notions. For example, in~\cite{BelPerXu01} traditional multi-agent \stit logics were extended with traditional deontic operators (e.g. ``It is obligatory that,'' ``It is permissible that,'' and ``It is forbidden that''~\cite{Aqv84}), in~\cite{Hor01,Mur04} utilitarian deontic operators were discussed, and in~\cite{BerLyo19b,BerLyo21} an assortment of non-utilitarian deontic operators and their interrelations were analyzed. We make use of an agent-specific (non-utilitarian deontic) \emph{obligation} operator\index{Obligation operator} $\Oi$, interpreted to mean that `it ought to be the case for agent $i$ that' (some proposition is realized). For example, the formula $\Oi \mathtt{door\_Open}$ expresses that `it ought to be the case for agent $i$ that the apartment door is open', meaning that in an ideal world for agent $i$ the apartment door would be open. 

\stit logics and extensions thereof continue to receive considerable attention and have found a range of applications, being applied in epistemic~\cite{Bro11}, temporal~\cite{BerLyo19b}, and legal~\cite{LorSar15} reasoning. Additionally, such logics have proven fruitful in the clarification of philosophical principles (e.g.~\cite{BerLyo21}) and in the verification of autonomous systems~\cite{SheAbb20}.

We note that the deontic \stit logics introduced in this section are closely related to those of Murakami~\cite{Mur04}. Yet, rather than supplying a utilitarian, branching-time semantics that orders states of affairs according to an agent's preferences (as is done in~\cite{Mur04}), we supply a relational semantics indicating which possible worlds are optimal for an agent (similar to what is done in~\cite{BerLyo19b,BerLyo21}). This difference necessitates a proof of strong completeness for our axiom systems relative to the new relational semantics, which we prove via a canonical model construction (cf.~\cite{BlaRijVen01}) in \sect~\ref{subsec:completeness-dsn}. First, however, we introduce our deontic \stit logics and associated preliminary concepts (\sect~\ref{subsec:logical-prelims-dsn}).

\subsection{Logical Preliminaries}\label{subsec:logical-prelims-dsn}

The deontic \stit logics\index{Deontic \stit logics} we consider employ a variety of modal operators to formalize reasoning about multi-agent choice making and agent dependent obligations. First off, such logics make use of a \emph{settledness} operator\index{Settledness operator} $\Box$, which is prefixed to formulae expressing that a proposition is `settled true' at a specific moment. As mentioned above, the language of each logic also includes a choice operator $\agbox$ expressing that `agent $i$ sees to it that' and an obligation operator $\Oi$ expressing that `it ought to be the case for agent $i$ that'. The operators $\Diamond$, $\agdia$, and $\ODi$ are the duals of $\Box$, $\agbox$, and $\Oi$, respectively.


\begin{definition}[The Language $\langdsn$] We let $\ag := \{0,1, \ldots ,n\}$ be our set of agents\index{Agent}. The multi-agent language $\langdsn$ (with $n = |\ag| \in \mathbb{N}$)\index{Language!for deontic \stit logics} is defined via the following grammar in BNF:
$$
\phi ::= p \ | \ \negnnf p \ | \ (\phi \lor \phi) \ | \ (\phi \land \phi) \ | \ \Diamond \phi \ | \ \Box \phi \ | \ \agdia \phi \ | \ \agbox \phi \ | \ \ODi\phi \ | \ \Oi \phi
$$
where $p$ is among a denumerable set of propositional variables $\prop = \{p, q, r, \ldots\}$ and $i \in \ag$. We use $\phi$, $\psi$, $\chi$, \etc (occasionally with subscripts) to denote formulae in $\langdsn$, and refer to formulae of the form $p$ and $\negnnf{p}$ (with $p \in \prop$) as \emph{literals}.
\end{definition}

As in the previous two sections, we introduce a measure on our logical formulae called the \emph{complexity}\index{Formula complexity!for deontic \stit logics}, which counts the number of binary connectives and modalities occurring in a formula. We define this measure on logical formulae from $\langdsn$ in the usual manner:

\begin{definition}[Complexity of a $\langdsn$ Formula] We define the \emph{complexity} $\fcomp{\phi}$ of a formula $\phi \in \langdsn$ inductively as follows:
\begin{itemize}

\item[$\li$] $\fcomp{p} = \fcomp{\neg p} = 0$ 

\item[$\li$] $\fcomp{\psi \land \chi} = \fcomp{\psi \lor \chi}  = max\{\fcomp{\psi},\fcomp{\chi}\} + 1$

\item[$\li$] $\fcomp{\Box \psi} = \fcomp{\Diamond \psi} = \fcomp{[i] \psi} = \fcomp{\agdia \psi} = \fcomp{\Oi\psi} =  \fcomp{\ODi \psi}  = \fcomp{\psi} + 1$

\end{itemize}
\end{definition}

Our formulae in $\langdsn$ are in negation normal form\index{Negation normal form!for deontic \stit logics}, which---as in the case of grammar logics (\sect~\ref{SEC:Grammar-Logics})---lets us simplify the sequents employed in our calculi and reduce the number of cases we need to consider when proving certain proof-theoretic results (see \sect~\ref{sec:lab-calc-dsn}). We stipulate that $p$ and $\negnnf{p}$ (for $p \in \prop$), $\land$ and $\lor$, $\Box$ and $\Diamond$, $\agbox$ and $\agdia$ (for $i \in \ag$), and $\Oi$ and $\ODi$ (for $i \in \ag$) are duals of one another. By utilizing the notion of duality\index{Duality!for deontic \stit logics}, we may define the negation $\negnnf \phi$ of a formula $\phi \in \langdsn$ to be the formula obtained by replacing each literal and logical operator with its corresponding dual. For example, given that $\phi$ is the formula $\Box (\agdia \negnnf{p} \land \Oi q)$, $\negnnf{\phi}$ would be equal to $\Diamond (\agbox p \lor \ODi \negnnf{q})$. The formal definition of negation is given below:

\begin{definition}[Negation of a $\langdsn$ Formula]\label{def:negation-dsn} We define the \emph{negation} $\negnnf{\phi}$ of a formula $\phi \in \langkm{\albet}$ inductively as follows:
\begin{multicols}{2}
\begin{itemize}

\item[$\li$] If $\phi = p$, then $\negnnf{\phi} = \neg p$

\item[$\li$] If $\phi = \neg p$, then $\negnnf{\phi} = p$

\item[$\li$] If $\phi = \psi \land \chi$, then $\negnnf{\phi} = \negnnf{\psi} \lor \negnnf{\chi}$

\item[$\li$] If $\phi = \psi \lor \chi$, then $\negnnf{\phi} = \negnnf{\psi} \land \negnnf{\chi}$

\item[$\li$] If $\phi = \Box \psi$, then $\negnnf{\phi} = \Diamond \negnnf{\psi}$

\item[$\li$] If $\phi = \Diamond \psi$, then $\negnnf{\phi} = \Box \negnnf{\psi}$

\item[$\li$] If $\phi = \agbox \psi$, then $\negnnf{\phi} = \agdia \negnnf{\psi}$

\item[$\li$] If $\phi = \agdia \psi$, then $\negnnf{\phi} = \agbox \negnnf{\psi}$

\item[$\li$] If $\phi = \Oi \psi$, then $\negnnf{\phi} = \ODi \negnnf{\psi}$

\item[$\li$] If $\phi = \ODi \psi$, then $\negnnf{\phi} = \Oi \negnnf{\psi}$

\end{itemize}
\end{multicols}
\end{definition}

Occasionally, we may also use the following abbreviations, where $p \in \prop$ is a fixed propositional variable:
$$
\top := p \lor \negnnf{p}, \quad \bot := p \land \negnnf{p}, \quad \phi \cimp \psi := \negnnf{\phi} \lor \psi, \quad \text{and} \quad \phi \cbiimp \psi := (\phi \cimp \psi) \land (\psi \cimp \phi).
$$
Due to the fact that our logic concerns instantaneous decision making, we make use of single moments in time where choices are made, as in~\cite{BalHerTro08,BerLyo21,LyoBer19}. Therefore, we forgo the use of the (more complex) traditional branching time structures often employed in atemporal \stit logics~\cite{BelPerXu01,Hor01}. Making use of this simplified semantics has the benefit that it simplifies our proof calculi (given in \sect~\ref{sec:lab-calc-dsn}) and associated automated reasoning procedures (given in \sect~\ref{sec:applicationsI}).

\begin{definition}[Frames and Models for Deontic STIT Logics]\label{def:frames-models-dsn} For each $i \in \ag$, we let $R_{[i]}(w) := \{v\in W \ | \ (w,v) \in R_{[i]}\}$. A \emph{$\dsn$-frame}\index{$\dsn$-frame} is defined to be a tuple $F = (W, \{R_{[i]} \ | \ i \in \ag\}, \{ \opt_{\Oi} \ | \ i \in \ag\} )$, where $W$ is a non-empty set of worlds $w$, $u$, $v$, \etc and the following hold:

\emph{\partcond} \quad For all $i\in \ag$, $R_{[i]} \subseteq W \times  W$ is an equivalence relation.\\
\emph{\ioacond} \quad For all $u_{1},...,u_{n} \in W$, $\bigcap_{i \in \ag} R_{[i]}(u_{i}) \neq \emptyset$.\\
\emph{\choicecond} \quad Let $k > 0$. For all $i \in \ag$ and $w_{0}, w_{1}, \cdots, w_{k} \in W$, $$\displaystyle{ \bigvee_{0 \leq m \leq k-1 \text{, } m+1 \leq j \leq k} R_{\agbox} w_{m}w_{j}}.$$\\
\emph{\donecond} \quad For all $i\in \ag$, $\opt_{\Oi}\subseteq W$.\\
\emph{\dtwocond} \quad For all $i\in \ag$, $\opt_{\Oi} \neq \emptyset$. \\
\emph{\dthreecond} \quad For all $i \in \ag$ and $w, v \in W$, if $w \in \opt_{\Oi}$ and $v \in R_{\agbox}(w)$, then $v \in \opt_{\Oi}$.\\

A \emph{$\dsn$-model}\index{$\dsn$-model} is a tuple $M = (F,V)$ where $F$ is a frame and $V : \prop \to 2^{W}$ is a valuation function\index{Valuation function!for deontic \stit logics} mapping propositional variables to subsets of $W$. 
\end{definition}

As in~\cite{BalHerTro08,LyoBer19}, $W$ represents a single moment of possible worlds where agents from $\ag$ are making decisions. Also, it should be noted that the parameters $n$ and $k$ (in $\dsn$) represent the number of agents in $\ag$ and the maximum number of choices available to an agent in $\ag$ at any given moment, respectively. Condition \partcond ensures that each relation $R_{\agbox}$ is a binary relation on $W$ and partitions $W$ into equivalence classes called \emph{choice-cells}\index{Choice-cell}, which represents a set of possible worlds that an agent can realize via their choice at a moment. Condition \ioacondns---referred to as the \emph{independence of agents} condition~\cite{BelPerXu01}\index{Independence of agents condition}---ensures that all choices made by agents are consistent, i.e., regardless of which choices are made, some state of affairs is realized. Condition \choicecond limits the number of choices available to an agent at a moment to a maximum of $k$, though we stipulate that if $k = 0$, then condition \choicecond is not enforced at all, and agents may have any number of choices available at a moment.  The condition \donecond ensures that all ideal worlds at the present moment for any agent $i$ (which are those worlds contained in $\opt_{\Oi}$) are in fact possible worlds. Condition \dtwocond ensures that at least one ideal world exists for an agent $i$ at a moment, and \dthreecond states that every ideal world extends to an ideal choice; in other words, \dtwocond and \dthreecond together ensure the existence of an ideal choice for each agent $i$ at a given moment.

Below, we define \emph{satisfaction}, \emph{global truth}, and \emph{validity} relative to these $\dsn$-frames and -models:

\begin{definition}[Satisfaction, Global Truth, Validity]\label{def:semantics-dsn} Let $M = (W, \{R_{[i]} \ | \ i \in \ag\}, \{ \opt_{\Oi} \ | \ i \in \ag\} ),V)$ be a $\dsn$-model and let $w\in W$. We define the \emph{satisfaction}\index{Formula satisfaction!for deontic \stit logics} of a formula $\phi\in \langdsn$ on $M$ at $w$ (written $M,w \Vdash \phi$) inductively as follows:
\begin{itemize}
\item[$\li$] $M, w \Vdash p$ \ifandonlyif $w \in V(p)$

\item[$\li$] $M, w \Vdash \negnnf p$ \ifandonlyif $w \not\in V(p)$

\item[$\li$] $M, w \Vdash \phi \wedge \psi$ \ifandonlyif $M, w \Vdash \phi$ and $M, w \Vdash \psi$

\item[$\li$] $M, w \Vdash \phi \vee \psi$ \ifandonlyif $M, w \Vdash \phi$ or $M, w \Vdash \psi$

\item[$\li$] $M, w \Vdash \Box \phi$ \ifandonlyif for all $u \in W$, $M, u \Vdash \phi$

\item[$\li$] $M, w \Vdash \Diamond \phi$ \ifandonlyif for some $u \in W$, $M, u \Vdash \phi$

\item[$\li$] $M, w \Vdash [i] \phi$ \ifandonlyif for all $u \in R_{[i]}(w)$, $M, u \Vdash \phi$

\item[$\li$] $M, w \Vdash \lb i \rb \phi$ \ifandonlyif for some $u \in R_{[i]}(w)$, $M, u \Vdash \phi$

\item[$\li$] $M, w \Vdash \Oi \phi$ \ifandonlyif for all $u \in \opt_{\Oi}$, $M, u \Vdash \phi$

\item[$\li$] $M, w \Vdash \ODi \phi$ \ifandonlyif for some $u \in \opt_{\Oi}$, $M, u \Vdash \phi$
\end{itemize}
A formula $\phi$ is \emph{globally true}\index{Global truth!for deontic \stit logics} on a model $M$ (written $M \Vdash \phi$) \ifandonlyif $M,u \Vdash \phi$ for all $u \in W$. A formula $\phi$ is \emph{$\dsn$-valid}\index{$\dsn$-valid} (written $\Vdash_{\dsn} \phi$) \ifandonlyif it is globally true on every $\dsn$-model. Last, we say that $\premises$ semantically implies\index{Semantically implies} $\phi$ (written $\premises\Vdash\phi$) \ifandonlyif for all models $M$ and worlds $w$ in $M$, if $M, w \Vdash \psi$ for all $\psi\in\premises$, then $M, w\Vdash \phi$.
\end{definition}


It is useful to observe that all modalities used in $\langdsn$ are normal and that the conditions \partcondns--\choicecond and \donecondns--\dthreecond fall within the class of \emph{Kracht formulae}\index{Kracht formula} (cf.~\cite[\dfn~3.58]{BlaRijVen01}). This implies that each (first-order) condition \partcondns--\choicecond and \donecondns--\dthreecond corresponds to a modal axiom in the Sahlqvist\index{Sahlqvist axiom} class that is canonical for that property (cf.~\cite[\thm~3.59]{BlaRijVen01}). We use this insight to provide sound and complete axiomatizations $\h\dsn$ for each logic $\dsn$:

\begin{definition}[The Axiomatization $\h\dsn$]\label{def:axiomatization-dsn} The axiomatization $\h\dsn$\index{$\h\dsn$} consists of all axioms and inference rules below, for all $i \in \ag$.
\begin{center}
\begin{itemize}

\item[A0] All instances of (classical) propositional tautologies




\item[A1] $\Box (\phi \rightarrow \psi) \rightarrow (\Box \phi \rightarrow \Box \psi)$

\item[A2] $\agbox{} (\phi \rightarrow \psi) \rightarrow (\agbox{} \phi \rightarrow \agbox{} \psi)$

\item[A3] $\Oi (\phi \rightarrow \psi) \rightarrow (\Oi \phi \rightarrow \Oi \psi)$




\item[A4] $\Box \phi \rightarrow [i] \phi$

\item[A5] $\Box \phi \rightarrow \Oi \phi$

\item[A6] $\Box \phi \rightarrow \phi$

\item[A7] $\Diamond \phi \rightarrow \Box \Diamond \phi$

\item[A8] $\agbox{} \phi \rightarrow \phi$

\item[A9] $\lb i \rb \phi \rightarrow \agbox{} \lb i \rb \phi$

\item[A10] $\Oi \phi \rightarrow \ODi \phi$

\item[A11] $\Diamond \Oi \phi \rightarrow \Box \Oi \phi$

\item[A12] $\Oi \phi \rightarrow \Oi \agbox \phi$

\item[A13] $\Diamond [0] \phi_{0} \land \cdots \land \Diamond [n] \phi_{n} \rightarrow \Diamond ([0] \phi_{0} \land \cdots \land [n] \phi_{n})$

\item[A14] $\Diamond \agbox \phi_{1} \land \Diamond (\negnnf{\phi}_{1} \land \agbox \phi_{2}) \land \cdots \land \Diamond (\negnnf{\phi}_{1} \land \cdots \land \negnnf{\phi}_{k-1} \land \agbox \phi_{k}) \rightarrow \phi_{1} \lor \cdots \lor \phi_{k}$

\item[R0] \AxiomC{$\phi$}
\AxiomC{$\phi \rightarrow \psi$}
\BinaryInfC{$\psi$}
\DisplayProof

\item[R1] \AxiomC{$\phi$}
\UnaryInfC{$\Box \phi$}
\DisplayProof
\end{itemize}
\end{center}
We define the \emph{logic $\dsn$}\index{$\dsn$} to be the smallest set of formulae from $\langdsn$ that is closed under substitutions of the axioms and applications of the inference rules in $\h\dsn$. We say that a formula $\phi \in \langdsn$ is a \emph{$\dsn$-theorem}\index{$\dsn$-theorem} (written $\vdash_{\dsn} \phi$) \ifandonlyif $\phi \in \dsn$.
\end{definition}

Axiom A0 ensures that each logic $\dsn$ is an extension of classical propositional logic, and R0 is the familiar rule modus ponens. Axioms A1 - A5 and rule R1 ensure that all modalities are normal. 
 Furthermore, the bridge axioms---A4 and A5---correspond to the fact that $R_{[i]} \subseteq W \times W$ in condition \partcond and that $\opt_{\Oi} \subseteq W$ in condition \textbf{(D\textsubscript{1})}, respectively. Axioms A6 - A9 encode the $\mathsf{S5}$ behavior of the $\Box$, $\Diamond$, $\agbox$, and $\agdia$ modalities. Within the deontic logic community, axiom A10 is referred to as the \emph{principle of deontic consistency}~\cite{Wri51}\index{Principle of deontic consistency}, whereas in the philosophical literature it is commonly called the \emph{ought implies logical possibility} principle~\cite{Vra07}\index{Ought implies logical possibility}. The axiom corresponds to condition \dtwocond and establishes that whatever is obligatory is possible (i.e. consistent). Axiom A11 stipulates that whatever is obligatory from the perspective of a world in a model, is obligatory from the perspective of all worlds in the model, thus making every obligation settled-true at any given moment. Axiom A12 corresponds to condition \dthreecond and secures that if a state of affairs is obligatory, then there is a choice available to the agent that realizes the obligatory state of affairs. Axiom A13 is the well-known \emph{independence of agents} axiom~\cite{BelPerXu01}\index{Independence of agents axiom}, which coincides with condition \ioacond and encodes the fact that any collection of choices made by the agents are consistent with one another. Last, axiom A14 corresponds to condition \choicecond and limits the number of choices available to each agent at a moment to a maximum of $k$, though when $k = 0$, we omit this axiom from the axiomatization, thus placing no limit on the number of choices available to each agent. That is to say, when $k > 0$, axiom A14 is included in our axiomatization for each $i \in \ag$ and semantically corresponds to each agent being limited to a maximum of $k$ choices at a moment, and when $k = 0$, no upper bound on choices is imposed.


\begin{definition}[$\dsn$-derivable]\label{def:dsn-derivation} We say that \emph{$\phi$ is $\dsn$-derivable\index{$\dsn$-derivable} from a set of premises $\premises$}, written $\premises \vdash_{\dsn} \phi$, \ifandonlyif there exist $\psi_{1}, \ldots, \psi_{m} \in \premises$ \suchthat $\psi_{1} \land \cdots \land \psi_{m} \rightarrow \phi \in \dsn$.
\end{definition}

It is quick to verify that the notion of a formula being an element of a deontic \stit logic is equivalent to being derivable from an empty set of premises:

\begin{proposition}
For any formula $\phi \in \langdsn$, $\vdash_{\dsn} \phi$ \ifandonlyif $\emptyset \vdash_{\dsn} \phi$. 
\end{proposition}

A nice feature of our logics $\dsn$ is that the notion of being derivable from a set of premises (\dfn~\ref{def:dsn-derivation}) coincides with the notion of being semantically implied by a set of premises (\dfn~\ref{def:semantics-dsn}), that is, our logics are (strongly) sound and complete (cf.~\cite[\dfn~4.10]{BlaRijVen01}). Soundness is straightforward, and is shown below. Completeness, on the other hand, requires more work, and is proven in the following section (\sect~\ref{subsec:completeness-dsn}) using a slight variation canonical model technique for normal modal logics~\cite[\cptr~4]{BlaRijVen01}.

\begin{theorem}[Soundness]\label{thm:soundness-hilbert-dsn}
Let $\phi\in\langdsn$ and $\premises \subseteq \langdsn$. If $\premises \vdash_{\dsn} \phi$, then $ \premises \Vdash \phi$.
\end{theorem}

\begin{proof} It is straightforward to show that if $\vdash_{\dsn} \phi$, then $\Vdash_{\dsn} \phi$ by showing that each axiom is $\dsn$-valid and that each inference rule of $\h\dsn$ preserves $\dsn$-validity. We use this fact to prove our claim and suppose that $\premises \vdash_{\dsn} \phi$. Then, by \dfn~\ref{def:dsn-derivation}, we know that there exist $\psi_{1}, \ldots, \psi_{m} \in \premises$ \suchthat $\psi_{1} \land \cdots \land \psi_{m} \rightarrow \phi \in \dsn$, i.e. $\vdash_{\dsn} \psi_{1} \land \cdots \land \psi_{m} \rightarrow \phi$. To prove the desired conclusion, let $M$ be an arbitrary $\dsn$-model with $w \in W$ of $M$, and assume that for all $\chi \in \premises$, $M,w \Vdash \chi$. It follows that $M,w \Vdash \psi_{1} \land \cdots \land \psi_{m}$, implying that $M,w \Vdash \phi$ (since $\vdash_{\dsn} \psi_{1} \land \cdots \land \psi_{m} \rightarrow \phi$, which implies that $\Vdash_{\dsn} \psi_{1} \land \cdots \land \psi_{m} \rightarrow \phi$). Hence, $ \premises \Vdash \phi$.
\end{proof}


\subsection{Completeness via Canonical Models}\label{subsec:completeness-dsn}

In this section, we prove the completeness of each logic $\dsn$ by adapting and adjusting the canonical model construction for normal modal logics from~\cite{BlaRijVen01} to our setting. We use the following notation throughout the course of the section: if $\premises$ is a given set of formulae, then $\bigwedge \premises$ represents a conjunction of all formulae (assuming $\premises$ is finite) and $\heartsuit \premises$ is taken to be the set of all formulae in $\premises$ prefixed with a $\heartsuit$ modality (for $\heartsuit \in \{\Box, \Diamond, \agbox, \agdia, \Oi, \ODi\}$). 



As is typical when proving completeness via canonical models for modal logics, we first define a notion of a \emph{maximally consistent set} of formulae, and then show that such sets possess advantageous properties~\cite[\cptr~4]{BlaRijVen01}. We define such sets below and detail their properties in \lem~\ref{lem:MCS-prop}. Note that such sets will be used as worlds in our canonical models, and, since such sets are inherently founded upon syntactic notions (e.g. being $\dsn$-derivable~\dfn~\ref{def:dsn-derivation}), they will serve as a point of connection between our syntactically defined logics $\dsn$ and the semantics. Ultimately, this connection will be leveraged to prove completeness.

\begin{definition}[$\dsn$-CS, $\dsn$-MCS]\label{def:dsn-cs-mcs} A set $\premises \subset \langdsn$ is a \emph{$\dsn$ consistent set ($\dsn$-CS)}\index{$\dsn$-consistent set} \ifandonlyif $\premises \not\vdash_{\dsn} \bot$. We call a set $\premises \subset \langdsn$ a \emph{$\dsn$-maximally consistent set ($\dsn$-MCS)}\index{$\dsn$-maximally consistent set} \ifandonlyif $\premises$ is a $\dsn$-CS and for any set $\premises'$ such that $\premises \subset \premises'$, $\premises' \vdash_{\dsn} \bot$.
\end{definition}

The following is similar to~\cite[\prp~4.16]{BlaRijVen01}.

\begin{lemma}\label{lem:MCS-prop} Let $\premises$ be a $\dsn$-MCS. Then, the following hold:
\begin{enumerate}

\item[(i)] $\premises \vdash_{\dsn} \phi$ \ifandonlyif $\phi \in \premises$;

\item[(ii)] $\phi \in \premises$ \ifandonlyif $\negnnf{\phi} \not\in \premises$.

\end{enumerate}
\end{lemma}

\begin{proof} (i) For the forward direction, assume for a contradiction that $\premises \vdash_{\dsn} \phi$ and $\phi \not\in \premises$. By the latter assumption, $\premises \cup \{\phi\} \vdash_{\dsn} \bot$ by \dfn~\ref{def:dsn-cs-mcs}, from which it follows that $\premises \vdash_{\dsn} \negnnf{\phi}$, contradicting our other assumption. The backward direction is trivial.

(ii) For the forward direction, assume that $\phi \in \premises$ and $\negnnf{\phi} \in \premises$. Then, $\premises \vdash_{\dsn} \bot$, contradicting the fact that $\premises$ is consistent. For the backward direction, assume that $\negnnf{\phi} \not\in \premises$ and $\phi \not\in \premises$. By the first assumption, $\premises \cup \{\negnnf{\phi}\} \vdash_{\dsn} \bot$, implying that $\premises \vdash_{\dsn} \phi$, which by part (i), means that $\phi \in \premises$, contradicting our second assumption and proving the desired claim.
\end{proof}

Using standard techniques, it is straightforward to show that every $\dsn$-CS can be extended to a $\dsn$-MCS. This fact will be frequently used.

\begin{lemma}[Lindenbaum's Lemma]\label{Lindenbaum}
Every $\dsn$-CS can be extended to a $\dsn$-MCS.
\end{lemma}

\begin{proof}
Similar to~\cite[\lem~4.17]{BlaRijVen01}.
\end{proof}

By using $\dsn$-MCS's as building blocks, we define our canonical models \emph{relative to a $\dsn$-MCS $\premises$}. The fact that our canonical models are defined \emph{relatively}, is what contrasts our approach with the standard approach of building canonical models for normal modal logics~\cite[\cptr~4]{BlaRijVen01}. Typically, the models for normal modal logics integrate an explicit relation for each modality occurring in the language (e.g. $R_{\agbox}$ is used to interpret $\agbox$ and $\ideal$ is used to interpret $\Oi$ in $\langdsn$). However, as explained in the previous section, we employ a simplified semantics for our $\dsn$ logics, which forgoes the introduction of a relation for the $\Box$ and $\Diamond$ modalities. Instead, these modalities are interpreted relative to the set of worlds in a $\dsn$-model, which has two consequences. First, this interpretation of $\Box$ and $\Diamond$ causes the modalities to exhibit $\mathsf{S5}$ behavior, though, such behavior is not only acceptable, but desirable, as these operators are usually interpreted as $\mathsf{S5}$-type modalities in \stit logics~\cite{BelPerXu01,BalHerTro08,BerLyo19b,Mur04}. Second, since the interpretation of $\Box$ and $\Diamond$ depends on the set of worlds, and not on an explicitly associated relation, the standard canonical model definition---which defines a relation over the set of maximally consistent sets, used to interpret the corresponding modalities---is rendered insufficient. Therefore, we slightly adjust the canonical model definition given in~\cite[\cptr~4]{BlaRijVen01} to account for our simplified semantics.


\begin{definition}[Canonical Model]\label{def:canonical-model} Let $\premises$ be a $\dsn$-MCS. We define the \emph{canonical model (relative to $\premises$)}\index{Canonical model} to be the tuple $\mcan = (\wcan, \{\rcan \ | \ i \in \ag\}, \{\ican \ | \ i \in \ag \}, \vcan)$ such that:

\begin{itemize}

\item[$\li$] $\wcan := \{w \ | \ \text{$w$ is a $\dsn$-MCS and for all $\Box \phi \in \langdsn$, if $\Box \phi \in \premises$, then $\phi \in w$.}\}$;

\item[$\li$] For all $w, u \in \wcan$, $u \in \rcan(w)$ \ifandonlyif for all $\agbox \phi \in w$, $\phi \in u$;

\item[$\li$] For all $w, u \in \wcan$, $w \in \ican$ \ifandonlyif for all $\Oi \phi \in u$, $\phi \in w$;

\item[$\li$] $\vcan (p) = \{w \in \wcan \ | \ p \in w\}$.

\end{itemize}

\end{definition}

We now show that our canonical models possess favorable properties, which will be used in establishing both the Containment Lemma (\lem~\ref{lem:canonical-mod-is-DSn-mod}) and Truth Lemma (\lem~\ref{lem:truth-lemma})---both of which are crucial in verifying completeness (\thm~\ref{thm:completeness-hilbert}).

\begin{lemma}\label{lem:relations-diamonds} Let $w$ and $u$ be $\dsn$-MCS's. (i) $w \in \wcan$ \ifandonlyif for all $\phi \in \langdsn$, if $\phi \in w$, then $\Diamond \phi \in \premises$, (ii) For all $u,w \in \wcan$, $u \in \rcan(w)$ \ifandonlyif $\phi \in \langdsn$, if $\phi \in u$, then $\agdia \phi \in w$, and (iii) For all $w, u \in \wcan$, $w \in \ican$ \ifandonlyif $\phi \in \langdsn$, if $\phi \in w$, then $\ODi \phi \in u$.
\end{lemma}

\begin{proof} Let $w$ and $u$ be $\dsn$-MCS's. We prove claim (i); claims (ii) and (iii) are similar. (i) For the forward direction, suppose that $w \in \wcan$ and $\phi \in w$. Since $w$ is a $\dsn$-MCS by \dfn~\ref{def:canonical-model}, our supposition implies that $\Box \negnnf{\phi} \not\in \premises$ by \dfn~\ref{def:canonical-model} as well. It follows that $\negnnf{\Box \negnnf{\phi}} \in \premises$, yielding that $\Diamond \phi \in \premises$. For the other direction, we assume that for all $\phi \in \langdsn$, if $\phi \in w$, then $\Diamond \phi \in \premises$, and further suppose that $\Box \psi \in \premises$ for an arbitrary $\psi$. It follows that $\negnnf{\Box \psi} \not\in \premises$, implying that $\Diamond \negnnf{\psi} \not\in \premises$. Therefore, $\negnnf{\psi} \not\in w$, meaning that $\psi \in w$, showing that $w \in \wcan$.
\end{proof}

\begin{lemma}[Existence Lemma\index{Existence lemma}]\label{Existence_Lemma} (i) For any world $w \in \wcan$, if $\Diamond \phi \in w$, then there exists a world $u \in \wcan$ such that and $\phi \in u$. (ii) For any world $w \in \wcan$, if $\agdia \phi \in w$, then there exists a world $u \in \wcan$ such that $u \in \rcan(w)$ and $\phi \in u$. (iii) For any world $w \in \wcan$, if $\ODi \phi \in w$, then there exists a world $u \in \wcan$ such that $u \in \ican$ and $\phi \in u$.
\end{lemma}

\begin{proof}
Similar to \cite[\lem~4.20]{BlaRijVen01}.
\end{proof}

\begin{lemma}\label{lem:S5-Properties-of-W}
For all $w, u \in \wcan$, $\Diamond \phi \in w$ \ifandonlyif $\Diamond \phi \in u$. 
\end{lemma}

\begin{proof} For the forward direction, suppose that $\Diamond \phi \in w$. Then, by \lem~\ref{lem:relations-diamonds}-(i), we have $\Diamond \Diamond \phi \in \premises$, which implies that $\Box \Diamond \phi \in \premises$ since $\premises$ is a $\dsn$-MCS. We therefore have that $\Diamond \phi \in u$. For the backward direction, the proof is almost identical.
\end{proof}

The following lemma ensures that any given canonical model is \emph{contained} in our class of $\dsn$-models. 

\begin{lemma}[Containment Lemma\index{Containment lemma}]\label{lem:canonical-mod-is-DSn-mod}
Let $\premises$ be a $\dsn$-MCS. The canonical model $\mcan$ is a $\dsn$-model.
\end{lemma}

\begin{proof} Let $\premises$ be a $\dsn$-MCS. Observe that $\{\phi \ | \ \Box \phi \in \premises\} \subseteq \premises$ due to the fact that $\premises$ is a $\dsn$-MCS and $\Box \phi \rightarrow \phi$ is an axiom of $\h\dsn$. This fact and \dfn~\ref{def:canonical-model} ensures that $\premises \in \wcan$, thus proving $\wcan$ non-empty. We now show that $\mcan$ satisfies properties \partcondns--\choicecond and \donecondns--\dthreecondns. We confirm below that each property holds:

\partcond The fact that $\rcan \subseteq \wcan \times \wcan$ follows from the definition of $\rcan$. To prove that $\rcan$ is an equivalence relation, we must show that (i) $\rcan$ is reflexive, and (ii) $\rcan$ is Euclidean. For (i), let $w$ be an arbitrary world in $\wcan$, and assume that $\agbox \phi \in w$. Since $w$ is a $\dsn$-MCS and $\agbox \phi \rightarrow \phi$ is an axiom, we know that $w \vdash_{\dsn} \phi$, which implies that $\phi \in w$ by \lem~\ref{lem:MCS-prop}-(i). By the definition of $\rcan$, this implies that $w \in \rcan(w)$. For (ii), let $w$, $u$, and $v$ be arbitrary worlds in $\wcan$, and assume that $u, v \in \rcan(w)$. We aim to show that $v \in \rcan(u)$. Suppose that $\phi \in v$. It follows that $\agdia \phi \in w$ by \lem~\ref{lem:relations-diamonds}-(ii), and since $\agdia \phi \rightarrow \agbox \agdia \phi$ is an axiom, we know that $w \vdash_{\dsn} \agbox \agdia \phi$, meaning that $\agbox \agdia \phi \in w$ by \lem~\ref{lem:MCS-prop}-(i). Due to the fact that $u \in \rcan(w)$, and by the definition of $\rcan$ (\dfn~\ref{def:canonical-model}), we have $\agdia \phi \in u$. Thus, by \lem~\ref{lem:relations-diamonds}-(ii), it follows that $v \in \rcan(u)$.

\ioacond Let $u_{0}, \ldots, u_{n} \in \wcan$. We aim to show that there exists a $v \in \wcan$ such that $v \in \bigcap_{i \in \ag} \rcan(u_{i})$. Let $v' := \bigcup_{i\in \ag} \{\phi \ | \ \agbox \phi \in u_{i}\} \cup \{\psi \ | \ \Box \psi \in \premises\}$, and suppose that $v'$ is inconsistent to derive a contradiction. Due to the inconsistency of $v'$, we know that there exist $\phi_{1}, \ldots, \phi_{k} \in \bigcup_{i\in \ag} \{\phi \ | \ \agbox \phi \in u_{i}\}$ and $\psi_{1}, \ldots, \psi_{m} \in \{\psi \ | \ \Box \psi \in \premises\}$ such that $\vdash_{\dsn} \phi_{1} \land \cdots \land \phi_{k} \rightarrow \negnnf{\psi}_{1} \lor \cdots \lor \negnnf{\psi}_{m}$. Define $\Phi_{i} := \{\phi_{l} \ | \ \agbox \phi_{l} \in u_{i}\} \cap \{\phi_{1}, \ldots, \phi_{k}\}$, and observe that for each $i \in \ag$, $\bigwedge \agbox \Phi_{i} \in u_{i}$, implying that $\agbox \bigwedge \Phi_{i} \in u_{i}$, by modal reasoning and \lem~\ref{lem:MCS-prop}. Since $\agbox \bigwedge \Phi_{i} \rightarrow \Diamond \agbox \bigwedge \Phi_{i}$ is a theorem of $\h\dsn$, it follows that $\Diamond \agbox \bigwedge \Phi_{i} \in u_{i}$ for all $i \in \ag$. Consequently, by \lem~\ref{lem:S5-Properties-of-W}, for any $i \in \ag$, we have $\Diamond [0] \bigwedge \Phi_{0}, \ldots, \Diamond [n] \bigwedge \Phi_{n} \in u_{i}$, implying  that $\Diamond [0] \bigwedge \Phi_{0} \land \cdots \land \Diamond [n] \bigwedge \Phi_{n} \in u_{i}$ since $u_{i}$ is a $\dsn$-MCS. By the independence of agents axiom, $\Diamond ([0] \bigwedge \Phi_{0} \land \cdots \land [n] \bigwedge \Phi_{n}) \in u_{i}$ for any $i \in \ag$. By \lem~\ref{Existence_Lemma}-(i), it follows that there exists a world $u' \in \wcan$ such that $[0] \bigwedge \Phi_{0} \land \cdots \land [n] \bigwedge \Phi_{n} \in u'$, which further implies that $\bigwedge \Phi_{1} \land \cdots \land \bigwedge \Phi_{n} \in u'$ by modal reasoning and the fact that $\agbox \bigwedge \Phi_{i} \rightarrow \bigwedge \Phi_{i}$ is an axiom instance for each $i \in \ag$. Hence, $\bigwedge \Phi_{1} \land \cdots \land \bigwedge \Phi_{n} \in u'$, so by the definition of each $\Phi_{i}$ and the fact that $\vdash_{\dsn} \phi_{1} \land \cdots \land \phi_{k} \rightarrow \negnnf{\psi}_{1} \lor \cdots \lor \negnnf{\psi}_{m}$, we have that $u' \vdash_{\dsn} \negnnf{\psi}_{1} \lor \cdots \lor \negnnf{\psi}_{m}$. Since $u' \in \wcan$, we know that $\{\psi \ | \ \Box \psi \in \premises\} \subseteq u'$, which implies that $u' \vdash_{\dsn} \psi_{1} \land \cdots \land \psi_{m}$, giving a contradiction. It follows that $v'$ is a $\dsn$-CS. By \lem~\ref{Lindenbaum}, $v'$ can be extended to a $\dsn$-MCS $v$ such that $\{\psi \ | \ \Box \psi \in \premises\} \subseteq v$. Therefore, $v \in \wcan$ and for all $[i] \phi \in u_{i}$, $\phi \in v$, so, by \dfn~\ref{def:canonical-model}, $v \in \rcan(u_{i})$ for all $i \in \ag$, which implies the desired result.

\choicecond For a contradiction, suppose that $\wcan$ contains $k' > k$ choice-cells, this is, there exist worlds $w_{0}, \ldots w_{k'} \in \wcan$ such that $w_{j} \not\in \rcan (w_{m})$ for $0 \leq m \leq k'-1$ and $m+1 \leq j \leq k'$. Let us first consider the world $w_{0}$. By our assumption, we have that $w_{j} \not\in \rcan (w_{0})$ for $1 \leq j \leq k'$. This, in conjunction with \dfn~\ref{def:canonical-model}, implies that there exists formulae $\agbox\phi_{0,1}, \ldots, \agbox\phi_{0,k'} \in w_{0}$ such that $\phi_{0,1} \not\in w_{1}$, $\ldots$, $\phi_{0,k'} \not\in w_{k'}$. Let us define $\bigwedge \Phi_{0,1 \leq j \leq k'} := \phi_{0,1} \land \cdots \land \phi_{0,k'}$. By \dfn~\ref{lem:MCS-prop}, the fact that $w_{0}$ is a $\dsn$-MCS, by modal reasoning, and by the fact that $\phi_{0,j} \not\in w_{j}$ for $1 \leq j \leq k'$ (entailing $\negnnf{\phi_{0,j}} \in w_{j}$ for $1 \leq j \leq k'$), we know that:
$$
\agbox(\bigwedge \Phi_{0,1 \leq j \leq k'}) \in w_{0}
\text{ and }
\negnnf{\bigwedge \Phi_{0,1 \leq j \leq k'}} \in w_{j} \text{ for } 1 \leq j \leq k'.
$$ 
We continue in the above fashion for the worlds $w_{m}$ with $1 \leq m \leq k$. By our assumptions, $\agbox \phi_{m,j} \in w_{m}$ and $\phi_{m,j} \not\in w_{j}$ for $m + 1 \leq j \leq k'$. Furthermore, we let $\bigwedge \Phi_{m,m+1 \leq j \leq k'} := \phi_{m,m+1} \land \cdots \land \phi_{m,k'}$. Similar to the $w_{0}$ case above, \dfn~\ref{lem:MCS-prop}, the fact that $w_{m}$ is a $\dsn$-MCS, by modal reasoning, and by the fact that  $\phi_{m,j} \not\in w_{j}$ for $m + 1 \leq j \leq k'$ (meaning that $\negnnf{\phi_{m,j}} \in w_{j}$ for $m + 1 \leq j \leq k'$), we have for each $m$ (with $1 \leq m \leq k$) that:
$$
\agbox(\bigwedge \Phi_{m,m+1 \leq j \leq k'}) \in w_{m}
\text{ and }
\negnnf{\bigwedge \Phi_{m,m+1 \leq j \leq k'}} \in w_{j} \text{ for } 1 \leq j \leq k'.
$$
Without loss of generality, we now consider the world $w_{k'}$. By \lem~\ref{lem:S5-Properties-of-W} and the above facts, the following large conjunction is an element of $w_{k'}$:
\begin{eqnarray*}
& & \Diamond \agbox(\bigwedge \Phi_{0,1 \leq j \leq k'}) \land\\
& & \Diamond\Big(\negnnf{\bigwedge \Phi_{0,1 \leq j \leq k'}} \land \agbox\Big(\bigwedge \Phi_{1,2 \leq j \leq k'}) \Big) \land \\
& & \Diamond\Big(\negnnf{\bigwedge \Phi_{0,1 \leq j \leq k'}} \land \negnnf{\bigwedge \Phi_{1,2 \leq j \leq k'}} \land \agbox\Big(\bigwedge \Phi_{2,3 \leq j \leq k'}) \Big) \land \\
& & \vdots\\
& & \Diamond\Big(\negnnf{\bigwedge \Phi_{0,1 \leq j \leq k'}} \land \cdots \land \negnnf{\bigwedge \Phi_{k-1,k \leq j \leq k'}} \land \agbox\big(\bigwedge \Phi_{k,k \leq j \leq k'}) \Big)
\end{eqnarray*}
Therefore, by axiom A14, we have the following:
$$
\bigwedge \Phi_{0,1 \leq j \leq k'} \lor \cdots \lor \bigwedge \Phi_{k,k \leq j \leq k'} \in w_{k'}
$$
Observe that $\phi_{m,k'} \not\in w_{k'}$ for $0 \leq m \leq k$, implying that $\negnnf{\phi_{m,k'}} \in w_{k'}$ for $0 \leq m \leq k$. This fact, in conjunction with the definition of $\bigwedge \Phi_{m, m+1 \leq j \leq k'}$ for $0 \leq m \leq k$, by \lem~\ref{lem:MCS-prop}, and by modal reasoning, it follows that $\negnnf{\bigwedge \Phi_{m, m+1 \leq j \leq k'}} \in w_{k'}$ for $0 \leq m \leq k$. Hence, by modal reasoning,
$$
\negnnf{\bigwedge \Phi_{0,1 \leq j \leq k'}} \land \cdots \land \negnnf{\bigwedge \Phi_{k,k \leq j \leq k'}} \in w_{k'},
$$
which gives a contradiction. Therefore, it must be the case that the number of choice-cells in $\mcan$ is a maximum of $k$.

\donecond Follows from the definition of $\ican$ (\dfn~\ref{def:canonical-model}).

\dtwocond Fix an $i \in \ag$. We want to show that there exists a $v \in\wcan$ such that $v \in \ican$. Let $v' := \bigcup_{w \in \wcan} \{\psi \ | \ \Oi \psi \in w\} \cup \{\phi \ | \ \Box \phi \in \premises\}$ and assume that $v'$ is inconsistent to derive a contradiction. Since $v'$ is inconsistent, we know that $\vdash_{\dsn} \chi_{1} \land \cdots \land \chi_{m} \rightarrow \bot$. Define $\Phi := \{\chi_{1}, \ldots, \chi_{m}\} \cap \{\phi \ | \ \Box \phi \in \premises\}$, and $\Psi_{w} := \{\chi_{1}, \ldots, \chi_{m}\} \cap \{\psi \ | \ \Oi \psi \in w\}$. It follows that $\vdash_{\dsn} \bigwedge \Phi \land \bigwedge \Psi_{w_{1}} \land \cdots \land \bigwedge \Psi_{w_{k}} \rightarrow \bot$ for some $w_{1}, \ldots, w_{k} \in \wcan$. Using modal reasoning, we obtain the following:
\begin{eqnarray}
& \vdash_{\dsn} & \Oi \bigwedge \Phi \land \Oi \bigwedge \Psi_{w_{1}} \land \cdots \land \Oi \bigwedge \Psi_{w_{k}} \rightarrow \Oi \bot \\
& \vdash_{\dsn} & \Box \bigwedge \Phi \land \Oi \bigwedge \Psi_{w_{1}} \land \cdots \land \Oi \bigwedge \Psi_{w_{k}} \rightarrow \Oi \bot \\
& \vdash_{\dsn} & \Box \bigwedge \Phi \land \Oi \bigwedge \Psi_{w_{1}} \land \cdots \land \Oi \bigwedge \Psi_{w_{k}} \rightarrow \ODi \bot \\
& \vdash_{\dsn} & \Box \bigwedge \Phi \land \Oi \bigwedge \Psi_{w_{1}} \land \cdots \land \Oi \bigwedge \Psi_{w_{k}} \rightarrow \bot \\
& \vdash_{\dsn} & \Box \Box \bigwedge \Phi \land \Box \Oi \bigwedge \Psi_{w_{1}} \land \cdots \land \Box \Oi \bigwedge \Psi_{w_{k}} \rightarrow \Box \bot \\
& \vdash_{\dsn} & \Box \bigwedge \Phi \land \Box \Oi \bigwedge \Psi_{w_{1}} \land \cdots \land \Box \Oi \bigwedge \Psi_{w_{k}} \rightarrow \bot \\
& \vdash_{\dsn} & \bigwedge \Box \Phi \land \bigwedge \Box \Oi \Psi_{w_{1}} \land \cdots \land \bigwedge \Box \Oi \Psi_{w_{k}} \rightarrow \bot \\
& \vdash_{\dsn} & \bigwedge \Box \Phi \land \bigwedge \Diamond \Oi \Psi_{w_{1}} \land \cdots \land \bigwedge \Diamond \Oi \Psi_{w_{k}} \rightarrow \bot
\end{eqnarray}
We explain each of the step in the above deduction: 2.2 follows from 2.1 by the bridge axiom A5 (i.e. $\Box \phi \rightarrow \Oi \phi$), 2.3 follows from 2.2 by the deontic consistency axiom A10 (i.e. $\Oi \phi \rightarrow \ODi \phi$), 2.4 and 2.5 follow from 2.3 by modal reasoning, 2.6 follows from 2.5 by modal reasoning and the fact that $\Box$ is an $\mathsf{S5}$ modality (by axioms A6 and A7), 2.7 follows from 2.6 by modal reasoning, and last, 2.8 follows from 2.7 by axiom A11 (i.e. $\Diamond \Oi \phi \rightarrow \Box \Oi \phi$). Now, observe that $\Box \Phi, \Diamond \Oi \Psi_{w_{1}}, \ldots, \Diamond \Oi \Psi_{w_{k}} \subseteq \premises$ (by the definition of $v'$ and \lem~\ref{lem:relations-diamonds}-(i)), which implies that $\premises \vdash_{\dsn} \bot$. Since $\premises$ is a $\dsn$-MCS, this gives a contradiction, and shows that $v'$ is consistent. By \lem~\ref{Lindenbaum}, $v'$ may be extended to a $\dsn$-MCS $v$. Since $\{\phi \ | \ \Box \phi \in \premises\} \subseteq v$, we know that $v \in \wcan$, and further, because $\{\psi \ | \ \Oi \psi \in w\} \subseteq v$ for each $w \in \wcan$, it follows that $v \in \ican$.

\dthreecond Let $w,u \in \wcan$ and $i \in \ag$. Assume that $w \in \ican$ and $u \in \rcan(w)$. We aim to show that $u \in \ican$. To prove this, we let $v$ be an arbitrary world in $\wcan$ and suppose that $\Oi \phi \in v$, with the aim of showing that $\phi \in u$. Since $v$ is a $\dsn$-MCS and $\Oi \phi \rightarrow \Oi \agbox \phi$ is an axiom instance, it follows that $\Oi \agbox \phi \in v$. By \dfn~\ref{def:canonical-model} and the assumption that $w \in \ican$, we have that for all $v' \in \wcan$ and each $\Oi \psi \in v'$, $\psi \in w$. This fact, in conjunction with our supposition above, implies that $\agbox \phi \in w$. By \dfn~\ref{def:canonical-model} and the assumption that $u \in \rcan(w)$, we have that $\phi \in u$, which entails the desired result.
\end{proof}

The following is similar to~\cite[\lem~4.21]{BlaRijVen01}.

\begin{lemma}[Truth Lemma\index{Truth lemma}]\label{lem:truth-lemma} For any $\phi \in \langdsn$, $\mcan , w \Vdash \phi$ \ifandonlyif $\phi \in w$.
\end{lemma}

\begin{proof} We prove the result by induction on the complexity of $\phi$. The base case trivially follows from the definition of $\vcan$ (\dfn~\ref{def:canonical-model}), so we focus on the inductive step. We show the $\lor$, $\Box$, $\agbox$, and $\Oi$ cases since the dual cases are similar.

\textit{Inductive step.} We consider each connective in turn.

\textbf{($\lor$)} $\mcan, w \Vdash \psi \lor \chi$ \ifandonlyif $\mcan, w \Vdash \psi$ or $\mcan, w \Vdash \chi$ (by \dfn~\ref{def:semantics-dsn}) \ifandonlyif $\psi \in w$ or $\chi \in w$ (by IH) \ifandonlyif $\psi \lor \chi \in w$ (by the fact that $w$ is a $\dsn$-MCS).


\textbf{($\Box$)} For the forward direction, suppose that $\mcan, w \Vdash \Box \psi$. It follows that for all $u \in \wcan$, $\mcan, u \Vdash \psi$ (by \dfn~\ref{def:semantics-dsn}), which implies that for all $u \in \wcan$, $\psi \in u$ (by IH). By \lem~\ref{Existence_Lemma}-(i), $\Box \psi \in w$. For the backward direction, assume that $\Box \psi \in w$ and let $u \in \wcan$. By \lem~\ref{lem:relations-diamonds}-(i), $\psi \in u$, which implies that $\mcan, u \Vdash \psi$ by IH. Since $u$ was arbitrary, it follows that $\mcan, w \Vdash \Box \psi$ by \dfn~\ref{def:semantics-dsn}.

\textbf{($\agbox$)} For forward direction, assume that $\mcan, w \Vdash \agbox \psi$. By \dfn~\ref{def:semantics-dsn} and IH, for all $u \in \wcan$, if $u \in R_{[i]}(w)$, then $\psi \in u$. By \lem~\ref{Existence_Lemma}-(ii), $\agbox \psi \in w$. For the backward direction, suppose that $\agbox \psi \in w$ and let $u \in \wcan$ such that $u \in R_{[i]}(w)$. By \dfn~\ref{def:canonical-model}, $\psi \in u$, which implies that $\mcan, u \Vdash \psi$ by IH. We may conclude that $\mcan, w \Vdash \agbox \psi$.

\textbf{($\Oi$)} For the forward direction, assume that $\mcan, w \Vdash \Oi \psi$. By \dfn~\ref{def:semantics-dsn} and IH, for all $u \in \ican$, $\psi \in u$. By \lem~\ref{Existence_Lemma}-(iii), $\Oi \psi \in w$. For the backward direction, suppose that $\Oi \psi \in w$ and let $u \in \ican$. By \dfn~\ref{def:canonical-model}, $\psi \in u$, implying that $\mcan, u \Vdash \psi$ by IH. Therefore, $\mcan, w \Vdash \Oi \psi$ since $u$ was an arbitrary world in $\wcan$.
\end{proof}

\begin{theorem}[Completeness]\label{thm:completeness-hilbert}
Let $\phi\in\langdsn$ and $\premises \subseteq \langdsn$. If $ \premises \Vdash \phi$, then $\premises \vdash_{\dsn} \phi$.
\end{theorem}

\begin{proof} We prove the claim by contraposition. Suppose that $\premises \not\vdash_{\dsn} \phi$. Then, it follows that $\premises \cup \{\negnnf{\phi}\}$ is consistent. By \lem~\ref{Lindenbaum}, we may extend this set to a $\dsn$-MCS $\premises'$. By \lem~\ref{lem:truth-lemma} and the fact that $\negnnf{\phi} \in \premises'$, we have that $M^{\premises'}, \premises' \Vdash \psi$ for all $\psi \in \premises$, and $M^{\premises'}, \premises' \Vdash \negnnf{\phi}$. Hence, $ \premises \not\Vdash \phi$ by \dfn~\ref{def:semantics-dsn}.
\end{proof}

\chapter{Labelled Systems for Modal and Constructive Logics}
\label{CPTR:Labelled} 

In this chapter, we introduce labelled sequent systems for context-free grammar logics with converse (\sect~\ref{sec:lab-calc-kms}), for first-order intuitionistic logics (\sect~\ref{sec:lab-calc-intFO}), and for deontic \stit logics (\sect~\ref{sec:lab-calc-dsn}). In each of these sections, we define the sequents and structures used in each class of labelled calculi, prove that such systems possess desirable properties (e.g. hp-invertibility of rules and syntactic cut-elimination), and confirm soundness and completeness relative to the classes of logics we are considering. In the first section (\sect~\ref{SECT:prelims-related-work-labelled}), we discuss preliminary concepts of labelled sequent systems, as well as touch on the history of such systems and significant results obtained in the labelled paradigm.

\section{Preliminaries and Related Work}\label{SECT:prelims-related-work-labelled}

Labelled sequent systems extend the structure of Gentzen-style systems by explicitly incorporating labels and semantic elements into the syntax of sequents. This idea---of integrating labels directly into the syntax of sequents---stretches back to the work of Kanger~\cite{Kan57}, who introduced sequent systems for modal logics based on \emph{spotted formulae}. Since then, numerous labelled sequent systems have been provided for large classes of modal and constructive logics~\cite{CasSma02,DycNeg12,Gab96,KusOka03,Min97,Neg05,Vig00}. 

The paradigm of labelled sequents has proven itself 
successful in producing modular calculi that uniformly cover extensive classes of logics, even when other proof-theoretic formalisms fail to do so, that is, through the addition or deletion of structural rules, a labelled sequent system for one logic may be converted into a labelled sequent system for another logic within a given class. In addition to uniform coverage and modularity, general results exist (\cite{DycNeg12,Neg05,Sim94,Vig00}) showing that labelled sequent systems share the same fundamental properties; e.g. height-preserving admissibility of contractions and syntactic cut-elimination. Perhaps one of the most beneficial characteristics however, is the ease with which labelled calculi are constructed. In essence, the semantic clauses and frame properties within a logic's relational semantics are transformed into inference rules, yielding a sound and complete calculus for the associated logic, which precisely demonstrates the semantic reasoning utilized in deriving a theorem. As was explained in the introduction (\cptr~\ref{CPTR:Intro}), this characteristic of the labelled paradigm, when composed with the process of refinement, effectively supplies a method for transforming the semantics of a logic into a calculus of economical structure that is also in possession of desirable properties and is well-suited for applications. After briefly introducing Vigan\`o's labelled sequent formalism below (which is predated by Simpson's labelled sequent formalism for modal intuitionistic logics~\cite{Sim94}), we will illustrate how semantic information is transformed into inference rules by considering the semantic clause for the standard $\Box$ modality.

In~\cite{Vig00}, Vigan{\`o} demonstrated the utility of the labelled formalism and provided labelled sequent calculi for a substantially broad class of modal, constructive, and relevance logics. The logics Vigan{\`o} considers are founded upon relational semantics, and allow for the accessibility relation(s) to satisfy any number of Horn formulae\index{Horn formula} of the form:
$$ 
\uc \Big( R_{i}t_{1}^{1} \ldots t_{n}^{1} \land \cdots \land R_{i}t_{1}^{m} \ldots t_{n}^{m} \cimp R_{i}t_{1}^{0} \ldots t_{n}^{0} \Big)
$$
where $R_{i}$ is an $n$-ary relational symbol and each term $t_{j}^{k}$ is built from the variables in $\vec{x}$ along with function symbols~\cite[p.~61]{Vig00}. Vigan{\`o}'s systems employ two types of labelled sequents: (i) labelled sequents of the form $\rel, \Gamma \sar \Delta$, where $\rel$ is a multiset of \emph{relational atoms} of the form $R_{i}t_{1} \ldots t_{n}$, and $\Gamma$ and $\Delta$ are multisets of \emph{labelled formulae} of the form $w : \phi$, and (ii) labelled sequents of the form $\rel \sar R_{i}t_{1} \ldots t_{n}$, where $\rel$ is a multiset of relational atoms. The relational atoms encode information about accessibility relations, and labelled formulae encode the satisfaction relation (i.e. $w : \phi$ can be interpreted as $M, w \Vdash \phi$). The separation between the two types of sequents is based on work by Gabbay~\cite{Gab96}, whereby labelled proof systems are divided into two parts: a \emph{base calculus} that corresponds to a base logic (with inference rules operating on the sequents of the form $\rel, \Gamma \sar \Delta$), and a \emph{relational algebra} that is used to reason about relational atoms, i.e. the properties of accessibility relations (with inference rules operating on the sequents of the form $\rel \sar R_{i}t_{1} \ldots t_{n}$)~\cite[p.~7]{Vig00}. In essence, the base calculus is a proof system consisting of rules for deriving logical formulae, and the relational algebra is a proof system consisting of rules that allow the derivation of relational properties. Vigan{\`o} argues that the modularity of his labelled sequent calculi is partially due to this division, which allows for a labelled sequent calculus to be constructed for a logic by taking the base calculus and equipping it with the proper relational algebra encoding the properties of that logic's accessibility relation(s). If another logic's accessibility relations possess distinct characteristics, then by modifying the relational algebra accordingly, a new labelled system for the logic can be obtained, showing that the formalism is modular.

To exhibit how a semantic clause can be transformed into a set of inference rules in the labelled formalism, let us consider the standard semantic clause for the $\Box$ modality in the modal logic $\mathsf{K}$ (cf.~\cite{BlaRijVen01}), which is similar to the semantic clause for the $\charabox$ modality used in grammar logics (see \sect~\ref{def:semantics-kms}). 
\begin{center}
$M,w \Vdash \Box \phi$ \ifandonlyif for all $u \in W$, if $Rwu$, then $M, u \Vdash \phi$.
\end{center}
The above semantic clause can be transformed into the following two clauses by means of classical reasoning---one of which expresses when $\Box \phi$ is satisfied, and another which expresses when $\Box \phi$ is unsatisfied (in a relational model $M = (W,R,V)$).
\begin{itemize}



\item[(i)] If for all $u \in W$, if $(w,u) \in R$, then $M, u \Vdash \phi$, then $M,w \Vdash \Box \phi$.

\item[(ii)] If there exists a $u \in W$ such that $(w,u) \in R$ and $M, u \not\Vdash \phi$, then $M,w \not\Vdash \Box \phi$.

\end{itemize}
Clause (i) can be written as the inference rule shown top-left below, which 
 states that if for any $u$, if $(w,u) \in R$, then $M, u \Vdash \phi$ holds, then $M,w \Vdash \Box \phi$ holds. Putting this into sequent notation, we can use ``$Rwu \sar u : \phi$'' to represent ``if $(w,u) \in R$, then $M, u \Vdash \phi$ holds'', and ``$\sar w : \Box \phi$'' to represent ``$M,w \Vdash \Box \phi$''. 
 This notational substitution gives the inference rule shown top-middle below. To obtain the final labelled sequent rule that corresponds to the first clause, we add in contexts $\rel$, $\Gamma$, and $\Delta$, which gives the final form of the rule shown top-right below. We note that a side condition must be imposed on the rule once contexts are added, i.e. $u$ is not allowed to occur in the conclusion, which ensures that the label is implicitly universally quantified. The inference rule corresponding to clause (ii) is obtained in a similar fashion, and its derivation is shown second below.
\begin{center}
\begin{tabular}{c c c c c}
\AxiomC{if $(w,u) \in R$, then $M, u \Vdash \phi$}
\UnaryInfC{$M,w \Vdash \Box \phi$}
\DisplayProof

&

$\leadsto$

&

\AxiomC{$Rwu \sar u : \phi$}
\UnaryInfC{$\sar w : \Box \phi$}
\DisplayProof

&

$\leadsto$

&

\AxiomC{$\rel, Rwu, \Gamma \sar u : \phi, \Delta$}
\UnaryInfC{$\rel, \Gamma \sar w : \Box \phi, \Delta$}
\DisplayProof
\end{tabular}
\end{center}

\begin{center}
\resizebox{\columnwidth}{!}{
\begin{tabular}{c c c}
\AxiomC{$(w,u) \in R$ and $M, u \not\Vdash \phi$}
\RightLabel{$\leadsto$}
\UnaryInfC{$M,w \not\Vdash \Box \phi$}
\DisplayProof

&

\AxiomC{$\sar Rwu$}
\AxiomC{$u : \phi \sar$}
\RightLabel{$\leadsto$}
\BinaryInfC{$w : \Box \phi \sar$}
\DisplayProof

&

\AxiomC{$\rel \sar Rwu$}
\AxiomC{$\rel, u : \phi, \Gamma \sar \Delta$}
\BinaryInfC{$\rel, w : \Box \phi, \Gamma \sar \Delta$}
\DisplayProof
\end{tabular}
}
\end{center}


Building off of Vigan{\`o}'s~\cite{Vig00} and Simpson's~\cite{Sim94} work, Dyckhoff and Negri~\cite{DycNeg12,Neg05}, showed that cut-free labelled sequent systems could be provided for an even broader class of modal and constructive logics than those provided by Vigan{\`o}. In~\cite{DycNeg12,Neg05}, the authors showed that labelled sequent calculi could be provided for any extension of intuitionistic logic $\int$, and any extension of the modal logic $\mathsf{K}$, with the accessibility relation satisfying any number of \emph{geometric formulae}~\cite{Sim94}\index{Geometric formula}, that is, formulae of the form shown below left. These classes of logics go beyond those considered by Vigan{\`o}, as Vigan{\`o} only considers extensions of $\int$ and $\mathsf{K}$ where the accessibility relation satisfies Horn formulae. 
\begin{center}
\begin{tabular}{c c}
$\displaystyle{\uc \Big( \bigwedge_{i = 1}^{n} \phi_{i} \cimp \ec (\bigvee_{j = 1}^{m} \bigwedge_{k = 1}^{l_{j}} \psi_{j,k}) \Big)}$

&

\AxiomC{$\rel, \overline{\phi}, \overline{\psi}_{1}, \Gamma \sar \Delta$}
\AxiomC{$\cdots$}
\AxiomC{$\rel, \overline{\phi}, \overline{\psi}_{m}, \Gamma \sar \Delta$}
\TrinaryInfC{$\rel, \overline{\phi}, \Gamma \sar \Delta$}
\DisplayProof
\end{tabular}
\end{center}
In the geometric formula (above left), each $\phi_{i}$ and $\psi_{j,k}$ is taken to be an atomic formula. Each geometric formula is equivalent to a \emph{geometric structural rule}\index{Geometric structural rule} (introduced in Simpson's PhD thesis~\cite{Sim94}) of the form shown above right, where $\overline{\phi} := \phi_{1}, \ldots, \phi_{n}$, $\overline{\psi}_{j} := \psi_{j,1}, \ldots, \psi_{j,l_{j}}$, and each variable from $\vec{y}$ is an \emph{eigenvariable}, meaning that it does not occur in the conclusion of the rule. Interestingly, the addition of any number of geometric structural rules to a base (labelled) calculus for $\int$ or $\mathsf{K}$ yields a sound and complete labelled sequent calculus for that extension of $\int$ or $\mathsf{K}$ with the corresponding geometric formulae.

Another distinction between the work of Vigan{\`o}~\cite{Vig00} and the work in~\cite{DycNeg12,Neg05}, is that the latter abandons the division between a base calculus and the relational algebra, and instead, only employs labelled sequents of the form $\rel, \Gamma \sar \Delta$ (as in Simpson's PhD thesis~\cite{Sim94}). Consequently, reasoning about relational atoms (i.e. applying inference rules that manipulate relational atoms) is done directly on labelled sequents of the form $\rel, \Gamma \sar \Delta$, instead of on labelled sequents of the form $\rel \sar R_{i}t_{1} \ldots t_{n}$. As will be seen in the succeeding chapters (\cptr~\ref{CPTR:Refinment-Modal} and~\ref{CPTR:Refinment-Constructive}), this difference is conducive to refining our labelled calculi since it allows for the interaction of rules that would have otherwise been kept separate in Vigan{\`o}'s formalism. Therefore, we will build labelled sequent calculi for grammar logics, first-order intuitionistic logics, and deontic \stit logics within the formalism of~\cite{DycNeg12,Neg05,Sim94}, thus setting the stage for refinement in \cptr~\ref{CPTR:Refinment-Modal} and~\ref{CPTR:Refinment-Constructive}.

Last, we note that although our labelled sequent systems admit syntactic cut-elimination, a \emph{strict} form of analyticity (i.e. the subformula property) fails to hold---as is typical of labelled systems. That is to say, cut-elimination does not immediately imply that every formula occurring within a cut-free derivation is a subformula of the conclusion. The blame for the loss of (a strict notion of) analyticity lies with the incorporation of structural rules that delete relational atoms from premise to conclusion. Nevertheless, since labelled sequent systems regularly include such rules, they are often qualified as analytic if every labelled formula occurring in a derivation is a subformula of some labelled formula in the conclusion~\cite{Neg05,Vig00}. It will be seen that this weaker notion of analyticity holds for our labelled sequent systems; yet, interestingly, the systems obtained via refinement (in the following chapters) will either possess a strict form of the subformula property or a higher degree of the property.


\subsubsection{Terminology and Notation.} We introduce notation and terminology which is uniformly applicable to the wide variety of labelled sequent systems we consider. First off, \emph{labelled sequents} are syntactic objects of the form $\Lambda := \rel, \Gamma \sar \Delta$. The multiset $\rel$ of relational atoms may be empty, or may consist of relational atoms of the form $R_{\chara}wu$, $\unda \in D_{w}$, $w \leq u$, or $R_{\agbox}wu$ depending on if the labelled calculus is for a grammar logic, first-order intuitionistic logic, or deontic \stit logic. 
 The labels $w$ and $u$ are taken to be among a denumerable set of labels $\lab$, and the multisets $\Gamma$ and $\Delta$ occurring in a labelled sequent may be empty, or may consist of labelled formulae of the form $w : \phi$, where $\phi \in \langkm{\albet}$, $\phi \in \langintfo$, or $\phi \in \langdsn$, depending on the logic under consideration. In each of the following sections, we will formally define the set of labelled sequents used. Also, we define the following useful operations on multisets:

\begin{definition}\label{def:restirction-prepend-labelled} If $\Gamma$ is a multiset of labelled formulae, then we define $\Gamma \restriction w$\index{$\Gamma \restriction w$} to be the multiset $\{\phi \ | \ w : \phi \in \Gamma\}$, and if $\Gamma := \phi_{1}, \ldots, \phi_{k}$ is a multiset of formulae, then we define $w : \Gamma := w : \phi_{1}, \ldots, w : \phi_{k}$\index{$w : \Gamma$}.
\end{definition}

A \emph{labelled calculus} is a set of \emph{inference rules}, which are objects of the following form:
\begin{center}
\AxiomC{$\Lambda_{1}$}
\AxiomC{$\cdots$}
\AxiomC{$\Lambda_{n}$}
\RightLabel{$\ru$}
\TrinaryInfC{$\Lambda$}
\DisplayProof
\end{center}
Inference rules are schematic, and when used in practice they will be instantiated with concrete labelled sequents, thus allowing for an instance of $\Lambda$ (the \emph{conclusion}) to be derived from instances of $\Lambda_{1}$, $\ldots$, $\Lambda_{n}$ (the \emph{premises}). Also, if $n = 0$ in an inference rule, that is, the inference rule does not contain premises, then we say that the inference rule is \emph{initial}\index{Initial rule} and each labelled sequent given as a conclusion of the rule is an \emph{initial sequent}\index{Initial sequent}.

Aside from initial rules, there are two other types of rules that our systems make use of: \emph{logical rules}\index{Logical rule} and \emph{structural rules}\index{Structural rule}. An inference rule is a \emph{logical rule} \ifandonlyif the rule explicitly introduces a complex logical formula, called the \emph{principal formula}\index{Principal formula}, from less complex logical formulae, called the \emph{auxiliary formulae}\index{Auxiliary formula} (all other formulae in the inference are called \emph{side formulae} or \emph{parametric formulae}). Furthermore, we refer to the principal and auxiliary formulae of an inference as \emph{active}\index{Active formula}, more generally. For example, the following rules are logical rules since they introduce the complex logical formulae $\phi \land \psi$ and $\Box \phi$, respectively.
\begin{center}
\begin{tabular}{c c}
\AxiomC{$\rel, \Gamma, w : \phi, w : \psi \sar \Delta$}
\UnaryInfC{$\rel, \Gamma, w : \phi \land \psi \sar \Delta$}
\DisplayProof

&

\AxiomC{$\rel, Rwu, \Gamma \sar u : \phi, \Delta$}
\UnaryInfC{$\rel, \Gamma \sar w : \Box \phi, \Delta$}
\DisplayProof
\end{tabular}
\end{center}
In the above rules, the formulae $w : \phi \land \psi$ and $w : \Box \phi$ are principal, $w : \phi, w : \psi$ and $Rwu, u : \phi$ are auxiliary, and all such formulae are active. The formulae in $\rel, \Gamma$, and $\Delta$ are side (or, parametric) formulae.

In contrast to logical rules, our labelled calculi also contain \emph{structural rules} which are rules that do not make explicit reference to a logical connective being introduced (as in the case of the logical rules), but rather, directly manipulate labelled sequents in a manner consistent with the associated logic's meta-logical properties. For example, the following two rules are labelled structural rules:
\begin{center}
\begin{tabular}{c c}
\AxiomC{$\rel, w \leq w, \Gamma \sar \Delta$}
\UnaryInfC{$\rel, \Gamma \sar \Delta$}
\DisplayProof

&

\AxiomC{$\rel, \Gamma \sar w :\phi, w : \phi, \Delta$}
\UnaryInfC{$\rel, \Gamma \sar w : \phi, \Delta$}
\DisplayProof
\end{tabular}
\end{center}
The top-left rule deletes a relational atom, and the top-right rule contracts the labelled formula $w : \phi$ from two copies into one copy. 

A \emph{proof} is defined to be a tree of labelled sequents such that 
each labelled sequent is the conclusion of an inference rule with premises also in the proof, and all leaves of the proof are initial sequents. The root of the proof is called the \emph{end sequent}, and the \emph{height} of the proof is the maximal number of labelled sequents that occur from the end sequent to an initial sequent.

We say that a rule $\ru$ is \emph{admissible} (\emph{height-preserving (hp-) admissible})\index{Hp-admissible rule}\index{Admissible rule} in a labelled calculus \ifandonlyif derivability of the premise(s) implies derivability of the conclusion with an $\ru$-free proof (whose height is less than or equal to the height(s) of the derivation(s) of the premise(s)). A rule $\ru$ is \emph{eliminable}\index{Eliminable rule} in a labelled calculus \ifandonlyif there exists an algorithm that transforms any proof in the labelled calculus into an $\ru$-free proof (observe that eliminability implies admissibility). Last, we say that a rule is \emph{invertible}\index{Hp-invertible rule}\index{Invertible rule} (\emph{height-preserving (hp-) invertible}) in a labelled calculus \ifandonlyif if an instance of the conclusion is derivable, then the premise(s) is (are) derivable (with height(s) less than or equal to the height of the derivation of the conclusion).

All of the above proof-theoretic terminology is fairly standard and can be found in most introductory texts on proof theory (e.g.~\cite{Bus98,Tak13}). Moreover, we extend such terminology to systems built within other proof theoretic formalisms (e.g. nested sequents). 

If a rule is (hp-)admissible or eliminable, then this implies that given a proof $\Pi$ of the premise of the rule, one can find a proof $\Pi'$ of the conclusion of the rule. Similarly, if a rule is (hp-)invertible, then this implies that given a proof $\Pi$ of the conclusion of the rule, one can find a proof $\Pi'$ of the premise(s) of the rule. When applying an (hp-)admissible, eliminable, or (hp-)invertible rule $\ru$ in a derivation, we will use a dashed line (as shown below left) to represent that the conclusion is not obtained via an inference rule, but rather, is obtained via the (hp-)admissibility, eliminability, or (hp-)invertibility property of the rule. Also, we will use a dashed line to indicate that a labelled sequent is known to be derivable by an assumption or by some result. Last, we use dotted lines (as shown below right) to indicate that two labelled sequents are identical to one another.
\begin{center}
\begin{tabular}{c c} 
\AxiomC{$\rel, \Gamma \sar \Delta$}
\RightLabel{$\ru$}
\dashedLine
\UnaryInfC{$\rel', \Gamma' \sar \Delta'$}
\DisplayProof

&

\AxiomC{$\rel, \Gamma \sar \Delta$}
\RightLabel{=}
\dottedLine
\UnaryInfC{$\rel', \Gamma' \sar \Delta'$}
\DisplayProof
\end{tabular}
\end{center}

In order to understand the data structures encoded by labelled sequents, we will transform labelled sequents into graphs. Therefore, it will be useful to define the notion of a (sub)graph, various types of graphs (e.g. trees, DAGs, etc.), and isomorphisms between graphs. In the following chapters, we will see that refining our labelled calculi necessitates a ``simplification'' in the data structures encoded by labelled sequents in derivations.

\begin{definition}[Graph, Induced Subgraph]\label{def:graph-subgraph} A \emph{graph}\index{Graph} is an object of the form $G = (V,E,L)$, where $V$ is a set of vertices, the set of edges $E \subseteq V \times V$ is a binary relation on $V$ (we also allow the set of edges $E \subseteq V \times V \times S$ to be indexed with elements of a set $S$), and $L$ is a labelling function mapping elements from $V$ to elements in a set $S'$.

We define an \emph{induced subgraph}\index{Induced Subgraph} $G' = (V',E',L')$ of a graph $G = (V,E,L)$ to be a graph such that $V' \subseteq V$, $E' = E \restriction V'$, and $L' = L \restriction V'$.
\end{definition}

\begin{definition}[Tree]\label{def:tree} A graph $G = (V,E,L)$ is a \emph{tree}\index{Tree} \ifandonlyif there exists a vertex $w \in V$ (called the \emph{root})\index{Root} such that there is a unique directed path from $w$ to every other vertex $u \in V$. Equivalently, a tree is a graph that is connected, free of directed cycles, and contains no backward branching.
\end{definition}

\begin{definition}[Polytree]\label{def:polytree} We say that a graph $G = (V,E,L)$ is a \emph{polytree}\index{Polytree} \ifandonlyif it is connected, free of directed cycles, and free of undirected cycles.
\end{definition}

\begin{definition}[Forest]\label{def:forest} A graph $G = (V,E,L)$ is a \emph{forest}\index{Forest} \ifandonlyif it consists of a disjoint union of trees. We refer to each root of a tree in the disjoint union as a \emph{root} of the forest.
\end{definition}

\begin{definition}[Directed Acyclic Graph (DAG)]\label{def:DAG} A graph $G = (V,E,L)$ is a \emph{directed acyclic graph (DAG)}\index{Directed acyclic graph (DAG)} \ifandonlyif it is free of directed cycles. We refer to a vertex $v \in V$ as a \emph{root} of a DAG given that no vertices $u \in V$ exist such that there is an edge from $u$ to $v$.
\end{definition}

\begin{definition}[Isomorphism, Isomorphic]\label{def:isomorphism} Let $G = (V,E,L)$ and $G' = (V',E',L')$ be two graphs. An \emph{isomorphism}\index{Isomorphism} $f : V \mapsto V'$ from $G$ to $G'$ is a function such that: 

\begin{itemize}

\item[$\li$] $f$ is bijective,

\item[$\li$] $(x,y) \in E$ \ifandonlyif $(fx,fy) \in E'$ (or, $(x,y,s) \in E$ \ifandonlyif $(fx,fy,s) \in E'$, if edges are indexed with elements from a set $S$), and

\item[$\li$] $L(w) = L'(fw)$.
\end{itemize}
We say $G$ and $G'$ are \emph{isomorphic}\index{Isomorphic} (and write $G \iso G'$) \ifandonlyif there exists an isomorphism between them.
\end{definition}




\section{Labelled Calculi for Grammar Logics}\label{sec:lab-calc-kms}

Although grammar logics were introduced in 1988~\cite{CerPen88}, the proof theory for such logics did not come about until a decade later~\cite{BalGioMar98}. In~\cite{BalGioMar98}, Baldoni \textit{et al.} formulated prefixed analytic tableaux for grammar logics and leveraged these systems to show the decidability of right linear grammar logics and the undecidability of context-free grammar logics. In subsequent years, alternative tableaux were proposed for classes of grammar logics, such as (non-prefixed) tableau with automaton-labelled formulae for regular grammar logics~\cite{GorNgu05} and tableaux for regular grammar logics with converse~\cite{NguSza09,NguSza11}. The former were used to provide EXPTIME decision procedures for regular grammar logics as well as to give an effective Craig interpolation lemma, while the latter was used to provide optimal EXPTIME decision procedures for regular grammar logics with converse.

Beyond the tableau formalism, nested sequent calculi were given for context-free grammar logics with converse~\cite{TiuIanGor12}. In that paper, the authors define proof-search and counter-model extraction algorithms, which are of EXPSPACE complexity when termination holds. Interestingly, the method of refinement (discussed in the next chapter) yields (slight variants of) the nested calculi of~\cite{TiuIanGor12} when applied to the labelled sequent calculi for context-free grammar logics with converse introduced in this section.

Below, we adapt the formalism and methodology of~\cite{DycNeg12,Neg05,Sim94,Vig00} to introduce labelled sequent systems for grammar logics. Our calculi consist of rules that manipulate \emph{labelled sequents}, that is, formulae extending the language $\langkm{\albet}$ and which are used to derive theorems. We define these as follows: 

\begin{definition}[Labelled Sequents for Grammar Logics]
\emph{Labelled sequents}\index{Labelled sequent!for grammar logics} for grammar logics are defined to be syntactic objects of the form $\Lambda := \rel \sar \Gamma$, where $\rel$ (the \emph{antecedent}) and $\Gamma$ (the \emph{consequent}) are defined via the following grammars in BNF:
$$
\rel ::= \epsilon \ | \ R_{\chara}wu \ | \ \rel, \rel \qquad \Gamma ::= \epsilon \ | \ w : \phi \ | \ \Gamma, \Gamma
$$
with $\chara \in \albet$, $\phi \in \langkm{\albet}$, and where $w$ and $u$ are among a denumerable set of labels $\lab := \{w, u, v, \ldots\}$. We refer to formulae of the form $R_{\chara}wu$ as \emph{relational atoms}\index{Relational atom!for grammar logics}, to formulae of the form $w : \phi$ as \emph{labelled formulae}\index{Labelled formula!for grammar logics}, and let $\lab(\Lambda)$, $\lab(\rel)$, and $\lab(\Gamma)$ represent the set of labels in $\Lambda$, $\rel$, and $\Gamma$, respectively.
\end{definition}

We use $\Lambda$, $\Lambda'$, \etc (occasionally with subscripts) to denote labelled sequents, $\rel$, $\rel'$, \etc (occasionally with subscripts) to denote multisets of relational atoms, and $\Gamma$, $\Gamma'$, \etc (occasionally with subscripts) to denote multisets of labelled formulae. Hence, we take the comma operator to be both associative and commutative; for example, we identify $R_{a}wu,R_{b}uv$ with $R_{b}uv,R_{a}wu$ and $w:\phi, u:\psi, v:\chi$ with $v:\chi, w:\phi, u:\psi$. This interpretation of comma is what lets us view strings $\rel$ and $\Gamma$ as multisets. The symbol $\epsilon$ is taken to represent the \emph{empty string}\index{Empty string} that acts as an identity element for the comma operator (e.g. $R_{a}wu, \epsilon, R_{b}uv$ is identified with $R_{a}wu, R_{b}uv$); 
hence, $\epsilon$ will typically be implicit in labelled sequents. 

\begin{figure}[t]
\noindent\hrule

\begin{center}
\begin{tabular}{c c}
\AxiomC{}
\RightLabel{$\id$\index{$\id$}}
\UnaryInfC{$\rel \sar w : p, w : \negnnf{p}, \Gamma$}
\DisplayProof

&

\AxiomC{$\rel \sar w : \phi, w : \psi, \Gamma$}
\RightLabel{$\disr$\index{$\disr$}}
\UnaryInfC{$\rel \sar w : \phi \lor \psi, \Gamma$}
\DisplayProof
\end{tabular}
\end{center}

\begin{center}
\begin{tabular}{c c}
\AxiomC{$\rel \sar w : \phi, \Gamma$}
\AxiomC{$\rel \sar w : \psi, \Gamma$}
\RightLabel{$\conr$\index{$\conr$}}
\BinaryInfC{$\rel \sar w : \phi \land \psi, \Gamma$}
\DisplayProof

&

\AxiomC{$\rel, R_{\chara}wu \sar w : \charadia \phi, u : \phi, \Gamma$}
\RightLabel{$\charadiar$\index{$\charadiar$}}
\UnaryInfC{$\rel, R_{\chara}wu \sar w : \charadia \phi, \Gamma$}
\DisplayProof
\end{tabular}
\end{center}

\begin{center}
\begin{tabular}{c c c}
\AxiomC{$\rel, R_{\chara}wu \sar u : \phi, \Gamma$}
\RightLabel{$\charaboxr^{\dag}$\index{$\charaboxr$}}
\UnaryInfC{$\rel \sar w : \charabox \phi, \Gamma$}
\DisplayProof

&

\AxiomC{$\rel, R_{\stra}wu, R_{\chara}wu \sar \Gamma$}
\RightLabel{$\psr$\index{$\psr$}}
\UnaryInfC{$\rel, R_{\stra}wu \sar \Gamma$}
\DisplayProof

&

\AxiomC{$\rel, R_{\chara}wu, R_{\conv{\chara}}uw \sar \Gamma$}
\RightLabel{$\convr$\index{$\convr$}}
\UnaryInfC{$\rel, R_{\chara}wu \sar \Gamma$}
\DisplayProof
\end{tabular}
\end{center}

\hrulefill
\caption{The labelled calculus $\gtkms$\index{$\gtkms$} for the grammar logic $\kms$. We have an $\charaboxr$, $\charadiar$, and $\convr$ rule for each $\chara \in \albet$, and for each production rule $\chara \pto s \in \thuesys$, we have a corresponding $\psr$ rule (see \dfn~\ref{def:production-to-structural-rule-kms} below for a description of the correspondence between production rules and structural rules). The side condition $\dag$ states that the associated rule can be applied only if the label $u$ is an \emph{eigenvariable}, i.e. it does not occur in the conclusion of the rule. Last, we assume that each calculus satisfies the closure condition (given in \dfn~\ref{def:closure-condition-kms}).}
\label{fig:G3Km(S)}
\end{figure}

Labelled sequents serve as an abstraction of a $\albet$-model (\dfn~\ref{def:frames-models-kms}), where the labels from $\lab$ represent the worlds, the relational atoms represent the accessibility relations, and the labelled formulae determine at which world a formula $\phi \in \langkm{\albet}$ is (un)satisfied. Intuitively, we interpret a labelled sequent $\rel \sar \Gamma$ as saying, `if all relational atoms in $\rel$ hold, then some formula in $\Gamma$ holds.' See \dfn~\ref{def:sequent-semantics-kms} for the formal definition of a labelled sequent interpretation, which also spells out the exact relationship between labelled sequents and $\albet$-models.

Our labelled calculi for grammar logics $\kms$ are displayed in \fig~\ref{fig:G3Km(S)}, and the derivability relation is defined below. Moreover, examples of concrete derivations in a calculus $\gtkms$ can be found in the proof of \thm~\ref{thm:completness-gtkms} below.

\begin{definition}\label{def:derives-gtkms} We write $\gtkmsder \Lambda$ to indicate that a labelled sequent $\Lambda$ is derivable in a calculus $\gtkms$. (NB. We will use the notation $\vdash_{\mathsf{X}}$ throughout the thesis to denote that a sequent is derivable in the calculus $\mathsf{X}$.) 
\end{definition}


The rules of the calculi are obtained from the notion of a $\albet$-model (\dfn~\ref{def:frames-models-kms}), and from the semantic clauses for the various connectives (\dfn~\ref{def:semantics-kms}). The $\id$ rule encodes the fact that in a $\albet$-model the valuation function will either make a propositional variable $p$ true at a world, or it will not. The rules $\conr$, $\disr$, $\charaboxr$, and $\charadiar$ are obtained from the semantic clauses of the respective logical connectives, and each rule $\psr$ lets us derive theorems based on the frame property obtained from the corresponding production rule in $\thuesys$ (see \dfn~\ref{def:production-to-structural-rule-kms} for the correspondence between production rules and frame properties). Each $\convr$ rule encodes the fact that in a $\albet$-model, the relation $R_{\chara}$ is the converse of the relation $R_{\conv{\chara}}$ for any $\chara \in \albet$, thus corresponding to the converse condition $\convcondns$ (\dfn~\ref{def:frames-models-kms}). 


 
As mentioned previously, a nice feature of the labelled paradigm is that it is relatively straightforward to transform the relational semantics of a logic into a labelled calculus.\footnote{As shown in~\cite{CiaMafSpe13}, the process of constructing labelled calculi for intermediate logics can be automated.} In addition to ease of construction, labelled calculi typically possess desirable proof-theoretic properties, such as (hp-)admissibility of structural rules (e.g. those presented in \fig~\ref{fig:struc-rules-kms} below) and hp-invertibility of rules~\cite{DycNeg12,Neg05,Sim94,Vig00}. In~\cite{Neg05}, the author shows that such properties are preserved under any extension of the labelled base calculus $\gtk$ (for the minimal modal logic $\mathsf{K}$) with geometric structural rules (mentioned into the introduction to this chapter). By analogy, let us think of the set of rules $\{\id, \conr, \disr, \charaboxr, \charadiar \ | \ \chara \in \albet\}$ as being a base calculus akin to $\gtk$. Despite the fact that~\cite{Neg05} does not address the construction and properties of labelled calculi in the multi-modal setting (and considers only the uni-modal setting), if we think of the aforementioned set as our base calculus, and because the rules $\psr$ and $\convr$ fit within the geometric rule scheme, it is straightforward to adapt the methods and results from~\cite{Neg05} to our setting. Below, we will explain how this is done.


Although we may think of the set of rules $\{\id, \conr, \disr, \charaboxr, \charadiar \ | \ \chara \in \albet\}$ as a base calculus in order to draw an analogy with the method of~\cite{Neg05} and transfer results to our setting, the set of rules $\{\id, \conr, \disr, \charaboxr, \charadiar, \convr \ | \ \chara \in \albet\}$ serves as a 
 sound and complete calculus for the base grammar logic $\km = \km(\emptyset)$ relative to an alphabet $\albet$ (see \thm~\ref{thm:soundness-gtkms} and~\ref{thm:completness-gtkms}). To obtain a calculus for an extension of a minimal grammar logic $\km$ with a \cfcst system $\thuesys$ (defined in \dfn~\ref{def:CFCST-kms}), i.e. for the logic $\kms$, we add a structural rule $\psr$ for each production rule in $\thuesys$. The correspondence between structural and production rules is detailed in the definition below:

\begin{definition}[Production and Structural Rules]\label{def:production-to-structural-rule-kms} Let $\thuesys$ be a \cfcst system, and let $\chara \pto \stra \in \thuesys$. If $\stra = \chara_{1} \cate \cdots \cate \chara_{n} \in \albetstr - \{\empstr\}$, we define $R_{\stra}wu := R_{\chara_{1}}wv_{1}, \ldots, R_{\chara_{n}}v_{n-1}u$, and if $\stra = \empstr$, then we define $R_{\stra}wu := \seqempstr$. Depending on if $\stra = \empstr$ or not, the structural rule corresponding to the production rule $\chara \pto \stra$ is defined as follows:

\begin{center}
\begin{tabular}{|c|c|}
\hline
$\stra \neq \empstr$

&

$\stra = \empstr$\\
\hline
\AxiomC{}
\noLine
\UnaryInfC{$\rel, R_{\stra}wu, R_{\chara}wu \sar \Gamma$}
\RightLabel{$\psr$}
\UnaryInfC{$\rel, R_{\stra}wu \sar \Gamma$}
\noLine
\UnaryInfC{}
\DisplayProof

&

\AxiomC{}
\noLine
\UnaryInfC{$\rel, R_{\chara}ww \sar \Gamma$}
\RightLabel{$\psr$}
\UnaryInfC{$\rel \sar \Gamma$}
\noLine
\UnaryInfC{}
\DisplayProof\\
\hline
\end{tabular}
\end{center}
In the rule top right, observe that $R_{\stra}wu$ does not occur because $\stra = \empstr$, meaning that $R_{\stra}wu$ is the empty string $\seqempstr$. 
\end{definition}

Although labelled calculi generally permit the hp-admissibility of relational atoms contractions $\ctrrel$ (see \fig~\ref{fig:struc-rules-kms}), there is an important caveat. When labelled calculi are extended with geometric structural rules of a specific shape, additional rules must be added to the system to ensure the hp-admissibility of $\ctrrel$~\cite{Neg05}. To motivate and explain the need for such rules, we consider an example: 
 suppose we have a \cfcst system $\thuesys := \{b \pto a \cate a, \conv{b} \pto \conv{a} \cate \conv{a} \}$. The first production rule gives rise to the $\psrp{b}{a \cate a}$ structural rule shown below left. (NB. Since the structural rule $\psrp{b}{a \cate a}$ is obtained from the first production rule, we let $b$ instantiate $\chara$ and $a \cate a$ instantiate $\stra$ in the schema $\psr$ as explained in \dfn~\ref{def:production-to-structural-rule-kms} above.) To explain why the hp-admissibility of relational atom contractions $\ctrrel$ is not immediately permitted, let us assume that we are given a derivation ending with the following two inferences shown below right, where an instance of $\psrp{b}{a \cate a}$ (with the label $w$ substituted for $v$ and $u$) is followed by an instance of $\ctrrel$:

\begin{center}
\begin{tabular}{c c}
\AxiomC{$\rel, R_{a}wv, R_{a}vu, R_{b}wu \sar \Gamma$}
\RightLabel{$\psrp{b}{a \cate a}$}
\UnaryInfC{$\rel, R_{a}wv, R_{a}vz \sar \Gamma$}
\DisplayProof

&

\AxiomC{$\rel, R_{a}ww, R_{a}ww, R_{b}ww \sar \Gamma$}
\RightLabel{$\psrp{b}{a \cate a}$}
\UnaryInfC{$\rel, R_{a}ww, R_{a}ww \sar \Gamma$}
\RightLabel{$\ctrrel$}
\UnaryInfC{$\rel, R_{a}ww \sar \Gamma$}
\DisplayProof
\end{tabular}
\end{center}

Typically, when proving a rule (such as $\ctrrel$) hp-admissible, one shows that the rule can be permuted above any rule of the calculus and can be deleted at initial sequents. Notice though, if we attempt to permute $\ctrrel$ above $\psrp{b}{a \cate a}$ and apply the rule directly to the top sequent, then we obtain the labelled sequent $\rel, R_{a}ww, R_{b}ww \sar \Gamma$. This labelled sequent is no longer in a form matching the premise of $\psrp{b}{a \cate a}$, and so the rule is inapplicable; the other production rule of $\thuesys$ does not give rise to a structural rule that could be used to delete $R_{b}ww$ (in order to derive the desired end sequent) either.

Nevertheless, this obstacle can be overcome by stipulating that our labelled calculi must adhere to the so-called \emph{closure condition}~\cite{Neg05}. Essentially, the closure condition states that if a substitution of labels within a structural rule causes a duplication of \emph{principal} relational atoms (i.e. those occurring in the conclusion), then another instance of the rule with those relational atoms contracted must be added to the calculus. Since the closure condition is only enforced when a duplication of principal relational atoms arises in the conclusion of the structural rule, this condition need not be enforced for the $\convr$ structural rules, which only contain a \emph{single} principal relational atom in the conclusion. Howbeit, each $\psr$ rule may contain more than a single principal relational atom, so we define the closure condition for such rules below: 

\begin{definition}[Closure Condition]\label{def:closure-condition-kms} 
Let $\thuesys$ be a \cfcst system, and let $\ruone$ be either a structural rule $\psr$ obtained from a production rule $\chara \pto \stra \in \thuesys$, or a structural rule $\psrcc$ obtained from the closure condition. (NB. We use a $\ddag$ subscript to indicate rules obtained via the closure condition.) For all such rules $\ruone$, if the following is an instance of $\ruone$:
\begin{center}
\AxiomC{$\rel, R_{\strb}wv, R_{\charb}vz,R_{\charb}vz, R_{\strc}zu, R_{\chara}wu \sar \Gamma$}
\RightLabel{$\ruone$}
\UnaryInfC{$\rel, R_{\strb}wv, R_{\charb}vz, R_{\charb}vz, R_{\strc}zu \sar \Gamma$}
\DisplayProof
\end{center}
then our calculus satisfies the \emph{closure condition}\index{Closure condition} \ifandonlyif it also contains the following instance of the rule (with $R_{\charb}vz$ contracted in both premise and conclusion):
\begin{center}
\AxiomC{$\rel, R_{\strb}wv, R_{\charb}vz, R_{\strc}zu, R_{\chara}wu \sar \Gamma$}
\RightLabel{$\rutwo$}
\UnaryInfC{$\rel, R_{\strb}vu,R_{\charb}vz, R_{\strc}zu \sar \Gamma$}
\DisplayProof
\end{center}
As stated in \fig~\ref{fig:G3Km(S)}, we assume that each $\gtkms$ calculus satisfies the closure condition.
\end{definition}

For any given structural rule $\psr$ only a finite number of instances with duplications can arise, implying that the closure condition only adds a finite number of rules for each $\psr$ rule. Hence, the above recursively defined condition terminates, meaning that the imposition of the closure condition is unproblematic. 

Now that we have defined each calculus $\gtkms$, we show how to interpret labelled sequents on $\albet$-models, giving rise to a notion of validity for sequents. We then utilize this notion to show that our calculi are sound (see \thm~\ref{thm:soundness-gtkms} below).

\begin{definition}[$\gtkms$ Semantics]\label{def:sequent-semantics-kms} Let $M = (W, \{R_{\chara} \ | \ \chara \in \albet\}, V)$ be a $\albet$-model satisfying the \cfcst system $\thuesys$ with $I : \ \lab \mapsto W$ an \emph{interpretation function}\index{Interpretation function!for grammar logics} mapping labels to worlds.  We define the \emph{satisfaction}\index{Satisfaction!for $\gtkms$} of a relational atom $R_{\chara}wu$ (written $M, I \models_{\kms} R_{\chara}wu$) and labelled formula $w : \phi$ (written $M, I \models_{\kms} w : \phi$) as follows:
\begin{itemize}

\item[$\li$] $M, I \models_{\kms} R_{\chara}wu$ \ifandonlyif $(I(w),I(u)) \in R_{\chara}$

\item[$\li$] $M, I \models_{\kms} w : \phi$ \ifandonlyif $M, I(w) \Vdash \phi$

\end{itemize}
We say that a multiset of relational atoms $\rel$ is \emph{satisfied} in $M$ with $I$ (written $M, I \models_{\kms} \rel$) \ifandonlyif $M, I \models_{\kms} R_{\chara}wu$ for all $R_{\chara}wu \in \rel$, and we say that a multiset of labelled formulae $\Gamma$ is \emph{satisfied} in $M$ with $I$ (written $M, I \models_{\kms} \Gamma$) \ifandonlyif $M, I \models_{\kms} w : \phi$ for all $w : \phi \in \Gamma$. 

A labelled sequent $\Lambda = \rel \sar \Gamma$ is \emph{satisfied} in $M$ with $I$ (written, $M,I \models_{\kms} \Lambda$) \ifandonlyif if $M, I \models_{\kms} \rel$, then $M, I \models_{\kms} \Gamma$. Also, we say that a labelled sequent $\Lambda$ is \emph{falsified} in $M$ with $I$ \ifandonlyif $M, I \not\models_{\kms} \Lambda$, that is, $\Lambda$ is not satisfied by $M$ with $I$.

Last, a labelled sequent $\Lambda$ is \emph{$\kms$-valid}\index{$\kms$-valid} (written $\models_{\kms} \Lambda$) \ifandonlyif it is satisfiable in every $\albet$-model $M$ satisfying $\thuesys$ with every interpretation function $I$. We say that a labelled sequent $\Lambda$ is \emph{$\kms$-invalid}\index{$\kms$-invalid} \ifandonlyif $\not\models_{\kms} \Lambda$, i.e. $\Lambda$ is not $\kms$-valid.
\end{definition}

\begin{theorem}[Soundness]\label{thm:soundness-gtkms}
If $\vdash_{\gtkms} \Lambda$, then $\models_{\kms} \Lambda$
\end{theorem}

\begin{proof} Let $\thuesys$ be a \cfcst system. We argue that each  
 rule preserves validity by contraposition, that is, we show that if the conclusion of the rule is $\kms$-invalid, then at least one premise of the rule is $\kms$-invalid. We only consider the $\charaboxr$, $\charadiar$, $\psr$, and $\convr$ cases.




\emph{$\charaboxr$} Suppose that the conclusion of $\charaboxr$ is $\kms$-invalid, that is, there exists a $\albet$-model $M$ satisfying $\thuesys$ and an interpretation function $I$ such that $M,I \not\models_{\kms} \rel \sar w : \charabox \phi, \Gamma$. It follows that $M, I(w) \not\Vdash \charabox \phi$, which further implies that there exists a world $v$ such that $(I(w),v) \in R_{\chara}$ and $M,v \not\Vdash \phi$. We define a new interpretation function $I'$ where $I'(z) = I(z)$ if $z \neq u$ and $I'(u) = v$ otherwise. Consequently, $M, I'(u) \not\Vdash \phi$, and since $u$ is an eigenvariable and $(I'(w),I'(u)) \in R_{\chara}$, we have that the premise is falsified in $M$ with $I'$.

\emph{$\charadiar$} Suppose that the conclusion of an $\charadiar$ inference in $\kms$-invalid, i.e. there exists a $\albet$-model $M$ satisfying $\thuesys$ and an interpretation function $I$ such that $M, I \not\models_{\kms} \rel, R_{\chara}wu \sar w : \charadia \phi, \Gamma$. It follows that $(I(w),I(u)) \in R_{\chara}$ and $M,I(w) \not\Vdash \charadia \phi$, implying that $M,I(u) \not\Vdash \phi$. This shows that the premise is falsified in $M$ and $I$.

\emph{$\psr$} Let $\chara \pto \stra$ be the production rule corresponding to $\psr$. We have two cases to consider: either (i) $\stra = \empstr$ or (ii) $\stra \neq \empstr$. For case (i), suppose that the conclusion of $\psr$ is $\kms$-invalid. Then, there exists a $\albet$-model $M$ satisfying $\thuesys$ and an interpretation function $I$ such that $M, I \not\models_{\kms} \rel, R_{\stra}wu \sar \Gamma$. Since our $\albet$-model satisfies all production rules in $S$, and $\chara \pto \stra \in S$, we know that $R_{\stra} \subseteq R_{\chara}$ by \dfn~\ref{def:production-rule-sat-kms}. This implies that $(I(w),I(u)) \in R_{\chara}$, showing that the premise is falsified by $M$ with $I$. For case (ii), suppose that there exists a $\albet$-model $M$ satisfying $\thuesys$ and an interpretation function $I$ such that $M, I \not\models_{\kms} \rel \sar \Gamma$. Since our $\albet$-model satisfies $\thuesys$, which includes the production rule $\chara \pto \empstr$, we know that $R_{\empstr} \subseteq R_{\chara}$ by \dfn~\ref{def:production-rule-sat-kms}. By \dfn~\ref{def:generalized-relations-kms}, $(I(w),I(w)) \in R_{\empstr} \subseteq R_{\chara}$, implying that the premise is falsified by $M$ with $I$. Also, note that the soundness of any rule $\psrcc$ obtained from $\psr$ via the closure condition holds as well.

\emph{$\convr$} Suppose the conclusion of $\convr$ is $\kms$-invlaid, that is, there exists a $\albet$-model $M$ and interpretation function $I$ such that $M, I \not\models_{\kms} \rel, R_{\chara}wu \sar \Gamma$. It follows that $(I(w),I(u)) \in R_{\chara}$, and since $M$ satisfies the converse condition \convcondns, we have $(I(u),I(w)) \in R_{\conv{\chara}}$, showing that the premise is falsified in $M$ with $I$.
\end{proof}



As mentioned previously, we may adapt the work from~\cite{Neg05} to our setting and confirm that each calculus $\gtkms$ possesses desirable proof theoretic properties. Not only are these properties useful in refining our labelled calculi, thus leading to applications---e.g. proving that all grammar logics based on \cfcst systems have the effective Lyndon interpolation property (\sect~\ref{sec:applicationsII})---but also help us confirm the completeness of our calculi (\thm~\ref{thm:completness-gtkms}). Before listing the properties of each calculus $\gtkms$, we follow \cite{DycNeg12,Neg05} and define the notion of a \emph{label substitution}, which is an operation fundamental in establishing the results that follow.

\begin{definition}[Label Substitution]\label{def:label-sub-kms} We define a \emph{label substitution}\index{Label substitution!for grammar logics} $(w/u)$ for $w, u \in \lab$ on individual relational atoms and labelled formulae as follows:
\begin{itemize}

\item[$\li$] $(R_{\chara}vz)(w/u) =
  \begin{cases}
                                   R_{\chara}wz & \text{if $u = v$ and $v \neq z$} \\
                                   R_{\chara}ww & \text{if $u = v$ and $v = z$} \\
                                   R_{\chara}uw & \text{if $u = z$ and $v \neq z$} \\
                                   R_{\chara}vz & \text{otherwise}
  \end{cases}
$

\item[$\li$] $(v : \phi)(w/u) =
  \begin{cases}
                                   w : \phi & \text{if $u = v$} \\
                                   v : \phi & \text{otherwise}
  \end{cases}
$
\end{itemize}
We define the label substitution $(w/u)$ on a multiset of relational atoms $\rel$ and a multiset of labelled formulae $\Gamma$ to be the multiset obtained by applying $(w/u)$ to each element of the multiset.
\end{definition}

\begin{figure}[t]
\noindent\hrule

\begin{center}
\begin{tabular}{c c c}
\AxiomC{$\rel \sar \Gamma$}
\RightLabel{$\wk$}
\UnaryInfC{$\rel, \rel' \sar \Gamma', \Gamma$}
\DisplayProof

&

\AxiomC{$\rel \sar \Gamma$}
\RightLabel{$\lsub$}
\UnaryInfC{$\rel(w/u) \sar \Gamma(w/u)$}
\DisplayProof

&

\AxiomC{$\rel, R_{a}wu, R_{a}wu \sar \Gamma$}
\RightLabel{$\ctrrel$}
\UnaryInfC{$\rel, R_{a}wu \sar \Gamma$}
\DisplayProof
\end{tabular}
\end{center}

\begin{center}
\begin{tabular}{c c}
\AxiomC{$\rel \sar w : \phi, w: \phi, \Gamma$}
\RightLabel{$\ctrr$}
\UnaryInfC{$\rel \sar w : \phi, \Gamma$}
\DisplayProof

&

\AxiomC{$\rel \sar w : \phi, \Gamma$}
\AxiomC{$\rel \sar w : \negnnf{\phi}, \Gamma$}
\RightLabel{$\cut$}
\BinaryInfC{$\rel \sar \Gamma$}
\DisplayProof
\end{tabular}
\end{center}

\hrule
\caption{The set $\strucsetkms$ of structural rules\index{Structural rule!for grammar logics} consists of all rules shown above.}\label{fig:struc-rules-kms}
\end{figure}

We are now in a position to leverage the results of~\cite{Neg05} in order to confirm that each calculus $\gtkms$ possesses useful characteristics. As mentioned previously, the work of~\cite{Neg05} holds for uni-modal logics, that is, the language makes use of the standard modalities $\Box$ and $\Diamond$ (cf.~\cite{BlaRijVen01}) as opposed to a variety of modalities like the ones we are using in the current multi-modal setting. Nevertheless, if we inspect the proofs of certain results in~\cite{Neg05}, we can see that by identifying $\charabox$ with $\Box$ and $\charadia$ with $\Diamond$, all proofs go through. Moreover, only minor modifications to the proofs are needed to account for the different logical signatures and the fact that our formulae are in negation normal form. (NB. In~\cite{Neg05}, the logical connectives consist of $\{\bot, \lor, \land, \cimp, \Box, \Diamond\}$; by contrast, our language uses $\{\neg, \lor, \land, \charabox, \charadia \ | \ \chara \in \albet\}$.) In what follows, we state the various properties possessed by each calculus $\gtkms$, and mention in the corresponding proof what results from~\cite{Neg05} are exploited.

\begin{lemma}\label{lem:general-id-kms}
For all $\phi \in \langkm{\albet}$, $\vdash_{\gtkms} \rel \sar w : \phi, w : \negnnf{\phi}, \Gamma$.
\end{lemma}

\begin{proof} The claim is shown by induction on the complexity $\fcomp{\phi}$ of $\phi$.
\end{proof}

\begin{lemma}\label{lem:struc-rules-admiss-kms}
The $\lsub$, $\wk$, $\ctrrel$, and $\ctrr$ rules are hp-admissible.
\end{lemma}

\begin{proof} All claims are shown by induction on the height of the given derivation and are similar to the proofs of \lem~4.3, \prp~4.4, and \thm~4.12 of~\cite{Neg05}.
\end{proof}

\begin{lemma}\label{lem:hp-invert-kms}
All rules of $\gtkms$ are hp-invertible.
\end{lemma}

\begin{proof} Hp-invertibility of the $\charadiar$, $\psr$, and $\convr$ rules follows from the hp-admissibility of $\wk$. The remaining rules are shown hp-invertible by induction on the height of the given derivation and are similar to the proof of \prp~4.11 of~\cite{Neg05}.
\end{proof}

\begin{theorem}\label{thm:cut-admiss-kms}
The rule $\cut$ is eliminable in $\gtkms$. 
\end{theorem}

\begin{proof} The result is shown by induction on the lexicographic ordering of pairs $(\fcomp{\phi},h_{1} + h_{2})$, where $\fcomp{\phi}$ is the complexity of the cut formula $\phi$, $h_{1}$ is the height of the derivation of the left premise of $\cut$, and $h_{2}$ is the height of the derivation of the right premise of $\cut$. The proof is similar to the proof of \thm~4.13 of~\cite{Neg05}.
\end{proof}

Let us now harness the above results to prove each calculus $\gtkms$ complete relative to $\kms$. Prior to proving full completeness (i.e. completeness relative to $\kms$), we confirm that the rules $\id$, $\disr$, and $\conr$ are enough to confirm \emph{classical completeness}, that is, all instances of classical propositional tautologies are derivable in each $\gtkms$ calculus. This is needed as it ensures that the axioms A0 (see \dfn~\ref{def:axiomatization-km} for each grammar logic axiomatization $\h\kms$) are satisfied. Although this result is needed to confirm full completeness of our calculi, we defer the proof to the appendix (Appendix B on p.~\pageref{app:classical-completeness}) to simplify presentation.

\begin{lemma}[Classical Completeness]\label{lem:classical-completeness-kms}
All instances of classical propositional tautologies in $\langkm{\albet}$ are derivable in $\gtkms$.
\end{lemma}

\begin{proof}
See Appendix B (p.~\pageref{app:classical-completeness}) for details.
\end{proof}

\begin{theorem}[Completeness]\label{thm:completness-gtkms}
If $\vdash_{\kms} \phi$, then $\vdash_{\gtkms} \seqempstr \sar w : \phi$.
\end{theorem}

\begin{proof} By \lem~\ref{lem:classical-completeness-kms}, we know that all instances of classical propositional tautologies in $\langkm{\albet}$ are derivable in $\gtkms$, ensuring that A0 holds. Therefore, we need only show that the axioms A1 -- A4 and the rules R0 and R1 (see \dfn~\ref{def:axiomatization-km} for the axiomatization of $\kms$) are derivable in $\gtkms$ to show completeness.

\emph{Axiom A1.}

\begin{center}
\AxiomC{$\Pi_{1}$}

\AxiomC{$\Pi_{2}$}

\BinaryInfC{$R_{\chara}wu \sar w : \charadia (\phi \lor \negnnf{\psi}), u : \negnnf{\phi} \lor \psi, w : \charadia \negnnf{\phi}, u : \psi$}
\RightLabel{$\charadiar$}
\UnaryInfC{$R_{\chara}wu \sar w : \charadia (\phi \lor \negnnf{\psi}), w : \charadia \negnnf{\phi}, u : \psi$}
\RightLabel{$\charaboxr$}
\UnaryInfC{$\sar w : \charadia (\phi \lor \negnnf{\psi}), w : \charadia \negnnf{\phi}, w : \charabox \psi$}
\RightLabel{$\dis \times 2$}
\UnaryInfC{$\sar w : \charadia (\phi \lor \negnnf{\psi}) \lor (\charadia \negnnf{\phi} \lor \charabox \psi)$}
\RightLabel{=}
\dottedLine
\UnaryInfC{$\sar w : \charabox (\phi \cimp \psi) \cimp (\charabox \phi \cimp \charabox \psi)$}
\DisplayProof
\end{center}

\begin{center}
\begin{tabular}{c c c}
$\Pi_{1}$

&

$:= \Bigg \{$

&

\AxiomC{}
\RightLabel{Lem.~\ref{lem:general-id-kms}}
\dashedLine
\UnaryInfC{$R_{\chara}wu \sar w : \charadia (\negnnf{\phi} \lor \psi), u : \phi, w : \charadia \negnnf{\phi}, u : \negnnf{\phi}, u : \psi$}
\RightLabel{$\charadiar$}
\UnaryInfC{$R_{\chara}wu \sar w : \charadia (\negnnf{\phi} \lor \psi), u : \phi, w : \charadia \negnnf{\phi}, u : \psi$}
\DisplayProof
\end{tabular}
\end{center}

\begin{center}
\begin{tabular}{c c c}
$\Pi_{2}$

&

$:= \Bigg \{$

&

\AxiomC{}
\RightLabel{Lem.~\ref{lem:general-id-kms}}
\dashedLine
\UnaryInfC{$R_{\chara}wu \sar w : \charadia (\negnnf{\phi} \lor \psi), u : \negnnf{\psi}, w : \charadia \negnnf{\phi}, u : \psi$}
\DisplayProof
\end{tabular}
\end{center}

\emph{Rule R0.} In the derivation of R0, the use of $\cut$ is eliminable due to \thm~\ref{thm:cut-admiss-kms}.

\begin{center}
\AxiomC{$\sar w : \phi \cimp \psi$}
\RightLabel{=}
\dottedLine
\UnaryInfC{$\sar w : \negnnf{\phi} \lor \psi$}
\RightLabel{\lem~\ref{lem:hp-invert-kms}}
\dashedLine
\UnaryInfC{$\sar w : \negnnf{\phi}, w : \psi$}

\AxiomC{$\sar w : \phi$}
\RightLabel{$\cut$}
\dashedLine
\BinaryInfC{$\sar w : \psi$}
\DisplayProof
\end{center}

\emph{Axiom A2.}

\begin{center}
\AxiomC{}
\RightLabel{Lem.~\ref{lem:general-id-kms}}
\dashedLine
\UnaryInfC{$R_{\chara}wu, R_{\conv{\chara}}uw \sar w : \negnnf{\phi}, u : \charadiac \phi, w : \phi$}

\RightLabel{$\charadiar$}
\UnaryInfC{$R_{\chara}wu, R_{\conv{\chara}}uw \sar w : \negnnf{\phi}, u : \charadiac \phi$}

\RightLabel{$\convr$}
\UnaryInfC{$R_{\chara}wu \sar w : \negnnf{\phi}, u : \charadiac \phi$}
\RightLabel{$\charaboxr$}
\UnaryInfC{$\sar w : \negnnf{\phi}, w : \charabox \charadiac \phi$}
\RightLabel{$\dis$}
\UnaryInfC{$\sar w : \negnnf{\phi} \lor \charabox \charadiac \phi$}
\RightLabel{=}
\dottedLine
\UnaryInfC{$\sar w : \phi \rightarrow \charabox \charadiac \phi$}
\DisplayProof
\end{center}

\emph{Axiom A3 and rule R1.} The use of $\lsub$ in the proof of R1 is admissible due to the use of \lem~\ref{lem:struc-rules-admiss-kms}.

\begin{center}
\begin{tabular}{c c}
\AxiomC{}
\RightLabel{Lem.~\ref{lem:general-id-kms}}
\dashedLine
\UnaryInfC{$R_{\stra}wu, R_{\chara}wu \sar u : \negnnf{\phi}, w : \charadia \phi, u : \phi$}
\RightLabel{$\charadiar$}
\UnaryInfC{$R_{\stra}wu, R_{\chara}wu \sar u : \negnnf{\phi}, w : \charadia \phi$}
\RightLabel{$\psr$}
\UnaryInfC{$R_{\stra}wu \sar u : \negnnf{\phi}, w : \charadia \phi$}
\RightLabel{$\charaboxr \times |\stra|$}
\UnaryInfC{$\sar w : [ \stra ] \negnnf{\phi}, w : \charadia \phi$}
\RightLabel{$\dis$}
\UnaryInfC{$\sar w : [ \stra ] \negnnf{\phi} \lor \charadia \phi$}
\RightLabel{=}
\dottedLine
\UnaryInfC{$\sar w : \langle \stra \rangle \phi \cimp \charadia \phi$}
\DisplayProof

&

\AxiomC{$\sar w : \phi$}
\RightLabel{$\wk$}
\dashedLine
\UnaryInfC{$R_{\chara}uw \sar w : \psi$}
\RightLabel{$\charabox$}
\UnaryInfC{$\sar u : \charabox \psi$}
\RightLabel{$\lsub$}
\dashedLine
\UnaryInfC{$\sar w : \charabox \psi$}
\DisplayProof
\end{tabular}
\end{center}


\end{proof}

An interesting question to ask of labelled calculi is: what structures are necessary to ensure completeness? Once we \emph{refine} our labelled calculi (in the following chapter), we will see that the labelled sequents needed to ensure completeness in the refined versions of our calculi are more minimalistic than those used in the $\gtkms$ calculi. To make this notion of ``structure'' precise, we define \emph{sequent graphs} (for grammar logics) below:

\begin{definition}[Sequent Graph for Grammar Logics]\label{def:sequent-graph-kms} Let $\Lambda := \rel \sar \Gamma$ be a labelled sequent. We define the \emph{sequent graph}\index{Sequent graph!for grammar logics} of $\Lambda$, $\seqgraph(\Lambda) = (V,E,L)$, as follows:
\begin{itemize}

\item[$\li$] $V = \lab(\Lambda)$

\item[$\li$] $E = \{(w,u,\chara) \ | \ R_{\chara}wu \in \rel\}$

\item[$\li$] $L(w) = \Gamma \restriction w$

\end{itemize}
(NB. Recall that $\Gamma \restriction w$ was defined in \dfn~\ref{def:restirction-prepend-labelled}.)
\end{definition}

\begin{example}\label{ex:sequent-graph-examples-kms} Below, we give an example of a labelled tree sequent $\Lambda$ (see \dfn~\ref{def:tree-sequent-kms}) and its corresponding sequent graph $\seqgraph(\Lambda)$.

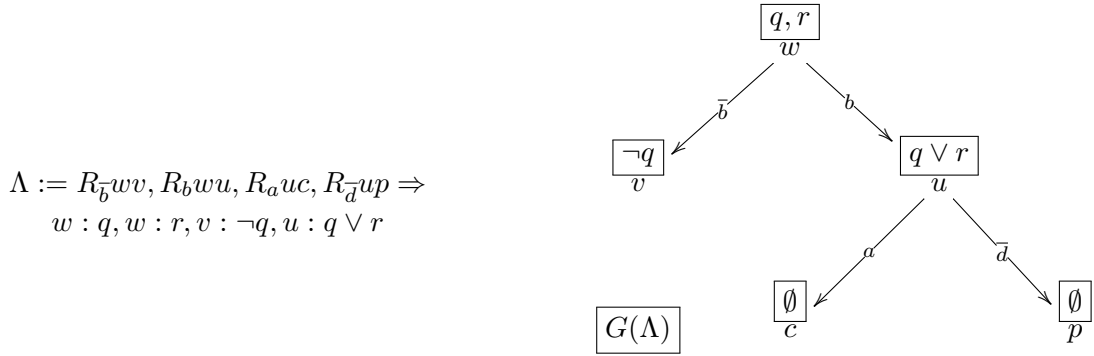
\begin{minipage}[t]{.33\textwidth}
\begin{tabular}{@{\hskip -.05em} c}
\vspace*{1 em}
\ \\
\AxiomC{ }
\noLine
\UnaryInfC{}
\noLine
\UnaryInfC{}
\noLine
\UnaryInfC{$\Lambda := R_{\conv{b}}wv, R_{b}wu, R_{a}uc, R_{\conv{d}}up \sar $}
\noLine
\UnaryInfC{$w : q, w : r, v : \negnnf{q}, u : q \lor r$}
\noLine
\UnaryInfC{}
\noLine
\UnaryInfC{ }
\noLine
\UnaryInfC{}
\noLine
\UnaryInfC{}
\DisplayProof
\end{tabular}
\end{minipage}
\begin{minipage}[t]{.15\textwidth}
\ 
\end{minipage}
\begin{minipage}[t]{.33\textwidth}
\begin{tabular}{c}
\xymatrix{
   & \overset{\boxed{q,r}}{w} \ar[dl]|-{\conv{b}} \ar[dr]|-{b} & &  \\
  \overset{\boxed{\negnnf{q}}}{v} & & \overset{\boxed{q \lor r}}{u} \ar[dl]|-{a}\ar[dr]|-{\conv{d}} & \\ 
\boxed{\seqgraph(\Lambda)}  & \overset{\boxed{\emptyset}}{c} &  & \overset{\boxed{\emptyset}}{p}
}
\end{tabular}
\end{minipage}

\end{example}

In the following chapter, for each calculus $\gtkms$, we will obtain a refined variant which makes use of labelled sequents whose graphs possess less structure, namely, \emph{labelled tree sequents}~\cite{GorRam12}. Such sequents are essentially the same as the labelled nested sequents presented in~\cite{Pim18}. We define labelled tree sequents here and provide a theorem below which shows that restricting to such sequents invalidates the completeness of each calculus $\gtkms$, that is, each calculus $\gtkms$ requires labelled sequents with a more liberal syntactic structure to ensure completeness.

\begin{definition}[Labelled Tree Sequent for Grammar Logics]\label{def:tree-sequent-kms} A labelled sequent $\Lambda$ is a \emph{labelled tree sequent}\index{Labelled tree sequent!for grammar logics} \ifandonlyif $\seqgraph(\Lambda) = (V,E,L)$ is a \emph{tree}.


\end{definition}

\begin{definition}[Labelled Tree Proof, Fixed Root Property for Grammar Logics]\label{def-tree-proof-kms} We say that a proof is a \emph{labelled tree proof}\index{Labelled tree proof!for grammar logics} \ifandonlyif it consists solely of labelled tree sequents.

Also, we say that a labelled tree proof has the \emph{fixed root property}\index{Fixed root property!for grammar logics} \ifandonlyif the every labelled tree sequent in the proof has the same root.
\end{definition}

As specified in the theorem below, if we restricted $\gtkms$ to only allow labelled tree derivations, then there would be theorems of $\kms$ that would no longer be provable. In such a situation, $\gtkms$ would be \emph{incomplete}\index{Incomplete} relative to $\kms$. We will refer to a calculus as \emph{incomplete} when it does not derive all theorems of its intended, associated logic.

\begin{theorem}\label{thm:sequent-structure-gtkms}
Let $\thuesys$ be a \cfcst system. The calculus $\gtkms$ is incomplete relative to labelled tree derivations.
\end{theorem}

\begin{proof} Since $\albet$ is non-empty by definition, we know there exists an $\chara \in \albet$. The claim follows by considering the proof of $p \rightarrow \charabox \charadiac p$ (an instance of axiom A3), which requires sequents that are not labelled tree sequents.
\end{proof}

\section{Labelled Calculi for First-Order Intuitionistic Logics}\label{sec:lab-calc-intFO}

In this section, we provide labelled sequent calculi for first-order intuitionistic logic with non-constant domains (also called, \emph{first-order intuitionistic logic proper}), and first-order intuitionistic logic with constant domains (introduced by Grzegorczyk in~\cite{Grz64}). A sequent calculus for the former has been in existence since the initiation of the sequent formalism by Gentzen~\cite{Gen35a,Gen35b}. Yet, sequent calculi for the latter were not provided until nearer the close of the century~\cite{KasShi94,FioMig99}. In~\cite{Fit14}, Fitting showed that both logics could be uniformly captured within the prefixed tableau and nested sequent formalisms, and also defined translations between the distinct calculi, thus showing their deductive equivalence.\footnote{In 1983~\cite{Fit83}, Fitting sketched a prefixed tableau calculus for \emph{propositional} intuitionistic logic, which stood as a basis for his work in the first-order setting.}

As it so happens, the labelled sequent calculi defined in this section have a natural relationship with Fitting's nested calculi. In~\cite{Lyo21} it was shown that the labelled calculi for the first-order intuitionistic logics $\intfond$ and $\intfocd$ can be transformed into Fitting's nested calculi for the logics via the refinement process (with some ``additional adjustments''). Also, both~\cite{Lyo20a,Pim18} show how labelled calculi for propositional intuitionistic logic can be transformed into nested calculi that are identical to, or equivalent to, Fitting's nested calculus for the logic, respectively. In the next chapter, we will show how refining the labelled calculi for first-order intuitionistic logics begets nested calculi that are related to Fitting's. 
 As will be commented on, these nested variants allow for a higher degree of modularity than Fitting's nested calculi 
 via the modification of side conditions imposed on certain logical rules.

The labelled calculi presented in this section are defined by extending the labelled calculus $\mathsf{G3I}$ for propositional intuitionistic logic of Dyckhoff and Negri~\cite{DycNeg12} with quantifier rules and rules concerning domains. Some of these quantifier and domain rules can be found in Vigan{\`o}~\cite{Vig00} and in~\cite{NegPla11}, where the authors provide labelled sequent calculi for first-order modal logics. These works however do not provide rules for the intuitionistic (i.e. strong) universal quantifier or the first-order monotonicity condition \moncondns, and so, we present rules for these in this section. Since our labelled sequent calculi are based on the work and formalisms of~\cite{DycNeg12,NegPla11,Sim94,Vig00}, so are our labelled sequents; these are defined as follows:

\begin{definition}[Labelled Sequents for First-Order Intuitionistic Logics]\label{def:labelled-sequents-FO-Int} A \emph{labelled sequent} for first-order intuitionistic logics\index{Labelled sequent!for first-order intuitionistic logics} is a syntactic object of the form $\Lambda := L_{1} \sar L_{2}$, where $L_{1}$ (the \emph{antecedent}) and $L_{2}$ (the \emph{consequent}) are defined via the following grammars in BNF:
\begin{center}
\begin{tabular}{c @{\hskip 2em} c}
$L_{1} ::= \seqempstr \ |\ w : \phi \ | \ \unda \in D_{w} \ | \ w \leq u \ | \ L_{1},L_{1}$

&

$L_{2} ::= \seqempstr \ |\ w : \phi \ | \ L_{2},L_{2}$
\end{tabular}
\end{center}
with $\phi \in \langintfo$, $\unda$ among a denumerable set of \emph{parameters} $\para = \{\unda, \undb, \undc, \ldots\}$, and $w$, $u$ among a denumerable set of labels $\lab = \{w, u, v, \ldots \}$. We refer to formulae of the forms $w \leq u$ and $\unda \in D_{w}$ as \emph{relational atoms}\index{Relational atom!for first-order intuitionistic logics} (with formulae of the form $\unda \in D_{w}$ referred to as \emph{domain atoms}\index{Domain atom}, more specifically) and refer to formulae of the form $w : \phi$ as \emph{labelled formulae}\index{Labelled formula!for first-order intuitionistic logics}. 
\end{definition}

We use $\Lambda$, $\Lambda'$, \etc (occasionally annotated) to denote labelled sequents as a whole, and due to the two types of formulae occurring in a labelled sequent, we use $\rel$, $\rel'$, \etc (occasionally annotated) to denote multisets of relational atoms, and $\Gamma$, $\Gamma'$, $\Delta$, $\Delta'$, \etc (occasionally annotated) to denote multisets of labelled formulae, thus distinguishing between the two. We therefore take the comma operator to be commutative and associative; for example, we identify the labelled sequent $w \leq u, w : \phi, \unda \in D_{w} \sar v : \psi, u : \chi$ with $\unda \in D_{w}, w \leq u, w : \phi \sar u : \chi, v : \psi$. This means that we may write a labelled sequent $\Lambda$ in a general form as $\rel, \Gamma \sar \Delta$, where we separate the relational atoms from the labelled formulae in the antecedent. Moreover, our interpretation of comma is what lets us view the antecedent $\rel, \Gamma$ and the consequent $\Delta$ of a labelled sequent $\rel, \Gamma \sar \Delta$ as multisets. We use $\lab(\Lambda)$, $\lab(\rel)$, and $\lab(\Gamma)$ to denote the sets of labels that occur in a labelled sequent $\Lambda$, a multiset $\rel$ of relational atoms, and a multiset $\Gamma$ of labelled formulae, respectively. Also, as in the previous section, we use $\seqempstr$ to denote the \emph{empty string}\index{Empty string} which acts as the identity element for the comma operator (e.g. we identify $w \leq v, \seqempstr, v : \psi$ with $w \leq v, v : \psi$). 
Consequently, $\seqempstr$ will often be implicit in labelled sequents. 

We syntactically distinguish between \emph{bound variables} $\{x, y, z, \ldots\}$ and \emph{free variables}, which are replaced with \emph{parameters}\index{Parameter} $\{\unda, \undb, \undc, \ldots\}$, to avoid clashes between the two categories (cf.~\cite[Sect.~8]{Fit14}). Instead of making use of formulae from the first-order language $\langintfo$, we use formulae from the first-order language where each freely occurring variable $x$ has been replaced by a distinct parameter $\unda$. 
For example, we would make use of the labelled formula $w : (\forall x) p(\unda,x) \lor q(\unda,\undb)$ instead of $w : (\forall x) p(y,x) \lor q(y,z)$ in a labelled sequent. 

We use the notation $\phi(\unda_{1}, \ldots, \unda_{n})$, with $n \in \mathbb{N}$, to denote that the parameters $\unda_{1}, \ldots, \unda_{n}$ are all parameters occurring in the formula $\phi$. (NB. If $n=0$, then we assume that the formula $\phi$ does not contain any parameters). We write $\phi(\vec{\unda})$ as shorthand for $\phi(\unda_{1}, \ldots, \unda_{n})$ and $\vec{\unda} \in D_{w}$ as shorthand for $\unda_{1} \in D_{w}, \ldots, \unda_{n} \in D_{w}$. Furthermore, the notation $\phi(\unda/x)$ represents the formula obtained by substituting the parameter $\unda$ for each free occurrence of $x$ in $\phi$. Substitutions of the form $(\unda/x)$ are formally defined as in \dfn~\ref{def:substitutions-FO-Int} in \sect~\ref{SEC:FO-Int-Logics}.

\begin{figure}[t]
\noindent\hrule

\begin{center}
\begin{tabular}{c @{\hskip 1em} c} 

\AxiomC{}
\RightLabel{$\idfo$\index{$\idfo$}}
\UnaryInfC{$\rel,w \leq u, \vec{\unda} \in D_{w},\Gamma, w :p(\vec{\unda}) \sar u :p(\vec{\unda}), \Delta$}
\DisplayProof

&

\AxiomC{}
\RightLabel{$\botl$\index{$\botl$}}
\UnaryInfC{$\rel, \Gamma, w :\bot \sar \Delta$}
\DisplayProof
\end{tabular}
\end{center}

\begin{center}
\begin{tabular}{c @{\hskip 1em} c @{\hskip 1em} c}

\AxiomC{$\rel, \Gamma, w :\phi, w :\psi \sar \Delta$}
\RightLabel{$\conl$\index{$\conl$}}
\UnaryInfC{$\rel, \Gamma, w :\phi \wedge \psi \sar \Delta$}
\DisplayProof

&

\AxiomC{$\rel, \Gamma \sar w :\phi, \Delta$}
\AxiomC{$\rel, \Gamma \sar w :\psi, \Delta$}
\RightLabel{$\conr$\index{$\conr$}}
\BinaryInfC{$\rel, \Gamma \sar w :\phi \wedge \psi, \Delta$}
\DisplayProof

\end{tabular}
\end{center}

\begin{center}
\begin{tabular}{c @{\hskip 1em} c}
\AxiomC{$\rel, \Gamma, w :\phi \sar \Delta$}
\AxiomC{$\rel, \Gamma, w :\psi \sar \Delta$}
\RightLabel{$\disl$\index{$\disl$}}
\BinaryInfC{$\rel, \Gamma, w :\phi \vee \psi \sar \Delta$}
\DisplayProof

&

\AxiomC{$\rel,w \leq w, \Gamma \sar \Delta$}
\RightLabel{$\refl$\index{$\refl$}}
\UnaryInfC{$\rel,\Gamma \sar \Delta$}
\DisplayProof
\end{tabular}
\end{center}

\begin{center}
\begin{tabular}{c}
\AxiomC{$\rel,w \leq u, \Gamma, w :\phi \imp \psi \sar \Delta, u :\phi$}
\AxiomC{$\rel,w \leq u, \Gamma, w :\phi \imp \psi, u :\psi \sar \Delta$}
\RightLabel{$\impl$\index{$\impl$}}
\BinaryInfC{$\rel,w \leq u, \Gamma, w :\phi \imp \psi \sar \Delta$}
\DisplayProof 
\end{tabular}
\end{center}

\begin{center}
\begin{tabular}{c c}
\AxiomC{$\rel, w \leq u, \Gamma, u :\phi \sar u :\psi, \Delta$}
\RightLabel{$\impr^{\dag_{1}}$\index{$\impr$}}
\UnaryInfC{$\rel, \Gamma \sar w :\phi \imp \psi, \Delta$}
\DisplayProof

&

\AxiomC{$\rel, \Gamma \sar w :\phi, w :\psi, \Delta$}
\RightLabel{$\disr$\index{$\disr$}}
\UnaryInfC{$\rel, \Gamma \sar w :\phi \vee \psi, \Delta$}
\DisplayProof
\end{tabular}
\end{center}

\begin{center}
\begin{tabular}{c @{\hskip 1em} c}
\AxiomC{$\rel,w \leq u, u \leq v, w \leq v, \Gamma \sar \Delta$}
\RightLabel{$\trans$\index{$\trans$}}
\UnaryInfC{$\rel,w \leq u, u \leq v, \Gamma \sar \Delta$}
\DisplayProof

&

\AxiomC{$\rel, \unda \in D_{w}, \Gamma \sar \Delta$}
\RightLabel{$\ned^{\dag_{2}}$\index{$\ned$}}
\UnaryInfC{$\rel,\Gamma \sar \Delta$}
\DisplayProof
\end{tabular}
\end{center}

\begin{center}
\resizebox{\columnwidth}{!}{
\begin{tabular}{c c} 
\AxiomC{$\rel, w \leq u, \unda \in D_{u}, \Gamma \sar  u : \phi(\unda/x), \Delta$}
\RightLabel{$\allr^{\dag_{3}}$\index{$\allr$}}
\UnaryInfC{$\rel, \Gamma \sar w : \forall x \phi, \Delta$}
\DisplayProof

&

\AxiomC{$\rel, \unda \in D_{w}, \Gamma \sar w: \phi(\unda/x), w: \exists x \phi, \Delta$}
\RightLabel{$\existsr$\index{$\existsr$}}
\UnaryInfC{$\rel, \unda \in D_{w}, \Gamma \sar w: \exists x \phi, \Delta$}
\DisplayProof

\end{tabular}
}
\end{center}

\begin{center}
\resizebox{\columnwidth}{!}{
\begin{tabular}{c c} 

\AxiomC{$\rel, \unda \in D_{w}, \Gamma, w: \phi(\unda/x) \sar \Delta$}
\RightLabel{$\existsl^{\dag_{2}}$\index{$\existsl$}}
\UnaryInfC{$\rel, \Gamma, w : \exists x \phi \sar \Delta$}
\DisplayProof

&

\AxiomC{$\rel, w \leq u, \unda \in D_{u}, \Gamma, u : \phi(\unda/x), w : \forall x \phi \sar \Delta$}
\RightLabel{$\alll$\index{$\alll$}}
\UnaryInfC{$\rel, w \leq u, \unda \in D_{u}, \Gamma, w : \forall x \phi \sar \Delta$}
\DisplayProof

\end{tabular}
}
\end{center}

\begin{center}
\begin{tabular}{c @{\hskip 1em} c}

\AxiomC{$\rel, w \leq u, \unda \in D_{w}, \unda \in D_{u}, \Gamma \sar \Delta$}
\RightLabel{$\nd$\index{$\nd$}}
\UnaryInfC{$\rel, w \leq u, \unda \in D_{w}, \Gamma \sar \Delta$}
\DisplayProof

&

\AxiomC{$\rel, u \leq w, \unda \in D_{u}, \unda \in D_{w}, \Gamma \sar \Delta$}
\RightLabel{$\cd$\index{$\cd$}}
\UnaryInfC{$\rel, u \leq w, \unda \in D_{w}, \Gamma \sar \Delta$}
\DisplayProof

\end{tabular}
\end{center}

\hrule
\caption{The labelled calculus $\gtintfond$\index{$\gtintfond$} for $\intfond$ consists of all rules minus the $\cd$ rule, and all rules give the calculus $\gtintfocd$\index{$\gtintfocd$} for $\intfocd$. The side condition $\dag_{1}$ states that the variable $u$ does not occur in the conclusion, $\dag_{2}$ states that $\unda$ does not occur in the conclusion, and $\dag_{3}$ states that neither $\unda$ nor $u$ occurs in the conclusion. As usual, labels and parameters restricted from occurring in the conclusion of an inference are called \emph{eigenvariables}.}
\label{fig:labelled-calculi-FO-Int}
\end{figure}

Our labelled calculi are given in Fig.~\ref{fig:labelled-calculi-FO-Int}, and are obtained from the models and semantic clauses of $\intfond$ and $\intfocd$. The derivability relations for our calculi are defined as follows:

\begin{definition} We write $\vdash_{\gtintfond} \Lambda$ and $\vdash_{\gtintfocd} \Lambda$ to indicate that the labelled sequent $\Lambda$ is derivable in $\gtintfond$ and $\gtintfocd$, respectively.
\end{definition}

The rules for $\botl$, $\conl$, $\conr$, $\disl$, $\disr$, $\impl$, $\impr$, $\existsl$, $\existsr$, $\alll$, and $\allr$ are rule representations of the semantic clauses given in \dfn~\ref{def:semantics-intfo}. The $\refl$ and $\trans$ rules allow inferences arising from the fact that frames (\dfn~\ref{def:IntFO-frame-model}) are reflexive and transitive, whereas $\ned$, $\nd$, and $\cd$ allow inferences based on the fact that the domains of frames are always inhabited (i.e. non-empty), satisfy the nested domain condition \ndcondns, and satisfy the constant domain condition \cdcondns, respectively. The rule $\idfo$ encodes the monotonicity condition \moncond imposed on models (\dfn~\ref{def:IntFO-frame-model}). Let us also comment on the new $\alll$ and $\allr$ rules for introducing universal quantifiers: since in the first-order intuitionistic setting the satisfaction of a universally quantified formula is determined not only by logical information holding at the present world, but at all future worlds, the corresponding logical rules require information at future labels to be considered when applying the rule. 

\begin{remark}\label{rem:G3I} We note that when the predicate $p$ is of arity $0$ (i.e. $p$ is a propositional atom), the $\idfo$ rule is of the following form:
\begin{center}
\AxiomC{}
\RightLabel{$\idfo$}
\UnaryInfC{$\rel,w \leq u, \Gamma, w :p \sar u :p, \Delta$}
\DisplayProof
\end{center}
The calculus obtained by taking $\botl$, $\conl$, $\conr$, $\disl$, $\disr$, $\impl$, $\impr$, $\refl$, $\trans$, and $\idfo$ restricted to the use of propositional atoms (i.e. the form of $\idfo$ shown above) is referred to as \textbf{G3I}, and is a sound and complete labelled calculus for propositional intuitionistic logic~\cite{DycNeg12}. The above instance of our $\idfo$ rule generates the initial sequents of \textbf{G3I} when we restrict ourselves to the propositional setting~\cite{DycNeg12}. Thus, our initial sequents are first-order generalizations of the initial sequents utilized for propositional intuitionistic logic.

Furthermore, the rules $\existsl$, $\existsr$, $\nd$, $\cd$, and $\ned$ appear in Vigan{\`o}~\cite[\cptr~6]{Vig00} and in~\cite[\cptr~12]{NegPla11}; however, the $\idfo$, $\alll$, and $\allr$ rules are (to the best of the author's knowledge) new.
\end{remark}

In fact, not only can the above calculi be seen as proof-theoretic encodings of semantic clauses (\dfn~\ref{def:semantics-intfo}) and frame properties (\dfn~\ref{def:IntFO-frame-model}) for $\intfond$ and $\intfocd$, but labelled sequents can be seen as abstractions of $\intfond$- and $\intfocd$-models. The relationship between the syntactic structures present in labelled sequents (\dfn~\ref{def:labelled-sequents-FO-Int}) and the semantics for the logics $\intfond$ and $\intfocd$ is spelled out formally in the definition below (\dfn~\ref{def:sequent-semantics-FO-Int}). Moreover, we also leverage the semantics for labelled sequents to define a notion of validity for labelled sequents, which is used in the subsequent theorem (\thm~\ref{thm:soundness-FO-Int}) to prove our calculi sound.

\begin{definition}[$\gtintfond$ and $\gtintfocd$ Semantics]\label{def:sequent-semantics-FO-Int} Let $\mathsf{IntX} \in \{\intfond, \intfocd\}$ and $M = (W, \leq, D, V)$ be an $\mathsf{IntX}$-model with $\overalldom = \bigcup_{w \in W} D_{w}$. We let $I : \ \lab \cup \para \mapsto W \cup \overalldom$ be an \emph{interpretation function}\index{Interpretation function!for first-order intuitionistic logics} mapping labels to worlds and parameters to elements of $\overalldom$.

We define the \emph{satisfaction}\index{Satisfaction!for $\gtintfond$ and $\gtintfocd$} of a relational atom $w \leq u$ or $\unda \in D_{w}$ (written $M, I \models_{\mathsf{IntX}} w \leq u$ and $M, I \models_{\mathsf{IntX}} \unda \in D_{w}$, \resp), and labelled formula $w : \phi$ (written $M, I \models_{\mathsf{IntX}} w : \phi$) as follows:
\begin{itemize}

\item[$\li$] $M, I \models_{\mathsf{IntX}} w \leq u$ \ifandonlyif $I(w) \leq I(u)$

\item[$\li$] $M, I \models_{\mathsf{IntX}} \unda \in D_{w}$ \ifandonlyif $I(\unda) \in D_{I(w)}$

\item[$\li$] $M, I \models_{\mathsf{IntX}} w : \phi$ \ifandonlyif $M, I(w) \Vdash \phi$

\end{itemize}
We say that a multiset of relational atoms $\rel$ is \emph{satisfied} in $M$ with $I$ (written $M, I \models_{\mathsf{IntX}} \rel$) \ifandonlyif $M, I \models_{\mathsf{IntX}} w \leq u$ and $M, I \models_{\mathsf{IntX}} \unda \in D_{w}$ for all $w \leq u, \unda \in D_{w} \in \rel$, and we say that a multiset of labelled formulae $\Gamma$ is \emph{satisfied} in $M$ with $I$ (written $M, I \models_{\mathsf{IntX}} \Gamma$) \ifandonlyif $M, I \models_{\mathsf{IntX}} w : \phi$ for all $w : \phi \in \Gamma$. 

A labelled sequent $\Lambda = \rel, \Gamma \sar \Delta$ is \emph{satisfied} in $M$ with $I$ (written, $M,I \models_{\mathsf{IntX}} \Lambda$) \ifandonlyif if $M, I \models_{\mathsf{IntX}} \rel$ and $M, I \models_{\mathsf{IntX}} \Gamma$, then $M, I \models_{\mathsf{IntX}} \Delta$. Also, we say that a labelled sequent $\Lambda$ is \emph{falsified} in $M$ with $I$ \ifandonlyif $M, I \not\models_{\mathsf{IntX}} \Lambda$, that is, $\Lambda$ is not satisfied by $M$ with $I$.

Last, a labelled sequent $\Lambda$ is \emph{$\mathsf{IntX}$-valid}\index{$\intfond$-valid}\index{$\intfocd$-valid} (written $\models_{\mathsf{IntX}} \Lambda$) \ifandonlyif it is satisfiable in every $\mathsf{IntX}$-model $M$ with every interpretation function $I$. We say that a labelled sequent $\Lambda$ is \emph{$\mathsf{IntX}$-invalid}\index{$\intfond$-invalid}\index{$\intfocd$-invalid} \ifandonlyif $\not\models_{\mathsf{IntX}} \Lambda$, i.e. $\Lambda$ is not $\mathsf{IntX}$-valid.
\end{definition}

By making use of the above notion of $\intfond$- and $\intfocd$-validity for labelled sequents, we prove our calculi $\gtintfond$ and $\gtintfocd$ sound.

\begin{theorem}[Soundness]\label{thm:soundness-FO-Int}
Let $\mathsf{X} \in \{\nnn, \ccc\}$. If $\vdash_{\mathsf{G3IntX}} \Lambda$, then $\models_{\mathsf{IntX}} \Lambda$.
\end{theorem}

\begin{proof} Let $\mathsf{IntX} \in \{\intfond, \intfocd\}$. With the exception of the $\idfo$, $\alll$, and $\allr$ rules, we can leverage the work of~\cite{DycNeg12,NegPla11,Vig00} to confirm that all other rules are sound. Therefore, we show that $\idfo$ produces only valid sequents, and that $\alll$ and $\allr$ preserve validity.

$\idfo$ Let $M = (W,\leq,D,V)$ be an arbitrary $\mathsf{IntX}$-model and $I$ be an arbitrary interpretation function. Suppose that the antecedent $\rel, w \leq u, \vec{\unda} \in D_{w}, w : p(\vec{\unda}), \Gamma$ is satisfied in $M$ with $I$. Then, $I(w) \leq I(u)$, $I(\unda_{1}) \in D_{I(w)}$, $\ldots$, $I(\unda_{n}) \in D_{I(w)}$, and $M, I(w) \Vdash p(\vec{\unda})$. These facts, along with the monotonicity condition \moncondns, implies that $(I(\unda_{1}), \ldots, I(\unda_{n})) \in V(p,I(w)) \subseteq V(p,I(u))$. Hence, $M, I(u) \Vdash p(\vec{\unda})$, showing that any sequent generated by the $\idfo$ rule is $\mathsf{IntX}$-valid.

$\alll$ Let $M$ be an $\mathsf{IntX}$-model and $I$ be an interpretation function such that $M, I \not\models \rel, w \leq u, \unda \in D_{u}, \Gamma, w : \forall x \phi \sar \Delta$. Thus, $I(w) \leq I(u)$, $I(\unda) \in D_{I(u)}$, and $M, I(w) \Vdash \forall x \phi$. It follows that $M, I(u) \Vdash \phi(\unda/x)$, showing that the premise is falsified by $M$ with $I$.

$\allr$ Suppose that $M$ is an $\mathsf{IntX}$-model and $I$ is an interpretation function such that $M, I \not\models \rel, \Gamma \sar w : \forall x \phi, \Delta$. Then, $M, I(w) \not\Vdash \forall x \phi$, which implies that there exists a world $u'$ and an $a \in D_{u'}$ such that $I(w) \leq u'$ and $M, u' \not\Vdash \phi(\unda/x)$. Let us define $I'(\undb) = I(\undb)$ for $\undb \neq \unda$ and $I'(\unda) = a$, and $I'(v) = I(v)$ for $v \neq u'$ and $I'(u) = u'$. Therefore, $I'(w) \leq I'(u)$, $I'(\unda) \in D_{I'(u)}$, and $M, I'(u) \not\Vdash \phi(\unda/x)$. Since $u$ and $\unda$ are eigenvariable, it follows that the premise is falsified by $M$ with $I'$.
\end{proof}

We now go on to show that the labelled calculi possess proof-theoretic properties such as the hp-admissibility of substitutions and structural rules (e.g. $\psub$ and $\wk$), hp-invertibility of all rules, and syntactic cut-elimination. The hp-admissible and eliminable structural rules are presented in \fig~\ref{fig:lab-struc-rules-FO-Int}. Ultimately, these properties will let us prove the completeness of our calculi (\thm~\ref{thm:completness-FO-Int}). While proving our results, we adapt proofs and results from~\cite{DycNeg12,NegPla11}, though, since these works do not explicitly consider first-order intuitionistic logic with (non-)constant domains, we still consider a few cases in our proofs.

Before moving on to showing that our calculi possess useful properties, we define the notion of a \emph{label substitution}\index{Label substitution!for first-order intuitionistic logics} and \emph{parameter substitution}\index{Parameter substitution}. Intuitively, a label substitution $(w/u)$ on a multiset of relational atoms or labelled formulae replaces all occurrences of the label $u$ with the label $w$, and a parameter substitution replaces all occurrences of the parameter $\undb$ with the parameter $\unda$. We formally define these operations below:

\begin{definition}[Label and Parameter Substitution]\label{lem:label-param-sub-FO-Int} We define a \emph{label substitution} $(w/u)$ for $w, u \in \lab$ on individual relational atoms and labelled formulae as follows:
\begin{itemize}

\item[$\li$] $(v \leq z)(w/u) =
  \begin{cases}
                                   w \leq z & \text{if $u = v$ and $v \neq z$} \\
                                   w \leq w & \text{if $u = v$ and $v = z$} \\
                                   u \leq w & \text{if $u = z$ and $v \neq z$} \\
                                   v \leq z & \text{otherwise}
  \end{cases}
$

\item[$\li$] $(\unda \in D_{v})(w/u) =
  \begin{cases}
                                   \unda \in D_{w} & \text{if $u = v$} \\
                                   \unda \in D_{v} & \text{otherwise}
  \end{cases}
$

\item[$\li$] $(v : \phi)(w/u) =
  \begin{cases}
                                   w : \phi & \text{if $u = v$} \\
                                   v : \phi & \text{otherwise}
  \end{cases}
$
\end{itemize}
We define the label substitution $(w/u)$ on a multiset of relational atoms $\rel$ and a multiset of labelled formulae $\Gamma$ to be the multiset obtained by applying $(w/u)$ to each element of the multiset.

We define a \emph{parameter substitution} $(\unda/\undb)$ for $\unda, \undb \in \para$ on individual relational atoms and labelled formulae as follows:
\begin{itemize}

\item[$\li$] $(w \leq u)(\unda/\undb) := w \leq u$

\item[$\li$] $(\undc \in D_{v})(\unda/\undb) := \unda \in D_{v}$ if $\undb = \undc$, and $(\undc \in D_{v})(\unda/\undb) := \undc \in D_{v}$ otherwise.

\item[$\li$] $(w : \phi)(\unda/\undb) := w : \phi(\unda/\undb)$

\end{itemize}
Note that for the last displayed parameter substitution (on a labelled formula) we make use of the notion of a parameter substitution on a formula, which is defined as in \dfn~\ref{def:substitutions-FO-Int}.

We define the parameter substitution $(\unda/\undb)$ on a multiset of relational atoms $\rel$ and a multiset of labelled formulae $\Gamma$ to be the multiset obtained by applying $(\unda/\undb)$ to each element of the multiset.
\end{definition}

\begin{figure}[t]
\noindent\hrule
\begin{center}
\begin{tabular}{c @{\hskip 1em} c}
\AxiomC{$\rel,\Gamma \sar \Delta$}
\RightLabel{$\lsub$}
\UnaryInfC{$\R(w/u),\Gamma(w/u) \sar \Delta(w/u)$}
\DisplayProof

&

\AxiomC{$\rel,\Gamma \sar \Delta$}
\RightLabel{$\psub$}
\UnaryInfC{$\R(\unda/\undb),\Gamma(\unda/\undb) \sar \Delta(\unda/\undb)$}
\DisplayProof
\end{tabular}
\end{center}

\begin{center}
\begin{tabular}{c @{\hskip 1em} c @{\hskip 1em} c}
\AxiomC{$\rel,\Gamma \sar \Delta$}
\RightLabel{$\wk$}
\UnaryInfC{$\rel',\rel,\Gamma',\Gamma \sar \Delta',\Delta$}
\DisplayProof

&

\AxiomC{$\rel,\rel',\rel',\Gamma \sar \Delta$}
\RightLabel{$\ctrrel$}
\UnaryInfC{$\rel,\rel',\Gamma \sar \Delta$}
\DisplayProof

&

\AxiomC{$\rel,w:\phi,w:\phi,\Gamma \sar \Delta$}
\RightLabel{$\ctrl$}
\UnaryInfC{$\rel,w:\phi,\Gamma \sar \Delta$}
\DisplayProof
\end{tabular}
\end{center}

\begin{center}
\begin{tabular}{c @{\hskip 1em} c}
\AxiomC{$\rel,\Gamma \sar w:\phi, w:\phi, \Delta$}
\RightLabel{$\ctrr$}
\UnaryInfC{$\rel,\Gamma \sar w:\phi, \Delta$}
\DisplayProof

&

\AxiomC{$\rel,\Gamma \sar \Delta, w :A$}
\AxiomC{$\rel,w :A,\Gamma \sar \Delta$}
\RightLabel{$\cut$}
\BinaryInfC{$\rel,\Gamma \sar \Delta$}
\DisplayProof
\end{tabular}
\end{center}

\hrule
\caption{The set $\strucsetint$ of structural rules\index{Structural rule!for first-order intuitionistic logics} consists of all rules shown above.}
\label{fig:lab-struc-rules-FO-Int}
\end{figure}

The first property we prove below is fundamental in securing completeness (\thm~\ref{thm:completness-FO-Int}), and ensures that all \emph{instances} of the axioms in $\h\intfond$ and $\h\intfocd$ (\dfn~\ref{def:axiomatization-IntFO}) are derivable in $\gtintfond$ and $\gtintfocd$, respectively.

\begin{lemma}\label{lem:general-id-FO-Int}
Let $\mathsf{G3X} \in \{\gtintfond, \gtintfocd\}$.

(i) For all $\phi \in \langintfo$, $\vdash_{\mathsf{G3X}} \rel,w \leq u, \vec{\unda} \in D_{w}, w : \phi(\vec{\unda}), \Gamma \sar u : \phi(\vec{\unda}), \Delta$.

(ii) For all $\phi \in \langintfo$, $\vdash_{\mathsf{G3X}} \rel, \vec{\unda} \in D_{w}, w:\phi(\vec{\unda}),\Gamma \sar \Delta, w :\phi(\vec{\unda})$.
\end{lemma}

\begin{proof} We prove claims (i) and (ii) for both calculi $\gtintfond$ and $\gtintfocd$ simultaneously by mutual induction on the complexity of $\phi$.

\textit{Base case for (i).} The base cases for claim (i) trivially follow from the $\idfo$ rule for atomic formulae and propositional variables, and from the $\botl$ rule for $\bot$. 

\textit{Inductive step for (i).} We provide the cases for when $\phi(\vec{\unda})$ is of the form $\exists x \psi(\vec{\unda})$ and $\forall x \psi(\vec{\unda})$, and omit the remaining cases which follow from~\cite[\lem~1]{DycNeg12}. In the $\exists x \psi(\vec{\unda})$ case, we let $\Delta' := u : \exists x \psi(\vec{a}), \Delta$, and in the $\forall x \psi(\vec{\unda})$ case, we let $\rel' = \rel, w \leq u, u \leq v, \vec{\unda} \in D_{w}$.

\begin{center}
\begin{tabular}{c}
\AxiomC{}
\RightLabel{\ih}
\dashedLine
\UnaryInfC{$\rel, w \leq u, \undb \in D_{w}, \vec{\unda} \in D_{w}, w : \psi(\vec{\unda})(\undb / x), \Gamma \sar u : \psi(\vec{\unda})(\undb / x), \Delta'$}
\RightLabel{$\existsr$}
\UnaryInfC{$\rel, w \leq u, \undb \in D_{w}, \vec{\unda} \in D_{w}, w : \psi(\vec{a})(\undb / x), \Gamma \sar u : \exists x \psi(\vec{a}), \Delta$}
\RightLabel{$\existsl$}
\UnaryInfC{$\rel, w \leq u, \vec{a} \in D_{w}, w : \exists x \psi(\vec{a}), \Gamma \sar u : \exists x \psi(\vec{a}), \Delta$}
\DisplayProof
\end{tabular}
\end{center}
\begin{center}
\begin{tabular}{c}
\AxiomC{}
\RightLabel{\ih}
\dashedLine
\UnaryInfC{$\rel', w \leq v, \undb \in D_{v}, v : \psi(\vec{\unda})(\undb / x), w : \forall x \psi(\vec{\unda}), \Gamma \sar v : \psi(\vec{\unda})(\undb / x), \Delta$}
\RightLabel{$\alll$}
\UnaryInfC{$\rel', w \leq v, \undb \in D_{v}, w : \forall x \psi(\vec{\unda}), \Gamma \sar v : \psi(\vec{\unda})(\undb / x), \Delta$}
\RightLabel{$\trans$}
\UnaryInfC{$\rel', \undb \in D_{v}, w : \forall x \psi(\vec{\unda}), \Gamma \sar v : \psi(\vec{\unda})(\undb / x), \Delta$}
\RightLabel{$\allr$}
\UnaryInfC{$\rel, w \leq u, \vec{\unda} \in D_{w}, w : \forall x \psi(\vec{a}), \Gamma \sar u : \forall x \psi(\vec{a}), \Delta$}
\DisplayProof
\end{tabular}
\end{center}

\textit{Base case for (ii).} The base case for atomic formulae is shown below (the base case for propositional variables is similar), and the case for when $\phi$ is of the form $\bot$ is omitted as it is simple to verify using the $\botl$ rule.

\begin{center}
\begin{tabular}{c}
\AxiomC{}
\RightLabel{$\idfo$}
\UnaryInfC{$\rel, w \leq w, \vec{\unda} \in D_{w}, w : p(\vec{\unda}), \Gamma \sar w : p(\vec{\unda}), \Delta$}
\RightLabel{$\refl$}
\UnaryInfC{$\rel, \vec{\unda} \in D_{w}, w : p(\vec{\unda}), \Gamma \sar w : p(\vec{\unda}), \Delta$}
\DisplayProof
\end{tabular}
\end{center}

\textit{Inductive step for (ii).}  We only show the case when $\phi(\vec{\unda})$ is of the form $\forall x \psi(\vec{\unda})$, as all other cases follow from~\cite[\lem~1]{DycNeg12} and \cite[\lem~12.1]{NegPla11}.

\begin{center}
\begin{tabular}{c}
\AxiomC{}
\RightLabel{\ih}
\dashedLine
\UnaryInfC{$\rel, w \leq u, \undb \in D_{u}, \vec{\unda} \in D_{w}, u : \psi(\vec{\unda})(\undb/x), w : \forall x \psi(\vec{\unda}), \Gamma \sar u : \psi(\vec{\unda})(\undb/x), \Delta$}
\RightLabel{$\alll$}
\UnaryInfC{$\rel, w \leq u, \undb \in D_{u}, \vec{\unda} \in D_{w}, w : \forall x \psi(\vec{a}), \Gamma \sar u : \psi(\vec{\unda})(\undb/x), \Delta$}
\RightLabel{$\allr$}
\UnaryInfC{$\rel, \vec{\unda} \in D_{w}, w : \forall x \psi(\vec{\unda}), \Gamma \sar w : \forall x \psi(\vec{\unda}), \Delta$}
\DisplayProof
\end{tabular}
\end{center}
\end{proof}

\begin{lemma}\label{lem:lsub-admiss-FO-Int}
The rule $\lsub$ is hp-admissible in $\gtintfond$ and $\gtintfocd$.
\end{lemma}

\begin{proof} We prove the result by induction on the height of the given derivation.

\textit{Base case.} The base case is easily resolved as any application of $\lsub$ to $\idfo$  or $\botl$ yields another instance of the rule.

\textit{Inductive step.} We need only consider the $\alll$ and $\allr$ cases, as all other cases follow from~\cite[\lem~3]{DycNeg12} and \cite[\lem~12.4]{NegPla11}. The $\alll$ case is handled by applying \ih and then the corresponding rule. The non-trivial $\allr$ case arises when the substitution introduces the eigenvariable of the $\allr$ inference. We show how to resolve this non-trivial case below, and omit the trivial case as it follows by applying \ih and then the corresponding rule. In what follows, we let $z$ be a fresh label.

\begin{flushleft}
\begin{tabular}{c c}
\AxiomC{$\rel, w \leq u, \unda \in D_{u}, \Gamma \sar  u : \phi(\unda/x), \Delta$}
\RightLabel{$\allr$}
\UnaryInfC{$\rel, \Gamma \sar w : \forall x \phi, \Delta$}
\RightLabel{$\lsub$}
\UnaryInfC{$\rel(u/v), \Gamma(u/v) \sar (w : \forall x \phi)(u/v), \Delta(u/v)$}
\DisplayProof

&

$\leadsto$
\end{tabular}
\end{flushleft}
\begin{flushright}
\AxiomC{$\rel, w \leq u, \unda \in D_{u}, \Gamma \sar  u : \phi(\unda/x), \Delta$}
\RightLabel{\ih}
\dashedLine
\UnaryInfC{$\rel(z/u), (w \leq u)(z/u), (\unda \in D_{u})(z/u), \Gamma(z/u) \sar  (u : \phi(\unda/x))(z/u), \Delta(z/u)$}
\RightLabel{=}
\dottedLine
\UnaryInfC{$\rel, w \leq z, \unda \in D_{z}, \Gamma \sar  z : \phi(\unda/x), \Delta$}
\RightLabel{\ih}
\dashedLine
\UnaryInfC{$\rel(u/v), (w \leq z)(u/v), (\unda \in D_{z})(u/v), \Gamma(u/v) \sar  (z : \phi(\unda/x))(u/v), \Delta(u/v)$}
\RightLabel{=}
\dottedLine
\UnaryInfC{$\rel(u/v), w \leq z, \unda \in D_{z}, \Gamma(u/v) \sar  z : \phi(\unda/x), \Delta(u/v)$}
\RightLabel{$\allr$}
\UnaryInfC{$\rel(u/v), \Gamma(u/v) \sar (w : \forall x \phi)(u/v), \Delta(u/v)$}
\DisplayProof
\end{flushright}
\end{proof}

\begin{lemma}\label{lem:psub-admiss-FO-Int}
The rule $\psub$ is hp-admissible in $\gtintfond$ and $\gtintfocd$. 
\end{lemma}

\begin{proof} We prove the claim by induction on the height of the given derivation.

\textit{Base case.} The base is trivial as any application of $\lsub$ to an instance of $\id$ or $\botl$ yields another instance of the rule.

\textit{Inductive step.} With the exception of the $\existsl$, $\allr$, and $\ned$ cases, all other cases follow by applying \ih and then the corresponding rule. The $\existsl$ and $\ned$ cases are handled as explained in~\cite[\lem~12.4]{NegPla11}. The non-trivial $\allr$ case arises when the parameter substitution $(\unda/\undb)$ introduces the eigenvariable of the $\allr$ inference. We show how to resolve the non-trivial case below, and omit the trivial case as it follows from invoking \ih followed by the corresponding rule. In the case below, we let $\undc$ be a fresh parameter.

\begin{flushleft}
\begin{tabular}{c c}
\AxiomC{$\rel, w \leq u, \unda \in D_{u}, \Gamma \sar  u : \phi(\unda/x), \Delta$}
\RightLabel{$\allr$}
\UnaryInfC{$\rel, \Gamma \sar w : \forall x \phi, \Delta$}
\RightLabel{$\psub$}
\UnaryInfC{$\R(\unda/\undb), \Gamma(\unda/\undb) \sar (w : \forall x \phi)(\unda/\undb), \Delta(\unda/\undb)$}
\DisplayProof

&

$\leadsto$
\end{tabular}
\end{flushleft}
\begin{flushright}
\AxiomC{$\rel, w \leq u, \unda \in D_{u}, \Gamma \sar  u : \phi(\unda/x), \Delta$}
\RightLabel{\ih}
\dashedLine
\UnaryInfC{$\rel(\undc/\unda), (w \leq u)(\undc/\unda), (\unda \in D_{u})(\undc/\unda), \Gamma(\undc/\unda) \sar (u : \phi(\unda/x))(\undc/\unda), \Delta(\undc/\unda)$}
\RightLabel{=}
\dottedLine
\UnaryInfC{$\rel, w \leq u, \undc \in D_{u}, \Gamma \sar u : \phi(\undc/x), \Delta$}
\RightLabel{\ih}
\dashedLine
\UnaryInfC{$\rel(\unda/\undb), (w \leq u)(\unda/\undb), (\undc \in D_{u})(\unda/\undb), \Gamma(\unda/\undb) \sar (u : \phi(\undc/x))(\unda/\undb), \Delta(\unda/\undb)$}
\RightLabel{=}
\dottedLine
\UnaryInfC{$\R(\unda/\undb), w \leq u, \undc \in D_{u}, \Gamma(\unda/\undb) \sar u : \phi(\undc/x)(\unda/\undb), \Delta(\unda/\undb)$}
\RightLabel{$\allr$}
\UnaryInfC{$\R(\unda/\undb), \Gamma(\unda/\undb) \sar (w : \forall x \phi)(\unda/\undb), \Delta(\unda/\undb)$}
\DisplayProof
\end{flushright}
\end{proof}

\begin{lemma}\label{lem:wk-admiss-FO-Int}
The rule $\wk$ is hp-admissible in $\gtintfond$ and $\gtintfocd$. 
\end{lemma}

\begin{proof} We prove the result by induction on the height of the given derivation.

\textit{Base case.} The base case is straightforward since any application of $\wk$ to $\idfo$ or $\botl$ yields another instance of the rule.

\textit{Inductive step.} The only non-trivial cases concern $\impr$, $\existsl$, $\allr$, and $\ned$, and arise when $\wk$ introduces a parameter or label identical to an eigenvariable of one of the aforementioned inferences. The $\impr$ case is resolved as in~\cite[\prp~1]{DycNeg12}, and the $\existsl$ and $\ned$ cases are resolved as in~\cite[\thm~12.5]{NegPla11}. Therefore, we only show how to resolve non-trivial $\allr$ case below. We let $\undc$ and $z$ be a fresh parameter and label, respectively.

\begin{flushleft}
\begin{tabular}{c c}
\AxiomC{$\rel, w \leq u, \unda \in D_{u}, \Gamma \sar  u : \phi(\unda/x), \Delta$}
\RightLabel{$\allr$}
\UnaryInfC{$\rel, \Gamma \sar w : \forall x \phi, \Delta$}
\RightLabel{$\wk$}
\UnaryInfC{$\rel, \rel', \Gamma, \Gamma' \sar w : \forall x \phi, \Delta, \Delta'$}
\DisplayProof

&

$\leadsto$
\end{tabular}
\end{flushleft}
\begin{flushright}
\AxiomC{$\rel, w \leq u, \unda \in D_{u}, \Gamma \sar  u : \phi(\unda/x), \Delta$}
\RightLabel{$\lsub$}
\dashedLine
\UnaryInfC{$\rel(z/u), (w \leq u)(z/u), (\unda \in D_{u})(z/u), \Gamma(z/u) \sar  (u : \phi(\unda/x))(z/u), \Delta(z/u)$}
\RightLabel{=}
\dottedLine
\UnaryInfC{$\rel, w \leq z, \unda \in D_{z}, \Gamma \sar  z : \phi(\unda/x), \Delta$}
\RightLabel{$\psub$}
\dashedLine
\UnaryInfC{$\rel(\undc/\unda), (w \leq z)(\undc/\unda), (\unda \in D_{z})(\undc/\unda), \Gamma(\undc/\unda) \sar  (z : \phi(\unda/x))(\undc/\unda), \Delta(\undc/\unda)$}
\RightLabel{=}
\dottedLine
\UnaryInfC{$\rel, w \leq z, \undc \in D_{z}, \Gamma \sar  z : \phi(\undc/x), \Delta$}
\RightLabel{\ih}
\dashedLine
\UnaryInfC{$\rel, \rel', w \leq z, \undc \in D_{z}, \Gamma, \Gamma' \sar  z : \phi(\undc/x), \Delta, \Delta'$}
\RightLabel{$\allr$}
\UnaryInfC{$\rel, \rel', \Gamma, \Gamma' \sar  w : \forall x \phi, \Delta, \Delta'$}
\DisplayProof
\end{flushright}
\end{proof}

\begin{lemma}\label{lem:invert-FO-Int}
All rules are hp-invertible in $\gtintfond$ and $\gtintfocd$. 
\end{lemma}

\begin{proof} The result is shown by induction on the height of the given derivation, and is proven similarly to~\cite[\prp~2]{DycNeg12} and~\cite[\thm~12.7]{NegPla11}.
\end{proof}

\begin{lemma}\label{lem:ctr-admiss-FO-Int}
The rules $\ctrrel$, $\ctrl$, and $\ctrr$ are hp-admissible in $\gtintfond$ and $\gtintfocd$. 
\end{lemma}

\begin{proof} The result is shown by induction on the height of the given derivation, and is proven similarly to~\cite[\thm~3]{DycNeg12} and~\cite[\thm~12.8]{NegPla11}.
\end{proof}

\begin{theorem}\label{thm:cut-admiss-FO-Int}
The rule $\cut$ is eliminable in $\gtintfond$ and $\gtintfocd$. 
\end{theorem}

\begin{proof} The result is shown by induction on the lexicographic ordering of pairs $(\fcomp{\phi},h_{1}+h_{2})$, where $\fcomp{\phi}$ is the complexity of the cut formula $\phi$, $h_{1}$ is the height of the derivation of the left premise of $\cut$, and $h_{2}$ is the height of the derivation of the right premise of $\cut$. We may assume w.l.o.g. that $\cut$ is the last rule used in our given derivation, and that no other instances of $\cut$ appear in the given derivation. The general result follows by repeatedly applying the procedure given in the proof to successively eliminate topmost instances of $\cut$ until the derivation is free of such inferences.

Due to the modularity of the labelled formalism, we may leverage the results~\cite[\thm~4]{DycNeg12} and~\cite[\thm~12.9]{NegPla11} to cover all cases with the exception of the case where the cut formula is of the form $\forall x \psi$ and is principal in both premises of $\cut$. This case is resolved as shown below:

\begin{flushleft}
\begin{tabular}{c c c}
$\Pi_{1}$

&

$= \Bigg \{$

&

\AxiomC{$\R, w \leq v, w \leq u, \unda \in D_{v}, \undb \in D_{u}, \Gamma \Rightarrow \Delta, u : \psi(\undb/x)$}
\RightLabel{$\allr$}
\UnaryInfC{$\R, w \leq v, \unda \in D_{v}, \Gamma \Rightarrow \Delta, w : \forall x \psi$}
\DisplayProof
\end{tabular}
\end{flushleft}

\begin{flushleft}
\begin{tabular}{c c c}
$\Pi_{2}$

&

$= \Bigg \{$

&

\AxiomC{$\R, w \leq v, \unda \in D_{v}, v : \psi(\unda / x), w : \forall x \psi, \Gamma \Rightarrow \Delta$}
\RightLabel{$\alll$}
\UnaryInfC{$\R, w \leq v, \unda \in D_{v}, w : \forall x \psi, \Gamma \Rightarrow \Delta$}
\DisplayProof
\end{tabular}
\end{flushleft}

\begin{center}
\begin{tabular}{c c c}
\AxiomC{$\Pi_{1}$}

\AxiomC{$\Pi_{2}$}

\RightLabel{$\cut$}
\BinaryInfC{$\R, w \leq v, \unda \in D_{v}, \Gamma \Rightarrow \Delta$}

\DisplayProof

&

$\leadsto$

&

\AxiomC{$\Pi_{1}'$}
\AxiomC{$\Pi_{2}'$}
\RightLabel{\ih}
\dashedLine
\BinaryInfC{$\R, w \leq v, \unda \in D_{v}, \Gamma \Rightarrow \Delta$}
\DisplayProof
\end{tabular}
\end{center}

\begin{center}
\begin{tabular}{c c c}
$\Pi_{1}'$

&

$= \Bigg \{$

&

\AxiomC{$\R, w \leq u, \undb \in D_{u}, \Gamma \Rightarrow \Delta, u : \psi(\undb / x)$}
\RightLabel{$\lsub$}
\dashedLine
\UnaryInfC{$\R, w \leq v, \undb \in D_{v}, \Gamma \Rightarrow \Delta, v : \psi(\undb / x)$}
\RightLabel{$\psub$}
\dashedLine
\UnaryInfC{$\R, w \leq v, \unda \in D_{v}, \Gamma \Rightarrow \Delta, v : \psi(\unda / x)$}
\DisplayProof
\end{tabular}
\end{center}

\begin{center}
\resizebox{\columnwidth}{!}{
\begin{tabular}{c c c}
$\Pi_{2}'$

&

$= \Bigg \{$

&

\AxiomC{$\R, w \leq v, w \leq u, \unda \in D_{v}, \undb \in D_{u}, \Gamma \Rightarrow \Delta, u : \psi(\undb / x)$}
\RightLabel{$\wk$}
\dashedLine
\UnaryInfC{$\R, w \leq v, w \leq u, \unda \in D_{v}, \undb \in D_{u}, v : \psi(\unda / x), \Gamma \Rightarrow \Delta, u : \psi(\undb / x)$}
\RightLabel{$\allr$}
\UnaryInfC{$\R, w \leq v, \unda \in D_{v}, v : \psi(\unda / x), \Gamma \Rightarrow \Delta, w : \forall x \psi$}

\AxiomC{$\Pi_{3}'$}

\RightLabel{\ih}
\dashedLine
\BinaryInfC{$\R, w \leq v, \unda \in D_{v}, v : \psi(\unda / x), \Gamma \Rightarrow \Delta$}
\DisplayProof
\end{tabular}
}
\end{center}
$$
\Pi_{3}' = \Bigg \{ \quad \R, w \leq v, \unda \in D_{v}, v : \psi(\unda / x), w : \forall x \psi, \Gamma \Rightarrow \Delta
$$
Note that we may invoke \ih in $\Pi_{2}'$ since the sum of the heights $h_{1} + h_{2}$ is one less than the original $\cut$, and we may invoke the second use of \ih since the cut formula $v : \psi(\unda / x)$ is of a smaller complexity, that is $\fcomp{\psi(\unda/x)} < \fcomp{\forall x \psi} = \fcomp{\phi}$.
\end{proof}

\begin{theorem}[Completeness]\label{thm:completness-FO-Int} Let $\phi \in \langintfo$, and let $\phi(\vec{\unda}) := \phi(\unda_{1}/x_{1})\cdots(\unda_{n}/x_{n})$, that is, $\phi(\vec{\unda})$ is the formula $\phi$ where the parameters $\vec{\unda} = a_{1}, \ldots, a_{n}$ have been substituted for each respective free variable $x_{1}, \ldots, x_{n}$ in $\phi$.

(i) If $\vdash_{\intfond} \phi$, then $\vdash_{\gtintfond} \vec{\unda} \in D_{w} \sar w : \phi(\vec{\unda})$.

(ii) If $\vdash_{\intfocd} \phi$, then $\vdash_{\gtintfocd} \vec{\unda} \in D_{w} \sar w : \phi(\vec{\unda})$.
\end{theorem}

\begin{proof} By \rem~\ref{rem:G3I}, we know that our calculi $\gtintfond$ and $\gtintfocd$ can derive all axioms A0 -- A8, that is, the axioms for propositional intuitionistic logic (cf.~\dfn~\ref{def:axiomatization-IntFO} and \cite[p.~6]{GabSheSkv09}). Therefore, we need only show that they can derive the axioms A9 -- A12 and A9 -- A13, respectively, as well as simulate the inference rules R0 and R1 (see \dfn~\ref{def:axiomatization-IntFO} for the axioms of $\intfond$ and $\intfocd$). We show this below, and note that the only proof that makes use of the constant domain rule $\cd$ is the proof of axiom A13 (the constant domain axiom). The conclusion of each derivation displays the axiom derived.


\emph{Axiom A9.} We present the case below where the universal quantifier is not vacuous, and note that the case where the universal quantifier is vacuous is similar.

\begin{center}
\resizebox{\columnwidth}{!}{
\begin{tabular}{c} 
\AxiomC{}
\RightLabel{\lem~\ref{lem:general-id-FO-Int}}
\dashedLine
\UnaryInfC{$w \leq u, u \leq u, \vec{\unda} \in D_{w}, \unda \in D_{w}, \unda \in D_{u}, \vec{\unda} \in D_{u}, u : \forall x \phi(\vec{\unda},x), u :\phi(\vec{\unda},\unda) \Rightarrow u : \phi(\vec{\unda},\unda)$}
\RightLabel{$\alll$}
\UnaryInfC{$w \leq u, u \leq u, \vec{\unda} \in D_{w}, \unda \in D_{w}, \unda \in D_{u}, \vec{\unda} \in D_{u}, u : \forall x \phi(\vec{\unda},x) \Rightarrow u : \phi(\vec{\unda},\unda)$}
\RightLabel{$\nd \times n$}
\UnaryInfC{$w \leq u, u \leq u, \vec{\unda} \in D_{w}, \unda \in D_{w}, u : \forall x \phi(\vec{\unda},x) \Rightarrow u : \phi(\vec{\unda},\unda)$}
\RightLabel{$\refl$}
\UnaryInfC{$w \leq u, \vec{\unda} \in D_{w}, \unda \in D_{w}, u : \forall x \phi(\vec{\unda},x) \Rightarrow u : \phi(\vec{\unda},\unda)$}
\RightLabel{$\impr$}
\UnaryInfC{$\vec{\unda} \in D_{w}, \unda \in D_{w} \Rightarrow w : \forall x \phi(\vec{\unda},x) \imp \phi(\vec{\unda},\unda)$}
\DisplayProof
\end{tabular}
}
\end{center}

\emph{Axiom A10.} Similar to the previous case, we show the case where the existential quantifier is not vacuous. The case where the quantifier is vacuous is proven similarly.

\begin{center}
\resizebox{\columnwidth}{!}{
\begin{tabular}{c}
\AxiomC{}
\RightLabel{\lem~\ref{lem:general-id-FO-Int}}
\dashedLine
\UnaryInfC{$w \leq u, \vec{\unda} \in D_{w}, \unda \in D_{w}, \vec{\unda} \in D_{u}, \unda \in D_{u}, u : \phi(\vec{\unda},\unda) \Rightarrow u : \phi(\vec{\unda},\unda), u : \exists x \phi(\vec{\unda},x)$}
\RightLabel{$\existsr$}
\UnaryInfC{$w \leq u, \vec{\unda} \in D_{w}, \unda \in D_{w}, \vec{\unda} \in D_{u}, \unda \in D_{u}, u : \phi(\vec{\unda},\unda) \Rightarrow u : \exists x \phi(\vec{\unda},x)$}
\RightLabel{$\nd \times n$}
\UnaryInfC{$w \leq u, \vec{\unda} \in D_{w}, \unda \in D_{w},  u : \phi(\vec{\unda},\unda) \Rightarrow u : \exists x \phi(\vec{\unda},x)$}
\RightLabel{$\impr$}
\UnaryInfC{$\vec{\unda} \in D_{w}, \unda \in D_{w} \Rightarrow w : \phi(\vec{\unda},\unda) \imp \exists x \phi(\vec{\unda},x)$}
\DisplayProof
\end{tabular}
}
\end{center}

\emph{Axiom A11.} We show how to derive the axiom A11 below when the universal quantifier is not vacuous, and omit the proof of the axiom when the universal quantifier is vacuous as the proof is similar. 
To save space and improve readability, we let $\R := w \leq v, v \leq u, u \leq z, \vec{\unda} \in D_{w}, \vec{\undb} \in D_{w}, \unda \in D_{z}$, $\R' := \R, v \leq z, z \leq z$, and $\Gamma := u : \psi(\vec{\undb}) \imp \phi(\vec{\unda},\unda), v : \forall x (\psi(\vec{\undb}) \imp \phi(\vec{\unda},x))$. Moreover, we suppose that $k_{1}$ is equal to the number of domain atoms $\vec{\undb} \in D_{w} = \undb_{1} \in D_{w}, \ldots, \undb_{k_{1}} \in D_{w}$, and $k_{2}$ is equal to the number of domain atoms $\vec{\unda} \in D_{w} = \unda_{1} \in D_{w}, \ldots, \unda_{k_{2}} \in D_{w}$.

\begin{center}
\begin{tabular}{c c c}
$\Pi_{1}$

&

$= \Bigg \{$

&

\AxiomC{}
\RightLabel{\lem~\ref{lem:general-id-FO-Int}}
\dashedLine
\UnaryInfC{$\R', \vec{\undb} \in D_{v}, \vec{\undb} \in D_{u}, \Gamma, u : \psi(\vec{\undb}) \Rightarrow z : \psi(\vec{\undb}), z : \phi(\vec{\unda},\unda)$}
\RightLabel{$\nd \times 2 k_{1}$}
\UnaryInfC{$\R', \Gamma, u : \psi(\vec{\undb}) \Rightarrow z : \psi(\vec{\undb}), z : \phi(\vec{\unda},\unda)$}
\DisplayProof
\end{tabular}
\end{center}

\begin{center}
\begin{tabular}{c c c}
$\Pi_{2}$

&

$= \Bigg \{$

&

\AxiomC{}
\RightLabel{\lem~\ref{lem:general-id-FO-Int}}
\dashedLine
\UnaryInfC{$\R', \vec{\unda} \in D_{v}, \vec{\unda} \in D_{u}, \vec{\unda} \in D_{z}, \Gamma, u : \psi(\vec{\undb}), z : \phi(\vec{\unda},\unda) \Rightarrow z : \phi(\vec{\unda},\unda)$}
\RightLabel{$\nd \times 3 k_{2}$}
\UnaryInfC{$\R', \Gamma, u : \psi(\vec{\undb}), z : \phi(\vec{\unda},\unda) \Rightarrow z : \phi(\vec{\unda},\unda)$}
\DisplayProof
\end{tabular}
\end{center}

\begin{center}
\begin{tabular}{c}
\AxiomC{$\Pi_{1}$}
\AxiomC{$\Pi_{2}$}
\RightLabel{$\impl$}
\BinaryInfC{$\R, v \leq z, z \leq z, z : \psi(\vec{\undb}) \imp \phi(\vec{\unda},\unda), v : \forall x (\psi(\vec{\undb}) \imp \phi(\vec{\unda},x)), u : \psi(\vec{\undb}) \Rightarrow z : \phi(\vec{\unda},\unda)$}
\RightLabel{$\refl$}
\UnaryInfC{$\R, v \leq z, z : \psi(\vec{\undb}) \imp \phi(\vec{\unda},\unda), v : \forall x (\psi(\vec{\undb}) \imp \phi(\vec{\unda},x)), u : \psi(\vec{\undb}) \Rightarrow z : \phi(\vec{\unda},\unda)$}
\RightLabel{$\alll$}
\UnaryInfC{$\R, v \leq z, v : \forall x (\psi(\vec{\undb}) \imp \phi(\vec{\unda},x)), u : \psi(\vec{\undb}) \Rightarrow z : \phi(\vec{\unda},\unda)$}
\RightLabel{$\trans$}
\UnaryInfC{$\R, v : \forall x (\psi(\vec{\undb}) \imp \phi(\vec{\unda},x)), u : \psi(\vec{\undb}) \Rightarrow z : \phi(\vec{\unda},\unda)$}
\RightLabel{$\allr$}
\UnaryInfC{$w \leq v, v \leq u, \vec{\unda} \in D_{w}, \vec{\undb} \in D_{w}, v : \forall x (\psi(\vec{\undb}) \imp \phi(\vec{\unda},x)), u : \psi(\vec{\undb}) \Rightarrow u : \forall x \phi(\vec{\unda},x)$}
\RightLabel{$\impr$}
\UnaryInfC{$w \leq v, \vec{\unda} \in D_{w}, \vec{\undb} \in D_{w}, v : \forall x (\psi(\vec{\undb}) \imp \phi(\vec{\unda},x)) \Rightarrow v : \psi(\vec{\undb}) \imp \forall x \phi(\vec{\unda},x)$}
\RightLabel{$\impr$}
\UnaryInfC{$\vec{\unda} \in D_{w}, \vec{\undb} \in D_{w} \Rightarrow w : \forall x (\psi(\vec{\undb}) \imp \phi(\vec{\unda},x)) \imp (\psi(\vec{\undb}) \imp \forall x \phi(\vec{\unda},x))$}
\DisplayProof
\end{tabular}
\end{center}

\emph{Axiom A12.} We show how to derive axiom A12 below and only consider the case where the quantifiers are non-vacuous, since the case where the quantifiers vacuously quantify $\psi$ is similar. To save space and improve readability, we let $\R := w \leq v, v \leq u, \vec{\unda} \in D_{w}, \vec{\undb} \in D_{w}, \unda \in D_{u}$ and $\Gamma := u : \psi(\vec{\undb},\unda) \imp \phi(\vec{\unda}), v : \forall x (\psi(\vec{\undb},x) \imp \phi(\vec{\unda}))$. Also, we let $k_{1}$ be equal to the number of domain atoms $\vec{\undb} \in D_{w} = \undb_{1} \in D_{w}, \ldots, \undb_{k_{1}} \in D_{w}$, and $k_{2}$ be equal to the number of domain atoms $\vec{\unda} \in D_{w} = \unda_{1} \in D_{w}, \ldots, \unda_{k_{2}} \in D_{w}$.

\begin{center}
\begin{tabular}{c c c}
$\Pi_{1}$

&

$= \Bigg \{$

&

\AxiomC{}
\RightLabel{\lem~\ref{lem:general-id-FO-Int}}
\dashedLine
\UnaryInfC{$\R, u \leq u, \vec{\undb} \in D_{v}, \vec{\undb} \in D_{u}, \Gamma, u : \psi(\vec{\undb},\unda) \sar u : \psi(\vec{\undb},\unda), u : \phi(\vec{\unda})$}
\RightLabel{$\nd \times 2 k_{1}$}
\UnaryInfC{$\R, u \leq u, \Gamma, u : \psi(\vec{\undb},\unda) \Rightarrow u : \psi(\vec{\undb},\unda), u : \phi(\vec{\unda})$}
\DisplayProof
\end{tabular}
\end{center}

\begin{center}
\begin{tabular}{c c c}
$\Pi_{2}$

&

$= \Bigg \{$

&

\AxiomC{}
\RightLabel{\lem~\ref{lem:general-id-FO-Int}}
\dashedLine
\UnaryInfC{$\R, u \leq u, \vec{\unda} \in D_{v}, \vec{\unda} \in D_{u}, \Gamma, u : \psi(\vec{\undb},\unda), u : \phi(\vec{\unda}) \sar  u : \phi(\vec{\unda})$}
\RightLabel{$\nd \times 2 k_{2}$}
\UnaryInfC{$\R, u \leq u, \Gamma, u : \psi(\vec{\undb},\unda), u : \phi(\vec{\unda}) \sar u : \phi(\vec{\unda})$}
\DisplayProof
\end{tabular}
\end{center}

\begin{center}
\begin{tabular}{c}
\AxiomC{$\Pi_{1}$}
\AxiomC{$\Pi_{2}$}
\RightLabel{$\impl$}
\BinaryInfC{$\R, u \leq u, u : \psi(\vec{\undb},\unda) \imp \phi(\vec{\unda}), v : \forall x (\psi(\vec{\undb},x) \imp \phi(\vec{\unda})), u : \psi(\vec{\undb},\unda) \sar u : \phi(\vec{\unda})$}
\RightLabel{$\refl$}
\UnaryInfC{$\R, u : \psi(\vec{\undb},\unda) \imp \phi(\vec{\unda}), v : \forall x (\psi(\vec{\undb},x) \imp \phi(\vec{\unda})), u : \psi(\vec{\undb},\unda) \sar u : \phi(\vec{\unda})$}
\RightLabel{$\alll$}
\UnaryInfC{$\R, v : \forall x (\psi(\vec{\undb},x) \imp \phi(\vec{\unda})), u : \psi(\vec{\undb},\unda) \sar u : \phi(\vec{\unda})$}
\RightLabel{$\existsl$}
\UnaryInfC{$w \leq v, v \leq u, \vec{\unda} \in D_{w}, \vec{\undb} \in D_{w}, v : \forall x (\psi(\vec{\undb},x) \imp \phi(\vec{\unda})), u : \exists x \psi(\vec{\undb},x) \sar u : \phi(\vec{\unda})$}
\RightLabel{$\impr$}
\UnaryInfC{$w \leq v, \vec{\unda} \in D_{w}, \vec{\undb} \in D_{w}, v : \forall x (\psi(\vec{\undb},x) \imp \phi(\vec{\unda})) \sar v : \exists x \psi(\vec{\undb},x) \imp \phi(\vec{\unda})$}
\RightLabel{$\impr$}
\UnaryInfC{$\vec{\unda} \in D_{w}, \vec{\undb} \in D_{w} \sar w : \forall x (\psi(\vec{\undb},x) \imp \phi(\vec{\unda})) \imp (\exists x \psi(\vec{\undb},x) \imp \phi(\vec{\unda}))$}
\DisplayProof
\end{tabular}
\end{center}

\emph{Axiom A13.} Below, we derive the constant domain axiom A13 and consider the case where the universal quantifier non-vacuously quantifiers over $\phi$; the case where the universal quantifier is vacuous is similar. Note that the derivation requires an application of the constant domain rule $\cd$, showing that the axiom is derivable in $\gtintfocd$, but not $\gtintfond$ (see proof $\Pi_{1}$ below). To save space and improve readability, we let $\rel := w \leq v, v \leq u, \vec{\unda} \in D_{w}, \vec{\undb} \in D_{w}, \unda \in D_{u}$.

\begin{center}
\resizebox{\columnwidth}{!}{
\begin{tabular}{c c c}
$\Pi_{1}$

&

$= \Bigg \{$

&

\AxiomC{}
\RightLabel{\lem~\ref{lem:general-id-FO-Int}}
\dashedLine
\UnaryInfC{$\R, \unda \in D_{u}, v \leq v, \vec{\unda} \in D_{v}, v : \phi(\vec{\unda},\unda), v : \forall x (\phi(\vec{\unda},x) \lor \psi(\vec{\undb})) \Rightarrow u : \phi(\vec{\unda},\unda), v : \psi(\vec{\undb})$}
\RightLabel{$\nd \times k_{1}$}
\UnaryInfC{$\R, \unda \in D_{u}, v \leq v, v : \phi(\vec{\unda},\unda), v : \forall x (\phi(\vec{\unda},x) \lor \psi(\vec{\undb})) \Rightarrow u : \phi(\vec{\unda},\unda), v : \psi(\vec{\undb})$}
\DisplayProof
\end{tabular}
}
\end{center}

\begin{center}
\resizebox{\columnwidth}{!}{
\begin{tabular}{c c c}
$\Pi_{2}$

&

$= \Bigg \{$

&

\AxiomC{}
\RightLabel{\lem~\ref{lem:general-id-FO-Int}}
\dashedLine
\UnaryInfC{$\R, \unda \in D_{u}, v \leq v, \vec{\undb} \in D_{v}, v : \psi(\vec{\undb}), v : \forall x (\phi(\vec{\unda},x) \lor \psi(\vec{\undb})) \Rightarrow u : \phi(\vec{\unda},\unda), v : \psi(\vec{\undb})$}
\RightLabel{$\nd \times k_{2}$}
\UnaryInfC{$\R, \unda \in D_{u}, v \leq v, v : \psi(\vec{\undb}), v : \forall x (\phi(\vec{\unda},x) \lor \psi(\vec{\undb})) \Rightarrow u : \phi(\vec{\unda},\unda), v : \psi(\vec{\undb})$}
\DisplayProof
\end{tabular}
}
\end{center}

\begin{center}
\begin{tabular}{c}
\AxiomC{$\Pi_{1}$}
\AxiomC{$\Pi_{2}$}
\RightLabel{$\disl$}
\BinaryInfC{$\R, \unda \in D_{u}, v \leq v, v : \phi(\vec{\unda},\unda) \lor \psi(\vec{\undb}), v : \forall x (\phi(\vec{\unda},x) \lor \psi(\vec{\undb})) \Rightarrow u : \phi(\vec{\unda},\unda), v : \psi(\vec{\undb})$}
\RightLabel{$\alll$}
\UnaryInfC{$\R, \unda \in D_{u}, v \leq v, v : \forall x (\phi(\vec{\unda},x) \lor \psi(\vec{\undb})) \Rightarrow u : \phi(\vec{\unda},\unda), v : \psi(\vec{\undb})$}
\RightLabel{$\cd$}
\UnaryInfC{$\R, v \leq v, v : \forall x (\phi(\vec{\unda},x) \lor \psi(\vec{\undb})) \Rightarrow u : \phi(\vec{\unda},\unda), v : \psi(\vec{\undb})$}
\RightLabel{$\refl$}
\UnaryInfC{$\rel, v : \forall x (\phi(\vec{\unda},x) \lor \psi(\vec{\undb})) \sar u : \phi(\vec{\unda},\unda), v : \psi(\vec{\undb})$}
\RightLabel{$\allr$}
\UnaryInfC{$w \leq v, \vec{\unda} \in D_{w}, \vec{\undb} \in D_{w}, v : \forall x (\phi(\vec{\unda},x) \lor \psi(\vec{\undb})) \Rightarrow v : \forall x \phi(\vec{\unda},x), v : \psi(\vec{\undb})$}
\RightLabel{$\disr$}
\UnaryInfC{$w \leq v, \vec{\unda} \in D_{w}, \vec{\undb} \in D_{w}, v : \forall x (\phi(\vec{\unda},x) \lor \psi(\vec{\undb})) \Rightarrow v : \forall x \phi(\vec{\unda},x) \lor \psi(\vec{\undb})$}
\RightLabel{$\impr$}
\UnaryInfC{$\vec{\unda} \in D_{w}, \vec{\undb} \in D_{w} \Rightarrow w : \forall x (\phi(\vec{\unda},x) \lor \psi(\vec{\undb})) \imp \forall x \phi(\vec{\unda},x) \lor \psi(\vec{\undb})$}
\DisplayProof
\end{tabular}
\end{center}

\emph{Rule R0.} To show that modus ponens can be simulated, we let $\vec{a}$ be all parameters occurring in $\phi$, $\vec{b}$ be all parameters occurring in $\psi$, and let $\vec{c}$ consist of all parameters occurring in $\psi$, but not $\phi$. The last inference consists of a sequence of $k$ $\ned$ applications that delete all domains atoms containing parameters from $\phi$, but not $\psi$.

\begin{center}
\resizebox{\columnwidth}{!}{
\begin{tabular}{c c c}
$\Pi$

&

$= \Bigg \{$

&

\AxiomC{$\vec{\unda} \in D_{w}, \vec{\undc} \in D_{w} \sar w : \phi(\vec{\unda}) \imp \psi(\vec{\undb})$}
\RightLabel{\lem~\ref{lem:invert-FO-Int}}
\dashedLine
\UnaryInfC{$w \leq u, \vec{\unda} \in D_{w}, \vec{\undc} \in D_{w}, u : \phi(\vec{\unda}) \sar u : \psi(\vec{\undb})$}
\RightLabel{$\lsub$}
\dashedLine
\UnaryInfC{$(w \leq u)(w/u), (\vec{\unda} \in D_{w})(w/u), (\vec{\undc} \in D_{w})(w/u), (u : \phi(\vec{\unda}))(w/u) \sar (u : \psi(\vec{\undb}))(w/u)$}
\RightLabel{=}
\dottedLine
\UnaryInfC{$w \leq w, \vec{\unda} \in D_{w}, \vec{\undc} \in D_{w}, w : \phi(\vec{\unda}) \sar w : \psi(\vec{\undb})$}
\RightLabel{$\refl$}
\UnaryInfC{$\vec{\unda} \in D_{w}, \vec{\undc} \in D_{w}, w : \phi(\vec{\unda}) \sar w : \psi(\vec{\undb})$}
\DisplayProof
\end{tabular}
}
\end{center}

\begin{center}
\begin{tabular}{c}
\AxiomC{$\vec{\unda} \in D_{w} \sar w : \phi(\vec{\unda})$}
\dashedLine
\RightLabel{$\wk$}
\dashedLine
\UnaryInfC{$\vec{\unda} \in D_{w}, \vec{\undc} \in D_{w} \sar w : \phi(\vec{\unda})$}

\AxiomC{$\Pi$}

\RightLabel{$\cut$}
\dashedLine
\BinaryInfC{$\vec{\unda} \in D_{w}, \vec{\undc} \in D_{w} \sar w : \psi(\vec{\undb})$}
\RightLabel{$\ned \times k$}
\UnaryInfC{$\vec{\undb} \in D_{w} \sar w : \psi(\vec{\undb})$}
\DisplayProof
\end{tabular}
\end{center}

\emph{Rule R1.} Last, we show that the generalization rule R1 can be simulated. We assume there are $k$ domains atoms $\vec{\unda} \in D_{w} := \vec{\unda}_{1} \in D_{1}, \ldots, \vec{\unda}_{k} \in D_{k}$.

\begin{center}
\begin{tabular}{c}
\AxiomC{$\vec{\unda} \in D_{w}, \unda \in D_{w} \Rightarrow w : \phi(\vec{\unda},\unda)$}
\RightLabel{$\wk$}
\dashedLine
\UnaryInfC{$u \leq w, \vec{\unda} \in D_{u}, \vec{\unda} \in D_{w}, \unda \in D_{w} \Rightarrow w : \phi(\vec{\unda},\unda)$}
\RightLabel{$\nd \times k$}
\UnaryInfC{$u \leq w, \vec{\unda} \in D_{u}, \unda \in D_{w} \Rightarrow w : \phi(\vec{\unda},\unda)$}
\RightLabel{$\allr$}
\UnaryInfC{$\vec{\unda} \in D_{u} \Rightarrow u : \forall x \phi(\vec{\unda},x)$}
\RightLabel{$\lsub$}
\dashedLine
\UnaryInfC{$\vec{\unda} \in D_{w} \Rightarrow w : \forall x \phi(\vec{\unda},x)$}
\DisplayProof
\end{tabular}
\end{center}

\end{proof}

As in the previous section, we define sequent graphs for labelled sequents, which uncovers the underlying data structures inherent in our sequents. This theoretical tool will assist us while refining our calculi in the next chapter.

\begin{definition}[Sequent Graph for First-Order Intuitionistic Logics]\label{def:sequent-graph-FO-Int} Let $\Lambda := \rel, \Gamma \sar \Delta$ be a labelled sequent for first-order intuitionistic logics. We define the \emph{sequent graph}\index{Sequent graph!for first-order intuitionistic logics} of $\Lambda$, $\seqgraph(\Lambda) = (V,E,L)$, as follows:
\begin{itemize}

\item[$\li$] $V = \lab(\Lambda)$

\item[$\li$] $E = \{(w,u) \ | \ w \leq u \in \rel\}$

\item[$\li$] $L(w) = \Gamma \restriction w \sar \Delta \restriction w$

\end{itemize}
Last, we define \emph{labelled tree sequents}, \emph{labelled tree proofs}, and the \emph{fixed root property}\index{Labelled tree sequent!for first-order intuitionistic logics}\index{Labelled tree proof!for first-order intuitionistic logics}\index{Fixed root property!for first-order intuitionistic logics} analogous to \dfn~\ref{def:tree-sequent-kms} and~\ref{def-tree-proof-kms}.
\end{definition}

\begin{example}\label{ex:sequent-graph-example-IntFO} We provide an example of a labelled tree sequent $\Lambda_{1}$ along with its corresponding sequent graph.

\begin{minipage}[t]{.33\textwidth}
\begin{tabular}{@{\hskip -.05em} c}
\vspace*{1 em}
\ \\
\AxiomC{ }
\noLine
\UnaryInfC{}
\noLine
\UnaryInfC{}
\noLine
\UnaryInfC{$\Lambda_{1} = w \leq u, u \leq z, u \leq v ,$}
\noLine
\UnaryInfC{$w : p, z : q, z : q \sar w : p, z : r, v : \bot$}
\noLine
\UnaryInfC{}
\noLine
\UnaryInfC{ }
\noLine
\UnaryInfC{}
\noLine
\UnaryInfC{}
\DisplayProof
\end{tabular}
\end{minipage}
\begin{minipage}[t]{.15\textwidth}
\ 
\end{minipage}
\begin{minipage}[t]{.33\textwidth}
\begin{tabular}{c}
\xymatrix{
&  \overset{\boxed{p \sar p}}{w} \ar[d] & & \\
 & \overset{\boxed{\seqempstr \sar \seqempstr}}{u} \ar[dr]\ar[dl]& & \\
\overset{\boxed{q,q \sar r}}{z} &  & \overset{\boxed{\seqempstr \sar \bot}}{v} & 
}
\end{tabular}
\end{minipage}

\end{example}

\begin{theorem}\label{thm:tree-incompleteness-FO-Int}
Each calculus $\gtintfond$ and $\gtintfocd$ is incomplete relative to labelled tree derivations.
\end{theorem}

\begin{proof} The result follows by considering the proof of $(p \imp q) \imp ((q \imp r) \imp (p \imp r))$, which requires the use of the $\trans$ rule, thus breaking the labelled tree property.
\end{proof}

\section{Labelled Calculi for Deontic STIT Logics}\label{sec:lab-calc-dsn}

Although \stit logics were introduced three decades ago by Belnap and Perloff~\cite{BelPer90}, the proof theory for such logics has only recently been developed. Tableau calculi for traditional (non-deontic) multi-agent \stit logics were introduced by Wansing in 2006~\cite{Wan06}, with extended versions 
 introduced in 2018~\cite{OlkWan18}. The first cut-free labelled sequent calculi for traditional multi-agent \stit logics were presented in 2019~\cite{BerLyo19a}, along with cut-free labelled sequent calculi for the \emph{temporal} \stit logics: $\mathsf{TSTIT}$~\cite{Lor13} and $\mathsf{XSTIT}$~\cite{Bro11,Bro11b}. While the papers~\cite{BerLyo19a,LyoBer19} provided labelled calculi for traditional, multi-agent \stit logics based on \emph{Kripke semantics}~\cite{BalHerTro08}, an equivalent set of labelled calculi were given for the same class of logics in~\cite{NegPav20} based on \emph{branching-time semantics}~\cite{BelPerXu01}. 
 Furthermore, the proof theory for a class of \emph{deontic} \stit logics was addressed as well, with labelled calculi being introduced in~\cite{BerLyo21} and employed to map out the interrelationships between certain deontic \stit logics and \emph{Ought-implies-Can} principles. The most relevant work for our purposes however is~\cite{LyoBer19}, where a simplified semantics for multi-agent \stit logics was exploited to design labelled calculi amenable to the method of refinement. These calculi were then refined and used to provide the first proof-search and counter-model construction algorithms for \stit logics. 

In this section, labelled sequent calculi are given for deontic \stit logics based on the author's joint work in~\cite{BerLyo19a,BerLyo21,LyoBer19}. It will be shown that these calculi possess useful proof-theoretic properties, and refined versions of the calculi (obtained in \cptr~\ref{CPTR:Refinment-Modal}) will be applied to provide proof-search and counter-model extraction algorithms for a class of deontic \stit logics (in \cptr~\ref{CPTR:Applications}).


\begin{definition}[Labelled Sequents for Deontic \stit Logics] We define labelled sequents for deontic \stit logics\index{Labelled sequent!for deontic \stit logics} to be syntactic objects of the form $\Lambda := \rel \sar \Gamma$, where $\rel$ (the \emph{antecedent}) and $\Gamma$ (the \emph{consequent}) are defined via the following grammars in BNF:
\begin{center}
$\rel ::= \seqempstr \ | \ R_{[i]}wu \ | \ \opt_{\Oi}w \ | \ \rel, \rel$ \qquad $\Gamma ::= \seqempstr \ | \ w:\phi \ | \ \Gamma, \Gamma$
\end{center}
with $i \in \ag$, $\phi \in \langdsn$, and $w, u$ elements from a denumerable set of labels $\lab := \{ w, u, v, ... \}$. We refer to formulae of the form $R_{[i]}wu$ and $\opt_{\Oi}w$ as \emph{relational atoms}\index{Relational atom!for deontic \stit logics}, and refer to formulae of the form $w : \phi$ as \emph{labelled formulae}\index{Labelled formula!for deontic \stit logics}.
\end{definition}
 
We use $\Lambda$, $\Lambda'$, \etc (occasionally annotated) to denote labelled sequents, $\rel$, $\rel'$, \etc (occasionally annotated) to denote multisets of relational atoms, and $\Gamma$, $\Gamma'$, \etc (occasionally annotated) to denote multisets of labelled formulae. We therefore take the comma operator to commute and associate in $\rel$ and $\Gamma$; therefore, we would identify $R_{[1]}wu, R_{[2]}wv, \opt_{\Oi}v$ with $R_{[2]}wv, \opt_{\Oi}v, R_{[1]}wu$, and $v:[1]\phi, u : \psi, w:[1]\phi$ with $u : \psi, w : [1]\phi, v : [1]\phi$. This interpretation of comma is what lets us view strings $\rel$ and $\Gamma$ as multisets. As in the previous two sections, we let $\seqempstr$ represent the \emph{empty string}\index{Empty string}, which acts as an identity element for comma (e.g. we identify $R_{[1]}wu, \seqempstr, R_{[2]}wv$ with $R_{[1]}wu, R_{[2]}wv$). 
Therefore, the empty string $\seqempstr$\index{Empty string} will usually be implicit in labelled sequents. Last, we use the notation $\lab(\rel)$, $\lab(\Gamma)$, and $\lab(\rel \sar \Gamma)$ to represent the set of labels contained in $\rel$, $\Gamma$, and $\rel \sar \Gamma$ respectively (e.g., $Lab(R_{[2]}wv \sar u : p) = \{w,v,u\}$).

\begin{figure}[t]
\noindent\hrule
\begin{small}
\begin{center}
\begin{tabular}{c c} 

\AxiomC{ }
\RightLabel{$\id$\index{$\id$}}
\UnaryInfC{$\rel \sar w : p, w : \negnnf{p}, \Gamma$}
\DisplayProof

&

\AxiomC{$\rel \sar w :  \phi, \Gamma$}
\AxiomC{$\rel \sar w :  \psi, \Gamma$}
\RightLabel{$\conr$\index{$\conr$}}
\BinaryInfC{$\rel \sar w :  \phi \wedge \psi, \Gamma$}
\DisplayProof

\end{tabular}
\end{center}

\begin{center}
\begin{tabular}{c c}
\AxiomC{$\rel \sar w :  \phi, w : \psi, \Gamma$}
\RightLabel{$\disr$\index{$\disr$}}
\UnaryInfC{$\rel \sar w :  \phi \vee \psi, \Gamma$}
\DisplayProof

&

\AxiomC{$\rel, R_{[i]}wu, R_{[i]}wv, R_{[i]}uv \sar \Gamma$}
\RightLabel{$\eucli$\index{$\eucli$}}
\UnaryInfC{$\rel, R_{[i]}wu, R_{[i]}wv \sar \Gamma$}
\DisplayProof
\end{tabular}
\end{center}

\begin{center}
\begin{tabular}{c @{\hskip 1em} c @{\hskip 1em} c}
\AxiomC{$\rel \sar u :  \phi, \Gamma$}
\RightLabel{$\boxr^{\dag}$\index{$\boxr$}}
\UnaryInfC{$\rel \sar w :  \Box \phi, \Gamma$}
\DisplayProof

&

\AxiomC{$\rel \sar w :  \Diamond \phi, u :  \phi, \Gamma$}
\RightLabel{$\diar$\index{$\diar$}}
\UnaryInfC{$\rel \sar w :  \Diamond \phi, \Gamma$}
\DisplayProof

&

\AxiomC{$\rel, R_{[i]}ww \sar \Gamma$}
\RightLabel{$\refli$\index{$\refli$}}
\UnaryInfC{$\rel \sar \Gamma$}
\DisplayProof
\end{tabular}
\end{center}

\begin{center}
\begin{tabular}{c c}
\AxiomC{$\rel, \opt_{\Oi}u \sar w : \ominus_{i} \phi, u : \phi, \Gamma$}
\RightLabel{$\ODir$\index{$\ODir$}}
\UnaryInfC{$\rel, \opt_{\Oi}u \sar w : \ominus_{i} \phi, \Gamma$}
\DisplayProof

&

\AxiomC{$\rel, R_{[i]}wu \sar w :  \agdia \phi, u : \phi, \Gamma$}
\RightLabel{$\agdiar$\index{$\agdiar$}}
\UnaryInfC{$\rel, R_{[i]}wu \sar w :  \agdia \phi, \Gamma$}
\DisplayProof
\end{tabular}
\end{center}

\begin{center}
\begin{tabular}{c c c}
\AxiomC{$\rel, R_{[i]}wu \sar u : \phi, \Gamma$}
\RightLabel{$\agboxr^{\dag}$\index{$\agboxr$}}
\UnaryInfC{$\rel \sar w :  [i] \phi, \Gamma$}
\DisplayProof

&

\AxiomC{$\rel, \opt_{\Oi}u \sar u : \phi, \Gamma$}
\RightLabel{$\Oir^{\dag}$\index{$\Oir$}}
\UnaryInfC{$\rel \sar w : \Oi \phi, \Gamma$}
\DisplayProof

&

\AxiomC{$\rel, \opt_{\Oi}u \sar \Gamma$}
\RightLabel{$\dtwoir^{\dag}$\index{$\dtwoir$}}
\UnaryInfC{$\rel \sar \Gamma$}
\DisplayProof
\end{tabular}
\end{center}
\end{small}

\begin{small}
\begin{center}
\begin{tabular}{c @{\hskip 1em} c}
\AxiomC{$\rel,R_{[0]}w_{0}u, ..., R_{[n]}w_{n}u \sar \Gamma$}
\RightLabel{$\ioa^{\dag}$\index{$\ioa$}}
\UnaryInfC{$\rel \sar \Gamma$}
\DisplayProof

&

\AxiomC{$\rel, R_{[i]}wu, \opt_{\Oi}w, \opt_{\Oi}u \sar \Gamma$}
\RightLabel{$\dthreeir$\index{$\dthreeir$}}
\UnaryInfC{$\rel, R_{[i]}wu, \opt_{\Oi}w \sar \Gamma$}
\DisplayProof
\end{tabular}
\end{center}
\end{small}

\begin{center}
\begin{tabular}{c}
\AxiomC{$\Big\{ \R, R_{\agbox}w_{m}w_{j} \sar \Gamma \ \Big| \ 0 \leq m \leq k-1 \text{, } m+1 \leq j \leq k \Big\}$}
\RightLabel{$\choicer$\index{$\choicer$}}
\UnaryInfC{$\R \sar \Gamma$}
\DisplayProof
\end{tabular}
\end{center}

\hrule
\caption{The calculi $\gtdsn$\index{$\gtdsn$} (with $|\ag| = n + 1$ and $n,k \in \mathbb{N}$). The superscript $\dag$ on $\boxr$, $\agboxr$, $\Oir$, $\ioa$, and $(\mathsf{D2}_{i})$ indicates that $u$ is a eigenvariable, i.e. it does not occur in the conclusion. We have $\agboxr$, $\agdiar$, $\Oir$, $\ODir$, $\refli$, $\eucli$, $\dtwoir$, $\dthreeir$, and $\choicer$ rules for each $i \in \ag$. We stipulate that if $k = 0$, then $\choicer$ is omitted from the calculus.
}
\label{fig:base-Gcalculus}
\end{figure}

The calculus $\gtdsn$ for the logic $\dsn$ (with $|\ag| = n + 1$, $k$ the maximum number of choices available to agents, and $n,k \in \mathbb{N}$) is shown in \fig~\ref{fig:base-Gcalculus} with the derivability relation defined as follows:

\begin{definition} We write $\vdash_{\gtdsn} \Lambda$ to indicate that a labelled sequent $\Lambda$ is derivable in $\gtdsn$. 
\end{definition}

The $\id$ rule encodes that fact that in a $\dsn$-model, either a propositional atom holds at a world, or it does not. The $\disr$, $\conr$, $\diar$, $\boxr$, $\ODir$, $\Oir$, $\agdiar$, and $\agboxr$ rules are obtained from the corresponding semantic clauses (\dfn~\ref{def:semantics-dsn}). The structural rules $\refli$ and $\eucli$ encode the fact that choice-cells in a $\dsn$-model are equivalence classes, that is each $R_{\agbox}$ relation is both reflexive and Euclidean as dictated by condition \partcondns. The other conditions \ioacondns, \dtwocondns, \dthreecondns, and \choicecond imposed on a $\dsn$-model are encoded by the rules, $\ioar$, $\dtwoir$, $\dthreeir$, and $\choicer$, respectively. Note that when the maximum number of choices $k > 0$, the $\choicer$ rule contains $k(k+1)/2$ premises, and each sequent $\R, R_{\agbox}x_{m}x_{j} \sar \Gamma$ (for $0 \leq m \leq k-1$ and $m+1 \leq j \leq k$) represents a different premise of the rule. For example, if $k = 1$ or $k = 2$, then the form of the $\choicer$ rule is as shown below top and below bottom, respectively:
\begin{center}
\AxiomC{$\R, R_{\agbox}w_{0}w_{1} \sar \Gamma$}
\RightLabel{$(APC^{1}_{i})$}
\UnaryInfC{$\R \sar \Gamma$}
\DisplayProof
\end{center}
\begin{center}
\AxiomC{$\R, R_{\agbox}w_{0}w_{1} \sar \Gamma$}
\AxiomC{$\R, R_{\agbox}w_{0}w_{2} \sar \Gamma$}
\AxiomC{$\R, R_{\agbox}w_{1}w_{2} \sar \Gamma$}
\RightLabel{$(APC^{2}_{i})$}
\TrinaryInfC{$\R \sar \Gamma$}
\DisplayProof
\end{center}


As is characteristic of labelled systems, our labelled sequents represent an abstraction of a model (\emph{viz.} a $\dsn$-model) with the labels denoting worlds, the relational atoms expressing the relations of the model, and labelled formulae representing that formulae are (un)satisfied at a particular world. To make the correspondence between the syntax of our labelled sequents and the content of a $\dsn$-model, we define how to interpret labelled sequents below. In addition, our sequent semantics will allow us to prove each $\gtdsn$ calculus sound relative to the associated logic $\dsn$, which we show after defining the sequent semantics.

\begin{definition}[$\gtdsn$ Semantics]\label{def:sequent-semantics-dsn} Let $M = (W, \{\R_{[i]} \ | \ i \in \ag\}, \{\opt_{\Oi} \ | \ i \in \ag\}, V)$ be a $\dsn$-model with $I$ an \emph{interpretation function}\index{Interpretation function!for deontic \stit logics} mapping labels to worlds, i.e. $I : \ \lab \mapsto W$. We define the \emph{satisfaction}\index{Satisfaction!for $\gtdsn$} of a relational atom $R_{\agbox}wu$ or $\ideal w$ (written $M, I \models_{\dsn} R_{\agbox}wu$ and $M, I \models_{\dsn} \ideal w$, \resp) and labelled formula $w : \phi$ (written $M, I \models_{\dsn} w : \phi$) as follows:
\begin{itemize}

\item[$\li$] $M, I \models_{\dsn} R_{\agbox}wu$ \ifandonlyif $(I(w),I(u)) \in R_{\agbox}$

\item[$\li$] $M, I \models_{\dsn} \ideal w$ \ifandonlyif $I(w) \in \ideal$

\item[$\li$] $M, I \models_{\dsn} w : \phi$ \ifandonlyif $M, I(w) \Vdash \phi$

\end{itemize}
We say that a multiset of relational atoms $\rel$ is \emph{satisfied} in $M$ with $I$ (written $M, I \models_{\dsn} \rel$) \ifandonlyif $M, I \models_{\dsn} R_{\agbox}wu$ and $M, I \models_{\dsn} \ideal w$ for all $R_{\agbox}wu, \ideal w \in \rel$, and we say that a multiset of labelled formulae $\Gamma$ is \emph{satisfied} in $M$ with $I$ (written $M, I \models_{\dsn} \Gamma$) \ifandonlyif $M, I \models_{\dsn} w : \phi$ for all $w : \phi \in \Gamma$. 

A labelled sequent $\Lambda := \rel \sar \Gamma$ is \emph{satisfied} in $M$ with $I$ (written, $M,I \models_{\dsn} \Lambda$) \ifandonlyif if $M, I \models_{\dsn} \rel$, then $M, I \models_{\dsn} \Gamma$. Also, we say that a labelled sequent $\Lambda$ is \emph{falsified} in $M$ with $I$ \ifandonlyif $M, I \not\models_{\dsn} \Lambda$, that is, $\Lambda$ is not satisfied by $M$ with $I$.

Last, a labelled sequent $\Lambda$ is \emph{$\dsn$-valid}\index{$\dsn$-valid} (written $\models_{\dsn} \Lambda$) \ifandonlyif it is satisfiable in every $\dsn$ model $M$ with every interpretation function $I$. We say that a labelled sequent $\Lambda$ is \emph{$\dsn$-invalid}\index{$\dsn$-invalid} \ifandonlyif $\not\models_{\dsn} \Lambda$, i.e. $\Lambda$ is not $\dsn$-valid.
\end{definition}

Intuitively, the above definition expresses that a sequent $\Lambda := \rel \sar \Gamma$ is satisfied in a $\dsn$-model \ifandonlyif some labelled formula $w : \phi \in \Gamma$ holds, given that all relational atoms in $\rel$ hold. Therefore, in traditional fashion, we interpret the comma in the antecedent conjunctively, and the comma in the consequent disjunctively, with the sequent arrow $\sar$ representing an implication. Using the above definition, we prove our calculi sound:

\begin{theorem}[Soundness]\label{thm:soundness-gtdsn}
If $\vdash_{\gtdsn} \Lambda$, then $\models_{\dsn} \Lambda$.
\end{theorem}

\begin{proof} The result is proven by showing that each initial sequent generated by $\id$ is $\dsn$-valid, and that each additional rule of $\gtdsn$ preserves validity. We prove the latter by contraposition, and show that if the conclusion of the rule is $\dsn$-invalid, then at least one of the premises (or, the premise) of the rule is $\dsn$-invalid. We omit consideration of the $\id$, $\disr$, and $\conr$ cases, as they are simple. 




$\boxr$ Suppose that $M, I \not\models_{\dsn} \rel \sar w : \Box \phi, \Gamma$. It follows that $M, I(w) \not\Vdash \Box \phi$, implying that there exists some $u'$ such that $M, u' \not\Vdash \phi$. Let us define $I'(v) = I(v)$ if $v \neq u$ and $I'(u) = u'$ otherwise. By definition, we have that $M, I'(u) \not\Vdash \phi$. Since $u$ is an eigenvariable, it follows that the premise is falsified by $M$ with $I'$.

$\agboxr$ Assume that $M, I \not\models_{\dsn} \rel \sar w : \agbox \phi, \Gamma$. Then, $M, I(w) \not\Vdash \agbox \phi$, implying that there exists a world $u'$ such that $(I(w),u') \in R_{\agbox}$ and $M, u' \not\Vdash \phi$. We let $I'(v) = I(v)$ if $v \neq u$ and $I'(u) = u'$ otherwise. By definition, $(I'(w),I'(u)) \in R_{\agbox}$ and $M, I'(u) \not\Vdash \phi$. Since the label $u$ is an eigenvariable, it follows that the premise is falsified by $M$ with $I'$.

$\Oir$ Assume that $M, I \not\models_{\dsn} \rel \sar w : \Oi \phi$. It follows that $M, I(w) \not\Vdash w : \Oi \phi$, implying that there exists a world $u'$ such that $u' \in \ideal$ and $M, u' \not\Vdash \phi$. We define $I'(v) = I(v)$ if $v \neq u$ and $I'(u) = u'$ otherwise. By definition, $I'(u) \in \ideal$ and $M, I'(u) \not\Vdash \phi$, and since $u$ is an eigenvariable, it follows that the premise is falsified by $M$ with $I'$.

$\diar$ Assume that $M, I \not\models_{\dsn} \rel \sar w : \Diamond \phi, \Gamma$. Then, $M, I(w) \not\Vdash \Diamond \phi$, meaning that for all $v \in W$, $M, v \not\Vdash \phi$. Since $I(u) \in W$, this implies that $M, I(u) \not\Vdash \phi$, which shows that the premise is falsified by $M$ with $I$.

$\agdiar$ Assume that $M, I \not\models_{\dsn} \rel, R_{\agbox}wu \sar w : \agdia \phi, \Gamma$. Then, $M, I(w) \not\Vdash \agbox \phi$, which implies that for all $u'$, if $(I(w),u') \in R_{\agbox}$, then $M, u' \not\Vdash \phi$. Since $(I(w),I(u)) \in R_{\agbox}$, it follows that $M, I(u) \not\Vdash \phi$, which shows that the premise is falsified by $M$ with $I$.

$\ODir$ Let us suppose that $M, I \not\models_{\dsn} \rel, \ideal u \sar w : \ODi \phi, \Gamma$. It follows that $M, I(w) \not\Vdash \ODi \phi$, implying that for all $u'$, if $u' \in \ideal$, then $M, u' \not\Vdash \phi$. Since $I(u) \in \ideal$, we have that $M, I(u) \not\Vdash \phi$, which shows that the premise is falsified by $M$ with $I$.

$\refli$ Let us suppose that $M, I \not\models_{\dsn} \rel \sar \Gamma$. Since the relation $R_{\agbox}$ is reflexive, it follows that $(I(w),I(w)) \in R_{\agbox}$, showing that the premise is falsified by $M$ with $I$.

$\eucli$ Let us suppose that $M, I \not\models_{\dsn} \rel, R_{\agbox}wu, R_{\agbox}uv \sar \Gamma$. Therefore, $(I(w),I(u)) \in R_{\agbox}$ and $(I(u), I(v)) \in R_{\agbox}$. Since the relation $R_{\agbox}$ is Euclidean, it follows that $(I(w),I(v)) \in R_{\agbox}$, which shows that the premise is falsified by $M$ with $I$.

$\dtwoir$ Let us suppose that $M, I \not\models_{\dsn} \rel \sar \Gamma$. Since $\ideal$ satisfies condition \dtwocondns, we know there exists some $u' \in W$ such that $u' \in \ideal$. Let us define $I'(v) = I(v)$ if $v \neq u$, and $I'(u) = u'$ otherwise. By definition, $I'(u) \in \ideal$, and since $u$ is an eigenvariable, it follows that the premise is falsified by $M$ with $I'$.

$\dthreeir$ Suppose that $M, I \not\models_{\dsn} \rel, R_{\agbox}wu, \ideal w \sar \Gamma$. This entails that $(I(w),I(u)) \in R_{\agbox}$ and $I(w) \in \ideal$. Since our model $M$ satisfies the condition \dthreecondns, it follows that $I(u) \in \ideal$, which shows that the premise is falsified by $M$ with $I$.

$\ioa$ Suppose that $M, I \not\models_{\dsn} \rel \sar \Gamma$. Since our model $M$ satisfies the \ioacond condition, we know that for $I(w_{0}), \ldots, I(w_{n}) \in W$, there exists some $u'$ such that $(I(w_{i}),u') \in R_{\agbox}$ for $i \in \{0, \ldots, n\}$. Let us define $I'(v) = I(v)$ if $v \neq u$ and $I'(u) = u'$ otherwise. Then, it follows that $(I(w_{i}),I(u)) \in R_{\agbox}$ for $i \in \{0, \ldots, n\}$. Because $u$ is an eigenvariable, we have that the premise is falsified by $M$ with $I'$.

$\choicer$ Suppose that $M, I \not\models_{\dsn} \rel \sar \Gamma$. Then, since our model $M$ satisfies the \choicecond condition, we know there exist worlds $I(w_{0}), \ldots, I(w_{k}) \in W$ such that 
$$
\bigvee_{0 \leq m \leq k-1, m+1 \leq j \leq k} (I(w_{m}),I(w_{j})) \in R_{\agbox}
$$
holds. Hence, there must exist an $m$ and $j$ (with $0 \leq m \leq k-1, m+1 \leq j \leq k$) such that $(I(w_{m}),I(w_{j})) \in R_{\agbox}$ holds, showing that one of the premises is falsified by $M$ with $I$.
\end{proof}


Our labelled calculi for deontic \stit logics possess fundamental proof-theoretic properties such as the hp-admissibility of label substitutions $\lsub$, weakening $\wk$, and contractions $\ctrrel$, $\ctrr$. Moreover, all rules of the calculi are hp-invertible, and each calculus admits syntactic cut-elimination. All results are shown below, with the methods of proof based upon the work in~\cite{Neg05,Sim94,Vig00}. 

The aforementioned properties are useful not only in establishing the completeness of our calculi (\thm~\ref{thm:completness-gtdsn}), but are useful in applications such as decidability via proof-search (see \sect~\ref{sec:applicationsI}). Before moving on to establish these results, we define the notion of a \emph{label substitution}, which is an operation that is fundamental for establishing our results.

\begin{definition}[Label Substitution]\label{lem:label-sub-dsn} We define a \emph{label substitution}\index{Label substitution!for deontic \stit logics} $(w/u)$ for $w, u \in \lab$ on individual relational atoms and labelled formulae as follows:
\begin{itemize}

\item[$\li$] $(R_{\agbox}vz)(w/u) =
  \begin{cases}
                                   R_{\agbox}wz & \text{if $u = v$ and $v \neq z$} \\
                                   R_{\agbox}ww & \text{if $u = v$ and $v = z$} \\
                                   R_{\agbox}uw & \text{if $u = z$ and $v \neq z$} \\
                                   R_{\agbox}vz & \text{otherwise}
  \end{cases}
$

\item[$\li$] $(\ideal v)(w/u) =
  \begin{cases}
                                   \ideal w & \text{if $u = v$} \\
                                   \ideal v & \text{otherwise}
  \end{cases}
$

\item[$\li$] $(v : \phi)(w/u) =
  \begin{cases}
                                   w : \phi & \text{if $u = v$} \\
                                   v : \phi & \text{otherwise}
  \end{cases}
$
\end{itemize}
We define the label substitution $(w/u)$ on a multiset of relational atoms $\rel$ and a multiset of labelled formulae $\Gamma$ to be the multiset obtained by applying $(w/u)$ to each element of the multiset.
\end{definition}

\begin{figure}[t]
\noindent\hrule
\begin{small}
\begin{center}
\begin{tabular}{c c c} 
\AxiomC{$\rel \sar \Gamma$}
\RightLabel{$\lsub$}
\UnaryInfC{$\rel(w/u) \sar \Gamma(w/u)$}
\DisplayProof

&

\AxiomC{$\rel \sar \Gamma$}
\RightLabel{$\wk$}
\UnaryInfC{$\rel,\rel' \sar \Gamma,\Gamma'$}
\DisplayProof

&

\AxiomC{$\rel,\rel',\rel' \sar \Gamma$}
\RightLabel{$\ctrrel$}
\UnaryInfC{$\rel,\rel' \sar \Gamma$}
\DisplayProof
\end{tabular}
\end{center}
\begin{center}
\begin{tabular}{c c}
\AxiomC{$\rel \sar w:\phi, w:\phi,\Gamma$}
\RightLabel{$\ctrr$}
\UnaryInfC{$\rel \sar w:\phi,\Gamma$}
\DisplayProof

&

\AxiomC{$\rel \sar w : \phi,\Gamma$}
\AxiomC{$\rel \sar w : \negnnf{\phi},\Gamma$}
\RightLabel{$\cut$}
\BinaryInfC{$\rel \sar \Gamma$}
\DisplayProof
\end{tabular}
\end{center}
\end{small}
\hrule
\caption{The set $\strucsetdsn$ of structural rules\index{Structural rule!for deontic \stit logics} consists of the rules above.}
\label{fig:structural-rules}
\end{figure}

\begin{lemma}\label{lem:general-id-dsn}
For all $\phi \in \langdsn$, $\vdash_{\gtdsn} \rel \sar w : \phi, w : \negnnf{\phi}, \Gamma$.
\end{lemma}

\begin{proof} We prove the result by induction on the complexity of $\phi$.

\textit{Base case.} If $\fcomp{\phi} = 0$, then $\phi$ is of the form $p$ or $\negnnf{p}$. In either case, the desired result is obtained as an instance of $\id$. (NB. If $\phi$ is of the form $\negnnf{p}$, then observe that the sequent $\rel \sar w : \negnnf{p}, w : \negnnf{\negnnf{p}}, \Gamma$ is identical to $\rel \sar w : \negnnf{p}, w : p, \Gamma$ by the definition of negation (\dfn~\ref{def:negation-dsn}), which is an instance of $\id$.)

\textit{Inductive step.} We show the cases where $\phi$ is of the form $\Box \psi$, $\agdia \psi$, and $\Oi \psi$; the other cases when $\phi$ is of the form $\psi \lor \chi$, $\psi \land \chi$, $\Diamond \psi$, $\agbox \psi$, or $\ODi \psi$ are simple or are shown similarly.

\begin{center}
\begin{tabular}{c c}
\AxiomC{}
\RightLabel{\ih}
\dashedLine
\UnaryInfC{$\rel \sar u : \phi, w : \Diamond \negnnf{\psi}, u : \negnnf{\psi}, \Gamma$}
\RightLabel{$\diar$}
\UnaryInfC{$\rel \sar u : \phi, w : \Diamond \negnnf{\psi}, \Gamma$}
\RightLabel{$\boxr$}
\UnaryInfC{$\rel \sar w : \Box \psi, w : \Diamond \negnnf{\psi}, \Gamma$}
\DisplayProof

&

\AxiomC{}
\RightLabel{\ih}
\dashedLine
\UnaryInfC{$\rel, R_{\agbox}wu \sar w : \agdia \psi, u : \psi, u : \negnnf{\psi}, \Gamma$}
\RightLabel{$\agdiar$}
\UnaryInfC{$\rel, R_{\agbox}wu \sar w : \agdia \psi, u : \negnnf{\psi}, \Gamma$}
\RightLabel{$\agboxr$}
\UnaryInfC{$\rel \sar w : \agdia \psi, w : \agbox \negnnf{\psi}, \Gamma$}
\DisplayProof
\end{tabular}
\end{center}

\begin{center}
\AxiomC{}
\RightLabel{\ih}
\dashedLine
\UnaryInfC{$\rel, \ideal u \sar u : \psi, u : \negnnf{\psi}, w : \ODi \negnnf{\psi}, \Gamma$}
\RightLabel{$\ODir$}
\UnaryInfC{$\rel, \ideal u \sar u : \psi, w : \ODi \negnnf{\psi}, \Gamma$}
\RightLabel{$\Oir$}
\UnaryInfC{$\rel \sar w : \Oi \psi, w : \ODi \negnnf{\psi}, \Gamma$}
\DisplayProof
\end{center}

\end{proof}

\begin{lemma}\label{lem:lsub-admiss-dsn}
The rule $\lsub$ is hp-admissible in $\gtdsn$. 
\end{lemma}

\begin{proof} We prove the result by induction on the height of the given derivation.

\textit{Base case.} Any application of $\lsub$ to an instance of $\id$ yields another instance of $\id$, which solves the base case.

\textit{Inductive step.} With the exception of the $\boxr$, $\agboxr$, $\Oir$, $\dtwoir$, and $\ioar$ rules, all cases are resolved by applying \ih and then the corresponding rule. Concerning the $\boxr$, $\agboxr$, $\Oir$, $\dtwoir$, and $\ioar$ cases, the non-trivial cases occur when the label substituted in is identical to the eigenvariable. We show how to resolve these non-trivial cases below and note that all other cases (i.e. the trivial cases) are resolved by applying \ih followed by the corresponding rule. Below, we assume that $z$ is a fresh label.

\begin{center}
\begin{tabular}{c c c}
\AxiomC{$\rel \sar u :  \phi, \Gamma$}
\RightLabel{$\boxr$}
\UnaryInfC{$\rel \sar w :  \Box \phi, \Gamma$}
\RightLabel{$\lsub$}
\UnaryInfC{$\rel(u/v) \sar (w :  \Box \phi)(u/v), \Gamma(u/v)$}
\DisplayProof

&

$\leadsto$

&

\AxiomC{$\rel \sar u :  \phi, \Gamma$}
\RightLabel{\ih}
\dashedLine
\UnaryInfC{$\rel(z/u) \sar (u : \phi)(z/u), \Gamma(z/u)$}
\RightLabel{=}
\dottedLine
\UnaryInfC{$\rel \sar z :  \phi, \Gamma$}
\RightLabel{\ih}
\dashedLine
\UnaryInfC{$\rel(u/v) \sar (z :  \phi)(u/v), \Gamma(u/v)$}
\RightLabel{=}
\dottedLine
\UnaryInfC{$\rel(u/v) \sar z :  \phi, \Gamma(u/v)$}
\RightLabel{$\boxr$}
\UnaryInfC{$\rel(u/v) \sar (w : \Box \phi)(u/v), \Gamma(u/v)$}
\DisplayProof
\end{tabular}
\end{center}

\begin{flushleft}
\begin{tabular}{c c c}
\AxiomC{$\rel, R_{[i]}wu \sar u : \phi, \Gamma$}
\RightLabel{$\agboxr$}
\UnaryInfC{$\rel \sar w : [i] \phi, \Gamma$}
\RightLabel{$\lsub$}
\UnaryInfC{$\rel(u/v) \sar (w : [i] \phi)(u/v), \Gamma(u/v)$}
\DisplayProof

&

$\leadsto$
\end{tabular}
\end{flushleft}
\begin{flushright}
\AxiomC{$\rel, R_{[i]}wu \sar u :  \phi, \Gamma$}
\RightLabel{\ih}
\dashedLine
\UnaryInfC{$\rel(z/u), (R_{[i]}wu)(z/u) \sar (u : \phi)(z/u), \Gamma(z/u)$}
\RightLabel{=}
\dottedLine
\UnaryInfC{$\rel, R_{[i]}wz \sar z :  \phi, \Gamma$}
\RightLabel{\ih}
\dashedLine
\UnaryInfC{$\rel(u/v), (R_{[i]}wu)(u/v) \sar (z : \phi)(u/v), \Gamma(u/v)$}
\RightLabel{=}
\dottedLine
\UnaryInfC{$\rel(u/v), (R_{[i]}wu)(u/v) \sar z : \phi, \Gamma(u/v)$}
\RightLabel{$\agboxr$}
\UnaryInfC{$\rel(u/v) \sar (w : \agbox \phi)(u/v), \Gamma(u/v)$}
\DisplayProof
\end{flushright}

\begin{flushleft}
\begin{tabular}{c c c}
\AxiomC{$\rel, \opt_{\Oi}u \sar u : \phi, \Gamma$}
\RightLabel{$\Oir$}
\UnaryInfC{$\rel \sar w : \Oi \phi, \Gamma$}
\RightLabel{$\lsub$}
\UnaryInfC{$\rel(u/v) \sar (w : \Oi \phi)(u/v), \Gamma(u/v)$}
\DisplayProof

&

$\leadsto$

\end{tabular}
\end{flushleft}
\begin{flushright}
\AxiomC{$\rel, \opt_{\Oi}u \sar u : \phi, \Gamma$}
\RightLabel{\ih}
\dashedLine
\UnaryInfC{$\rel(z/u), (\opt_{\Oi}u)(z/u) \sar (u :  \phi)(z/u), \Gamma(z/u)$}
\RightLabel{=}
\dottedLine
\UnaryInfC{$\rel, \ideal z \sar z :  \phi, \Gamma$}
\RightLabel{\ih}
\dashedLine
\UnaryInfC{$\rel(u/v), (\ideal z)(u/v) \sar (z :  \phi)(u/v), \Gamma(u/v)$}
\RightLabel{=}
\dottedLine
\UnaryInfC{$\rel(u/v), \ideal z \sar z :  \phi, \Gamma(u/v)$}
\RightLabel{$\Oir$}
\UnaryInfC{$\rel(u/v) \sar (w : \Oi \phi)(u/v), \Gamma(u/v)$}
\DisplayProof
\end{flushright}

\begin{center}
\begin{tabular}{c c c}
\AxiomC{$\rel, \opt_{\Oi}u \sar \Gamma$}
\RightLabel{$\dtwoir$}
\UnaryInfC{$\rel \sar \Gamma$}
\RightLabel{$\lsub$}
\UnaryInfC{$\rel(u/v) \sar \Gamma(u/v)$}
\DisplayProof

&

$\leadsto$

&

\AxiomC{$\rel, \opt_{\Oi}u \sar \Gamma$}
\RightLabel{\ih}
\dashedLine
\UnaryInfC{$\rel(z/u), (\opt_{\Oi}u)(z/u) \sar \Gamma(z/u)$}
\RightLabel{=}
\dottedLine
\UnaryInfC{$\rel, \ideal z \sar \Gamma$}
\RightLabel{\ih}
\dashedLine
\UnaryInfC{$\rel(u/v), (\ideal z)(u/v) \sar \Gamma(u/v)$}
\RightLabel{=}
\dottedLine
\UnaryInfC{$\rel(u/v), \ideal z \sar \Gamma(u/v)$}
\RightLabel{$\dtwoir$}
\UnaryInfC{$\rel(u/v) \sar \Gamma(u/v)$}
\DisplayProof
\end{tabular}
\end{center}

\begin{flushleft}
\begin{tabular}{c c c}
\AxiomC{$\rel,R_{[0]}w_{0}u, ..., R_{[n]}w_{n}u \sar \Gamma$}
\RightLabel{$\ioa$}
\UnaryInfC{$\rel \sar \Gamma$}
\RightLabel{$\lsub$}
\UnaryInfC{$\rel(u/v) \sar \Gamma(u/v)$}
\DisplayProof

&

$\leadsto$

\end{tabular}
\end{flushleft}
\begin{flushright}
\AxiomC{$\rel,R_{[0]}w_{0}u, ..., R_{[n]}w_{n}u \sar \Gamma$}
\RightLabel{\ih}
\dashedLine
\UnaryInfC{$\rel(z/u), (R_{[0]}w_{0}u, ..., R_{[n]}w_{n}u)(z/u) \sar \Gamma(z/u)$}
\RightLabel{=}
\dottedLine
\UnaryInfC{$\rel, R_{[0]}w_{0}z, ..., R_{[n]}w_{n}z \sar \Gamma$}
\RightLabel{\ih}
\dashedLine
\UnaryInfC{$\rel(u/v), (R_{[0]}w_{0}u, ..., R_{[n]}w_{n}u)(u/v) \sar \Gamma(u/v)$}
\RightLabel{$\ioar$}
\UnaryInfC{$\rel(u/v) \sar \Gamma(u/v)$}
\DisplayProof
\end{flushright}
\end{proof}

\begin{lemma}\label{lem:wk-admiss-dsn}
The rule $\wk$ is hp-admissible in $\gtdsn$. 
\end{lemma}

\begin{proof} We prove the result by induction on the height of the given derivation.

\textit{Base case.} The base case is easily resolved as any application of $\wk$ to an instance of $\id$ gives another instance of $\id$.

\textit{Inductive step.} With the exception of the $\boxr$, $\agboxr$, $\Oir$, $\dtwoir$, and $\ioar$ cases, all cases are resolved by invoking \ih followed by the corresponding rule. The non-trivial $\boxr$, $\agboxr$, $\Oir$, $\dtwoir$, and $\ioar$ cases arise when a relational atom or labelled formula is weakened in containing the eigenvariable of the inference. We show how to resolve these cases below and omit the other (trivial) cases as they follow by invoking \ih and then applying the corresponding rule. We let $z$ be a fresh label below.

\begin{center}
\begin{tabular}{c c c}
\AxiomC{$\rel \sar u :  \phi, \Gamma$}
\RightLabel{$\boxr$}
\UnaryInfC{$\rel \sar w :  \Box \phi, \Gamma$}
\RightLabel{$\wk$}
\UnaryInfC{$\rel, \rel' \sar w :  \Box \phi, \Gamma, \Gamma'$}
\DisplayProof

&

$\leadsto$

&

\AxiomC{$\rel \sar u :  \phi, \Gamma$}
\RightLabel{$\lsub$}
\dashedLine
\UnaryInfC{$\rel(z/u) \sar u :  \phi(z/u), \Gamma(z/u)$}
\RightLabel{=}
\dottedLine
\UnaryInfC{$\rel \sar z :  \phi, \Gamma$}
\RightLabel{\ih}
\dashedLine
\UnaryInfC{$\rel, \rel' \sar z :  \phi, \Gamma, \Gamma'$}
\RightLabel{$\boxr$}
\UnaryInfC{$\rel \sar w :  \Box \phi, \Gamma$}
\DisplayProof
\end{tabular}
\end{center}

\begin{center}
\begin{tabular}{c c c}
\AxiomC{$\rel, R_{\agbox}wu \sar u : \phi, \Gamma$}
\RightLabel{$\agboxr$}
\UnaryInfC{$\rel \sar w :  \agbox \phi, \Gamma$}
\RightLabel{$\wk$}
\UnaryInfC{$\rel, \rel' \sar w :  \agbox \phi, \Gamma, \Gamma'$}
\DisplayProof

&

$\leadsto$

&

\AxiomC{$\rel, R_{[i]}wu \sar u : \phi, \Gamma$}
\RightLabel{$\lsub$}
\dashedLine
\UnaryInfC{$\rel(z/u), R_{[i]}wu(z/u) \sar u :  \phi(z/u), \Gamma(z/u)$}
\RightLabel{=}
\dottedLine
\UnaryInfC{$\rel, R_{[i]}wz \sar z : \phi, \Gamma$}
\RightLabel{\ih}
\dashedLine
\UnaryInfC{$\rel, \rel', R_{[i]}wz \sar z :  \phi, \Gamma, \Gamma'$}
\RightLabel{$\agboxr$}
\UnaryInfC{$\rel, \rel' \sar w :  \agbox \phi, \Gamma, \Gamma'$}
\DisplayProof
\end{tabular}
\end{center}

\begin{center}
\begin{tabular}{c c c}
\AxiomC{$\rel, \opt_{\Oi}u \sar u : \phi, \Gamma$}
\RightLabel{$\Oir$}
\UnaryInfC{$\rel \sar w : \Oi \phi, \Gamma$}
\RightLabel{$\wk$}
\UnaryInfC{$\rel, \rel' \sar w : \Oi \phi, \Gamma, \Gamma'$}
\DisplayProof

&

$\leadsto$

&

\AxiomC{$\rel, \opt_{\Oi}u \sar u : \phi, \Gamma$}
\RightLabel{$\lsub$}
\dashedLine
\UnaryInfC{$\rel(z/u), \ideal u(z/u) \sar u :  \phi(z/u), \Gamma(z/u)$}
\RightLabel{=}
\dottedLine
\UnaryInfC{$\rel, \ideal z \sar z : \phi, \Gamma$}
\RightLabel{\ih}
\dashedLine
\UnaryInfC{$\rel, \rel', \ideal z \sar z :  \phi, \Gamma, \Gamma'$}
\RightLabel{$\Oir$}
\UnaryInfC{$\rel, \rel' \sar w : \Oi \phi, \Gamma, \Gamma'$}
\DisplayProof
\end{tabular}
\end{center}

\begin{center}
\begin{tabular}{c c c}
\AxiomC{$\rel, \opt_{\Oi}u \sar \Gamma$}
\RightLabel{$\dtwoir$}
\UnaryInfC{$\rel \sar \Gamma$}
\RightLabel{$\wk$}
\UnaryInfC{$\rel, \rel' \sar \Gamma, \Gamma'$}
\DisplayProof

&

$\leadsto$

&

\AxiomC{$\rel, \opt_{\Oi}u \sar \Gamma$}
\RightLabel{$\lsub$}
\dashedLine
\UnaryInfC{$\rel(z/u), \ideal u(z/u) \sar \Gamma(z/u)$}
\RightLabel{=}
\dottedLine
\UnaryInfC{$\rel, \ideal z \sar \Gamma$}
\RightLabel{\ih}
\dashedLine
\UnaryInfC{$\rel, \rel', \ideal z \sar \Gamma, \Gamma'$}
\RightLabel{$\Oir$}
\UnaryInfC{$\rel, \rel' \sar \Gamma, \Gamma'$}
\DisplayProof
\end{tabular}
\end{center}

\begin{flushleft}
\begin{tabular}{c c c}
\AxiomC{$\rel,R_{[0]}w_{0}u, ..., R_{[n]}w_{n}u \sar \Gamma$}
\RightLabel{$\ioa$}
\UnaryInfC{$\rel \sar \Gamma$}
\RightLabel{$\wk$}
\UnaryInfC{$\rel,\rel' \sar \Gamma, \Gamma'$}
\DisplayProof

&

$\leadsto$
\end{tabular}
\end{flushleft}
\begin{flushright}
\AxiomC{$\rel,R_{[0]}w_{0}u, ..., R_{[n]}w_{n}u \sar \Gamma$}
\RightLabel{$\lsub$}
\dashedLine
\UnaryInfC{$\rel(z/u),R_{[0]}w_{0}u(z/u), ..., R_{[n]}w_{n}u(z/u) \sar \Gamma(z/u)$}
\RightLabel{=}
\dottedLine
\UnaryInfC{$\rel,R_{[0]}w_{0}z, ..., R_{[n]}w_{n}z \sar \Gamma$}
\RightLabel{\ih}
\dashedLine
\UnaryInfC{$\rel, \rel', R_{[0]}w_{0}z, ..., R_{[n]}w_{n}z \sar \Gamma, \Gamma'$}
\RightLabel{$\ioar$}
\UnaryInfC{$\rel, \rel' \sar \Gamma, \Gamma'$}
\DisplayProof
\end{flushright}

\end{proof}

\begin{lemma}\label{lem:invert-dsn}
All rules  in $\gtdsn$ are hp-invertible. 
\end{lemma}

\begin{proof} The hp-invertibility of $\diar$, $\agdiar$, $\ODir$, $\refli$, $\eucli$, $\ioa$, $\dtwoir$, and $\dthreeir$ follow from \lem~\ref{lem:wk-admiss-dsn}. The remaining rules are shown to be hp-invertible by induction on the height of the given derivation, and the proof is similar to the proof of \cite[\prp~4.11]{Neg05}.
\end{proof}

\begin{lemma}\label{lem:ctrr-admiss-dsn}
The rules $\ctrrel$ and $\ctrr$ is hp-admissible in $\gtdsn$. 
\end{lemma}

\begin{proof} The result is shown by induction on the height of the given derivation. The hp-admissibility of $\ctrrel$ is straightforward, as any application of $\ctrrel$ to an initial sequent yields another initial sequent, and each case of the inductive step follows by applying \ih and then the corresponding rule. Therefore, we focus solely on the hp-admissibility proof of $\ctrr$.

\textit{Base case.} Any application of $\ctrr$ to an instance of $\id$ yields another instance of $\id$, resolving the base case.

\textit{Inductive step.} For the inductive step, we assume that the given derivation ends with an application of $\ru$ followed by an application of $\ctrr$. If the principal formula of $\ru$ is not active in $\ctrr$, then the active formulae of $\ctrr$ occur in the premise(s) of $\ru$. Hence, the desired conclusion is obtained by applying \ih to the premise(s) of $\ru$, followed by an application of $\ru$. We may therefore assume that the principal formula of $\ru$ is active in $\ctrr$. If the rule $\ru$ is an instance of $\diar$, $\agdiar$, or $\ODir$ then since the principal formula occurs in the premise of the rule, the case is resolved by applying \ih to the premise, followed by an application of the respective rule. Also, note that $\ru$ cannot be $\refli$, $\eucli$, $\dtwoir$, $\dthreeir$, $\ioar$, or $\choicer$ by our assumption, since it does not have a principal formula on the right. This leaves the $\disr$, $\conr$, $\boxr$, $\agboxr$, and $\Oir$ cases; we show the latter three cases below, as the first two are resolved as in~\cite[\thm~4.12]{Neg05}.

\begin{center}
\begin{tabular}{c c c}
\AxiomC{$\rel \sar u : \phi, w : \Box \phi, \Gamma$}
\RightLabel{$\boxr$}
\UnaryInfC{$\rel \sar w : \Box \phi, w : \Box \phi, \Gamma$}
\RightLabel{$\ctrr$}
\UnaryInfC{$\rel \sar w : \Box \phi, \Gamma$}
\DisplayProof

&

$\leadsto$

&

\AxiomC{$\rel \sar u : \phi, w : \Box \phi, \Gamma$}
\RightLabel{\lem~\ref{lem:invert-dsn}}
\dashedLine
\UnaryInfC{$\rel \sar u : \phi, v : \phi, \Gamma$}
\RightLabel{$\lsub$}
\dashedLine
\UnaryInfC{$\rel(u/v) \sar (u : \phi)(u/v), (v : \phi)(u/v), \Gamma(u/v)$}
\RightLabel{=}
\dottedLine
\UnaryInfC{$\rel \sar u : \phi, u : \phi, \Gamma$}
\RightLabel{\ih}
\dashedLine
\UnaryInfC{$\rel \sar u : \phi, \Gamma$}
\RightLabel{$\boxr$}
\UnaryInfC{$\rel \sar w : \Box \phi, \Gamma$}
\DisplayProof
\end{tabular}
\end{center}

\begin{flushleft}
\begin{tabular}{c c}
\AxiomC{$\rel, R_{\agbox}wu \sar u : \phi, w : \agbox \phi, \Gamma$}
\RightLabel{$\agboxr$}
\UnaryInfC{$\rel \sar w : \agbox \phi, w : \agbox \phi, \Gamma$}
\RightLabel{$\ctrr$}
\UnaryInfC{$\rel \sar w : \agbox \phi, \Gamma$}
\DisplayProof

&

$\leadsto$
\end{tabular}
\end{flushleft}
\begin{flushright}
\AxiomC{$\rel, R_{\agbox}wu \sar u : \phi, w : \agbox \phi, \Gamma$}
\RightLabel{\lem~\ref{lem:invert-dsn}}
\dashedLine
\UnaryInfC{$\rel, R_{\agbox}wu, R_{\agbox}wv \sar u : \phi, v : \phi, \Gamma$}
\RightLabel{$\lsub$}
\dashedLine
\UnaryInfC{$\rel(u/v), (R_{\agbox}wu)(u/v), (R_{\agbox}wv)(u/v) \sar (u : \phi)(u/v), (v : \phi)(u/v), \Gamma(u/v)$}
\RightLabel{=}
\dottedLine
\UnaryInfC{$\rel, R_{\agbox}wu, R_{\agbox}wu \sar u : \phi, u : \phi, \Gamma$}
\RightLabel{$\ctrrel$ + \ih}
\dashedLine
\UnaryInfC{$\rel, R_{\agbox}wu \sar u : \phi, \Gamma$}
\RightLabel{$\agboxr$}
\UnaryInfC{$\rel \sar w : \agbox \phi, \Gamma$}
\DisplayProof
\end{flushright}

\begin{flushleft}
\begin{tabular}{c c}
\AxiomC{$\rel, \ideal u \sar u : \phi, w : \Oi \phi, \Gamma$}
\RightLabel{$\Oir$}
\UnaryInfC{$\rel \sar w : \Oi \phi, w : \Oi \phi, \Gamma$}
\RightLabel{$\ctrr$}
\UnaryInfC{$\rel \sar w : \Oi \phi, \Gamma$}
\DisplayProof

&

$\leadsto$
\end{tabular}
\end{flushleft}
\begin{flushright}
\AxiomC{$\rel, \ideal u \sar u : \phi, w : \Oi \phi, \Gamma$}
\RightLabel{\lem~\ref{lem:invert-dsn}}
\dashedLine
\UnaryInfC{$\rel, \ideal u, \ideal v \sar u : \phi, v : \phi, \Gamma$}
\RightLabel{$\lsub$}
\dashedLine
\UnaryInfC{$\rel(u/v), (\ideal u)(u/v), (\ideal v)(u/v) \sar (u : \phi)(u/v), (v : \phi)(u/v), \Gamma(u/v)$}
\RightLabel{=}
\dottedLine
\UnaryInfC{$\rel, \ideal u, \ideal u \sar u : \phi, u : \phi, \Gamma$}
\RightLabel{$\ctrrel$ + \ih}
\dashedLine
\UnaryInfC{$\rel, \ideal u \sar u : \phi, \Gamma$}
\RightLabel{$\Oir$}
\UnaryInfC{$\rel \sar w : \Oi \phi, \Gamma$}
\DisplayProof
\end{flushright}
\end{proof}

\begin{theorem}\label{lem:cut-admiss-dsn}
The rule $\cut$ is eliminable in $\gtdsn$. 
\end{theorem}

\begin{proof} We assume w.l.o.g. that we are given a derivation where $\cut$ is the last rule used, and that no other instances of $\cut$ appear in the given derivation. The general result follows by repeatedly applying the algorithm described below to successively eliminate topmost instances of $\cut$ until the derivation is free of $\cut$ instances. We prove the result by induction on the lexicographic ordering of pairs $(\fcomp{\phi},h_{1}+h_{2})$, where $\fcomp{\phi}$ is the complexity of the cut formula $\phi$, $h_{1}$ is the height of the derivation of the left premise of $\cut$, and $h_{2}$ is the height of the derivation of the right premise of $\cut$. As is typical when proving cut elimination, there are a large number of cases, and so, we explicitly write the assumptions being made in each case for clarity.

\textit{1. One of the premises of $\cut$ is an instance of $\id$.}

\quad \textit{1.1 The left premise of $\cut$ is an instance of $\id$.} Then, the left premise of $\cut$ is of the form $\rel \sar w : p, w : \negnnf{p}, \Delta$, and there are three subcases to consider:

\qquad \textit{1.1.1 The cut formula is $w : p$.} It follows that the right premise of $\cut$ is of the form $\rel \sar w : \negnnf{p}, w :\negnnf{p}, \Gamma$, and the conclusion of $\cut$ is of the form $\rel \sar w : \negnnf{p}, \Gamma$. Observe that applying the hp-admissiblity of $\ctrr$ (\lem~\ref{lem:ctrr-admiss-dsn}) to the proof of the right premise yields a proof of the desired conclusion.

\qquad \textit{1.1.2 The cut formula is $w : \negnnf{p}$.} This case is similar to the previous case 1.1.1.

\qquad \textit{1.1.3 The cut formula is in $\Gamma$.} Then, the conclusion of $\cut$ is of the form $\rel \sar w : p, w : \negnnf{p}, \Gamma'$, which is an instance of $\id$.

\quad \textit{1.2 The right premise of $\cut$ is an instance of $\id$.} This case is shown similarly to the previous case 1.1.

\textit{2. None of the premises of $\cut$ are an instance of $\id$.} In this case, we let $\rel \sar w : \phi, \Gamma$ be the left premise of $\cut$ and $\rel \sar w : \negnnf{\phi}, \Gamma$ by the right premise of $\cut$.

\quad \textit{2.1 The cut formula $w : \phi$ is not principal in the left premise of $\cut$.} We show the $\Oir$, $\diar$, and $\ioar$ cases, as the remaining cases are similar. In each case below we may invoke \ih as the sum of the heights $h_{1} + h_{2}$ has decreased.

\qquad \textit{2.1.1 The left premise of $\cut$ is derived with $\Oir$.}

\begin{flushleft}
\begin{tabular}{c c}
\AxiomC{$\rel, \opt_{\Oi}u \sar u : \psi, w : \phi, \Gamma$}
\RightLabel{$\Oir$}
\UnaryInfC{$\rel \sar v : \Oi \psi, w : \phi, \Gamma$}

\AxiomC{$\rel \sar v : \Oi \psi, w : \negnnf{\phi}, \Gamma$}
\RightLabel{$\cut$}
\BinaryInfC{$\rel \sar v : \Oi \psi, \Gamma$}
\DisplayProof

&

$\leadsto$
\end{tabular}
\end{flushleft}

\begin{flushright}
\AxiomC{$\rel, \opt_{\Oi}u \sar u : \psi, w : \phi, \Gamma$}
\AxiomC{$\rel \sar v : \Oi \psi, w : \negnnf{\phi}, \Gamma$}
\RightLabel{\lem~\ref{lem:invert-dsn}}
\dashedLine
\UnaryInfC{$\rel, \ideal u \sar u : \psi, w : \negnnf{\phi}, \Gamma$}
\RightLabel{\ih}
\dashedLine
\BinaryInfC{$\rel, \ideal u \sar u : \psi, \Gamma$}
\RightLabel{$\Oir$}
\UnaryInfC{$\rel \sar v : \Oi \psi, \Gamma$}
\DisplayProof
\end{flushright}

\qquad \textit{2.1.2 The left premise of $\cut$ is derived with $\diar$.}

\begin{flushleft}
\begin{tabular}{c c}
\AxiomC{$\rel \sar v : \Diamond \psi, u : \psi, w : \phi, \Gamma$}
\RightLabel{$\diar$}
\UnaryInfC{$\rel \sar v : \Diamond \psi, w : \phi, \Gamma$}

\AxiomC{$\rel \sar v : \Diamond \psi, w : \negnnf{\phi}, \Gamma$}
\RightLabel{$\cut$}
\BinaryInfC{$\rel \sar v : \Diamond \psi, \Gamma$}
\DisplayProof

&

$\leadsto$
\end{tabular}
\end{flushleft}

\begin{flushright}
\AxiomC{$\rel \sar v : \Diamond \psi, u : \psi, w : \phi, \Gamma$}

\AxiomC{$\rel \sar v : \Diamond \psi, w : \negnnf{\phi}, \Gamma$}
\RightLabel{\lem~\ref{lem:invert-dsn}}
\dashedLine
\UnaryInfC{$\rel \sar v : \Diamond \psi, u : \psi, w : \negnnf{\phi}, \Gamma$}
\RightLabel{\ih}
\dashedLine
\BinaryInfC{$\rel \sar v : \Diamond \psi, u : \psi, \Gamma$}
\RightLabel{$\diar$}
\UnaryInfC{$\rel \sar v : \Diamond \psi, \Gamma$}
\DisplayProof
\end{flushright}

\qquad \textit{2.1.3 The left premise of $\cut$ is derived with $\ioar$.}

\begin{flushleft}
\begin{tabular}{c c}
\AxiomC{$\rel,R_{[0]}w_{0}u, ..., R_{[n]}w_{n}u \sar w : \phi, \Gamma$}
\RightLabel{$\ioa$}
\UnaryInfC{$\rel \sar w : \phi, \Gamma$}

\AxiomC{$\rel \sar w : \negnnf{\phi}, \Gamma$}
\RightLabel{$\cut$}
\BinaryInfC{$\rel \sar \Gamma$}
\DisplayProof

&

$\leadsto$
\end{tabular}
\end{flushleft}

\begin{flushright}
\AxiomC{$\rel,R_{[0]}w_{0}u, ..., R_{[n]}w_{n}u \sar w : \phi, \Gamma$}

\AxiomC{$\rel \sar w : \negnnf{\phi}, \Gamma$}
\RightLabel{\lem~\ref{lem:invert-dsn}}
\dashedLine
\UnaryInfC{$\rel, R_{[0]}w_{0}u, ..., R_{[n]}w_{n}u \sar w : \negnnf{\phi}, \Gamma$}
\RightLabel{\ih}
\dashedLine
\BinaryInfC{$\rel, R_{[0]}w_{0}u, ..., R_{[n]}w_{n}u \sar \Gamma$}
\RightLabel{$\ioar$}
\UnaryInfC{$\rel \sar \Gamma$}
\DisplayProof
\end{flushright}

\quad \textit{2.2 The cut formula $w : \phi$ is principal in the left premise of $\cut$ only.} This case is proven in a similar manner to the previous case 2.1.

\quad \textit{2.3 The cut formula $w : \phi$ is principal both premises of $\cut$.} We omit the cases where $\phi$ is of the form $\psi \lor \chi$, $\psi \land \chi$, $\agbox \psi$, and $\agdia \psi$ as these are handled in a similar fashion to the cut-elimination proof~\cite[\thm~4.13]{Neg05} for extensions of the modal logic $\mathsf{K}$. Without loss of generality, we show the cases where $\phi$ is of the form $\Box \psi$ and $\Oi \psi$.

\qquad \textit{2.3.1 The cut formula is of the form $\Box \psi$.}

\begin{flushleft}
\begin{tabular}{c c}
\AxiomC{$\rel \sar v : \psi, \Gamma$}
\RightLabel{$\boxr$}
\UnaryInfC{$\rel \sar w : \Box \psi, \Gamma$}

\AxiomC{$\rel \sar w : \Diamond \negnnf{\psi}, u : \negnnf{\psi}, \Gamma$}
\RightLabel{$\diar$}
\UnaryInfC{$\rel \sar w : \Diamond \negnnf{\psi},\Gamma$}

\RightLabel{$\cut$}
\BinaryInfC{$\rel \sar \Gamma$}
\DisplayProof

&

$\leadsto$
\end{tabular}
\end{flushleft}

\begin{flushright}
\AxiomC{$\Pi$}

\AxiomC{$\rel \sar v : \psi, \Gamma$}
\RightLabel{$\lsub$}
\dashedLine
\UnaryInfC{$\rel(u/v) \sar (v : \psi)(u/v), \Gamma(u/v)$}
\RightLabel{=}
\dottedLine
\UnaryInfC{$\rel \sar u : \psi, \Gamma$}
\RightLabel{\ih}
\dashedLine
\BinaryInfC{$\rel \sar \Gamma$}
\DisplayProof
\end{flushright}

\begin{center}
\begin{tabular}{c c c}
$\Pi$

&

$= \Bigg \{$

&

\AxiomC{$\rel \sar v : \psi, \Gamma$}
\RightLabel{$\wk$}
\dashedLine
\UnaryInfC{$\rel \sar v : \psi, u : \negnnf{\psi}, \Gamma$}
\RightLabel{$\boxr$}
\UnaryInfC{$\rel \sar w : \Box \psi, u : \negnnf{\psi}, \Gamma$}

\AxiomC{$\rel \sar w : \Diamond \negnnf{\psi}, u : \negnnf{\psi}, \Delta$}
\RightLabel{\ih}
\dashedLine
\BinaryInfC{$\rel \sar u : \negnnf{\psi}, \Gamma$}
\DisplayProof
\end{tabular}
\end{center}

\qquad \textit{2.3.2 The cut formula is of the form $\Oi \psi$.}

\begin{flushleft}
\begin{tabular}{c c}
\AxiomC{$\rel, \opt_{\Oi}v \sar v : \phi, \Gamma$}
\RightLabel{$\Oir$}
\UnaryInfC{$\rel \sar w : \Oi \phi, \Gamma$}

\AxiomC{$\rel, \opt_{\Oi}u \sar w : \ODi \negnnf{\phi}, u : \negnnf{\phi}, \Gamma$}
\RightLabel{$\ODir$}
\UnaryInfC{$\rel, \opt_{\Oi}u \sar w : \ODi \negnnf{\phi},\Gamma$}

\RightLabel{$\cut$}
\BinaryInfC{$\rel, \opt_{\Oi}u \sar \Gamma$}
\DisplayProof

&

$\leadsto$
\end{tabular}
\end{flushleft}

\begin{flushright}
\AxiomC{$\Pi$}

\AxiomC{$\rel, \opt_{\Oi}v \sar v : \phi, \Gamma$}
\RightLabel{$\lsub$}
\dashedLine
\UnaryInfC{$\rel(u/v), (\opt_{\Oi}v)(u/v) \sar (u : \phi)(u/v), \Gamma(u/v)$}
\RightLabel{=}
\dottedLine
\UnaryInfC{$\rel, \opt_{\Oi}u \sar u : \phi, \Gamma$}
\RightLabel{\ih}
\dashedLine
\BinaryInfC{$\rel, \opt_{\Oi}u \sar \Gamma$}
\DisplayProof
\end{flushright}

\begin{center}
\begin{tabular}{c c c}
$\Pi$

&

$= \Bigg \{$

&

\AxiomC{$\rel, \opt_{\Oi}v \sar v : \phi, \Gamma$}
\RightLabel{$\wk$}
\dashedLine
\UnaryInfC{$\rel, \opt_{\Oi}u, \opt_{\Oi}v \sar v : \phi, u : \negnnf{\phi}, \Gamma$}
\RightLabel{$\Oir$}
\UnaryInfC{$\rel, \opt_{\Oi}u \sar w : \Oi \phi, u : \negnnf{\phi}, \Gamma$}

\AxiomC{$\rel, \opt_{\Oi}u \sar w : \ODi \negnnf{\phi}, u : \negnnf{\phi}, \Delta$}
\RightLabel{\ih}
\dashedLine
\BinaryInfC{$\rel, \opt_{\Oi}u \sar u : \negnnf{\phi}, \Gamma, \Delta$}
\DisplayProof
\end{tabular}
\end{center}
\end{proof}

As for the calculi \sect~\ref{sec:lab-calc-kms}, we need to show that each calculus $\gtdsn$ is complete relative to classical propositional logic, that is, all classical propositional tautologies are derivable in each calculus $\gtdsn$. This ensures that axioms given via A0 in $\h\dsn$ (\dfn~\ref{def:axiomatization-dsn}) are derivable in $\gtdsn$, which can then be harnessed to show full completeness relative to $\dsn$. We state the classical completeness lemma below, and defer the proof to the appendix (Appendix B on p.~\pageref{app:classical-completeness}) to simplify presentation.

\begin{lemma}[Classical Completeness]\label{lem:classical-completeness-dsn}
All instances of classical propositional tautologies in $\langdsn$ are derivable in $\gtdsn$.
\end{lemma}

\begin{proof}
See Appendix B (p.~\pageref{app:classical-completeness}) for details.
\end{proof}

\begin{theorem}[Completeness]\label{thm:completness-gtdsn}
If $\vdash_{\dsn} \phi$, then $\vdash_{\gtdsn} w : \phi$. 
\end{theorem}

\begin{proof} By \lem~\ref{lem:classical-completeness-dsn}, we know that A0 holds. Moreover, the derivations of axioms A1 -- A3 are similar to the derivation of axiom A1 in the proof of \thm~\ref{thm:completness-gtkms}, 
 R0 is derived similar to the derivation of R0 in the proof of \thm~\ref{thm:completness-gtkms}, and R1 is derived similar to the derivation of R1 in the proof of \thm~\ref{thm:completness-gtkms}. We therefore show how to derive axioms A4 -- A14 below:

\emph{Axioms A4 and A5.}

\begin{center}
\begin{tabular}{c c}
\AxiomC{}
\RightLabel{\lem~\ref{lem:general-id-dsn}}
\dashedLine
\UnaryInfC{$R_{\agbox}wu \sar w : \Diamond \negnnf{\phi}, u : \negnnf{\phi}, u : \phi$}
\RightLabel{$\diar$}
\UnaryInfC{$R_{\agbox}wu \sar w : \Diamond \negnnf{\phi}, u : \phi$}
\RightLabel{$\agboxr$}
\UnaryInfC{$\sar w : \Diamond \negnnf{\phi}, w : \agbox \phi$}
\RightLabel{$\disr$}
\UnaryInfC{$\sar w : \Diamond \negnnf{\phi} \lor \agbox \phi$}
\RightLabel{=}
\dottedLine
\UnaryInfC{$\sar w : \Box \phi \cimp \agbox \phi$}
\DisplayProof

&

\AxiomC{}
\RightLabel{\lem~\ref{lem:general-id-dsn}}
\dashedLine
\UnaryInfC{$\ideal u \sar w : \Diamond \negnnf{\phi}, u : \negnnf{\phi}, u : \phi$}
\RightLabel{$\diar$}
\UnaryInfC{$\ideal u \sar w : \Diamond \negnnf{\phi}, u : \phi$}
\RightLabel{$\Oir$}
\UnaryInfC{$\sar w : \Diamond \negnnf{\phi}, w : \Oi \phi$}
\RightLabel{$\disr$}
\UnaryInfC{$\sar w : \Diamond \negnnf{\phi} \lor \Oi \phi$}
\RightLabel{=}
\dottedLine
\UnaryInfC{$\sar w : \Box \phi \cimp \Oi \phi$}
\DisplayProof
\end{tabular}
\end{center}

\emph{Axioms A6 and A7.}

\begin{center}
\begin{tabular}{c c}
\AxiomC{}
\RightLabel{\lem~\ref{lem:general-id-dsn}}
\dashedLine
\UnaryInfC{$\sar w : \Diamond \negnnf{\phi}, w : \negnnf{\phi}, w : \phi$}
\RightLabel{$\diar$}
\UnaryInfC{$\sar w : \Diamond \negnnf{\phi}, w : \phi$}
\RightLabel{$\disr$}
\UnaryInfC{$\sar w : \Diamond \negnnf{\phi} \lor \phi$}
\RightLabel{=}
\dottedLine
\UnaryInfC{$\sar w : \Box \phi \cimp \phi$}
\DisplayProof

&

\AxiomC{}
\RightLabel{\lem~\ref{lem:general-id-dsn}}
\dashedLine
\UnaryInfC{$\sar u : \negnnf{\phi}, v : \Diamond \phi, u : \phi$}
\RightLabel{$\diar$}
\UnaryInfC{$\sar u : \negnnf{\phi}, v : \Diamond \phi$}
\RightLabel{$\boxr \times 2$}
\UnaryInfC{$\sar w : \Box \negnnf{\phi}, w : \Box \Diamond \phi$}
\RightLabel{$\disr$}
\UnaryInfC{$\sar w : \Box \negnnf{\phi} \lor \Box \Diamond \phi$}
\RightLabel{=}
\dottedLine
\UnaryInfC{$\sar w : \Diamond \phi \cimp \Box \Diamond \phi$}
\DisplayProof
\end{tabular}
\end{center}

\emph{Axiom A8}

\begin{center}
\begin{tabular}{c c}
\AxiomC{}
\RightLabel{\lem~\ref{lem:general-id-dsn}}
\dashedLine
\UnaryInfC{$R_{\agbox}wu \sar w : \agdia \negnnf{\phi}, u : \negnnf{\phi}, u : \phi$}
\RightLabel{$\agdiar$}
\UnaryInfC{$R_{\agbox}wu \sar w : \agdia \negnnf{\phi}, w : \phi$}
\RightLabel{$\refli$}
\UnaryInfC{$\sar w : \agdia \negnnf{\phi}, w : \phi$}
\RightLabel{$\disr$}
\UnaryInfC{$\sar w : \agdia \negnnf{\phi} \lor \phi$}
\RightLabel{=}
\dottedLine
\UnaryInfC{$\sar w : \agbox \phi \cimp \phi$}
\DisplayProof
\end{tabular}
\end{center}

\emph{Axiom A9.}

\begin{center}
\AxiomC{}
\RightLabel{\lem~\ref{lem:general-id-dsn}}
\dashedLine
\UnaryInfC{$R_{\agbox}wu, R_{\agbox}wv, R_{\agbox}vu \sar u : \negnnf{\phi}, v : \agdia \phi, u : \phi$}
\RightLabel{$\agdiar$}
\UnaryInfC{$R_{\agbox}wu, R_{\agbox}wv, R_{\agbox}vu \sar u : \negnnf{\phi}, v : \agdia \phi$}
\RightLabel{$\eucli$}
\UnaryInfC{$R_{\agbox}wu, R_{\agbox}wv \sar u : \negnnf{\phi}, v : \agdia \phi$}
\RightLabel{$\agboxr \times 2$}
\UnaryInfC{$\sar w : \agbox \negnnf{\phi}, w : \agbox \agdia \phi$}
\RightLabel{$\disr$}
\UnaryInfC{$\sar w : \agbox \negnnf{\phi} \lor \agbox \agdia \phi$}
\RightLabel{=}
\dottedLine
\UnaryInfC{$\sar w : \agdia \phi \cimp \agbox \agdia \phi$}
\DisplayProof
\end{center}

\emph{Axiom A10.}
 
\begin{center}
\begin{tabular}{c}
\AxiomC{}
\RightLabel{\lem~\ref{lem:general-id-dsn}}
\dashedLine
\UnaryInfC{$\opt_{\Oi}u \sar u : \phi, u : \negnnf{\phi}, w : \ODi \negnnf{\phi}, w : \ODi \phi$}
\RightLabel{$\ODir$}
\UnaryInfC{$\opt_{\Oi}u \sar u : \negnnf{\phi}, w : \ODi \negnnf{\phi}, w : \ODi \phi$}
\RightLabel{$\ODir$}
\UnaryInfC{$\opt_{\Oi}u \sar w : \ODi \negnnf{\phi}, w : \ODi \phi$}
\RightLabel{$\dtwoir$}
\UnaryInfC{$\sar w : \ODi \negnnf{\phi}, w : \ODi \phi$}
\RightLabel{$\disr$}
\UnaryInfC{$\sar w : \ODi \negnnf{\phi} \lor \ODi \phi$}
\RightLabel{=}
\dottedLine
\UnaryInfC{$\sar w : \Oi \phi \cimp \ODi \phi$}
\DisplayProof
\end{tabular}
\end{center}

\emph{Axiom A11.}

\begin{center}
\AxiomC{}
\RightLabel{\lem~\ref{lem:general-id-dsn}}
\dashedLine
\UnaryInfC{$\opt_{\Oi}v \sar v : \negnnf{\phi}, v : \phi, u : \ODi \negnnf{\phi}, v : \Oi \phi$}
\RightLabel{$\Oir$}
\UnaryInfC{$\opt_{\Oi}v \sar v : \phi, u : \ODi \negnnf{\phi}, v : \Oi \phi$}
\RightLabel{$\Oir$}
\UnaryInfC{$\sar u : \ODi \negnnf{\phi}, v : \Oi \phi$}
\RightLabel{$\boxr \times 2$}
\UnaryInfC{$\sar w : \Box \ODi \negnnf{\phi}, w : \Box \Oi \phi$}
\RightLabel{$\disr$}
\UnaryInfC{$\sar w : \Box \ODi \negnnf{\phi} \lor \Box \Oi \phi$}
\RightLabel{=}
\dottedLine
\UnaryInfC{$\sar w : \Diamond \Oi \phi \cimp \Box \Oi \phi$}
\DisplayProof
\end{center}

\emph{Axiom A12.}

\begin{center}
\AxiomC{}
\RightLabel{\lem~\ref{lem:general-id-dsn}}
\dashedLine
\UnaryInfC{$R_{[i]}uv, \opt_{\Oi}u, \opt_{\Oi}v \sar w : \ODi \negnnf{\phi}, v : \negnnf{\phi}, v : \phi$}
\RightLabel{$\ODir$}
\UnaryInfC{$R_{[i]}uv, \opt_{\Oi}u, \opt_{\Oi}v \sar w : \ODi \negnnf{\phi}, v : \phi$}
\RightLabel{$\dthreeir$}
\UnaryInfC{$R_{[i]}uv, \opt_{\Oi}u \sar w : \ODi \negnnf{\phi}, v : \phi$}
\RightLabel{$\agboxr$}
\UnaryInfC{$\opt_{\Oi}u \sar w : \ODi \negnnf{\phi}, u : \agbox \phi$}
\RightLabel{$\Oir$}
\UnaryInfC{$\sar w : \ODi \negnnf{\phi}, w : \Oi \agbox \phi$}
\RightLabel{$\disr$}
\UnaryInfC{$\sar w : \ODi \negnnf{\phi} \lor \Oi \agbox \phi$}
\RightLabel{=}
\dottedLine
\UnaryInfC{$\sar w : \Oi \phi \cimp \Oi \agbox \phi$}
\DisplayProof
\end{center}

\emph{Axiom A13.} Let $i \in \ag = \{0, \ldots, n\}$, $\rel := R_{[0]}u_{0}v, \ldots, R_{[n]}u_{n}v$, and $\Gamma := u_{0} : \lb 0 \rb \negnnf{\phi}_{0}, \ldots, u_{n}: \lb n \rb \negnnf{\phi}_{n}, w :  \Diamond (\bigwedge_{i \in \ag} [i] \phi_{i})$. Also, we define $\rel_{i}$ be equal to $\rel$ minus the relational atom $R_{[i]}u_{i}v$, and $\Gamma_{i}$ be equal to $\Gamma$ minus the labelled formula $u_{i} : \negnnf{\phi}$. The independence of agents axiom A13 is derived as follows:

\begin{center}
\begin{tabular}{c}

\AxiomC{$\Pi_{1}$}
\AxiomC{$\cdots$}
\AxiomC{$\Pi_{n}$}
\RightLabel{$\conr \times (n - 1)$}
\TrinaryInfC{$\R \sar u_{0} : \lb 0 \rb \negnnf{\phi}_{0}, \ldots, u_{n}: \lb n \rb \negnnf{\phi}_{n}, w :  \Diamond (\bigwedge_{i \in \ag} [i] \phi_{i}), v: \bigwedge_{i \in \ag} [i] \phi_{i}$}

\RightLabel{$\diar$}

\UnaryInfC{$\R \sar u_{0} : \lb 0 \rb \negnnf{\phi}_{0}, \ldots, u_{n}: \lb n \rb \negnnf{\phi}_{n}, w :  \Diamond (\bigwedge_{i \in \ag} [i] \phi_{i})$}
\RightLabel{$\ioar$}
\UnaryInfC{$\sar u_{0} : \lb 0 \rb \negnnf{\phi}_{0}, \ldots, u_{n}: \lb n \rb \negnnf{\phi}_{n}, w :  \Diamond (\bigwedge_{i \in \ag} [i] \phi_{i})$}
\RightLabel{$\boxr \times n$}
\UnaryInfC{$\sar w : \Box \lb 0 \rb \negnnf{\phi}_{0}, \ldots, w :  \Box \lb n \rb \negnnf{\phi}_{n}, w :  \Diamond (\bigwedge_{i \in \ag} [i] \phi_{i})$}
\RightLabel{$\disr \times (n-1)$}
\UnaryInfC{$\sar w : \Box \lb 0 \rb \negnnf{\phi}_{0} \vee \cdots \vee \Box \lb n \rb \negnnf{\phi}_{n} \vee \Diamond (\bigwedge_{i \in \ag} [i] \phi_{i})$}
\RightLabel{=}
\dottedLine
\UnaryInfC{$\sar w : \Diamond [ 0 ] \phi_{0} \land \cdots \land \Diamond [n] \phi_{n} \cimp \Diamond (\bigwedge_{i \in \ag} [i] \phi_{i})$}
\DisplayProof
\end{tabular}
\end{center}

\begin{center}
\begin{tabular}{c c @{\hskip 1em} c}
$\Pi_{i}$

&

$= \Bigg \{$

&

\AxiomC{}
\RightLabel{\lem~\ref{lem:general-id-dsn}}
\dashedLine
\UnaryInfC{$\R_{i}, R_{[i]}vu,R_{[i]}u_{i}v, R_{[i]}u_{i}u \sar \Gamma_{i}, u_{i} : \lb i \rb \negnnf{\phi}_{i}, u : \negnnf{\phi}_{i},  u : \phi_{i}$}
\RightLabel{$\agdiar$}
\UnaryInfC{$\R_{i}, R_{[i]}vu,R_{[i]}u_{i}v, R_{[i]}u_{i}u_{i}, \R_{[i]}vu_{i}, \R_{[i]}u_{i}u \sar \Gamma_{i}, u_{i} : \lb i \rb \negnnf{\phi}_{i}, u : \phi_{i}$}
\RightLabel{$\eucli$}
\UnaryInfC{$\R_{i}, R_{[i]}vu,R_{[i]}u_{i}v, R_{[i]}u_{i}u_{i}, \R_{[i]}vu_{i} \sar \Gamma_{i}, u_{i} : \lb i \rb \negnnf{\phi}_{i}, u : \phi_{i}$}
\RightLabel{$\eucli$}
\UnaryInfC{$\R_{i}, R_{[i]}vu,R_{[i]}u_{i}v, R_{[i]}u_{i}u_{i} \sar \Gamma_{i}, u_{i} : \lb i \rb \negnnf{\phi}_{i}, u : \phi_{i}$}
\RightLabel{$\refli$}
\UnaryInfC{$\R_{i}, R_{[i]}vu,R_{\agbox}u_{i}v \sar \Gamma_{i}, u_{i} : \lb i \rb \negnnf{\phi}_{i}, u : \phi_{i}$}
\RightLabel{$\agbox$}
\UnaryInfC{$\R_{i}, R_{[i]}u_{i}v \sar \Gamma_{i}, u_{i} : \lb i \rb \negnnf{\phi}_{i}, v: [i] \phi_{i}$}
\DisplayProof
\end{tabular}
\end{center}

\emph{Axiom A14.} For the $\Pi_{0,j}$ derivations below, we let $1 \leq j \leq k$, and for the $\Pi_{m,j}$ derivations, we let $0 < m \leq k-1$ and $m+1 \leq j \leq n$. We let
$$
\Gamma := w_{0}:\phi_1, \ldots, w_{0}: \phi_k
$$
and
$\Gamma_{0,j}$ be the multiset
$$
w_{1} : \agdia \negnnf{\phi_{1}}, w_{2} : \phi_{1}, w_{2} : \agdia \negnnf{\phi_2}, \ldots, w_{k}: \phi_{1}, \ldots, w_{k} : \phi_{k-1}, w_{k} : \agdia \negnnf{\phi_{k}},
 w_{0}:\phi_1, \ldots, w_{0}: \phi_k
$$
minus $w_{0} : \phi_{j}, w_{j} : \agdia \negnnf{\phi_{j}}$. Last, we let $\Gamma_{m,j}$ be the multiset
$$
w_{1} : \agdia \negnnf{\phi_{1}}, w_{2} : \phi_{1}, w_{2} : \agdia \negnnf{\phi_2}, \ldots, w_{k}: \phi_{1}, \ldots, w_{k} : \phi_{k-1}, w_{k} : \agdia \negnnf{\phi_{k}},
 w_{0}:\phi_1, \ldots, w_{0}: \phi_k
$$
minus $w_{k}: \agdia \negnnf{\phi_{k}}, w_{j}: \phi_{k}$ and $q = \frac{k^{2}-k}{2}$. The axiom is derived as follows:

\begin{center}
\begin{tabular}{c c c}
$\Pi_{0,j}$

&

$= \Bigg \{$

&

\AxiomC{$R_{\agbox}w_{0}w_{0}, R_{\agbox}w_{0}w_{j}, R_{\agbox}w_{j}w_{0} \sar w_{0}: \phi_{j}, w_{j}: \agdia \negnnf{\phi_{j}}, w_{0} : \negnnf{\phi_{j}}, \Gamma_{0,j}$}
\RightLabel{$\agdia$}
\UnaryInfC{$R_{\agbox}w_{0}w_{0}, R_{\agbox}w_{0}w_{j}, R_{\agbox}w_{j}w_{0}, \ldots, w_{0}: \phi_{j} , w_{j}: \agdia \negnnf{\phi_{j}}, \Gamma_{0,j}$}
\RightLabel{$\eucli$}
\UnaryInfC{$R_{\agbox}w_{0}w_{0}, R_{\agbox}w_{0}w_{j}, \ldots, w_{0}: \phi_{j} , w_{j}: \agdia \negnnf{\phi_{j}}, \Gamma_{0,j}$}
\RightLabel{$\refli$}
\UnaryInfC{$R_{\agbox}w_{0}w_{j} \sar w_{0}: \phi_{j} , w_{j}: \agdia \negnnf{\phi_{j}}, \Gamma_{0,j}$}
\DisplayProof
\end{tabular}
\end{center}

\begin{center}
\begin{tabular}{c c c}
$\Pi_{m,j}$

&

$= \Bigg \{$

&

\AxiomC{$R_{\agbox}w_{k}w_{j} \sar w_{k}: \agdia \negnnf{\phi_{k}}, w_{j} : \negnnf{\phi_{k}}, w_{j}: \phi_{k}, \Gamma_{m,j}$}
\RightLabel{$\agdia$}
\UnaryInfC{$R_{\agbox}w_{k}w_{j} \sar w_{k}: \agdia \negnnf{\phi_{k}}, w_{j}: \phi_{k}, \Gamma_{m,j}$}
\DisplayProof
\end{tabular}
\end{center}

\begin{center}
\resizebox{\columnwidth}{!}{
\begin{tabular}{c}

\AxiomC{$\Big\{ \Pi_{m,j} \ \Big| \ 0 \leq m \leq k-1 \text{, } m+1 \leq j \leq k \Big\}$}


\RightLabel{$\choicer$}

\UnaryInfC{$\sar w_{1} : \agdia \negnnf{\phi_{1}}, w_{2} : \phi_{1}, w_{2} : \agdia \negnnf{\phi_2}, \ldots, w_{k}: \phi_{1}, \ldots, w_{k} : \phi_{k-1}, w_{k} : \agdia \negnnf{\phi_{k}},
 \Gamma$}
\RightLabel{$\boxr \times k + \disr \times q$}
\UnaryInfC{$\sar w_{0}:\Box\langle i \rangle \negnnf{\phi_1}, w_{0}: \Box (\phi_1\lor\langle i\rangle \negnnf{\phi_2}), \ldots, w_{0}:\Box (\phi_{1} \lor \cdots \phi_{k-1} \lor \agdia \negnnf{\phi_{k}}), \Gamma$}
\RightLabel{$\disr \times (2k - 1)$}
\UnaryInfC{$\sar w_{0}:\Box\langle i \rangle \negnnf{\phi_1} \lor \Box (\phi_1\lor\langle i\rangle \negnnf{\phi_2}) \lor \cdots \lor \Box (\phi_{1} \lor \cdots \phi_{k-1} \lor \agdia \negnnf{\phi_{k}}) \lor\phi_1\lor \cdots \lor \phi_k$}
\RightLabel{=}
\dottedLine
\UnaryInfC{$\sar w : \Diamond \agbox \phi_{1} \land \Diamond (\negnnf{\phi}_{1} \land \agbox \phi_{2}) \land \cdots \land \Diamond (\negnnf{\phi}_{1} \land \cdots \land \negnnf{\phi}_{k-1} \land \agbox \phi_{k}) \rightarrow \phi_{1} \lor \cdots \lor \phi_{k}$}
\DisplayProof 
\end{tabular}
}
\end{center}
\end{proof}

Similar to the previous two sections, we define sequent graphs for our labelled sequents as well as special types of labelled sequents called \emph{labelled forest sequents} and \emph{labelled DAG sequents}. These reduced structures will naturally emerge in derivations after we refined our deontic \stit calculi (\cptr~\ref{CPTR:Refinment-Modal}).

\begin{definition}[Sequent Graphs for Deontic \stit Logics]\label{def:sequent-graph-kms} Let $\Lambda := \rel \sar \Gamma$ be a labelled sequent for deontic \stit logics. We define the \emph{sequent graph}\index{Sequent graph!for deontic \stit logics} of $\Lambda$, $\seqgraph(\Lambda) = (V,E,L)$, as follows:
\begin{itemize}

\item[$\li$] $V = \lab(\Lambda)$

\item[$\li$] $E = \{(w,u,\agbox) \ | \ R_{\agbox}wu \in \rel\}$

\item[$\li$] $L(w) = \big(\{i \in \ag \ | \ \ideal w \in \rel\}; \{\phi \ | \ w : \phi \in \Gamma\}\big)$

\end{itemize}
\end{definition}

\begin{definition}[Labelled Forest/DAG sequent]\label{def:tree-DAG-sequent-dsn} A labelled sequent for deontic \stit logics $\Lambda$ is a \emph{labelled forest sequent}\index{Labelled forest sequent} (\emph{labelled DAG sequent}\index{Labelled DAG sequent}) \ifandonlyif $\seqgraph(\Lambda) = (V,E,L)$ is a \emph{forest} ($\seqgraph(\Lambda) = (V,E,L)$ is a \emph{DAG}, \resp).
\end{definition}

To provide additional intuition regarding such labelled sequents, we give an example of a labelled forest sequent and a labelled DAG sequent below:


\begin{example}\label{ex:sequent-graph-examples-dsn} Below, we give an example of a labelled forest sequent $\Lambda_{1}$ along with its corresponding sequent graph $\seqgraph(\Lambda_{1})$, and an example of a labelled DAG sequent $\Lambda_{2}$ along with its corresponding sequent graph $\seqgraph(\Lambda_{2})$. The labels $w$ and $u$ serve as the roots in $\Lambda_{1}$, and the label $w$ and $v$ serve as the roots in $\Lambda_{2}$.

\begin{center}
$\Lambda_{1} := R_{[1]}wx, R_{[2]}wy, R_{[1]}wz, R_{[1]}uv, \opt_{\otimes_{1}}x, \opt_{\otimes_{1}}u, \opt_{\otimes_{2}}u, \opt_{\otimes_{1}}v, \opt_{\otimes_{3}}v \sar \Gamma_{1}$
\end{center}
$$\Gamma_{1} := w : p, w : q, x : \negnnf{p}, z : p \land q, z : \otimes_{1} r, u : q$$

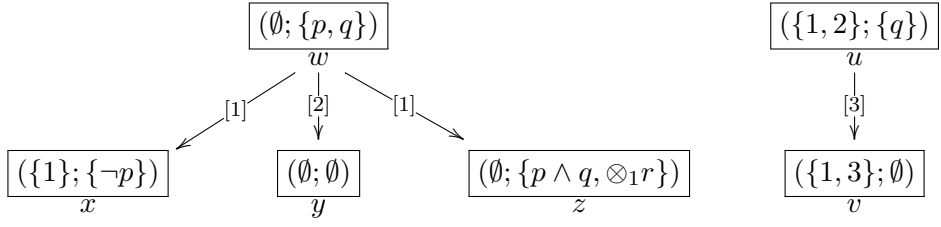
\begin{center}
\begin{tabular}{c}
\xymatrix{
&  \overset{\boxed{(\emptyset ; \{p,q\})}}{w} \ar[dl]|-{[1]}\ar[d]|-{[2]}\ar[dr]|-{[1]} & &  \overset{\boxed{(\{1,2\} ; \{q\})}}{u}\ar[d]|-{[3]} \\
\overset{\boxed{(\{1\} ; \{\negnnf{p}\})}}{x} & \overset{\boxed{(\emptyset ; \emptyset)}}{y} & \overset{\boxed{(\emptyset ; \{p \land q, \otimes_{1} r\})}}{z} & \overset{\boxed{(\{1,3\} ; \emptyset)}}{v}
}
\end{tabular}
\end{center}

\begin{center}
$\Lambda_{2} := R_{[2]}wu, R_{[2]}wz, R_{[2]}uz, R_{[3]}vy, R_{[1]}vx, \opt_{\otimes_{1}}w, \opt_{\otimes_{1}}u, \opt_{\otimes_{2}}u, \opt_{\otimes_{3}}u, \opt_{\otimes_{1}}y \sar \Gamma_{2}$
\end{center}
$$\Gamma_{2} := w : q, w : q, u : p, u : \negnnf{p}, v : \Box p \lor q, x : r$$

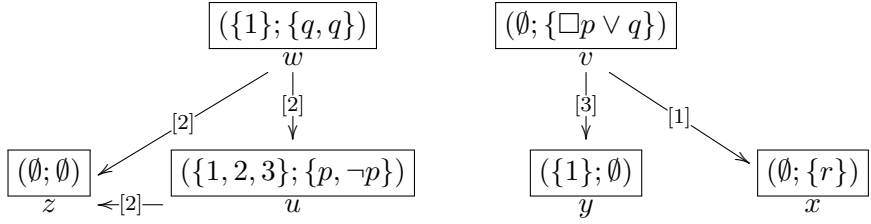
\begin{center}
\begin{tabular}{c}
\xymatrix{
&  \overset{\boxed{(\{1\} ; \{q,q\})}}{w} \ar[dl]|-{[2]}\ar[d]|-{[2]} &  \overset{\boxed{(\emptyset ; \{\Box p \lor q\})}}{v}\ar[dr]|-{[1]}\ar[d]|-{[3]} & \\
\overset{\boxed{(\emptyset ; \emptyset)}}{z} &    \overset{\boxed{(\{1,2,3\} ; \{p,\negnnf{p}\})}}{u} \ar[l]|-{[2]} & \overset{\boxed{(\{1\} ; \emptyset)}}{y} & \overset{\boxed{(\emptyset ; \{r\})}}{x}
}
\end{tabular}
\end{center}
\end{example}

After refining our class of calculi for deontic \stit logics, we will find that certain subclasses of calculi utilize derivations of a certain shape. In the refined setting, when we set $n = 0$, that is, the number of agents $|\ag| = 1$, we will find that completeness is preserved if we restrict ourselves to the use of labelled forest sequents, and if we set $k = 0$, meaning that no upper bound is imposed on the number of choices available to our agents, then we may restrict ourselves to the use of labelled DAG sequents in derivations. Below, we explicitly define such derivations, and also introduce the \emph{rooted property}. In essence, a labelled forest or DAG derivation possesses the rooted property if, reading the derivation in a bottom-up manner, whenever a label is introduced as a root in the sequent graph of a labelled sequent, it remains a root in all labelled sequents higher up in the derivation. The significance of the rooted property is twofold: first, confirming that the rooted property holds for labelled forest or DAG derivations lets us relate our refined labelled calculi (derived in the next chapter) to the nested and indexed-nested sequent formalisms, where proofs effectively have the rooted property. Second, the rooted property has practical value, as it tells us that information `flows' from certain points when inference rules are applied to labelled forest or DAG sequents in reverse, i.e. during proof-search, trees and DAGs `grow' from roots in sequent graphs of labelled sequents. This observation is useful in establishing termination, and will be seen in \sect~\ref{sec:applicationsI} on proof-search and decidability for deontic \stit logics.

\begin{definition}[Labelled Forest/DAG Proof, Rooted Property]\label{def-tree-DAG-proof-dsn} We say that a proof is a \emph{labelled forest proof}\index{Labelled forest proof} (\emph{labelled DAG proof}\index{Labelled DAG sequent}\index{Labelled DAG proof}) \ifandonlyif it consists solely of labelled forest sequents (it consists solely of labelled DAG sequents, \resp).

Also, we say that a labelled forest (DAG) proof $\Pi$ has the \emph{rooted property}\index{Rooted property} \ifandonlyif 
 for all labelled sequents $\Lambda_{1}$ and $\Lambda_{2}$ occurring in $\Pi$, if $\Lambda_{1}$ occurs below $\Lambda_{2}$ and $w_{1}$, $\ldots$, $w_{m}$ are the roots of all connected induced subgraphs in $\seqgraph(\Lambda_{1})$, then they are all roots of connected induced subgraphs in $\seqgraph(\Lambda_{2})$.
\end{definition}

Last, we prove a theorem below showing that each calculus $\gtdsn$ is incomplete relative to labelled forest and DAG derivations. This theorem is useful in comparing our labelled and refined labelled calculi. In fact, we will confirm that refinement produces systems which permit less underlying structure in their labelled sequents, that is to say, refined labelled systems for certain deontic \stit logics allow for completeness relative to labelled forest and DAG derivations. This topic will be discussed in \sect~\ref{SECT:Refine-STIT} on refining labelled calculi for deontic \stit logics.

\begin{theorem}\label{thm:sequent-structure-gtdsn}
Let $n, k \in \mathbb{N}$. The calculus $\gtdsn$ is incomplete relative to labelled forest and labelled DAG derivations.
\end{theorem}

\begin{proof} The theorem follows by considering the proof(s) of $\agbox p \cimp p = \agdia \negnnf{p} \lor p$ (an instance of axiom A11), which requires the use of $\refli$ as shown below:

\begin{center}
\AxiomC{}
\RightLabel{$\id$}
\UnaryInfC{$R_{\agbox}ww \sar w : \agdia \negnnf{p}, w : \negnnf{p}, w : p$}
\RightLabel{$\agdiar$}
\UnaryInfC{$R_{\agbox}ww \sar w : \agdia \negnnf{p}, w : p$}
\RightLabel{$\refli$}
\UnaryInfC{$\seqempstr \sar w : \agdia \negnnf{p}, w : p$}
\RightLabel{$\disr$}
\UnaryInfC{$\seqempstr \sar w : \agdia \negnnf{p} \lor p$}
\RightLabel{=}
\dottedLine
\UnaryInfC{$\seqempstr \sar w : \agbox p \cimp p$}
\DisplayProof
\end{center}

 When applied bottom-up, $\refli$ adds a loop to the sequent graph of a labelled sequent, which breaks both the labelled tree and labelled DAG property. Moreover, a quick inspection of the rules of $\gtdsn$ will show that $\refli$ is necessary to derive the above theorem, showing the incompleteness of the calculus relative to labelled forest and DAG derivations.
\end{proof}



\chapter{The Method of Refinement: Modal Propositional Logics}
\label{CPTR:Refinment-Modal} 






We introduce the method of refinement\index{Method of refinement}---a means by which labelled calculi may be simplified through the introduction of propagation rules and the elimination of structural rules---and apply the method in the context of grammar and deontic \stit logics. The next chapter will apply the method of refinement in the context of first-order intuitionistic logics. The central mechanism behind this process is \emph{structural rule elimination}\index{Structural rule elimination}, whereby structural rules encoding frame properties---e.g. $\psr$ in $\gtkms$ (see \fig~\ref{fig:G3Km(S)}), $\nd$ in $\gtintfond$ (see \fig~\ref{fig:labelled-calculi-FO-Int}), and $\dtwoir$ in $\gtdsn$ (see \fig~\ref{fig:base-Gcalculus})---are permuted upward in a derivation and are either deleted at initial sequents or absorbed into certain logical rules. In order to carry out the elimination of structural rules, we introduce \emph{propagation rules}\index{Propagation rule} (see, e.g.~\cite{CasCerGasHer97,CiaLyoRamTiu20,Fit72,LyoBer19,Sim94,TiuIanGor12}), or a generalization thereof (which we call, \emph{reachability rules}), to our calculi. We note that we will refer to such rules---that allow for the elimination of structural rules to go through---as \emph{conducive}\index{Conducive rule}. 
 
 The addition of conducive rules (i.e. propagation and reachability rules in our setting) and the elimination of structural rules yields \emph{refined} labelled proof systems that are more economical in the sense that they consist of less rules, produce shorter proofs (compared to the original labelled systems from which they were derived), and need only utilize labelled sequents of a reduced form (e.g. labelled sequents whose sequent graphs are trees; see \dfn~\ref{def:tree-sequent-kms}). We will discuss the consequences of these effects along with additional practical consequences of refinement below. However, before discussing the advantages of refinement as well as related work, we give the reader a brief introduction to the methodology. First, we introduce and provide an example of propagation rules, followed by an example of structural rule elimination, and last, give an example of the type of (nested) proof system that is commonly produced via refinement.

Propagation rules have a unique operation with bottom-up applications corresponding to the propagation of a formula along a path within the data structure (or, graph) encoded by the sequent. For example, we might have a propagation rule such as $(Pr_{\adia}^{1})$ shown below left, which makes use of the labelled formula $w : \adia \phi$ to bottom-up introduce the labelled formula $w : \phi$ (observe that there is a path of length $0$ between the label of the principal and auxiliary formula, i.e. the labels are identical). Another example of a propagation rule is $(Pr_{\adia}^{2})$, which is shown below right, and makes use of the labelled formula $w : \adia \phi$ and the path of relational atoms $R_{a}wv, R_{a}vu$ to bottom-up introduce the labelled formula $u : \phi$ associated with the label $u$ at the end of the relational path of length $2$.

\begin{center}
\begin{tabular}{c c}
\AxiomC{$\rel \sar w : \adia \phi, w : \phi, \Gamma$}
\RightLabel{$(Pr_{\adia}^{1})$}
\UnaryInfC{$\rel \sar w : \adia \phi, \Gamma$}
\DisplayProof

&

\AxiomC{$\rel, R_{a}wv, R_{a}vu \sar w : \adia \phi, u : \phi, \Gamma$}
\RightLabel{$(Pr_{\adia}^{2})$}
\UnaryInfC{$\rel, R_{a}wv, R_{a}vu \sar w : \adia \phi, \Gamma$}
\DisplayProof
\end{tabular}
\end{center}

The use of propagation rules goes at least as far back as 1972, where Fitting employed rules of a similar functionality within the context of prefixed tableaux for normal modal logics~\cite{Fit72}.\footnote{The term \emph{propagation rule} was not used in Fitting's paper~\cite{Fit72} and seems to have been coined in~\cite{CasCerGasHer97}.} Such rules have been integrated into numerous proof-theoretic frameworks for diverse classes of logics, such as in the labelled sequent formalism for intuitionistic modal logics~\cite{Sim94}, tense logics~\cite{CiaLyoRamTiu20}, and \stit logics~\cite{LyoBer19}; in the nested sequent formalism for grammar logics~\cite{TiuIanGor12}; and in the context of prefixed tableaux for normal modal logics~\cite{CasCerGasHer97,Fit72}. While propagation rules function by propagating formulae along paths within a sequent (when applied bottom-up), applications of reachability rules (which will be discussed in \cptr~\ref{CPTR:Refinment-Constructive}) additionally depend on the (non-)existence of data occurring along paths within a sequent.

To provide the reader with intuition concerning the process of structural rule elimination and the utility of propagation rules, we consider an example derivation in the labelled calculus $\gtkms$ (see \fig~\ref{fig:G3Km(S)}) for the grammar logic $\kms$ (see \dfn~\ref{def:axiomatization-km}) where $\thuesys := \{a \pto \empstr, \conv{a} \pto \empstr\}$ and $\albet := \{a,\conv{a}\}$ (i.e. the alphabet is the set of characters $\{a,\conv{a}\}$). The calculus consists of the rules shown below, where we have a $(p^{\chara}_{\empstr})$, $\convr$, $\charadiar$, and $\charaboxr$ rule for each $\chara \in \albet$. Also, as usual, the label $u$ is an eigenvariable in the $\charaboxr$ rule.

\begin{center}
\begin{tabular}{c c c}
\AxiomC{}
\RightLabel{$\id$}
\UnaryInfC{$\rel \sar w : p, w : \negnnf{p}, \Gamma$}
\DisplayProof

&

\AxiomC{$\rel \sar w : \phi, w : \psi, \Gamma$}
\RightLabel{$\disr$}
\UnaryInfC{$\rel \sar w : \phi \lor \psi, \Gamma$}
\DisplayProof

&

\AxiomC{$\rel, R_{\chara}ww \sar \Gamma$}
\RightLabel{$(p^{\chara}_{\empstr})$}
\UnaryInfC{$\rel \sar \Gamma$}
\DisplayProof
\end{tabular}
\end{center}

\begin{center}
\begin{tabular}{c c}
\AxiomC{$\rel \sar w : \phi, \Gamma$}
\AxiomC{$\rel \sar w : \psi, \Gamma$}
\RightLabel{$\conr$}
\BinaryInfC{$\rel \sar w : \phi \land \psi, \Gamma$}
\DisplayProof

&

\AxiomC{$\rel, R_{\chara}wu \sar w : \charadia \phi, u : \phi, \Gamma$}
\RightLabel{$\charadiar$}
\UnaryInfC{$\rel, R_{\chara}wu \sar w : \charadia \phi, \Gamma$}
\DisplayProof
\end{tabular}
\end{center}

\begin{center}
\begin{tabular}{c c c}
\AxiomC{$\rel, R_{\chara}wu \sar u : \phi, \Gamma$}
\RightLabel{$\charaboxr$}
\UnaryInfC{$\rel \sar w : \charabox \phi, \Gamma$}
\DisplayProof

&

\AxiomC{$\rel, R_{\chara}wu, R_{\conv{\chara}}uw \sar \Gamma$}
\RightLabel{$\convr$}
\UnaryInfC{$\rel, R_{\chara}wu \sar \Gamma$}
\DisplayProof
\end{tabular}
\end{center}

Let us now consider the elimination of the structural rule $\psrp{a}{\empstr}$ from a given derivation. We assume that we are given the derivation shown top-left below, and make use of the propagation rule $(Pr_{\adia}^{1})$, which was introduced above, to permute the $\psrp{a}{\empstr}$ rule upward at the first step. Thus, the $\adiar$ and $\psrp{a}{\empstr}$ inferences are transformed into a $\psrp{a}{\empstr}$ inference followed by a $(Pr_{\adia}^{1})$ inference. Since the conclusion of the $\psrp{a}{\empstr}$ inference in the top-right derivation below is an instance of $\id$, we may delete the $\psrp{a}{\empstr}$ instance altogether, giving the output derivation shown below bottom.

\begin{tabular}{c c c}
\AxiomC{}
\RightLabel{$\id$}
\UnaryInfC{$R_{a}ww \sar w : \adia p, w : p, w : \negnnf{p}$}
\RightLabel{$\adiar$}
\UnaryInfC{$R_{a}ww \sar w : \adia p, w : \negnnf{p}$}
\RightLabel{$\psrp{a}{\empstr}$}
\UnaryInfC{$\seqempstr \sar w : \adia p, w : \negnnf{p}$}
\DisplayProof

&

$\leadsto$

&

\AxiomC{}
\RightLabel{$\id$}
\UnaryInfC{$R_{a}ww \sar w : \adia p, w : p, w : \negnnf{p}$}
\RightLabel{$\psrp{a}{\empstr}$}
\UnaryInfC{$\seqempstr \sar w : \adia p, w : p, w : \negnnf{p}$}
\RightLabel{$(Pr_{\adia}^{1})$}
\UnaryInfC{$\seqempstr \sar w : \adia p, w : \negnnf{p}$}
\DisplayProof
\end{tabular}

\begin{center}
\begin{tabular}{c c}
$\leadsto$

&

\AxiomC{}
\RightLabel{$\id$}
\UnaryInfC{$\seqempstr \sar w : \adia p, w : p, w : \negnnf{p}$}
\RightLabel{$(Pr_{\adia}^{1})$}
\UnaryInfC{$\seqempstr \sar w : \adia p, w : \negnnf{p}$}
\DisplayProof
\end{tabular}
\end{center}

Typically, structural rule elimination has the effect of simplifying the data structure underlying labelled sequents in a proof. This fact can be seen in the example above: observe that the sequent graphs (\dfn~\ref{def:sequent-graph-kms}) of the top two sequents in the input (i.e. top-left) derivation contain a loop due to the presence of the relational atom $R_{a}ww$, whereas the sequent graphs of all sequents in the output derivation consist of a single point. As explained above, introducing propagation or reachability rules to a calculus, which permits the elimination of structural rules, produces labelled calculi that employ simpler syntactic structures in their proofs and---interestingly---tend to be notational variants of \emph{nested sequent systems}~\cite{Bul92,Kas94,Str13,TiuIanGor12}. It should be noted that refinement does not always produce refined labelled calculi that are notational variants of nested systems, as will be discussed and demonstrated in \sect~\ref{SECT:Refine-STIT}; similar to the systems in~\cite{MarStr14} for (intuitionistic) modal logics, our calculi for deontic \stit logics may contain a mixture of propagation rules and structural rules since it is not known if propagation rules exist which allow for certain structural rules to be eliminated. Still, as will be discussed in that section as well, the data structures underlying labelled sequents in a proof are nevertheless reduced after refinement has taken place.

The creation of the nested sequent formalism is often attributed to Bull~\cite{Bul92} and Kashima~\cite{Kas94} (and is equivalent to the prefixed tableaux formalism introduced much earlier by Fitting~\cite{Fit72}). The formalism is characterized by the use of generalized versions of Gentzen-style sequents, which utilize nesting constructors to organize Gentzen-style sequents (or, multisets of formulae) into trees. To give the reader an idea of nested sequent systems and an idea of the output of the refinement process, we introduce nested sequents for the grammar logic $\kms$ with $\thuesys := \{a \pto \empstr, \conv{a} \pto \empstr\}$ and $\albet := \{a,\conv{a}\}$ (just introduced above), and show the nested calculus that results from refining the labelled calculus $\gtkms$ (which was just introduced above as well).

The following grammar in BNF defines nested sequents for the grammar logic $\kms$ considered above:
$$
\na ::= \seqempstr \ | \ \phi \ | \ \na, \na \ | \ (a)\nbbl \na \nbbr \ | \ (\conv{a})\nbbl \na \nbbr
$$
where $\phi \in \langkm{\albet}$ and $\seqempstr$ is the empty sequent. One can readily check that such syntactic structures encode trees; for example, the sequent graph encoded by the nested sequent $p, (\conv{a})\{\negnnf{q}\}, (a)\{q \lor r, (a)\{\seqempstr\}\}$ is shown below left and the sequent graph encoded by the nested sequent $q,r, (a)\{\abox p, (a)\{\seqempstr\}, (\conv{a})\{\seqempstr\}\}$ is shown below right. Both graphs are trees in the sense of \dfn~\ref{def:tree}.

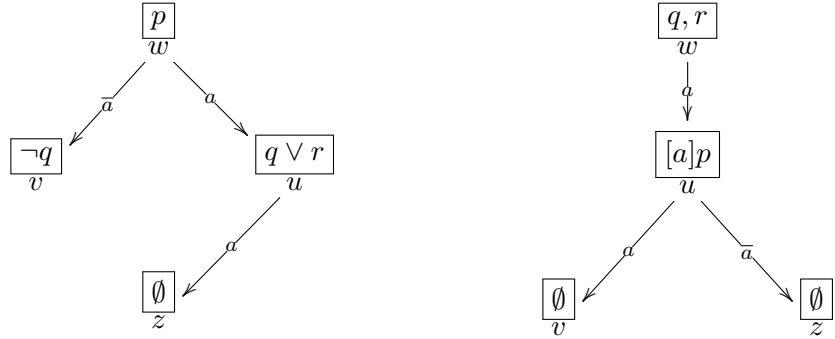
\begin{center}
\begin{tabular}{c c}
\xymatrix{
   & \overset{\boxed{p}}{w} \ar[dl]|-{\conv{a}} \ar[dr]|-{a} & &  \\
  \overset{\boxed{\negnnf{q}}}{v} & & \overset{\boxed{q \lor r}}{u} \ar[dl]|-{a}& \\ 
  & \overset{\boxed{\emptyset}}{z} &  &
}

&

\xymatrix{
   & & \overset{\boxed{q,r}}{w} \ar[d]|-{a} &  \\
   & & \overset{\boxed{\abox p}}{u} \ar[dl]|-{a}\ar[dr]|-{\conv{a}} & \\ 
  & \overset{\boxed{\emptyset}}{v} &  & \overset{\boxed{\emptyset}}{z}
}
\end{tabular}
\end{center}

If we were to eliminate the structural rules $(p^{\chara}_{\empstr})$ and $\convr$ for each $\chara \in \albet = \{a, \conv{a}\}$ from the labelled calculus $\gtkms$ introduced above (by expanding the calculus with sufficient propagation rules), then after switching from labelled to nested notation, we would obtain the nested calculus shown below, where we have a $\charaboxr$ and $\prcharadiar$ rule for each $\chara \in \albet$. (NB. As will be discussed in \sect~\ref{SECT:Refine-Grammar} below, it is sometimes advantageous to work with labelled notation as opposed to nested notation. Therefore, translating refined labelled calculi into nested calculi is an \emph{optional} last step in the refinement process.) The notation $X [ Y ]$ and $X [ Y ] [ Z ]$ is used to represent that $Y$ and/or $Z$ occur at some level of the nestings in the nested sequent $X$. For example, if our nested sequent $X$ is $p, (a)\{q, (\conv{a}) \{ r \land \negnnf{r}\}\}$, then $X[p]$, $X[q, (\conv{a}) \{ r \land \negnnf{r}\}]$, and $X[r \land \negnnf{r}]$ would all be correct representations of $X$ in this notation. 
 Last, we note that the propagation rule $\prcharadiar$ possesses a side condition $\dag$ stipulating how to use $\charadia \phi$ to bottom-up propagate the formula $\phi$ to the point witnessing $Z$ in the nested sequent $X$. We omit the exact description of this side condition here as it is rather complex, and note that such rules will be defined in the succeeding section (\sect~\ref{SECT:Refine-Grammar}). 

\begin{center}
\begin{tabular}{c c c}
\AxiomC{}
\RightLabel{$\id$}
\UnaryInfC{$X[p, \negnnf{p}, Y]$}
\DisplayProof

&

\AxiomC{$X[\phi, \psi, Y]$}
\RightLabel{$\disr$}
\UnaryInfC{$X[\phi \lor \psi, Y]$}
\DisplayProof

&

\AxiomC{$X[\phi, Y]$}
\AxiomC{$X[\psi, Y]$}
\RightLabel{$\conr$}
\BinaryInfC{$X[\phi \land \psi, Y]$}
\DisplayProof
\end{tabular}
\end{center}

\begin{center}
\begin{tabular}{c c}
\AxiomC{$X[Y, (\chara)\{\phi\}]$}
\RightLabel{$\charaboxr$}
\UnaryInfC{$X[Y, \charabox \phi]$}
\DisplayProof

&

\AxiomC{$X[\charadia \phi, Y][\phi, Z]$}
\RightLabel{$\prcharadiar^{\dag}$}
\UnaryInfC{$X[\charadia \phi, Y][Z]$}
\DisplayProof
\end{tabular}
\end{center}

Refining labelled calculi yields a variety of advantages: first, the replacement of structural rules with propagation or reachability rules leads to a compression in the size of proofs, a decrease in the total number of inference rules, and a reduction in the complexity of the data structure underlying labelled sequents in a proof. This can lead to a savings in space and an increase in efficiency for implementations or automated reasoning algorithms (e.g. proof-search, counter-model extraction, effective interpolation). Second, the removal of certain semantic elements and structures---e.g. domain atoms $\unda \in D_{w}$ (see \dfn~\ref{def:labelled-sequents-FO-Int}) used in labelled sequents---from the syntax of sequents, makes the calculi easier to understand and handle. Third, proving decidability for logics via (unrefined) labelled calculi is often complicated and/or uses \emph{ad hoc} methods that are logic dependent~\cite{Neg05,Vig00}. By contrast, if we employ more refined sequents (e.g. trees, forests, DAGs) in our calculi, then analyzing and demonstrating the termination of proof-search procedures becomes easier and has proven to be \emph{uniform}; for example, terminating proof-search with nested sequent calculi is shown uniformly for a class of grammar logics in~\cite{TiuIanGor12}. Fourth, the refinement method has been applied to translate between the labelled paradigm and non-labelled proof-theoretic formalisms (e.g.~\cite{CiaLyoRamTiu20}), thus allowing for results to be transferred between the distinct settings, for formalisms to be changed when one is better suited for the task at hand, and elucidating the semantic information inherent in syntactic structures of the non-labelled formalism (which has explanatory value). Last, the calculi we obtain are \emph{modular} with respect to their associated classes of logics, that is, by the alteration of side conditions or the deletion/addition of rules, any refined calculus can be transformed into another refined calculus with the same properties (e.g. invertibility of rules, cut-admissibility, etc.), but for another logic in the considered class. 
 
 Another benefit of refinement concerns how the method may be composed with methods and results from the labelled sequent paradigm. Although refinement is intimately tied to simplification, the method can be seen as part of a more extensive method whereby the semantics of a logic may be transformed into a proof calculus possessing desirable properties and which is suited for certain applications. As explained in \cptr~\ref{CPTR:Intro} and~\ref{CPTR:Labelled}, an advantage of the labelled formalism is that it allows for the straightforward extraction of (labelled) proof systems from semantics. However, such calculi involve complicated structures, violate the subformula property to a high degree, and tend to be unwieldy. By composing the construction process for labelled calculi with the simplification process obtained from structural rule elimination, we obtain a strategy for transforming the semantics of a logic into a refined calculus in possession of `nice' properties (e.g. such systems satisfy the desiderata of~\cite{Wan94} to a large extent).\footnote{See the introduction (\cptr~\ref{CPTR:Intro}) for a discussion of Wansing's desiderata, used to characterize `nice' proof systems.} 


The idea of refinement is foreshadowed by works studying the relationship between labelled and alternative proof systems. Such works include~\cite{GorRam12}, where translations between labelled and tree-hypersequent/nested sequent calculi were introduced for G\"odel-L\"ob provability logic; the papers~\cite{CiaLyoRam18,CiaLyoRamTiu20}, which provided translations between labelled, display, and nested calculi for tense logics; and the paper~\cite{Pim18}, which provided translations between labelled and nested calculi for intuitionistic, normal, and non-normal modal propositional logics. In each case, the labelled calculi were shown equivalent to calculi within a reduced proof-theoretic formalism, and made use of structural rule elimination arguments to obtain the more economical calculi. The refinement methodology abstracts from such works and gives a more general strategy for simplifying labelled calculi within a wide range of settings. A significant departure from the aforementioned works concerns the use of grammar theoretic machinery (\emph{viz.} \cfcst systems; see \dfn~\ref{def:CFCST-kms}) in defining propagation and reachability rules. As will be seen in \sect~\ref{SECT:Refine-Grammar} below and the following chapter (\cptr~\ref{CPTR:Refinment-Constructive}), the use of such machinery allows for relatively broad classes of propagation and reachability rules to be defined, thus permitting the elimination of structural rules in a large number of scenarios. This allows us to provide the first results showing that labelled calculi for grammar, deontic STIT, and first-order intuitionistic logics can be `simplified' and transformed into nested calculi in many cases.\footnote{It should be noted that labelled calculi for traditional \stit logics and first-order intuitionistic logics were refined in~\cite{LyoBer19} and~\cite{Lyo20a,Lyo21}, essentially yielding nested systems, so similar results have been shown in more restricted settings.}

This chapter is based on work from~\cite{BerLyo21,CiaLyoRam18,CiaLyoRamTiu20,LyoBer19} and is organized as follows: In the first section (\sect~\ref{SECT:Refine-Grammar}), we show that each labelled calculus for a grammar logic can be algorithmically transformed into a refined labelled calculus with structural rules replaced by propagation rules (\thm~\ref{thm:gtkms-to-kmsl-kms}) and where each derivation of a theorem only makes use of labelled tree sequents (\thm~\ref{thm:tree-derivations-kmsl}). We additionally prove that all refined labelled calculi are notational variants of (slight reformulations of) the nested calculi from~\cite{TiuIanGor12} (\thm~\ref{thm:kmsl-to-dkms} and~\ref{thm:dkms-to-kmsl}). The section ends with a discussion on the relationship between refined labelled calculi and the shallow nested (i.e. display) calculi from~\cite{TiuIanGor12}. In the second section (\sect~\ref{SECT:Refine-STIT}), we refine the labelled calculi for our deontic \stit logics, demonstrating that structural rules may be eliminated in the presence of certain propagation rules (\thm~\ref{thm:gtdsn-to-dsnl}), begetting systems that are complete relative to labelled DAG derivations (\thm~\ref{thm:DAG-proofs-dsn}) or labelled forest derivations (\cor~\ref{thm:forest-proofs-dsn}) depending on the deontic \stit logic considered. In the following chapter (\cptr~\ref{CPTR:Refinment-Constructive}), we discuss and apply refinement in the first-order setting, which, due to its more complex nature, will require us to improve upon and expand the refinement methodology. Details of how the methodology is augmented in a first-order context will be discussed there.






\section{Refining Labelled Calculi for Grammar Logics}\label{SECT:Refine-Grammar}


We apply the refinement method to each labelled calculus $\gtkms$ for the grammar logic $\kms$ to obtain a new calculus $\kmsl$. Before giving formal proofs confirming the extraction of the latter from the former, we motivate and explain how the conductive (i.e. propagation) rules (shown in \fig~\ref{fig:Refined-Calculus-i-kms}) of $\kmsl$ arise naturally when attempting to eliminate structural rules---\emph{viz.} $\psr$ and $\convr$ (see \fig~\ref{fig:G3Km(S)})---from $\gtkms$. The goal, therefore, is not simply to claim that refinement can be done, but to demonstrate the type of analysis sufficient to allow for refinement to be performed. After deriving each calculus $\kmsl$ from its parent calculus $\gtkms$, we show that each refined calculus is a notational variant of a nested calculus $\dkms$ from~\cite{TiuIanGor12}. The insight that refinement generates (slight variants of) known nested calculi (discovered independently of the method), suggests a naturalness to the method of refinement and of the associated nested calculi. 

\fig~\ref{fig:-translation-relationships-kms} summarizes the main transformations and translations between the calculi studied in this section. We use solid arrows to represent \emph{transformations}\index{Transformation}, which preserve the language of the calculus (e.g. labelled sequents) while algorithmically mapping proofs between the two systems (as in Thm.~\ref{thm:gtkms-to-kmsl-kms} and~\ref{thm:kmsl-to-gtkms-kms}), and we use dotted arrows to represent \emph{translations}\index{Translation}, which not only algorithmically map proofs between the two systems, but also change the language in the process (as in Thm.~\ref{thm:kmsl-to-dkms} and~\ref{thm:dkms-to-kmsl}). The dotted arrows are annotated with the symbols $\lnkms$ and $\nlkms$, which are the translation functions introduced in \dfn~\ref{def:ln-kms} and~\ref{def:nl-kms}, respectively.

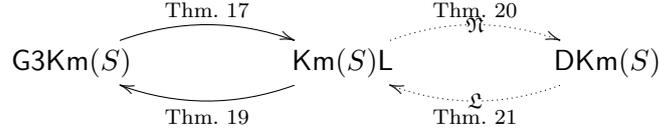
\begin{figure}[t]
\noindent\hrule
\begin{center}
\begin{tabular}{c}
\xymatrix{
	 \gtkms\ar@/^1.2pc/@{->}[rr]^{\text{Thm.~\ref{thm:gtkms-to-kmsl-kms}}}	& &   
	 \kmsl\ar@/^1.2pc/@{.>}[rr]|{\lnkms}^{\text{Thm.~\ref{thm:kmsl-to-dkms}}}\ar@/^1.2pc/@{->}[ll]^{\text{Thm.~\ref{thm:kmsl-to-gtkms-kms}}} & & 
	 \dkms\ar@/^1.2pc/@{.>}[ll]|{\nlkms}^{\text{Thm.~\ref{thm:dkms-to-kmsl}}} \\
}
\end{tabular}
\end{center}
\hrule
\caption{Transformations and translations between grammar logic calculi.}\label{fig:-translation-relationships-kms}
\end{figure}


\subsection{Extracting $\kmsl$ from $\gtkms$}

We begin by analyzing the elimination of a $\psr$ rule in a labelled calculus $\gtkms$, and observe under what conditions $\psr$ cannot be permuted upward in a derivation. Such situations will motivate and suggest the addition of propagation rules to $\gtkms$ that allow for the permutation to go through. In the labelled setting, such rules propagate formulae along paths of relational atoms occurring in a labelled sequent (when read bottom-up). The rules introduced here are based on the propagation rules for nested and labelled sequents introduced in~\cite{CiaLyoRamTiu20,LyoBer19,GorPosTiu11,TiuIanGor12}. Before defining our propagation rules (which are provided in \fig~\ref{fig:Refined-Calculus-i-kms}), we will explain how such rules arise naturally when considering the elimination of a $\psr$ rule.

For the sake of illustration, let us suppose that we have the following \cfcst system $\thuesys := \{a \pto a \cate b, \conv{a} \pto \conv{a} \cate \conv{b}\}$. We will attempt to prove the admissibility of the structural rule $\psrp{a}{a \cate b}$ by showing that the rule can always be permuted upward in a derivation. The reader can verify that the $\psrp{a}{a \cate b}$ rule is permutable with every rule in $\gtkms$ with the exception of the $\charadiar$ and $\convr$ rules. We fix the characters $a, b \in \albet$ and focus on permuting the $\psrp{a}{a \cate b}$ rule above the $\adiar$ rule, as analyzing this case will be sufficient to motivate and explain the definition of propagation rules and their addition to $\gtkms$. Furthermore, we will not consider the permutation of $\psrp{a}{a \cate b}$ above any $\convr$ structural rule for the following reason: since our aim is to prove the $\psr$ and $\convr$ rules admissible via an elimination algorithm, we can consider topmost occurrences of rules in a given derivation and successively eliminate them via their respective elimination procedures; therefore, we never need to consider the permutation of one rule above the other. As stated previously, our analysis will motivate the definition of a propagation rule, and thus provide the reader with insight into how such rules are discovered and defined via the process of structural rule elimination.

Let us assume that we have a derivation ending with an $\adiar$ inference followed by a $\psrp{a}{a \cate b}$ inference. We want to show that the rule can always be moved upward in a derivation and still allow for the same conclusion to be derived.
\begin{center}
\AxiomC{$\rel, R_{a}wv, R_{b}vu, R_{a}wu \sar w : \adia \phi, u : \phi, \Gamma$}
\RightLabel{$\adiar$}
\UnaryInfC{$\rel, R_{a}wv, R_{b}vu, R_{a}wu \sar w : \adia \phi, \Gamma$}
\RightLabel{$\psrp{a}{a \cate b}$}
\UnaryInfC{$\rel,R_{a}wv, R_{b}vu \sar w : \adia \phi, \Gamma$}
\DisplayProof


\end{center}
Since our aim is to eliminate the $\psrp{a}{a \cate b}$ rule, we will apply the rule directly to the top sequent shown above with the goal of using other rules in $\gtkms$ to derive the same conclusion. Applying $\psrp{a}{a \cate b}$ directly to the top sequent shown above gives us the following:
\begin{center}
\AxiomC{$\rel,R_{a}wv, R_{b}vu, R_{a}wu \sar w : \adia \phi, u : \phi, \Gamma$}
\RightLabel{$\psrp{a}{a \cate b}$}
\UnaryInfC{$\rel,R_{a}wv, R_{b}vu \sar w : \adia \phi, u : \phi, \Gamma$}
\DisplayProof
\end{center}
In order to derive the desired conclusion, we need to find a set of rules in $\gtkms$ that lets us delete $u : \phi$; however, a quick glance at the rules of $\gtkms$ will demonstrate that no rules are applicable allowing for $u : \phi$ to be deleted. Nevertheless, notice that if we extend our calculus with the $\ruone$ rule below, then the desired conclusion can be derived by applying the rule directly to the conclusion of the inference above.
\begin{center}
\AxiomC{$\rel,R_{a}wv, R_{b}vu \sar w : \adia \phi, u : \phi, \Gamma$}
\RightLabel{$\ruone$}
\UnaryInfC{$\rel,R_{a}wv, R_{b}vu \sar w : \adia \phi, \Gamma$}
\DisplayProof
\end{center}
The rule $\ruone$ essentially states that if there is a sequence of relational atoms of the form $R_{a}wv, R_{b}vu$ between a label $w$ and a label $u$ in a sequent, with the labelled formulae $w : \adia \phi$ and $u : \phi$ also occurring in the sequent, then $u : \phi$ may be deleted from premise to conclusion. If we consider the sequent semantics (\dfn~\ref{def:sequent-semantics-kms}), we can see that this rule is sound since if a $\kms$-model satisfies $R_{a}wv, R_{b}vu$, then because $a \pto a \cate b \in \thuesys$, the model will also satisfy $R_{a}wu$ as $R_{a \cate b} \subseteq R_{a}$. This last fact implies that we can think of the relational atom $R_{a}wu$ as being \emph{implicit} in both premise and conclusion of $\ruone$, showing that the rule is essentially a special instance of the (sound) $\adiar$ rule.

In accordance with what has been said, we can see that if we extend our calculus $\gtkms$ with the rule $\ruone$, then the $\psrp{a}{a \cate b}$ rule can be permuted above $\adiar$. Still, this does not immediately imply that $\psrp{a}{a \cate b}$ can be eliminated from any given derivation, as we now have to check if the $\psrp{a}{a \cate b}$ rule can be permuted above $\ruone$. We will consider this below, and observe the problematic case:
\begin{center}
\AxiomC{$\rel,R_{a}wz, R_{b}zv, R_{a}wv, R_{b}vu \sar w : \adia \phi, u : \phi, \Gamma$}
\RightLabel{$\ruone$}
\UnaryInfC{$\rel, R_{a}wz, R_{b}zv, R_{a}wv, R_{b}vu \sar w : \adia \phi, \Gamma$}
\RightLabel{$\psrp{a}{a \cate b}$}
\UnaryInfC{$\rel, R_{a}wz, R_{b}zv, R_{b}vu \sar w : \adia \phi, \Gamma$}
\DisplayProof
\end{center}
Again, since our goal is to eliminate the $\psrp{a}{a \cate b}$ rule, we apply it directly to the top sequent above (giving the inference below), and then aim to derive the same conclusion using other rules of $\gtkms + \ruone$.
\begin{center}
\AxiomC{$\rel,R_{a}wz, R_{b}zv, R_{a}wv, R_{b}vu \sar w : \adia \phi, u : \phi, \Gamma$}
\RightLabel{$\psrp{a}{a \cate b}$}
\UnaryInfC{$\rel, R_{a}wz, R_{b}zv, R_{b}vu \sar w : \adia \phi, u : \phi, \Gamma$}
\DisplayProof
\end{center}
As in the case of permuting $\psrp{a}{a \cate b}$ above $\adiar$, no rules are present in $\gtkms + \ruone$ that will allow us to derive the desired conclusion. We could, as before, add a new rule to our calculus that will bring about the desired permutation:
\begin{center}
\AxiomC{$\rel, R_{a}wz, R_{b}zv, R_{b}vu \sar w : \adia \phi, u : \phi, \Gamma$}
\RightLabel{$\rutwo$}
\UnaryInfC{$\rel, R_{a}wz, R_{b}zv, R_{b}vu \sar w : \adia \phi, \Gamma$}
\DisplayProof
\end{center}
It is easy to see that applying this rule to the conclusion of the former $\psrp{a}{a \cate b}$ inference allows for us to derive the desired conclusion, thus showing that $\psrp{a}{a \cate b}$ can be permuted above $\ruone$ in the calculus $\gtkms + \{\ruone,\rutwo\}$. As the reader might have noticed, although the addition of $\rutwo$ to our calculus allows for $\psrp{a}{a \cate b}$ to be permuted above $\ruone$, we now have to check if $\psrp{a}{a \cate b}$ can be permuted above $\rutwo$. If such a situation is analyzed, then one would observe that in order for the $\psrp{a}{a \cate b}$ rule to be permuted above $\rutwo$, a new rule $\ruthree$ would have to be added to our calculus, which would necessitate the addition of another rule $\rufour$ to allow for $\psrp{a}{a \cate b}$ to be permuted above $\ruthree$---this phenomenon would continue ad infinitum. Although we could add the whole infinite lot of such rules to our calculus in order to secure the elimination of $\psrp{a}{a \cate b}$, there is a more elegant solution to our problem.

Observe that the active relational atoms in $\ruone$ are of the form $R_{a}wv, R_{b}vu = R_{a \cate b}wu$ and the active relational atom in $\rutwo$ are of the form $R_{a}wz, R_{b}zv, R_{b}vu = R_{a \cate b \cate b}wu$. Recall that the production rule $a \pto a \cate b$ is element of our \cfcst system $\thuesys$, implying that $a \dtoann a \cate b$ and $a \dtoann a \cate b \cate b$. A quick comparison of the strings derived from $a$ and the active relational atoms of $\ruone$ and $\rutwo$ shows that the strings derived from $a$ serve as the indices in the active relational atoms. In fact, if one observes the infinite set of rules generated by trying to prove the eliminability of $\psrp{a}{a \cate b}$, they would find that the active relational atoms of all such rules correspond to strings derivable from $a$ in $\thuesys$. In other words, the indices correspond to strings in the language $\thuesyslang{a}$. Therefore, instead of adding an infinite number of rules to $\gtkms$ to allow for the elimination of $\psrp{a}{a \cate b}$, we could instead add a single rule of the form:
\begin{center}
\AxiomC{$\rel \sar w : \adia \phi, u : \phi, \Gamma$}
\RightLabel{$\pradiar$}
\UnaryInfC{$\rel \sar w : \adia \phi, \Gamma$}
\DisplayProof
\end{center}
where we impose a side condition stating that a sequence of relational atoms must exist in $\rel$ `corresponding to' strings in the language $\thuesyslang{a}$. The exact meaning of `corresponding to' relies on the formulation of a \emph{propagation graph}\index{Propagation graph!for $\kmsl$} and \emph{propagation path}\index{Propagation path}, which are formally defined in \dfn~\ref{def:propagation-graph-kms} and \dfn~\ref{def:propagation-path-kms} below, respectively, and are based on the work in~\cite{CiaLyoRamTiu20,GorPosTiu11}. The first notion transforms a labelled sequent into a graph, and the second notion defines a correspondence between paths in the graph and strings in a language generated by a \cfcst system. After giving these definitions, examples of a propagation graph, path, and rule are provided for clarity, followed by \fig~\ref{fig:Refined-Calculus-i-kms} which defines the propagation rules $\prcharadiar$.

\begin{definition}[Propagation Graphs for $\kmsl$]\label{def:propagation-graph-kms} Let $\Lambda = \rel \sar \Gamma$ be a labelled sequent for grammar logics. We define the \emph{propagation graph} $\prgr{\Lambda} = (\prgrdom, \prgredges)$ to be the directed graph such that
\begin{itemize}

\item[$\li$] $\prgrdom := \lab(\Lambda)$;

\item[$\li$] $\prgredges := \{(w,u,a), (u,w,\overline{a}) \ | \ R_{a}wu \in \rel \text{ or } R_{\conv{a}}uw \in \rel \}$.

\end{itemize}
We will often write $w \in \prgr{\Lambda}$ to mean $w \in \prgrdom$, and $(w,u,\chara) \in \prgr{\Lambda}$ to mean $(w,u,\chara) \in \prgredges$, for $\chara \in \albet$.
\end{definition}

\begin{definition}[Propagation Path for $\kmsl$]\label{def:propagation-path-kms} Let $\Lambda$ be a labelled sequent with $\prgr{\Lambda} = (\prgrdom, \prgredges)$. We define a \emph{propagation path from $w_{1}$ to $w_{n}$ in $\prgr{\Lambda}$} to be an alternating sequence of vertices $w_{1}, \ldots, w_{n} \in \prgrdom$ and characters $\chara_{1}, \ldots, \chara_{n-1} \in \albet$ of the form:
$$
\ppath(w_{1},w_{n}) := w_{1}, \chara_{1}, w_{2}, \chara_{2}, \ldots , \chara_{n-1}, w_{n}
$$
such that $(w_{1}, w_{2}, \chara_{1}) , (w_{2}, w_{3}, \chara_{2}), \ldots, (w_{n-1}, w_{n}, \chara_{n-1}) \in \prgredges$. Given a propagation path $\ppath(w_{1},w_{n}) = w_{1}, \chara_{1}, w_{2}, \chara_{2}, \ldots , \chara_{n-1}, w_{n}$, we let $\stra_{\ppath}(w_{1},w_{n}) = \chara_{1} \cate \chara_{2} \cate \cdots \cate \chara_{n-1}$ denote the \emph{string of the propagation path from $w_{1}$ to $w_{n}$}. 

We define the \emph{converse of a propagation path} as follows:
\begin{center}
$\overline{\ppath}(w_{n},w_{1}) := w_{n}, \overline{\chara}_{n}, w_{n-1}, \ldots, w_{2}, \overline{\chara}_{1}, w_{1}$ \ifandonlyif $\ppath(w_{1},w_{n}) := w_{1}, \chara_{1}, w_{2}, \ldots, w_{n-1}, \chara_{n-1}, w_{n}$.
\end{center}
Also, the \emph{converse string of a propagation path} as defined follows:
\begin{center}
$\overline{\stra}_{\ppath}(w_{n},w_{1}) := \overline{\chara}_{n} \cate \overline{\chara}_{n-1} \cate \cdots \cate \overline{\chara}_{1}$ \ifandonlyif $\stra_{\ppath}(w_{1},w_{n}) = \chara_{1} \cate \chara_{2} \cate \cdots \cate \chara_{n-1}$.
\end{center}

Last, we let $\emppath(w,w) := w$ represent the \emph{empty path}\index{Empty path} that holds between any vertex $w \in V$ and itself, with the string of the empty path defined as follows: $\stra_{\emppath}(w,w) := \empstr$.
\end{definition}

We are now in a position to properly define our propagation rules. \fig~\ref{fig:Refined-Calculus-i-kms} defines a propagation rule $\prcharadiar$ for each $\chara \in \albet$ and gives the side condition dictating its (in)application. For the moment, let us refer to the premise of the propagation rule $\prcharadiar$ as $\Lambda$ and its conclusion as $\Lambda'$. The side condition `$\exists \ppath ( \stra_{\ppath}(w,u) \in \thuesyslang{\chara} )$' should be read as stating that `there exists a propagation path $\ppath(w,u)$ in the propagation graph $\prgr{\Lambda} = \prgr{\Lambda'}$ such that $\stra_{\ppath}(w,u) \in \thuesyslang{\chara}$.' Observe that since the propagation graph of the premise is identical to the propagation graph of the conclusion, that is $\prgr{\Lambda} = \prgr{\Lambda'}$, propagation rules can be just as easily applied bottom-up as they can be applied top-down. We will use a similar notation to denote the side conditions of propagation (and reachability) rules in \cptr~\ref{CPTR:Refinment-Constructive} as well. Last, each refined labelled calculus $\kmsl$ for each grammar logic $\kms$ is displayed in \fig~\ref{fig:Refined-Calculus-i-kms} with the derivability relation for each calculus defined as follows: 

\begin{definition}\label{def:terminology-kmsl} We write $\vdash_{\kmsl} \Lambda$ to indicate that a labelled sequent $\Lambda$ is derivable in a calculus $\kmsl$.
\end{definition}

To supply the reader with additional intuition regarding propagation graphs, paths, and rules, an example unifying all such concepts is provided below:

\begin{example}\label{ex:propagation-graph-path} We give a pictorial representation of the propagation graph $\prgr{\Lambda}$, where $\Lambda$ is the labelled sequent defined below. Although it is not technically an aspect of the definition of a propagation graph, we also decorate the vertices of the propagation graph to show which formulae are associated with what vertices.

\begin{minipage}[t]{.5\textwidth}
\xymatrix{
 & & & \overset{\boxed{\emptyset}}{u} \ar@/^1pc/@{.>}[dl]|-{\conv{b}} \\
  \overset{\boxed{\adia p}}{w} \ar@/^1pc/@{.>}[rr]|-{a} &   & \overset{\boxed{\emptyset}}{v} \ar@/^1pc/@{.>}[dr]|-{\conv{c}} \ar@/^1pc/@{.>}[ll]|-{\overline{a}}\ar@/^1pc/@{.>}[ur]|-{b} &  \\
  &  & & \overset{\boxed{p,q}}{z} \ar@/^1pc/@{.>}[ul]|-{c}   
}
\end{minipage}
\begin{minipage}[t]{.5\textwidth}
\vspace{2.5em}
\begin{tabular}{c}
\AxiomC{ }
\noLine
\UnaryInfC{$\Lambda := R_{a}wv, R_{\conv{b}}uv, R_{c}zv \sar w : \adia p, z : p, z : q$}
\DisplayProof
\end{tabular}
\end{minipage}

Let us suppose that our \cfcst system is $\thuesys := \{a \pto a \cate b \conv{b} \cate \conv{c}, \conv{a} \pto \conv{a} \cate \conv{b} \cate b \cate c \}$. Observe that the path $\ppath(w,z) := w, a, v, b, u, \conv{b}, v, \conv{c}, z$ exists between $w$ and $z$ with $\adia p$ occurring at the starting vertex $w$, and $p$ occurring at the terminal vertex $z$. Since $\stra_{\ppath}(w,z) = a b \conv{b} \conv{c} \in \thuesyslang{a}$ (due to the first production rule of $\thuesys$), we can apply the propagation rule $\pradiar$ to $\Lambda$ to delete the labelled formula $z : p$, giving $R_{a}wv, R_{\conv{b}}uv, R_{c}zv \sar w : \adia p, z : q$ as the conclusion.
\end{example}

\begin{figure}[t]
\noindent\hrule

\begin{center}
\AxiomC{}
\RightLabel{$\id$}
\UnaryInfC{$\rel \sar w : p, w : \negnnf{p}, \Gamma$}
\DisplayProof
\end{center}

\begin{center}
\begin{tabular}{c c}
\AxiomC{$\rel \sar w : \phi, w : \psi, \Gamma$}
\RightLabel{$\disr$}
\UnaryInfC{$\rel \sar w : \phi \lor \psi, \Gamma$}
\DisplayProof

&

\AxiomC{$\rel \sar w : \phi, \Gamma$}
\AxiomC{$\rel \sar w : \psi, \Gamma$}
\RightLabel{$\conr$}
\BinaryInfC{$\rel \sar w : \phi \land \psi, \Gamma$}
\DisplayProof
\end{tabular}
\end{center}

\begin{center}
\begin{tabular}{c c}
\AxiomC{$\rel, R_{\chara}wu \sar u : \phi, \Gamma$}
\RightLabel{$\charabox^{\dag_{1}}$}
\UnaryInfC{$\rel \sar w : \charabox \phi, \Gamma$}
\DisplayProof

&

\AxiomC{$\rel \sar w : \charadia \phi, u : \phi, \Gamma$}
\RightLabel{$\prcharadiar^{\dag_{2}}$\index{$\prcharadiar$}}
\UnaryInfC{$\rel \sar w : \charadia \phi, \Gamma$}
\DisplayProof
\end{tabular}
\end{center}

\hrulefill
\caption{The refined calculus $\kmsl$\index{$\kmsl$} for the grammar logic $\kms$. We have a $\charabox$ and $\prcharadiar$ rule for each $\chara \in \albet$. The side condition $\dag_{1}$ states that the associated rule can be applied only if the label $u$ is an eigenvariable. The side condition $\dag_{2}$ states that $\exists \ppath ( \stra_{\ppath}(w,u) \in \thuesyslang{\chara} )$.}
\label{fig:Refined-Calculus-i-kms}
\end{figure}

\begin{lemma}\label{lem:diamond-is-propagation-rule-ksm}
For each $\chara \in \albet$, the rule $\charadiar$ is an instance of $\prcharadiar$.
\end{lemma}

\begin{proof} Let $\chara \in \albet$. The premise of $\charadiar$ is of the form $\rel, R_{\chara}wu \sar w : \charadia \phi, u : \phi, \Gamma$ and its propagation graph contains the propagation path $w, \chara, u$ (due to the occurrence of $R_{\chara}wu$). Since $\chara \in \thuesyslang{\chara}$, the side condition of the propagation rule $\prcharadiar$ is satisfied, allowing for the conclusion $\rel, R_{\chara}wu \sar w : \charadia \phi, \Gamma$ to be derived.
\end{proof}

The above lemma tells us that each propagation rule $\prcharadiar$ subsumes $\charadiar$, meaning that we need only consider the permutation of a $\psr$ rule above $\prcharadiar$ and not $\charadiar$. Therefore, since the only non-trivial case of proving the eliminability of $\psr$ concerns its permutation with $\charadiar$, if we are able to prove that the structural rule $\psr$ can always be permuted above the propagation rule $\prcharadiar$, then we will have successfully solved the problem discussed at the onset of this section, namely, the problem of eliminating each $\psr$ rule from a given derivation. 
 In the following lemma, we show that the propagation rules are in fact sufficient to allow for the upward mobility of each $\psr$ rule in a derivation; after showing this result, we give a concrete example showing how to permute a $\psr$ rule above a $\prcharadiar$ rule for clarity.

\begin{lemma}\label{lem:permuting-psr-padiar-kms} Let $\thuesys$ be a \cfcst system with $\chara, \chara_{0}, \ldots, \chara_{n}, \charb \in \albet$, and define the following:
\begin{itemize}

\item[$\li$] $R_{\stra}wu := R_{\chara_{0}}wu_{1}, \ldots R_{\chara_{n}}u_{n}u$

\item[$\li$] $\Lambda := \rel, R_{\stra}wu, R_{\chara}wu \sar v : \charbdia \phi, z:\phi, \Gamma$

\item[$\li$] $\Lambda' := \rel, R_{\stra}wu \sar v: \charbdia \phi, z:\phi, \Gamma$

\end{itemize}
Moreover, let $\ppath(v,z)$ be a propagation path between $v$ and $z$ occurring in $\prgr{\Lambda}$.

Suppose we are given a derivation in $\gtkms + \{\prcharadiar \ | \ \chara \in \albet\}$ ending with the following inferences:
\begin{center}
\begin{tabular}{c}
\AxiomC{$\rel, R_{\stra}wu, R_{\chara}wu \sar v : \charbdia \phi, z:\phi, \Gamma$}
\RightLabel{$\prcharbdiar$}
\UnaryInfC{$\rel, R_{\stra}wu, R_{\chara}wu \sar v : \charbdia \phi, \Gamma$}
\RightLabel{$\psrp{\chara}{\stra}$}
\UnaryInfC{$\rel, R_{\stra}wu \sar v : \charbdia \phi, \Gamma$}
\DisplayProof
\end{tabular}
\end{center}
where $\stra_{\ppath}(v,z) \in \thuesyslang{\charb}$. 
Then, there exists a path $\ppath'(v,z)$ in $\prgr{\Lambda'}$ such that $\stra_{\ppath'}(v,z) \in \thuesyslang{\charb}$, that is to say, the $\psrp{\chara}{\stra}$ rule may be permuted above the $\prcharbdiar$ rule to derive the same end sequent as shown below:
\begin{center}
\AxiomC{$\rel, R_{\stra}wu, R_{\chara}wu \sar v: \charbdia \phi, z:\phi, \Gamma$}
\RightLabel{$\psrp{\chara}{\stra}$}
\UnaryInfC{$\rel, R_{\stra}wu \sar v: \charbdia \phi, z:\phi, \Gamma$}
\RightLabel{$\prcharbdiar$}
\UnaryInfC{$\rel, R_{\stra}wu \sar v: \charbdia \phi, \Gamma$}
\DisplayProof
\end{center}
Note that $\psr$ may represent a structural rule obtained via the closure condition.
\end{lemma}

\begin{proof} Let our assumptions be those expressed in the statement of the lemma above. We consider two cases: either (i) the relational atom $R_{\chara}wu$ is not active in the $\prcharbdiar$ inference, or (ii) the relational atom $R_{\chara}wu$ is active in the $\prcharbdiar$ inference.

(i) Our assumption implies that the propagation path $\ppath(v,z)$ does not depend on the relational atom $R_{\chara}wu$, but rather, only depends on the relational atoms $\rel, R_{\stra}wu$. Therefore, we may take the desired propagation path $\ppath'(v,z)$ in $\prgr{\Lambda'}$ such that $\stra_{\ppath'}(v,z) \in \thuesyslang{\charb}$ to be the propagation path $\ppath(v,z)$, thus showing that the two rules are permutable.

(ii) For the second case, suppose that the relational atom $R_{\chara}wu$ is active in the $\prcharbdiar$ inference. To prove the claim we need to show the existence of a propagation path $\ppath'(v,z)$ in $\prgr{\Lambda'}$ such that $\stra_{\ppath'}(v,z) \in \thuesyslang{\charb}$. We construct such a propagation path by simultaneously performing the following replacements on the propagation path $\ppath(v,z)$:
\begin{itemize}

\item[$\li$] Replace each occurrence of $w, \chara, u$ in $\ppath(u,z)$ with $w, \chara_{0}, u_{1}, \ldots, u_{n}, \chara_{n}, u$, and

\item[$\li$] replace each occurrence of $u, \overline{\chara}, w$ in $\ppath(u,z)$ with $u, \overline{\chara}_{n}, u_{n}, \ldots, u_{1}, \overline{\chara}_{0}, w$.
\end{itemize}
We call the resulting path, after the above replacements on $\ppath(v,z)$ have been performed, $\ppath'(v,z)$. First, observe that the former propagation paths correspond to the edges $(w, u_{1}, \chara_{0}), \ldots, (u_{n}, u, \chara_{n}) \in \prgr{\Lambda'}$ obtained from the relational atoms $R_{\stra}wu$ that occur in $\Lambda'$ (by \dfn~\ref{def:propagation-graph-kms}). Second, observe that the latter propagation paths correspond to the edges $(u, u_{n}, \overline{\chara}_{n}), \ldots, (u_{1}, w, \overline{\chara}_{0}) \in \prgr{\Lambda'}$ which are also obtained from the relational atoms $R_{\stra}wu$ that occur in $\Lambda'$ (by \dfn~\ref{def:propagation-graph-kms}). Therefore, since the only difference between $\prgr{\Lambda}$ and $\prgr{\Lambda'}$ is that the former contains the edges $(w,u,\chara)$ and $(u,w,\overline{\chara})$ corresponding the relational atom $R_{\chara}wu$, whereas the latter does not, and because $\ppath'(v,z)$ omits use of paths $w, \chara, u$ and $u, \overline{\chara}, w$ (corresponding to edges $(w,u,\chara)$ and $(u,w,\overline{\chara})$, \resp), it follows that $\ppath'(v,z)$ occurs in $\prgr{\Lambda'}$.  

To finish the proof, we need to show that $\stra_{\ppath'}(v,z) \in \thuesyslang{\charb}$. By assumption, we know that $\stra_{\ppath}(v,z) \in \thuesyslang{\charb}$, which by \dfn~\ref{def:derivation-relation-language-kms}, implies that $\charb \dtoann \stra_{\ppath}(v,z)$. Moreover, since $\psr$ is in our calculus, as explained in \fig~\ref{fig:G3Km(S)}, there must exist a corresponding production rule $\chara \pto \stra \in \thuesys$. Also, since $\thuesys$ is a \cfcst system, we know that $\overline{\chara} \pto \overline{\stra} \in \thuesys$ by the closure condition \thueclcondns. Now, by the assumption of case (ii), we know that either $\chara$ or $\overline{\chara}$ occurs in $\stra_{\ppath}(v,z)$. Observe that applying $\chara \pto \stra$ to each occurrence of $\chara$ in $\stra_{\ppath}(v,z)$ (obtained from the relational atom $R_{\chara}wu$), and $\overline{\chara} \pto \overline{\stra}$ to each occurrence of $\overline{\chara}$ in $\stra_{\ppath}(v,z)$ (obtained from the relational atom $R_{\chara}wu$), yields the string $\stra_{\ppath'}(v,z)$. Hence, $\charb \dtoann \stra_{\ppath'}(v,z)$, which implies that $\stra_{\ppath'}(v,z) \in \thuesyslang{\charb}$ by \dfn~\ref{def:derivation-relation-language-kms}.
\end{proof}


\begin{example}\label{ex:psr-permutation} Let $\albet := \{a, \conv{a}, b, \conv{b}, c, \conv{c}\}$ and $\stra := \conv{b} \cate c \cate a \in \albet^{*}$. We give an example of permuting the structural rule $\psrp{\conv{b}}{\stra}$ above a propagation rule $\pradiar$. Let our \cfcst system be:
$$
\thuesys := \{a \pto b \cate \overline{b}, \overline{b} \pto \overline{b} \cate c \cate a, \overline{a} \pto a \cate \overline{b}, b \pto \overline{a} \cate \overline{c} \cate b\}.
$$
In the derivation below, we assume that $\pradiar$ is applied due to the occurrence of the propagation path $\ppath(w,u) = w, b, v, \overline{b}, u$ based on the relational atoms $R_{\conv{b}}vw, R_{\overline{b}}vu$, which further implies that $\stra_{\ppath}(w,u)  = b \cate \overline{b} \in \thuesyslang{a}$. This holds since the production rule $a \pto b \cate \overline{b}$ occurs in $\thuesys$. Concerning the $\psrp{\conv{b}}{\stra}$ rule, we assume that it is obtained from the production rule $\overline{b} \pto \overline{b} \cate c \cate a \in \thuesys$, and that the relational atom $R_{\conv{b}}vu$ is deleted from premise to conclusion due to the existence of the relational atoms $R_{\conv{b}}vw,R_{c}wz,R_{a}zu$. The propagation graph of the top sequent of the derivation is shown below right:

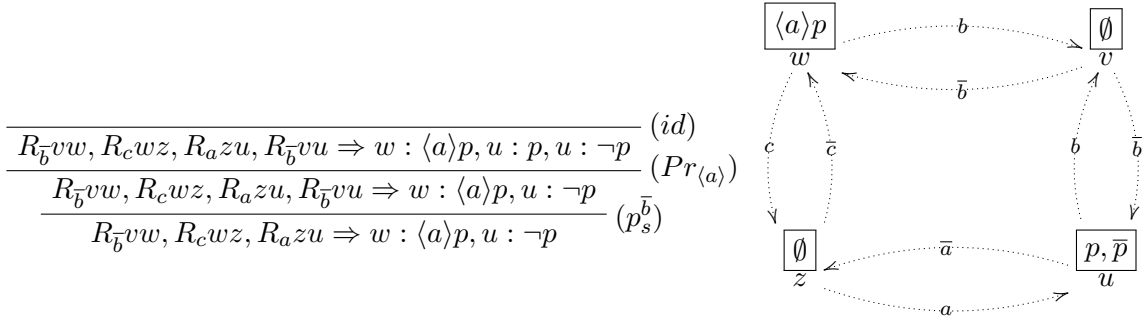
\begin{minipage}[t]{.33\textwidth}
\begin{tabular}{@{\hskip -.05em} c}
\vspace*{1 em}
\ \\
\AxiomC{}
\RightLabel{$\id$}
\UnaryInfC{$R_{\conv{b}}vw,R_{c}wz,R_{a}zu,R_{\conv{b}}vu \sar w:\adia p, u:p,u:\negnnf{p}$}
\RightLabel{$\pradiar$}
\UnaryInfC{$R_{\conv{b}}vw,R_{c}wz,R_{a}zu,R_{\conv{b}}vu \sar w:\adia p, u:\negnnf{p}$}
\RightLabel{$\psrp{\conv{b}}{\stra}$}
\UnaryInfC{$R_{\conv{b}}vw,R_{c}wz,R_{a}zu \sar w:\adia p,u:\negnnf{p}$}
\DisplayProof
\end{tabular}
\end{minipage}
\begin{minipage}[t]{.15\textwidth}
\ 
\end{minipage}
\begin{minipage}[t]{.33\textwidth}
\begin{tabular}{c}
\xymatrix{
  & &  \overset{\boxed{\adia p}}{w} \ar@/^-1pc/@{.>}[dd]|-{c} \ar@/^1pc/@{.>}[rrr]|-{b}	&   &   & \overset{\boxed{\emptyset}}{v}\ar@/^1pc/@{.>}[dd]|-{\overline{b}} \ar@/^1pc/@{.>}[lll]|-{\overline{b}}	\\
 & & & & & \\
 & &  \overset{\boxed{\emptyset}}{z} \ar@/^-1pc/@{.>}[uu]|-{\overline{c}}\ar@/^-1pc/@{.>}[rrr]|-{a}  &   	&	& \overset{\boxed{p, \overline{p}}}{u}\ar@/^1pc/@{.>}[uu]|-{b}\ar@/^-1pc/@{.>}[lll]|-{\overline{a}}
}
\end{tabular}
\end{minipage}

Let us now apply the $\psrp{\conv{b}}{\stra}$ to the top sequent of the derivation above in attempt to permute the two rules. If we apply $\psrp{\conv{b}}{\stra}$ to the top sequent, then we obtain the derivation shown below left. Below right, we provide the propagation graph of the conclusion of the derivation:

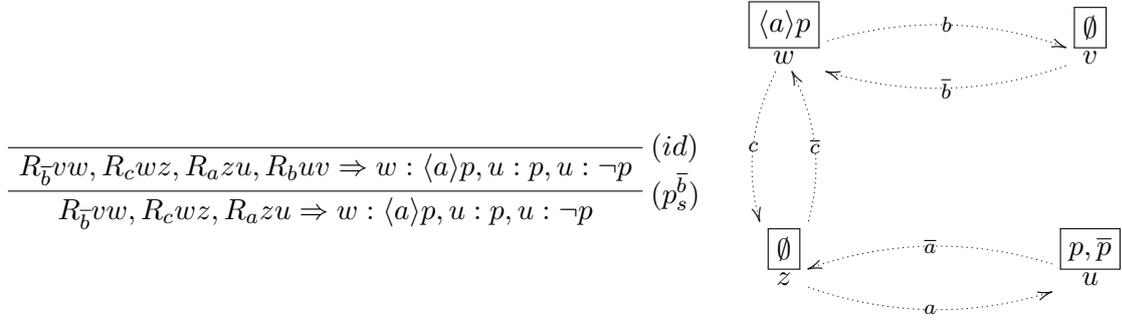
\begin{minipage}[t]{.33\textwidth}
\begin{tabular}{c}
\vspace*{1 em}
\ \\
\AxiomC{}
\RightLabel{$\id$}
\UnaryInfC{$R_{\conv{b}}vw,R_{c}wz,R_{a}zu,R_{b}uv \sar w:\adia p, u:p,u:\negnnf{p}$}
\RightLabel{$\psrp{\conv{b}}{\stra}$}
\UnaryInfC{$R_{\conv{b}}vw,R_{c}wz,R_{a}zu \sar w:\adia p, u : p, u:\negnnf{p}$}
\DisplayProof
\end{tabular}
\end{minipage}
\begin{minipage}[t]{.15\textwidth}
\ 
\end{minipage}
\begin{minipage}[t]{.33\textwidth}
\begin{tabular}{c}
\xymatrix{
  & &  \overset{\boxed{\adia p}}{w} \ar@/^-1pc/@{.>}[dd]|-{c} \ar@/^1pc/@{.>}[rrr]|-{b}	&   &   & \overset{\boxed{\emptyset}}{v} \ar@/^1pc/@{.>}[lll]|-{\overline{b}}	\\
 & & & & & \\
 & &  \overset{\boxed{\emptyset}}{z} \ar@/^-1pc/@{.>}[uu]|-{\overline{c}}\ar@/^-1pc/@{.>}[rrr]|-{a}  &   	&	& \overset{\boxed{p, \overline{p}}}{u}\ar@/^-1pc/@{.>}[lll]|-{\overline{a}}
}
\end{tabular}
\end{minipage}

In order to apply the propagation rule $\pradiar$, we need to show that there exists a propagation path $\ppath'(w,u)$ in the propagation graph of the conclusion such that $\stra_{\ppath'}(w,u) \in \thuesyslang{a}$. To show such a propagation path exists, we apply the procedure explained in the proof of \lem~\ref{lem:permuting-psr-padiar-kms}. In the initial derivation, we had the propagation path $\ppath(w,u) = w, b, v, \overline{b}, u$. Applying our procedure, we replace each occurrence of $u, b, v$ with $u, \overline{a}, z, \overline{c}, w, b, v$ and each occurrence of $v, \overline{b}, u$ with $v, \overline{b}, w, c, z, a, u$. Since no path $u, b, v$ 
 exists in $\ppath(w,u)$, only the latter replacement of $v, \conv{b}, u$ is applicable, and yields the propagation path $\ppath'(w,u) := w, b, v, \overline{b}, w, c, z, a, u$, which indeed exists in the propagation graph of the conclusion. Moreover, the derivation
$$
a \pto b \cate \overline{b} \pto b \cate \overline{b} \cate c \cate a
$$
in $\thuesys$ confirms that $\stra_{\ppath'}(w,u) = b \cate \overline{b} \cate c \cate a \in \thuesyslang{a}$. Therefore, we may apply $\pradiar$ to the conclusion above to derive the desired sequent and successfully permute the two rules.
\end{example}

We are now in a position to prove two critical lemmata, which respectively claim that each structural rule $\psr$ and $\convr$ is eliminable in $\gtkms + \{\prcharadiar \ | \ \chara \in \albet\}$. Note that when proving $\psr$ elimination, we need not consider permuting the rule above $\convr$, or vice-versa. As explained previously, given a derivation in $\gtkms + \{\prcharadiar \ | \ \chara \in \albet\}$, we may consider topmost occurrences of either a $\psr$ or $\convr$ rule, and successively delete such topmost occurrences via the elimination algorithms provided in the following two lemmata. Processing the input derivation in this way has the consequence that a $\psr$ rule will never be permuted above a $\convr$ rule, or vice-versa, since topmost occurrences are eliminated first.

Up until this point, we have primarily focused on securing the elimination of each $\psr$ rule. Coincidentally, although our analysis of $\psr$ elimination gave rise to the definition of a propagation rule, it just so happens that propagation rules are sufficient for proving the elimination of each $\convr$ rule as well. This is primarily based on the fact that propagation rules rely on the notion of a propagation graph, which `builds in' the information of each $\convr$ rule; the specifics of how each $\convr$ rule is `built into' the definition of a propagation graph will be explained after the following lemma:

\begin{lemma}\label{lem:psr-elimination-kms}
Let $\thuesys$ be a \cfcst system. For each $\chara \pto \stra \in \thuesys$, the rule $\psr$ is eliminable in the calculus $\gtkms + \{\prcharadiar \ | \ \chara \in \albet\} - \{\convr \ | \ \chara \in \albet\}$.
\end{lemma}

\begin{proof} We prove the result by induction on the height of the given derivation, and assume that our derivation contains one application of the $\psr$ rule, which is the last inference of the derivation. The general result follows by successively applying the procedure given below to delete topmost occurrences of $\psr$ rules until the derivation is free of such inferences.

\textit{Base case.} The base case follows from the fact that any application of a $\psr$ rule to an instance of $\id$ yields another instance of $\id$.

\textit{Inductive step.} By \lem~\ref{lem:diamond-is-propagation-rule-ksm}, we need not consider the $\charadiar$ case. Also, with the exception of the $\prcharadiar$ case, all cases are resolved by invoking \ih followed by an application of the corresponding rule. The $\prcharadiar$ case follows from \lem~\ref{lem:permuting-psr-padiar-kms} above.
\end{proof}

Having confirmed that each $\psr$ rule is eliminable, we now focus on showing $\convr$ elimination. As mentioned above, the information inherent in the $\convr$ rule, namely, that any relations $R_{\chara}$ and $R_{\conv{\chara}}$ in a $\albet$-model are converses of one another, is incorporated into the definition of a propagation graph. If one observes the definition of a propagation graph $\prgr{\Lambda} = (V,E)$ of a labelled sequent $\Lambda$ (\dfn~\ref{def:propagation-graph-kms}), they will notice that the edges $(w,u,\chara)$ and $(u,w,\conv{\chara})$ are added to the set of edges $E$ regardless of if the relational atom $R_{\chara}wu$ or the relational atom $R_{\conv{\chara}}uw$ occurs in $\Lambda$. Therefore, since the $\convr$ rule transforms a labelled sequent of the form $\rel, R_{\chara}wu, R_{\conv{\chara}}uw \sar \Gamma$ into a labelled sequent of the form $\rel, R_{\chara}wu \sar \Gamma$, the edges $(w,u,\chara)$ and $(u,w,\conv{\chara})$ in the propagation graph of the former will still be present in the propagation graph of the latter because of the continued presence of $R_{\chara}wu$. To put it another way, the propagation graph of the premise of a $\convr$ rule is identical to the propagation graph of the conclusion of the rule. 
As shown below, the fact that propagation graphs are invariant under applications of $\convr$ 
allows for the permutation of $\convr$ above an propagation rule. We will make use of this insight to prove $\convr$ elimination below:

\begin{lemma}\label{lem:convr-elim-kms}
Let $\thuesys$ be a \cfcst system with alphabet $\albet$. For each $\chara \in \albet$, the $\convr$ rule is eliminable in $\gtkms + \{\prcharadiar \ | \ \chara \in \albet\} - \{\psr \ | \ \chara \pto \stra \in \thuesys\}$.
\end{lemma}

\begin{proof} We prove the result by induction on the height of the given derivation, and assume that only one instance of $\convr$ occurs as the last inference of the derivation. The general result follows by successively deleting topmost occurrences of $\convr$ in a derivation, until it is free of such inferences.

\textit{Base case.} The base case follows from the fact that any application of $\convr$ to an instance of $\id$ yields another instance of $\id$.

\textit{Inductive step.} By \lem~\ref{lem:diamond-is-propagation-rule-ksm}, we need not consider the permutation of $\convr$ above $\charadiar$. With the exception of the $\prcharbdiar$ rule, all cases trivially follow by invoking \ih followed by the corresponding rule. Let us now consider the case of permuting $\convr$ above a $\prcharbdiar$ inference.

Suppose that our derivation ends with a propagation inference $\prcharbdiar$ followed by a $\convr$ inference as shown below left. Also, let $\Lambda := \rel, R_{\chara}wu, R_{\conv{\chara}}uw \sar v : \charbdia \phi, z : \phi, \Gamma$ with $\prgr{\Lambda} = (V,E)$, and take note that $(w,u,\chara),(u,w,\conv{\chara}) \in E$. Due to the side condition of the propagation rule, we know there exists a propagation path $\ppath(v,z)$ in $\prgr{\Lambda}$ such that $\stra_{\ppath}(v,z) \in \thuesyslang{\charb}$.  If we invoke \ih and delete $R_{\conv{\chara}}uw$ from $\Lambda$, then since the relational atom $R_{\chara}wu$ is still present, the edges $(w,u,\chara),(u,w,\conv{\chara})$ are still in $E$. In fact, the propagation graph of the conclusion $\rel, R_{\chara}wu \sar v : \charbdia \phi, z : \phi, \Gamma$ is identical to $\prgr{\Lambda}$. Therefore, the side condition for $\prcharbdiar$ is still satisfied, meaning the rule may be applied.

\begin{flushleft}
\begin{tabular}{c c}
\AxiomC{$\rel, R_{\chara}wu, R_{\conv{\chara}}uw \sar v : \charbdia \phi, z : \phi, \Gamma$}
\RightLabel{$\prcharbdiar$}
\UnaryInfC{$\rel, R_{\chara}wu, R_{\conv{\chara}}uw \sar v : \charbdia \phi, \Gamma$}
\RightLabel{$\convr$}
\UnaryInfC{$\rel, R_{\chara}wu \sar v : \charadia \phi, \Gamma$}
\DisplayProof

&

$\leadsto$
\end{tabular}
\end{flushleft}

\begin{flushright}
\AxiomC{$\rel, R_{\chara}wu, R_{\conv{\chara}}uw \sar v : \charbdia \phi, z : \phi, \Gamma$}
\RightLabel{\ih}
\dashedLine
\UnaryInfC{$\rel, R_{\chara}wu \sar v : \charbdia \phi, z : \phi, \Gamma$}
\RightLabel{$\prcharbdiar$}
\UnaryInfC{$\rel, R_{\chara}wu \sar v : \charbdia \phi, \Gamma$}
\DisplayProof
\end{flushright}
\end{proof}

The two previous lemmata establish the following theorem:

\begin{theorem}\label{thm:gtkms-to-kmsl-kms}
Every derivation in $\gtkms$ can be algorithmically transformed into a derivation in $\kmsl$.
\end{theorem}

\begin{proof} The result follows from \lem~\ref{lem:psr-elimination-kms} and \ref{lem:convr-elim-kms} by successively eliminating topmost occurrences of each $\psr$ and $\convr$ rule.
\end{proof}

By considering the elimination of structural rules, we have found that  each calculus $\gtkms$ can be transformed into a calculus $\kmsl$ that contains fewer rules overall and which are purely \emph{formula driven rules}\index{Formula driven rule}, i.e. each rule's (bottom-up) applicability is solely determined by the occurrence of a complex logical formula (from $\langkm{\albet}$) in the conclusion. Each calculus $\gtkms$ is different in this regard, since structural rules such as $\psr$ and $\convr$ are not formula driven in the above sense, as bottom-up applications of the rules introduce relational atoms independent of which labelled formulae occur. In addition, an analysis of the proofs of \lem~\ref{lem:psr-elimination-kms} and~\ref{lem:convr-elim-kms} shows that the proofs of $\kmsl$ are linearly compressed versions of the proofs from $\gtkms$. Moreover, as the following theorem establishes, our refined calculi only require the use of labelled tree sequents in derivations (\dfn~\ref{def:tree-sequent-kms}), whereas each $\gtkms$ calculus requires the use of more complex structures (\thm~\ref{thm:sequent-structure-gtkms}); this is significant as it shows that through structural rule elimination we have essentially obtained nested calculi for grammar logics (which we discuss in more detail in \sect~\ref{subsect:relationship-nested}  below). In these respects, each $\kmsl$ calculus is simpler than the $\gtkms$ calculus it was derived from; of course, the trade-off is that each $\charadiar$ rule has absorbed each $\psr$ and $\convr$ rule, thus generating the $\prcharadiar$ rule that possesses a more complex functionality.

\begin{theorem}\label{thm:tree-derivations-kmsl}
Let $\thuesys$ be a \cfcst system. Every derivation in $\kmsl$ of a labelled formula $w : \phi$ is a labelled tree derivation with the fixed root property.\footnote{The \emph{fixed root property} is defined in \dfn~\ref{def-tree-proof-kms} on p.~\pageref{def-tree-proof-kms}.}
\end{theorem}

\begin{proof} To see that the derivation of $w :\phi$ is a labelled tree derivation with the fixed root property (see \dfn~\ref{def-tree-proof-kms}) observe that applying inference rules of $\kmsl$ bottom-up either preserve relational structure or add forward relational structure (with the $\charaboxr$ rule), thus constructing a tree emanating from $w$.
\end{proof}

Although we have derived $\kmsl$ from $\gtkms$, we have not confirmed whether such calculi possess similar proof-theoretic properties. One course of action could be to prove the (hp-)admissibility of the rules $\strucsetkms$ as well as the (hp-)invertibility of all rules in $\kmsl$ (as we did for each $\gtkms$ calculus). However, we opt for another course of action, and show that every derivation in a $\kmsl$ calculus can be transformed into a derivation in the corresponding $\gtkms$ calculus. This backward transformation kills two birds with one stone: First, it demonstrates the deductive correspondence between $\gtkms$ and $\kmsl$, and second, it establishes that $\kmsl$ inherits proof-theoretic properties from $\gtkms$. Regarding the second point, leveraging both transformations (\thm~\ref{thm:gtkms-to-kmsl-kms} and~\ref{thm:kmsl-to-gtkms-kms}), lets us establish the soundness and completeness of $\kmsl$ (see \cor~\ref{cor:sound-comp-kmsl} below), and shows that the refined calculi possess similar admissibility and invertiblity properties (detailed in \cor~\ref{cor:struc-admiss-kmsl} and~\ref{cor:invert-kmsl} below) that are useful in applications; e.g. the extraction of interpolants (see \sect~\ref{sec:applicationsII}).

Before showing that any derivation in a calculus $\kmsl$ can be transformed into a derivation in $\gtkms$, we first need to show a lemma relating derivations in a \cfcst system to what is provable in $\gtkms$.

\begin{lemma}\label{lem:deleting-relational-atoms-kms}
Let $\Lambda := \rel \sar \Gamma$. Suppose $\rel, R_{\chara}wu \sar \Gamma$ is derivable in $\gtkms$, and that $\ppath(w,u)$ is a propagation path (potentially empty) in $\prgr{\Lambda}$ such that $\chara \dtoann \stra_{\ppath}(w,u)$. Then, $\Lambda = \rel \sar \Gamma$ is derivable in $\gtkms$.
\end{lemma}

\begin{proof} We prove the result by induction on the length (\dfn~\ref{def:derivation-relation-language-kms}) of the derivation $\chara \dtoann \stra_{\ppath}(w,u)$.

\textit{Base case.} For the base case, we consider (i) a derivation of length $0$, meaning that $\chara \dtoann \chara$, and (ii) a derivation of length $1$, meaning that $\chara \pto \stra_{\ppath}(w,u) \in \thuesys$. Both cases are respectively shown below, where $\rel := \rel', R_{\chara}wu$ in the first case, and $\rel := \rel', R_{\stra}wu$ with $\stra := \stra_{\ppath}(w,u)$ in the second case.

\begin{center}
\begin{tabular}{c c c}
\AxiomC{$\Pi_{1}$}
\UnaryInfC{$\rel', R_{\chara}wu, R_{\chara}wu \sar \Gamma$}
\DisplayProof

&

$\leadsto$

&

\AxiomC{$\Pi_{1}$}
\UnaryInfC{$\rel', R_{\chara}wu, R_{\chara}wu \sar  \Gamma$}
\RightLabel{$\ctrrel$}
\dashedLine
\UnaryInfC{$\rel', R_{\chara}wu \sar \Gamma$}
\DisplayProof
\end{tabular}
\end{center}

\begin{center}
\begin{tabular}{c c c}
\AxiomC{$\Pi_{2}$}
\UnaryInfC{$\rel', R_{\stra}wu, R_{\chara}wu \sar \Gamma$}
\DisplayProof

&

$\leadsto$

&

\AxiomC{$\Pi_{2}$}
\UnaryInfC{$\rel', R_{\stra}wu, R_{\chara}wu \sar  \Gamma$}
\RightLabel{$\psrp{\stra}{\chara}$}
\UnaryInfC{$\rel', R_{\stra}wu \sar \Gamma$}
\DisplayProof
\end{tabular}
\end{center}

\textit{Inductive step.} Let $\Lambda := \rel \sar \Gamma$ and assume that we have a proof $\Pi$ of $\rel, R_{\chara}wu \sar \Gamma$. Suppose our derivation $\chara \dtoann \stra_{\ppath}(w,u)$ is of length $n+1$, that is, it consists of a derivation $\chara \dtoann \strb$ of length $n$ followed by a one-step derivation $\strb \dto_{\thuesys} \stra_{\ppath}(w,u)$. Hence, there exist strings $\strc_{0}, \strc_{1} \in \albet^{*}$ and a production rule $\charb \pto \strc \in \thuesys$ such that $\strb = \strc_{0} \cate \charb \cate \strc_{1}$ and $\strc_{0} \cate \strc \cate \strc_{1} = \stra_{\ppath}(w,u)$. This implies the existence of a propagation path $\ppath_{\strc_{0}}(w,v), \ppath_{\strc}(v,z), \ppath_{\strc_{1}}(z,u)$ in $\prgr{\Lambda}$, where $v,z \in \lab(\Lambda)$. We give a proof below showing how to derive the desired result.

First, we apply hp-admissibility of $\wk$ (\lem~\ref{lem:struc-rules-admiss-kms}) to the proof $\Pi$ of $\rel, R_{\chara}wu \sar \Gamma$, introducing the relational atom $R_{\charb}vz$ and giving a proof of the labelled sequent $\Lambda' := \rel, R_{\charb}vz, R_{\chara}wu \sar \Gamma$. Observe that the propagation path $\ppath_{\strc_{0}}(w,v), \ppath_{\charb}(v,z), \ppath_{\strc_{1}}(z,u)$ exists in $\prgr{\Lambda'}$. Also, by assumption, we know that $\chara \dtoann \strb = \strc_{0} \cate \charb \cate \strc_{1}$ with a derivation of length $n$. Thus, we may invoke \ih to derive the sequent $\Lambda'' := \rel, R_{\charb}vz \sar \Gamma$. Last, by our assumption that $\charb \pto \strc \in \thuesys$, and the fact that $\ppath_{\strc}(v,z)$ occurs in $\prgr{\Lambda''}$, we know that there exists a structural rule $\psrp{\charb}{\stra}$ in $\gtkms$ that is applicable to $\Lambda''$. Applying this rule lets us derive the desired conclusion.

\begin{center}
\AxiomC{$\Pi$}
\UnaryInfC{$\rel, R_{\chara}wu \sar \Gamma$}
\RightLabel{$\wk$}
\dashedLine
\UnaryInfC{$\rel, R_{\charb}vz, R_{\chara}wu \sar \Gamma$}
\RightLabel{\ih}
\dashedLine
\UnaryInfC{$\rel, R_{\charb}vz \sar \Gamma$}
\RightLabel{$\psrp{\charb}{\stra}$}
\UnaryInfC{$\rel \sar \Gamma$}
\DisplayProof
\end{center}
\end{proof}

\begin{theorem}\label{thm:kmsl-to-gtkms-kms}
Every derivation in $\kmsl + \strucsetkms$ can be algorithmically transformed into a derivation in $\gtkms$.
\end{theorem}

\begin{proof} We prove the result by induction on the height of the given derivation.

\textit{Base case.} Any instance of $\id$ in $\kmsl + \strucsetkms$ is an instance of $\id$ in $\gtkms$.

\textit{Inductive step.} As usual, we prove the inductive step by a case-distinction on the last rule applied in the given derivation. If the last rule applied is a rule in the set $\strucsetkms$, then we invoke the (hp-)admissibility result of the corresponding rule (\lem~\ref{lem:struc-rules-admiss-kms} and \thm~\ref{thm:cut-admiss-kms}). With the exception of $\prcharadiar$, all other cases are handled by invoking \ih and then applying the same rule. We show how to resolve the case when the last inference is $\prcharadiar$.
\begin{center}
\begin{tabular}{c c c}
\AxiomC{$\rel \sar w : \charadia \phi, u : \phi, \Gamma$}
\RightLabel{$\prcharadiar$}
\UnaryInfC{$\rel \sar w : \charadia \phi, \Gamma$}
\DisplayProof

&

$\leadsto$

&

\AxiomC{}
\RightLabel{\ih}
\dashedLine
\UnaryInfC{$\rel \sar w : \charadia \phi, u : \phi, \Gamma$}
\RightLabel{$\wk$}
\dashedLine
\UnaryInfC{$\rel, R_{\chara}wu \sar w : \charadia \phi, u : \phi, \Gamma$}
\RightLabel{$\charadiar$}
\UnaryInfC{$\rel, R_{\chara}wu \sar w : \charadia \phi, \Gamma$}
\RightLabel{\lem~\ref{lem:deleting-relational-atoms-kms}}
\dashedLine
\UnaryInfC{$\rel \sar w : \charadia \phi, \Gamma$}
\DisplayProof
\end{tabular}
\end{center}
By the side condition imposed on the $\prcharadiar$ rule, we know that there exists a propagation path $\ppath(w,u)$ in $\prgr{\rel \sar w : \charadia \phi, u : \phi, \Gamma}$ such that $\stra_{\ppath}(w,u) \in \thuesyslang{\chara}$. The applicability of \lem~\ref{lem:deleting-relational-atoms-kms} follows from the fact that $\prgr{\rel \sar w : \charadia \phi, u : \phi, \Gamma} = \prgr{\rel \sar w : \charadia \phi, \Gamma}$, implying that $\ppath(w,u)$ exists in $\prgr{\rel \sar w : \charadia \phi, \Gamma}$.
\end{proof}

We may now utilize our translations between $\gtkms$ and $\kmsl$ (\thm~\ref{thm:gtkms-to-kmsl-kms} and~\ref{thm:kmsl-to-gtkms-kms}) to show that our refined calculi inherit proof-theoretic properties from their parent labelled calculi.

\begin{corollary}\label{cor:struc-admiss-kmsl}
The rules in $\strucsetkms$ are admissible in $\kmsl$.
\end{corollary}

\begin{proof} Follows from the (hp-)admissibility of the rules $\strucsetkms$ in $\gtkms$ (\lem~\ref{lem:struc-rules-admiss-kms} and \thm~\ref{thm:cut-admiss-kms}), as well as \thm~\ref{thm:gtkms-to-kmsl-kms} and \ref{thm:kmsl-to-gtkms-kms}.
\end{proof}

\begin{corollary}\label{cor:invert-kmsl}
All rules of  $\kmsl$ are invertible.
\end{corollary}

\begin{proof} The invertibility of $\prcharadiar$ follows from the admissibility of $\wk$ (\cor~\ref{cor:struc-admiss-kmsl} above). The invertibility of $\disr$, $\conr$, and $\charaboxr$ is argued as follows: Let $\Lambda$ be an instance of a conclusion of $\disr$, $\conr$, or $\charaboxr$, and assume $\Lambda$ is derivable in $\kmsl$. Then, by \thm~\ref{thm:kmsl-to-gtkms-kms} we know that $\Lambda$ is derivable in $\gtkms$, and so, by the (hp-)invertibility of the rules in $\gtkms$ (\lem~\ref{lem:hp-invert-kms}), we know that the corresponding premise(s) $\Lambda'$ (and $\Lambda''$) of the rule is (are) derivable. Hence, by \thm~\ref{thm:gtkms-to-kmsl-kms} $\Lambda'$ (and $\Lambda''$) is (are) derivable in $\kmsl$.
\end{proof}

\begin{corollary}[Soundness and Completeness of $\kmsl$]\label{cor:sound-comp-kmsl} Let $\thuesys$ be a \cfcst system.

(i) If $\vdash_{\kmsl} \Lambda$, then $\models_{\kms} \Lambda$.

(ii) If $\vdash_{\kms} \phi$, then $\vdash_{\kmsl} \seqempstr \sar w : \phi$.
\end{corollary}

\begin{proof} Follows from the soundness of $\gtkms$ (\thm~\ref{thm:soundness-gtkms}), the completeness of $\gtkms$ (\thm~\ref{thm:completness-gtkms}), as well as \thm~\ref{thm:gtkms-to-kmsl-kms} and \ref{thm:kmsl-to-gtkms-kms}.
\end{proof}


\subsection{Relationship to Nested Sequent Formalism}\label{subsect:relationship-nested}

In this section, we prove that each refined labelled calculus $\kmsl$ is a notational variant of a deep nested calculus $\dkms$ (displayed in \fig~\ref{fig:DKm(S)} below). It should be noted that the deep nested calculi presented here are slight reformulations of the nested calculi for grammar logics introduced in~\cite{TiuIanGor12}. Therefore, due to the minute differences between the nested calculi introduced here and those given in~\cite{TiuIanGor12}, we assign our calculi the same names and refer to each calculus (for a \cfcst system $\thuesys$) as $\dkms$.

The observation that refinement yields variants of known nested systems (as will be shown for first-order intuitionistic logics as well), suggests that the discovery of such systems was---in some sense---not accidental, and that refinement is a natural procedure connecting the relational semantics of a logic to a nested calculus for the logic. In order to establish the equivalence between refined labelled and nested systems, we will first introduce the $\dkms$ nested calculi below, and then define translations permitting us to switch between labelled and nested notation. Ultimately, these translations will be used to construct (relatively simple) algorithms that allow for proofs to be translated between the refined labelled and nested calculi.

Let us now define the building blocks for our $\dkms$ calculi: \emph{nested sequents} for grammar logics~\cite{TiuIanGor12}.\footnote{Note that the notation used for nested sequents in \cite{TiuIanGor12} was obtained from the notation put forth in Kashima's paper~\cite{Kas94} and uses brackets of the form $\{$ and $\}$ as opposed to brackets of the form $[$ and $]$ for nesting data. This is the reverse of the notation used by many writers who opt to nest data with the latter brackets; e.g.~\cite{Bru09,Bul92,Fit14,Str13}.}

\begin{definition}[Nested Sequents for Grammar Logics~\cite{TiuIanGor12}]\label{def:nested-sequents-kms} \emph{Nested sequents for grammar logics}\index{Nested sequent!for grammar logics} are syntactic objects $\na$ defined via the following grammar in BNF:
$$
\na ::= \seqempstr \ | \ \phi \ | \ \na, \na \ | \ (\chara)\nbbl \na \nbbr
$$
where $\chara \in \albet$ and $\phi \in \langkm{\albet}$.
\end{definition}

We use $\na$, $\nb$, $\nc$, $\ldots$ (possibly annotated) to denote nested sequents, and we let $\seqempstr$ represent the \emph{empty string} (i.e. empty nested sequent). As with labelled sequents, $\seqempstr$ is an identity element for comma, and comma associates and commutes. We use the notation $\na \nbl \nb \nbr$ (and $\na \nbl \nb \nbr \nbl \nc \nbr$) to mean that $\nb$ ($\nb$ and $\nc$, \resp) occurs (occur, \resp) at some depth in the nestings of $\na$. For example, if the nested sequent $\na$ is $p, (a) \nbbl q, (b) \nbbl \negnnf{p}, r \nbbr \nbbr$, then $\na \nbl p \nbr$, $\na \nbl q, (b) \nbbl \negnnf{p}, r \nbbr \nbr$, $\na \nbl q \nbr \nbl \negnnf{p}, r \nbr$, etc. are all valid representations of $\na$.

\subsubsection{The Value of Switching Notation.} Before we proceed, it is interesting to wonder if switching from labelled to nested notation possesses any utility in its own right, that is, 
does the option of converting notation offer any advantage? Two reasons come to mind which justify a positive answer. First, switching from labelled to nested notation ensures that the language of our calculi enjoys a degree of parsimony, that is, the language does not allow for syntactic structures that go too far beyond what is needed for completeness. Second, switching to a more restrictive notation 
 has practical value as it provides a priori knowledge about the structure of proofs; e.g. one can be certain that all sequents within a given proof encode trees prior to observing any proof within the associated system. To ground this second point further, we consider a concrete case: in~\cite{LyoBer19} the first proof-search procedures for \stit logics were introduced using refined labelled calculi, and to ensure the correctness of each proof-search algorithm, it was necessary to prove that all labelled sequents generated throughout the course of proof-search were labelled forest sequents (see \dfn~\ref{def:tree-DAG-sequent-dsn}). Had the notation of the refined labelled calculi been translated to a \emph{rigid} notation only allowing for sequents of a forest shape (e.g. like how nested sequents only allow for sequents of a tree shape) prior to the writing of the proof-search procedure, then confirming that all sequents generated throughout the course of proof-search were of a forest shape would have been rendered unnecessary. Due to the rigidity of the notation used, which---by its very definition---only allows for certain structures, it would have been known in advance that all sequents possessed the desired shape. Therefore, \emph{rigidifying} notation (i.e. switching from a more liberal notation like that of labelled sequents to a more restrictive notation like that of nested sequents) appears to be of value.

Furthermore, translating in the opposite direction---from nested to labelled notation---appears to be worthwhile as well. For instance, in~\cite{LyoTiuGorClo20} a purely syntactic method of proving interpolation for logics via their nested sequent calculi was introduced. In that paper, the nested sequent calculi were translated into labelled notation as the notation was easier to work with and allowed for simpler definitions. Hence, translations 
 provide one with the freedom to switch proof-theoretic formalisms when one is better suited for a particular task.

In order to translate between labelled and nested sequents, we introduce \emph{sequent graphs} for nested sequents, which are objects intermediary in the translation.

\begin{definition}[Sequent Graph of a Nested Sequent for $\kms$]\label{def:sequent-graph-nested-kms} We define the \emph{sequent graph of a nested sequent $\na$}\index{Sequent graph!of a nested sequent for $\kms$} inductively on the depth of the nestings of $\na$ as shown below. Notice that we make use of sequences of natural numbers
$$
\numseq \in \bigcup_{n \in \mathbb{N}} \mathbb{N}^{n}
$$
to represent our vertices similar to the prefixes used in prefixed tableaux (cf.~\cite{Fit72}). We represent sequences as $n_{1} . n_{2} . \ldots . n_{k-1} . n_{k}$ and use $\numseq$ (possible annotated) to denote them. Our inductive definition of $\seqgraph(X) := \seqgraph_{0}(X)$ is as follows:
\begin{itemize}

\item[$\li$] If $\na = \phi_{1}, \ldots, \phi_{k}$, then $G_{\numseq}(\na) := (V_{\numseq},E_{\numseq},L_{\numseq})$, where
$$
(i) \quad V_{\numseq} := \{\numseq\} \qquad (ii) \quad E_{\numseq} := \emptyset \qquad (iii) \quad L_{\numseq} := \{(\numseq, \{\phi_{1}, \ldots, \phi_{k}\})\}
$$

\item[$\li$] Let $\na := \phi_{1}, \ldots, \phi_{k}, (\chara_{1}) \nbbl \nc_{1} \nbbr, \ldots, (\chara_{n}) \nbbl \nc_{n} \nbbr$ and suppose that each $G_{\numseq  . i}(\nc_{i}) = ( V_{ \numseq . i}, E_{\numseq . i}, L_{\numseq . i} )$ (with $i \in \{1, \ldots n\}$ and $n \in \mathbb{N}$) is already defined. We define $G_{\numseq}(\na) := ( V_{\numseq}, E_{\numseq}, L_{\numseq} )$ as shown below:
\begin{flushleft}
$\li \quad V_{\numseq} := \{\numseq\} \cup \displaystyle{\bigcup_{1 \leq i \leq n} V_{\numseq . i}}$\\
$\li \quad E_{\numseq} := \{(\numseq,\numseq . i,\chara_{i}) \ | \ 1 \leq i \leq n \} \cup \displaystyle{\bigcup_{1 \leq i \leq n} E_{\numseq . i}}$\\
$\li \quad L_{\numseq} := \{(\numseq, \{\phi_{1}, \ldots, \phi_{k}\})\} \cup \displaystyle{\bigcup_{1 \leq i \leq n} L_{\numseq . i}}
$
\end{flushleft}
\end{itemize}
Note that when $k = 0$, the multiset $\phi_{1}, \ldots, \phi_{k}$ is taken to be the empty string $\seqempstr$. Also, we will often use $w$, $u$, $v$, $\ldots$ to represent vertices as opposed to sequences of natural numbers.
\end{definition}

Even though we may use the notation $\na \nbl \nb \nbr$ or $\na \nbl \nb \nbr \nbl \nc \nbr$ to denote a nested sequent $\na$, we also allow for the notation $\na \nbl \nb \nbr_{w}$ and $\na \nbl \nb \nbr_{w} \nbl \nc \nbr_{u}$ to denote that $\nb$ is associated with the vertex $w$ (meaning that $L(w) = \nb$ in $\seqgraph(X) = (V,E,L)$), and to denote that $\nb$ and $\nc$ are associated with the vertices $w$ and $u$ (meaning that $L(w) = \nb$ and $L(u) = \nc$ in $\seqgraph(X) = (V,E,L)$), respectively.

Since each calculus $\dkms$ is a notational variant of a refined labelled calculus $\kmsl$, each nested calculus likewise employs a set of propagation rules. Therefore, it is necessary to define propagation graphs for nested sequents. Still, the definition of a propagation path and the string of a propagation path, along with their converses,  remains the same in the nested setting and are as given in \dfn~\ref{def:propagation-path-kms}. This follows from the fact that propagation graphs for nested sequents are identical to propagation graphs for labelled (tree) sequents.

\begin{definition}[Propagation Graphs for $\dkms$]\label{def:propagation-graph-nested-kms} Let $\na$ be a nested sequent for grammar logics with sequent graph $\seqgraph(\na) = (V,E,L)$. We define the \emph{propagation graph}\index{Propagation graph!for $\dkms$} $\prgr{\na} = (\prgrdom', \prgredges')$ to be the directed graph such that
\begin{itemize}

\item[$\li$] $\prgrdom' := V$;

\item[$\li$] $\prgredges' := \{(w,u,\chara), (u,w,\overline{\chara}) \ | \ (w,u,\chara) \in E \text{ or } (u,w,\overline{\chara}) \in E \}$.

\end{itemize}
We will often write $w \in \prgr{\na}$ to mean $w \in \prgrdom'$, and $(w,u,\chara) \in \prgr{\na}$ to mean $(w,u,\chara) \in \prgredges'$, for $\chara \in \albet$.
\end{definition}

Each nested calculus $\dkms$ for a given \cfcst system $\thuesys$ is displayed in \fig~\ref{fig:DKm(S)}. The derivability relation for each calculus is defined as follows:

\begin{definition}
We write $\vdash_{\dkms} X$ to indicate that the nested sequent $X$ is derivable in $\dkms$.
\end{definition}

 With the exception of each propagation rule $\prcharadiar$ (for $\chara \in \albet$), all rules are identical to those given in~\cite{TiuIanGor12}. The distinguishing feature between the propagation rules presented here and the propagation rules presented in~\cite{TiuIanGor12} concerns the side condition imposed. The side condition of a $\prcharadiar$ rule states that $\exists \ppath (\stra_{\ppath}(w,u) \in \thuesyslang{\chara})$ must hold if the rule is applied, i.e. there must exist a propagation path $\ppath(w,u)$ in the propagation graph of the premise (or, equivalently, the conclusion) such that $\stra_{\ppath}(w,u) \in \thuesyslang{\chara}$. By contrast, the side condition of a propagation rule from~\cite{TiuIanGor12} checks if a certain string is both accepted by a certain automaton and within a certain language~\cite[p.~523]{TiuIanGor12}. Nevertheless, due to the equivalence between formal grammars and automata, it turns out that both side conditions are equivalent. 

\begin{figure}[t]
\noindent\hrule

\begin{center}
\begin{tabular}{c c c}
\AxiomC{}
\RightLabel{$\id$}
\UnaryInfC{$X[p, \negnnf{p}, Y]$}
\DisplayProof

&

\AxiomC{$X[\phi, \psi, Y]$}
\RightLabel{$\disr$}
\UnaryInfC{$X[\phi \lor \psi, Y]$}
\DisplayProof

&

\AxiomC{$X[\phi, Y]$}
\AxiomC{$X[\psi, Y]$}
\RightLabel{$\conr$}
\BinaryInfC{$X[\phi \land \psi, Y]$}
\DisplayProof
\end{tabular}
\end{center}

\begin{center}
\begin{tabular}{c c}
\AxiomC{$X[Y, (\chara)\{\phi\}]$}
\RightLabel{$\charaboxr$}
\UnaryInfC{$X[Y, \charabox \phi]$}
\DisplayProof

&

\AxiomC{$X[\charadia \phi, Y]_{w}[\phi, Z]_{u}$}
\RightLabel{$\prcharadiar^{\dag}$\index{$\prcharadiar$}}
\UnaryInfC{$X[\charadia \phi, Y]_{w}[Z]_{u}$}
\DisplayProof
\end{tabular}
\end{center}

\hrulefill
\caption{The nested calculus $\dkms$\index{$\dkms$}, which is a variant 
 of a calculus from~\cite{TiuIanGor12} of the same name. The calculus has a $\charaboxr$ and $\prcharadiar$ rule for each $\chara \in \albet$. The side condition $\dag$ states that the rule is applicable only if $\exists \ppath (\stra_{\ppath}(w,u) \in \thuesyslang{\chara})$.}
\label{fig:DKm(S)}
\end{figure}

Nested sequents encode a tree structure by definition, with the nestings representing edges occurring within the tree. In this regard, nested sequents differ from labelled sequents, which encode edges via their relational atoms. In addition, the nested notation is more rigid than the labelled notation, since it only allows for trees to be represented, whereas labelled sequents can represent arbitrary graphs. 
 To provide more intuition regarding the tree that corresponds to a nested sequent, we give an example of a nested sequent and the tree it encodes below; an example of the propagation graph of the nested sequent is also provided.

\begin{example} The nested sequent $\na := q,r, (\conv{b}) \nbbl \negnnf{q} \nbbr, (b)\nbbl q \lor r, (a) \nbbl \seqempstr \nbbr, (\conv{d}) \nbbl \seqempstr \nbbr \nbbr$ encodes the tree $\seqgraph(X)$ shown below left and has the propagation graph $\prgr{X}$ shown below right:
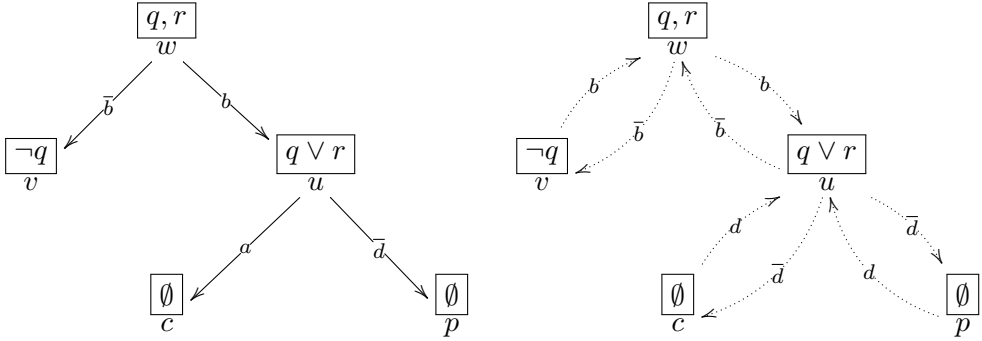
\begin{center}
\begin{tabular}{c c}
\xymatrix{
   & \overset{\boxed{q,r}}{w} \ar[dl]|-{\conv{b}} \ar[dr]|-{b} & &  \\
  \overset{\boxed{\negnnf{q}}}{v} & & \overset{\boxed{q \lor r}}{u} \ar[dl]|-{a}\ar[dr]|-{\conv{d}} & \\ 
  & \overset{\boxed{\emptyset}}{c} &  & \overset{\boxed{\emptyset}}{p}
}

&

\xymatrix{
   & \overset{\boxed{q,r}}{w} \ar@/^1pc/@{.>}[dl]|-{\conv{b}} \ar@/^1pc/@{.>}[dr]|-{b} & &  \\
  \overset{\boxed{\negnnf{q}}}{v}\ar@/^1pc/@{.>}[ur]|-{b} & & \overset{\boxed{q \lor r}}{u} \ar@/^1pc/@{.>}[dr]|-{\conv{d}} \ar@/^1pc/@{.>}[dl]|-{\conv{d}}\ar@/^1pc/@{.>}[ul]|-{\conv{b}} & \\ 
  & \overset{\boxed{\emptyset}}{c}\ar@/^1pc/@{.>}[ur]|-{d} &  & \overset{\boxed{\emptyset}}{p}\ar@/^1pc/@{.>}[ul]|-{d}
}
\end{tabular}
\end{center}
\end{example}

Perhaps the reader will find the above sequent graph $\seqgraph(X)$ familiar, and in fact it is. The graph is identical to the sequent graph $\seqgraph(\Lambda)$ of the labelled tree sequent $\Lambda$ given in \exmpl~\ref{ex:sequent-graph-examples-kms} of \sect~\ref{sec:lab-calc-kms}. This demonstrates a natural correspondence between labelled tree sequents and nested sequents. We now define the translation $\lnkms$ mapping labelled tree sequents to nested sequents, and the translation $\nlkms$ mapping nested sequents to labelled tree sequents---both of which rely on the notion of a sequent graph. After defining these two translations, we introduce an algorithm that converts derivations in $\kmsl$ to derivations in $\dkms$ (\thm~\ref{thm:kmsl-to-dkms}), and an algorithm that translates derivations in $\dkms$ to derivations in $\kmsl$ (\thm~\ref{thm:dkms-to-kmsl}).



\begin{definition}[Downward Closure]\label{def:downward-closure-kms} Let $\na$ be a nested sequent and $\Lambda$ be a labelled tree sequent. Also, let $\seqgraph(\na) := (V,E,L)$ and $\seqgraph(\Lambda) := (V,E,L)$ with $w \in V$. (NB. The ambiguity in notation is of no consequence and simplifies presentation.) We define the \emph{$w$-downward closure}\index{Downward closure} $\seqgraph_{w}(X) = (V',E',L')$ and $\seqgraph_{w}(\Lambda) = (V',E',L')$ to be the smallest induced subgraph (\dfn~\ref{def:graph-subgraph}) of $G(X)$ and $G(\Lambda)$, respectively, such that $w \in V'$ and
\begin{itemize}

\item[$\blacktriangleright$] if $v \in V'$ and $(v,u) \in E$, then $u \in V'$;

\item[$\blacktriangleright$] $E' = E \restriction V'$ 

\item[$\blacktriangleright$] $L' = L \restriction V'$

\end{itemize}
\end{definition}

\begin{remark}
If $X$ is a nested sequent and $\Lambda$ is a labelled tree sequent with $w$ the root of the sequent graph $\seqgraph(X)$ and $\seqgraph(\Lambda)$, then $\seqgraph_{w}(X) = \seqgraph(X)$ and  $\seqgraph_{w}(\Lambda) = \seqgraph(\Lambda)$, respectively.
\end{remark}

\begin{definition}[The Translation $\lnkms$]\label{def:ln-kms} Let $\Lambda := \rel \sar \Gamma$ be the a labelled tree sequent with $\seqgraph(\Lambda) = (V,E,L)$ and $w \in V$ the root. We define the translation $\lnt(\Lambda) := \lnt(\seqgraph_{w}(\Lambda))$\index{Translation $\lnt$!for grammar logics} inductively as follows:
\begin{itemize}

\item[$\li$] If $\seqgraph(\Lambda) = (V,E,L)$ with $V = \{w\}$, $E = \emptyset$, and $L = \{(w, \{\phi_{1}, \ldots, \phi_{k}\})\}$, then
$$
\lnt(\seqgraph_{w}(\Lambda)) := \phi_{1}, \ldots, \phi_{k}
$$

\item[$\li$] If $\seqgraph(\Lambda) = (V,E,L)$ with $w, u_{1}, \ldots, u_{n} \in V$, $(w,u_{i},\chara_{i}) \in E$ (for $i \in \{1, \ldots, n\}$), then
$$
\lnt(\seqgraph_{w}(\Lambda)) := L(w), (\chara_{1})\nbbl \lnt(\seqgraph_{u_{1}}(\Lambda)) \nbbr, \ldots, (\chara_{n})\nbbl \lnt(\seqgraph_{u_{n}}(\Lambda)) \nbbr
$$
\end{itemize}
\end{definition}

We demonstrate the operation of the translation function $\lnkms$ with an example:

\begin{example} We show how to translate the labelled tree sequent
$$
\Lambda:= R_{a}wu, R_{a}uv, R_{\conv{b}}wz \sar w : q, w : \bdia q, u : p \land \negnnf{p}, v : q, v :r, z : c, z :p
$$
into a nested sequent via the computation below:
\begin{eqnarray*}
& \lnkms(\Lambda) & := \lnkms(\seqgraph_{w}(\Lambda))\\
&  & = q, \bdia q, (a)\{\lnkms(\seqgraph_{u}(\Lambda))\}, (\conv{b})\{\lnkms(\seqgraph_{z}(\Lambda))\}\\
&  & = q, \bdia q, (a)\{p \land \negnnf{p}, (a) \{\lnkms(\seqgraph_{v}(\Lambda))\}\}, (\conv{b})\{c,p\}\\
&  & = q, \bdia q, (a)\{p \land \negnnf{p}, (a) \{q,r\}\}, (\conv{b})\{c,p\}\\
\end{eqnarray*}
\end{example}

In order to define the reverse translation from nested to labelled notation, it is helpful to introduce the notion of a \emph{sequent composition}\index{Sequent composition}. Since this concept will be used in the sequel, we define the notion for labelled sequents in general.

\begin{definition}[Labelled Sequent Composition] Given two labelled sequents $\Lambda_{1} := \rel_{1}, \Gamma_{1} \sar \Delta_{1}$ and $\Lambda_{2} := \rel_{2}, \Gamma_{2} \sar \Delta_{2}$, we define the \emph{sequent composition} as follows:
$$
\Lambda_{1} \seqcomp \Lambda_{2} := \rel_{1}, \rel_{2}, \Gamma_{1}, \Gamma_{2} \sar \Delta_{1}, \Delta_{2}
$$
\end{definition}

\begin{definition}[The Translation $\nlt$]\label{def:nl-kms} Let $\na$ be the a nested sequent. We define the translation $\nlt(\na) := \nlt(\seqgraph_{0}(\na))$\index{Translation $\nlt$!for grammar logics} inductively as follows:
\begin{itemize}

\item[$\li$] If $\seqgraph_{\numseq}(\na) = (V,E,L)$ with $V = \{\numseq\}$, $E = \emptyset$, and $L = \{(\numseq, Y)\}$, then
$$
\nlt(\seqgraph_{\numseq}(\na)) := \seqempstr \sar w_{\numseq} : \nb
$$

\item[$\li$] If $\seqgraph_{\numseq}(\na) = (V,E,L)$ with $\numseq, \numseq . 1, \ldots, \numseq . n \in V$, $(\numseq, \numseq . 1, \chara_{1}), \ldots, (\numseq, \numseq . n, \chara_{n}) \in E$, and $(\numseq, \nb) \in L$, then
\end{itemize}
$$
\nlt(\seqgraph_{\numseq}(\na)) := (R_{\chara_{1}}w_{\numseq}w_{\numseq . 1}, \ldots, R_{\chara_{m}}w_{\numseq}w_{\numseq . n} \sar w_{\numseq} : \nb) \seqcomp \nlt(\seqgraph_{\numseq . 1}(\na)) \seqcomp \cdots \seqcomp \nlt(\seqgraph_{\numseq . n}(\na))
$$
In practice, we will often use labels such as $w$, $u$, $v$, $\ldots$ as opposed to labels indexed with sequences of natural numbers for simplicity.
\end{definition}

Similar to before, we demonstrate how the translation function works with an example:

\begin{example} Below, we show how to translate the nested sequent
$$
X := q, \bdia q, (a)\{p \land \negnnf{p}, (a) \{q,r\}\}, (\conv{b})\{c,p\}
$$
into a labelled (tree) sequent:
\begin{eqnarray*}
& \nlkms(X) & := \nlkms(\seqgraph_{0}(X))\\
&  & = (R_{a}wu, R_{\conv{b}}wz \sar w : q, w : \bdia q) \seqcomp \nlkms(\seqgraph_{u}(X)) \seqcomp \nlkms(\seqgraph_{z}(X)) \\
&  & = (R_{a}wu, R_{\conv{b}}wz \sar w : q, w : \bdia q) \seqcomp (R_{a}uv \sar u : p \land \negnnf{p}) \seqcomp  \\
& & \textcolor{white}{=}\nlkms(\seqgraph_{v}(X)) \seqcomp (\seqempstr \sar z : c, z :p)\\
&  & = (R_{a}wu, R_{\conv{b}}wz \sar w : q, w : \bdia q) \seqcomp (R_{a}uv \sar u : p \land \negnnf{p}) \seqcomp \\
& & \textcolor{white}{=} (\seqempstr \sar v : q, v : r) \seqcomp (\seqempstr \sar z : c, z :p)\\
&  & = R_{a}wu, R_{a}wz, R_{\conv{b}}uv \sar w : q, w : \bdia q, u : p \land \negnnf{p}, v : q, v : r, z : c, z : p
\end{eqnarray*}
\end{example}

Based on the above definitions, it is not hard to see that the sequent graph of a labelled tree sequent $\Lambda$ is isomorphic to the sequent graph of $\lnkms(\Lambda)$ or that the sequent graph of a nested sequent $X$ is isomorphic to the sequent graph of $\nlkms(X)$. (NB. The relation of being \emph{isomorphic} is defined in \dfn~\ref{def:isomorphism} in \cptr~\ref{CPTR:Labelled}.) Thus, we have the following lemma:

\begin{lemma}\label{lem:iso-translation-kms} Let $\Lambda$ be a labelled tree sequent and $X$ be a nested sequent (for grammar logics). Then,

(i) $G(\Lambda) \iso G(\lnkms(\Lambda))$

(ii) $G(X) \iso G(\nlkms(X))$
\end{lemma}

Let us now leverage our translation functions to show that for any \cfcst system $\thuesys$, the calculi $\kmsl$ and $\dkms$ are notational variants of one another. Translating proofs between refined labelled and nested systems for grammar logics is straightforward and merely consists of changing notation.

\begin{theorem}\label{thm:kmsl-to-dkms}
Let $\thuesys$ be a \cfcst system. Every derivation in $\kmsl$ is algorithmically translatable to a derivation in $\dkms$.
\end{theorem}

\begin{proof} We prove the result by induction on the height of the given derivation.

\textit{Base case.} The base case is straightforward as the principal formulae $w:p, w:\negnnf{p}$ will be included together in the output nested sequent, thus ensuring that it is initial. We let $Y := \Gamma \restriction w$ in the nested sequent $X$ below.

\begin{center}
\begin{tabular}{c c c}
\AxiomC{ }
\RightLabel{$\id$}
\UnaryInfC{$\rel \sar w : p, w : \negnnf{p}, \Gamma$}
\DisplayProof

&

$\leadsto$

&

\AxiomC{ }
\RightLabel{$\id$}
\UnaryInfC{$\lnt(\rel \sar w : p, w : \negnnf{p}, \Gamma)$}
\RightLabel{=}
\dottedLine
\UnaryInfC{$\na \nbl p, \negnnf{p}, Y \nbr_{w}$}
\DisplayProof
\end{tabular}
\end{center}

\textit{Inductive step.} Each case of the inductive step is straightforward and is resolved as shown below. In all cases, we let $Y := \Gamma \restriction w$, and in the $\prcharadiar$ case we let $Z := \Gamma \restriction u$. Also, note that the side condition in the $\prcharadiar$ case holds due to \lem~\ref{lem:iso-translation-kms} above, which ensures that $\prgr{\rel \sar w : \charadia \phi, u : \phi, \Gamma} \iso \prgr{X[\charadia \phi, Y]_{w}[\phi, Z]_{u}}$.

\begin{center}
\begin{tabular}{c c c}
\AxiomC{$\rel \sar w : \phi, w : \psi, \Gamma$}
\RightLabel{$\disr$}
\UnaryInfC{$\rel \sar w : \phi \lor \psi, \Gamma$}
\DisplayProof

&

$\leadsto$

&

\AxiomC{$\lnt(\rel \sar w : \phi, w : \psi, \Gamma)$}
\RightLabel{=}
\dottedLine
\UnaryInfC{$\na \nbl \phi, \psi, Y \nbr_{w}$}
\RightLabel{$\disr$}
\UnaryInfC{$\na \nbl \phi \lor \psi, Y \nbr_{w}$}
\dottedLine
\RightLabel{=}
\UnaryInfC{$\lnt(\rel \sar w : \phi \lor \psi, \Gamma)$}
\DisplayProof
\end{tabular}
\end{center}

\begin{center}
\begin{tabular}{c @{\hskip -1em} c}
\AxiomC{$\rel \sar w : \phi, \Gamma$}
\AxiomC{$\rel \sar w : \psi, \Gamma$}
\RightLabel{$\conr$ \ $\leadsto$}
\BinaryInfC{$\rel \sar w : \phi \land \psi, \Gamma$}
\DisplayProof

&

\AxiomC{$\lnt(\rel \sar w : \phi, \Gamma)$}
\RightLabel{=}
\dottedLine
\UnaryInfC{$\na \nbl \phi, \nb \nbr_{w}$}

\AxiomC{$\lnt(\rel \sar w : \psi, \Gamma)$}
\RightLabel{=}
\dottedLine
\UnaryInfC{$\na \nbl \psi, \nb \nbr_{w}$}

\RightLabel{$\conr$}
\BinaryInfC{$\na \nbl \phi \land \psi, \nb \nbr_{w}$}
\dottedLine
\RightLabel{=}
\UnaryInfC{$\lnt(\rel \sar w : \phi \land \psi, \Gamma)$}
\DisplayProof
\end{tabular}
\end{center}

\begin{center}
\begin{tabular}{c c c}
\AxiomC{$\rel, R_{\chara}uw \sar u : \phi, \Gamma$}
\RightLabel{$\charaboxr$}
\UnaryInfC{$\rel \sar w : \charabox \phi, \Gamma$}
\DisplayProof

&

$\leadsto$

&

\AxiomC{$\lnt(\rel, R_{\chara}uw \sar u : \phi, \Gamma)$}
\RightLabel{=}
\dottedLine
\UnaryInfC{$X[Y, (\chara)\nbbl \phi \nbbr]_{w}$}
\RightLabel{$\charaboxr$}
\UnaryInfC{$X[Y, \charabox \phi ]_{w}$}
\RightLabel{=}
\dottedLine
\UnaryInfC{$\lnt(\rel \sar w : \charabox \phi, \Gamma)$}
\DisplayProof
\end{tabular}
\end{center}

\begin{center}
\begin{tabular}{c c c}
\AxiomC{$\rel \sar w : \charadia \phi, u : \phi, \Gamma$}
\RightLabel{$\prcharadiar$}
\UnaryInfC{$\rel \sar w : \charadia \phi,\Gamma$}
\DisplayProof

&

$\leadsto$

&

\AxiomC{$\lnkms(\rel \sar w : \charadia \phi, u : \phi, \Gamma)$}
\RightLabel{=}
\dottedLine
\UnaryInfC{$X[\charadia \phi, Y]_{w}[\phi, Z]_{u}$}
\RightLabel{$\prcharadiar$}
\UnaryInfC{$X[\charadia \phi, Y]_{w}[Z]_{u}$}
\RightLabel{=}
\dottedLine
\UnaryInfC{$\lnkms(\rel \sar w : \charadia \phi, u : \phi, \Gamma)$}
\DisplayProof
\end{tabular}
\end{center}
\end{proof}

\begin{theorem}\label{thm:dkms-to-kmsl}
Let $\thuesys$ be a \cfcst system. Every derivation in $\dkms$ is algorithmically translatable to a derivation in $\kmsl$.
\end{theorem}

\begin{proof} The theorem is proven by induction on the height of the given derivation and is similar to the proof of the previous theorem (\thm~\ref{thm:kmsl-to-dkms}). In essence, the proof puts forth a simple algorithm showing that one can switch the notation of a given derivation from nested to labelled through applications of the $\nlkms$ function.
\end{proof}

An example is provided below that demonstrates the correspondence between proofs in a refined labelled calculus $\kmsl$ and nested calculus $\dkms$.

\begin{example} The derivation on the left translates via $\lnkms$ to the derivation on the right, and the derivation on the right translates via $\nlkms$ to the derivation on the left (up to a change of labels).
\begin{center}

\begin{minipage}[t]{.33\textwidth}
\AxiomC{}
\RightLabel{$\id$}
\UnaryInfC{$R_{\chara}wu \sar w : \negnnf{p}, w : p, u : \charadiac p$}
\RightLabel{$\prcharadiar$}
\UnaryInfC{$R_{\chara}wu \sar w : \negnnf{p}, u : \charadiac p$}
\RightLabel{$\charaboxr$}
\UnaryInfC{$\sar w : \negnnf{p}, w : \charabox \charadiac p$}
\RightLabel{$\disr$}
\UnaryInfC{$\sar w : \negnnf{p} \lor \charabox \charadiac p$}
\DisplayProof
\end{minipage}
\begin{minipage}[t]{.15\textwidth}
\
\end{minipage}
\begin{minipage}[t]{.33\textwidth}
\AxiomC{}
\RightLabel{$\id$}
\UnaryInfC{$\negnnf{p}, p, (\chara)\{ \charadiac p\}$}
\RightLabel{$\prcharadiar$}
\UnaryInfC{$\negnnf{p}, (\chara)\{ \charadiac p\}$}
\RightLabel{$\charaboxr$}
\UnaryInfC{$\negnnf{p}, \charabox \charadiac p$}
\RightLabel{$\disr$}
\UnaryInfC{$\negnnf{p} \lor \charabox \charadiac p$}
\DisplayProof
\end{minipage}

\end{center}
\end{example}


\subsection{A Note on the Relationship between Refined Labelled and Display Calculi} 

Before concluding this section, we mention an interesting property that holds for each $\kmsl$ calculus, and is largely reminiscent of the \emph{display}/\emph{residuation} rules\index{Display rule}\index{Residuation rule} employed in \emph{display}/\emph{shallow nested} calculi\index{Display calculus}\index{Shallow nested calculus}~\cite{Bel82,GorPosTiu11,Kas94,LyoIttEckGra17,TiuIanGor12,Wan02}.\footnote{The shallow nested calculi of~\cite{GorPosTiu11,Kas94,TiuIanGor12} can be seen as one-sided versions of display calculi (cf.~\cite{Bel82,Wan02}).} Display calculi were introduced by Belnap in~\cite{Bel82}, and their one-sided variants were introduced by Kashima in~\cite{Kas94}. Such calculi generalize Gentzen's sequent calculus formalism by extending the formalism with additional structural connectives that go beyond the comma utilized in Gentzen-style sequents. Moreover, a significant feature of the display calculus/shallow-nested sequent calculus formalism is that it admits a general cut-elimination theorem stating that any calculus built within the formalism, which also satisfies eight easily verifiable syntactic conditions, allows for cut-elimination~\cite{Bel82}. This property has proven the formalism useful in uniformly providing cut-free calculi for large classes of logics, independent of the logical connectives or semantics employed.

The paper~\cite{TiuIanGor12} not only introduced (slight reformulations of) the $\dkms$ calculi presented above, but also introduced shallow nested calculi $\skms$ for context-free grammar logics with converse. Such calculi make use of the same nested sequents, but restrict applications of logical rules (\emph{viz.} $\disr$, $\conr$, $\charaboxr$) to the root, or top-level, of a nested sequent. Furthermore, propagation rules are replaced by two sets of rules: rules that introduce $\charadia$ formulae, much like the $\charadiar$ rules, and structural rules, much like the $\psr$ rules. We do not introduce such rules here since they are unneeded in our analysis of $\kmsl$. The interested reader is referred to~\cite[p.~521]{TiuIanGor12} for a formal introduction of shallow nested calculi $\skms$ for context-free grammar logics with converse. The essential characteristic of shallow nested calculi however, is the incorporation of the following residuation rule in each calculus:
\begin{center}
\AxiomC{$X, (\chara)\{Y\}$}
\RightLabel{$(r)$}
\UnaryInfC{$(\conv{\chara})\{X\},Y$}
\DisplayProof
\end{center}

The residuation rule $(r)$ allows for a nesting to be `flipped' (much like turning the page of a book) from one structure $Y$ within the nested sequent to another structure $X$, so long as the index is changed from $\chara$ to $\conv{\chara}$ in the process. Keeping this in mind, we will prove the following proposition concerning $\kmsl$, which verifies the validity of `flipping' relational atoms, and afterwards, will briefly discuss the relevance of this property with respect to shallow nested calculi and `display-style reasoning'.

\begin{proposition}\label{prop:display-prop-kmsl} Let $\thuesys$ be a \cfcst system. Every proof of a sequent $\rel, R_{\chara}wu \sar \Gamma$ in $\kmsl$ is algorithmically transformable to a proof of $\rel, R_{\conv{\chara}}uw \sar \Gamma$ with the same height or less, and vice-versa.
\end{proposition}

\begin{proof} We prove the result by induction on the height of the given derivation.

\textit{Base case.} The base case is resolved as shown below:

\begin{center}
\begin{tabular}{c c c}
\AxiomC{}
\RightLabel{$\id$}
\UnaryInfC{$\rel, R_{\chara}wu \sar v : p, v : \negnnf{p}, \Gamma$}
\DisplayProof

&

$\leadsto$ 

&

\AxiomC{}
\RightLabel{$\id$}
\UnaryInfC{$\rel, R_{\conv{\chara}}uw \sar v : p, v : \negnnf{p}, \Gamma$}
\DisplayProof
\end{tabular}
\end{center}

\textit{Inductive step.} We show how to resolve the case when the last rule is a $\prcharbdiar$ inference, as all other cases are resolved by applying \ih and then the corresponding rule. We solve and explain the $\prcharbdiar$ case below:
\begin{center}
\begin{tabular}{c c c}
\AxiomC{$\rel, R_{\chara}wu \sar v : \charbdia \phi, z : \phi, \Gamma$}
\RightLabel{$\prcharbdiar$}
\UnaryInfC{$\rel, R_{\chara}wu \sar v : \charbdia \phi, \Gamma$}
\DisplayProof

&

$\leadsto$

&

\AxiomC{$\rel, R_{\conv{\chara}}wu \sar v : \charbdia \phi, z : \phi, \Gamma$}
\RightLabel{$\prcharbdiar$}
\UnaryInfC{$\rel, R_{\conv{\chara}}wu \sar v : \charbdia \phi, \Gamma$}
\DisplayProof

\end{tabular}
\end{center}
Let $\Lambda := \rel, R_{\chara}wu \sar v : \charbdia \phi, z : \phi, \Gamma$ and $\Lambda' := \rel, R_{\conv{\chara}}wu \sar v : \charbdia \phi, z : \phi, \Gamma$. The side condition of the $\prcharbdiar$ inference on the left implies that there exists a path $\ppath(v,z)$ in $\prgr{\Lambda}$ such that $\stra_{\ppath}(v,z) \in \thuesyslang{\chara}$. By \dfn~\ref{def:propagation-graph-kms}, $\prgr{\Lambda} = \prgr{\Lambda'}$, implying that the path $\ppath(v,z)$ exists in $\prgr{\Lambda'}$. Therefore, the derivation on the left may be transformed into the derivation on the right.
\end{proof}

The above proposition is interesting in that it implies the hp-admissibility of a labelled version of the residuation rule $(r)$, namely:
\begin{center}
\AxiomC{$\rel, R_{\chara}wu \sar \Gamma$}
\RightLabel{$(l)$}
\UnaryInfC{$\rel, R_{\conv{\chara}}uw \sar \Gamma$}
\DisplayProof
\end{center}
The hp-admissibility of the above rule suggests that our refined labelled calculi are capable of simulating the display-style reasoning inherent in the shallow nested calculi of \cite{TiuIanGor12}. In fact, by the work done in this section, and the work in~\cite{TiuIanGor12}, we know that for any \cfcst system $\thuesys$, every derivation in $\kmsl$ can be algorithmically transformed into a derivation in the shallow nested calculus $\skms$, and vice-versa. This conclusion is justified by the fact that each calculus $\kmsl$ is a notational variant of a nested calculus $\dkms$, and in~\cite{TiuIanGor12} it is shown that (a slight variant of) each nested calculus $\dkms$ can be algorithmically transformed into a shallow nested calculus $\skms$. 

What is more noteworthy perhaps, is the interplay between the more liberal notation of labelled sequents and the display-style reasoning engendered by the hp-admissible $(l)$ rule. The residuation rule $(r)$ transforms one nested sequent into another nested sequent, effectively transforming an object encoding a tree into a new object encoding a tree. 
 By contrast, the flexibility of labelled notation allows for $(l)$ to not only transform a labelled tree sequent into another labelled tree sequent (thus simulating the behavior of $(r)$ in the nested setting), but applications of $(l)$ allow for the production of intermediate labelled sequents outside the class of labelled tree sequents as well (which do not correspond to nested sequents via the translation $\lnkms$). That is to say, the labelled notation coupled with the labelled residuation rule $(l)$ allows for a more unrestrained version of display-style reasoning. To make this point concrete, an example comparing applications of $(l)$ to $(r)$ is provided below.

\begin{example} In the first part of the example, we consider a derivation of a labelled tree sequent $\Lambda_{2}$ (defined below) that consists of two applications of the $(l)$ rule to a labelled tree sequent $\Lambda_{1}$ (also defined below). This derivation translates (in its entirety) to a derivation of $\lnkms(\Lambda_{2})$ from $\lnkms(\Lambda_{1})$ consisting of two applications of the $(r)$ rule. In addition, we display the sequent graphs of the labelled tree and nested sequents to illustrate how applications of $(l)$ and $(r)$ manipulate the underlying data structure of the associated sequents. 
 In the second part of the example, we give a derivation of $\Lambda_{2}$ from $\Lambda_{1}$ that uses alternative instances of the $(l)$ rule and does not translate via $\lnkms$ (in its entirety) to a derivation in the nested setting since it includes a labelled sequent $\Lambda$ which is not a labelled tree sequent. This observation demonstrates the additional methods of proof offered in the labelled setting.

\begin{center}
\begin{tabular}{c c}
\AxiomC{$\Lambda_{1} :=$}
\noLine
\UnaryInfC{\textcolor{white}{$R_{a}$}}
\noLine
\UnaryInfC{$\Lambda_{2} :=$}
\DisplayProof

&

\AxiomC{$R_{\conv{a}}uw, R_{b}wv, R_{\conv{c}}wz \sar u : q \lor r, w : p, w : \negnnf{p}, z : \negnnf{q}$}
\RightLabel{$(l)$}
\UnaryInfC{$R_{\conv{a}}uw, R_{b}wv, R_{c}zw \sar u : q \lor r, w : p, w : \negnnf{p}, z : \negnnf{q}$}
\RightLabel{$(l)$}
\UnaryInfC{$R_{\conv{a}}wu, R_{b}wv, R_{c}zw \sar u : q \lor r, w : p, w : \negnnf{p}, z : \negnnf{q}$}
\DisplayProof
\end{tabular}
\end{center}

\begin{center}
\AxiomC{$\lnkms(R_{\conv{a}}uw, R_{b}wv, R_{\conv{c}}wz \sar u : q \lor r, w : p, w : \negnnf{p}, z : \negnnf{q})$}
\RightLabel{=}
\dottedLine
\UnaryInfC{$q \lor r, (\conv{a}) \{p, \negnnf{p}, (b) \{\seqempstr\}, (\conv{c})\{\negnnf{q}\} \}$}
\RightLabel{$(r)$}
\UnaryInfC{$(a) \{ q \lor r \}, p, \negnnf{p}, (b) \{\seqempstr\}, (\conv{c})\{\negnnf{q}\}$}
\RightLabel{$(r)$}
\UnaryInfC{$(c)\{ (a) \{ q \lor r \}, p, \negnnf{p}, (b) \{\seqempstr\} \}, \negnnf{q}$}
\dottedLine
\RightLabel{=}
\UnaryInfC{$\lnkms(R_{\conv{a}}wu, R_{b}wv, R_{c}zw \sar u : q \lor r, w : p, w : \negnnf{p}, z : \negnnf{q})$}
\DisplayProof
\end{center}

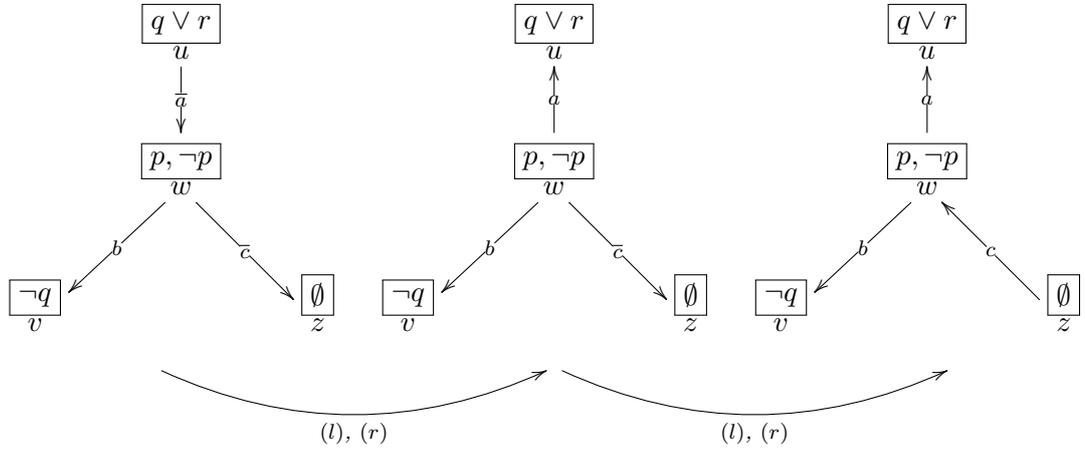
\begin{center}
\begin{tabular}{c c c}
\xymatrix{
& \overset{\boxed{q \lor r}}{u}\ar[d]|-{\conv{a}} & \\
   & \overset{\boxed{p,\negnnf{p}}}{w} \ar[dl]|-{b}\ar[dr]|-{\conv{c}} &  \\
  \overset{\boxed{\negnnf{q}}}{v} & & \overset{\boxed{\emptyset}}{z}
}

&

\xymatrix{
& \overset{\boxed{q \lor r}}{u} & \\
   & \overset{\boxed{p,\negnnf{p}}}{w} \ar[dl]|-{b}\ar[dr]|-{\conv{c}}\ar[u]|-{a} &  \\
  \overset{\boxed{\negnnf{q}}}{v} & & \overset{\boxed{\emptyset}}{z}
}

&

\xymatrix{
& \overset{\boxed{q \lor r}}{u} & \\
   & \overset{\boxed{p,\negnnf{p}}}{w} \ar[dl]|-{b}\ar[u]|-{a} &  \\
  \overset{\boxed{\negnnf{q}}}{v} & & \overset{\boxed{\emptyset}}{z}\ar[ul]|-{c}
}
\end{tabular}
\begin{tabular}{@{\hskip -1.5em} c}
\xymatrix{
 & & \ar@/^-1.5pc/@{->}[rrrrr]_{\text{$(l)$, $(r)$}}
 & & & &  
 & \ar@/^-1.5pc/@{->}[rrrrr]_{\text{$(l)$, $(r)$}} &  & & & & &
}
\end{tabular}
\end{center}

In the above derivation the root of $\Lambda_{1}$ (and $\lnkms(\Lambda_{1})$) switches from $u$ to $w$, and then from $w$ to $z$. This behavior is not exhibited in the alternative proof of $\Lambda_{2}$ from $\Lambda_{1}$, which includes an intermediary labelled sequent $\Lambda$ whose sequent graph is not a tree, but rather, is a polytree (\dfn~\ref{def:polytree}). This derivation and the corresponding sequent graphs are shown below:

\begin{center}
\begin{tabular}{c c}
\AxiomC{$\Lambda_{1} :=$}
\noLine
\UnaryInfC{$\Lambda_{\textcolor{white}{2}} :=$}
\noLine
\UnaryInfC{$\Lambda_{2} :=$}
\DisplayProof

&

\AxiomC{$R_{\conv{a}}uw, R_{b}wv, R_{\conv{c}}wz \sar u : q \lor r, w : p, w : \negnnf{p}, z : \negnnf{q}$}
\RightLabel{$(l)$}
\UnaryInfC{$R_{a}wu, R_{b}wv, R_{\conv{c}}wz \sar u : q \lor r, w : p, w : \negnnf{p}, z : \negnnf{q}$}
\RightLabel{$(l)$}
\UnaryInfC{$R_{a}wu, R_{b}wv, R_{c}zw \sar u : q \lor r, w : p, w : \negnnf{p}, z : \negnnf{q}$}
\DisplayProof
\end{tabular}
\end{center}

\begin{center}
\begin{tabular}{c c c}
\xymatrix{
& \overset{\boxed{q \lor r}}{u}\ar[d]|-{\conv{a}} & \\
   & \overset{\boxed{p,\negnnf{p}}}{w} \ar[dl]|-{b}\ar[dr]|-{\conv{c}} &  \\
  \overset{\boxed{\negnnf{q}}}{v} & & \overset{\boxed{\emptyset}}{z}
}

&

\xymatrix{
& \overset{\boxed{q \lor r}}{u}\ar[d]|-{\conv{a}} & \\
   & \overset{\boxed{p,\negnnf{p}}}{w} \ar[dl]|-{b} &  \\
  \overset{\boxed{\negnnf{q}}}{v} & & \overset{\boxed{\emptyset}}{z}\ar[ul]|-{c}
}

&

\xymatrix{
& \overset{\boxed{q \lor r}}{u} & \\
   & \overset{\boxed{p,\negnnf{p}}}{w} \ar[dl]|-{b}\ar[u]|-{a} &  \\
  \overset{\boxed{\negnnf{q}}}{v} & & \overset{\boxed{\emptyset}}{z}\ar[ul]|-{c}
}
\end{tabular}
\begin{tabular}{@{\hskip -1.5em} c}
\xymatrix{
 & & \ar@/^-1.5pc/@{->}[rrrrr]_{\text{$(l)$}}
 & & & &  
 & \ar@/^-1.5pc/@{->}[rrrrr]_{\text{$(l)$}} &  & & & & &
}
\end{tabular}
\end{center}
\end{example}

The above example demonstrates the added flexibility embedded within labelled notation, which not only explains why labelled systems can simulate other systems (e.g. nested and display) with relative ease~\cite{CiaLyoRam18,CiaLyoRamTiu20,GorRam12,Pim18,Res06}, but perhaps explains why labelled systems possess stronger proof-theoretic properties in relation to other systems (which use a more rigid or restrictive notation). For example, certain linear nested sequent systems (e.g.~\cite{KuzLel18,Lyo20b}) only allow for the \emph{admissibility} of contractions as opposed to the \emph{height-preserving admissibility} of contractions, which seems to be due to the fact that the manipulation of data must occur within a linear structure. 

Last, the paper~\cite{CiaLyoRamTiu20} discusses the relationship between labelled, refined labelled, nested, and shallow nested (i.e. display) calculi for a class of tense logics. The logics considered in that paper can be seen as a restricted class of grammar logics that use an alphabet of the form $\albet = \{a, \conv{a}\}$. The translations in~\cite{CiaLyoRamTiu20} may be generalized to define mutual translations between (refined) labelled and shallow nested calculi for context-free grammar logics with converse.






\section{Refining Labelled Calculi for Deontic STIT Logics}\label{SECT:Refine-STIT}

In this section, we will refine each deontic \stit calculus $\gtdsn$ (see \fig~\ref{fig:base-Gcalculus} in \sect~\ref{sec:lab-calc-dsn}).  
 Unlike in the previous section, we forgo a detailed analysis of structural rule elimination as such an analysis would proceed along the same lines---observe which permutations of structural rules in $\gtdsn$ cannot be performed, and extract propagation rules that allow for the permutations to go through. Instead, we simply state which propagation rules were obtained from such an analysis. After defining such rules, we show that each calculus $\gtdsn$ can be refined, yielding a deductively equivalent calculus $\dsnl$ that inherits the properties of $\gtdsn$. Last, we show that derivations in $\dsnl$ need only use labelled forest or DAG sequents (see \dfn~\ref{def:tree-DAG-sequent-dsn}), depending on the values of $n,k \in \mathbb{N}$, thus demonstrating that the refined calculi employ simpler structures in deriving theorems relative to the labelled calculi $\gtdsn$.

One of the more interesting observations made in this section concerns the conditions under which refinement can be performed. It will be shown that the structural rules $\refli$, $\eucli$, $\dtwoir$, and $\dthreeir$ (see \fig~\ref{fig:base-Gcalculus} in \sect~\ref{sec:lab-calc-dsn}) are eliminable from any derivation in a calculus $\gtdsn$, given that the calculus is extended with certain conducive rules. However, we will not achieve complete structural rule elimination as we did with the grammar logic 
 calculi since the rules $\ioa$ and $\choicer$ will still be present in our refined labelled calculi, showing that refinement may still be performed in the presence of certain structural rules. In our situation, this holds true because both the $\ioa$ and $\choicer$ rules lack active relational atoms in their conclusion, and so, any structural rule may be permuted above them. This observation suggests that other rules of such a shape would allow for refinement as well.

Due to the presence of $\ioa$ and $\choicer$ in our refined labelled systems $\dsnl$ (shown in \fig~\ref{fig:refined-calculus-dsn}), such systems will not be notational variants of nested systems in general. Even so, if we unbound the number of choices available to an agent (i.e. $k = 0$), or limit ourselves to the use of a single agent (i.e. $n = 0$), then the refined labelled calculi $\dsnlkz$ and $\dsnlnz$ only require labelled DAG proofs and labelled forest proofs (\dfn~\ref{def:tree-DAG-sequent-dsn}), respectively, showing that behind the scenes a reduction in structure is still present. The former class of deontic \stit calculi (\emph{viz.} $\dsnlkz$) can be seen as close relatives of indexed nested sequent calculi~\cite{Fit15,MarStr17}, which generalize the nested sequent notation to encode directed acyclic graphs, while the latter class of calculi (\emph{viz.} $\dsnlnz$) can be viewed as proper nested calculi. 

Another distinguishing feature of our refined calculi for deontic \stit logics is that we replace the use of \cfcst systems in our propagation rules with the use of \emph{undirected paths} (cf.~\cite{Lyo21}). Intuitively, an undirected path within a labelled sequent is a sequence of relational atoms of the form $R_{\agbox}vz$ connecting two labels $w$ and $u$, but where we ignore the `orientation' of the relational atoms. For example, the path $R_{\agbox}wv, R_{\agbox}vu$ moves forward from $w$ to $v$ and then forward from $v$ to $u$, yielding in path from $w$ to $u$, but $R_{\agbox}vw, R_{\agbox}vu$ gives an (undirected) path from $w$ to $u$ as well, since we first move backward from $w$ to $v$, and then forward from $v$ to $u$. We make use of such structures and forgo the use of our grammar theoretic machinery as it simplifies the presentation and application of our propagation rules. Still, such simplicity comes at a cost since we can no longer `plug in' a new \cfcst system to obtain a refined labelled calculus for a new logic (as was done in the previous section). Although this is a shortcoming in a sense, it is actually not a drawback in the current setting since deontic \stit logics were specifically designed to possess certain properties with the goal of modeling multi-agent normative reasoning. Hence, we do not desire to change the properties of our deontic \stit logics, as this would damage their correspondence with the underlying philosophical reasons for having imposed such properties in the first place. Therefore, we are justified in trading modularity for simplicity in the current setting.

Let us now define undirected paths, and afterward, we will discuss and define our propagation rules that make use of such objects. 

\begin{definition}[Undirected $i$-Path]\label{def:i-path-dsn} Let $w \sim_{i} u \in \{R_{\agbox}wu, R_{\agbox}uw\}$ and $\Lambda = \rel \sar \Gamma$. An \emph{undirected $i$-path}\index{Undirected $i$-path} of relational atoms from a label $w$ to $u$ occurs in $\rel$ (and therefore, in $\Lambda$), written $w \sim^{\rel}_{i} u$, \ifandonlyif $w = u$, $w \sim_{i} u$, or there exist labels $v_{j} \in \lab(\Lambda)$ (with $j \in \{1,\ldots,m\}$) such that $w \sim_{i} v_{1}, \ldots, v_{m} \sim_{i} u$ occurs in $\rel$.
\end{definition}

\begin{lemma}\label{lem:uipath-equiv-relation-dsn}
Let $\Lambda := \rel \sar \Gamma$ be a labelled sequent. The undirected $i$-path relation $\uipathrel$ is an equivalence relation over the set of labels $\lab(\Lambda)$.
\end{lemma}

\begin{proof}
Follows from \dfn~\ref{def:i-path-dsn} above.
\end{proof}

\begin{figure}[t]
\noindent\hrule

\begin{center}
\begin{tabular}{c c} 

\AxiomC{ }
\RightLabel{$\id$}
\UnaryInfC{$\rel \sar w:p, w:\negnnf{p}, \Gamma$}
\DisplayProof

&

\AxiomC{$\rel \sar w: \phi, \Gamma$}
\AxiomC{$\rel \sar w: \psi, \Gamma$}
\RightLabel{$\conr$}
\BinaryInfC{$\rel \sar w: \phi \wedge \psi, \Gamma$}
\DisplayProof

\end{tabular}
\end{center}

\begin{center}
\begin{tabular}{c @{\hskip 2em} c}
\AxiomC{$\rel, \opt_{\Oi}u \sar u : \phi, \Gamma$}
\RightLabel{$\Oir^{\dag_{1}}$}
\UnaryInfC{$\rel \sar w : \Oi \phi, \Gamma$}
\DisplayProof

&

\AxiomC{$\rel, R_{[0]}w_{0}u, ..., R_{[n]}w_{n}u \sar \Gamma$}
\RightLabel{$\ioa^{\dag_{1}}$}
\UnaryInfC{$\rel \sar \Gamma$}
\DisplayProof

\end{tabular}
\end{center}

\begin{center}
\begin{tabular}{c c c} 
\AxiomC{$\rel \sar w: \phi, w : \psi, \Gamma$}
\RightLabel{$\disr$}
\UnaryInfC{$\rel \sar w: \phi \vee \psi, \Gamma$}
\DisplayProof

&

\AxiomC{$\rel \sar w: \Diamond \phi, u : \phi, \Gamma$}
\RightLabel{$\diar$}
\UnaryInfC{$\rel \sar w: \Diamond \phi, \Gamma$}
\DisplayProof

&

\AxiomC{$\rel, R_{\agbox}wu \sar u : \phi, \Gamma$}
\RightLabel{$\agboxr^{\dag_{1}}$}
\UnaryInfC{$\rel \sar w: \agbox \phi, \Gamma$}
\DisplayProof
\end{tabular}
\end{center}

\begin{center}
\begin{tabular}{c c} 

\AxiomC{$\rel \sar u : \agdia \phi, v : \phi, \Gamma$}
\RightLabel{$\pragdiar^{\dag_{2}}$\index{$\pragdiar$}}
\UnaryInfC{$\rel \sar u : \agdia \phi, \Gamma$}
\DisplayProof

&

\AxiomC{$\rel, \opt_{\Oi}u  \sar w : \ominus_{i} \phi, v : \phi, \Gamma$}
\RightLabel{$\prODirone^{\dag_{2}}$\index{$\prODirone$}}
\UnaryInfC{$\rel, \opt_{\Oi}u \sar w : \ominus_{i} \phi, \Gamma$}
\DisplayProof
\end{tabular}
\end{center}

\begin{center}
\begin{tabular}{c c c} 
\AxiomC{$\rel \sar u : \phi, \Gamma$}
\RightLabel{$\boxr^{\dag_{1}}$}
\UnaryInfC{$\rel \sar w: \Box \phi, \Gamma$}
\DisplayProof

&

\AxiomC{$\rel, \opt_{\Oi}u \sar w : \ominus_{i} \phi, u : \phi, \Gamma$}
\RightLabel{$\prODirtwo^{\dag_{1}}$\index{$\prODirtwo$}}
\UnaryInfC{$\rel \sar w : \ominus_{i} \phi, \Gamma$}
\DisplayProof
\end{tabular}
\end{center}

\begin{center}
\begin{tabular}{c}
\AxiomC{$\Big\{ \R, R_{\agbox}w_{m}w_{j} \sar \Gamma \ \Big| \ 0 \leq m \leq k-1 \text{, } m+1 \leq j \leq k \Big\}$}
\RightLabel{$\choicer$}
\UnaryInfC{$\R \sar \Gamma$}
\DisplayProof
\end{tabular}
\end{center}

\hrule
\caption{The calculi $\dsnl$\index{$\dsnl$} (with $|\ag| = n + 1$ and $n, k \in \mathbb{N}$). The side condition $\dag_{1}$ on $\boxr$, $\agboxr$, $\Oir$, $\ioa$, and $\prODirtwo$ indicates that $u$ is a eigenvariable, and the side condition $\dag_{2}$ states that $u \sim^{\rel}_{i} v$. Last, we have $\agboxr$, $\pragdiar$, $\Oir$, $\prODirone$, $\prODirtwo$, and $\choicer$ rules for each $i \in \ag$. We stipulate that if $k = 0$, then $\choicer$ is omitted from the calculus.}
\label{fig:refined-calculus-dsn}
\end{figure}

Our refined labelled calculi are displayed in \fig~\ref{fig:refined-calculus-dsn} and employ three new rules: $\pragdiar$, $\prODirone$, and $\prODirtwo$. The propagation rule $\pragdiar$ is obtained from the $\agdiar$, $\refli$, and $\eucli$ rules, the $\prODirone$ rule is obtained from the $\ODir$, $\refli$, $\eucli$, and $\dthreeir$ rules, and last, the $\prODirtwo$ rule is obtained from $\prODirone$ and $\dtwoir$. The $\pragdiar$ and $\prODirone$ rules make use of undirected $i$-paths in their side conditions, which arise naturally when analyzing the permutation of $\refli$ and $\eucli$ upwards in a $\gtdsn$ derivation. Moreover, undirected $i$-paths can be seen as syntactic encodings of the \partcond condition imposed on $\dsn$-frames (\dfn~\ref{def:frames-models-dsn}), which forces the relations $\{R_{\agbox} \ | \ i \in \ag\}$ to be equivalence classes, or (equivalently) to satisfy the properties of reflexivity and Euclideanity. Since we will repeatedly make reference to the $\pragdiar$, $\prODirone$, and $\prODirtwo$ rules while refining our calculi, we collect them into a set defined as follows:

\begin{definition}[The Set $\prset$] 
 Let $\ag = \{0, \ldots, n\} \subseteq \mathbb{N}$ be our set of agents. We define our set of conducive rules as follows:
$$
\prset := \{\pragdiar, \prODirone, \prODirtwo \ | \ i \in \ag\}
$$
\end{definition}

As with the propagation rules we saw in the previous section, the side condition imposed on $\pragdiar$ and $\prODirone$ can be determined whether we are applying the rules top-down or bottom-up. Regarding the $\pragdiar$ rule, this follows from the fact that the side condition only requires data from $\rel$ and the principal formula $u : \agdia \phi$ (see \fig~\ref{fig:refined-calculus-dsn}) to determine the rules (in)applicability, and all such data occurs in both the premise and conclusion---the same holds for the $\prODirone$ rule. This feature renders our propagation rules suitable for proof-search methods which apply inference rules in reverse. To provide more intuition regarding the functionality of $\pragdiar$ and $\prODirone$ we provide an example below, showing how the rules may be applied bottom-up as would be done during proof-search.

\begin{example} We examine applications of propagation rules to the labelled sequent:
$$
\Lambda := \opt_{\otimes_{1}} u, \opt_{\otimes_{2}} v, R_{[2]}wv, R_{[2]}uv, R_{[1]}uz \sar w : \ominus_{1} q, u : \langle 2 \rangle p
$$
whose sequent graph $\seqgraph(\Lambda)$ (\dfn~\ref{def:sequent-graph-kms}) is shown below:
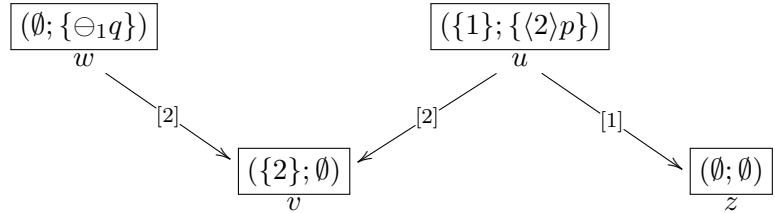
\begin{center}
\begin{tabular}{c}
\xymatrix{
\overset{\boxed{(\emptyset ; \{\ominus_{1} q\})}}{w} \ar[dr]|-{[2]} & & \overset{\boxed{(\{1\} ; \{\langle 2 \rangle p\})}}{u} \ar[dl]|-{[2]}\ar[dr]|-{[1]} & \\
 & \overset{\boxed{(\{2\} ; \emptyset)}}{v} & &
\overset{\boxed{(\emptyset ; \emptyset)}}{z}\\
}
\end{tabular}
\end{center}
Due to the undirected $2$-path $R_{[2]}uv$, the $(Pr_{\langle 2 \rangle})$ rule may be applied (bottom-up) to propagate the propositional variable $p$ to $v$, effectively adding $v : p$ to our labelled sequent $\Lambda$. In addition, we may make use of the undirected $2$ path $R_{[2]}uv, R_{[2]}wv$ from $u$ to $w$ to (bottom-up) apply $(Pr_{\langle 2 \rangle})$ another time and propagate the propositional atom $p$ to $w$, i.e. adding $w : p$ to our labelled sequent $\Lambda$. Due to the occurrence of $\opt_{\otimes_{1}} u$ and $w : \ominus_{1} q$ in $\Lambda$, and because there is an undirected $1$-path from $u$ to itself, and an undirected $1$-path $R_{[1]}uz$ from $u$ to $z$, we may apply $(Pr_{\ominus_{1}}^{1})$ (bottom-up) two times to add a $q$ at $u$ and a $q$ at $z$, effectively adding $u : q, z : q$ to $\Lambda$. After performing these four inferences, we obtain the following labelled sequent:
$$
\Lambda' := \opt_{\otimes_{1}} u, \opt_{\otimes_{2}} v, R_{[2]}wv, R_{[2]}uv, R_{[1]}uz \sar w : \ominus_{1} q, u : \langle 2 \rangle p, w : p, v : p, u : q, z : q
$$
whose sequent graph $\seqgraph(\Lambda')$ is as follows:
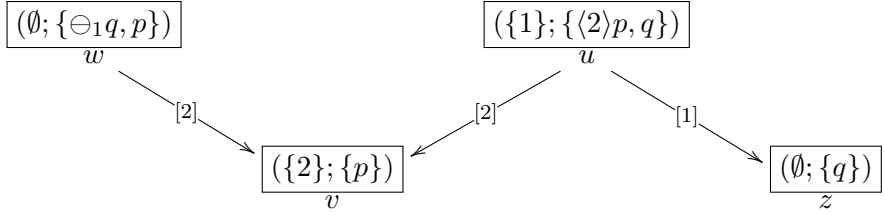
\begin{center}
\begin{tabular}{c}
\xymatrix{
\overset{\boxed{(\emptyset ; \{\ominus_{1} q, p\})}}{w} \ar[dr]|-{[2]} & & \overset{\boxed{(\{1\} ; \{\langle 2 \rangle p, q\})}}{u} \ar[dl]|-{[2]}\ar[dr]|-{[1]} & \\
 & \overset{\boxed{(\{2\} ; \{p\})}}{v} & &
\overset{\boxed{(\emptyset ; \{q\})}}{z}\\
}
\end{tabular}
\end{center}
\end{example}

We now move on to showing that $\refli$, $\eucli$, $\dtwoir$, and $\dthreeir$ are eliminable in $\gtdsn + \prset$. As in the last section, when proving a rule eliminable, we need not consider its permutation above the other structural rules we aim to eliminate; e.g. when considering $\refli$ elimination, we do not consider permuting the rule above $\eucli$, $\dtwoir$, or $\dthreeir$. The reason being, if we consider any derivation in $\gtdsn + \prset$, we can always successively eliminate topmost occurrences of $\refli$, $\transi$, $\dtwoir$, and $\dthreeir$, yielding a proof free of such inferences and without any of the rules ever being permuted above each other in the process. We first show that our new rules $\pragdiar$ and $\prODirone$ subsume $\agdiar$ and $\ODir$, respectively, as this will let us simplify proofs of subsequent results:

\begin{lemma}\label{lem:dia-is-propagation-dsn}
The $\agdiar$ and $\ODir$ rules are instances of the $\pragdiar$ and $\prODirone$ rules, respectively.
\end{lemma}

\begin{proof} Observe that the premise of the $\pragdiar$ inference below contains the relational atom $R_{[i]}wu$. Hence, there exists an undirected $i$-path from $w$ to $u$, meaning that the side condition of $\pragdiar$ holds. Regarding the $\prODirone$ inference, let $\rel' := \rel, \opt_{\Oi}u$. The application of the rule is valid as well since $u \uipath^{\rel'} u$ by \dfn~\ref{def:i-path-dsn}, that is, the side condition holds.
\begin{center}
\begin{tabular}{c c}
\AxiomC{$\rel, R_{[i]}wu \sar w :  \agdia \phi, u : \phi, \Gamma$}
\RightLabel{$\pragdiar$}
\UnaryInfC{$\rel, R_{[i]}wu \sar w :  \agdia \phi, \Gamma$}
\DisplayProof

&

\AxiomC{$\rel, \opt_{\Oi}u \sar w : \ominus_{i} \phi, u : \phi, \Gamma$}
\RightLabel{$\prODirone$}
\UnaryInfC{$\rel, \opt_{\Oi}u \sar w : \ominus_{i} \phi, \Gamma$}
\DisplayProof
\end{tabular}
\end{center}
\end{proof}

\begin{lemma}\label{lem:refli-elim-dsn}
For all $i \in \ag$, the $\refli$ rule is eliminable in the calculus $\gtdsn + \prset - \{\eucli, \dtwoir, \dthreeir \ | \ i \in \ag\}$.
\end{lemma}

\begin{proof} We prove the claim by induction on the height of the given derivation.

\textit{Base case.} Suppose that we have an instance of $\id$ followed by an instance of $\refli$ as shown below left. Then, the conclusion is an instance of $\id$ as shown below right. Also, note that $w$ may or may not be identical to the label $u$.
\begin{center}
\begin{tabular}{c c c}
\AxiomC{}
\RightLabel{$\id$}
\UnaryInfC{$\rel, R_{\agbox}uu \sar w : p, w : \negnnf{p},\Gamma$}
\RightLabel{$\refli$}
\UnaryInfC{$\rel, \sar w : p, w : \negnnf{p},\Gamma$}
\DisplayProof

&

$\leadsto$

&

\AxiomC{}
\RightLabel{$\id$}
\UnaryInfC{$\rel \sar w : p, w : \negnnf{p},\Gamma$}
\DisplayProof
\end{tabular}
\end{center}

\textit{Inductive step.} 
 By \lem~\ref{lem:dia-is-propagation-dsn}, we need not consider permutations above the $\agdiar$ or $\ODir$ rules. With the exception of $\pragdiar$ and $\prODirone$, the rule $\refli$ freely permutes above all other rules in our calculus. Below, we consider the $\pragdiar$ case as the $\prODirone$ case is similar, and suppose our derivation ends with the following:
\begin{center}
\begin{tabular}{c}
\AxiomC{$\rel, R_{\agbox}vv \sar w : \agdia \phi, u : \phi,\Gamma$}
\RightLabel{$\pragdiar$}
\UnaryInfC{$\rel, R_{\agbox}vv \sar w : \agdia \phi, \Gamma$}
\RightLabel{$\refli$}
\UnaryInfC{$\rel \sar w : \agdia \phi, \Gamma$}
\DisplayProof
\end{tabular}
\end{center}
Let $\rel' := \rel, R_{\agbox}vv$. Since the $\pragdiar$ rule was applied, we know that there exists an undirected $i$-path $w \uipath^{\rel'} u$. We have two cases to consider: either $R_{\agbox}vv$ occurs in the undirected $i$-path, or it does not. In the latter case, we may freely permute the two rules, so we focus on the former case. 

Suppose that $R_{\agbox}vv$ occurs in $w \uipath^{\rel'} u$. Observe that by deleting each occurrence of $R_{\agbox}vv$ in $w \uipath^{\rel'} u$, we obtain an alternative undirected $i$-path $w \uipath^{\rel} u$ that does not rely on $R_{\agbox}vv$. Therefore, we may apply $\refli$ to the top sequent above, and use this path to satisfy the side condition, thus allowing for the two rules to be permuted.
\end{proof}

\begin{lemma}\label{lem:eucli-elim-dsn}
For all $i \in \ag$, the $\eucli$ rule is eliminable in the calculus $\gtdsn + \prset - \{\refli, \dtwoir, \dthreeir \ | \ i \in \ag\}$.
\end{lemma}

\begin{proof} We prove the claim by induction on the height of the given derivation. 

\textit{Base case.} Any application of $\eucli$ to an instance of $\id$ yields another instance of $\id$, which resolves the base case.

\textit{Inductive step.} By \lem~\ref{lem:dia-is-propagation-dsn}, we need not consider permutations above the $\agdiar$ or $\ODir$ rules. With the exception of $\pragdiar$ and $\prODirone$, the $\eucli$ rule freely permutes above all other rules in our calculus. We consider only the $\prODirone$ case below as the $\pragdiar$ case is similar, and assume our derivation ends with the following:
\begin{center}
\begin{tabular}{c}
\AxiomC{$\rel, R_{\agbox}wu, R_{\agbox}wv, R_{\agbox}uv, \ideal y \sar  z : \ODi \phi, x : \phi, \Gamma$}
\RightLabel{$\prODirone$}
\UnaryInfC{$\rel, R_{\agbox}wu, R_{\agbox}wv, R_{\agbox}uv, \ideal y \sar  z : \ODi \phi, \Gamma$}
\RightLabel{$\eucli$}
\UnaryInfC{$\rel, R_{\agbox}wu, R_{\agbox}wv, \ideal y \sar  z : \ODi \phi, \Gamma$}
\DisplayProof
\end{tabular}
\end{center}
Let $\rel' := \rel, R_{\agbox}wu, R_{\agbox}wv, R_{\agbox}uv, \ideal y$. Since $\prODirone$ was applied, we know that the side condition holds, implying that there exists an undirected $i$-path $y \uipath^{\rel'} x$ in the premise of the inference. There are two possible cases, either $R_{\agbox}uv$ occurs in the aforementioned $i$-path, or it does not. In the latter case, the two rules may be freely permuted, so we focus on the former case. We replace each occurrence of $R_{\agbox}uv$ in $y \uipath^{\rel'} x$ with $R_{\agbox}wu, R_{\agbox}wv$. Observe that this yields an undirected $i$-path from $y$ to $x$ that does not rely on $R_{\agbox}uv$. Using this path, we may apply $\eucli$ first, and then $\prODirone$, since the side condition will be satisfied.
\end{proof}

\begin{lemma}\label{lem:dtwoi-elim-dsn}
For all $i \in \ag$, the $\dtwoir$ rule is eliminable in the calculus $\gtdsn + \prset - \{\refli, \eucli, \dthreeir \ | \ i \in \ag\}$.
\end{lemma}

\begin{proof} We prove the result by induction on the height of the given derivation.

\textit{Base case.} The base case follows from the fact that any application of $\dtwoir$ to an instance of $\id$ begets another instance of $\id$.

\textit{Inductive step.} By \lem~\ref{lem:dia-is-propagation-dsn}, we need not consider permutations above the $\agdiar$ or $\ODir$ rules. With the exception of the $\prODirone$ rule, $\dtwoir$ permutes above all other rules in our calculus. We show how to resolve the non-trivial case of permuting $\dtwoir$ above $\prODirone$ below:
\begin{center}
\begin{tabular}{c c c}
\AxiomC{$\rel, \ideal u \sar w : \ODi \phi, u : \phi, \Gamma$}
\RightLabel{$\prODirone$}
\UnaryInfC{$\rel, \ideal u \sar w : \ODi \phi, \Gamma$}
\RightLabel{$\dtwoir$}
\UnaryInfC{$\rel \sar w : \ODi \phi, \Gamma$}
\DisplayProof

&

$\leadsto$

&

\AxiomC{$\rel, \ideal u \sar w : \ODi \phi, u : \phi, \Gamma$}
\RightLabel{$\prODirtwo$}
\UnaryInfC{$\rel \sar w : \ODi \phi, \Gamma$}
\DisplayProof
\end{tabular}
\end{center}
\end{proof}

\begin{lemma}\label{lem:dthreei-elim-dsn}
For all $i \in \ag$, the $\dthreeir$ rule is eliminable in the calculus $\gtdsn + \prset - \{\refli, \eucli, \dtwoir \ | \ i \in \ag\}$.
\end{lemma}

\begin{proof} We prove the result by induction on the height of the given derivation.

\textit{Base case.} The base case is resolved by observing that any application of $\dthreeir$ to $\id$ gives another instance of $\id$.

\textit{Inductive step.} By \lem~\ref{lem:dia-is-propagation-dsn}, we need not consider permutations above the $\agdiar$ or $\ODir$ rules. With the exception of $\prODirone$, the $\dthreeir$ rule permutes above all other rules in our calculus. We show how to resolve the case of permuting $\dthreeir$ above $\prODirone$, and assume our derivation ends with the derivation shown top left.
\begin{flushleft}
\begin{tabular}{c c}
\AxiomC{$\rel, R_{\agbox}wu, \ideal w, \ideal u \sar v : \ODi \phi, z : \phi, \Gamma$}
\RightLabel{$\prODirone$}
\UnaryInfC{$\rel, R_{\agbox}wu, \ideal w, \ideal u \sar v : \ODi \phi, \Gamma$}
\RightLabel{$\dthreeir$}
\UnaryInfC{$\rel, R_{\agbox}wu, \ideal w \sar v : \ODi \phi, \Gamma$}
\DisplayProof

&

$\leadsto$
\end{tabular}
\end{flushleft}
\begin{flushright}
\AxiomC{$\rel, R_{\agbox}wu, \ideal w, \ideal u \sar v : \ODi \phi, z : \phi, \Gamma$}
\RightLabel{$\dthreeir$}
\UnaryInfC{$\rel, R_{\agbox}wu, \ideal w \sar v : \ODi \phi, z : \phi, \Gamma$}
\RightLabel{$\prODirone$}
\UnaryInfC{$\rel, R_{\agbox}wu, \ideal w \sar v : \ODi \phi, \Gamma$}
\DisplayProof
\end{flushright}
Let $\rel' := \rel, R_{\agbox}wu$. We have two cases to consider: either $\ideal u$ is active in the $\prODirone$ inference, or it is not. In the latter case, the two rules may be freely permuted. In the former case, if $\ideal u$ is active in the $\prODirone$ inference (in the above left derivation), then we know that there is an undirected $i$-path $u \sim^{\rel'}_{i} z$ due to the side condition of $\prODirone$. Observe that the side condition of $\prODirone$ continues to hold after $\dthreeir$ is applied since $\ideal w$ occurs in the multiset of relational atoms (as can be seen in the lower right derivation above) and by combining $R_{\agbox}wu$ with the undirected $i$-path $u \sim^{\rel'}_{i} z$, we obtain an undirected $i$-path $w \sim^{\rel'}_{i} z$.
\end{proof}

\begin{theorem}\label{thm:gtdsn-to-dsnl}
Every derivation in $\gtdsn$ can be algorithmically transformed into a derivation in $\dsnl$.
\end{theorem}

\begin{proof} The result follows from \lem~\ref{lem:dia-is-propagation-dsn} -- \ref{lem:dthreei-elim-dsn} above.
\end{proof}

At this stage, we could opt to show that our refined labelled calculi possess proof-theoretic properties (e.g. hp-admissibility of $\ctrr$, hp-invertibility of rules, or syntactic $\cut$ elimination) by carrying out proofs as we did for all $\gtdsn$ calculi. We opt for a different method of proof however, and instead, establish a correspondence between each $\dsnl$ and $\gtdsn$ calculus as this uncovers the relationship between derivations in the different settings. We have already shown that every derivation in a calculus $\gtdsn$ can be transformed into a derivation in $\dsnl$ (\thm~\ref{thm:gtdsn-to-dsnl} above), so to establish our correspondence, we show that proof transformations can be performed in the opposite direction as well. We will then leverage these proof transformations to show that each $\dsnl$ calculus inherits favorable proof-theoretic properties from its associated $\gtdsn$ calculus (\cor~\ref{cor:admissible-strucset-refined-dsn}--\ref{cor:soundness-completeness-refined-dsn}). We first prove a valuable lemma (\lem~\ref{lem:delete-relations-dsn} below), which states that if an undirected $i$-path $w \uipathrel u$ occurs in the relational atoms $\rel$ of a derivable labelled sequent $\rel, R_{\agbox}wu \sar \Gamma$, then the relational atom $R_{\agbox}wu$ is superfluous, and so, the labelled sequent $\rel \sar \Gamma$ may be derived on its own. After proving this lemma, we provide an algorithm showing that proofs in $\dsnl$ can be transformed into proofs in $\gtdsn$.

\begin{lemma}\label{lem:delete-relations-dsn}
If $\rel, R_{\agbox}wu \sar \Gamma$ is derivable in $\gtdsn$ and $w \uipathrel u$, then $\rel \sar \Gamma$ is derivable in $\gtdsn$.
\end{lemma}

\begin{proof} We prove the claim by induction on the length of the undirected $i$-path between $w$ and $u$ in $\rel$, and let $\Pi$ be the derivation of $\rel, R_{\agbox}wu \sar \Gamma$ in $\gtdsn$.

\textit{Base case.} If the undirected $i$-path between $w$ and $u$ is of length $0$, then $w = u$, and the case is resolved as shown below left. If the undirected $i$-path between $w$ and $u$ is of length $1$, then the case is resolved as shown below right and makes use of the hp-admissibility of $\ctrrel$ (\lem~\ref{lem:ctrr-admiss-dsn}). Note that in the second case, $\rel = \rel', R_{\agbox}wu$.
\begin{center}
\begin{tabular}{c c}
\AxiomC{$\Pi$} 
\UnaryInfC{$\rel, R_{\agbox}ww \sar \Gamma$}
\RightLabel{$\refli$}
\UnaryInfC{$\rel \sar \Gamma$}
\DisplayProof

&

\AxiomC{$\Pi$} 
\UnaryInfC{$\rel, R_{\agbox}wu \sar \Gamma$}
\RightLabel{=}
\UnaryInfC{$\rel', R_{\agbox}wu, R_{\agbox}wu \sar \Gamma$}
\RightLabel{$\ctrrel$}
\dashedLine
\UnaryInfC{$\rel', R_{\agbox}wu \sar \Gamma$}
\RightLabel{$=$}
\dottedLine
\UnaryInfC{$\rel \sar \Gamma$}
\DisplayProof
\end{tabular}
\end{center}

\textit{Inductive step.} We assume that the undirected $i$-path $w \uipathrel u$ between $w$ and $u$ is of length $m + 1$. Therefore, there exists a $v$ such that there is an undirected $i$-path $w \uipathrel v$ between $w$ and $v$ of length $m$ and an undirected $i$-path between $v$ and $u$ of length $1$. Since the length of the undirected $i$-path $w \uipathrel v$ between $w$ and $v$ is $m$, we may invoke \ih (in both cases) below to delete the relational atom $R_{\agbox}wv$. Additionally, since there is an undirected $i$-path between $v$ and $u$ of length $1$, either $\rel = \rel', R_{\agbox}uv$ or $\rel = \rel', R_{\agbox}vu$ by Def.~\ref{def:i-path-dsn}. The first case is resolved as shown below top, and the second case is resolved as shown below bottom. Also, we indicate active relational atoms in the premise of a $\eucli$ inference with an asterisk $*$ to improve comprehensibility.
\begin{center}
\begin{tabular}{c}
\AxiomC{$\Pi$} 
\UnaryInfC{$\rel', R_{\agbox}uv, R_{\agbox}wu, \sar \Gamma$}
\RightLabel{$\wk$}
\dashedLine
\UnaryInfC{$\rel', R_{\agbox}uv, R_{\agbox}wv, R_{\agbox}ww, R_{\agbox}^{*}uu, R_{\agbox}vv, R_{\agbox}vw, R_{\agbox}vu, R_{\agbox}^{*}uw, R_{\agbox}^{*}wu \sar \Gamma$}
\RightLabel{$\eucli$}
\UnaryInfC{$\rel', R_{\agbox}uv, R_{\agbox}wv, R_{\agbox}ww, R_{\agbox}uu, R_{\agbox}vv, R_{\agbox}^{*}vw, R_{\agbox}^{*}vu, R_{\agbox}^{*}uw \sar \Gamma$}
\RightLabel{$\eucli$}
\UnaryInfC{$\rel', R_{\agbox}^{*}uv, R_{\agbox}wv, R_{\agbox}ww, R_{\agbox}^{*}uu, R_{\agbox}vv, R_{\agbox}vw, R_{\agbox}^{*}vu \sar \Gamma$}
\RightLabel{$\eucli$}
\UnaryInfC{$\rel', R_{\agbox}uv, R_{\agbox}^{*}wv, R_{\agbox}^{*}ww, R_{\agbox}uu, R_{\agbox}vv, R_{\agbox}^{*}vw \sar \Gamma$}
\RightLabel{$\eucli$}
\UnaryInfC{$\rel', R_{\agbox}uv, R_{\agbox}wv, R_{\agbox}ww, R_{\agbox}uu, R_{\agbox}vv \sar \Gamma$}
\RightLabel{$\refli \times 3$}
\UnaryInfC{$\rel', R_{\agbox}uv, R_{\agbox}wv \sar \Gamma$}
\RightLabel{IH}
\dashedLine
\UnaryInfC{$\rel', R_{\agbox}uv \sar \Gamma$}
\RightLabel{=}
\dottedLine
\UnaryInfC{$\rel \sar \Gamma$}
\DisplayProof
\end{tabular}
\end{center}
\begin{center}
\begin{tabular}{c}
\AxiomC{$\Pi$}
\UnaryInfC{$\rel', R_{\agbox}vu, R_{\agbox}wu \sar \Gamma$}
\RightLabel{$\wk$}
\dashedLine
\UnaryInfC{$\rel', R_{\agbox}^{*}vu, R_{\agbox}^{*}vw, R_{\agbox}^{*}wu,  R_{\agbox}wv,  R_{\agbox}ww \sar \Gamma$}
\RightLabel{$\eucli$}
\UnaryInfC{$\rel', R_{\agbox}vu, R_{\agbox}^{*}vw,  R_{\agbox}^{*}wv,  R_{\agbox}^{*}ww \sar \Gamma$}
\RightLabel{$\eucli$}
\UnaryInfC{$\rel', R_{\agbox}vu,  R_{\agbox}wv,  R_{\agbox}ww \sar \Gamma$}
\RightLabel{$\refli$}
\UnaryInfC{$\rel', R_{\agbox}vu,  R_{\agbox}wv \sar \Gamma$}
\RightLabel{IH}
\dashedLine
\UnaryInfC{$\rel', R_{\agbox}vu \sar \Gamma$}
\RightLabel{=}
\dottedLine
\UnaryInfC{$\rel \sar \Gamma$}
\DisplayProof
\end{tabular}
\end{center}
\end{proof}

\begin{theorem}\label{thm:dsnl-to-gtdsn}
Every derivation in $\dsnl + \strucsetdsn$ can be algorithmically transformed into a derivation in $\gtdsn$.
\end{theorem}

\begin{proof} We prove the result by induction on the height of the given derivation in $\dsnl + \strucsetdsn$.

\textit{Base case.} Any instance of $\id$ in $\dsnl + \strucsetdsn$ is an instance of $\id$ in $\gtdsn$, which resolves the base case.

\textit{Inductive step.} We prove the inductive step by considering the last rule applied in $\dsnl + \strucsetdsn$. All rules in $\strucsetdsn$ are handled by applying \ih followed by the hp-admissibility or eliminability of the rule (\lem~\ref{lem:lsub-admiss-dsn}, \ref{lem:wk-admiss-dsn}, \ref{lem:ctrr-admiss-dsn}, and  \thm~\ref{lem:cut-admiss-dsn}). With the exception of the $\pragdiar$, $\prODirone$, and $\prODirtwo$ rules, all other cases are easily resolved since each is a rule in $\gtdsn$. We show the $\pragdiar$, $\prODirone$, and $\prODirtwo$ cases below, and note that in the $\pragdiar$ case the side condition $w \uipathrel u$ holds, and in the $\prODirone$ case the side condition $u \uipathrel v$ holds, which allows for the invocation of \lem~\ref{lem:delete-relations-dsn}.
\begin{center}
\begin{tabular}{c c c}
\AxiomC{$\rel \sar w : \agdia \phi, u : \phi, \Gamma$}
\RightLabel{$\pragdiar$}
\UnaryInfC{$\rel \sar w : \agdia \phi, \Gamma$}
\DisplayProof

&

$\leadsto$

&

\AxiomC{ } 
\RightLabel{\ih}
\dashedLine
\UnaryInfC{$\rel \sar w : \agdia \phi, u : \phi, \Gamma$}
\RightLabel{$\wk$}
\dashedLine
\UnaryInfC{$\rel, R_{[i]}wu \sar w : \agdia \phi, u : \phi, \Gamma$}
\RightLabel{$\agdiar$}
\UnaryInfC{$\rel, R_{[i]}wu \sar w : \agdia \phi, \Gamma$}
\RightLabel{Lem.~\ref{lem:delete-relations-dsn}}
\dashedLine
\UnaryInfC{$\rel \sar w : \agdia \phi, \Gamma$}
\DisplayProof
\end{tabular}
\end{center}

\begin{flushleft}
\begin{tabular}{c c}
\AxiomC{$\rel, \ideal u \sar w : \ODi \phi, v : \phi, \Gamma$}
\RightLabel{$\prODirone$}
\UnaryInfC{$\rel, \ideal u \sar w : \ODi \phi, \Gamma$}
\DisplayProof

&

$\leadsto$
\end{tabular}
\end{flushleft}
\begin{flushright}
\AxiomC{ } 
\RightLabel{\ih}
\dashedLine
\UnaryInfC{$\rel, \ideal u \sar w : \ODi \phi, v : \phi, \Gamma$}
\RightLabel{$\wk$}
\dashedLine
\UnaryInfC{$\rel,  R_{\agbox}uv, \ideal u, \ideal v \sar w : \ODi \phi, v : \phi, \Gamma$}
\RightLabel{$\ODir$}
\UnaryInfC{$\rel,  R_{\agbox}uv, \ideal u, \ideal v \sar w : \ODi \phi, \Gamma$}
\RightLabel{$\dthreeir$}
\UnaryInfC{$\rel, R_{\agbox}uv, \ideal u \sar w : \ODi \phi, \Gamma$}
\RightLabel{Lem.~\ref{lem:delete-relations-dsn}}
\dashedLine
\UnaryInfC{$\rel, \ideal u \sar w : \ODi \phi, \Gamma$}
\DisplayProof
\end{flushright}

\begin{center}
\begin{tabular}{c c c}
\AxiomC{$\rel, \ideal u \sar w : \ODi \phi, u : \phi, \Gamma$}
\RightLabel{$\prODirtwo$}
\UnaryInfC{$\rel \sar w : \ODi \phi, \Gamma$}
\DisplayProof

&

$\leadsto$

&

\AxiomC{ } 
\RightLabel{\ih}
\dashedLine
\UnaryInfC{$\rel, \ideal u \sar w : \ODi \phi, u : \phi, \Gamma$}
\RightLabel{$\ODir$}
\dashedLine
\UnaryInfC{$\rel, \ideal u \sar w : \ODi \phi, \Gamma$}
\RightLabel{$\dtwoir$}
\UnaryInfC{$\rel \sar w : \ODi \phi, \Gamma$}
\DisplayProof
\end{tabular}
\end{center}
\end{proof}

\begin{example} The top derivation below is in $\dsnl$ and the bottom derivation below is in $\gtdsn$. The first $\pragdiar$ instance in the top derivation translates to the first $\agdiar$ instance and $\eucli$ instance in the bottom derivation, the second $\pragdiar$ inference in the top derivation translates to the second $\agdiar$ inference along with the $\refli$ instance in the bottom derivation, and the $\prODirtwo$ instance translates to the $\ODir$ and $\dtwoir$ instances in the bottom derivation. To transform the bottom derivation into the top derivation, the $\eucli$, $\refli$, and $\dtwoir$ inferences are eliminated as explained in \lem~\ref{lem:eucli-elim-dsn}, \ref{lem:refli-elim-dsn}, and \ref{lem:dtwoi-elim-dsn}, respectively. Last, we let $\phi :=  \agbox \negnnf{p} \lor \agbox \agdia \agdia p$ to save space.

\begin{center}
\AxiomC{}
\RightLabel{$\id$}
\UnaryInfC{$\ideal w, R_{\agbox}wu, R_{\agbox}wv \sar z : \ODi \phi, v : \agdia \agdia p, v : \agdia p, u : p, u : \negnnf{p}$}
\RightLabel{$\pragdiar$}
\UnaryInfC{$\ideal w, R_{\agbox}wu, R_{\agbox}wv \sar z : \ODi \phi, v : \agdia \agdia p, v : \agdia p, u : \negnnf{p}$}
\RightLabel{$\pragdiar$}
\UnaryInfC{$\ideal w, R_{\agbox}wu, R_{\agbox}wv \sar z : \ODi \phi, v : \agdia \agdia p, u : \negnnf{p}$}
\RightLabel{$\agboxr \times 2$}
\UnaryInfC{$\ideal w \sar z : \ODi \phi, w : \agbox \negnnf{p}, w : \agbox \agdia \agdia p$}
\RightLabel{$\disr$}
\UnaryInfC{$\ideal w  \sar z : \ODi \phi, w : \agbox \negnnf{p} \lor \agbox \agdia \agdia p$}
\RightLabel{=}
\dottedLine
\UnaryInfC{$\ideal w  \sar z : \ODi (\agbox \negnnf{p} \lor \agbox \agdia \agdia p), w : \agbox \negnnf{p} \lor \agbox \agdia \agdia p$}
\RightLabel{$\prODirtwo$}
\UnaryInfC{$\seqempstr \sar z : \ODi (\agbox \negnnf{p} \lor \agbox \agdia \agdia p)$}
\DisplayProof
\end{center}

\begin{center}
\AxiomC{}
\RightLabel{$\id$}
\UnaryInfC{$\ideal w, R_{\agbox}wu, R_{\agbox}wv, R_{\agbox}vu, R_{\agbox}vv \sar z : \ODi \phi, v : \agdia \agdia p, v : \agdia p, u : p, u : \negnnf{p}$}
\RightLabel{$\agdiar$}
\UnaryInfC{$\ideal w, R_{\agbox}wu, R_{\agbox}wv, R_{\agbox}vu, R_{\agbox}vv \sar z : \ODi \phi, v : \agdia \agdia p, v : \agdia p, u : \negnnf{p}$}
\RightLabel{$\eucli$}
\UnaryInfC{$\ideal w, R_{\agbox}wu, R_{\agbox}wv, R_{\agbox}vv \sar z : \ODi \phi, v : \agdia \agdia p, v : \agdia p, u : \negnnf{p}$}
\RightLabel{$\agdiar$}
\UnaryInfC{$\ideal w, R_{\agbox}wu, R_{\agbox}wv, R_{\agbox}vv \sar z : \ODi \phi, v : \agdia \agdia p, u : \negnnf{p}$}
\RightLabel{$\refli$}
\UnaryInfC{$\ideal w, R_{\agbox}wu, R_{\agbox}wv \sar z : \ODi \phi, v : \agdia \agdia p, u : \negnnf{p}$}
\RightLabel{$\agboxr \times 2$}
\UnaryInfC{$\ideal w \sar z : \ODi \phi, w : \agbox \negnnf{p}, w : \agbox \agdia \agdia p$}
\RightLabel{$\disr$}
\UnaryInfC{$\ideal w  \sar z : \ODi \phi, w : \agbox \negnnf{p} \lor \agbox \agdia \agdia p$}
\RightLabel{=}
\dottedLine
\UnaryInfC{$\ideal w  \sar z : \ODi (\agbox \negnnf{p} \lor \agbox \agdia \agdia p), w : \agbox \negnnf{p} \lor \agbox \agdia \agdia p$}
\RightLabel{$\ODir$}
\UnaryInfC{$\ideal w  \sar z : \ODi (\agbox \negnnf{p} \lor \agbox \agdia \agdia p)$}
\RightLabel{$\dtwoir$}
\UnaryInfC{$\seqempstr \sar z : \ODi (\agbox \negnnf{p} \lor \agbox \agdia \agdia p)$}
\DisplayProof
\end{center}

\end{example}

We may now leverage our proof-theoretic transformations to show that our refined labelled calculi possess admissibility and invertibility properties, and in addition, are sound and complete.

\begin{corollary}\label{cor:admissible-strucset-refined-dsn}
The rules in $\strucsetdsn$ are admissible in $\dsnl$.
\end{corollary}

\begin{proof} Follows from \thm~\ref{thm:dsnl-to-gtdsn} and \thm~\ref{thm:gtdsn-to-dsnl}.
\end{proof}

\begin{corollary}\label{cor:invertible-rules-refined-dsn}
All rules of $\dsnl$ are invertible.
\end{corollary}

\begin{proof} Follows from \thm~\ref{thm:dsnl-to-gtdsn} and \thm~\ref{thm:gtdsn-to-dsnl}, as well as \lem~\ref{lem:invert-dsn}.
\end{proof}

\begin{corollary}[Soundness and Completeness of $\dsnl$]\label{cor:soundness-completeness-refined-dsn} Let $n,k \in \mathbb{N}$. Then, \\
(i) If $\vdash_{\dsnl} \Lambda$, then $\models_{\dsn} \Lambda$.\\
(ii) If $\vdash_{\dsn} \phi$, then $\vdash_{\dsnl} \seqempstr \sar w : \phi$.
\end{corollary}

\begin{proof} Follows from \thm~\ref{thm:soundness-gtdsn}, \ref{thm:completness-gtdsn}, \ref{thm:gtdsn-to-dsnl}, and \thm~\ref{thm:dsnl-to-gtdsn}.
\end{proof}

Before concluding the section, we discuss the reduction in sequential structure that arises through refinement, and recall that each $\gtdsn$ calculus is incomplete relative to labelled DAG and forest derivations (\thm~\ref{thm:sequent-structure-gtdsn}). If we consider the calculi in the class $\{\dsnlkz \ | \ n \in \mathbb{N}\}$, where the maximum number of choices is unrestricted (i.e. $k = 0$), then it is relatively straightforward to show that any derivation of a labelled theorem $w : \phi$ in such a calculus is a labelled DAG derivation. The reason why derivations for such calculi require labelled DAG sequents, as opposed to more minimalistic structures (e.g. labelled forest sequents), is due to the presence of the $\ioa$ rule, which introduces relational atoms that `converge' to the same point when applied bottom-up. This is illustrated in the graphic below, which shows the kind of structure that would appear in the sequent graph of a labelled sequent if $\ioa$ was applied in reverse, introducing the relational atoms $R_{[0]}w_{0}u, R_{[1]}w_{1}u, \ldots, R_{[n-1]}w_{n-1}u, R_{[n]}w_{n}u$.

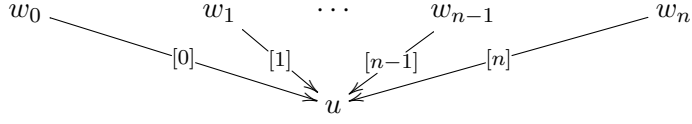
\begin{center}
\begin{tabular}{c}
\xymatrix{
w_{0}\ar[drrr]|-{[0]} & & w_{1}\ar[dr]|-{[1]} &  \cdots & w_{n-1}\ar[dl]|-{[n-1]} & & w_{n}\ar[dlll]|-{[n]} \\
& & & u & & & 
}
\end{tabular}
\end{center}


Therefore, as it currently stands, each refined labelled calculus $\dsnlkz$ (with $n > 0$) appears to require labelled DAG derivations as the $\ioa$ rule is vital for completeness.

\begin{theorem}\label{thm:DAG-proofs-dsn}
Let $n \in \mathbb{N}$ and $k = 0$. Every derivation in $\dsnlkz$ of a labelled formula $w : \phi$ is a labelled DAG derivation with the rooted property.
\end{theorem}

\begin{proof} First, we recall that if $k = 0$, then $\choicer$ is omitted from the calculus for each $i \in \ag$ (see \fig~\ref{fig:refined-calculus-dsn}). To prove the result, let us consider a derivation in $\dsnlkz$ of the labelled sequent $\seqempstr \sar w : \phi$ in a bottom-up manner. The only rules that add new labels (i.e. eigenvariables)---and therefore, change the set of vertices or edges in the sequent graph of a labelled sequent---are the $\agboxr$, $\Oir$, $\prODirtwo$, and $\ioa$ rules. If the $\agboxr$ rule is applied bottom-up to a labelled sequent $\Lambda$, then it will introduce a forward edge to the sequent graph of $\Lambda$. If the $\Oir$ or $\prODirtwo$ rule is applied bottom-up to a labelled sequent $\Lambda$, then it will introduce a new, disconnected vertex. If the $\ioa$ rule is applied bottom-up to a labelled sequent $\Lambda$, then it introduces $|\ag| = n+1$ new edges all converging to a fresh vertex in the sequent graph of $\Lambda$. Therefore, one can see that applying rules in reverse to the labelled sequent $\seqempstr \sar w : \phi$ cannot introduce a cycle since new edges always connect to fresh vertices in sequent graphs. Furthermore, when new labels are introduced via bottom-up applications of $\agboxr$, $\Oir$, $\prODirtwo$, or $\ioa$, either forward edges are added to sequent graphs, or new disconnected vertices are added; this shows that if a label serves as a root, then no edges can be introduced pointing to that label, and hence, the labelled derivation possesses the rooted property.
\end{proof}

Last, we note that in the single-agent setting, i.e. for the class of calculi $\{\dsnlnz \ | \ k \in \mathbb{N}\}$, completeness only requires labelled forest derivations. This minimalism is partly due to the fact that the $\ioa$ rule is redundant when $|\ag| = 1$, implying that convergent structures---such as the example presented above---need not be introduced. Still, the $\choicer$  rule could break the property of being a labelled forest sequent when applied bottom-up, but it turns out that applications of $\choicer$ can be restricted to ensure that the forest structure of a labelled sequent is preserved. We do not prove that each $\dsnlnz$ calculus only requires labelled forest derivations (with the rooted property) here, as the proof requires more sophisticated methods, which will be introduced in \sect~\ref{sec:applicationsI}, and follows as a corollary from our work on proof-search for deontic \stit logics. 



\chapter{The Method of Refinement: First-Order Intuitionistic Logics}
\label{CPTR:Refinment-Constructive} 






We demonstrate how refining the labelled calculi $\gtintfond$ and $\gtintfocd$ (see \fig~\ref{fig:labelled-calculi-FO-Int}) for the first-order intuitionistic logics $\intfond$ and $\intfocd$ (see \dfn~\ref{def:axiomatization-IntFO}) yields new nested calculi for the logics (which we refer to as $\nintfond$ and $\nintfocd$, respectively; see \fig~\ref{fig:nested-calculi-FOInt} on p.~\pageref{fig:nested-calculi-FOInt}). The characteristic feature of these nested calculi is the employment of \emph{nested sequents} in deriving theorems, which---in the current context---are formulae encoding trees of two-sided Gentzen-style sequents. For example, the nested sequent:
$$
p, \bot \sar p, \{q \imp q, q \imp q \sar \seqempstr\}, \{q \lor r \sar q \land p, \{\seqempstr \sar p\}, \{\seqempstr \sar \seqempstr\}\}
$$
encodes the following tree of sequents:
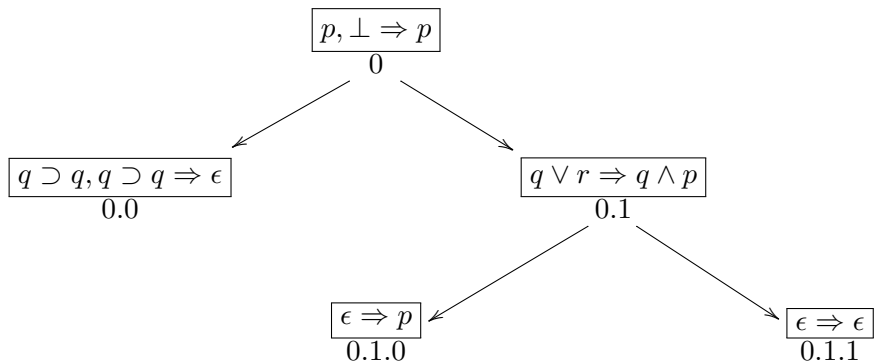
\begin{center}
\begin{tabular}{c c}
\xymatrix{
   & \overset{\boxed{p, \bot \sar p}}{0} \ar[dl] \ar[dr] & &  \\
  \overset{\boxed{q \imp q, q \imp q \sar \seqempstr}}{0.0} & & \overset{\boxed{q \lor r \sar q \land p}}{0.1} \ar[dl]\ar[dr]& \\ 
  & \overset{\boxed{\seqempstr \sar p}}{0.1.0} &  & \overset{\boxed{\seqempstr \sar \seqempstr}}{0.1.1}
}
\end{tabular}
\end{center}

The nested calculi we obtain via the method of refinement 
 are distinct from the existing nested calculi for $\intfond$ and $\intfocd$ provided by Fitting in~\cite{Fit14}, which have been included in Appendix A (see p.~\pageref{app:fittings-nested-calculi}). As will be argued below, the new nested systems $\nintfond$ and $\nintfocd$ obtained via refinement possess certain qualities that improve upon Fitting's nested calculi in a number of ways. Perhaps the most significant and advantageous distinction concerns the theoretical apparatus underpinning inference rules in $\nintfond$ and $\nintfocd$; we make use of \cfcst systems (see \dfn~\ref{def:CFCST-kms}) to define inference rules for $\imp$, $\exists$, and $\forall$ and in the definition of initial sequents. Therefore, such rules are parameterized by formal grammars dictating (in)applications of each rule---just as was done with propagation rules in \sect~\ref{SECT:Refine-Grammar}. 
 To provide the reader with intuition concerning how formal grammars play a role in our nested systems, we give a concrete example of how an initial sequent is determined in $\nintfond$ and $\nintfocd$:
$$
X := r(\undb,\undc) \sar \seqempstr, \nbbl \forall x q(x) \sar \seqempstr, \nbbl q(\undb), p \sar r(\undb,\undc)  \nbbr \nbbr
$$
Determining that $X$ is an initial sequent consists of two steps: \emph{first}, we transform the nested sequent into a \emph{propagation graph} (much like the propagation graphs introduced in the previous chapter; see \dfn~\ref{def:propagation-graph-nested-kms}), which is analogous to a finite deterministic automaton.\footnote{See~\cite[\cptr~1]{Sip12} for an introduction to finite deterministic automata.} A pictorial representation of the propagation graph of $X$ is shown below:
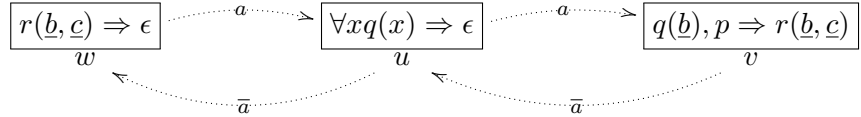
\begin{center}
\begin{tabular}{c}
\xymatrix{
\overset{\boxed{r(\undb,\undc) \sar \seqempstr}}{w}\ar@/^1.5pc/@{.>}[rr]|-{a} & &
 \overset{\boxed{\forall x q(x) \sar \seqempstr}}{u}\ar@/^1.5pc/@{.>}[ll]|-{\conv{a}}\ar@/^1.5pc/@{.>}[rr]|-{a}
 & & 
 \overset{\boxed{q(\undb), p \sar r(\undb,\undc)}}{v}\ar@/^1.5pc/@{.>}[ll]|-{\conv{a}} \\
}
\end{tabular}
\end{center}
For the propagation graph of a nested sequent, we make use of the character $a$ to encode a \emph{forward} move, that is, a transition to a (two-sided) Gentzen-style sequent that is \emph{deeper} in the nestings, and make use of its converse $\conv{a}$ to encode a \emph{backward} move to a \emph{shallower} Gentzen-style sequent. 
 Labeling the edges of the propagation graph with characters is useful because sequences of edges can be thought of as strings generated by a \cfcst system encoding how formulae should be correctly positioned (or, correctly propagated---in the case of inference rules for $\imp$, $\exists$, and $\forall$) throughout the graph of the given nested sequent; this brings us to our \emph{second} step in determining that $X$ is initial, which relies on matching strings generated by a \cfcst system to paths in the propagation graph of $X$. 
 
For reasons that will be made clear in the subsequent section (\sect~\ref{subsect:refinement-part-one-FOInt}), a nested sequent is considered initial in $\nintfond$ and $\nintfocd$ so long as there is a sequence of edges in its propagation graph from a Gentzen-style sequent of the form $Y_{1}, p(\vec{\unda}) \sar Z_{1}$ to another of the form $Y_{2} \sar p(\vec{\unda}), Z_{2}$ spelling a string from the set $\{\empstr, a, a \cate a, a \cate a \cate a, \ldots\}$, where $Y_{1}$, $Y_{2}$, $Z_{1}$, and $Z_{2}$ are multisets of formulae from $\langintfo$ (see \dfn~\ref{def:langintfo}) and $p(\vec{\unda})$ is an atomic formula. We can re-phrase this condition in grammar-theoretic terms by stating that a nested sequent is initial so long as there is a sequence of edges in its propagation graph from a Gentzen-style sequent $Y_{1}, p(\vec{\unda}) \sar Z_{1}$ to another of the form $Y_{2} \sar p(\vec{\unda}), Z_{2}$ that spells a string generated by $a$ with the following \cfcst system:
$$
S := \{a \pto \empstr, a \pto a \cate a, \conv{a} \pto \empstr, \conv{a} \pto \conv{a} \cate \conv{a}\},
$$
that is, the string is in the language $\thuesyslang{a}$ (see \dfn~\ref{def:derivation-relation-language-kms}). Note that $\thuesyslang{a} = \{\empstr, a, a \cate a, a \cate a \cate a, \ldots\}$ and that strings within the language can be seen as encoding a notion of \emph{reachability} in the propagation graph of a nested sequent with $\empstr$ meaning that there is a forward path of length $0$ between $Y_{1}, p(\vec{\unda}) \sar Z_{1}$ and $Y_{2} \sar p(\vec{\unda}), Z_{2}$, with $a$ meaning that there is a forward path of length $1$ between $Y_{1}, p(\vec{\unda}) \sar Z_{1}$ and $Y_{2} \sar p(\vec{\unda}), Z_{2}$, with $a \cate a$ meaning that there is a forward path of length $2$ between $Y_{1}, p(\vec{\unda}) \sar Z_{1}$ and $Y_{2} \sar p(\vec{\unda}), Z_{2}$, and so on. In terms of the semantics of $\intfond$ and $\intfocd$, this condition on initial sequents is equivalent to the monotonicity condition \moncond (see \dfn~\ref{def:IntFO-frame-model}) imposed on $\intfond$- and $\intfocd$-models, which states that if an atomic formula $p(\vec{\unda})$ is true at a world, then it is true at all future worlds. One can easily verify that the nested sequent $X$ is initial via this condition since there is a sequence of (two) forward edges from $w$ to $v$ spelling the string $a \cate a$---which can also be generated by $a$ with the second production rule in $S$---where the initial node $w$ is associated with $r(\undb,\undc) \sar \seqempstr$ and the terminal node $v$ is associated with $q(\undb), p \sar r(\undb,\undc)$.



The method of defining rules that make reference to formal grammars produces interesting and beneficial consequences. First, it allows us to recognize a class of rules that generalize the behavior of propagation rules, which we dub \emph{reachability rules}\index{Reachability rule}. While propagation rules function by propagating a formula to the terminal node of a path, reachability rules function by (additionally) checking if data exists at the end of a (potentially separate) path. It is easy to imagine this behavior being generalized even further with reachability rules propagating formulae along an arbitrary number of paths conditional on the existence of data found along some number of paths, albeit we will only consider rather simple versions of such rules here. A second outcome of our grammar/language theoretical approach to defining rules is a substantial increase in the modularity of our nested systems compared to the systems put forth by Fitting. Fitting employs a rule---referred to as \emph{lift} (see Appendix A on p.~\pageref{app:fittings-nested-calculi})---which encodes the transitivity and monotonicity properties of intuitionistic logics. We note that the \emph{lift} rule is a reformulation of the propagation rules introduced by Postniece in~\cite{Post09,Post10}, which were used to provide a deep-inference nested sequent system for bi-intuitionistic logic amenable to proof-search. In the approach presented here, monotonicity is encoded into an initial, reachability rule, and transitivity is encoded in a propagation rule for the $\imp$ connective and a reachability rule for the $\forall$ connective. By comparison then, our approach is more fine-grained than Fitting's as it distinguishes the monotonicity and transitivity properties as opposed to unifying them in a single rule. Consequently, this appears to bring about an increase in modularity as transforming our systems into systems for sub-intuitionistic logics~\cite{Cor87,Res94} should follow by the modification of our reachability and/or propagation rules. In fact, by `plugging-in' alternative formal grammars and/or adding certain rules for new logical connectives, it appears that we can uniformly obtain proof calculi for a sizable class of propositional and first-order (sub-)(bi-)intuitionistic logics (such as those discussed in~\cite{Cor87,IshKik07,PinUus18,Rau80,Res94}), though such investigations are left to future work as they go beyond the logics considered in this thesis. Last, connecting our proof systems (and therefore, our logics) to formal grammars opens up the possibility of transferring results (e.g. (un)decidability results) between formal language theory and logic.


As mentioned at the onset, the primary goal of this chapter is to show how (new) nested calculi can be obtained for $\intfond$ and $\intfocd$ via the method of refinement. In contrast to the propositional setting where refinement primarily proceeds via structural rule elimination, in the first-order setting we must also consider the removal of domain atoms (e.g. $\unda \in D_{w}$) from labelled syntax. Thus, structural rule elimination will beget the \emph{quasi-refined} labelled calculi\index{Quasi-refined labelled calculus} $\intfondlq$ and $\intfocdlq$ containing rules whose applicability still depends on the occurrence of domain atoms. After obtaining these calculi, we then redefine our rules so that their applicability is independent of the occurrence of domain atoms, producing our \emph{(fully) refined} labelled calculi $\intfondl$ and $\intfocdl$, from which the nested calculi $\nintfond$ and $\nintfocd$ may be obtained. To make our strategy for deriving $\nintfond$ and $\nintfocd$ via refinement explicit, we clearly list the steps involved in the process, which also explains the main content of each of the following four sections:


\begin{enumerate}

\item \sect~\ref{subsect:refinement-part-one-FOInt}: Discover through analyzing the elimination of structural rules, what conducive rules (i.e propagation and reachability rules) permit the elimination of all structural rules from the \emph{labelled} calculi $\gtintfond$ and $\gtintfocd$ (defined in \fig~\ref{fig:labelled-calculi-FO-Int}).\footnote{Recall that \emph{conducive rules} are rules that permit the elimination of structural rules when added to our calculi.}

\item \sect~\ref{subsect:refinement-part-two-FOInt}: Derive the \emph{quasi-refined} labelled calculi $\intfondlq$ and $\intfocdlq$ (defined in \dfn~\ref{def:quasi-refined-calculi}) from the labelled calculi $\gtintfond$ and $\gtintfocd$ via structural rule elimination. 

\item \sect~\ref{subsect:refinement-part-three-FOInt} (Part 1): Remove the domain atoms from the syntax of $\intfondlq$ and $\intfocdlq$ and re-define relevant inference rules so that applications of the rules are independent of the existence of domain atoms to obtain the \emph{(fully) refined} labelled calculi $\intfondl$ and $\intfocdl$ (defined in \fig~\ref{fig:refined-calculi-FO-Int}).

\item \sect~\ref{subsect:refinement-part-three-FOInt} (Part 2): Show that the refined labelled calculi $\intfondl$ and $\intfocdl$ possess favorable proof-theoretic properties.

\item \sect~\ref{subsect:translation-to-nested-FOInt}: Show that the \emph{nested} calculi $\nintfond$ and $\nintfocd$ (defined in \fig~\ref{fig:nested-calculi-FOInt}) can be obtained from, and inherit the properties of, the refined labelled calculi $\intfondl$ and $\intfocdl$.

\end{enumerate}

The relationships established between the various proof calculi considered are displayed in \fig~\ref{fig:relationships}. Solid arrows indicate that proofs in one calculus may be transformed into proofs in another calculus (which preserves the language of the calculus) and dotted arrows signify that proofs in one calculus may be translated into proofs in another calculus (which changes the language of the calculus). The symbol $\subset$ is used to indicate that one calculus is a restricted version of 
another; e.g. $\gtintfond$ is a restricted version of $\gtintfocd$ as the former omits the $\cd$ rule and the latter includes it. Furthermore, arrows are annotated with the name of the result that establishes the transformation or translation, and the dotted arrows are additionally annotated with the $\lnint$ and $\nlint$ symbols, indicating the functions that translate from labelled to nested notation and the opposite, respectively, which are introduced in \sect~\ref{subsect:translation-to-nested-FOInt}. 

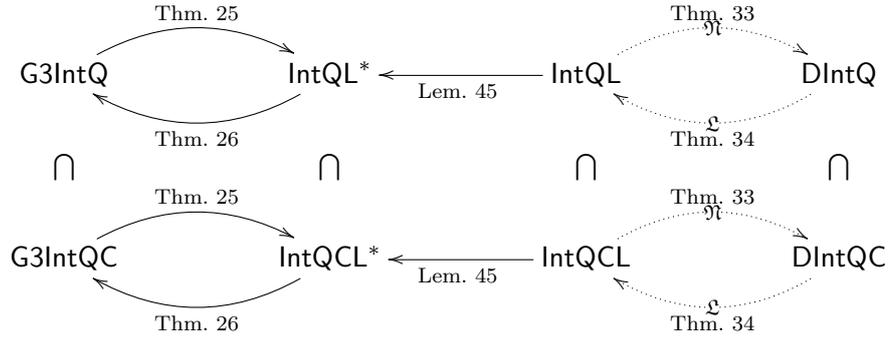
\begin{figure}[t]
\noindent\hrule
\begin{center}
\begin{tabular}{c}
\xymatrix{
	 \gtintfond\ar@/^1.5pc/@{->}[rr]^{\text{Thm.~\ref{thm:admissible-rules-G3-Calc}}}\ar@{}[dd]|{\bigcap}	& &   
	 \intfondlq\ar@/^1.5pc/@{->}[ll]^{\text{Thm.~\ref{thm:reach-prop-admissible-rules-G3-Calc}}}\ar@{}[dd]|{\bigcap} & & \intfondl\ar@/^1.5pc/@{.>}[rr]|{\lnint}^{\text{Thm.~\ref{thm:Refined-to-Nested-FOInt}}}\ar@{->}[ll]^{\text{\lem~\ref{lem:refined-to-quasi-refined-FOInt}}}\ar@{}[dd]|{\bigcap} & & 
	 \nintfond\ar@/^1.5pc/@{.>}[ll]|{\nlint}^{\text{Thm.~\ref{thm:Nested-to-Refined-FOInt}}}\ar@{}[dd]|{\bigcap}
	 \\
	 & & & & \\
	 \gtintfocd\ar@/^1.5pc/@{->}[rr]^{\text{Thm.~\ref{thm:admissible-rules-G3-Calc}}}	& &   
	 \intfocdlq\ar@/^1.5pc/@{->}[ll]^{\text{Thm.~\ref{thm:reach-prop-admissible-rules-G3-Calc}}} & & \intfocdl\ar@/^1.5pc/@{.>}[rr]|{\lnint}^{\text{Thm.~\ref{thm:Refined-to-Nested-FOInt}}}\ar@{->}[ll]^{\text{\lem~\ref{lem:refined-to-quasi-refined-FOInt}}} & & 
	 \nintfocd\ar@/^1.5pc/@{.>}[ll]|-{\nlint}^{\text{Thm.~\ref{thm:Nested-to-Refined-FOInt}}}
	 \\
}
\end{tabular}
\end{center}
\hrule
\caption{Transformations and translations between the intuitionistic calculi considered.}\label{fig:relationships}
\end{figure}


\section{Refinement Part I: Analysis}\label{subsect:refinement-part-one-FOInt}


We analyze the issues that arise when attempting to perform structural rule elimination in $\gtintfond$ and $\gtintfocd$ (see \fig~\ref{fig:labelled-calculi-FO-Int}). This analysis will lead to the discovery of (and motivate the definition of) conducive rules that permit the elimination of $\refl$, $\trans$, $\nd$, $\cd$, and $\ned$ (see \fig~\ref{fig:labelled-calculi-FO-Int}) from our labelled calculi. 
 Our analysis of structural rule elimination will consist of three parts: (i) Considering the permutation of $\refl$ and $\trans$ above the $\idfo$ and $\impl$ rules, (ii) Considering the permutation of $\nd$ above the $\idfo$ and $\existsr$ rules, and (iii) Considering the permutation of $\ned$ above the $\alll$ rule. Despite the fact that the case of permuting $\refl$ and $\trans$ above $\alll$, permuting $\cd$ (with $\nd$) above $\idfo$, $\existsr$, and $\alll$, and permuting $\ned$ above $\existsr$ are all interesting cases, the understanding forged from considering the cases in (i) -- (iii) above will be sufficient to motivate the definition of our reachability and propagation rules. Before going through our analysis, we introduce some helpful terminology: 

\begin{definition}[Principal, Auxiliary Label] We refer to the label of the principal formula in a labelled inference rule as the \emph{principal label}\index{Principal label}. Also, the label(s) of the auxiliary formula(e) in a labelled inference rule is (are) referred to as the \emph{auxiliary label(s)}\index{Auxiliary label}. Principal and auxiliary labels are referred to as \emph{active labels}\index{Active label}, more generally.
\end{definition}

Let us begin part (i) of our analysis. We first consider the problematic cases that arise when attempting to permute $\refl$ above $\idfo$ and $\impl$, and after, look at the problematic cases that arise when attempting to permute $\trans$ above the two rules. The $\refl$ cases we consider are as follows:

\begin{center}
\AxiomC{}
\RightLabel{$\idfo$}
\UnaryInfC{$\rel,w \leq w, \vec{\unda} \in D_{w},\Gamma, w :p(\vec{\unda}) \sar w :p(\vec{\unda}), \Delta$}
\RightLabel{$\refl$}
\UnaryInfC{$\rel, \vec{\unda} \in D_{w},\Gamma, w :p(\vec{\unda}) \sar w :p(\vec{\unda}), \Delta$}
\DisplayProof
\end{center}
\begin{center}
\AxiomC{$\rel,w \leq w, \Gamma, w :\phi \imp \psi \sar \Delta, w :\phi$}
\AxiomC{$\rel,w \leq w, \Gamma, w :\phi \imp \psi, w :\psi \sar \Delta$}
\RightLabel{$\impl$}
\BinaryInfC{$\rel,w \leq w, \Gamma, w :\phi \imp \psi \sar \Delta$}
\RightLabel{$\refl$}
\UnaryInfC{$\rel, \Gamma, w :\phi \imp \psi \sar \Delta$}
\DisplayProof 
\end{center}

In order to successfully eliminate $\refl$ in the above $\idfo$ case, the end sequent must be derivable without the use of $\refl$. However, by inspection of $\gtintfond$ and $\gtintfocd$, it is apparent that none of the rules can be used to derive the end sequent without the use of $\refl$. A similar issue arises when attempting to permute the $\refl$ rule upwards in the $\impl$ derivation, as shown below:

\begin{center}
\begin{tabular}{c c}
\AxiomC{$\rel,w \leq w, \Gamma, w :\phi \imp \psi \sar \Delta, w :\phi$}
\RightLabel{$\refl$}
\UnaryInfC{$\rel, \Gamma, w :\phi \imp \psi \sar \Delta, w :\phi$}
\DisplayProof 

&

\AxiomC{$\rel,w \leq w, \Gamma, w :\phi \imp \psi, w :\psi \sar \Delta$}
\RightLabel{$\refl$}
\UnaryInfC{$\rel, \Gamma, w :\phi \imp \psi, w :\psi \sar \Delta$}
\DisplayProof 
\end{tabular}
\end{center}

In the above cases, no rule exists within $\gtintfond$ or $\gtintfocd$ permitting us to derive the desired conclusion of the original proof. Nevertheless, if we add the following (sound) rules to our calculi, then the permutations with $\refl$ may be accomplished:

\begin{center}
\AxiomC{}
\RightLabel{$(id_{0})$}
\UnaryInfC{$\rel, \vec{\unda} \in D_{w},\Gamma, w :p(\vec{\unda}) \sar w :p(\vec{\unda}), \Delta$}
\DisplayProof
\end{center}

\begin{center}
\AxiomC{$\rel, \Gamma, w :\phi \imp \psi \sar \Delta, w :\phi$}
\AxiomC{$\rel, \Gamma, w :\phi \imp \psi, w :\psi \sar \Delta$}
\RightLabel{$(\imp_{l}^{0})$}
\BinaryInfC{$\rel, \Gamma, w :\phi \imp \psi \sar \Delta$}
\DisplayProof 
\end{center}

Notice that both of the above rules share a common characteristic. In the $(id_{0})$ rule, what was the (principal) labelled formula $u : p(\vec{\unda})$ on the right of the sequent arrow $\sar$ has now become the labelled formula $w : p(\vec{\unda})$, thus matching the principal formula on the left. By comparison, in the $(\imp_{l}^{0})$ rule, the auxiliary formulae $u : \phi$ (in the left premise) and $u : \psi$ (in the right premise) have also \emph{propagated} backward and are associated with the label $w$ with which the principal formula $w : \phi \imp \psi$ is associated. This observation suggests that for the elimination of $\refl$, we require new versions of the $\idfo$, $\impl$, and $\alll$ rules (notice that all such rules include a relational atom of the form $w \leq u$ in their conclusion, which causes them to `interact' with $\refl$) where the auxiliary labels are reachable with a path of length zero from the principal label.\footnote{Recall that we are omitting consideration of certain cases (e.g. $\alll$) in our analysis to keep it concise. In spite of this fact, for the curious reader, the new version of the $\alll$ rule would be as follows:
\begin{center}
\AxiomC{$\rel, \unda \in D_{w}, \Gamma, w : \phi(\unda/x), w : \forall x \phi \sar \Delta$}
\RightLabel{$(\forall_{l}^{0})$}
\UnaryInfC{$\rel, \unda \in D_{w}, \Gamma, w : \forall x \phi \sar \Delta$}
\DisplayProof
\end{center}
} We will keep this fact in mind as we continue our analysis and observe the problematic case of permuting $\trans$ above $\id$ and $\impl$. These cases are as follows:

\begin{center}
\AxiomC{}
\RightLabel{$\idfo$}
\UnaryInfC{$\rel, w \leq v, v \leq u, w \leq u, \vec{\unda} \in D_{w},\Gamma, w :p(\vec{\unda}) \sar u :p(\vec{\unda}), \Delta$}
\RightLabel{$\trans$}
\UnaryInfC{$\rel, w \leq v, v \leq u, \vec{\unda} \in D_{w},\Gamma, w :p(\vec{\unda}) \sar u :p(\vec{\unda}), \Delta$}
\DisplayProof
\end{center}

$$
\Lambda_{1} := \rel,w \leq v, v \leq u, w \leq u, \Gamma, w :\phi \imp \psi \sar \Delta, u :\phi
$$
$$
\Lambda_{2} := \rel,w \leq v, v \leq u, w \leq u, \Gamma, w :\phi \imp \psi, u :\psi \sar \Delta
$$
\begin{center}
\AxiomC{$\Lambda_{1}$}
\AxiomC{$\Lambda_{2}$}
\RightLabel{$\impl$}
\BinaryInfC{$\rel,w \leq v, v \leq u, w \leq u, \Gamma, w :\phi \imp \psi \sar \Delta$}
\RightLabel{$\trans$}
\UnaryInfC{$\rel, w \leq v, v \leq u, \Gamma, w :\phi \imp \psi \sar \Delta$}
\DisplayProof 
\end{center}

As before, the end sequent of the first proof is not derivable without the use of $\trans$ in $\gtintfond$ and $\gtintfocd$, suggesting that the end sequent should be taken as an initial sequent in its own right. Also, permuting the $\trans$ rule upwards in the $\impl$ proof yields proofs ending with the following two inferences:

\begin{center}
\AxiomC{$\rel,w \leq v, v \leq u, w \leq u, \Gamma, w :\phi \imp \psi \sar \Delta, u :\phi$}
\RightLabel{$\trans$}
\UnaryInfC{$\rel, w \leq v, v \leq u, \Gamma, w :\phi \imp \psi \sar \Delta, u :\phi$}
\DisplayProof 
\end{center}
\begin{center}
\AxiomC{$\rel, w \leq v, v \leq u, w \leq u, \Gamma, w :\phi \imp \psi, u :\psi \sar \Delta$}
\RightLabel{$\trans$}
\UnaryInfC{$\rel, w \leq v, v \leq u, \Gamma, w :\phi \imp \psi, u :\psi \sar \Delta$}
\DisplayProof 
\end{center}

Since no rule is applicable to the conclusions above that would allow us to derive the desired conclusion, the $\impl$ case also suggests the need of a new $\impl$ rule. The new versions of $\idfo$ and $\impl$ suggested by our analysis are as shown below, with $\rel_{2} := \rel, w \leq v, v \leq u$. It is crucial to observe that a directed path of length two (due to the $w \leq v, v \leq u$ relational atoms) occurs between the principal and auxiliary labels in the rules below:

\begin{center}
\AxiomC{}
\RightLabel{$(id_{2})$}
\UnaryInfC{$\rel_{2}, \vec{\unda} \in D_{w},\Gamma, w :p(\vec{\unda}) \sar u :p(\vec{\unda}), \Delta$}
\DisplayProof
\end{center}

\begin{center}
\AxiomC{$\rel_{2}, \Gamma, w :\phi \imp \psi \sar \Delta, u :\phi$}
\AxiomC{$\rel_{2}, \Gamma, w :\phi \imp \psi, u :\psi \sar \Delta$}
\RightLabel{$(\imp_{l}^{2})$}
\BinaryInfC{$\rel_{2}, \Gamma, w :\phi \imp \psi \sar \Delta$}
\DisplayProof 
\end{center}

Adding the above rules to our calculi would permit the permutation of $\trans$ above $\idfo$ and $\impl$, yet, if such rules are added to our calculi, then we must consider permuting $\trans$ above them as well. Let us examine the problematic cases that arise when attempting to permute $\trans$ above $(id_{2})$ and $(\imp_{l}^{2})$, and after, we will summarize and discuss the conducive rules suggested by part (i) of our analysis.

\begin{center}
\AxiomC{}
\RightLabel{$(id_{2})$}
\UnaryInfC{$\rel, w \leq v, v \leq z, z \leq u, w \leq u, \vec{\unda} \in D_{w},\Gamma, w :p(\vec{\unda}) \sar u :p(\vec{\unda}), \Delta$}
\RightLabel{$\trans$}
\UnaryInfC{$\rel, w \leq v, v \leq z, z \leq u, \vec{\unda} \in D_{w},\Gamma, w :p(\vec{\unda}) \sar u :p(\vec{\unda}), \Delta$}
\DisplayProof
\end{center}

$$
\Lambda_{1} := \rel,w \leq v, v \leq z, z \leq u, w \leq u, \Gamma, w :\phi \imp \psi \sar \Delta, u :\phi
$$
$$
\Lambda_{2} := \rel,w \leq v, v \leq z, z \leq u, w \leq u, \Gamma, w :\phi \imp \psi, u :\psi \sar \Delta
$$
\begin{center}
\AxiomC{$\Lambda_{1}$}
\AxiomC{$\Lambda_{2}$}
\RightLabel{$\impl$}
\BinaryInfC{$\rel,w \leq v, v \leq z, z \leq u, w \leq u, \Gamma, w :\phi \imp \psi \sar \Delta$}
\RightLabel{$\trans$}
\UnaryInfC{$\rel, w \leq v, v \leq z, z \leq u, \Gamma, w :\phi \imp \psi \sar \Delta$}
\DisplayProof 
\end{center}

At this point, the reader might have guessed that the above cases cannot be resolved within our calculi, unless new versions of the $(id_{2})$ and $(\imp_{l}^{2})$ rules are added. Such rules would have the following form, with $\rel_{3} := \rel, w \leq v, v \leq z, z \leq u$, thus requiring the existence of a (directed) relational path of length three between principal and auxiliary labels. 

\begin{center}
\AxiomC{}
\RightLabel{$(id_{3})$}
\UnaryInfC{$\rel_{3}, \vec{\unda} \in D_{w},\Gamma, w :p(\vec{\unda}) \sar u :p(\vec{\unda}), \Delta$}
\DisplayProof
\end{center}

\begin{center}
\AxiomC{$\rel_{3}, \Gamma, w :\phi \imp \psi \sar \Delta, u :\phi$}
\AxiomC{$\rel_{3}, \Gamma, w :\phi \imp \psi, u :\psi \sar \Delta$}
\RightLabel{$(\imp_{l}^{3})$}
\BinaryInfC{$\rel_{3}, \Gamma, w :\phi \imp \psi \sar \Delta$}
\DisplayProof 
\end{center}

Of course, extending our calculi with the above rules would require considering their permutability with $\trans$, thus necessitating the addition of new rules $(id_{4})$ and $(\imp_{l}^{4})$ (having relational paths of length four between their auxiliary labels and principal labels), which would further require the addition of new rules to permit their permutation with $\trans$, ad infinitum. 

A pattern has emerged from our analysis: The rules $(id_{0})$ and $(\imp_{l}^{0})$ require relational paths of length zero between auxiliary labels and principal labels, the original rules $\idfo$ and $\impl$ require relational paths of length one (i.e. a single relational atom) between auxiliary labels and principal labels, the $(id_{2})$ and $(\imp_{l}^{2})$ rules require relational paths of length two (i.e. there are two relational atoms $w \leq v, v \leq u$) between auxiliary labels and principal labels, etc. Similar to \sect~\ref{SECT:Refine-Grammar}, it appears that what is needed to allow for the elimination of $\refl$ and $\trans$ is the addition of propagation rules to our calculi, where, the side condition permits an application of the rule given that the auxiliary labels are \emph{reachable} (with a directed path) from the principal label. There are a variety of ways in which we could define such rules, but since we have \cfcst systems at our disposal, we will make use of such objects to define our propagation rules, analogous to what was done in \sect~\ref{SECT:Refine-Grammar} for grammar logics. (NB. This approach differs from the author's approach in~\cite{Lyo20a,Lyo21} where applications of rules rely on the notion of a \emph{path}.) 

Recall that propagation rules view labelled sequents as propagation graphs, and allow for formulae to be introduced at terminal nodes of propagation paths corresponding to strings generated by a \cfcst system. By part (i) of our analysis, we found that we should employ rules that propagate formulae to \emph{reachable} labels. Hence, the question arises; how do we encode a notion of reachability in a \cfcst system? Answering this question will tell us what side conditions to impose on our propagation rules and will dictate the definition of a propagation graph (given below).

As previously implied, a label $u$ in a labelled sequent is reachable from another label $w$ \ifandonlyif there exists a sequence of relational atoms $w \leq v_{1}, \ldots, v_{n} \leq u$ of length zero or greater from $w$ to $u$ in the labelled sequent. Let us fix the minimal alphabet $\albet := \{a, \conv{a}\}$ for the remainder of the next four sections (\sect~\ref{subsect:refinement-part-one-FOInt} -- \ref{subsect:translation-to-nested-FOInt}), i.e. any reference to $\albet$ will be a reference to the alphabet $\{a, \conv{a}\}$ in \sect~\ref{subsect:refinement-part-one-FOInt} -- \ref{subsect:translation-to-nested-FOInt}. If we think of the character $a$ as encoding a single relational atom of the form $w \leq u$, then all of the strings $\empstr, a, a \cate a, a \cate a \cate a, \ldots \in (\albet^{+})^{*}$ can be thought of as encoding sequences of relational atoms. Note that all such sequences are generated by the following \cfcst system:\footnote{The name of the \cfcst system $\thuesysi$ arises because the first two production rules correspond to the properties of reflexivity and transitivity, respectively, as shown in \fig~\ref{fig:frame-conditions-production-rules}. Since the modal logic $\mathsf{S4}$ is the minimal normal modal logic that is sound and complete relative to reflexive and transitive Kripke frames, the name $\thuesysi$ for the above \cfcst systems is apt.}
$$
\thuesysi := \{a \pto \empstr, a \pto a \cate a, \conv{a} \pto \empstr, \conv{a} \pto \conv{a} \cate \conv{a}\}
$$
Observe that the above \cfcst system is sufficient to encode a notion of reachability as $\thuesysilang{a} = (\albet^{+})^{*}$. Furthermore, since we are thinking of the character $a$ as encoding a relational atom of the form $w \leq u$, we will use $a$ to index forward edges in our propagation graphs, and $\conv{a}$ to index backward edges, as described in the definition below:

\begin{definition}[Propagation Graphs for $\intfondl$ and $\intfocdl$]\label{def:propagation-graph-FOInt} Let $\Lambda = \rel, \Gamma \sar \Delta$ be a labelled sequent for first-order intuitionistic logics. We define the \emph{propagation graph}\index{Propagation graph!for $\intfondl$ and $\intfocdl$} $\prgr{\Lambda} = (\prgrdom, \prgredges)$ to be the directed graph such that
\begin{itemize}

\item[$\li$] $\prgrdom := \lab(\Lambda)$;

\item[$\li$] $\prgredges := \{(w,u,a), (u,w,\overline{a}) \ | \ w \leq u \in \rel \}$.

\end{itemize}
We will often write $w \in \prgr{\Lambda}$ to mean $w \in \prgrdom$, and $(w,u,\chara) \in \prgr{\Lambda}$ to mean $(w,u,\chara) \in \prgredges$, for $\chara \in \albet$.
\end{definition}

Propagation paths, strings of propagation paths, and their converses are defined as in \dfn~\ref{def:propagation-path-kms}, so we need not repeat the definitions here. However, we provide an example of all such concepts below for review, as well an an example of a propagation graph of a labelled sequent for first-order intuitionistic logics.

\begin{example} Let our labelled sequent be the following:
$$
\Lambda := w \leq u, w \leq v, v \leq z, \unda \in D_{w}, w : p(\unda), u : \forall x p(x), z : \negnnf{r} \sar w : q, u : q \imp q, v : q
$$
The propagation graph $\prgr{\Lambda}$ is shown below:
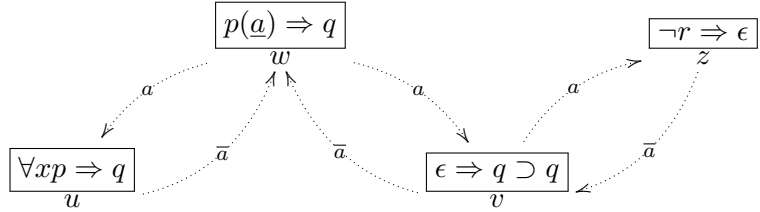
\begin{center}
\begin{tabular}{c}
\xymatrix{
 & \overset{\boxed{p(\unda) \sar q}}{w}\ar@/^1.5pc/@{.>}[dr]|-{a} \ar@/^-1.5pc/@{.>}[dl]|-{a} &  &
 \overset{\boxed{\negnnf{r} \sar \seqempstr}}{z}\ar@/^1.5pc/@{.>}[dl]|-{\conv{a}}\\
\overset{\boxed{\forall x p \sar q}}{u}\ar@/^-1.5pc/@{.>}[ur]|-{\conv{a}} & & \overset{\boxed{\seqempstr \sar q \imp q}}{v}\ar@/^1.5pc/@{.>}[ul]|-{\conv{a}}\ar@/^1.5pc/@{.>}[ur]|-{a} &
}
\end{tabular}
\end{center}
Examples of propagation paths include the propagation path $\ppath(w,z) = w, a, v, a, z$ and the propagation path $\ppath'(v,v) := v, \conv{a}, w, a, v$, with converses $\conv{\ppath}(z,w) = z, \conv{a}, v, \conv{a}, w$ and $\conv{\ppath}'(v,v) := v, \conv{a}, w, a, v$, respectively. The string of each propagation path is $\stra_{\ppath}(w,z) = a \cate a$ and $\stra_{\ppath'}(v,w) = \conv{a} \cate a$ with the converse strings $\conv{\stra}_{\ppath}(z,w) = \conv{a} \cate \conv{a}$ and $\conv{\stra}_{\ppath'}(w,v) = \conv{a} \cate a$, respectively.
\end{example}

\begin{figure}[t]
\noindent\hrule

\begin{center}
\begin{tabular}{c}
\AxiomC{}
\RightLabel{$\idfonca^{\dag_{1}({\norc})}$\index{$\idfonca$}}
\UnaryInfC{$\rel, \unda_{1} \in D_{u_{1}}, \ldots, \unda_{n} \in D_{u_{n}}, \Gamma, w : p(\vec{\unda}) \Rightarrow u : p(\vec{\unda}), \Delta$}
\DisplayProof
\end{tabular}
\end{center}

\begin{center}
\begin{tabular}{c}
\AxiomC{$\rel, \unda \in D_{u}, \Gamma \Rightarrow \Delta, w: \phi(\unda/x), w: \exists x \phi$}
\RightLabel{$\existsrnca^{\dag_{2}(\norc)}$\index{$\existsrnca$}}
\UnaryInfC{$\rel, \unda \in D_{u}, \Gamma \Rightarrow \Delta, w: \exists x \phi$}
\DisplayProof
\end{tabular}
\end{center}

\begin{center}
\begin{tabular}{c}
\AxiomC{$\rel, \unda \in D_{u}, \Gamma \Rightarrow w : \phi(\unda / x), w : \exists x \phi, \Delta$}
\RightLabel{$\existsrncia^{\dag_{3}({\norc})}$\index{$\existsrncia$}}
\UnaryInfC{$\rel, \Gamma \Rightarrow w : \exists x \phi, \Delta$}
\DisplayProof
\end{tabular}
\end{center}

\begin{center}
\begin{tabular}{c}
\AxiomC{$\rel, \unda \in D_{u}, w : \forall x \phi, v : \phi(\unda/x), \Gamma \Rightarrow \Delta$}
\RightLabel{$\alllnca^{\dag_{4}(\norc)}$\index{$\alllnca$}}
\UnaryInfC{$\rel, \unda \in D_{u}, w : \forall x \phi, \Gamma \Rightarrow \Delta$}
\DisplayProof
\end{tabular}
\end{center}

\begin{center}
\begin{tabular}{c}
\AxiomC{$\rel, \unda \in D_{u}, \Gamma, w : \forall x \phi, v : \phi(\unda / x) \sar \Delta$}
\RightLabel{$\alllncia^{\dag_{5}({\norc})}$\index{$\alllncia$}}
\UnaryInfC{$\rel, \Gamma, w : \forall x \phi \sar \Delta$}
\DisplayProof
\end{tabular}
\end{center}

\begin{center}
\AxiomC{$\rel, w : \phi \imp \psi, \Gamma \Rightarrow \Delta, u : \phi$}
\AxiomC{$\rel, w : \phi \imp \psi, u : \psi, \Gamma \Rightarrow \Delta$}
\RightLabel{$\primp^{\dag_{6}({\norc})}$\index{$\primp$}}
\BinaryInfC{$\rel, w : \phi \imp \psi, \Gamma \Rightarrow \Delta$}
\DisplayProof
\end{center}

\hrule
\caption{Reachability and propagation rules. The rules when $\mathsf{X} = \nnn$ will be added to $\gtintfond$ to refine the calculus, and the rules when $\mathsf{X} = \ccc$ will be added to $\gtintfocd$ to refine the calculus. The side conditions $\dag_{1}(\nnn)$ -- $\dag_{6}(\nnn)$ are defined in \fig~\ref{fig:side-conditionsi-FO-Int} and the side conditions $\dag_{1}(\ccc)$ -- $\dag_{6}(\ccc)$ are defined in \fig~\ref{fig:side-conditionsii-FO-Int}.}
\label{fig:reachability-propagation-rules-atoms-FO-Int}
\end{figure}

\begin{figure}[t]
\begin{center}
\bgroup
\def\arraystretch{1.5}
\begin{tabular}{| c | c |} 
\hline
Name & Side Condition \\ 
\hline
$\dag_{1}(\nnn)$\index{$\dag_{1}(\nnn)$} &  $\exists \ppath_{i}(\stra_{\ppath_{i}}(u_{i},w) \in \thuesysilang{a})$ for each  \\ 
 &  $i \in \{1, \ldots, n\}$, and $\exists \ppath(\stra_{\ppath}(w,u) \in \thuesysilang{a})$ \\ 
\hline
$\dag_{2}(\nnn)$\index{$\dag_{2}(\nnn)$} &  $\exists \ppath(\stra_{\ppath}(u,w) \in \thuesysilang{a})$  \\ 
\hline
$\dag_{3}(\nnn)$\index{$\dag_{3}(\nnn)$} &  $\unda$ is an eigenvariable  \\ 
 &  and $\exists \ppath(\stra_{\ppath}(u,w) \in \thuesysilang{a})$ \\ 
\hline
$\dag_{4}(\nnn)$\index{$\dag_{4}(\nnn)$} &  $\exists \ppath(\stra_{\ppath}(u,v) \in \thuesysilang{a})$  and $\exists \ppath(\stra_{\ppath}(w,v) \in \thuesysilang{a})$ \\ 
\hline
$\dag_{5}(\nnn)$\index{$\dag_{5}(\nnn)$} &  $\unda$ is an eigenvariable, $\exists \ppath(\stra_{\ppath}(u,v) \in \thuesysilang{a})$\\
 &  and $\exists \ppath(\stra_{\ppath}(w,v) \in \thuesysilang{a})$ \\ 
\hline
$\dag_{6}(\nnn)$\index{$\dag_{6}(\nnn)$} &  $\exists \ppath(\stra_{\ppath}(w,u) \in \thuesysilang{a})$  \\ 
\hline
\end{tabular}
\egroup
\end{center}
\caption{Side conditions for the reachability and propagation rules used to refine $\gtintfond$.}
\label{fig:side-conditionsi-FO-Int}
\end{figure}

We are now in a position to put forth the propagation rules resulting from part (i) of our analysis. The propagation rule $\primp$ which replaces the $\impl$ rule in our (quasi-)refined calculi is shown in \fig~\ref{fig:reachability-propagation-rules-atoms-FO-Int}. The propagation rule $(Pr_{id})$ for $\id$ resulting from part (i) of our analysis is as follows:
\begin{center}
\AxiomC{}
\RightLabel{$(Pr_{id})^{\dag}$}
\UnaryInfC{$\rel, \vec{\unda} \in D_{w}, \Gamma, w : p(\vec{\unda}) \Rightarrow u : p(\vec{\unda}), \Delta$}
\DisplayProof
\end{center}
where the side condition $\dag$ states that $\exists \ppath (\stra_{\ppath}(w,u) \in \thuesysilang{a})$, that is, there must exist a propagation path $\ppath(w,u)$ in the propagation graph of the labelled sequent such that $\stra_{\ppath}(w,u) \in \thuesysilang{a}$. It should be noted however, that the above rule will not occur in our quasi-refined labelled calculi for $\intfond$ and $\intfocd$. The reason being, the occurrence of the domain atoms $\vec{\unda} \in D_{w}$ cause the rule to interact with the $\nd$ and $\cd$ rules, and so, via step (ii) of our analysis the $(Pr_{id})$ rule will be generalized to a proper reachability rule, yielding the final version of the rule that will occur in our quasi-refined labelled calculi. With that being said, let us begin part (ii) of our analysis.

In part (ii) of our analysis, we consider the permutation of $\nd$ above the $\id$ and $\existsr$ rules. Notwithstanding, since we discovered in part (i) that our quasi-refined systems require the propagation rule $(Pr_{id})$ instead of $\id$ (which is subsumed by $(Pr_{id})$ as shown in \lem~\ref{lem:instance-of} below), we will analyze permutations of $\nd$ above $(Pr_{id})$ rather than above $\id$. To ease our analysis, we assume the existence of a single principal domain atom in $(Pr_{id})$. Below, we show the non-trivial cases of applying $\nd$ to $(Pr_{id})$ and $\existsr$, along with the result of attempting to permute $\nd$ above $\existsr$ by applying $\nd$ to the premise of $\existsr$.

\begin{center}
\AxiomC{}
\RightLabel{$(Pr_{id})$}
\UnaryInfC{$\rel, v \leq w,  \unda \in D_{v}, \unda \in D_{w}, \Gamma, w : p(\unda) \Rightarrow u : p(\unda), \Delta$}
\RightLabel{$\nd$}
\UnaryInfC{$\rel, v \leq w,  \unda \in D_{v}, \Gamma, w : p(\unda) \Rightarrow u : p(\unda), \Delta$}
\DisplayProof
\end{center}

\begin{flushleft}
\begin{tabular}{c c}
\AxiomC{$\rel, v \leq w,  \unda \in D_{v}, \unda \in D_{w}, \Gamma \sar w: \phi(\unda/x), w: \exists x \phi, \Delta$}
\RightLabel{$\existsr$}
\UnaryInfC{$\rel, v \leq w,  \unda \in D_{v}, \unda \in D_{w}, \Gamma \sar w: \exists x \phi, \Delta$}
\RightLabel{$\nd$}
\UnaryInfC{$\rel, v \leq w,  \unda \in D_{v}, \Gamma \sar w: \exists x \phi, \Delta$}
\DisplayProof

&
$\leadsto$
\end{tabular}
\end{flushleft}
\begin{flushright}
\AxiomC{$\rel, v \leq w,  \unda \in D_{v}, \unda \in D_{w}, \Gamma \sar w: \phi(\unda/x), w: \exists x \phi, \Delta$}
\RightLabel{$\nd$}
\UnaryInfC{$\rel, v \leq w,  \unda \in D_{v}, \Gamma \sar w: \phi(\unda/x), w: \exists x \phi, \Delta$}
\DisplayProof
\end{flushright}

By inspecting the rules of our calculi, one can confirm that the end sequent of the $(Pr_{id})$ proof cannot be derived by other means. The issue is that the domain atom $\unda \in D_{v}$ is associated with a label $v$ that is one step in the past of the label $w$ (due to the $v \leq w$ relational atom) from where it needs to occur. Similarly, no rule is applicable in the $\existsr$ case to allow for the permutation to go through, and the domain atom has also shifted one step into the past (after applying $\nd$ to the premise of $\existsr$). As we have typically done, we can add new rules to our calculi to allow for the permutation of $\nd$ above $(Pr_{id})$ and $\existsr$ to go through. The (sound) rules suggested by the above analysis are the following:

\begin{center}
\AxiomC{}
\RightLabel{$(Pr_{id}^{-1})$}
\UnaryInfC{$\rel, v \leq w,  \unda \in D_{v}, \Gamma, w : p(\unda) \Rightarrow u : p(\unda), \Delta$}
\DisplayProof
\end{center}

\begin{center}
\AxiomC{$\rel, v \leq w,  \unda \in D_{v}, \Gamma \sar w: \phi(\unda/x), w: \exists x \phi, \Delta$}
\RightLabel{$(\exists^{-1}_{r})$}
\UnaryInfC{$\rel, v \leq w,  \unda \in D_{v}, \Gamma \sar w: \exists x \phi, \Delta$}
\DisplayProof
\end{center}

Adding the above rules to our labelled calculi will allow for the permutation of $\nd$ above $(Pr_{id})$ and $\existsr$ to go through, however, their addition implies that we must consider their permutability with $\nd$. The non-trivial cases of applying $\nd$ to $(Pr_{id}^{-1})$ and $(\exists^{-1}_{r})$ are shown below, along with the application of $\nd$ to the premise of $(\exists^{-1}_{r})$ in attempt to permute the two rules.

\begin{center}
\AxiomC{}
\RightLabel{$(Pr_{id}^{-1})$}
\UnaryInfC{$\rel, z \leq v, v \leq w,  \unda \in D_{z}, \unda \in D_{v}, \Gamma, w : p(\unda) \Rightarrow u : p(\unda), \Delta$}
\RightLabel{$\nd$}
\UnaryInfC{$\rel, z \leq v, v \leq w,  \unda \in D_{z}, \Gamma, w : p(\unda) \Rightarrow u : p(\unda), \Delta$}
\DisplayProof
\end{center}

\begin{flushleft}
\begin{tabular}{c c}
\AxiomC{$\rel, z \leq v, v \leq w,  \unda \in D_{z}, \unda \in D_{v}, \Gamma \sar w: \phi(\unda/x), w: \exists x \phi, \Delta$}
\RightLabel{$(\exists^{-1}_{r})$}
\UnaryInfC{$\rel, z \leq v, v \leq w,  \unda \in D_{z}, \unda \in D_{v}, \Gamma \sar w: \exists x \phi, \Delta$}
\RightLabel{$\nd$}
\UnaryInfC{$\rel, z \leq v, v \leq w,  \unda \in D_{z}, \Gamma \sar w: \exists x \phi, \Delta$}
\DisplayProof

&
$\leadsto$
\end{tabular}
\end{flushleft}
\begin{flushright}
\AxiomC{$\rel, z \leq v, v \leq w,  \unda \in D_{z}, \unda \in D_{v}, \Gamma \sar w: \phi(\unda/x), w: \exists x \phi, \Delta$}
\RightLabel{$\nd$}
\UnaryInfC{$\rel, z \leq v, v \leq w,  \unda \in D_{z}, \Gamma \sar w: \phi(\unda/x), w: \exists x \phi, \Delta$}
\DisplayProof
\end{flushright}

Yet again, inspecting the rules of our labelled calculi will prove that the end sequent of the $(Pr_{id}^{-1})$ inference cannot be derived by other means, and that no rules permit the permutation of $\nd$ above $(\exists^{-1}_{r})$. Nonetheless, what is interesting is that both the end sequent of the $(Pr_{id}^{-1})$ derivation and the concluding sequent obtained from applying $\nd$ to the premise of $(\exists^{-1}_{r})$ contain domain atoms $\unda \in D_{z}$ associated with a label $z$ that is two steps in the past of the principal label $w$ (due to the relational atoms $z \leq v, v \leq w$). Therefore, similar to how permutations of the $\trans$ rule caused formulae to be propagated forward, we have found that permutations of the $\nd$ rule propagate domain atoms \emph{backward}. Likewise, an analysis of permuting the $\cd$ rule above $(Pr_{id})$ and $\existsr$ would have us conclude the \emph{forward} propagation of active domain atoms. In the constant domain setting then (where both $\nd$ and $\cd$ are present), successive permutations of $\nd$ and $\cd$ could cause the active domain atom(s) to propagate to any label within the labelled sequent (via a sequence of forward and backward `shifts' along relational atoms of the form $w \leq u$).

Such findings suggest that in the non-constant domain setting, we ought to impose a side condition on $(Pr_{id})$ and $\existsr$ stating that the principal label $w$ is reachable from the label(s) associated with the active domain atom(s) (via a directed path of relational atoms). Alternatively, in the constant domain setting, we ought to impose the side condition that the principal label $w$ is reachable from the label(s) associated with the active domain atom(s), \emph{but not necessarily with a directed path}. Imposing these side conditions yields the reachability rules $\idfonca$ and $\existsrnca$ shown in \fig~\ref{fig:reachability-propagation-rules-atoms-FO-Int}. The reachability rule $\alllnca$ is also displayed in that figure; the form of the rule and side conditions imposed were discovered through an analysis of permuting $\refl$, $\trans$, $\nd$ and $\cd$ above the $\alll$ rule, similar to what was done for $\id$, $\impl$, and $\existsr$ above.

\begin{figure}[t]
\begin{center}
\bgroup
\def\arraystretch{1.5}
\begin{tabular}{| c | c |} 
\hline
Name & Side Condition \\ 
\hline
$\dag_{1}(\ccc)$\index{$\dag_{1}(\ccc)$} &  $\exists \ppath_{i}(\stra_{\ppath_{i}}(u_{i},w) \in \thuesysiilang{a})$ for each \\
 &  $i \in \{1, \ldots, n\}$, and $\exists \ppath(\stra_{\ppath}(w,u) \in \thuesysilang{a})$\\
\hline
 $\dag_{2}(\ccc)$\index{$\dag_{2}(\ccc)$} &  $\exists \ppath(\stra_{\ppath}(u,w) \in \thuesysiilang{a})$\\
\hline
$\dag_{3}(\ccc)$\index{$\dag_{3}(\ccc)$} &  $\unda$ is an eigenvariable\\
 &  and $\exists \ppath(\stra_{\ppath}(u,w) \in \thuesysiilang{a})$ \\ 
\hline
$\dag_{4}(\ccc)$\index{$\dag_{4}(\ccc)$} &  $\exists \ppath(\stra_{\ppath}(u,v) \in \thuesysiilang{a})$ and $\exists \ppath(\stra_{\ppath}(w,v) \in \thuesysilang{a})$\\
\hline
$\dag_{5}(\ccc)$\index{$\dag_{5}(\ccc)$} &  $\unda$ is an eigenvariable, $\exists \ppath(\stra_{\ppath}(u,v) \in \thuesysiilang{a})$\\
 &  and $\exists \ppath(\stra_{\ppath}(w,v) \in \thuesysilang{a})$ \\ 
\hline
 $\dag_{6}(\ccc)$\index{$\dag_{6}(\ccc)$} &  $\exists \ppath(\stra_{\ppath}(w,u) \in \thuesysilang{a})$\\
\hline
\end{tabular}
\egroup
\end{center}
\caption{Side conditions for the reachability and propagation rules used to refine $\gtintfocd$.}
\label{fig:side-conditionsii-FO-Int}
\end{figure}

By what was said in part (i) of our analysis, we found that the \cfcst system $\thuesysi$ encodes a notion of \emph{directed} reachability. To encode a notion of \emph{undirected} reachability---used to formulate side conditions of rules in the constant domain setting---we make use of the following \cfcst system:\footnote{The name of $\thuesysii$ was chosen because the first two production rules correspond to the properties of reflexivity and Euclideanity, respectively, as shown in \fig~\ref{fig:frame-conditions-production-rules}. The name alludes to the logic $\mathsf{S5}$, which is the minimal normal modal logic sound and complete on all reflexive and Euclidean frames.}
$$
\thuesysii := \{a \pto \empstr, a \pto \conv{a} \cate a, \conv{a} \pto \empstr, \conv{a} \pto a \cate \conv{a}\}
$$
Recall that in the propagation graph of a labelled sequent (\dfn~\ref{def:propagation-graph-FOInt}) the character $a$ is used to encode a forward movement along a relational atom of the form $w \leq u$, and the converse character $\conv{a}$ is used to encode a reverse movement. Hence, two labels are reachable (\emph{in an undirected sense}) \ifandonlyif there is a propagation path whose string lies within the set $\albet^{*} = \{a,\conv{a}\}^{*}$. Since $\albet^{*} = \thuesysiilang{a}$, we know that the above \cfcst system properly encodes the notion of undirected reachability. Before moving onto step (iii) of our analysis, we consider an example of a concrete application of the $\alllna$ rule to demonstrate the functionality of a reachability rule in the current setting.

\begin{example} Let us consider the labelled sequent $\Lambda$ below, which has the propagation graph $\prgr{\Lambda}$ shown below bottom. Note that the vertices of the propagation graph have been decorated to make the corresponding formulae explicit.
$$
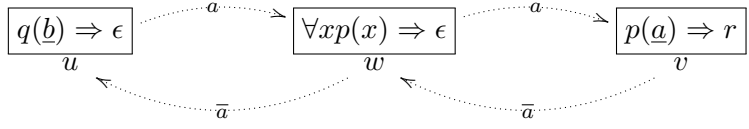
\Lambda :=  u \leq w, w \leq v, \unda \in D_{u}, u : q(\undb), w : \forall v p(x), v : p(\unda) \sar v : r
$$
\begin{center}
\begin{tabular}{c}
\xymatrix{
 \overset{\boxed{q(\undb) \sar \seqempstr}}{u}\ar@/^1.75pc/@{.>}[rr]|-{a} &  & \overset{\boxed{\forall x p(x) \sar \seqempstr}}{w}\ar@/^1.75pc/@{.>}[rr]|-{a} \ar@/^1.5pc/@{.>}[ll]|-{\conv{a}} &  & \overset{\boxed{p(\unda) \sar r}}{v}\ar@/^1.5pc/@{.>}[ll]|-{\conv{a}}
}
\end{tabular}
\end{center}
Since there exists a propagation path $\ppath(u,v) := u, a, w, a, v$ such that $\stra_{\ppath}(u,v) = a \cate a \in \thuesysilang{a}$ and a propagation path $\ppath'(w,v) := w, a, v$ such that $\stra_{\ppath}(w,v) = a \in \thuesysilang{a}$, we can apply the $\alllna$ rule to delete the labelled formula $v : p(\unda)$ and derive the following labelled sequent:
$$
\Lambda' :=  u \leq w, w \leq v, \unda \in D_{u}, u : q(\undb), w : \forall v p(x) \sar v : r
$$
\end{example}

A useful feature of our propagation and reachability rules (shown in \fig~\ref{fig:reachability-propagation-rules-atoms-FO-Int}) is that the propagation graph of the premise of the rule is identical to the propagation graph of the conclusion, meaning that the side condition can be checked regardless of if we are applying the rule top-down or bottom-up (the latter fact being useful for proof-search). Also, it should be noted that the side conditions are read in a similar manner to the side conditions of propagation rules for grammar logics introduced in \sect~\ref{SECT:Refine-Grammar}. For instance, the side condition `$\exists \ppath(\stra_{\ppath}(u,v) \in \thuesysiilang{a})$' states there exists a propagation path $\ppath(u,v)$ in the propagation graph of the premise or conclusion (depending on if the rule is being applied top-down or bottom-up) such that $\stra_{\ppath}(u,v) \in \thuesysiilang{a}$. Often times when discussing the side conditions of propagation or reachability rules we will omit mention of the propagation graph and take it for the granted that the propagation path exists in the propagation graph of the premise and/or conclusion of the rule being discussed.

Let us now move onto step (iii) of our analysis. We have determined that instead of employing the $\existsr$ and $\alll$ rules, we ought to make use of the reachability rules $\existsrnca$ and $\alllnca$, which subsume $\existsr$ and $\alll$, respectively (see \lem~\ref{lem:instance-of} below). It so happens that permuting $\ned$ above $\existsrnca$ and $\alllnca$ are the only non-trivial cases that occur when analyzing $\ned$ elimination. We analyze the non-trivial case that occurs when attempting to permute $\ned$ above $\alllnca$, which will motivate how both $\existsrnca$ and $\alllnca$ ought to be reformulated to allow for $\ned$ elimination.

\begin{center}
\AxiomC{$\rel, \unda \in D_{u}, w : \forall x \phi, v : \phi(\unda/x), \Gamma \Rightarrow \Delta$}
\RightLabel{$\alllnca$}
\UnaryInfC{$\rel, \unda \in D_{u}, w : \forall x \phi, \Gamma \Rightarrow \Delta$}
\RightLabel{$\ned$}
\UnaryInfC{$\rel, w : \forall x \phi, \Gamma \Rightarrow \Delta$}
\DisplayProof
\end{center}

The issue in the above derivation is that due to the existence of $v : \phi(\unda/x)$, which contains the parameter $\unda$, $\ned$ is inapplicable to the top sequent. The solution to our problem then, is to absorb the $\ned$ inference, yielding a new rule of the form:

\begin{center}
\AxiomC{$\rel, \unda \in D_{u}, w : \forall x \phi, v : \phi(\unda/x), \Gamma \Rightarrow \Delta$}
\RightLabel{$\alllncia$}
\UnaryInfC{$\rel, w : \forall x \phi, \Gamma \Rightarrow \Delta$}
\DisplayProof
\end{center}

\begin{figure}[t]
\begin{center}
\bgroup
\def\arraystretch{1.5}
\begin{tabular}{| c  c  c |}
\hline
Name & & Set of Rules\\
\hline
$\rulesna$\index{$\rulesna$} & $ := $ &  $\{\idfona, \existsrna, \existsrnia, \alllna, \alllnia, \primp\}$\\
$\rulesca$\index{$\rulesca$} & $ := $ &  $\{\idfoca, \existsrca, \existsrcia, \alllca, \alllcia, \primp\}$\\
\hline
\end{tabular}
\egroup
\end{center}
\caption{Sets of reachability and propagation rules.}
\label{fig:Rule-sets-FO-Int}
\end{figure}

We obtain the $\existsrncia$ rule in a similar fashion. Both of the reachability rules $\existsrncia$ and $\alllncia$ are shown in \fig~\ref{fig:reachability-propagation-rules-atoms-FO-Int}.

At this stage, we have completed our analysis, and move on to proving that the conducive rules shown in \fig~\ref{fig:reachability-propagation-rules-atoms-FO-Int} allow for the elimination of $\refl$, $\trans$, $\nd$, $\cd$, and $\ned$ in $\gtintfond$ and $\gtintfocd$. To simplify notation, the conducive rules are organized into two sets $\rulesna$ and $\rulesca$, shown in \fig~\ref{fig:Rule-sets-FO-Int}, with the former collecting the conducive rules for $\gtintfond$ and the latter collecting the conducive rules for $\gtintfocd$. Moreover, we use the symbol $\nnn$ to index rules and side conditions that are relevant to refining the calculus $\gtintfond$ and use the symbol $\ccc$ to index rules and side conditions that are relevant to refining the calculus $\gtintfocd$. When reference is made to a rule or side condition indexed with $\mathsf{X}$, it is taken to represent both the rule or side condition indexed with $\nnn$ and the rule or side condition indexed with $\ccc$ (e.g. $\idfonca \in \{\idfona, \idfoca\}$. Last, we present a formal definition for the \cfcst systems $\thuesysi$ and $\thuesysii$ for reference. 

\begin{definition}[The \cfcst Systems $\thuesysi$ and $\thuesysii$]\label{def:S4-S5-FOInt} We define the \cfcst systems $\thuesysi$ and $\thuesysii$ as follows:
$$
\thuesysi := \{a \pto \empstr, a \pto a \cate a, \conv{a} \pto \empstr, \conv{a} \pto \conv{a} \cate \conv{a}\}
$$
$$
\thuesysii := \{a \pto \empstr, a \pto \conv{a} \cate a, \conv{a} \pto \empstr, \conv{a} \pto a \cate \conv{a}\}
$$
\end{definition}




\section{Refinement Part II: Structural Rule Elimination}\label{subsect:refinement-part-two-FOInt}

In this section, we formally prove that the conducive rules shown in \fig~\ref{fig:reachability-propagation-rules-atoms-FO-Int} allow for the refinement of $\gtintfond$ and $\gtintfocd$, and ultimately, allow for the extraction of the nested calculi $\nintfond$ and $\nintfocd$ for $\intfond$ and $\intfocd$, respectively (discussed in \sect~\ref{subsect:translation-to-nested-FOInt}). Our first step is to show that the reachability and propagation rules $\idfonca$, $\primp$, $\existsrnca$, and $\alllnca$ subsume the rules $\idfo$, $\impl$, $\existsr$, and $\alll$. Afterward, we show that the structural rules (\emph{viz.} $\refl$, $\trans$, $\nd$, $\cd$, and $\ned$; see \fig~\ref{fig:labelled-calculi-FO-Int}) are eliminable in the labelled calculi $\gtintfond$ and $\gtintfocd$ when extended with the proper reachability and propagation rules (see \thm~\ref{thm:admissible-rules-G3-Calc} below), yielding the quasi-refined labelled calculi $\intfondlq$ and $\intfocdlq$ (defined in \dfn~\ref{def:quasi-refined-calculi} below). Last, we show that the quasi-refined labelled calculi $\intfondlq$ and $\intfocdlq$ are deductively equivalent to the labelled calculi $\gtintfond$ and $\gtintfocd$ (see \thm~\ref{thm:reach-prop-admissible-rules-G3-Calc} below), and therefore inherit proof-theoretic properties from their parental labelled calculi (see \thm~\ref{thm:properties-quasi-refined-FOInt} below). In the following section, we show how to rewrite the rules of the quasi-refined calculi $\intfondlq$ and $\intfocdlq$, rendering domain atoms expendable, and producing the fully refined labelled calculi $\intfondl$ and $\intfocdl$ from which the nested calculi $\nintfond$ and $\nintfocd$ may be easily extracted.

\begin{lemma}\label{lem:instance-of} \ 

(i) The rule $\idfo$ is an instance of $\idfona$ and $\idfoca$.

(ii) The $\impl$ rule is an instance of the $\primp$ rule.

(iii) The $\existsr$ rule is an instance of the rules $\existsrna$ and $\existsrca$.

(iv) The $\alll$ rule is an instance of the rules $\alllna$ and $\alllca$.

\end{lemma}

\begin{proof} We prove claim (i) below and defer the proofs of (ii)--(iv) to Appendix C (p.~\pageref{app:proofs}).

(i) We show that $\idfo$ is an instance of $\idfoca$ since showing that $\idfo$ is an instance of $\idfona$ is similar. By \dfn~\ref{def:S4-S5-FOInt}, we know that $\empstr \in \thuesysiilang{a}$ (due to the first production rule in $\thuesysii$). Moreover, by \dfn~\ref{def:propagation-path-kms}, we know that the empty path $\emppath(w,w) = w$ always holds between a label and itself, and has the string $\stra_{\emppath}(w,w) = \empstr$. Therefore, $\stra_{\emppath}(w,w) \in \thuesysiilang{a}$.

In the inference below, observe that the relational atom $w \leq u$ implies the existence of a propagation path $\ppath(w,u) := w, a, u$, where $\stra_{\ppath}(w,u) = a$. Since $a \in \thuesysilang{a}$, it follows that $\stra_{\ppath}(w,u) \in \thuesysilang{a}$. Therefore, by all that has been argued, we know that 
 the side condition $\dag_{1}(\ccc)$ of $\idfoca$ from \fig~\ref{fig:side-conditionsii-FO-Int} is indeed satisfied, and the following inference is an instance of $\idfoca$:
\begin{center}
\AxiomC{}
\RightLabel{$\idfoca$}
\UnaryInfC{$\rel, w \leq u, \unda_{1} \in D_{w}, \ldots, \unda_{n} \in D_{w}, \Gamma, w : p(\vec{\unda}) \Rightarrow u : p(\vec{\unda}), \Delta$}
\RightLabel{=}
\dottedLine
\UnaryInfC{$\rel,w \leq u, \vec{\unda} \in D_{w},\Gamma, w :p(\vec{\unda}) \sar u :p(\vec{\unda}), \Delta$}
\DisplayProof
\end{center}
\end{proof}

The above lemma confirms that our reachability and propagation rules are strengthened versions of the rules $\idfo$, $\impl$, $\existsr$, and $\alll$ from which they were extracted. Additionally, the above lemma will assist us in proving the elimination of structural rules from $\gtintfond$ and $\gtintfocd$ as less cases will need to be considered. We now show that all structural rules in our labelled calculi for first-order intuitionistic logics can be eliminated.

\begin{lemma}\label{lem:refl-elim-FOInt} \ 

(i) The rule $\refl$ is eliminable in $\gtintfond + \rulesna - \{\trans, \nd, \ned\}$.

(ii) The rule $\refl$ is eliminable in $\gtintfocd + \rulesca - \{\trans, \nd, \cd, \ned\}$. 
\end{lemma}

\begin{proof} We prove the result by induction on the height of the given derivation and show claim (ii) as it subsumes claim (i).

\textit{Base case.} The $\botl$ case is simple to verify, and we need not consider the $\idfo$ case due to \lem~\ref{lem:instance-of}. Hence, we consider the $\idfoca$ case, which is resolved as shown below:
\begin{flushleft}
\begin{tabular}{c c}
\AxiomC{}
\RightLabel{$\idfoca$}
\UnaryInfC{$\rel, v \leq v, \unda_{1} \in D_{u_{1}}, \ldots, \unda_{n} \in D_{u_{n}}, \Gamma, w : p(\vec{\unda}) \Rightarrow u : p(\vec{\unda}), \Delta$}
\RightLabel{$\refl$}
\UnaryInfC{$\rel, \unda_{1} \in D_{u_{1}}, \ldots, \unda_{n} \in D_{u_{n}}, \Gamma, w : p(\vec{\unda}) \Rightarrow u : p(\vec{\unda}), \Delta$}
\DisplayProof

&

$\leadsto$
\end{tabular}
\end{flushleft}
\begin{flushright}
\AxiomC{}
\RightLabel{$\idfoca$}
\UnaryInfC{$\rel, \unda_{1} \in D_{u_{1}}, \ldots, \unda_{n} \in D_{u_{n}}, \Gamma, w : p(\vec{\unda}) \Rightarrow u : p(\vec{\unda}), \Delta$}
\DisplayProof
\end{flushright}
If the relational atom $v \leq v$ that is active in $\refl$ occurs along one or more propagation paths $\ppath_{i}(u_{i},w)$ or along the propagation path $\ppath(w,u)$, then by replacing each occurrence of the path $v, a, v$ and $v, \conv{a}, v$ with the empty path $\emppath(v,v) = v$ in each such $\ppath_{i}(u_{i},w)$ and in $\ppath(w,u)$, we obtain propagation paths $\ppath_{i}'(u_{i},w)$ and a propagation path $\ppath'(w,u)$ that do not rely on the relational atom $v \leq v$. By assumption, we know that $\stra_{\ppath_{i}}(u_{i},w) \in \thuesysiilang{a}$ and $\stra_{\ppath}(w,u) \in \thuesysilang{a}$. By applying the production rules $a \pto \empstr, \conv{a} \pto \empstr \in \thuesysii$ and $a \pto \empstr, \conv{a} \pto \empstr \in \thuesysi$ to each $a$ and $\conv{a}$ obtained from the path $v, a, v$ and $v, \conv{a}, v$ (respectively) in $\stra_{\ppath_{i}}(u_{i},w)$ and $\stra_{\ppath}(w,u)$, we obtain the strings $\stra_{\ppath_{i}'}(u_{i},w)$ and $\stra_{\ppath'}(w,u)$, which, by the fact that they are derived from $\stra_{\ppath_{i}}(u_{i},w)$ and $\stra_{\ppath}(w,u)$ using the production rules $a \pto \empstr$ and $\conv{a} \pto \empstr$ ensures that they are elements of $\thuesysiilang{a}$ and $\thuesysilang{a}$, respectively.

\textit{Inductive step.} With the exception of the $\primp$, $\existsrca$, and $\alllca$ rules, all cases are handled by invoking \ih and then applying the corresponding rule. (NB. By \lem~\ref{lem:instance-of}, we may omit consideration of the $\impl$, $\existsr$, and $\alll$ rules.) We consider the $\primp$, $\existsrca$, and $\alllca$ cases.

$\primp$. Let $\ppath(w,u)$ be a propagation path in the premises of the top derivation below such that $\stra_{\ppath}(w,u) \in \thuesysilang{a}$. If the relational atom $v \leq v$ that is active in $\refl$ occurs along the propagation path $\ppath(w,u)$, then by replacing each occurrence of $v, a, v$ and $v, \conv{a}, v$ (obtained from the relational atom $v \leq v$) in $\ppath(w,u)$ with the empty path $\emppath(v,v) = v$, we obtain a new propagation path $\ppath'(w,u)$ that does not rely on the relational atom $v \leq v$. Since $a \pto \empstr, \conv{a} \pto \empstr \in \thuesysi$, we may apply the production rules to each $a$ and $\conv{a}$ in $\stra_{\ppath}(w,u)$ that was obtained from the path $v, a, v$ and the path $v, \conv{a}, v$ (respectively), yielding the string $\stra_{\ppath'}(w,u)$. Due to the fact that $\stra_{\ppath}(w,u) \in \thuesysilang{a}$ and $\stra_{\ppath}(w,u) \dto^{*}_{\thuesysi} \stra_{\ppath'}(w,u)$, we know that $\stra_{\ppath'}(w,u) \in \thuesysilang{a}$, thus showing that the side condition of the $\primp$ inference in the second derivation holds, allowing for the permutation to go through.
\begin{flushleft}
\begin{tabular}{c c}
\AxiomC{$\rel, v \leq v, w : \phi \imp \psi, \Gamma \Rightarrow \Delta, u : \phi$}
\AxiomC{$\rel, v \leq v, w : \phi \imp \psi, u : \psi, \Gamma \Rightarrow \Delta$}
\RightLabel{$\primp$}
\BinaryInfC{$\rel, v \leq v, w : \phi \imp \psi, \Gamma \Rightarrow \Delta$}
\RightLabel{$\refl$}
\UnaryInfC{$\rel, \Gamma, w :\phi \imp \psi \sar \Delta$}
\DisplayProof

&

$\leadsto$
\end{tabular}
\end{flushleft}
\begin{flushright}
\AxiomC{$\rel,v \leq v,, \Gamma, w :\phi \imp \psi \sar \Delta, u :\phi$}
\RightLabel{$\refl$}
\UnaryInfC{$\rel, \Gamma, w :\phi \imp \psi \sar \Delta, u :\phi$}

\AxiomC{$\rel,v \leq v,, \Gamma, w :\phi \imp \psi, u :\psi \sar \Delta$}
\RightLabel{$\refl$}
\UnaryInfC{$\rel, \Gamma, w :\phi \imp \psi, u :\psi \sar \Delta$}

\RightLabel{$\primp$}
\BinaryInfC{$\rel,\Gamma, w :\phi \imp \psi \sar \Delta$}
\DisplayProof
\end{flushright}

$\existsrca$. Let $\ppath(u,w)$ be a propagation path in the premise of the top left derivation below such that $\stra_{\ppath}(u,w) \in \thuesysiilang{a}$. If the relational atom $v \leq v$ that is active in the $\refl$ inference occurs along the propagation path $\ppath(u,w)$, then by replacing each occurrence of the path $v, a, v$ and $v, \conv{a}, v$ (obtained from the relational atom $v \leq v$ and where $\chara \in \albet$) with the empty path $\emppath(v,v) = v$, we obtain a new propagation path $\ppath'(u,w)$ that does not rely on the relational atom $v \leq v$. Because $a \pto \empstr, \conv{a} \pto \empstr \in \thuesysiilang{a}$, we can derive $\stra_{\ppath'}(u,w)$ from $\stra_{\ppath}(u,w)$ by applying the production rules to each $a$ and $\conv{a}$ in $\stra_{\ppath}(u,w)$ obtained from the paths $v, a, v$ and $v, \conv{a}, v$, respectively. It follows that $\stra_{\ppath'}(u,w) \in \thuesysiilang{a}$, implying that we may apply $\existsrca$ after applying $\refl$.

\begin{flushleft}
\begin{tabular}{c c}
\AxiomC{$\rel, v \leq v, \unda \in D_{u}, \Gamma \Rightarrow \Delta, w: \phi(\unda/x), w: \exists x \phi$}
\RightLabel{$\existsrca$}
\UnaryInfC{$\rel, v \leq v, \unda \in D_{u}, \Gamma \Rightarrow \Delta, w: \exists x \phi$}
\RightLabel{$\refl$}
\UnaryInfC{$\rel, \unda \in D_{u}, \Gamma \sar \Delta, w : \exists x \phi$}
\DisplayProof

&

$\leadsto$
\end{tabular}
\end{flushleft}
\begin{flushright}
\AxiomC{$\rel, v \leq v, \unda \in D_{u}, \Gamma \Rightarrow \Delta, w: \phi(\unda/x), w: \exists x \phi$}
\RightLabel{$\refl$}
\UnaryInfC{$\rel, \unda \in D_{u}, \Gamma \Rightarrow \Delta, w: \phi(\unda/x), w: \exists x \phi$}
\RightLabel{$\existsrca$}
\UnaryInfC{$\rel, \unda \in D_{u}, \Gamma \Rightarrow \Delta, w: \exists x \phi$}
\DisplayProof
\end{flushright}

$\alllca$. Let $\ppath_{1}(u,v)$ be the propagation path in the premise of the top left derivation below such that $\stra_{\ppath_{1}}(u,v) \in \thuesysiilang{a}$ and $\ppath_{2}(w,v)$ be the propagation path in the premise of the top left derivation below such that $\stra_{\ppath_{2}}(w,v) \in \thuesysilang{a}$. If the relational atom $z \leq z$ that is active in the $\refl$ inference occurs along the propagation path $\ppath_{1}(u,v)$ or $\ppath_{2}(w,v)$, then by replacing each path $z, a, z$ and $z, \conv{a}, z$ with the empty path $\emppath(z,z) = z$, we obtain new propagation paths $\ppath_{1}'(u,v)$ and $\ppath_{2}'(w,v)$ that do not rely on the relational atom $z \leq z$. Since $a \pto \empstr, \conv{a} \pto \empstr \in \thuesysii$ and $a \pto \empstr, \conv{a} \pto \empstr \in \thuesysi$ (\dfn~\ref{def:S4-S5-FOInt}), we may apply the production rules to each $a$ and $\conv{a}$ in $\stra_{\ppath_{1}}(u,v)$ and $\stra_{\ppath_{2}}(w,v)$ obtained from the paths $z, a, z$ and $z, \conv{a}, z$ (respectively), letting us derive the strings $\stra_{\ppath_{1}'}(u,v)$ and $\stra_{\ppath_{2}'}(w,v)$. It follows that $\stra_{\ppath_{1}'}(u,v) \in \thuesysiilang{a}$ and $\stra_{\ppath_{2}'}(w,v) \in \thuesysilang{a}$, implying that we may apply $\refl$ prior to $\alllca$.

\begin{flushleft}
\begin{tabular}{c c}
\AxiomC{$\rel, z \leq z, \unda \in D_{u}, w : \forall x \phi, v : \phi(\unda/x), \Gamma \Rightarrow \Delta$}
\RightLabel{$\alllca$}
\UnaryInfC{$\rel,  z \leq z, \unda \in D_{u}, w : \forall x \phi, \Gamma \Rightarrow \Delta$}
\RightLabel{$\refl$}
\UnaryInfC{$\rel, \unda \in D_{u}, w : \forall x \phi, \Gamma \Rightarrow \Delta$}
\DisplayProof

&

$\leadsto$
\end{tabular}
\end{flushleft}
\begin{flushright}
\AxiomC{$\rel, z \leq z, \unda \in D_{u}, w : \forall x \phi, v : \phi(\unda/x), \Gamma \Rightarrow \Delta$}
\RightLabel{$\refl$}
\UnaryInfC{$\rel, \unda \in D_{u}, w : \forall x \phi, v : \phi(\unda/x), \Gamma \Rightarrow \Delta$}
\RightLabel{$\alllca$}
\UnaryInfC{$\rel, \unda \in D_{u}, w : \forall x \phi, \Gamma \Rightarrow \Delta$}
\DisplayProof
\end{flushright}
\end{proof}

\begin{lemma}\label{lem:trans-elim-FOInt} \ 

(i) The rule $\trans$ is eliminable in $\gtintfond + \rulesna - \{\refl, \nd, \ned\}$.

(ii) The rule $\trans$ is eliminable in $\gtintfocd + \rulesca - \{\refl, \nd, \cd, \ned\}$. 
\end{lemma}

\begin{proof} We defer the proof to Appendix C (p.~\pageref{app:proofs}).
\end{proof}

\begin{lemma}\label{lem:nd-elim-FOInt} \ 

(i) The rule $\nd$ is eliminable in $\gtintfond + \rulesna - \{\refl, \trans, \ned\}$.

(ii) The rule $\nd$ is eliminable in $\gtintfocd + \rulesca - \{\refl, \trans, \cd, \ned\}$. 
\end{lemma}

\begin{proof} We prove the result by induction on the height of the given derivation and only show claim (ii) as it subsumes claim (i).

\textit{Base case.} The $\botl$ case is straightforward and we need not consider the $\idfo$ case by \lem~\ref{lem:instance-of}. If we apply the $\nd$ rule to $\idfoca$, and none of the principal domain atoms of $\idfoca$ are active in the application of $\nd$, then the conclusion is an instance of $\idfoca$. Let us suppose then that one of the principal domain atoms of $\idfoca$ is active in the application of $\nd$. We assume w.l.o.g. that the domain atom is $\unda_{1} \in D_{u_{1}}$, and argue how the case is resolved below:
\begin{flushleft}
\begin{tabular}{c c}
\AxiomC{}
\RightLabel{$\idfoca$}
\UnaryInfC{$\rel, v \leq u_{i}, \unda_{1} \in D_{v}, \unda_{1} \in D_{u_{1}}, \ldots, \unda_{n} \in D_{u_{n}}, \Gamma, w : p(\vec{\unda}) \Rightarrow u : p(\vec{\unda}), \Delta$}
\RightLabel{$\nd$}
\UnaryInfC{$\rel, v \leq u_{i}, \unda_{1} \in D_{v}, \ldots, \unda_{n} \in D_{u_{n}}, \Gamma, w : p(\vec{\unda}) \Rightarrow u : p(\vec{\unda}), \Delta$}
\DisplayProof

&

$\leadsto$
\end{tabular}
\end{flushleft}
\begin{flushright}
\AxiomC{}
\RightLabel{$\idfoca$}
\UnaryInfC{$\rel, v \leq u_{i}, \unda_{1} \in D_{v}, \ldots, \unda_{n} \in D_{u_{n}}, \Gamma, w : p(\vec{\unda}) \Rightarrow u : p(\vec{\unda}), \Delta$}
\DisplayProof
\end{flushright}
By the side condition of $\idfoca$, we know that there exists a propagation path $\ppath(u_{1},w)$ in the premise of the top left derivation such that $\stra_{\ppath}(u_{1},w) \in \thuesysiilang{a}$. By prefixing the propagation path $\ppath(u_{1},w)$ with $v, a, u_{i}$ (obtained from the relational atom $v \leq u_{i}$), we obtain a propagation path $\ppath'(v,w)$. If $\stra_{\ppath}(u_{1},w) \neq \empstr$, then $\ppath(u_{1},w)$ is either of the form $u_{1}, a, \ppath_{0}(z,w)$ or $u_{1}, \conv{a}, \ppath_{0}(z,w)$ for some label $z$ occurring in the premise of the top derivation above. As explained in \lem~\ref{lem:trans-elim-FOInt} above, $a \dto^{*}_{\thuesysii} a \cate a$, and by \dfn~\ref{def:S4-S5-FOInt}, $\conv{a} \pto a \cate \conv{a} \in \thuesysii$. If the leading character of $\stra_{\ppath}(u_{1},w)$ is $a$, then applying the former derivation to this occurrence of $a$ gives the string $\stra_{\ppath'}(v,w)$, and if the leading character of $\stra_{\ppath}(u_{1},w)$ is $\conv{a}$, then applying the aforementioned production rule gives the string $\stra_{\ppath'}(v,w)$. In either case, $\stra_{\ppath}(v,w) \in \thuesysiilang{a}$. If $\stra_{\ppath}(u_{1},w) = \empstr$, then $\stra_{\ppath'}(v,w) = a$, and since $a \in \thuesysiilang{a}$, it follows that $\stra_{\ppath'}(v,w) \in \thuesysiilang{a}$. Hence, the conclusion of the top left derivation is an instance of $\idfoca$ in its own right.

\textit{Inductive step.} By \lem~\ref{lem:instance-of}, we need not consider the $\impl$, $\existsr$, or $\alll$ cases. With the exception of the $\existsrca$ and $\alllca$ cases, all remaining cases are resolved by invoking \ih and then applying the corresponding rule. We consider the $\existsrca$ case as the $\alllca$ case is similar.

$\existsrca$. If the domain atom $\unda \in D_{u}$ deleted via the $\nd$ inference in the top left proof below is not active in the $\existsrca$ inference, then the two rules freely permute. Let us suppose then that the domain atom $\unda \in D_{u}$ deleted via the $\nd$ inference is active in the $\existsrca$ inference. By the side condition on $\existsrca$, we know that there exists a propagation path $\ppath(u,w)$ such that $\stra_{\ppath}(u,w) \in \thuesysiilang{a}$. By prefixing the propagation path $\ppath(u,w)$ with the path $v, a, u$ obtained from the relational atom $v \leq u$, we obtain a new propagation path $\ppath'(v,w)$. If $\stra_{\ppath}(u,w) \neq \empstr$, then $\ppath(u,w)$ is either of the form $u, a, \ppath_{0}(z,w)$ or $u, \conv{a}, \ppath_{0}(z,w)$ for some label $z$ occurring in the premise of the top derivation below. In \lem~\ref{lem:trans-elim-FOInt} above, we found that $a \dto^{*}_{\thuesysii} a \cate a$, and by \dfn~\ref{def:S4-S5-FOInt}, $\conv{a} \pto a \cate \conv{a} \in \thuesysii$. If the leading character of $\stra_{\ppath}(u,w)$ is $a$, then applying the former derivation to this occurrence of $a$ gives the string $\stra_{\ppath'}(v,w)$, and if the leading character of $\stra_{\ppath}(u,w)$ is $\conv{a}$, then applying the former production rule gives the string $\stra_{\ppath'}(v,w)$. Hence, regardless of which case holds $\stra_{\ppath'}(v,w) \in \thuesysiilang{a}$. If, on the other hand, $\stra_{\ppath}(u,w) = \empstr$, then $\stra_{\ppath'}(v,w) = a$, and since $a \in \thuesysiilang{a}$, we have that $\stra_{\ppath'}(v,w) \in \thuesysiilang{a}$. Hence, the $\nd$ rule may be applied before the $\existsrca$ rule, showing that the two rules may be permuted.
\begin{flushleft}
\begin{tabular}{c c}
\AxiomC{$\rel,v \leq u, \unda \in D_{v}, \unda \in D_{u}, \Gamma \Rightarrow \Delta, w: \phi(\unda/x), w: \exists x \phi$}
\RightLabel{$\existsrca$}
\UnaryInfC{$\rel, v \leq u, \unda \in D_{v}, \unda \in D_{u}, \Gamma \Rightarrow \Delta, w: \exists x \phi$}
\RightLabel{$\nd$}
\UnaryInfC{$\rel, v \leq u, \unda \in D_{v}, \Gamma \Rightarrow \Delta, w: \exists x \phi$}
\DisplayProof

&

$\leadsto$
\end{tabular}
\end{flushleft}
\begin{flushright}
\AxiomC{$\rel,v \leq u, \unda \in D_{v}, \unda \in D_{u}, \Gamma \Rightarrow \Delta, w: \phi(\unda/x), w: \exists x \phi$}
\RightLabel{$\nd$}
\UnaryInfC{$\rel,v \leq u, \unda \in D_{v}, \Gamma \Rightarrow \Delta, w: \phi(\unda/x), w: \exists x \phi$}
\RightLabel{$\existsrca$}
\UnaryInfC{$\rel,v \leq u, \unda \in D_{v}, \Gamma \Rightarrow \Delta, w: \phi(\unda/x)$}
\DisplayProof
\end{flushright}
\end{proof}

\begin{lemma}\label{lem:ned-elim-FOInt} \ 

(i) The rule $\ned$ is eliminable in $\gtintfond + \rulesna - \{\refl, \trans, \nd\}$.

(ii) The rule $\ned$ is eliminable in $\gtintfocd + \rulesca - \{\refl, \trans, \nd, \cd\}$. 
\end{lemma}

\begin{proof} We prove the result by induction on the height of the given derivation and show (ii) as it subsumes (i). 

\textit{Base case.} Any application of $\ned$ to $\botl$, $\idfo$, or $\idfoca$ yields another instance of rule.

\textit{Inductive step.} By \lem~\ref{lem:instance-of}, we need not consider the $\impl$, $\existsr$, or $\alll$ cases. With the exception of the $\existsrca$ and $\alllca$ cases, all remaining cases are resolved by applying \ih and then the corresponding rule. We show the non-trivial $\existsrca$ and $\alllca$ cases below:

\begin{flushleft}
\begin{tabular}{c c}
\AxiomC{$\rel, \unda \in D_{u}, \Gamma \Rightarrow \Delta, w: \phi(\unda/x), w: \exists x \phi$}
\RightLabel{$\existsrca$}
\UnaryInfC{$\rel, \unda \in D_{u}, \Gamma \Rightarrow \Delta, w: \exists x \phi$}
\RightLabel{$\ned$}
\UnaryInfC{$\rel, \Gamma \Rightarrow \Delta, w: \exists x \phi$}
\DisplayProof

&

$\leadsto$
\end{tabular}
\end{flushleft}
\begin{flushright}
\AxiomC{$\rel, \unda \in D_{u}, \Gamma \Rightarrow \Delta, w: \phi(\unda/x), w: \exists x \phi$}
\RightLabel{$\existsrcia$}
\UnaryInfC{$\rel, \Gamma \Rightarrow \Delta, w: \exists x \phi$}
\DisplayProof
\end{flushright}

\begin{flushleft}
\begin{tabular}{c c}
\AxiomC{$\rel, \unda \in D_{u}, w : \phi(\unda/x), w : \forall x \phi, \Gamma \Rightarrow \Delta$}
\RightLabel{$\alllca$}
\UnaryInfC{$\rel, \unda \in D_{u}, w : \forall x \phi, \Gamma \Rightarrow \Delta$}
\RightLabel{$\ned$}
\UnaryInfC{$\rel, w : \forall x \phi, \Gamma \Rightarrow \Delta$}
\DisplayProof

&

$\leadsto$
\end{tabular}
\end{flushleft}
\begin{flushright}
\AxiomC{$\rel, \unda \in D_{u}, w : \phi(\unda/x), w : \forall x \phi, \Gamma \Rightarrow \Delta$}
\RightLabel{$\alllca$}
\UnaryInfC{$\rel, w : \forall x \phi, \Gamma \Rightarrow \Delta$}
\DisplayProof
\end{flushright}
\end{proof}

\begin{lemma}\label{lem:cd-elim-FOInt}
The rule $\cd$ is eliminable in $\gtintfocd + \rulesca - \{\refl, \trans, \nd, \ned\}$.
\end{lemma}

\begin{proof} The proof is similar to the proof of \lem~\ref{lem:nd-elim-FOInt}.
\end{proof}

\begin{theorem}\label{thm:admissible-rules-G3-Calc} \ 

(i) The rules  $\{\idfo, \impl, \existsr, \alll, \refl, \trans, \nd, \ned\}$ are admissible in the calculus $\gtintfond + \rulesna$.

(ii) The rules $\{\idfo, \impl, \existsr, \alll, \refl, \trans, \nd, \ned, \cd\}$ are admissible in the calculus $\gtintfocd + \rulesca$. 
\end{theorem}

\begin{proof} Follows from \lem~\ref{lem:instance-of} -- \ref{lem:cd-elim-FOInt}.
\end{proof}

We have formally confirmed that our conducive rules $\rulesna$ and $\rulesca$ are sufficient to allow for structural rule elimination in $\gtintfond$ and $\gtintfocd$, respectively. Removing all structural rules, along with all unnecessary rules (e.g. $\impl$ which is subsumed by $\primp$), from $\gtintfond$ and $\gtintfocd$ gives us our quasi-refined labelled calculi, defined below:

\begin{definition}\label{def:quasi-refined-calculi}
We define the quasi-refined labelled calculi $\intfondlq$\index{$\intfondlq$} and $\intfocdlq$\index{$\intfocdlq$} as follows:
$$
\intfondlq := \{\idfona,\botl,\disl,\disr,\conl,\conr,\primp,\impr,\existsrna,\existsrnia,\alllna,\alllnia\}
$$
$$
\intfocdlq := \{\idfoca,\botl,\disl,\disr,\conl,\conr,\primp,\impr,\existsrca,\existsrcia,\alllca,\alllcia\}
$$
\end{definition}

We could, at this stage, directly prove that our quasi-refined labelled calculi $\intfondlq$ and $\intfocdlq$ possess desirable proof-theoretic properties, as we did for $\gtintfond$ and $\gtintfocd$ in \sect~\ref{sec:lab-calc-intFO} (i.e. without relying on proof transformations). However, we will take a separate route---similar to what was done in \sect~\ref{SECT:Refine-Grammar} for grammar logics---and show that proofs in our quasi-refined labelled calculi $\intfondlq$ and $\intfocdlq$ can be algorithmically transformed into proofs in $\gtintfond$ and $\gtintfocd$ (\resp), and vice-versa. This approach shows that our quasi-refined labelled calculi inherit proof-theoretic properties from their parental labelled calculi, and also establishes a relationship between proofs within the different settings. 
 To accomplish this aim, we first show the following lemma:

\begin{lemma}\label{lem:deleting-relational-atoms-FOInt}
Let $\Lambda := \rel, \Gamma \sar \Delta$. Suppose $\rel, w \leq u, \Gamma \sar \Delta$ is derivable in $\gtintfond$ or $\gtintfocd$, and that $\ppath(w,u)$ is a propagation path (potentially empty) in $\prgr{\Lambda}$ such that $\stra_{\ppath}(w,u) \in \thuesysilang{a}$. 
 Then, $\rel, \Gamma \sar \Delta$ is derivable in $\gtintfond$ or $\gtintfocd$, respectively.
\end{lemma}

\begin{proof} We prove the result for $\gtintfond$ as the proof for $\gtintfocd$ is similar. Since $\stra_{\ppath}(w,u) \in \thuesysilang{a}$, we know that $a \dto^{*}_{\thuesysi} \stra_{\ppath}(w,u)$. The lemma is shown by induction on the length (\dfn~\ref{def:derivation-relation-language-kms}) of the derivation $a \dto^{*}_{\thuesysi} \stra_{\ppath}(w,u)$.

\textit{Base case.} For the base case, we consider (i) a derivation of length $0$, meaning that $a \dtoann a$, and (ii) a derivation of length $1$, meaning that $a \pto \stra_{\ppath}(w,u) \in \thuesysi$, and therefore, either (ii.1) $a \pto \stra_{\ppath}(w,u) = a \pto \empstr$, or (ii.2) $a \pto \stra_{\ppath}(w,u) = a \pto a \cate a$. (NB. It cannot be the case that $a \pto \stra_{\ppath}(w,u) \in \thuesysi$ is the production rule $\conv{a} \pto \empstr$ or $\conv{a} \pto \conv{a} \cate \conv{a}$ since both have the character $\conv{a}$ as their head.) Cases (i), (ii.1), and (ii.2) are respectively shown below, where $\rel := \rel', w \leq u$ in case (i) and $\rel := \rel', w \leq u, u \leq v$ in case (ii.2).

\begin{center}
\begin{tabular}{c c c}
\AxiomC{$\Pi_{1}$}
\UnaryInfC{$\rel', w \leq u, w \leq u, \Gamma \sar \Delta$}
\DisplayProof

&

$\leadsto$

&

\AxiomC{$\Pi_{1}$}
\UnaryInfC{$\rel', w \leq u, w \leq u, \Gamma \sar  \Delta$}
\RightLabel{$\ctrrel$}
\dashedLine
\UnaryInfC{$\rel', w \leq u, \Gamma \sar \Delta$}
\DisplayProof
\end{tabular}
\end{center}

\begin{center}
\begin{tabular}{c c c}
\AxiomC{$\Pi_{2}$}
\UnaryInfC{$\rel, w \leq w, \Gamma \sar \Delta$}
\DisplayProof

&

$\leadsto$

&

\AxiomC{$\Pi_{2}$}
\UnaryInfC{$\rel, w \leq w, \Gamma \sar \Delta$}
\RightLabel{$\refl$}
\UnaryInfC{$\rel, \Gamma \sar \Delta$}
\DisplayProof
\end{tabular}
\end{center}

\begin{center}
\begin{tabular}{c c c}
\AxiomC{$\Pi_{3}$}
\UnaryInfC{$\rel', w \leq u, u \leq v, w \leq v, \Gamma \sar \Delta$}
\DisplayProof

&

$\leadsto$

&

\AxiomC{$\Pi_{2}$}
\UnaryInfC{$\rel', w \leq u, u \leq v, w \leq v, \Gamma \sar \Delta$}
\RightLabel{$\trans$}
\UnaryInfC{$\rel', w \leq u, u \leq v, \Gamma \sar \Delta$}
\DisplayProof
\end{tabular}
\end{center}

\textit{Inductive step.} Let $\Lambda := \rel, \Gamma \sar \Delta$ and assume that we have a proof $\Pi$ of $\rel, w \leq u, \Gamma \sar \Delta$. Suppose our derivation $a \dto^{*}_{\thuesysi} \stra_{\ppath}(w,u)$ is of length $n+1$. Recall that the length of a derivation is the minimal number of one-step derivations necessary to derive the output string from the input string in the \cfcst system (\dfn~\ref{def:derivation-relation-language-kms}). Since we are making use of the \cfcst system $\thuesysi$, it must be the case that our derivation consists of a sequence of production rules of the form $a \pto a \cate a$ starting with the forward character $a \in \albet^{+}$. Also, because our derivation is of length $n+1$, we know that it consists of a derivation $a \dto^{*}_{\thuesysi} \strb$ of length $n$ followed by a one-step derivation $\strb \dto_{\thuesysi} \stra_{\ppath}(w,u)$. Hence, there exist strings $\strc_{0}, \strc_{1} \in (\albet^{+})^{*}$ such that $\strb = \strc_{0} \cate a \cate \strc_{1}$ and $\strc_{0} \cate a \cate a \cate \strc_{1} = \stra_{\ppath}(w,u)$. This implies the existence of a propagation path of the form $\ppath_{\strc_{0}}(w,v_{1}), a, z, a, \ppath_{\strc_{1}}(v_{2},u)$ in $\prgr{\Lambda}$, where $v_{1}, v_{2}, z \in \lab(\Lambda)$. It follows that $\rel$ must be of the form $\rel', v_{1} \leq z, z \leq v_{2}$. Using this fact, we derive the desired conclusion as follows:
\begin{center}
\AxiomC{$\Pi$}
\UnaryInfC{$\rel', v_{1} \leq z, z \leq v_{2}, w \leq u, \Gamma \sar \Delta$}
\RightLabel{$\wk$}
\dashedLine
\UnaryInfC{$\rel', v_{1} \leq z, z \leq v_{2}, v_{1} \leq v_{2}, w \leq u, \Gamma \sar \Delta$}
\RightLabel{\ih}
\dashedLine
\UnaryInfC{$\rel', v_{1} \leq z, z \leq v_{2}, v_{1} \leq v_{2}, \Gamma \sar \Delta$}
\RightLabel{$\trans$}
\UnaryInfC{$\rel', v_{1} \leq z, z \leq v_{2}, \Gamma \sar \Delta$}
\RightLabel{=}
\dottedLine
\UnaryInfC{$\rel, \Gamma \sar \Delta$}
\DisplayProof
\end{center}
We know that a propagation path of the form $\ppath_{\strc_{0}}(w,v_{1}), a, z, a, \ppath_{\strc_{1}}(v_{2},u)$ exists in $\rel$, so after applying $\wk$ to introduce the relational atom $v_{1} \leq v_{2}$, we know that a propagation path of the form $\ppath'(w,u) := \ppath_{\strc_{0}}(w,v_{1}), a, \ppath_{\strc_{1}}(v_{2},u)$ exists in $\rel, v_{1} \leq v_{2}$ (since we can take a detour through the path $v_{1}, a, v_{w}$ instead of $v_{1}, a, z, a, v_{2}$). Observe that the string $\stra_{\ppath'}(w,u) = t$ has a derivation of length $n$, which means that we may invoke \ih to delete the relational atom $w \leq u$. The remaining steps of the proof are self-explanatory, and thus, we have resolved the inductive step.
\end{proof}

\begin{theorem}\label{thm:reach-prop-admissible-rules-G3-Calc} Let $\mathsf{X} \in \{\nnn, \ccc\}$. Every proof of a labelled sequent $\Lambda$ in $\mathsf{IntXL^{*}}$ can be algorithmically transformed into a proof in $\mathsf{G3IntX}$.


\end{theorem}

\begin{proof} We prove the result for $\intfocdlq$ and $\gtintfocd$ as the proof for $\intfondlq$ and $\gtintfond$ is similar. All rules common to both $\intfocdl$ and $\gtintfocd$ straightforwardly translate, so we only show how to translate each rule from $\rulesca$, and consider each in turn below.

$\idfoca$. Let us consider an instance of $\idfoca$:
\begin{center}
\AxiomC{}
\RightLabel{$\idfoca$}
\UnaryInfC{$\rel, \unda_{1} \in D_{u_{1}}, \ldots, \unda_{n} \in D_{u_{n}}, \Gamma, w : p(\vec{\unda}) \Rightarrow u : p(\vec{\unda}), \Delta$}
\DisplayProof
\end{center}
By the side condition imposed on $\idfoca$, we know that there exist propagation paths $\ppath_{i}(u_{i},w)$ for $i \in \{1, \ldots, n\}$ such that $\stra_{\ppath_{i}}(u_{i},w) \in \thuesysiilang{a}$ and a propagation path $\ppath(w,u)$ such that $\stra_{\ppath}(w,u) \in \thuesysilang{a}$. We may assume w.l.o.g. that the propagation paths $\ppath_{i}(u_{i},w)$ are \emph{minimal}, that is, no propagation paths $\ppath_{i}'(u_{i},w)$ exist such that $\lenstr{\stra_{\ppath_{i}'}(u_{i},w)} < \lenstr{\stra_{\ppath_{i}}(u_{i},w)}$. Since each propagation path $\ppath_{i}(u_{i},w)$ is minimal, we know that no label occurs more than once in each propagation path, which is important in applying the $\nd$ and $\cd$ rules below. For $i \in \{1, \ldots, n\}$, we let $\unda_{i} \in D_{\ppath_{i}} := \unda_{i} \in D_{u_{i}}, \unda_{i} \in D_{v_{1}^{i}}, \ldots, \unda_{i} \in D_{v_{k_{i}}^{i}}$, where $\ppath_{i}(u_{i},w) := u_{i}, \chara_{0}^{i}, v_{1}^{i}, \ldots, v_{k_{i}}^{i}, \chara_{k_{i}}^{i}, w$ and $\chara_{j}^{i} \in \albet = \{a,\conv{a}\}$ with $j \in \{1, \ldots, k_{i}\}$. Let $n_{i}(a)$ and $n_{i}(\conv{a})$ be the respective number of occurrences of $a$ and $\conv{a}$ in the string $\stra_{\ppath_{i}}(u_{i},w)$.  

To prove $\idfoca$ admissible in $\gtintfocd$, we first weaken in the domain atoms $\unda_{i} \in D_{\ppath_{i}}$ for each $i \in \{1, \ldots, n\}$ as shown in the derivation below. Then, we perform a \emph{domain atom deletion procedure} and successively delete domain atoms along the propagation paths $\ppath_{i}(u_{i},w)$ by first deleting the domain atom $\unda_{i} \in D_{w}$ at $w$ and working our way backward toward the initial node of the path $u_{i}$ through applications of $\nd$ and $\cd$. We let $(d_{i}) := \nd \times n_{i}(a) + \cd \times n_{i}(\conv{a})$ represent this sequence of $\nd$ and $\cd$ applications; note that by the definition of $n_{i}(a)$ and $n_{i}(\conv{a})$, there will be $n_{i}(a)$ applications of $\nd$ and $n_{i}(\conv{a})$ applications of $\cd$. Also, the minimality of each propagation path $\ppath_{i}(u_{i},w)$ is significant here as it implies the non-existence of repetitions of labels in $\ppath_{i}(u_{i},w)$, which, if present, could potentially block the deletion of domain atoms within the domain atom deletion procedure described above, thus causing the procedure to halt prematurely without outputting the desired result.\footnote{To demonstrate the significance of minimality, suppose for the sake of argument that a path $v, a, v', a, v$ (with a repetition of $v$) occurs within a propagation path $\ppath_{i}(u_{i},w)$. Then, at some stage of the domain atom deletion procedure, we will have derived a labelled sequent of the form $\Lambda := \rel', v \leq v', v' \leq v, \unda_{i} \in D_{v'}, \unda_{i} \in D_{v}, \Gamma' \sar \Delta'$. Since we delete domain atoms in reverse from the end node $w$ toward the initial node $u_{i}$ of the propagation path $\ppath_{i}(u_{i},w)$, the rule $\nd$ will be applied to $\Lambda$ deleting the domain atom $\unda_{i} \in D_{v}$, and yielding the labelled sequent $\Lambda' := \rel', v \leq v', v' \leq v, \unda_{i} \in D_{v'}, \Gamma' \sar \Delta'$ due to the path $v', a, v$ occurring in $\ppath_{i}(u_{i},w)$. At this point, following the domain atom deletion procedure, we should apply $\nd$ to delete the domain atom $\unda_{i} \in D_{v'}$ from $\Lambda'$ (as the path $v', a, v$ of $v, a, v', a, v$ in $\ppath_{i}(u_{i},w)$ was just processed, and so, we must process the path $v, a, v'$ of $v, a, v', a, v$ in $\ppath_{i}(u_{i},w)$), but since we are not guaranteed the existence of a domain atom $\unda_{i} \in D_{v}$ in $\Lambda'$ (as the only domain atom of such a shape and that was guaranteed to exist was just deleted), we are not necessarily permitted to apply the $\nd$ rule. Hence, the procedure halts without the desired result being obtained. 
} Last, we invoke \lem~\ref{lem:deleting-relational-atoms-FOInt} to delete the relational atom $w \leq u$, which is permitted due to the side condition of $\idfoca$.
\begin{center}
\AxiomC{}
\RightLabel{$\idfo$}
\UnaryInfC{$\rel,w \leq u, \unda_{1} \in D_{w}, \ldots, \unda_{n} \in D_{w}, \Gamma, w :p(\vec{\unda}) \sar u :p(\vec{\unda}), \Delta$}
\RightLabel{$\wk$}
\dashedLine
\UnaryInfC{$\rel,w \leq u, \unda_{1} \in D_{\ppath_{1}}, \ldots, \unda_{n} \in D_{\ppath_{n}}, \unda_{1} \in D_{w}, \ldots, \unda_{n} \in D_{w}, \Gamma, w :p(\vec{\unda}) \sar u :p(\vec{\unda}), \Delta$}
\RightLabel{$(d_{1})$}
\UnaryInfC{$\rel,w \leq u, \unda_{1} \in D_{u_{1}}, \ldots, \unda_{n} \in D_{\ppath_{n}}, \unda_{2} \in D_{w}, \ldots, \unda_{n} \in D_{w}, \Gamma, w :p(\vec{\unda}) \sar u :p(\vec{\unda}), \Delta$}
\UnaryInfC{$\vdots$}
\UnaryInfC{$\rel,w \leq u, \unda_{1} \in D_{u_{1}}, \ldots, \unda_{n} \in D_{\ppath_{n}}, \unda_{n} \in D_{w}, \Gamma, w :p(\vec{\unda}) \sar u :p(\vec{\unda}), \Delta$}
\RightLabel{$(d_{n})$}
\UnaryInfC{$\rel,w \leq u, \unda_{1} \in D_{u_{1}}, \ldots, \unda_{n} \in D_{u_{n}}, \Gamma, w :p(\vec{\unda}) \sar u :p(\vec{\unda}), \Delta$}
\RightLabel{\lem~\ref{lem:deleting-relational-atoms-FOInt}}
\dashedLine
\UnaryInfC{$\rel, \unda_{1} \in D_{u_{1}}, \ldots, \unda_{n} \in D_{u_{n}}, \Gamma, w :p(\vec{\unda}) \sar u :p(\vec{\unda}), \Delta$}
\DisplayProof
\end{center}

$\primp$. Suppose that the side condition of a $\primp$ instance holds, i.e. a propagation path $\ppath(w,u)$ exists such that $\stra_{\ppath}(w,u) \in \thuesysilang{a}$. Using this fact, we may invoke \lem~\ref{lem:deleting-relational-atoms-FOInt} below, allowing us to obtain our desired conclusion:
\begin{center}
\AxiomC{$\rel, w : \phi \imp \psi, \Gamma \Rightarrow \Delta, u : \phi$}
\RightLabel{$\wk$}
\dashedLine
\UnaryInfC{$\rel, w \leq u, w : \phi \imp \psi, \Gamma \Rightarrow \Delta, u : \phi$}

\AxiomC{$\rel, w : \phi \imp \psi, u : \psi, \Gamma \Rightarrow \Delta$}
\RightLabel{$\wk$}
\dashedLine
\UnaryInfC{$\rel, w \leq u, w : \phi \imp \psi, u : \psi, \Gamma \Rightarrow \Delta$}

\RightLabel{$\impl$}
\BinaryInfC{$\rel, w \leq u, w : \phi \imp \psi, \Gamma \Rightarrow \Delta$}
\dashedLine
\RightLabel{\lem~\ref{lem:deleting-relational-atoms-FOInt}}
\UnaryInfC{$\rel, w : \phi \imp \psi, \Gamma \Rightarrow \Delta$}
\DisplayProof
\end{center}

$\existsrca$. Suppose that the side condition of an $\existsrca$ inference holds, that is, there exists a propagation path $\ppath(u,w)$ such that $\stra_{\ppath}(u,w) \in \thuesysiilang{a}$. We assume w.l.o.g. that $\ppath(u,w)$ is minimal (as in the base case), let $\ppath(u,w) := u, \chara_{0}, v_{1}, \ldots v_{n}, \chara_{n}, w$ (with $\chara_{i} \in \albet = \{a,\conv{a}\}$ for $i \in \{1, \ldots, n\}$), and let $n(a)$ and $n(\conv{a})$ be the number of occurrences of $a$ and $\conv{a}$ in $\stra_{\ppath}(u,w)$, respectively. To prove the rule admissible in $\gtintfocd$, we first apply hp-admissibility of $\wk$ to the premise of the rule (as shown below) and add domain atoms along the path $\ppath(u,w)$. Then, we apply $\existsr$ a single time, and last apply the $\nd$ and $\cd$ rules to delete the weakened in domain atoms by starting at $w$ and working backward toward $u$ in $\ppath(u,w)$. The $\nd$ rule will be applied $n(a)$ times and the $\cd$ rule will be applied $n(\conv{a})$ times, and we use $(d) := \nd \times n(a) + \cd \times n(\conv{a})$ to denote these inferences.
\begin{center}
\AxiomC{$\R, \unda \in D_{u}, \Gamma \Rightarrow \Delta, w: \phi(\unda/x), w: \exists x \phi$}
\RightLabel{$\wk$}
\dashedLine
\UnaryInfC{$\R, \unda \in D_{u}, \unda \in D_{v_{1}}, \ldots, \unda \in D_{v_{n}}, \unda \in D_{w},  \Gamma \Rightarrow \Delta, w: \phi(\unda/x), w: \exists x \phi$}
\RightLabel{$\existsr$}
\UnaryInfC{$\R, \unda \in D_{u}, \unda \in D_{v_{1}}, \ldots, \unda \in D_{v_{n}}, \unda \in D_{w},  \Gamma \Rightarrow \Delta, w: \exists x \phi$}
\RightLabel{$(d)$}
\UnaryInfC{$\R, \unda \in D_{u}, \Gamma \Rightarrow \Delta, w: \exists x \phi$}
\DisplayProof
\end{center}

$\alllca$. Similar to the proof of $\existsrca$ above.

$\existsrcia$. Suppose that the side condition holds, i.e. there exists a propagation path $\ppath(u,w)$ such that $\stra_{\ppath}(u,w) \in \thuesysiilang{a}$. We assume w.l.o.g. that $\ppath(u,w)$ is minimal (as in the base case), let $\ppath(u,w) := u, \chara_{0}, v_{1}, \ldots, v_{n}, \chara_{n}, w$ (with $\chara_{i} \in \albet = \{a,\conv{a}\}$ for $i \in \{1, \ldots, n\}$), and let $n(a)$ and $n(\conv{a})$ be the number of occurrences of $a$ and $\conv{a}$ in $\stra_{\ppath}(u,w)$, respectively. To prove $\existsrcia$ admissible, we first apply hp-admissibility of $\wk$ to add in domain atoms along the propagation path $\ppath(u,w)$, and use the domain atom $\unda \in D_{w}$ to apply $\existsr$. After this, we successively delete each of the weakened in domain atoms by applying $\nd$ and $\cd$ starting at $w$ and working our way backward toward the label $u$ in the propagation path $\ppath(u,w)$. The rule $\nd$ will be applied $n(a)$ times and $\cd$ will be applied $n(\conv{a})$ times; we let $(d) := \nd \times n(a) + \cd \times n(\conv{a})$ represent the successive applications of $\nd$ and $\cd$ that are applied to delete the weakened in domain atoms. Last, we apply $\ned$ to derive the desired conclusion, thus showing the rule admissible in $\gtintfocd$. 
\begin{center}
\AxiomC{$\rel, \unda \in D_{u}, \Gamma \Rightarrow w : \phi(\unda / x), w : \exists x \phi, \Delta$}
\RightLabel{$\wk$}
\dashedLine
\UnaryInfC{$\R, \unda \in D_{u}, \unda \in D_{v_{1}}, \ldots, \unda \in D_{v_{n}}, \unda \in D_{w},  \Gamma \Rightarrow \Delta, w: A(\unda/x), w: \exists x A$}
\RightLabel{$\existsr$}
\UnaryInfC{$\R, \unda \in D_{u}, \unda \in D_{v_{1}}, \ldots, \unda \in D_{v_{n}}, \unda \in D_{w},  \Gamma \Rightarrow \Delta, w: \exists x A$}
\RightLabel{$(d)$}
\UnaryInfC{$\R, \unda \in D_{u}, \Gamma \Rightarrow \Delta, w: \exists x A$}
\RightLabel{$\ned$}
\UnaryInfC{$\rel, \Gamma \Rightarrow w : \exists x \phi, \Delta$}
\DisplayProof
\end{center}

$\alllcia$. Similar to the proof of $\existsrcia$ above.

\end{proof}

We may now leverage \thm~\ref{thm:admissible-rules-G3-Calc} and~\ref{thm:reach-prop-admissible-rules-G3-Calc} to show that our quasi-refined labelled calculi inherit proof-theoretic properties from their parental labelled calculi:

\begin{theorem}[Proof-theoretic Properties of $\intfondlq$ and $\intfocdlq$]\label{thm:properties-quasi-refined-FOInt} Let $\mathsf{X} \in \{\nnn, \ccc\}$.

(i) All rules from $\strucsetint$ are admissible in $\mathsf{IntXL}^{*}$.

(ii) All rules in $\mathsf{IntXL}^{*}$ are invertible.

(iii) If $\vdash_{\mathsf{IntXL}^{*}} \Lambda$, $\models_{\mathsf{IntX}} \Lambda$.

(iv) If $\vdash_{\mathsf{IntX}} \phi$, then $\vdash_{\mathsf{IntXL}^{*}} \seqempstr \sar w : \phi$.
\end{theorem}

\begin{proof} Claims (i) -- (iv) follow from \thm~\ref{thm:soundness-FO-Int}, \ref{thm:cut-admiss-FO-Int}, \ref{thm:completness-FO-Int}, \ref{thm:admissible-rules-G3-Calc},~\ref{thm:reach-prop-admissible-rules-G3-Calc}, and \lem~\ref{lem:lsub-admiss-FO-Int}, \ref{lem:psub-admiss-FO-Int}, \ref{lem:wk-admiss-FO-Int}, \ref{lem:invert-FO-Int}, and \ref{lem:ctr-admiss-FO-Int}.
\end{proof}


Before moving on to the next section on domain atom removal, we show that our quasi-refined labelled calculi are sound and complete relative to labelled tree derivations with the fixed root property. The proof of this result is stated below, and although the proof is sufficient to confirm that our quasi-refined labelled calculi are `simpler' than their parental labelled calculi---as the latter are incomplete relative to labelled tree derivations (\thm~\ref{thm:tree-incompleteness-FO-Int})---it does not explain \emph{why} structural rule elimination yields calculi that only require labelled tree sequents. 

To answer such a question, we recognize observations made in~\cite{CiaLyoRam18,Lyo20a,Lyo21}: The rules $\refl$ and $\trans$ allow for labelled sequents to occur in derivations which are not labelled tree sequents. To demonstrate this fact, observe the following derivation:
\begin{center}
\AxiomC{$w \leq v, v \leq v, v : p \sar v : p$}
\RightLabel{$\refl$}
\UnaryInfC{$w \leq v, v : p \sar v : p$}
\RightLabel{$\impr$}
\UnaryInfC{$\sar w : p \imp p$}
\DisplayProof
\end{center}
As we can see in the example above, the initial sequent is \emph{not} a labelled tree sequent due to the existence of the relational atom $v \leq v$ (one can check its sequent graph via \dfn~\ref{def:sequent-graph-FO-Int}). Once $\refl$ is applied however, a labelled tree sequent results. Most importantly, we can see that the derivation consists of two fragments: a top fragment that does not use labelled tree sequents (i.e. the initial sequent), and a bottom fragment that does use labelled tree sequents (i.e. the bottom two sequents).

It turns out that the above observation---i.e. that a labelled derivation in $\intfondlq$ or $\intfocdlq$ can be partitioned into a `treelike' and `non-treelike' fragment---always holds, if we consider derivations that derive a labelled formula $w : \phi$. Due to the fact that the end sequent of such a derivation is of the form $\seqempstr \sar w : \phi$, there will necessarily be a bottom fragment containing labelled tree sequents. If we consider the derivation in a bottom-up manner, then all rules---with the exception of $\refl$ and $\trans$---will either preserve relational atoms or add new ones (e.g. $\impr$ and $\allr$), which constructs a tree emanating from $w : \phi$ due to the eigenvariable condition. Yet, once an application of $\refl$ or $\trans$ is applied (bottom-up) either a loop or undirected cycle will be added, breaking the `treelike' structure and beginning the `non-treelike' fragment of the proof. Therefore, if one imagines permuting instances of $\refl$ and $\trans$ upward in a $\intfondlq$ or $\intfocdlq$ derivation (making use of the elimination algorithms explained in \lem~\ref{lem:refl-elim-FOInt} and~\ref{lem:trans-elim-FOInt}), then one can see that the bottom `treelike' fragment grows and the top `non-treelike' fragment diminishes. Once all $\refl$ and $\trans$ inferences have been removed, the derivation becomes a labelled tree derivation, explaining how we transition to a setting where only labelled tree derivations are required for completeness.


\begin{theorem}\label{thm:treelike-derivations-quasi-refined-FOInt}
Every derivation in $\intfondlq$ and $\intfocdlq$ of a labelled formula $w : \phi$ is a labelled tree derivation with the fixed root property.
\end{theorem}

\begin{proof} Consider a proof of $\seqempstr \sar w : \phi$ in $\intfondlq$ or $\intfocdlq$ in a bottom-up manner. The only rules that introduce relational atoms of the form $u \leq v$ are the $\impr$ and $\allr$. Hence, bottom-up applications of rules will either preserve relational structure or will construct a tree emanating from (the root) $w$.
\end{proof}


\section{Refinement Part III: Removal of Domain Atoms}\label{subsect:refinement-part-three-FOInt}

We introduce our refined labelled calculi $\intfondl$ and $\intfocdl$ (shown in \fig~\ref{fig:refined-calculi-FO-Int} below) which are obtained from the quasi-refined calculi $\intfondlq$ and $\intfocdlq$ derived in the previous section. The extraction of $\intfondl$ and $\intfocdl$ from $\intfondlq$ and $\intfocdlq$ (\resp) results from two modifications made to the quasi-refined labelled calculi: (i) principal and auxiliary domain atoms are removed from all relevant inference rules, and (ii) the $\existsrnca$ and $\existsrncia$ rules, and the $\alllnca$ and $\alllncia$ rules, are combined to make the single rules $\existsrnc$ and $\alllnc$ (given in \fig~\ref{fig:refined-calculi-FO-Int} below) with a more complex side condition (expressed in \fig~\ref{fig:refined-side-conditions-FO-Int}). This section will be devoted to showing that the refined labelled calculi $\intfondl$ and $\intfocdl$ possess fundamental proof-theoretic properties, and in the subsequent section, we will show that the systems are notational variants of nested systems.

The removal of principal and auxiliary domain atoms from $\idfonca$, $\existsl$, $\existsrnca$, $\existsrncia$, $\allr$, $\alllnca$, and $\alllncia$ (yielding the rules $\idfonc$, $\existslnc$, $\existsrnc$, $\allrnc$, and $\alllnc$ in \fig~\ref{fig:refined-calculi-FO-Int} below, \resp) is the central difference between our quasi-refined and refined labelled calculi. The rules $\idfonc$, $\existslnc$, and $\allrnc$ are straightforwardly obtained from $\idfonca$, $\existsl$, and $\allr$, respectively, by simply omitting the principal and auxiliary domain atoms. By contrast, since applications of the $\existsrnca$, $\existsrncia$, $\alllnca$, and $\alllncia$ rules depend on side conditions that further depend on the existence of domain atoms, the extraction of $\existsrnc$ and $\alllnc$ requires more insight. The key to transforming the $\existsrnca$ and $\existsrncia$ rules into $\existsrnc$, and the $\alllnca$ and $\alllncia$ rules into $\alllnc$, is to re-write the side condition so that it no longer relies on the existence of a domain atom $\unda \in D_{u}$, but rather, relies on the existence of a labelled formula $u : \psi(\vec{\unda})$ with $\unda$ occurring in $\vec{\unda}$. Breaking the dependence of the rules on the existence of domain atoms effectively renders such syntactic structures expendable---a fact which is formally established later on (\prp~\ref{prop:remove-domain-atoms-FOInt}).

Recognizing that the reliance on domain atoms can be replaced by a reliance on labelled formulae, came about by comparing the side conditions of the $\existsrnca$, $\existsrncia$, $\alllnca$, and $\alllncia$ rules with the side conditions imposed by Fitting on his $R\exists$ and $L\forall$ rules employed in his nested calculus for first-order intuitionistic logic proper~\cite[p.~59]{Fit14}. Whereas applications of our $\existsrnca$ and $\alllnca$ rules rely on the existence of a propagation path connecting the label $u$ of the active domain atom to the auxiliary label (see \fig~\ref{fig:side-conditionsi-FO-Int} and~\ref{fig:side-conditionsii-FO-Int}), Fitting's $R\exists$ and $L\forall$ rules rely (in part) on the existence of a path connecting a formula with a parameter $\unda$ to the auxiliary formula of the inference---when such a state of affairs holds, Fitting calls the parameter $\unda$ \emph{available}. We impose similar side conditions in our refined labelled calculi, where a rule's applicability is dictated by a parameter being \emph{$\thuesysi$-available} or \emph{$\thuesysii$-available}:

\begin{definition}[$\thuesysi$-available, $\thuesysii$-available]\label{def:S4-S5-available-FOInt} Let $\Lambda := \rel, \Gamma \sar \Delta$. We say that a parameter $\unda$ is \emph{$\thuesysi$-available}\index{$\thuesysi$-available} (\emph{$\thuesysii$-available}\index{$\thuesysii$-available}) in $\Lambda$ for a label $w$ \ifandonlyif there exists a labelled formula $u : \phi(\vec{\unda}) \in \Gamma, \Delta$ with $\unda$ in $\vec{\unda}$ such that $\exists \ppath(\stra_{\ppath}(u,w) \in \thuesysilang{a})$ ($\exists \ppath(\stra_{\ppath}(u,w) \in \thuesysiilang{a})$, \resp).
\end{definition}

When making use of the fact that a parameter $\unda$ is $\thuesysi$-available or $\thuesysii$-available, we will abuse notation and often denote the labelled formula $u : \phi(\vec{\unda}) \in \Gamma, \Delta$ with $\unda$ in $\vec{\unda}$---which gives rise to the $\thuesysi$-availability or $\thuesysii$-availability of $\unda$---as $u : \phi(\unda)$. This will simplify discussions and proofs that concern the side conditions of reachability rules. 

\begin{example} We give an example of $\thuesysi$- and $\thuesysii$-available labels in the labelled sequent $\Lambda$ shown below. Since such labels are determined by considering a labelled sequent's propagation graph, we also include a pictorial representation of $\prgr{\Lambda}$.
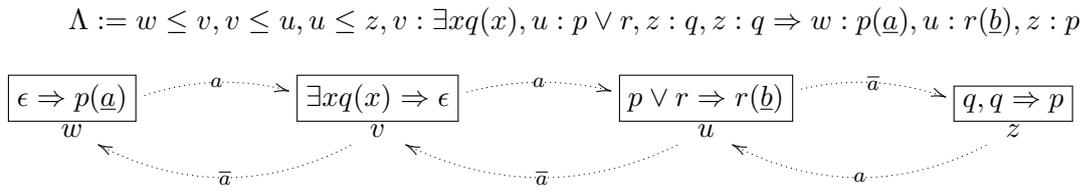
$$
\Lambda := w \leq v, v \leq u, u \leq z, v : \exists x q(x), u : p \lor r, z :q, z : q \sar w : p(\unda), u : r(\undb), z : p
$$
\begin{center}
\xymatrix{
  \overset{\boxed{\seqempstr \sar p(\unda)}}{w} \ar@/^1.5pc/@{.>}[rr]|-{a} &   & \overset{\boxed{\exists x q(x) \sar \seqempstr}}{v} \ar@/^1.5pc/@{.>}[ll]|-{\overline{a}}\ar@/^1.5pc/@{.>}[rr]|-{a} &  & \overset{\boxed{p \lor r \sar r(\undb)}}{u} \ar@/^1.5pc/@{.>}[ll]|-{\conv{a}}\ar@/^1.5pc/@{.>}[rr]|-{\conv{a}}  & 
 & \overset{\boxed{q, q \sar p}}{z} \ar@/^1.5pc/@{.>}[ll]|-{a}   
}
\end{center}
In the above example, the parameter $\unda$ is both $\thuesysi$- and $\thuesysii$-available for $w$, $v$, and $u$ since there exist propagation paths $\emppath(w,w) = w$, $\ppath_{1}(w,v) := w, a, v$, and $\ppath_{2}(w,u) := w, a, v, a, u$ such that $\stra_{\emppath}(w,w) \in \thuesysilang{a}$, $\stra_{\emppath}(w,w) \in \thuesysiilang{a}$, $\stra_{\ppath_{1}}(w,v) \in \thuesysilang{a}$, $\stra_{\ppath_{1}}(w,v) \in \thuesysiilang{a}$, $\stra_{\ppath_{2}}(w,u) \in \thuesysilang{a}$, and $\stra_{\ppath_{2}}(w,u) \in \thuesysiilang{a}$. However, note that $\unda$ is not $\thuesysi$-available for $z$ since none of the strings in $\thuesysilang{a}$ contain the converse character $\conv{a}$, and any propagation path from $w$ to $z$ must contain such a character. On the other hand, the label $z$ is $\thuesysii$-available since there exists a propagation path $\ppath_{3}(w,z) := w, a, v, a, u, \conv{a}, z$ such that $\stra_{\ppath_{3}}(w,z) \in \thuesysiilang{a}$.
\end{example}

Our refined labelled calculi $\intfondl$ and $\intfocdl$ are given in \fig~\ref{fig:refined-calculi-FO-Int} with the side conditions of the reachability and propagation rules given in \fig~\ref{fig:refined-side-conditions-FO-Int}. As explained previously, the calculi are obtained from $\intfondlq$ and $\intfocdlq$ by deleting principal and auxiliary domain atoms from rules as well as combining $\existsrnca$ and $\existsrncia$, and $\alllnca$ and $\alllncia$, yielding the rules $\existsrnc$ and $\alllnc$, respectively. Furthermore, the side conditions of the $\existsrnc$ and $\alllnc$ rules make use of the notion of $\thuesysi$- and $\thuesysii$-availability depending on if $\mathsf{X} = \nnn$ or $\mathsf{X} = \ccc$. As can be witnessed, none of the rules make reference to domain atoms, implying their superfluity. We formally establish this fact with the following proposition:

\begin{figure}[t]
\noindent\hrule

\begin{center}
\begin{tabular}{c @{\hskip 1em} c} 
\AxiomC{}
\RightLabel{$\idfonc^{\dag_{1}(\norc)}$\index{$\idfonc$}}
\UnaryInfC{$\rel, \Gamma, w : p(\vec{\unda}) \Rightarrow u : p(\vec{\unda}), \Delta$}
\DisplayProof

&

\AxiomC{}
\RightLabel{$\botl$}
\UnaryInfC{$\rel, \Gamma, w :\bot \sar \Delta$}
\DisplayProof
\end{tabular}
\end{center}

\begin{center}
\begin{tabular}{c @{\hskip 1em} c @{\hskip 1em} c}

\AxiomC{$\rel, \Gamma, w :\phi, w :\psi \sar \Delta$}
\RightLabel{$\conl$}
\UnaryInfC{$\rel, \Gamma, w :\phi \wedge \psi \sar \Delta$}
\DisplayProof

&

\AxiomC{$\rel, \Gamma \sar w :\phi, \Delta$}
\AxiomC{$\rel, \Gamma \sar w :\psi, \Delta$}
\RightLabel{$\conr$}
\BinaryInfC{$\rel, \Gamma \sar w :\phi \wedge \psi, \Delta$}
\DisplayProof
\end{tabular}
\end{center}

\begin{center}
\begin{tabular}{c}
\AxiomC{$\rel, \Gamma, w :\phi \sar \Delta$}
\AxiomC{$\rel, \Gamma, w :\psi \sar \Delta$}
\RightLabel{$\disl$}
\BinaryInfC{$\rel, \Gamma, w :\phi \vee \psi \sar \Delta$}
\DisplayProof
\end{tabular}
\end{center}

\begin{center}
\begin{tabular}{c c}
\AxiomC{$\rel, w \leq u, \Gamma, u :\phi \sar u :\psi, \Delta$}
\RightLabel{$\impr^{\dag_{2}(\norc)}$}
\UnaryInfC{$\rel, \Gamma \sar w :\phi \imp \psi, \Delta$}
\DisplayProof

&

\AxiomC{$\rel, \Gamma \sar w :\phi, w :\psi, \Delta$}
\RightLabel{$\disr$}
\UnaryInfC{$\rel, \Gamma \sar w :\phi \vee \psi, \Delta$}
\DisplayProof
\end{tabular}
\end{center}

\begin{center}
\AxiomC{$\rel, w : \phi \imp \psi, \Gamma \Rightarrow \Delta, u : \phi$}
\AxiomC{$\rel, w : \phi \imp \psi, u : \psi, \Gamma \Rightarrow \Delta$}
\RightLabel{$\primp^{\dag_{1}(\norc)}$\index{$\primp$}}
\BinaryInfC{$\rel, w : \phi \imp \psi, \Gamma \Rightarrow \Delta$}
\DisplayProof
\end{center}

\begin{center}
\begin{tabular}{c c}
\AxiomC{$\rel, \Gamma \Rightarrow \Delta, w: \phi(\unda/x), w: \exists x \phi$}
\RightLabel{$\existsrnc^{\dag_{3}(\norc)}$\index{$\existsrnc$}}
\UnaryInfC{$\rel, \Gamma \Rightarrow \Delta, w: \exists x \phi$}
\DisplayProof

&

\AxiomC{$\rel, w : \forall x \phi, v : \phi(\unda/x), \Gamma \Rightarrow \Delta$}
\RightLabel{$\alllnc^{\dag_{4}(\norc)}$\index{$\alllnc$}}
\UnaryInfC{$\rel, w : \forall x \phi, \Gamma \Rightarrow \Delta$}
\DisplayProof
\end{tabular}
\end{center}

\begin{center}
\begin{tabular}{c c} 
\AxiomC{$\rel, w \leq u, \Gamma \sar  u : \phi(\unda/x), \Delta$}
\RightLabel{$\allrnc^{\dag_{5}(\norc)}$\index{$\allrnc$}}
\UnaryInfC{$\rel, \Gamma \sar w : \forall x \phi, \Delta$}
\DisplayProof

&

\AxiomC{$\rel, \Gamma, w: \phi(\unda/x) \sar \Delta$}
\RightLabel{$\existslnc^{\dag_{6}(\norc)}$\index{$\existslnc$}}
\UnaryInfC{$\rel, \Gamma, w: \exists x \phi \sar \Delta$}
\DisplayProof

\end{tabular}
\end{center}

\hrule
\caption{The refined labelled calculus $\intfondl$\index{$\intfondl$} consists of all rules with $\mathsf{X} = \nnn$, and the refined labelled calculus $\intfocdl$\index{$\intfocdl$} consists of all rules with $\mathsf{X} = \ccc$. The side conditions $\dag_{1}(\mathsf{X})$ -- $\dag_{6}(\mathsf{X})$ are given in \fig~\ref{fig:refined-side-conditions-FO-Int}.}
\label{fig:refined-calculi-FO-Int}
\end{figure}

\begin{figure}[t]
\begin{center}
\bgroup
\def\arraystretch{1.25}
\begin{tabular}{| c | c | c | c |}
\hline
Name & Side Condition & Name & Side Condition\\
\hline
$\dag_{1}(\nnn)$\index{$\dag_{1}(\nnn)$} & $\exists\ppath(\stra_{\ppath}(w,u) \in \thuesysilang{a})$ &
$\dag_{1}(\ccc)$\index{$\dag_{1}(\ccc)$} & $\exists\ppath(\stra_{\ppath}(w,u) \in \thuesysilang{a})$\\
\hline
$\dag_{2}(\nnn)$\index{$\dag_{2}(\nnn)$} &  $u$ is an eigenvariable &
$\dag_{2}(\ccc)$\index{$\dag_{2}(\ccc)$} &  $u$ is an eigenvariable\\
\hline
$\dag_{3}(\nnn)$\index{$\dag_{3}(\nnn)$} & $\unda$ is $\thuesysi$-available for $w$ & $\dag_{3}(\ccc)$\index{$\dag_{3}(\ccc)$} &  $\unda$ $\thuesysii$-available for $w$\\
 &  or $\unda$ is an eigenvariable &  &  or $\unda$ is an eigenvariable\\
\hline
$\dag_{4}(\nnn)$\index{$\dag_{4}(\nnn)$} & $\unda$ is $\thuesysi$-available for $v$ & $\dag_{4}(\ccc)$\index{$\dag_{4}(\ccc)$} &  $\unda$ $\thuesysii$-available for $v$\\
 &  or $\unda$ is an eigenvariable, and  &  &  or $\unda$ is an eigenvariable, and \\
  &  $\exists\ppath(\stra_{\ppath}(w,v) \in \thuesysilang{a})$  &  &  $\exists\ppath(\stra_{\ppath}(w,v) \in \thuesysilang{a})$\\
 \hline
$\dag_{5}(\nnn)$\index{$\dag_{5}(\nnn)$} &  $\unda$ and $u$ are eigenvariables & $\dag_{5}(\ccc)$\index{$\dag_{5}(\ccc)$} &  $\unda$ and $u$ are eigenvariables\\
\hline
$\dag_{6}(\nnn)$\index{$\dag_{6}(\nnn)$} &  $\unda$ is an eigenvariable & $\dag_{6}(\ccc)$\index{$\dag_{6}(\ccc)$} &  $\unda$ is an eigenvariable\\
\hline
\end{tabular}
\egroup
\end{center}
\caption{Side conditions for rules in $\intfondl$ and $\intfocdl$.}
\label{fig:refined-side-conditions-FO-Int}
\end{figure}

\begin{proposition}\label{prop:remove-domain-atoms-FOInt}
If $\rel, \unda \in D_{w}, \Gamma \sar \Delta$ is derivable in $\intfondl$ or $\intfocdl$, then $\rel, \Gamma \sar \Delta$ is derivable in $\intfondl$ or $\intfocdl$, respectively.
\end{proposition}

\begin{proof} We prove the result by induction on the height of the given derivation and show the result for $\intfocdl$ as the proof for $\intfondl$ is similar.

\textit{Base case.} The base case follows from the fact that removing any domain atoms from an instance of $\idfonc$ or $\botl$ yields another instance of $\idfonc$ or $\botl$, respectively.

\textit{Inductive step.} All cases follow by invoking \ih and then applying the corresponding rule.
\end{proof}

Let us now establish the soundness of our refined labelled calculi. Although we could attempt to prove soundness by supplying a semantics for our calculi and then showing that each rule of $\intfondl$ and $\intfocdl$ preserves validity, we opt for another strategy and show that each derivation in $\intfondl$ and $\intfocdl$ can be transformed into a derivation in $\intfondlq$ and $\intfocdlq$, respectively. In order to transform derivations of the former systems into derivations in the latter systems, we introduce the \emph{domain closure}\index{Domain closure function} function $\domcl$, which ensures that each parameter $\unda$ occurring in any labelled formula $w : \phi(\vec{\unda})$ of a labelled sequent $\Lambda$ is accompanied by a corresponding domain atom $\unda \in D_{w}$ (specifying the domain in which the parameter is included). We define the domain closure function below, followed by an example to demonstrate its functionality.

\begin{definition}[The Function $\domcl)$] Let $\Lambda := \rel, \Gamma \sar \Delta$ be a labelled sequent and define
$$
\para(\Lambda,w) := \{\unda \ | \ \text{there exists a } w : \phi(\vec{\unda}) \in \Gamma, \Delta \text{ with $\unda$ in $\vec{\unda}$}\}.
$$
We define $\domcl(\Lambda) := \rel, \rel', \Gamma \sar \Delta$, where
$$
\rel' := \{\unda \in D_{w} \ | \ w \in \lab(\Lambda) \text{ and } \unda \in \para(\Lambda,w)\}.
$$
\end{definition}

\begin{example} Suppose we are given the following labelled sequent:
$$
\Lambda := w \leq u, w \leq v, v : p(\undc) \imp q(\undc), w : r(\unda) \sar w :\forall x p(x)
$$
An application of  $\domcl(\cdot)$ to $\Lambda$ would output the following:
$$
\domcl(\Lambda) := w \leq u, w \leq v, \unda \in D_{w}, \undc \in D_{v}, v : p(\undc) \imp q(\undc), w : r(\unda) \sar w :\forall x p(x)
$$
\end{example}

As discussed above, certain inference rules in $\intfondlq$ and $\intfocdlq$ depend on the existence of domain atoms, whereas all inference rules in $\intfondl$ and $\intfocdl$ are independent of domain atoms (see \prp~\ref{prop:remove-domain-atoms-FOInt}). Therefore, the significance of the domain closure function $\domcl$ is that it adds a sufficient number of domain atoms to inferences within an $\intfondl$ or $\intfocdl$ proof to ensure that the proof is a valid derivation in $\intfondlq$ or $\intfocdlq$, respectively. The following lemma demonstrates that the domain closure function permits the desired proof transformation (thus securing soundness, as will be subsequently proven).

\begin{lemma}\label{lem:refined-to-quasi-refined-FOInt} \ 

(i) Every proof of a labelled sequent $\Lambda$ in $\intfondl$ can be algorithmically transformed into a proof of $\domcl(\Lambda)$ in $\intfondlq$.

(ii) Every proof of a labelled sequent $\Lambda$ in $\intfocdl$ can be algorithmically transformed into a proof of $\domcl(\Lambda)$ in $\intfocdlq$.
\end{lemma}

\begin{proof} We defer the proof to Appendix C (p.~\pageref{app:proofs}).
\end{proof}


\begin{theorem}[Soundness of $\intfondl$ and $\intfocdl$] \ 

(i) If $\vdash_{\intfondl} \seqempstr \sar w : \phi$, then $\Vdash_{\intfond} \phi$.

(ii) If $\vdash_{\intfocdl} \seqempstr \sar w : \phi$, then $\Vdash_{\intfocd} \phi$.
\end{theorem}

\begin{proof}
Follows from \thm~\ref{thm:properties-quasi-refined-FOInt}, \lem~\ref{lem:refined-to-quasi-refined-FOInt}, and \dfn~\ref{def:sequent-semantics-FO-Int}.
\end{proof}

We now show that $\intfondl$ and $\intfocdl$ possess desirable proof-theoretic properties such as hp-admissibility of structural rules (e.g. $\wk$ and $\ctrr$), hp-invertibility of all rules, and syntactic cut-elimination. After establishing these results, we harness the proof-theoretic properties of $\intfondl$ and $\intfocdl$ to prove completeness. The completeness result also relies on the following lemma, which essentially states that generalized instances of $\idfonc$ are always derivable.

\begin{lemma}\label{lem:general-id-FOInt-refined}
Let $\mathsf{XL} \in \{\intfondl, \intfocdl\}$ and $\Lambda := \rel, \Gamma, w : \phi(\vec{\unda}) \sar u : \phi(\vec{\unda}), \Delta$. If there exists a propagation path $\ppath(w,u)$ in $\prgr{\Lambda}$ such that $\stra_{\ppath}(w,u) \in \thuesysilang{a}$, then $\vdash_{\mathsf{XL}} \Lambda$.
\end{lemma}

\begin{proof} We prove the result by induction on the complexity of $\phi(\vec{\unda})$. The base cases are trivial, so we only consider the inductive step.

\textit{Inductive step.} We only show the cases where $\phi(\vec{\unda})$ is of the form $\exists x \psi(\vec{\unda})$ and $\forall x \psi(\vec{\unda})$. Both cases are resolved as follows:

\begin{center}
\AxiomC{}
\RightLabel{\ih}
\dashedLine
\UnaryInfC{$\rel, \Gamma, w : \psi(\vec{\unda})(\undb/x) \sar u : \psi(\vec{\unda})(\undb/x), u : \exists x \psi(\vec{\unda}), \Delta$}
\RightLabel{$\existsrnc$}
\UnaryInfC{$\rel, \Gamma, w : \psi(\vec{\unda})(\undb/x) \sar u : \exists x \psi(\vec{\unda}), \Delta$}
\RightLabel{$\existslnc$}
\UnaryInfC{$\rel, \Gamma, w : \exists x \psi(\vec{\unda}) \sar u : \exists x \psi(\vec{\unda}), \Delta$}
\DisplayProof
\end{center}

\begin{center}
\AxiomC{}
\RightLabel{\ih}
\dashedLine
\UnaryInfC{$\rel, u \leq v, \Gamma, w : \forall x \psi(\vec{\unda}), w : \psi(\vec{\unda})(\undb/x) \sar v : \psi(\vec{\unda})(\undb/x), \Delta$}
\RightLabel{$\alllnc$}
\UnaryInfC{$\rel, u \leq v, \Gamma, w : \forall x \psi(\vec{\unda}) \sar v : \psi(\vec{\unda})(\undb/x), \Delta$}
\RightLabel{$\allrnc$}
\UnaryInfC{$\rel, \Gamma, w : \forall x \psi(\vec{\unda}) \sar u : \forall x \psi(\vec{\unda}), \Delta$}
\DisplayProof
\end{center}
\end{proof}

Next, we confirm the hp-admissibility of the substitution rules $\lsub$ and $\psub$, which serves as a foundation for all 
 remaining results. Recall that label and parameter substitutions were defined in \dfn~\ref{lem:label-param-sub-FO-Int}. Additionally, we note that all structural rules shown hp-admissible in this section are the same as those found in \fig~\ref{fig:lab-struc-rules-FO-Int}. Also, most of the remaining proofs of this section will be deferred to the Appendix C (starting on p.~\pageref{app:proofs}) to improve readability and because such results are similar to the hp-admissibility, hp-invertibility, and elimination results proven in \cptr~\ref{CPTR:Labelled}. Unless stated otherwise, all omitted proofs (i.e. proofs deferred to Appendix C) are shown by induction on the height of the given derivation.

\begin{lemma}\label{lem:lsub-admiss-FOInt-refined}
The rule $\lsub$ is hp-admissible in $\intfondl$ and $\intfocdl$.
\end{lemma}


\begin{lemma}\label{lem:psub-admiss-FOInt-refined}
The rule $\psub$ is hp-admissible in $\intfondl$ and $\intfocdl$. 
\end{lemma}


\begin{lemma}\label{lem:wk-admiss-FOInt-refined}
The rule $\wk$ is hp-admissible in $\intfondl$ and $\intfocdl$. 
\end{lemma}


\begin{lemma}\label{lem:dis-con-invert-FOInt-refined}
The $\conl$, $\conr$, $\disl$, and $\disr$ rules are hp-invertible in $\intfondl$ and $\intfocdl$. 
\end{lemma}


\begin{lemma}\label{lem:imp-invert-FOInt-refined}
The $\primp$ and $\impr$ rules are hp-invertible in $\intfondl$ and $\intfocdl$. 
\end{lemma}


\begin{lemma}\label{lem:exists-all-invert-FOInt-refined} \ 

(i) The $\existsln$, $\existsrn$, $\allln$, and $\allrn$ rules are hp-invertible in $\intfondl$. 

(ii) The $\existslc$, $\existsrc$, $\alllc$, and $\allrc$ rules are hp-invertible in $\intfocdl$. 
\end{lemma}


\begin{lemma}\label{lem:invert-FOInt-refined}
All rules are hp-invertible in $\intfondl$ and $\intfocdl$. 
\end{lemma}

\begin{proof} Follows from \lem~\ref{lem:dis-con-invert-FOInt-refined} -- \ref{lem:exists-all-invert-FOInt-refined} above.
\end{proof}

\begin{lemma}\label{lem:ctrrel-admiss-FOInt-refined}
The rule $\ctrrel$ is hp-admissible in $\intfondl$ and $\intfocdl$. 
\end{lemma}

\begin{proof} We proceed by induction on the height of the given derivation.

\textit{Base case.} Any application of $\ctrrel$ to $\idfonc$ or $\botl$ yields another instance of the rule, which confirms the base case.

\textit{Inductive step.} All cases are resolved by applying \ih followed by the corresponding rule.
\end{proof}

\begin{lemma}\label{lem:ctrl-admiss-FOInt-refined}
The rule $\ctrl$ is hp-admissible in $\intfondl$ and $\intfocdl$. 
\end{lemma}


\begin{lemma}\label{lem:ctrr-admiss-FOInt-refined}
The rule $\ctrr$ is hp-admissible in $\intfondl$ and $\intfocdl$. 
\end{lemma}

\begin{proof} The result is shown by induction on the height of the given derivation and is similar to the proof of \lem~\ref{lem:ctrl-admiss-FOInt-refined} above.
\end{proof}

\begin{lemma}\label{lem:refl-elim-refined-FOInt}
The $\refl$ and $\trans$ rules are elimimable in $\intfondl$ and $\intfocdl$.
\end{lemma}

\begin{proof}
The lemma is proven similarly to \lem~\ref{lem:refl-elim-FOInt} and~\ref{lem:trans-elim-FOInt}.
\end{proof}

\begin{lemma}\label{lem:deleting-relational-atoms-refined}
Let $\Lambda := \rel, \Gamma \sar \Delta$. Suppose $\rel, w \leq u, \Gamma \sar \Delta$ is derivable in $\intfondl$ or $\intfocdl$, and that $\ppath(w,u)$ is a propagation path (potentially empty) in $\prgr{\Lambda}$ such that $\stra_{\ppath}(w,u) \in \thuesysilang{a}$. Then, $\rel, \Gamma \sar \Delta$ is derivable in $\intfondl$ or $\intfocdl$, respectively.
\end{lemma}

\begin{proof}
Similar to the proof of \lem ~\ref{lem:deleting-relational-atoms-FOInt}, but relies on \lem~\ref{lem:refl-elim-refined-FOInt} above.
\end{proof}

\begin{theorem}\label{thm:cut-admiss-FOInt-refined}
The rule $\cut$ is eliminable in $\intfondl$ and $\intfocdl$. 
\end{theorem}

\begin{proof} The result is proven by induction on the lexicographic ordering of pairs $(\fcomp{\phi},h_{1}+h_{2})$, where $\fcomp{\phi}$ is the complexity of the cut formula $\phi$, $h_{1}$ is the height of the derivation of the left premise of $\cut$, and $h_{2}$ is the height of the derivation of the right premise of $\cut$. We defer the proof to Appendix C (starting on p.~\pageref{app:proofs}).
\end{proof}



We are now in a position to prove the completeness of our refined labelled calculi $\intfondl$ and $\intfocdl$. The proof is as follows:


\begin{theorem}[Completeness]\label{thm:completness-FOInt-refined}\ 

(i) If $\vdash_{\intfond} \phi$, then $\vdash_{\intfondl} \seqempstr \sar w : \phi$.

(ii) If $\vdash_{\intfocd} \phi$, then $\vdash_{\intfocdl} \seqempstr \sar w : \phi$.
\end{theorem}

\begin{proof} We show that our calculi $\intfondl$ and $\intfocdl$ can derive axioms A9 -- A12 and A9 -- A13, respectively, as well as simulate the inference rules R0 and R1 (see \dfn~\ref{def:axiomatization-IntFO} for the axioms of $\intfond$ and $\intfocd$). Proofs of axioms A0 -- A8, i.e. the axioms for propositional intuitionistic logic (cf.~\dfn~\ref{def:axiomatization-IntFO} and \cite[p.~6]{GabSheSkv09}), are straightforward. 

We provide derivations for axioms A9 -- A13 below, and note that the proof of axiom A13 (the constant domain axiom) requires the rule $\alllc$, i.e. the proof does not go through with the weaker rule $\allln$. For all other cases we let $\mathsf{X} \in \{\nnn,\ccc\}$ to prove the results uniformly.


\emph{Axiom A9.}

\begin{center}
\begin{tabular}{c} 
\AxiomC{}
\RightLabel{\lem~\ref{lem:general-id-FOInt-refined}}
\dashedLine
\UnaryInfC{$w \leq u, u : \forall x \phi, u : \phi(\unda/x) \sar u : \phi(\unda/x)$}
\RightLabel{$\alllnc$}
\UnaryInfC{$w \leq u, u : \forall x \phi \sar u : \phi(\unda/x)$}
\RightLabel{$\impr$}
\UnaryInfC{$\seqempstr \sar w : \forall x \phi \imp \phi(\unda/x)$}
\DisplayProof
\end{tabular}
\end{center}

\emph{Axiom A10.}

\begin{center}
\begin{tabular}{c}
\AxiomC{}
\RightLabel{\lem~\ref{lem:general-id-FOInt-refined}}
\dashedLine
\UnaryInfC{$w \leq u, u : \phi(\unda/x) \sar u : \exists x \phi, u : \phi(\unda/x)$}
\RightLabel{$\existsrnc$}
\UnaryInfC{$w \leq u, u : \phi(\unda/x) \sar u : \exists x \phi$}
\RightLabel{$\impr$}
\UnaryInfC{$\seqempstr \sar w : \phi(\unda/x) \imp \exists x \phi$}
\DisplayProof
\end{tabular}
\end{center}

\emph{Axiom A11.} To save space and improve readability, we let $\rel := w \leq u, u \leq v, v \leq z$. Also, by assumption we know that $x$ does not occur in $\psi$ (see \dfn~\ref{def:axiomatization-IntFO}).

\begin{flushleft}
\begin{tabular}{c c c}
$\Pi_{1}$

&

$= \Bigg \{$

&

\AxiomC{}
\RightLabel{\lem~\ref{lem:general-id-FOInt-refined}}
\dashedLine
\UnaryInfC{$\rel, u : \forall x (\psi \imp \phi), v :\psi, z : \psi \imp \phi(\unda/x) \sar z : \psi, z : \phi(\unda/x)$}
\DisplayProof
\end{tabular}
\end{flushleft}

\begin{flushleft}
\begin{tabular}{c c c}
$\Pi_{2}$

&

$= \Bigg \{$

&

\AxiomC{}
\RightLabel{\lem~\ref{lem:general-id-FOInt-refined}}
\dashedLine
\UnaryInfC{$\rel, u : \forall x (\psi \imp \phi), v :\psi, z : \psi \imp \phi(\unda/x), z : \phi(\unda/x) \sar z : \phi(\unda/x)$}
\DisplayProof
\end{tabular}
\end{flushleft}

\begin{center}
\begin{tabular}{c}
\AxiomC{$\Pi_{1}$}
\AxiomC{$\Pi_{2}$}
\RightLabel{$\primp$}
\BinaryInfC{$w \leq u, u \leq v, v \leq z, u : \forall x (\psi \imp \phi), v :\psi, z : \psi \imp \phi(\unda/x) \sar z : \phi(\unda/x)$}
\RightLabel{$\alllnc$}
\UnaryInfC{$w \leq u, u \leq v, v \leq z, u : \forall x (\psi \imp \phi), v :\psi \sar z : \phi(\unda/x)$}
\RightLabel{$\allrnc$}
\UnaryInfC{$w \leq u, u \leq v, u : \forall x (\psi \imp \phi), v :\psi \sar v : \forall x \phi$}
\RightLabel{$\impr$}
\UnaryInfC{$w \leq u, u : \forall x (\psi \imp \phi) \sar u : \psi \imp \forall x \phi$}
\RightLabel{$\impr$}
\UnaryInfC{$\seqempstr \sar w : \forall x (\psi \imp \phi) \imp (\psi \imp \forall x \phi)$}
\DisplayProof
\end{tabular}
\end{center}

\emph{Axiom A12.} To save space and improve readability, we let $\rel := w \leq u, u \leq v$. Also, by assumption we know that $x$ does not occur in $\phi$ (see \dfn~\ref{def:axiomatization-IntFO}).

\begin{flushleft}
\begin{tabular}{c c c}
$\Pi_{1}$

&

$= \Bigg \{$

&

\AxiomC{}
\RightLabel{\lem~\ref{lem:general-id-FOInt-refined}}
\dashedLine
\UnaryInfC{$\rel, u : \forall x (\psi \imp \phi), v : \psi(\unda/x), v : \psi(\unda/x) \imp \phi \sar v : \psi(\unda/x), v : \phi$}
\DisplayProof
\end{tabular}
\end{flushleft}

\begin{flushleft}
\begin{tabular}{c c c}
$\Pi_{2}$

&

$= \Bigg \{$

&

\AxiomC{}
\RightLabel{\lem~\ref{lem:general-id-FOInt-refined}}
\dashedLine
\UnaryInfC{$\rel, u : \forall x (\psi \imp \phi), v : \psi(\unda/x), v : \psi(\unda/x) \imp \phi, v : \phi \sar v : \phi$}
\DisplayProof
\end{tabular}
\end{flushleft}

\begin{center}
\begin{tabular}{c}
\AxiomC{$\Pi_{1}$}
\AxiomC{$\Pi_{2}$}
\RightLabel{$\primp$}
\BinaryInfC{$w \leq u, u \leq v, u : \forall x (\psi \imp \phi), v : \psi(\unda/x), v : \psi(\unda/x) \imp \phi \sar v : \phi$}
\RightLabel{$\alllnc$}
\UnaryInfC{$w \leq u, u \leq v, u : \forall x (\psi \imp \phi), v : \psi(\unda/x) \sar v : \phi$}
\RightLabel{$\existslnc$}
\UnaryInfC{$w \leq u, u \leq v, u : \forall x (\psi \imp \phi), v : \exists x \psi \sar v : \phi$}
\RightLabel{$\impr$}
\UnaryInfC{$w \leq u, u : \forall x (\psi \imp \phi) \sar u : \exists x \psi \imp \phi$}
\RightLabel{$\impr$}
\UnaryInfC{$\seqempstr \sar w : \forall x (\psi \imp \phi) \imp (\exists x \psi \imp \phi)$}
\DisplayProof
\end{tabular}
\end{center}

\emph{Axiom A13.}

\begin{flushleft}
\begin{tabular}{c c c}
$\Pi_{1}$

&

$= \Bigg \{$

&

\AxiomC{}
\RightLabel{\lem~\ref{lem:general-id-FOInt-refined}}
\dashedLine
\UnaryInfC{$w \leq u, u \leq v, u : \forall x (\phi \lor \psi), u : \psi \sar v : \phi(\unda/x), u : \psi$}
\DisplayProof
\end{tabular}
\end{flushleft}

\begin{flushleft}
\begin{tabular}{c c c}
$\Pi_{2}$

&

$= \Bigg \{$

&

\AxiomC{}
\RightLabel{\lem~\ref{lem:general-id-FOInt-refined}}
\dashedLine
\UnaryInfC{$w \leq u, u \leq v, u : \forall x (\phi \lor \psi), u : \phi(\unda/x) \sar v : \phi(\unda/x), u : \psi$}
\DisplayProof
\end{tabular}
\end{flushleft}

\begin{center}
\begin{tabular}{c}
\AxiomC{$\Pi_{1}$}
\AxiomC{$\Pi_{2}$}
\RightLabel{$\disl$}
\BinaryInfC{$w \leq u, u \leq v, u : \forall x (\phi \lor \psi), u : \phi(\unda/x) \lor \psi \sar v : \phi(\unda/x), u : \psi$}
\RightLabel{$\alllc$}
\UnaryInfC{$w \leq u, u \leq v, u : \forall x (\phi \lor \psi) \sar v : \phi(\unda/x), u : \psi$}
\RightLabel{$\allrc$}
\UnaryInfC{$w \leq u, u : \forall x (\phi \lor \psi) \sar u : \forall x \phi, u : \psi$}
\RightLabel{$\disr$}
\UnaryInfC{$w \leq u, u : \forall x (\phi \lor \psi) \sar u : \forall x \phi \lor \psi$}
\RightLabel{$\impr$}
\UnaryInfC{$\seqempstr \sar w : \forall x (\phi \lor \psi) \imp \forall x \phi \lor \psi$}
\DisplayProof
\end{tabular}
\end{center}

\emph{Rule R0.} We invoke \lem~\ref{lem:refl-elim-refined-FOInt} and \thm~\ref{thm:cut-admiss-FOInt-refined} below to apply the admissible rules $\refl$ and $\cut$.

\begin{center}
\begin{tabular}{c}
\AxiomC{$\seqempstr \sar w : \phi$}

\AxiomC{$\seqempstr \sar w : \phi \imp \psi$}
\RightLabel{\lem~\ref{lem:imp-invert-FOInt-refined}}
\dashedLine
\UnaryInfC{$w \leq u, u : \phi \sar u : \psi$}
\RightLabel{$\lsub$}
\dashedLine
\UnaryInfC{$(w \leq u)(w/u), (u : \phi)(w/u) \sar (u : \psi)(w/u)$}
\RightLabel{=}
\dottedLine
\UnaryInfC{$w \leq w, w : \phi \sar w : \psi$}
\RightLabel{$\refl$}
\dashedLine
\UnaryInfC{$w : \phi \sar w : \psi$}

\RightLabel{$\cut$}
\dashedLine
\BinaryInfC{$\seqempstr \sar w : \psi$}
\DisplayProof
\end{tabular}
\end{center}

\emph{Rule R1.}

\begin{center}
\begin{tabular}{c}
\AxiomC{$\seqempstr \sar w : \phi(\unda/x)$}
\RightLabel{$\wk$}
\dashedLine
\UnaryInfC{$u \leq w \sar w : \phi(\unda/x)$}
\RightLabel{$\allrnc$}
\UnaryInfC{$\seqempstr \sar u : \forall x \phi$}
\RightLabel{$\lsub$}
\dashedLine
\UnaryInfC{$\seqempstr \sar w : \forall x \phi$}
\DisplayProof
\end{tabular}
\end{center}

\end{proof}

Due to the similarity of the refined labelled calculi $\intfondl$ and $\intfocdl$, and the quasi-refined labelled calculi $\intfondlq$ and $\intfocdlq$, it should not be surprising that $\intfondl$ and $\intfocdl$ are complete relative to labelled tree derivations with the fixed root property. The following theorem confirms this fact. Yet, since we have removed all dependencies on domain atoms within our refined labelled calculi, such systems can be proven notational variants of nested sequent systems, which we demonstrate in the following section.

\begin{theorem}\label{thm:treelike-derivations-refined-FOInt}
Every derivation in $\intfondl$ and $\intfocdl$ of a labelled formula $w : \phi$ is a labelled tree derivation with the fixed root property.
\end{theorem}

\begin{proof} Similar to the proof of \thm~\ref{thm:treelike-derivations-quasi-refined-FOInt}.
\end{proof}


\section{Relationship to Nested Sequent Formalism}\label{subsect:translation-to-nested-FOInt}

We show that our refined labelled calculi $\intfondl$ and $\intfocdl$ can be viewed as nested systems. The systems we put forth are distinct from Fitting's nested calculi for $\intfond$ and $\intfocd$ (which have been provided in Appendix A on p.~\pageref{app:fittings-nested-calculi}), as we employ reachability and propagation rules such as $\idfonc$, $\primp$, $\existsrnc$, and $\alllnc$ (shown in \fig~\ref{fig:nested-calculi-FOInt} below), whereas Fitting's calculi utilize different versions of these rules and adds a rule called \emph{lift} (which is a reformulation of the propagation rules introduced by Postniece in~\cite{Post09,Post10}). As discussed previously, making use of reachability and propagation rules (which take a \cfcst system as a parameter) is advantageous, as it lets us easily transform our calculi into systems for new logics by taking alternative \cfcst systems as parameters. We first define nested sequents, and afterward, define their sequent graphs, which facilitate translations between labelled and nested notation.

\begin{definition}[Nested Sequents for First-Order Intuitionistic Logics] We define a \emph{nested sequent for first-order intuitionistic logics}\index{Nested sequent!for first-order intuitionistic logics} to be a syntactic object $X$ defined via the grammars in BNF shown below:
\begin{center}
\begin{tabular}{c @{\hskip 2em} c}
$X ::= Y \sar Y \ | \ Y \sar Y, \nbbl X \nbbr, \ldots, \nbbl X \nbbr$

&

$Y ::= \varepsilon \ | \ \phi \ | \ Y, Y$
\end{tabular}
\end{center}
where $\phi \in \langintfo$.
\end{definition}

We denote nested sequents with $\na$, $\nb$, $\nc$, $\ldots$ (possibly annotated), and as usual, $\seqempstr$ represents the empty string\index{Empty string}, which is an identity element for comma, where comma associates and commutes. As before, we use the notation $\na \nbl \nb \nbr$ (and $\na \nbl \nb \nbr \nbl \nc \nbr$) to mean that the nested sequent $\nb$ ($\nb$ and $\nc$, \resp) occurs (occur, \resp) at some depth in the nestings of $\na$. For example, if the nested sequent $\na$ is $p(\unda) \sar q, \nbbl \forall x p(x) \sar q \imp q\}, \nbbl \seqempstr \sar q \nbbr$, then $\na \nbl p(\unda) \sar q \nbr$, $\na \nbl p(\unda) \sar q, \nbbl \forall x p(x) \sar q \imp q \nbbr \nbr$, and $\na \nbl p(\unda) \sar q \nbr \nbl \seqempstr \sar q \nbr$ are all valid representations of $\na$.

\begin{definition}[Sequent Graph of a Nested Sequent for $\intfond$ and $\intfocd$] We define the \emph{sequent graph of a nested sequent $\na$}\index{Sequent graph!of a nested sequent for $\intfond$ and $\intfocd$} inductively on the depth of the nestings of $\na$ as shown below. As in \dfn~\ref{def:sequent-graph-nested-kms}, we make use of sequences of natural numbers, and represent such sequences as $n_{1} . n_{2} . \ldots . n_{k-1} . n_{k}$, using $\numseq$ (possible annotated) to denote them. Our inductive definition of $\seqgraph(X) := \seqgraph_{0}(X)$ is as follows:
\begin{itemize}

\item[$\li$] If $\na = \phi_{1}, \ldots, \phi_{n} \sar \psi_{1}, \ldots, \psi_{k}$, then $G_{\numseq}(\na) := (V_{\numseq},E_{\numseq},L_{\numseq})$, where
$$
(i) \quad V_{\numseq} := \{\numseq\} \qquad (ii) \quad E_{\numseq} := \emptyset \qquad (iii) \quad L_{\numseq} := \{(\numseq, \phi_{1}, \ldots, \phi_{n} \sar \psi_{1}, \ldots, \psi_{k})\}
$$

\item[$\li$] Let $\na := \phi_{1}, \ldots, \phi_{n} \sar \psi_{1}, \ldots, \psi_{k}, \nbbl \nb_{1} \nbbr, \ldots, \nbbl \nb_{m} \nbbr$ and suppose that each $G_{\numseq  . i}(\nb_{i}) := ( V_{ \numseq . i}, E_{\numseq . i}, L_{\numseq . i} )$ (with $i \in \{1, \ldots m\}$ and $m \in \mathbb{N}$) is already defined. We define $G_{\numseq}(\na) := ( V_{\numseq}, E_{\numseq}, L_{\numseq} )$ as shown below:
\begin{flushleft}
$\li \quad V_{\numseq} := \{\numseq\} \cup \displaystyle{\bigcup_{1 \leq i \leq m} V_{\numseq . i}}$\\
$\li \quad E_{\numseq} := \{(\numseq,\numseq . i, a) \ | \ 1 \leq i \leq m \} \cup \displaystyle{\bigcup_{1 \leq i \leq m} E_{\numseq . i}}$\\
$\li \quad L_{\numseq} := \{(\numseq, \phi_{1}, \ldots, \phi_{n} \sar \psi_{1}, \ldots, \psi_{k})\} \cup \displaystyle{\bigcup_{1 \leq i \leq m} L_{\numseq . i}}
$
\end{flushleft}
\end{itemize}
Note that when $n = 0$ or $k = 0$, the multiset $\phi_{1}, \ldots, \phi_{n}$ and $\psi_{1}, \ldots, \psi_{k}$ is taken to be the empty string $\seqempstr$. We will often use $w$, $u$, $v$, $\ldots$ to represent vertices as opposed to sequences of natural numbers.
\end{definition}

For a nested sequent $\na \nbl \nb \sar \nc \nbr$ or $\na \nbl \nb_{1} \sar \nc_{1} \nbr \nbl \nb_{2} \sar \nc_{2} \nbr$ , we also use the notation $\na \nbl \nb \sar \nc \nbr_{w}$ and $\na \nbl \nb_{1} \sar \nc_{1} \nbr_{w} \nbl \nb_{2} \sar \nc_{2} \nbr_{u}$ to denote that $\nb \sar \nc$ is associated with the vertex $w$ (meaning that $L(w) = \nb \sar \nc$ in $\seqgraph(X) = (V,E,L)$), and to denote that $\nb_{1} \sar \nc_{1}$ and $\nb_{2} \sar \nc_{2}$ are associated with the vertices $w$ and $u$ (meaning that $L(w) = \nb_{1} \sar \nc_{1}$ and $L(u) = \nb_{2} \sar \nc_{2}$ in $\seqgraph(X) = (V,E,L)$), respectively.

\begin{figure}[t]
\noindent\hrule

\begin{center}
\begin{tabular}{c c}
\AxiomC{}
\RightLabel{$\idfonc^{\dag_{1}(\norc)}$}
\UnaryInfC{$X \nbl Y_{1}, p(\vec{\unda}) \sar Z_{1} \nbr _{w} [ Y_{2} \sar p(\vec{\unda}), Z_{2} \nbr_{u}$}
\DisplayProof

&

\AxiomC{}
\RightLabel{$\botl$}
\UnaryInfC{$X \nbl  Y, \bot \sar Z \nbr $}
\DisplayProof
\end{tabular}
\end{center}

\begin{center}
\begin{tabular}{c c c}
\AxiomC{$X \nbl Y, \phi, \psi \sar Z \nbr  $}
\RightLabel{$\conl$}
\UnaryInfC{$X \nbl Y, \phi \land \psi \sar Z \nbr $}
\DisplayProof

&

\AxiomC{$X \nbl  Y \sar \phi, \psi, Z \nbr  $}
\RightLabel{$\disr$}
\UnaryInfC{$X \nbl  Y \sar \phi \lor \psi, Z \nbr $}
\DisplayProof

&

\AxiomC{$X \nbl  Y \sar Z, \nbbl \phi \sar \psi\nbbr \nbr $}
\RightLabel{$\impr$}
\UnaryInfC{$X \nbl Y \sar \phi \imp \psi, Z \nbr $}
\DisplayProof
\end{tabular}
\end{center}

\begin{center}
\begin{tabular}{c c}
\AxiomC{$X \nbl Y, \phi \sar Z \nbr $}
\AxiomC{$X \nbl Y, \psi \sar Z \nbr $}
\RightLabel{$\disl$}
\BinaryInfC{$X \nbl Y, \phi \lor \psi \sar Z \nbr $}
\DisplayProof

&

\AxiomC{$X \nbl Y \sar \phi, Z \nbr $}
\AxiomC{$X \nbl Y \sar \psi, Z \nbr $}
\RightLabel{$\conr$}
\BinaryInfC{$X \nbl Y \sar \phi \land \psi, Z \nbr $}
\DisplayProof

\end{tabular}
\end{center}

\begin{center}
\begin{tabular}{c}
\AxiomC{$X \nbl Y_{1}, \phi \imp \psi \sar  Z_{1} \nbr _{w} \nbl Y_{2} \sar \phi, Z_{2} \nbr _{u}$}
\AxiomC{$X \nbl Y_{1}, \phi \imp \psi \sar Z_{1} \nbr _{w} \nbl Y_{2}, \psi \sar  Z_{2} \nbr_{u}$}
\RightLabel{$\primp^{\dag_{1}(\norc)}$}
\BinaryInfC{$X \nbl Y_{1}, \phi \imp \psi \sar Z_{1} \nbr_{w} \nbl Y_{2} \sar  Z_{2} \nbr _{u}$}
\DisplayProof
\end{tabular}
\end{center}

\begin{center}
\begin{tabular}{c c}
\AxiomC{$X[ Y, \phi(\unda/x) \sar Z  \nbr $}
\RightLabel{$\existslnc^{\dag_{2}(\norc)}$}
\UnaryInfC{$X[ Y, \exists x \phi \sar Z \nbr $}
\DisplayProof

&

\AxiomC{$X \nbl Y_{1}, \forall x \phi \sar Z_{1} \nbr_{w} \nbl Y_{2}, \phi(\unda/x) \sar Z_{2} \nbr_{v}$}
\RightLabel{$\alllnc^{\dag_{4}(\norc)}$}
\UnaryInfC{$X \nbl Y_{1}, \forall x \phi \sar Z_{1} \nbr_{w} \nbl Y_{2} \sar Z_{2} \nbr_{v}$}
\DisplayProof
\end{tabular}
\end{center}

\begin{center}
\begin{tabular}{c c}
\AxiomC{$X[ Y \sar \phi(\unda/x), \exists x \phi, Z  \nbr _{w}$}
\RightLabel{$\existsrnc^{\dag_{3}(\norc)}$}
\UnaryInfC{$X[ Y \sar \exists x \phi, Z \nbr _{w}$}
\DisplayProof

&

\AxiomC{$X[ Y \sar Z, \nbbl \seqempstr \sar \phi(\unda/x)\nbbr   \nbr $}
\RightLabel{$\allrnc^{\dag_{2}(\norc)}$}
\UnaryInfC{$X[ Y \sar \forall x \phi, Z \nbr $}
\DisplayProof
\end{tabular}
\end{center}

\hrule
\caption{The nested calculus $\nintfond$\index{$\nintfond$} consists of all rules with $ \mathsf{X} = \nnn$. The nested calculus $\nintfocd$\index{$\nintfocd$} consists of all rules with $\mathsf{X} = \ccc$. The side conditions of the propagation and reachability rules are given in \fig~\ref{fig:side-conditions-Nested-FOInt} below.}
\label{fig:nested-calculi-FOInt}
\end{figure}

As with our refined labelled calculi, applications of the reachability and propagation rules depend on the notion of a propagation graph, $\thuesysi$-availability, and $\thuesysii$-availability. We define these concepts below, and subsequently provide an example of a sequent graph and propagation graph for a nested sequent, as well as give an example of a parameter which is $\thuesysi$-available and $\thuesysii$-available.

\begin{definition}[Propagation Graphs for $\nintfond$ and $\nintfocd$]\label{def:propagation-graph-nested-FOInt} Let $\na$ be a nested sequent for $\intfond$ and $\intfocd$\index{Propagation graph!for $\nintfond$ and $\nintfocd$} with the sequent graph $\seqgraph(\na) = (V,E,L)$. We define the \emph{propagation graph} $\prgr{\na} = (\prgrdom', \prgredges')$ to be the directed graph such that
\begin{itemize}

\item[$\li$] $\prgrdom' := V$;

\item[$\li$] $\prgredges' := \{(w,u,a), (u,w,\overline{a}) \ | \ (w,u,a) \in E \}$.

\end{itemize}
We will often write $w \in \prgr{\na}$ to mean $w \in \prgrdom'$, and $(w,u,\chara) \in \prgr{\na}$ to mean $(w,u,\chara) \in \prgredges'$, for $\chara \in \albet = \{a, \conv{a}\}$.
\end{definition}

\begin{definition}[$\thuesysi$-available, $\thuesysii$-available]\label{def:S4-S5-available-nested-FOInt} Let $X := X[Y_{1} \sar Z_{1}]_{u}[Y_{2} \sar Z_{2}]_{w}$ be a nested sequent. We say that a parameter $\unda$ is \emph{$\thuesysi$-available}\index{$\thuesysi$-available}\index{$\thuesysii$-available} (\emph{$\thuesysii$-available}) in $X$ for $w$ \ifandonlyif there exists a formula $\psi(\vec{\unda}) \in Y_{1}, Z_{1}$ with $\unda$ in $\vec{\unda}$ and a propagation path $\ppath(u,w)$ in $\prgr{X}$ such that $\stra_{\ppath}(u,w) \in \thuesysilang{a}$ ($\stra_{\ppath}(u,w) \in \thuesysiilang{a}$, \resp).
\end{definition}

Propagation paths and strings of propagation paths (along with their converses) are defined as in \dfn~\ref{def:propagation-path-kms}, so we do not repeat these definitions here. However, we do provide examples of these objects below along with our example:

\begin{example} Let our nested sequent be the following:
$$
\na := p(\unda) \sar q, \nbbl \forall x p(x) \sar q \imp q\}, \nbbl \seqempstr \sar q \nbbr
$$
The sequent graph $\seqgraph(\na)$ is obtained by deleting the dotted edges in the graph below, and the propagation graph $\prgr{\na}$ is obtained by deleting the solid edges.
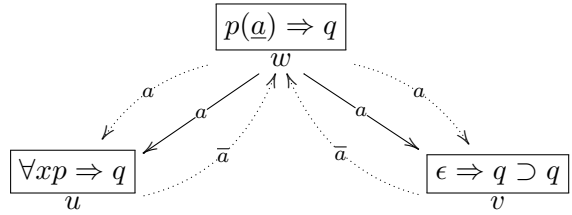
\begin{center}
\begin{tabular}{c}
\xymatrix{
 & \overset{\boxed{p(\unda) \sar q}}{w} \ar[dr]|-{a} \ar[dl]|-{a}\ar@/^1.5pc/@{.>}[dr]|-{a} \ar@/^-1.5pc/@{.>}[dl]|-{a} & \\
\overset{\boxed{\forall x p \sar q}}{u}\ar@/^-1.5pc/@{.>}[ur]|-{\conv{a}} & & \overset{\boxed{\seqempstr \sar q \imp q}}{v}\ar@/^1.5pc/@{.>}[ul]|-{\conv{a}}
}
\end{tabular}
\end{center}
Examples of propagation paths include the propagation path $\ppath(w,u) = w, a, u$ and the propagation path $\ppath'(u,w) := u, \conv{a}, w, a, v, \conv{a}, w$. The string of each propagation path is $\stra_{\ppath}(w,u) = a$ and $\stra_{\ppath'}(u,w) = \conv{a} \cate a \cate \conv{a}$, respectively. Also, since $p(\unda)$ occurs in the antecedent at $w$ and there exists a propagation path $\ppath(w,u) := w, a , u$ with $a \in \thuesysilang{a}$ and $a \in \thuesysiilang{a}$, we know that $\unda$ is both $\thuesysi$- and $\thuesysii$-available for $u$.
\end{example}

Making use of the above notions, we define our nested calculi $\nintfond$ and $\nintfocd$ for the first-order intuitionistic logics $\intfond$ and $\intfocd$, respectively, which are presented in \fig~\ref{fig:nested-calculi-FOInt} with the side conditions of reachability and propagation rules given in \fig~\ref{fig:side-conditions-Nested-FOInt}.

\begin{figure}[t]
\begin{center}
\bgroup
\def\arraystretch{1.25}
\begin{tabular}{| c | c | c | c |}
\hline
Name & Side Condition & Name & Side Condition\\
\hline
$\dag_{1}(\nnn)$ & $\exists\ppath(\stra_{\ppath}(w,u) \in \thuesysilang{a})$ &
$\dag_{1}(\ccc)$ & $\exists\ppath(\stra_{\ppath}(w,u) \in \thuesysilang{a})$\\
\hline
$\dag_{2}(\nnn)$ &  $\unda$ is an eigenvariable & $\dag_{2}(\ccc)$ &  $\unda$ is an eigenvariable\\
\hline
$\dag_{3}(\nnn)$ & $\unda$ is $\thuesysi$-available for $w$ & $\dag_{3}(\ccc)$ &  $\unda$ $\thuesysii$-available for $w$\\
 &  or $\unda$ is an eigenvariable &  &  or $\unda$ is an eigenvariable\\
\hline
$\dag_{4}(\nnn)$ & $\unda$ is $\thuesysi$-available for $v$ & $\dag_{4}(\ccc)$ &  $\unda$ $\thuesysii$-available for $v$\\
 &  or $\unda$ is an eigenvariable, and &  &  or $\unda$ is an eigenvariable, and\\
  &  $\exists\ppath(\stra_{\ppath}(w,v) \in \thuesysilang{a})$ &  &  $\exists\ppath(\stra_{\ppath}(w,v) \in \thuesysilang{a})$\\
\hline
\end{tabular}
\egroup
\end{center}
\caption{Side conditions for rules in $\intfondl$ and $\intfocdl$.}
\label{fig:side-conditions-Nested-FOInt}
\end{figure}

We now define our translations functions $\lnint$ and $\nlint$, which translate from labelled to nested notation, and from nested to labelled notation, respectively. The definition depends on the notation of the downward closure of a sequent graph, defined in \dfn~\ref{def:downward-closure-kms}. After presenting these definitions, we prove that $\intfondl$ and $\intfocdl$ are the respective notational variants of the calculi $\nintfond$ and $\nintfocd$, and give an example of translating derivations between the calculi.

\begin{definition}[The Translation $\lnkms$]\label{def:ln-kms} Let $\Lambda := \rel, \Gamma \sar \Delta$ be the a labelled tree sequent with $\seqgraph(\Lambda) = (V,E,L)$ and $w \in V$ the root. We define the translation $\lnt(\Lambda) := \lnt(\seqgraph_{w}(\Lambda))$\index{Translation $\lnt$!for first-order intuitionistic logics} inductively as follows:
\begin{itemize}

\item[$\li$] If $\seqgraph(\Lambda) = (V,E,L)$ with $V = \{w\}$, $E = \emptyset$, and $L = \{(w, \Gamma \restriction w \sar \Delta \restriction w)\}$, then
$$
\lnt(\seqgraph_{w}(\Lambda)) := \Gamma \restriction w \sar \Delta \restriction w
$$

\item[$\li$] If $\seqgraph(\Lambda) = (V,E,L)$ with $w, u_{1}, \ldots, u_{n} \in V$, $(w,u_{i}) \in E$ (for $i \in \{1, \ldots, n\}$), then
$$
\lnt(\seqgraph_{w}(\Lambda)) := L(w), \nbbl \lnt(\seqgraph_{u_{1}}(\Lambda)) \nbbr, \ldots, \nbbl \lnt(\seqgraph_{u_{n}}(\Lambda)) \nbbr
$$
\end{itemize}
\end{definition}

\begin{example} We show how to translate the labelled tree sequent
$$
\Lambda:= w \leq u, u \leq v, w \leq z, w : q,  v : p \imp q, v :r \sar , w : \forall x q, u : p \lor p, z : c, z :p
$$
into a nested sequent via the computation below:
\begin{eqnarray*}
& \lnkms(\Lambda) & := \lnkms(\seqgraph_{w}(\Lambda))\\
&  & = q \sar \forall x q, \nbbl \lnkms(\seqgraph_{u}(\Lambda))\nbbr, \nbbl \lnkms(\seqgraph_{z}(\Lambda))\nbbr\\
&  & = q \sar \forall x q, \nbbl \seqempstr \sar p \lor p, \nbbl\lnkms(\seqgraph_{v}(\Lambda))\nbbr\nbbr, \nbbl \seqempstr \sar c,p\nbbr\\
&  & = q \sar \forall x q, \nbbl \seqempstr \sar p \lor p, \nbbl p \imp q, r \sar \seqempstr\nbbr\nbbr, \nbbl\seqempstr \sar c,p\nbbr\\
\end{eqnarray*}
\end{example}

\begin{definition}[The Translation $\nlt$]\label{def:nl-kms} Let $\na$ be the a nested sequent. We define the translation $\nlt(\na) := \nlt(\seqgraph_{0}(\na))$\index{Translation $\nlt$!for first-order intuitionistic logics} inductively as follows:
\begin{itemize}

\item[$\li$] If $\seqgraph_{\numseq}(\na) = (V,E,L)$ with $V = \{\numseq\}$, $E = \emptyset$, and $L = \{(\numseq, Y \sar Z)\}$, then
$$
\nlt(\seqgraph_{\numseq}(\na)) := w_{\numseq} : Y \sar w_{\numseq} : Z
$$

\item[$\li$] If $\seqgraph_{\numseq}(\na) = (V,E,L)$ with $\numseq, \numseq . 1, \ldots, \numseq . n \in V$, $(\numseq, \numseq . 1), \ldots, (\numseq, \numseq . n) \in E$, and $(\numseq, Y \sar Z) \in L$, then
\end{itemize}
$$
\nlt(\seqgraph_{\numseq}(\na)) := (w_{\numseq} \leq w_{\numseq . 1}, \ldots, w_{\numseq} \leq w_{\numseq . n}, w_{\numseq} : Y \sar w_{\numseq} : Z) \seqcomp \nlt(\seqgraph_{\numseq . 1}(\na)) \seqcomp \cdots \seqcomp \nlt(\seqgraph_{\numseq . n}(\na))
$$
In practice, we will often use labels such as $w$, $u$, $v$, $\ldots$ as opposed to labels indexed with sequences of natural numbers for simplicity.
\end{definition}

\begin{example} Below, we show how to translate the nested sequent
$$
X := q \sar \forall x q, \nbbl \seqempstr \sar p \lor p, \nbbl p \imp q, r \sar \seqempstr \nbbr \nbbr, \nbbl \seqempstr \sar c,p \nbbr
$$
into a labelled (tree) sequent:
\begin{eqnarray*}
& \nlkms(X) & := \nlkms(\seqgraph_{0}(X))\\
&  & = (w \leq u, w \leq z, w : q \sar w : \forall x q) \seqcomp \nlkms(\seqgraph_{u}(X)) \seqcomp \nlkms(\seqgraph_{z}(X)) \\
&  & = (w \leq u, w \leq z, w : q \sar w : \forall x q) \seqcomp (u \leq v \sar u : p \lor p) \seqcomp  \\
& & \textcolor{white}{=}\nlkms(\seqgraph_{v}(X)) \seqcomp (\seqempstr \sar z : c, z :p)\\
&  & = (w \leq u, w \leq z, w : q \sar w : \forall x q) \seqcomp (u \leq v \sar u : p \lor p) \seqcomp \\
& & \textcolor{white}{=} (v : p \imp q, v : r \sar \seqempstr) \seqcomp (\seqempstr \sar z : c, z :p)\\
&  & = w \leq u, u \leq v, w \leq z, w : q,  v : p \imp q, v :r \sar , w : \forall x q, u : p \lor p, z : c, z :p
\end{eqnarray*}
\end{example}

As with our translations for grammar logics, it is not difficult to confirm that the sequent graph of a labelled tree sequent or nested sequent is isomorphic to the sequent graph of its translatee under $\lnint$ and $\nlint$, respectively. This fact is expressed in the following lemma:

\begin{lemma}\label{lem:iso-translation-FOInt} Let $\Lambda$ be a labelled tree sequent and $X$ be a nested sequent (for first-order intuitionistic logics). Then,

(i) $G(\Lambda) \iso G(\lnkms(\Lambda))$

(ii) $G(X) \iso G(\nlkms(X))$
\end{lemma}

\begin{theorem}\label{thm:Refined-to-Nested-FOInt}
Let $\mathsf{X} \in \{\nnn, \ccc\}$. Every derivation in $\mathsf{IntXL}$ is algorithmically translatable to a derivation in $\mathsf{DIntX}$.
\end{theorem}

\begin{proof} We prove the result by induction on the height of the given derivation and show the claim for $\intfocdl$ and $\nintfocd$, as translating from $\intfondl$ to $\nintfond$ is similar.

\textit{Base case.} Let the initial sequent shown below left be $\Lambda$, and assume that the nested sequent shown below right is $\lnint(\Lambda)$. Due to the application of $\idfonc$ (bottom left), we know that there exists a propagation path $\ppath(w,u)$ in $\prgr{\Lambda}$ such that $\stra_{\ppath}(w,u) \in \thuesysilang{a}$. By \lem~\ref{lem:iso-translation-FOInt} above, the same propagation path will exist in $\prgr{\lnint(\Lambda)}$, and hence, the application of the $\idfonc$ rule (bottom-right) deriving $\lnint(\Lambda)$ is valid.

\begin{center}
\begin{tabular}{c c c}
\AxiomC{}
\RightLabel{$\idfonc$}
\UnaryInfC{$\rel, \Gamma, w : p(\vec{\unda}) \sar u : p(\vec{\unda}), \Delta$}
\DisplayProof

&

$\leadsto$

&

\AxiomC{}
\RightLabel{$\idfonc$}
\UnaryInfC{$X \nbl Y_{1}, p(\vec{\unda}) \sar Z_{1} \nbr _{w} [ Y_{2} \sar p(\vec{\unda}), Z_{2} \nbr_{u}$}
\DisplayProof
\end{tabular}
\end{center}

\textit{Inductive step.} We consider the $\primp$, $\existslc$, and $\existsrc$ cases; all other cases are straightforward or similar.

$\primp$. Let the left premise of the $\primp$ inference shown below top be $\Lambda_{1}$ and the right premise be $\Lambda_{2}$. We assume that the left premise of the $\primp$ inference shown below bottom is $\lnint(\Lambda_{1})$ and the right premise is $\lnint(\Lambda_{2})$. Due to the application of $\primp$ shown below top, we know that there exists a propagation path $\ppath(w,u)$ in $\prgr{\Lambda_{1}}$ and $\prgr{\Lambda_{2}}$ such that $\stra_{\ppath}(w,u) \in \thuesysilang{a}$. By \lem~\ref{lem:iso-translation-FOInt}, we know that the propagation path exists in $\prgr{\lnint(\Lambda_{1})}$ and $\prgr{\lnint(\Lambda_{2})}$ as well, and so, $\primp$ may be applied to $\lnint(\Lambda_{1})$ and $\lnint(\Lambda_{2})$ as shown below bottom.

\begin{flushleft}
\begin{tabular}{c c}
\AxiomC{$\rel, w : \phi \imp \psi, \Gamma \Rightarrow \Delta, u : \phi$}
\AxiomC{$\rel, w : \phi \imp \psi, u : \psi, \Gamma \Rightarrow \Delta$}
\RightLabel{$\primp$}
\BinaryInfC{$\rel, w : \phi \imp \psi, \Gamma \Rightarrow \Delta$}
\DisplayProof

&

$\leadsto$
\end{tabular}
\end{flushleft}
\begin{flushright}
\AxiomC{$X \nbl Y_{1}, \phi \imp \psi \sar  Z_{1} \nbr _{w} \nbl Y_{2} \sar \phi, Z_{2} \nbr _{u}$}
\AxiomC{$X \nbl Y_{1}, \phi \imp \psi \sar Z_{1} \nbr _{w} \nbl Y_{2}, \psi \sar  Z_{2} \nbr_{u}$}
\RightLabel{$\primp$}
\BinaryInfC{$X \nbl Y_{1}, \phi \imp \psi \sar Z_{1} \nbr_{w} \nbl Y_{2} \sar  Z_{2} \nbr _{u}$}
\DisplayProof
\end{flushright}

$\existslc$. Let $\Lambda$ be the premise of the left inference below, and $\lnint(\Lambda)$ be the premise of the right inference. By assumption, we know that $\unda$ is an eigenvariable in $\Lambda$, and so, by \dfn~\ref{def:ln-kms}, $\unda$ is an eigenvariable in $\lnint(\Lambda)$. Hence, we may apply the $\existslc$ rule to $\lnint(\Lambda)$, giving the desired conclusion as shown below right.

\begin{center}
\begin{tabular}{c c c}
\AxiomC{$\rel, \Gamma, w: \phi(\unda/x) \sar \Delta$}
\RightLabel{$\existslc$}
\UnaryInfC{$\rel, \Gamma, w: \exists x \phi \sar \Delta$}
\DisplayProof

&

$\leadsto$

&

\AxiomC{$X[ Y, \phi(\unda/x) \sar Z  \nbr $}
\RightLabel{$\existslc$}
\UnaryInfC{$X[ Y, \exists x \phi \sar Z \nbr $}
\DisplayProof
\end{tabular}
\end{center}

$\existsrc$. Let the premise of the left inference be $\Lambda$ and the premise of the right inference be $\lnint(\Lambda)$. By the side condition on the left $\existslc$ inference, we know that either $\unda$ is an eigenvariable, or $\unda$ is $\thuesysii$-available for $w$. In the first case, by the definition of $\lnint$ (\dfn~\ref{def:ln-kms}), we know that $\unda$ will be an eigenvariable in $\lnint(\Lambda)$ as well. In the second case, by \lem~\ref{lem:iso-translation-FOInt}, $\unda$ will be $\thuesysii$-available. Regardless of the case then, the side condition holds, and we may apply the rule $\existslc$ to $\lnint(\Lambda)$ as shown below right, giving the desired conclusion.

\begin{center}
\begin{tabular}{c c c}
\AxiomC{$\rel, \Gamma \Rightarrow \Delta, w: \phi(\unda/x), w: \exists x \phi$}
\RightLabel{$\existsrc$}
\UnaryInfC{$\rel, \Gamma \Rightarrow \Delta, w: \exists x \phi$}
\DisplayProof

&

$\leadsto$

&

\AxiomC{$X[ Y \sar \phi(\unda/x), \exists x \phi, Z  \nbr _{w}$}
\RightLabel{$\existsrc$}
\UnaryInfC{$X[ Y \sar \exists x \phi, Z \nbr _{w}$}
\DisplayProof
\end{tabular}
\end{center}












\end{proof}

\begin{theorem}\label{thm:Nested-to-Refined-FOInt}
Let $\mathsf{X} \in \{\nnn, \ccc\}$. Every derivation in $\mathsf{DIntX}$ is algorithmically translatable to a derivation in $\mathsf{IntXL}$.
\end{theorem}

\begin{proof} The theorem is proven in a similar fashion as \thm~\ref{thm:Refined-to-Nested-FOInt} above. One proves by induction on the height of the given derivation that each rule in $\nintfond$ and $\nintfocd$ can be translated into an instance of the corresponding rule in $\intfondl$ and $\intfocdl$, respectively. Similar to the proof of \thm~\ref{thm:Refined-to-Nested-FOInt}, each rule straightforwardly translates (under the $\nlint$ function).
\end{proof}

\begin{example} If the top derivation below (in $\intfocdl$) is input into the translation algorithm of \thm~\ref{thm:Refined-to-Nested-FOInt}, then we obtain the derivation shown below bottom. On the other hand, the derivation below bottom (in $\nintfocd$) translates via \thm~\ref{thm:Nested-to-Refined-FOInt} to the derivation shown below top. 

Also, obverse that the $\alllc$ rule may be applied in the top derivation because $\unda$ is $\thuesysii$-available for $u$, that is, we have the labelled formula $v : p(\unda)$ and the propagation path $\ppath(v,u) := v, \conv{a}, u$ such that $\conv{a} \in \thuesysiilang{a}$. Similarly, the side condition holds in the $\nintfocd$ derivation, so the $\alllc$ rule may be applied there as well. It should be pointed out that in the non-constant domain setting, the theorem shown below (which is an instance of the constant domain axiom A13) is not derivable because the instance of $\alllc$ cannot be replaced by an instance of $\allln$. The inability to substitute $\allln$ for $\alllc$ is due to the fact that the string of every propagation path from $v$ to $u$ must contain at least one occurrence of the character $\conv{a}$, and so, the string will never be in $\thuesysilang{a}$.

\begin{flushleft}
$\Lambda_{1} := w \leq u, u \leq v, u : \forall x (p(x) \lor q), u : p(\unda) \sar v : p(\unda), u : q$
\end{flushleft}
\begin{flushleft}
$\Lambda_{2} := w \leq u, u \leq v, u : \forall x (p(x) \lor q), u : q \sar v : p(\unda), u : q$
\end{flushleft}

\begin{center}
\AxiomC{}
\RightLabel{$\idfonc$}
\UnaryInfC{$\Lambda_{1}$}

\AxiomC{}
\RightLabel{$\idfonc$}
\UnaryInfC{$\Lambda_{2}$}

\RightLabel{$\disl$}
\BinaryInfC{$w \leq u, u \leq v, u : \forall x (p(x) \lor q), u : p(\unda) \lor q \sar v : p(\unda), u : q$}
\RightLabel{$\alllc$}
\UnaryInfC{$w \leq u, u \leq v, u : \forall x (p(x) \lor q) \sar v : p(\unda), u : q$}
\RightLabel{$\allrc$}
\UnaryInfC{$w \leq u, u : \forall x (p(x) \lor q) \sar u : \forall x p(x), u : q$}
\RightLabel{$\disr$}
\UnaryInfC{$w \leq u, u : \forall x (p(x) \lor q) \sar u : \forall x p(x) \lor q$}
\RightLabel{$\impr$}
\UnaryInfC{$\seqempstr \sar w : \forall x (p(x) \lor q) \imp \forall x p(x) \lor q$}
\DisplayProof
\end{center}

\begin{flushleft}
$X_{1} := \seqempstr \sar \seqempstr , \nbbl \forall x (p(x) \lor q), p(\unda) \sar q, \nbbl \seqempstr \sar p(\unda) \nbbr \nbbr$
\end{flushleft}
\begin{flushleft}
$X_{2} := \seqempstr \sar \seqempstr , \nbbl \forall x (p(x) \lor q), q \sar q, \nbbl \seqempstr \sar p(\unda) \nbbr \nbbr$
\end{flushleft}

\begin{center}
\AxiomC{}
\RightLabel{$\idfonc$}
\UnaryInfC{$X_{1}$}

\AxiomC{}
\RightLabel{$\idfonc$}
\UnaryInfC{$X_{2}$}

\RightLabel{$\disl$}
\BinaryInfC{$\seqempstr \sar \seqempstr , \nbbl \forall x (p(x) \lor q), p(\unda) \lor q \sar q, \nbbl \seqempstr \sar p(\unda) \nbbr \nbbr$}
\RightLabel{$\alllc$}
\UnaryInfC{$\seqempstr \sar \seqempstr , \nbbl \forall x (p(x) \lor q) \sar q, \nbbl \seqempstr \sar p(\unda) \nbbr \nbbr$}
\RightLabel{$\allrc$}
\UnaryInfC{$\seqempstr \sar \seqempstr , \nbbl \forall x (p(x) \lor q) \sar \forall x p(x), q \nbbr$}
\RightLabel{$\disr$}
\UnaryInfC{$\seqempstr \sar \seqempstr , \nbbl \forall x (p(x) \lor q) \sar \forall x p(x) \lor q \nbbr$}
\RightLabel{$\impr$}
\UnaryInfC{$\seqempstr \sar \forall x (p(x) \lor q) \imp \forall x p(x) \lor q$}
\DisplayProof
\end{center}

\end{example}







\chapter{Applications: Decidability and Interpolation}
\label{CPTR:Applications} 



In this chapter, we put our refined labelled calculi for deontic \stit logics and grammar logics to work. In \sect~\ref{sec:applicationsI}, we build off of the author's work in~\cite{LyoBer19} and show how each calculus $\dsnsal$ for the single-agent deontic \stit logic $\dsnnz$ (with $n=0$, i.e. $\ag = \{0\}$) can be utilized in a proof-search procedure to decide the validity of formulae. As is typical when proving decidability via proof-search, we write an algorithm that applies the rules of $\dsnsal$ in reverse, attempting to construct a proof of the input formula. As a corollary, we will not only obtain decidability for each logic $\dsnnz$, but will also confirm that each logic $\dsnnz$ is in possession of the \emph{finite model property}\index{Finite model property}, i.e. any formula that is satisfiable, is satisfiable on a finite model. We note that we do not provide proof-search procedures for first-order intuitionistic logics since they are undecidable and do not provide proof-search algorithms for grammar logics either since such algorithms were already provided for a decidable sub-class of such logics in~\cite{TiuIanGor12}. (NB. Context-free grammar logics with converse are undecidable in general.)

In \sect~\ref{sec:applicationsII}, the syntactic method of interpolation for nested sequent systems introduced in~\cite{LyoTiuGorClo20} is presented and applied to show that each context-free grammar logic with converse possesses the effective Lyndon interpolation property. At the beginning of \sect~\ref{sec:applicationsII}, we define (Lyndon) interpolation, discuss the importance of the property, and also describe the state of the art in the proof-theoretic approach to confirming the property. Last, we note that the new results of the second section (i.e. \sect~\ref{sec:applicationsII}) generalize those of~\cite{GorNgu05} from the class of regular grammar logics to the class of context-free grammar logics with converse (and from effective Craig interpolation to effective Lyndon interpolation).

\section{Proof-Search and Decidability for Deontic STIT Logics}\label{sec:applicationsI}


Proof-search algorithms\index{Proof-search} are special procedures defined relative to proof systems that construct or find the proof of an input formula (or, sequent) by applying the rules of the proof system in reverse. 
 Such algorithms are often used to prove logics decidable, but are also indispensable in the domain of automated reasoning---having been used to decide knowledge representation languages~\cite{Rad12}, to provide complexity-optimal decision procedures~\cite{LelPim15,Vig00}, and to automate the extraction of counter-models~\cite{LyoBer19,TiuIanGor12}, among other things. 

The proof-search algorithms we introduce in this section leverage refined labelled calculi from the class $\{\dsnlnz \ | \ k \in \mathbb{N}\}$ (see \fig~\ref{fig:refined-calculus-dsn}). The algorithms operate by taking a labelled sequent $\Lambda$ as input, and check (i) if a labelled formula $w : \phi$ and its negation $w : \negnnf{\phi}$ occur in $\Lambda$, and if not, then check (ii) if a complex formula exists in $\Lambda$ that has not yet been analyzed. Regarding point (i), if a labelled formula and its negation occur in the input labelled sequent, then by \lem~\ref{lem:general-id-dsn} and \thm~\ref{thm:gtdsn-to-dsnl}, we know the sequent is provable, meaning that the algorithm no longer needs to search for a proof. Regarding point (ii), if a complex formula has yet to be analyzed, then this leaves open the possibility that a proof of the labelled sequent may be found by applying the appropriate rule (bottom-up) to the input labelled sequent which analyzes the complex formula into the simpler auxiliary formulae from which it may be formed. It is important to point out that in our setting, our proof-search algorithms are performing two tasks simultaneously---searching for a proof of the input labelled sequent and constructing a counter-model of the input labelled sequent in case a proof is not found. To demonstrate this dualistic functionality of our proof-search procedure, we consider a concrete example of proof-search using a refined labelled calculus for a deontic \stit logic.

Let us suppose for the sake of the example that we have a single-agent (i.e. $n = 0$ and $\ag := \{0\}$) and that the agent's choices are not bounded (that is, $k = 0$). We want to search for a proof or counter-model of the labelled sequent $\seqempstr \sar w : \agboxz \Oiz (p \lor \negnnf{q})$ using the calculus $\mathsf{DS}^{0}_{0}\mathsf{L}$. In attempt to construct a proof, we apply relevant rules to the sequent in reverse, yielding the derivation shown below. Also, note that when applying rules in reverse we preserve the analyzed (principal) formula of the inference into the premise forcing us to invoke the admissibility of $\ctrr$ (\cor~\ref{cor:admissible-strucset-refined-dsn}) before each inference. The reason for doing this is that if a counter-model is to be constructed from failed proof-search, then having the formula present simplifies the proof confirming that the counter-model of the end sequent is indeed a counter-model; this will become apparent when showing the correctness of our proof-search procedures in \thm~\ref{thm:correctness} later on.

\begin{center}
\AxiomC{$R_{\agboxz}wu, \idealz v \sar w : \agboxz \Oiz (p \lor \negnnf{q}), u : \Oiz (p \lor \negnnf{q}), v : p \lor \negnnf{q}, v : p, v : \negnnf{q}$}
\RightLabel{$\disr$}
\UnaryInfC{$R_{\agboxz}wu, \idealz v \sar w : \agboxz \Oiz (p \lor \negnnf{q}), u : \Oiz (p \lor \negnnf{q}), v : p \lor \negnnf{q}, v : p \lor \negnnf{q}$}
\RightLabel{$\ctrr$}
\dashedLine
\UnaryInfC{$R_{\agboxz}wu, \idealz v \sar w : \agboxz \Oiz (p \lor \negnnf{q}), u : \Oiz (p \lor \negnnf{q}), v : p \lor \negnnf{q}$}
\RightLabel{$\Oirz$}
\UnaryInfC{$R_{\agboxz}wu \sar w : \agboxz \Oiz (p \lor \negnnf{q}), u : \Oiz (p \lor \negnnf{q}), u : \Oiz (p \lor \negnnf{q})$}
\RightLabel{$\ctrr$}
\dashedLine
\UnaryInfC{$R_{\agboxz}wu \sar w : \agboxz \Oiz (p \lor \negnnf{q}), u : \Oiz (p \lor \negnnf{q})$}
\RightLabel{$\agboxrz$}
\UnaryInfC{$\seqempstr \sar w : \agboxz \Oiz (p \lor \negnnf{q}), w : \agboxz \Oiz (p \lor \negnnf{q})$}
\RightLabel{$\ctrr$}
\dashedLine
\UnaryInfC{$\seqempstr \sar w : \agboxz \Oiz (p \lor \negnnf{q})$}
\DisplayProof
\end{center}

Despite the fact that each principal formula is preserved upward, it happens to be sufficient (for building a proof or counter-model of the end sequent) to analyze (i.e. apply an inference rule bottom-up to) each complex labelled formula only once. By this fact, our proof-search example will terminate with the derivation above since each complex labelled formula occurring in the top sequent was analyzed at some point lower in the derivation. Since the above `derivation' is not a valid derivation in $\mathsf{DS}^{0}_{0}\mathsf{L}$, as the top sequent is not an initial sequent, we can extract a counter-model for the end sequent from the top seqeunt, which we now explain.

A nice feature of the (refined) labelled formalism is that its close association with the semantics of the logic allows labelled sequents to be readily converted into models. In the current example then, our aim is to transform the top sequent of the derivation above into a $\mathsf{DS}^{0}_{0}$-model $M = (W,R_{\agboxz},\idealz,V)$. We define $W$ to be the set of labels, i.e. $W := \{w,u,v\}$. To define $R_{\agboxz}$, we make use of the relational atom $R_{\agboxz}wu$ and the set of worlds $W$, forming a set $\{(w,w), (u,u), (v,v), (w,u)\}$ that contains a reflexive pair for each world in $W$ (i.e. each label in the top sequent), and a pair from each relevant relational atom, which in this case just provides $(w,u)$ due to the occurrence of $R_{\agboxz}wu$. We then take the transitive and symmetric closure of the set, thus giving $R_{\agboxz} := \{(w,w),(w,u),(u,w),(u,u), (v,v)\}$, which ensures that the relation is an equivalence relation that partitions $W$ as dictated by condition \partcond (\dfn~\ref{def:frames-models-dsn}). We define $\idealz$ to be all labels in the choice-cell (from $R_{\agboxz}$) of an `ideal label'; since the only `ideal label' is $v$ due to the $\idealz v$ occurring in the top sequent, and because the only label in the choice-cell of $v$ is $v$, we have that $\idealz := \{v\}$. Last, we define the valuation function $V$ so that it invalidates each literal at the world (i.e. label) that prefixes it; thus, we have $V(p) := \{w,u\}$ and $V(q) := \{v\}$. It is straightforward to verify that $M$ is a $\mathsf{DS}^{0}_{0}$-model and satisfies conditions \partcondns--\choicecond and \donecondns--\dthreecondns.

We can see that $M$ is a counter-model for $\agboxz \Oiz (p \lor \negnnf{q})$ via the following calculation: Since $v \not\in V(p) \text{ and } v \in V(q)$, it follows that $M,v \not\Vdash p \text{ and } M,v \not\Vdash \negnnf{q}$, which further implies that $M,v \not\Vdash p \lor \negnnf{q}$. This fact, in conjunction with the fact that $v \in \idealz$, implies that $M,u \not\Vdash \Oiz (p \lor \negnnf{q})$, which entails that $M,w \not\Vdash \agboxz \Oiz (p \lor \negnnf{q})$, when one takes into account that $u \in R_{\agboxz}(w)$. Through failed proof-search then, we have shown that the top sequent of the above derivation can be used to construct a counter-model of the end sequent. Let us now move on to show that such operations can be performed generally.


\subsection{The Proof-Search Algorithm \provedsks and its Corollaries}

We provide a correct and terminating proof-search procedure \provedsks (\alg~\ref{alg:ProveDSk-i} below), which harnesses each refined labelled calculus $\dsnsal$ to decide its associated logic $\dsnnz$. 
 We only consider the single-agent setting where $n=0$ (i.e. $\ag = \{0\}$) and use $0$ to denote the single-agent throughout the course of the section. Despite this restriction, we still allow for choice limitation, and therefore the proof-search algorithm \provedsks takes a parameter $k$ limiting the number of choices of the agent $0$ to a maximum of $k$ (when $k > 0$), and imposes no limitation when $k = 0$ (that is, the agent $0$ may have an arbitrary number of choices available when $k = 0$). The single-agent setting greatly simplifies proof-search as the independence of agents rule $\ioa$ becomes redundant and no longer needs to be considered during proof-search as demonstrated by the following lemma:\footnote{We note that proof-search algorithms can be given for the multi-agent case as well, but require more sophisticated methods. We only consider the single-agent case here as it simplifies our work and is closely related to the published results in~\cite{LyoBer19} on proof-search for single-agent (non-deontic) \stit logics.}

\begin{lemma}
For each $k \in \mathbb{N}$, the $\ioa$ rule is eliminable in $\dsnnz$.
\end{lemma}

\begin{proof} Let $\Pi$ be a derivation in $\dsnsal$ containing some number of $\ioa$ inferences. We show how to eliminate topmost occurrences of the $\ioa$ rule, which, through repeated application yields a proof free of such inferences. Therefore, let us consider a topmost occurrence of $\ioa$ as shown below left. We invoke admissibility of $\lsub$ (\cor~\ref{cor:admissible-strucset-refined-dsn}) as shown below right to replace the eigenvariable $u$ with the label $w$, followed by the eliminability of $(ref_{0})$ (\lem~\ref{lem:refli-elim-dsn}) to obtain a derivation without the $\ioa$ inference. It is important to note two things: (i) the admissibility of $\lsub$ and the eliminability of $(ref_{0})$ do not introduce instances of $\ioa$, and (ii) although the resulting derivation may increase in size, since the number of $\ioa$ inferences has decreased, through repeated application of the explained procedure we still obtain a proof free of $\ioa$ inferences.

\begin{center}
\begin{tabular}{c c c}
\AxiomC{$\rel, R_{[0]}wu \sar \Gamma$}
\RightLabel{$\ioa$}
\UnaryInfC{$\rel \sar \Gamma$}
\DisplayProof

&

$\leadsto$

&

\AxiomC{$\rel, R_{[0]}wu \sar \Gamma$}
\RightLabel{$\lsub$}
\dashedLine
\UnaryInfC{$\rel, R_{[0]}ww \sar \Gamma$}
\RightLabel{$(ref_{0})$}
\dashedLine
\UnaryInfC{$\rel \sar \Gamma$}
\DisplayProof
\end{tabular}
\end{center}
\end{proof}

As discussed at the end of \sect~\ref{SECT:Refine-STIT}, bottom-up applications of the $\ioa$ rule introduce edges in the sequent graph of a labelled sequent that all meet at a fresh vertex, ensuring that the resulting labelled sequent is not a labelled forest sequent. In the single-agent setting, such structures do not appear, and furthermore, the redundancy of the $\ioa$ rule implies that we need not consider the rule whatsoever during proof-search. 
 As we will see (in \lem~\ref{lem:forestlike-invariance} below), this has the desirable consequence that we are permitted to restrict ourselves to considering only labelled forest derivations during proof-search---an insight which proves advantageous in designing our proof-search procedure. 
 Since labelled forest sequents (which compose labelled forest derivations) play a crucial role in our proof-search methodology, we introduce the relevant concept of a \emph{choice-tree} and explain the semantic meaning of such an object in \rem~\ref{rmk:forest-structure-remark} below.

\begin{definition}[Choice-Tree] Let $\Lambda$ be a labelled forest sequent with sequent graph $G(\Lambda)= (V,E,L)$. We refer to each tree in $G(\Lambda)$ (which is a forest since $\Lambda$ is a labelled forest sequent) as a \emph{choice-tree}\index{Choice-tree} and if $w \in V$, we let $\ct(w)$ denote the choice-tree to which $w$ belongs. Last, we use $\lab(\ct(w))$ to represent the set of all vertices (i.e. labels) occurring in the choice-tree $\ct(w)$.
\end{definition}

\begin{example}\label{ex:choice-tree-dsn} We provide an example of a labelled forest sequent $\Lambda$ below along with a pictorial representation of its sequent graph $\seqgraph(\Lambda)$.
\begin{flushleft}
$
\Lambda := \idealz w_{2}, \idealz u, \idealz u_{1}, \idealz v, R_{\agboxz}ww_{1}, R_{\agboxz}ww_{2}, R_{\agboxz}uu_{1}, R_{\agboxz}vv_{1} \sar 
$
\end{flushleft}
\begin{flushright}
$
w : \Box r, w_{1} : p, u_{1} : q \lor \agdiaz q, v : p, v : \negnnf{q}, v_{1} : \Oiz q
$
\end{flushright}

\begin{center}
\begin{tabular}{c}
\xymatrix{
 & \overset{\boxed{\big(\emptyset,\{\Box r\}\big)}}{w}\ar[dl]|-{\agboxz}\ar[d]|-{\agboxz} &  \overset{\boxed{\big(\{0\},\emptyset\big)}}{u}\ar[d]|-{\agboxz} & \overset{\boxed{\big(\{0\},\{p, \negnnf{q}\}\big)}}{v}\ar[d]|-{\agboxz}  \\
\overset{\boxed{\big(\emptyset,\{p\}\big)}}{w_{1}} &  \overset{\boxed{\big(\{0\},\emptyset\big)}}{w_{2}} & \overset{\boxed{\big(\{0\},\{q \lor \agdiaz q\}\big)}}{u_{1}} & \overset{\boxed{\big(\emptyset,\{\Oiz q\}\big)}}{v_{1}}   \\ 
}
\end{tabular}
\end{center}
The labels $w$, $u$, and $v$ are the roots of their respective choice-trees that make up the forest. The choice-tree above left is equally denoted by $\ct(w)$, $\ct(w_{1})$, and $\ct(w_{2})$, the choice-tree above middle is equally denoted by $\ct(u)$ and $\ct(u_{1})$, and the choice-tree above right is equally denoted by $\ct(v)$ and $\ct(v_{1})$. Furthermore, we have $\lab(\ct(w)) = \lab(\ct(w_{1})) = \lab(\ct(w_{2})) = \{w,w_{1},w_{2}\}$, $\lab(\ct(u)) = \lab(\ct(u_{1})) = \{u,u_{1}\}$, and $\lab(\ct(v)) = \lab(\ct(v_{1})) = \{v,v_{1}\}$.
\end{example}

\begin{remark}\label{rmk:forest-structure-remark}
Each choice-tree that occurs in the sequent graph of a labelled forest sequent is a syntactic representation of a choice-cell for agent $\sa$, that is, it represents an equivalence class in $R_{\agboxz}$ in $\dsnnz$-model. This insight tells us that if agent $\sa$ is restricted to a maximum of $k > 0$ choices and there are $m > k$ choice-trees in the sequent graph of a labelled forest sequent, then at least two choice-trees must correspond to the same choice-cell. We use this observation to specify how $\choicerz$ is (bottom-up) applied in our proof-search algorithm.
\end{remark}


We now introduce blocking conditions\index{Blocking condition} (similar to those used in~\cite{TiuIanGor12}), which will be employed in our proof-search procedure. A na\"ive approach to proof-search would simply apply inference rules in reverse on an input formula in attempt to construct a proof. Yet, such a strategy often leads to a non-terminating proof-search algorithm, suggesting that our algorithm must be properly equipped to determine when a rule should and should not be applied; for example, the $\agdiarz$ rule can be bottom-up applied an infinite number of times to the labelled sequent $R_{\agboxz}wu \sar w : \agdiaz p$. Blocking conditions are essential in this regard, as such conditions allow for a (bottom-up) rule application when false, and prohibit such an application when true. Since each blocking condition is associated with the application of a single rule, once all such conditions are true, it becomes clear that none of the inference rules are bottom-up applicable, and proof-search may safely terminate. 

We introduce four sets of blocking conditions in \dfn~\ref{def:saturation-proof-search-dsn} -- \ref{def:dtwo-choicecond-sat} below. The first set of blocking conditions are the \emph{saturation conditions}, which determine if a derivable sequent has been reached during proof-search and if the $\disr$ and $\conr$ rules are bottom-up applicable. The second set of blocking conditions---the \emph{realization conditions}---dictate the applicability of the $\boxr$, $\agboxrz$, and $\Oirz$ rules, while the third set of \emph{propagation conditions} dictate the applicability of the $\diar$, $\pragdiarz$, and $\prODironez$ rules. Last, the notion of being \dtwocondns-satisfied and the notion of being \choicecondns-satisfied (composing the fourth set of blocking conditions below) govern bottom-up applications of the $\prODirtwoz$ and $\choicerz$ rules, respectively. Last, we note that the terminology and blocking conditions used here are based on and motivated by the terminology and blocking conditions of~\cite{TiuIanGor12}.

\begin{definition}[Saturation]\label{def:saturation-proof-search-dsn} Let $\Lambda := \rel \sar \Gamma$ be a labelled forest sequent with $w \in \lab(\Lambda)$. The label $w$ is \emph{saturated}\index{Saturated} \ifandonlyif (i) for all $w : \phi \in \Gamma$, $w : \negnnf{\phi} \not\in \Gamma$, (ii) for all $w:\phi \lor \psi \in \Gamma$, both $w:\phi \in \Gamma$ and $w:\psi \in \Gamma$, (iii) for all $w:\phi \land \psi \in \Gamma$, either $w:\phi \in \Gamma$ or $w:\psi \in \Gamma$.
\end{definition}




\begin{definition}[$\Box$-, $\agboxz$, $\Oiz$-realization]\label{def:realization-proof-search-dsn} Let $\Lambda := \rel \sar \Gamma$ be a labelled forest sequent with $w \in \lab(\Lambda)$.
\begin{itemize}

\item[$\li$] The label $w$ is \emph{$\Box$-realized}\index{$\Box$-realized} \ifandonlyif for every $w : \Box \phi \in \Gamma$, there exists a label $u \in \lab(\Lambda)$ such that $u:\phi \in \Gamma$. 

\item[$\li$] The label $w$ is \emph{$\agboxz$-realized}\index{$\agboxz$-realized} \ifandonlyif for every $w : \agboxz \phi \in \Gamma$, there exists a label $u \in \lab(\Lambda)$ such that $u \in \ct(w)$ and $u:\phi \in \Gamma$.

\item[$\li$] The label $w$ is \emph{$\Oiz$-realized}\index{$\Oiz$-realized} \ifandonlyif for every $w : \Oiz \phi \in \Lambda$, there exists a label $u \in \lab(\Lambda)$ such that $\idealz u \in \rel$ and $u:\phi \in \Gamma$.

\end{itemize}
\end{definition}

\begin{definition}[$\Diamond$-, $\agdiaz$-, $\ODiz$-propagated]\label{def:propagated-proof-search-dsn} Let $\Lambda$ be a labelled forest sequent with $w \in \lab(\Lambda)$.
\begin{itemize}

\item[$\li$] The label $w$ is \emph{$\Diamond$-propagated}\index{$\Diamond$-propagated} \ifandonlyif for every $w : \Diamond \phi \in \Gamma$, we have $u:\phi \in \Gamma$ for all $u \in \lab(\Lambda)$.

\item[$\li$] The label $w$ is \emph{$\agdiaz$-propagated}\index{$\agdiaz$-propagated} \ifandonlyif for every $w : \agdiaz \phi \in \Gamma$, we have $u:\phi \in \Gamma$ for all labels $u \in \lab(\Lambda)$ such that $w \uipathrelz u$.

\item[$\li$] The label $w$ is \emph{$\ODiz$-propagated}\index{$\ODiz$-propagated} \ifandonlyif for every $w : \ODiz \phi \in \Gamma$, we have $v:\phi \in \Gamma$ for all labels $v \in \lab(\Lambda)$ such that there exists a label $u \in \lab(\Lambda)$ with $\idealz u \in \rel$ and $u \uipathrelz v$.

\end{itemize}
\end{definition}

\begin{definition}[\dtwocondns-satisfied, \choicecondns-satisfied]\label{def:dtwo-choicecond-sat} Let $\Lambda := \rel \sar \Gamma$ be a labelled forest sequent with $n > 0$. We say that $\Lambda$ is \emph{\dtwocondns-satisfied}\index{\dtwocondns-satisfied} \ifandonlyif for all $w : \ODiz \phi \in \Upgamma$, there exists a label $u \in \lab(\Lambda)$ such that $\idealz u, u : \phi \in \rel, \Gamma$. We say that $\Lambda$ is \emph{\choicecondns-satisfied}\index{\choicecondns-satisfied} \ifandonlyif $\seqgraph(\Lambda)$ contains at most $n$-many choice-trees.
\end{definition}

As mentioned above, if a labelled sequent satisfies all blocking conditions, then the proof-search algorithm no longer needs to continue building a proof above that sequent and can terminate on that branch of the computation. 
 When a sequent generated during the course of proof-search enters such a state 
 we say that the sequent is \emph{stable} (cf.~\cite{TiuIanGor12}). The formal definition of stability is as follows:

\begin{definition}[Stability]\label{def:stability-proof-search-dsn} A forestlike labelled sequent $\Lambda$ is \emph{stable}\index{Stable} \ifandonlyif (i) all labels $w$ in $\Lambda$ are saturated, (ii) all labels are $\Box$-, $\agboxz$-, $\Oiz$-realized, (iii) all labels are $\Diamond$-, $\agdiaz$-, $\ODiz$-propagated,  and (iv) $\Lambda$ is \dtwocondns-satisfied, (v) $\Lambda$ is \choicecondns-satisfied.
\end{definition}

We have now laid the necessary groundwork to write our proof-search algorithm \provedsk, i.e. \alg~\ref{alg:ProveDSk-i} below. Due to the length of the algorithm, its instructions have been split into two parts occurring on different pages. The algorithm works by sequentially checking if an input labelled sequent satisfies each blocking condition, and if so, then the sequent will be stable by \dfn~\ref{def:stability-proof-search-dsn} above, and the procedure will return \texttt{False} for that input labelled sequent. If, on the other hand, the input labelled sequent does not satisfy some blocking condition, then the labelled sequent contains a complex labelled formula that has not yet been analyzed, leaving open the possibility that a proof of the initial input formula may still be found; 
 in such a situation, \provedsks effectively applies the relevant rule (bottom-up) yielding a new labelled sequent (or, new labelled sequents) that is (are) recursively input back into the proof-search algorithm. Note that due to instructions 10--20 and 49--57, more than one recursive call may be made, corresponding to the branching rules $\conr$ and $\choicerz$, and which attempts to build proofs for the premises of each rule instance. If all branches of the computation return \texttt{True}, then a proof of the input labelled sequent has been found, but if a single branch returns \texttt{False}, then the input labelled sequent is not provable as is shown in the correctness theorem (\thm~\ref{thm:correctness}) below. The reader should also take note of the following remark, which explains how to modify \provedsk \ when $k = 0$. 

\begin{remark}
If $k = 0$, meaning that the agent $0$'s choices are unbounded, then we omit instructions 49--57 from the proof-search algorithm. Recall from \fig~\ref{fig:refined-calculus-dsn} (introducing each refined labelled calculus $\dsnl$) that the choice-limitation rule $\choicer$ is omitted when $k = 0$. Since instructions 49--57 govern bottom-up applications of $\choicerz$ during proof-search, the instructions are unnecessary when $k = 0$, justifying their exclusion in this case.
\end{remark}

It is significant how \provedsk \ handles the $\choicerz$ rule (see instructions 49--57 in \alg~\ref{alg:ProveDSk-i}). As explained in \rem~\ref{rmk:forest-structure-remark} above, each choice-tree in the sequent graph of a labelled forest sequent is a syntactic representation of a choice-cell. If at some step of the computation \provedsk \ generates a labelled forest sequent whose sequent graph contains $m > k$ choice-trees, then we know that at least two of the choice-trees must represent the same choice-cell. The means by which we syntactically encode that two choice-trees represent the same choice-cell is by connecting them with an edge, and in fact, this is the operation performed by instructions 49--57. However, since we do not know which two choice-trees represent the same choice-cell we must try all possible connections as dictated by the $\choicerz$ rule when applied bottom-up. It is important to point out that in order to preserve the forest structure of a labelled sequent, we cannot simply allow for two choice-trees to be arbitrarily connected in \provedsk. Rather, to preserve the forest structure of labelled sequents throughout the computation of \provedsk, we connect two choice-trees by introducing an edge from the root of one to the root of the other, which begets a new tree combining the previous two. To provide more intuition regarding bottom-up applications of the $\choicerz$ rule (corresponding to instructions 49--57 in \provedsk), we present an example below showing how $(APC_{0}^{2})$ is bottom-up applied to a labelled forest sequent.

\begin{example}\label{ex:connecting-choice-trees-dsn} Let us consider the labelled forest sequent $\Lambda$ from \exmpl~\ref{ex:choice-tree-dsn} above. To improve readability and focus only on essential details, we omit the decoration of the vertices in our sequent graphs. The sequent graph $\seqgraph(\Lambda)$ (without decorated vertices) is shown below top-left. If we were to apply the $(APC_{0}^{2})$ rule bottom-up to $\Lambda$ (as dicatated by instructions 49--57 in \provedsk), then the other three labelled forest sequents (whose non-decorated sequent graphs are shown below) will result. Notice that the sequent graphs of the newly generated labelled forest sequents have two choice-trees instead of three, and that by connecting the root of one choice-tree to the root of another, we preserve the forest structure in the output.

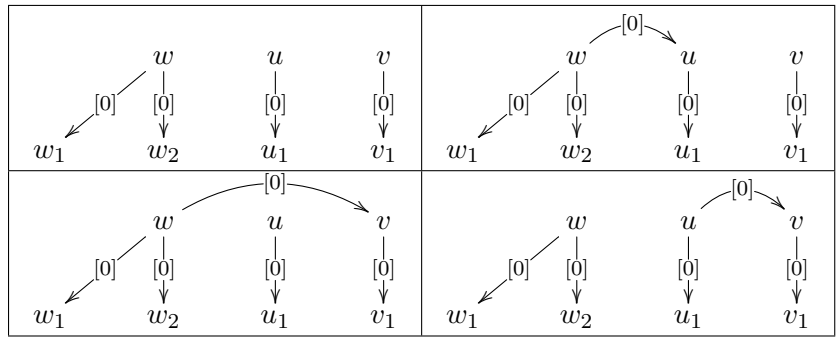
\begin{center}
\bgroup
\def\arraystretch{2}
\begin{tabular}{|c|c|}
\hline
\xymatrix{
 & w\ar[dl]|-{\agboxz}\ar[d]|-{\agboxz} &  u\ar[d]|-{\agboxz} & v\ar[d]|-{\agboxz}  \\
w_{1} &  w_{2} & u_{1} & v_{1}   \\
}

&

\xymatrix{
 & w\ar[dl]|-{\agboxz}\ar[d]|-{\agboxz}\ar@/^1.0pc/[r]|-{\agboxz} &  u\ar[d]|-{\agboxz} & v\ar[d]|-{\agboxz}  \\
w_{1} &  w_{2} & u_{1} & v_{1}   \\
}\\
\hline
\xymatrix{
 & w\ar[dl]|-{\agboxz}\ar[d]|-{\agboxz}\ar@/^1.2pc/[rr]|-{\agboxz} &  u\ar[d]|-{\agboxz} & v\ar[d]|-{\agboxz}  \\
w_{1} &  w_{2} & u_{1} & v_{1}   \\
}

&

\xymatrix{
 & w\ar[dl]|-{\agboxz}\ar[d]|-{\agboxz} &  u\ar[d]|-{\agboxz}\ar@/^1.0pc/[r]|-{\agboxz} & v\ar[d]|-{\agboxz}  \\
w_{1} &  w_{2} & u_{1} & v_{1}   \\
}\\
\hline
\end{tabular}
\egroup
\end{center}

\end{example}

Since bottom-up applications of $\choicerz$ rely on our input labelled sequent being a labelled forest sequent, it is necessary to confirm that all labelled sequents generated throughout the course of the computation of \provedsks are labelled forest sequents (since otherwise, instructions 49--57 will be nonsensical). We argue this result in \lem~\ref{lem:forestlike-invariance} below. Furthermore, this fact, in conjunction with the correctness and termination theorems below (\thm~\ref{thm:correctness} and~\ref{thm:termination-prove}, \resp), has the interesting consequence that each calculus $\dsnsal$ is complete relative to labelled forest derivations with the rooted property (\cor~\ref{thm:forest-proofs-dsn}). This result tells us that refinement has brought about a reduction in the underlying data structure employed in labelled sequents occurring in derivations (cf.~\thm~\ref{thm:sequent-structure-gtdsn}), though, this fact is observed and confirmed via our proof-search algorithm.

\begin{lemma}\label{lem:forestlike-invariance} Let $\phi \in \langdsn$. Every labelled sequent generated throughout the course of computing \provedsk$(\seqempstr \sar w : \phi)$ is a labelled forest sequent.
\end{lemma}

\begin{proof} By \dfn~\ref{def:tree-DAG-sequent-dsn}, we know that $w : \phi$ is a labelled forest sequent since it's graph is of the form $\seqgraph(\seqempstr \sar w : \phi) = (\{w\},\emptyset,\{(w,\{\phi\}\})$, which is a single point. Observe that only conditional statements in \provedsk \ that add new nodes or edges (when the condition holds true) to the sequent graph of the input labelled sequent are those given on lines 21--24, 25--28, 29--32, and 45--48 in the algorithm corresponding to bottom-up applications of the $\boxr$, $\agboxrz$, $\ODirz$, and $\prODirtwoz$ rules, respectively; with the exception of the instruction block 49--57, all other instructions preserve the forest structure inherent in the input labelled forest sequent because they only change the labeling function $L$ of the sequent graph of the input labelled sequent (i.e. such rules only redecorate vertices). Instructions 21--24, 29--32, and 45--48 add new disconnected vertices to the sequent graph of the input labelled sequent, which serve as new roots of new choice-trees, thus preserving the forest structure. 
 The instructions 25--28 add a new edge to a fresh vertex in a choice-tree of the sequent graph of the input labelled sequent, which preserves the forest structure because only an additional branch is added to some choice-tree. The instructions 49--57 connect the root of one choice-tree to the root of another choice-tree (as explained in \exmpl~\ref{ex:connecting-choice-trees-dsn} above), thus yielding a new choice-tree (that subsumes the original two from which it was formed) in the sequent graph of the input labelled sequent. Therefore, every labelled sequent generated throughout the course of \provedsk$(\seqempstr \sar w : \phi)$ will be a labelled forest sequent since all instructions preserve the forest structure of any input labelled forest sequent. 
\end{proof}

\begin{algorithm}
\KwIn{A Labelled Sequent: $\rel \sar \Gamma$}
\KwOut{A Boolean: \texttt{True}, \texttt{False}}

\If{$w : \phi, w :\negnnf{\phi} \in \Upgamma$}
     {\Return \texttt{True};}

\If{$\rel \sar \Upgamma$ is stable}
     {\Return \texttt{False};}
     
\If{there exists a $w : \phi \lor \psi \in \Upgamma$, but either $w : \phi \not\in \Upgamma$ or $w : \psi \not\in \Upgamma$}
     {Let $\Upgamma' := w :\phi, w :\psi, \Upgamma$;\\
     \Return \provedsk($\rel \sar \Upgamma'$);}
     
\If{there exists a $w : \phi \land \psi \in \Upgamma$, but $w : \phi, w : \psi \not\in \Upgamma$}
{
    Let $\Upgamma_{1} := w :\phi, \Upgamma$;\\
    Let $\Upgamma_{2} := w :\psi, \Upgamma$;\\
    \If{\provedsk($\rel \sar \Upgamma_{i}$) = \texttt{False} for some $i \in \{1,2\}$}
    {
    \Return \texttt{False};
    }\Else{
    \Return \texttt{True};
    }
}

\caption{\provedsk}\label{alg:ProveDSk-i}
\end{algorithm}
    
\begin{algorithm}
\setcounter{AlgoLine}{20}
\If{some $w \in \lab(\rel \sar \Upgamma)$ is not $\Box$-realized}
    {
     For a labelled formula $w : \Box \phi \in \Upgamma$ such that $u : \phi \not\in \Upgamma$ for all $u \in \lab(\rel \sar \Gamma)$, let $\Upgamma' := v :\phi, \Upgamma$ with $v$ fresh;\\
     \Return \provedsk($\rel \sar \Upgamma'$);
    }
    
\If{some $w \in \lab(\rel \sar \Upgamma)$ is not $\agboxz$-realized}
    {
    For a labelled formula $w : \agboxz \phi \in \Upgamma$ such that $R_{\agboxz}wu, u : \phi \not\in \rel, \Upgamma$ for all $u \in \lab(\rel \sar \Upgamma)$, let $\rel' := R_{\agboxz}wv$ and $\Upgamma' := v :\phi, \Upgamma$ with $v$ fresh;\\
    \Return \provedsk($\rel' \sar \Upgamma'$);
    }
    
\If{some $w \in \lab(\rel \sar \Upgamma)$ is not $\Oiz$-realized}
    {
     For a labelled formula $w : \Oiz \phi \in \Upgamma$ such that $\idealz u, u : \phi \not\in \rel, \Upgamma$ for all $u \in \lab(\rel \sar \Upgamma)$, let $\rel' := \idealz v, \rel$ and $\Upgamma' := v :\phi, \Upgamma$ with $v$ fresh;\\
     \Return \provedsk($\rel' \sar \Upgamma'$);
    }
    
\If{some $w \in \lab(\rel \sar \Upgamma)$ is not $\Diamond$-propagated}
    {
     Pick a label $u \in \lab(\rel \sar \Upgamma)$ such that $w : \Diamond \phi \in \Upgamma$ and $u : \phi \not\in \Upgamma$, and let $\Upgamma' := u :\phi, \Upgamma$;\\
     \Return \provedsk($\rel \sar \Upgamma'$);
    }
    
\If{some $w \in \lab(\rel \sar \Upgamma)$ is not $\agdiaz$-propagated}
    {
    Pick a label $u \in \lab(\rel \sar \Upgamma)$ such that $w : \agdiaz \phi \in \Upgamma$, $w \uipathrelz u$, $u : \phi \not\in \Upgamma$, and let $\Upgamma' := u :\phi, \Upgamma$;\\
     \Return \provedsk($\rel \sar \Upgamma'$);
    }
    
\If{some $w \in \lab(\rel \sar \Upgamma)$ is not $\ODiz$-propagated}
    {
     Pick labels $u, v \in \lab(\rel \sar \Upgamma)$ such that $w : \ODiz \phi \in \Upgamma$, $\idealz u \in \rel$, $u \uipathrelz v$, and $v : \phi \not\in \Upgamma$, and let $\Upgamma' := v :\phi, \Upgamma$;\\
     \Return \provedsk($\rel \sar \Upgamma'$);
    }

\If{$\rel \sar \Upgamma$ is not \dtwocondns-satisfied}
    {
    Pick a label $w \in \lab(\rel \sar \Upgamma)$ such that $w : \ODiz \phi \in \Upgamma$, and for all $u \in \lab(\rel \sar \Upgamma)$, $\idealz u, u : \phi \not\in \rel, \Upgamma$, and let $\rel' := \idealz v, \rel$ and $\Upgamma' := v :\phi, \Upgamma$ with $v$ fresh;\\
    \Return \provedsk($\rel' \sar \Upgamma'$);
    }
    
\If{$\rel \sar \Upgamma$ is not \choicecondns-satisfied}
    {
    Let $\rel_{m,j}, \rel := R_{\agboxz}w_{m}w_{j}$ (with $0 \leq m \leq k - 1$ and $m + 1 \leq j \leq k$) where $w_{m}$ and $w_{j}$ are distinct roots of choice-trees in $\rel \sar \Gamma$;\\
    \If{\provedsk($\rel_{m,j} \sar \Upgamma$) = \texttt{False} for some $m$ and $j$}
        {
        \Return \texttt{False};
        }\Else{
        \Return \texttt{True};
        }
    }
    
\end{algorithm}

We are now in a position to prove the correctness and termination of our proof-search algorithm \provedsk. Before showing each of these results, we prove a useful lemma relating choice-trees to undirected $0$-paths:

\begin{lemma}\label{lem:in-CT-iff-connected-dsn}
Let $\Lambda := \rel \sar \Gamma$ be a labelled forest sequent with $w,u \in \lab(\Lambda)$, then $u \in \ct(w)$ \ifandonlyif $w \uipathrelz u$.
\end{lemma}

\begin{proof} The forward direction is trivial since if $u \in \ct(w)$, then $w$ and $u$ occur in the same choice-tree, implying that there is some undirected $0$-path between the two labels. The backward direction is simple as well: if $\Lambda$ is a labelled forest sequent, then the only way there can be an undirected $0$-path between two labels $w$ and $u$ is if they occur in the same choice-tree.
\end{proof}

\begin{theorem}[Correctness]\label{thm:correctness} Let $\phi \in \langdsn$ and $k \in \mathbb{N}$.

(i) If \provedsk$(\seqempstr \sar w : \phi)$ returns \texttt{True}, then $\seqempstr \sar w:\phi$ is $\dsnsal$-provable.

(ii) If \provedsk$(\seqempstr \sar w : \phi)$ returns \texttt{False}, then $\seqempstr \sar w:\phi$ is not $\dsnsal$-provable and this is witnessed by a finite counter-model. 
\end{theorem}

\begin{proof} (i) Every recursive call in \provedsk \ is a bottom-up application of a rule from $\dsnsal$, and so, if \provedsk$(\seqempstr \sar w : \phi)$ returns \texttt{True}, then we obtain a proof of $\seqempstr \sar w:\phi$ where all top sequents are of the form $\rel \sar u : \psi, u : \negnnf{\psi}, \Gamma$. By \lem~\ref{lem:general-id-dsn} and \thm~\ref{thm:gtdsn-to-dsnl}, we know that all such labelled sequents have a derivation in $\dsnsal$. Also, it should be noted that because the principal formula of a bottom-up inference (corresponding to a recursive call) is preserved upwards, it may be necessary to apply hp-admissibility of $ \ctrr$ (\lem~\ref{lem:ctrr-admiss-dsn}) to obtain the final proof.

(ii) Suppose that \provedsk$(\seqempstr \sar w : \phi)$ returns \texttt{False}. This implies that a stable labelled forest sequent $\Lambda := \rel \sar \Gamma$ was generated. Let $\seqgraph(\Lambda) := (V,E,L)$. We use $\Lambda$ to define two counter-models $M = (W,R_{\agboxz}, \idealz,V)$ and $M' = (W',R_{\agboxz}', \idealz',V')$ for $\phi$, depending on if $\Lambda$ contains a labelled formula of the form $u : \ODiz \psi$, or it does not, respectively. The $\dsnnz$-model $M$ is defined as follows:
\begin{itemize}

\item[$\li$] $W := \lab(\Lambda)$

\item[$\li$] For all $u, v \in \lab(\Lambda)$, $v \in R_{\agboxz}(u)$ \ifandonlyif $u \uipathrelz v$

\item[$\li$] $\idealz := \{v \ | \ \text{exists }u \in \lab(\Lambda), \ \idealz u \in \rel \text{ and } u \uipathrelz v \}$

\item[$\li$] $V(u) := \{p \ | \ u : \negnnf{p} \in \Gamma \}$, for all $u \in \lab(\Lambda)$

\end{itemize}
For the $\dsnnz$-model $M'$, we let $W' := W$, $R_{\agboxz}' := R_{\agboxz}$, $\idealz' = \{u \ | \ w \uipathrelz u\}$, and $V' = V$, that is, the only difference is that $\idealz$ is set equal to the choice-cell of $w$, which must be non-empty (as it must contain $w$ at the very least). The reason for this is that we need to ensure that $\idealz\neq \emptyset$, i.e. the model $M'$ satisfies the condition \dtwocond (and is a $\dsnnz$-model). We let $\countmod \in \{M,M'\}$, $\countw \in \{W, W'\}$, $\countrel \in \{R_{\agboxz}, R_{\agboxz}'\}$, $\countideal \in \{\idealz, \idealz'\}$, and $\countval \in \{V, V'\}$ to prove our results uniformly, and only distinguish between $M$, $M'$, and their components when needed.

To complete the proof of claim (ii), we prove two things: (ii.1) $\countmod$ is a $\dsnnz$-model with $\ag = \{\sa\}$, and (ii.2) $\countmod, w \not\Vdash \phi$, i.e. $\countmod$ is a counter-model for $\phi$. Showing these two cases proves that $\seqempstr \sar w : \phi$ is not $\dsnsal$-provable by soundness (\thm~\ref{thm:soundness-gtdsn}).

(ii.1) We prove that $\countmod$ is a $\dsnnz$-model by showing that it satisfies conditions \partcondns--\choicecond and \donecondns--\dthreecondns. 

\partcond Follows from the definition of $\countrel$ and the fact that $\uipathz$ is an equivalence relation (\lem~\ref{lem:uipath-equiv-relation-dsn}).

\ioacond The \ioacond condition is trivially satisfied in the single-agent setting.

\choicecond Since the labelled sequent $\Lambda$ is stable, we know that it is \choicecondns-satisfied, meaning that $\seqgraph(\Lambda)$ contains at most $k$ many choice-trees. By \lem~\ref{lem:in-CT-iff-connected-dsn}, we know that $u \uipathz v$ \ifandonlyif $v \in \ct(u)$. This fact, in conjunction with the definition of $\countrel$ above, and the fact that $\uipathz$ is an equivalence relation (\lem~\ref{lem:uipath-equiv-relation-dsn}), implies that each choice-tree corresponds to a choice-cell in $\countrel$. Hence, there will be at most $k$ many choice cells.

\donecond Follows from the definition of $\countideal$ above.

\dtwocond There are two possible cases depending on if $\countmod = M$ or $\countmod = M'$, i.e. depending of if $\Lambda$ contains a labelled formula of the form $u : \ODiz \psi$, or if it does not. For the first case, since $\Lambda$ is \dtwocondns-satisfied and contains a labelled formula of the form $u : \ODiz \psi$, we know that there exists a label $u \in \lab(\Lambda)$ such that $\idealz u \in \rel$ by instructions 45--48. For the second case, the non-emptiness of $\idealz'$ follows by the definition of $\idealz'$. Hence, $\idealz$ and $\idealz'$ will be non-empty, that is, $\countideal \neq \emptyset$.

\dthreecond There are two possible cases depending on if $\countmod = M$ or $\countmod = M'$, i.e. depending of if $\Lambda$ contains a labelled formula of the form $u : \ODiz \psi$, or if it does not. The second case is trivial, as $\idealz'$ satisfies \dthreecond by definition, so we focus on the first case. Suppose that $u \in \idealz$ and $v \in R_{\agbox}(u)$. Then, there exists a label $z \in \lab(\Lambda)$ such that $\idealz z$ and $z \uipathrelz u$ holds by the definition of $\idealz$ above. By the definition of $R_{\agboxz}$ above, we also know that $u \uipathrelz v$ holds, which implies that $z \uipathrelz v$ by \lem~\ref{lem:uipath-equiv-relation-dsn}; therefore, $v \in \idealz$ by the definition of $\idealz$ above.

(ii.2) We prove that if $u : \psi \in \Gamma$, then $\countmod, u \not\Vdash \psi$ by induction on the complexity of $\psi$. The desired conclusion that $\countmod, w \not\Vdash \phi$ follows from this claim since $w : \phi \in \Gamma$.

\textit{Base case.} Suppose that $u : \psi$ is of the form $u : p$ or $u : \negnnf{p}$. In the former case, since $\Lambda$ is stable, we know that it is saturated, implying that $u : \negnnf{p} \not\in \Gamma$ because $u : p \in \Gamma$. By the definition of $\countval$ above, $p \not\in \countval(u)$, which lets us conclude that $\countmod, u \not\Vdash p$. In the latter case, by the definition of $\countval$, we know that $p \in \countval(u)$, entailing that $\countmod, u \not\Vdash \negnnf{p}$.

\textit{Inductive step.} We consider the main connective of $\psi$ below and show that in each case the desired result follows.

$\lor$. Assume that $\psi := \chi \lor \theta$ and $u : \chi \lor \theta \in \Gamma$. Then, since $\Lambda$ is saturated, we know that $u : \chi, u : \theta \in \Gamma$, which implies that $\countmod, u \not\Vdash \chi$ and $\countmod,u \not\Vdash \theta$ by the inductive hypothesis. Hence, $\countmod, u \not\Vdash \chi \lor \theta$.

$\land$. Assume that $\psi := \chi \land \theta$ and $u : \chi \land \theta \in \Gamma$. Then, since $\Lambda$ is saturated, we know that either $u : \chi \in \Gamma$ or $u : \theta \in \Gamma$, which implies that either $\countmod, u \not\Vdash \chi$ or $\countmod,u \not\Vdash \theta$ by the inductive hypothesis. Regardless of the case, $\countmod, u \not\Vdash \chi \land \theta$.

$\Diamond$. Assume that $\psi := \Diamond \chi$ and $u : \Diamond \chi \in \Gamma$. Then, since $\Lambda$ is saturated, we know that $\Lambda$ is $\Diamond$-propagated. Hence, $v : \chi \in \Gamma$ for all $v \in \lab(\Lambda)$. By \ih and the definition of $\countw$ then, $\countmod, v \not\Vdash \chi$ for all $v \in \countw$. Therefore, $\countmod, u \not\Vdash \Diamond \chi$.

$\agdiaz$. Assume that $\psi := \agdiaz \chi$ and $u : \agdiaz \chi \in \Gamma$. Then, since $\Lambda$ is saturated, we know that $\Lambda$ is $\agdiaz$-propagated. Hence, $v : \chi \in \Gamma$ for all $v \in \lab(\Lambda)$ such that $u \uipathrelz v$. By \ih and the definition of $\countrel$ then, $\countmod, v \not\Vdash \chi$ for all $v \in \countrel(u)$. Therefore, $\countmod, u \not\Vdash \agdiaz \chi$.

$\ODiz$. If $\Lambda$ does not contain a labelled formula of the form $u : \ODiz \chi$, then the case holds vacuously. Let us suppose then that $\Lambda$ contains such a labelled formula, meaning that $\countmod = M$. Assume that $\psi := \ODiz \chi$ and $u : \ODiz \chi \in \Gamma$. Then, since $\Lambda$ is saturated, we know that $\Lambda$ is $\ODiz$-propagated. Hence, $v : \chi \in \Gamma$ for all $v \in \lab(\Lambda)$ such that for some $z \in \lab(\Lambda)$ with $\idealz z \in \rel$ and $z \uipathrelz v$. By \ih and the definition of $\idealz$ then, we know that $M, v \not\Vdash \chi$ for all $v \in \idealz$, implying that $M, u \not\Vdash \ODiz \chi$.

$\Box$. Assume that $\psi := \Box \chi$ and $u : \Box \chi \in \Gamma$. Then, since $\Lambda$ is saturated, we know that $\Lambda$ is $\Box$-realized, meaning there exists a label $v \in \lab(\Lambda)$ such that $v : \chi \in \Gamma$. By the definition of $\countw$ we know that $v \in \countw$ and by \ih we have that $\countmod, v \not\Vdash \chi$. Therefore, $\countmod, u \not\Vdash \Box \chi$. 

$\agboxz$. Assume that $\psi := \agboxz \chi$ and $u : \agboxz \chi \in \Gamma$. Then, since $\Lambda$ is saturated, we know that $\Lambda$ is $\agboxz$-realized, meaning there exists a label $v \in \lab(\Lambda)$ such that $R_{\agboxz}uv$ and $v : \chi \in \Gamma$. By the definition of $\countrel$ we know that $v \in \countrel(u)$ and by \ih we have that $\countmod, v \not\Vdash \chi$. Therefore, $\countmod, u \not\Vdash \agboxz \chi$. 

$\Oiz$. Assume that $\psi := \Oiz \chi$ and $u : \Oiz \chi \in \Gamma$. Then, since $\Lambda$ is saturated, we know that $\Lambda$ is $\Oiz$-realized, meaning there exists a label $v \in \lab(\Lambda)$ such that $\idealz v \in \rel$ and $v : \chi \in \Gamma$. By the definition of $\countideal$ we know that $v \in \countideal$ and by \ih we have that $\countmod, v \not\Vdash \chi$. Therefore, $\countmod, u \not\Vdash \Oiz \chi$. 
\end{proof}

\begin{theorem}[Termination]\label{thm:termination-prove} For each $\phi \in \langdsnz$, \provedsk$(\seqempstr \sar w : \phi)$ terminates.
\end{theorem}

\begin{proof} Consider a run of \provedsk$(\seqempstr \sar w : \phi)$. The only rules that add new vertices in the sequent graphs of labelled forest sequents generated throughout the course of the computation are instructions 21--24, 25--28, 29--32, 45--48, which correspond to bottom-up applications of the rules $\boxr$, $\agboxrz$, $\Oirz$, and $\prODirtwoz$, respectively. The number of bottom-up applications of $\boxr$, $\agboxrz$, $\Oirz$, and $\prODirtwoz$ rules is restricted by the number of $\Box$, $\agboxz$, $\Oiz$, and $\ODiz$ subformulae of $\phi$, and hence can only be applied a finite number of times. 
 Since the aforementioned instructions/rules are the only ones that introduce new vertices in the sequent graphs of labelled forest sequents, only a finite number of vertices (i.e. labels) can be introduced throughout the computation of \provedsk$(\seqempstr \sar w : \phi)$. 

Instructions 7--10 and 11--20 correspond to bottom-up applications of the $\disr$ and $\conr$ rules, and once such instructions are applied to a disjunctive or conjunctive subformula of $\phi$, only formulae of a smaller complexity are introduced and the considered disjunctive or conjunctive subformula need not be considered again. Instructions 33--36, 37--40, 41--44 correspond to bottom-up applications of the $\diar$, $\agdiarz$, and $\ODirz$ rules, and ensure that for each $\Diamond \psi$, $\agdiaz \psi$, and $\ODiz \psi$ subformula of $\phi$,  $u : \psi$ (which is a formula of less complexity) is added only once with some label $u$. Since only a finite number of labels (i.e. vertices) can be introduced throughout the course of the computation (by what was said above), this implies that the $\diar$, $\agdiarz$, and $\ODirz$ rules can only be applied a finite number of times. 

Last, instructions 49--57 (corresponding to a bottom-up application of $\choicerz$) take a labelled forest sequent whose sequent graph has $m > k$ choice-trees as input and returns a new labelled forest sequent whose sequent graph contains $m - 1 \geq k$ choice-trees. Since only a finite number of labels (and relational atoms) can be introduced throughout the computation, only a finite number of choice-trees can be introduced throughout the computation; hence, a point will always be reached where the $\choicerz$ rule need not be applied.
\end{proof}

\begin{corollary}[Decidability and FMP]
For each $k \in \mathbb{N}$, the logic $\dsnsa$ is decidable and has the finite model property\index{Finite model property (FMP)}.
\end{corollary}

\begin{proof} Follows from \thm~\ref{thm:correctness} and~\ref{thm:termination-prove}.
\end{proof}

\begin{corollary}\label{thm:forest-proofs-dsn}
Let $k \in \mathbb{N}$ and $n = 0$. If a labelled formula $w : \phi$ is provable in $\dsnlnz$, then it has a labelled forest derivation with the rooted property.
\end{corollary}

\begin{proof} Follows from \lem~\ref{lem:forestlike-invariance}, \thm~\ref{thm:correctness}, \thm~\ref{thm:termination-prove}, and that fact that in the sequent graph of a labelled forest sequent, none of the instructions of \provedsk \ ever add an edge that points to the root of a choice-tree, ensuring that roots are preserved throughout the course of the computation, and hence, the output derivation has the rooted property. 
\end{proof}



\section{Interpolation for Context-free Grammar Logics with Converse}\label{sec:applicationsII}

The Lyndon interpolation property\index{Lyndon interpolation} for a logic $\mathsf{L}$ states that if $\phi \cimp \psi \in \mathsf{L}$, then there exists a formula $\chi$ (called an \emph{interpolant}) such that $\phi \cimp \chi, \chi \cimp \psi \in \mathsf{L}$ and $\chi$ is built from logical connectives and propositional variables common to both $\phi$ and $\psi$ that have the same polarity. A weaker version of Lyndon interpolation---referred to as \emph{Craig interpolation}\index{Craig interpolation}---only requires an interpolant to be built using propositional variables common to both $\phi$ and $\psi$, thus dropping the added condition that the propositional variables used in $\chi$ share the same polarity in $\phi$ and $\psi$ (therefore, Lyndon interpolation implies Craig interpolation). Interpolation has found a variety of applications in logic and computer science; e.g. interpolation is used in computer-aided verification to divide a problem involving an implication $\phi \cimp \psi$ into smaller problems involving $\phi \cimp \chi$ and $\chi \cimp \psi$~\cite{McM18}, to establish Beth definability~\cite{KihOno10}, and is used in knowledge representation to conceal or forget superfluous or private information in ontology querying~\cite{LutWol11}. Hence, interpolation is of practical consequence.

Both semantic or syntactic arguments may be and have been used to establish that a logic $\mathsf{L}$ possesses the Lyndon or Craig interpolation property. As one might expect, the semantic approach requires a semantics for the logic $\mathsf{L}$, whereas the syntactic approach requires a proof calculus for the logic. The latter, syntactic approach---initiated by Maehara~\cite{Mae60}, who utilized Gentzen-style sequent systems---constructs an interpolant $\chi$ for an implication $\phi \cimp \psi \in \mathsf{L}$ and has the advantage that the method transforms a proof of $\phi \cimp \psi$ into proofs of $\phi \cimp \chi$ and $\chi \cimp \psi$ witnessing that $\chi$ is an interpolant. To provide intuition regarding Maehara's interpolation methodology\index{Maehara's method}, which motivates the more general methodology we will employ for grammar logics, we explain the method below in the context of classical propositional logic. Afterward, we describe how Maehara's method for Gentzen-style sequent systems can be generalized and extended to nested/refined labelled systems.

As explained in Appendix B (p.~\pageref{app:classical-completeness}), the rules $\id$, $\disr$, and $\conr$ (from $\gtkms$) serve as a sound and complete calculus for classical propositional logic. Consequently, we use these three rules as our calculus to demonstrate Maehara's method, and note that by the work in the Appendix B (p.~\pageref{app:classical-completeness}), our calculus only needs sequents of the form $\seqempstr \sar \Gamma$ where every labelled formula in $\Gamma$ has the same label $w$. Since every labelled formula has the same label $w$, it is permissible to omit labels; hence, we omit the use of labels in the sequents of our example, meaning that we will utilize (one-sided) Gentzen-style sequents while demonstrating Maehara's method. 

If showing Lyndon interpolation, then Maehara's method is a means of transforming a proof of a sequent $\seqempstr \sar \Gamma_{1}, \Gamma_{2}$ into proofs of $\seqempstr \sar \Gamma_{1}, \chi$ and $\seqempstr \sar \negnnf{\chi}, \Gamma_{2}$, where $\chi$ only contains propositional variables common to both $\negnnf{\Gamma_{1}}$ and $\Gamma_{2}$ that have the same polarity.\footnote{If $\Gamma_{1} := \phi_{1}, \ldots, \phi_{n}$, we let $\negnnf{\Gamma_{1}} := \negnnf{\phi}_{1}, \ldots, \negnnf{\phi}_{n}$.} If such a transformation can be performed for each derivable sequent of the form $\seqempstr \sar \Gamma_{1}, \Gamma_{2}$, then it can be performed for any derivable sequent of the form $\seqempstr \sar \negnnf{\phi}, \psi$, which is equivalent to $\seqempstr \sar \phi \cimp \psi$. It follows that the proof of $\seqempstr \sar \negnnf{\phi}, \psi$ can be transformed into proofs of $\seqempstr \sar \negnnf{\phi}, \chi$ and $\seqempstr \sar \negnnf{\chi}, \psi$, which are equivalent to $\seqempstr \sar \phi \cimp \chi$ and $\seqempstr \sar \chi \cimp \psi$, respectively---yielding an interpolant $\chi$ for the implication $\phi \cimp \psi$.

The method works by constructing an interpolant $\chi$ by induction on the proof of the derivable sequent $\seqempstr \sar \Gamma_{1}, \Gamma_{2}$ we wish to interpolate. This inductive construction is encoded within inference rules that employ more expressive  sequents, specifying how the context of the  sequent is partitioned (with the left component of the partition representing the antecedent that implies the interpolant, and the right component of the partition representing the consequent that is implied by the interpolant) and the interpolant constructed thus far. The more expressive sequents we employ are of the form $\seqempstr \sar \Gamma_{1} \sep \Gamma_{2} \sepgi \chi$ denoting that the sequent $\seqempstr \sar \Gamma_{1}, \Gamma_{2}$ can be partitioned into $\seqempstr \sar \Gamma_{1}, \chi$ and $\seqempstr \sar \negnnf{\chi}, \Gamma_{2}$ with interpolant $\chi$. The inference rules we make use of are variants of the $\id$, $\disr$, and $\conr$ rules where we consider all possible `partitions' of the context and assign interpolants based on how the context is split. We first introduce the initial rules, and then introduce the variants of $\disr$ and $\conr$. 

\begin{center}
\begin{tabular}{c c}
\AxiomC{}
\RightLabel{$(id_{1})$}
\UnaryInfC{$\seqempstr \sar \Gamma_{1}, p, \negnnf{p} \sep \Gamma_{2} \sepgi \bot$}
\DisplayProof

&

\AxiomC{}
\RightLabel{$(id_{2})$}
\UnaryInfC{$\seqempstr \sar \Gamma_{1}, p \sep \negnnf{p}, \Gamma_{2} \sepgi \negnnf{p}$}
\DisplayProof
\end{tabular}
\end{center}
\begin{center}
\begin{tabular}{c c}
\AxiomC{}
\RightLabel{$(id_{3})$}
\UnaryInfC{$\seqempstr \sar \Gamma_{1}, \negnnf{p} \sep p, \Gamma_{2} \sepgi p$}
\DisplayProof

&

\AxiomC{}
\RightLabel{$(id_{4})$}
\UnaryInfC{$\seqempstr \sar \Gamma_{1} \sep p, \negnnf{p}, \Gamma_{2} \sepgi \top$}
\DisplayProof
\end{tabular}
\end{center}

As shown above, there are four possible interpolants, depending on how the context of an initial sequent is partitioned. The $(id_{1})$ rule tells us that the derivable  sequent $\seqempstr \sar \Gamma_{1}, p, \negnnf{p}, \Gamma_{2}$ can be split into the derivable sequents $\seqempstr \sar \Gamma_{1}, p, \negnnf{p}, \bot$ and $\seqempstr \sar \negnnf{\bot}, \Gamma_{2}$ with interpolant $\bot$. The other three initial rules are read in a similar fashion. Our disjunction and conjunction rules are as follows:

\begin{center}
\begin{tabular}{c c}
\AxiomC{$\seqempstr \sar \Gamma_{1}, \phi, \psi \sep \Gamma_{2} \sepgi \chi$}
\RightLabel{$(\lor_{r1})$}
\UnaryInfC{$\seqempstr \sar \Gamma_{1}, \phi \lor \psi \sep \Gamma_{2} \sepgi \chi$}
\DisplayProof

&

\AxiomC{$\seqempstr \sar \Gamma_{1}, \phi \sep \Gamma_{2} \sepgi \chi$}
\AxiomC{$\seqempstr \sar \Gamma_{1}, \psi \sep \Gamma_{2} \sepgi \theta$}
\RightLabel{$(\land_{r1})$}
\BinaryInfC{$\seqempstr \sar \Gamma_{1}, \phi \land \psi \sep \Gamma_{2} \sepgi \chi \lor \theta$}
\DisplayProof
\end{tabular}
\end{center}

\begin{center}
\begin{tabular}{c c}
\AxiomC{$\seqempstr \sar \Gamma_{1} \sep \phi, \psi, \Gamma_{2} \sepgi \chi$}
\RightLabel{$(\lor_{r2})$}
\UnaryInfC{$\seqempstr \sar \Gamma_{1} \sep \phi \lor \psi, \Gamma_{2} \sepgi \chi$}
\DisplayProof

&

\AxiomC{$\seqempstr \sar \Gamma_{1} \sep \phi, \Gamma_{2} \sepgi \chi$}
\AxiomC{$\seqempstr \sar \Gamma_{1} \sep \psi, \Gamma_{2} \sepgi \theta$}
\RightLabel{$(\land_{r2})$}
\BinaryInfC{$\seqempstr \sar \Gamma_{1} \sep \phi \land \psi, \Gamma_{2} \sepgi \chi \land \theta$}
\DisplayProof
\end{tabular}
\end{center}

Notice how each of the rules above constructs an interpolant of the conclusion by making use of the interpolant(s) of the premise(s). Maehara's method therefore works by assigning interpolants to all initial sequents in the given derivation, and then inductively constructing the interpolant of the end sequent via the above rules. We demonstrate this process with an example, and consider a derivation of an implication $p \land q \cimp p \land q$, which is represented as $(\negnnf{p} \lor \negnnf{q}) \lor (p \land q)$ since we are working with formulae negation normal form.

\begin{center}
\AxiomC{}
\RightLabel{$\id$}
\UnaryInfC{$\seqempstr \sar \negnnf{p}, \negnnf{q}, p$}

\AxiomC{}
\RightLabel{$\id$}
\UnaryInfC{$\seqempstr \sar \negnnf{p}, \negnnf{q}, q$}

\RightLabel{$\conr$}
\BinaryInfC{$\seqempstr \sar \negnnf{p}, \negnnf{q}, p \land q$}
\RightLabel{$\disr$}
\UnaryInfC{$\seqempstr \sar \negnnf{p} \lor \negnnf{q}, p \land q$}
\RightLabel{$\disr$}
\UnaryInfC{$\seqempstr \sar (\negnnf{p} \lor \negnnf{q}) \lor (p \land q)$}
\RightLabel{=}
\dottedLine
\UnaryInfC{$\seqempstr \sar p \land q \cimp p \land q$}
\DisplayProof
\end{center}

We transform the above derivation into a new derivation that makes use of the (relevant) inference rules introduced previously. In doing so, we only make use of the top portion of the above derivation down to the  sequent $\seqempstr \sar \negnnf{p} \lor \negnnf{q}, p \land q$ since Maehara's method demands the antecedent formula $\negnnf{p} \lor \negnnf{q}$ and consequent formula $p \land q$ be split from one another:

\begin{center}
\AxiomC{}
\RightLabel{$(id_{3})$}
\UnaryInfC{$\seqempstr \sar \negnnf{p}, \negnnf{q} \sep p \sepgi p$}
\AxiomC{}
\RightLabel{$(id_{3})$}
\UnaryInfC{$\seqempstr \sar \negnnf{p}, \negnnf{q} \sep q \sepgi q$}
\RightLabel{$(\land_{r2})$}
\BinaryInfC{$\seqempstr \sar \negnnf{p}, \negnnf{q} \sep p \land q \sepgi p \land q$}
\RightLabel{$(\lor_{r1})$}
\UnaryInfC{$\seqempstr \sar \negnnf{p} \lor \negnnf{q} \sep p \land q \sepgi p \land q$}
\DisplayProof
\end{center}

The derivation above may now be split into two derivations in the original calculus witnessing that $p \land q$ is the interpolant of $p \land q \cimp p \land q := (\negnnf{p} \lor \negnnf{q}) \lor (p \land q)$. The derivation below left is obtained by taking the data from the left components of the partitioned sequents together with the initial interpolants, and the derivation below right is obtained by taking the right components together with the negation of the initial interpolants. Observe that the use of $\conr$ in the derivation below left builds the interpolant $p \land q$, and the $\disr$ rule in the derivation below right builds the interpolant $\negnnf{p} \lor \negnnf{q}$. Also, note that it may be necessary to invoke the (hp-)admissibility of structural rules such as $\wk$ (as shown in the derivation below right) to obtain the desired result.

\begin{center}
\begin{tabular}{c c}
\AxiomC{}
\RightLabel{$\id$}
\UnaryInfC{$\seqempstr \sar \negnnf{p}, \negnnf{q}, p$}
\AxiomC{}
\RightLabel{$\id$}
\UnaryInfC{$\seqempstr \sar \negnnf{p}, \negnnf{q}, q$}
\RightLabel{$\conr$}
\BinaryInfC{$\seqempstr \sar \negnnf{p}, \negnnf{q}, p \land q$}
\RightLabel{$\disr$}
\UnaryInfC{$\seqempstr \sar \negnnf{p} \lor \negnnf{q}, p \land q$}
\DisplayProof

&

\AxiomC{}
\RightLabel{$\id$}
\UnaryInfC{$\seqempstr \sar \negnnf{p}, p$}
\RightLabel{$\wk$}
\dashedLine
\UnaryInfC{$\seqempstr \sar \negnnf{q}, \negnnf{p}, p$}
\AxiomC{}
\RightLabel{$\id$}
\UnaryInfC{$\seqempstr \sar \negnnf{q}, q$}
\RightLabel{$\wk$}
\dashedLine
\UnaryInfC{$\seqempstr \sar \negnnf{p}, \negnnf{q}, q$}
\RightLabel{$\conr$}
\BinaryInfC{$\seqempstr \sar \negnnf{p}, \negnnf{q}, p \land q$}
\RightLabel{$\disr$}
\UnaryInfC{$\seqempstr \sar \negnnf{p} \lor \negnnf{q}, p \land q$}
\DisplayProof
\end{tabular}
\end{center}

By applying $\disr$ to both end sequents, we obtain two derivations of the valid formula $(\negnnf{p} \lor \negnnf{q}) \lor (p \land q) := p \land q \cimp p \land q$, which witness that $p \land q$ is the interpolant $p \land q \cimp p \land q$. Note that although the implication $p \land q \cimp p \land q$ is rather simple to interpolate, the above example was chosen for its simplicity and because of the fact that it still gives a concrete demonstration of Maehara's method.

In Maehara's seminal paper~\cite{Mae60}, the above method was provided for Gentzen-style sequent calculi. However, if our aim is to proof-theoretically (and syntactically) demonstrate that our class of grammar logics possesses the Lyndon interpolation property, then we must figure out how to generalize Maehara's method to nested (or, refined labelled) sequent calculi, since we have such calculi at our disposal for grammar logics.\footnote{Recall that in the context of grammar logics, nested sequent and refined labelled sequent calculi are notational variants of one another (see \sect~\ref{SECT:Refine-Grammar}). Therefore, in the current context, we may use the two terms interchangeably.} Such a generalization was presented in~\cite{LyoTiuGorClo20}, where Craig interpolation was established for tense logics and bi-intuitionistic logic.\footnote{Although the method of~\cite{LyoTiuGorClo20} establishes Craig interpolation, strengthening the method to yield Lyndon interpolation is almost trivial, and follows from also considering the polarities of propositional variables.} Also, it should be noted that although the method of~\cite{LyoTiuGorClo20} was introduced for nested systems, refined labelled systems were leveraged in the methodology; as explained there, the labelled \emph{notation} tends to be easier to work with and allows one to avoid the introduction of certain auxiliary concepts. We therefore follow the same strategy, and utilize our refined labelled calculi $\kmsl$ as opposed to the nested calculi $\dkms$ while proving Lyndon interpolation for grammar logics. 

Extending Maehara's method to the framework of refined labelled (i.e. nested) sequents calls for generalizing two of the notions inherent in Maehara's method: (i) the notion of an interpolant, and (ii) the negation of an interpolant. We first motivate how interpolants ought to be extended in the refined labeled (i.e. nested) setting by considering a simple example, and then discuss (ii) afterward. 

For our simple example, let us consider the following instance of the $\id$ rule (recall that we are now working in the calculus $\kmsl$):
\begin{center}
\AxiomC{}
\RightLabel{$\id$}
\UnaryInfC{$R_{a}wu \sar u : p, u : \negnnf{p}$}
\DisplayProof
\end{center}
As we saw in the example of Maehara's method above, each rule of the given calculus is transformed into a set of rules employing more expressive sequents that partition the contexts and assign interpolants accordingly---\emph{where interpolants are proper logical formulae}. For the sake of the example, let us suppose that we partition the above initial sequent as follows: $R_{a}wu \sar u : p \sep u : \negnnf{p} \sepgi \gi$, with $\gi$ denoting the interpolant we aim to discover. In the current context, it is nonsensical to make use of proper logical formulae as interpolants because such formulae omit labels, which encode essential information in the refined labelled setting. At the very least then, our interpolant ought to make use of labels. If we pick the interpolant $u : \negnnf{p}$ as our interpolant $\gi$, then the aforementioned split sequent becomes $R_{a}wu \sar u : p \sep u : \negnnf{p} \sepgi u : \negnnf{p}$. If we read this sequent in a manner analogous to how we read partitioned sequents in the context of Maehara's method, then the partitioned sequent states that $R_{a}wu \sar u : p, u : \negnnf{p}$ and $R_{a}wu \sar u : \negnnf{\negnnf{p}}, u : \negnnf{p}$ are both derivable in $\kmsl$ with interpolant $u : \negnnf{p}$. Since $\negnnf{\negnnf{p}} := p$, this happens to be true. We may conclude that the generalized interpolants used in the refined labelled setting must make use of labelled formulae at the very least.

Drawing an analogy with Maehara's method, we must introduce inference rules to calculate interpolants for each inference rule of $\kmsl$, which includes the two premise rule $\conr$. In the context of classical propositional logic, we saw that the $(\land_{r1})$ and $(\land_{r2})$ rules constructed the interpolants $\chi \lor \theta$ and $\chi \land \theta$, respectively, from the interpolants of the premises. In the refined labelled setting with $\kmsl$, where we have already agreed that interpolants ought to include labels, what would happen if we had a conjunction inference where the interpolant of one premise was of the form $w : \chi$ and the interpolant in the second premise was of the form $u : \theta$? To resolve this issue, it happens to be sufficient to define interpolants to be sets of labelled sequents, and when a rule such as $\conr$ is applied, we union to the two sets to construct the new interpolant. 

The discussion thus far has led us to confirm that interpolants ought to be sets of labelled sequents when generalizing Maehara's method to the refined labelled setting. Nevertheless, recall that Maehara's method takes a derivable sequent $\seqempstr \sar \Gamma_{1}, \Gamma_{2}$ and returns a proof of $\seqempstr \sar \Gamma_{1}, \chi$ and $\seqempstr \sar \negnnf{\chi}, \Gamma_{2}$, where $\chi$ is the interpolant. By analogy, to show our grammar logics interpolate via $\kmsl$, we aim to show that given a derivable labelled sequent $\rel \sar \Gamma_{1}, \Gamma_{2}$, we can transform its proof into proofs of $\rel \sar \Gamma_{1}, \gi$ and $\rel \sar \negnnf{\gi}, \Gamma_{2}$, where $\gi$ is the interpolant. Yet, if $\gi$ is a set of labelled sequents, what should the negation $\negnnf{\gi}$ represent?

At this point, it behooves us to recognize a general pattern underlying the relationship between an interpolant and its negation in Maehara's method, namely, the empty sequent can be derived from an interpolant and its negation by making use of $\cut$ (we will also allow for structural rules such as $\wk$ and $\ctrr$ to be used in our more general setting); we say that an interpolant and its negation are \emph{orthogonal} if such a relationship holds. In the context of classical propositional logic and Gentzen-style sequent calculi, this relationship is easy to verify; if $\chi$ is an interpolant, then a single application of $\cut$ between $\seqempstr \sar \chi$ and $\seqempstr \sar \negnnf{\chi}$ yields the empty sequent $\seqempstr \sar \seqempstr$. Nonetheless, in the refined labelled setting, phrasing the relationship between an interpolant and its negation in such a general manner permits us to define the negation $\negnnf{\gi}$ of an interpolant $\gi$ as a set of labelled sequents orthogonal to $\gi$. By this definition, the negation of an interpolant is not necessarily unique, though 
 we will strengthen the definition in our interpolation work below to ensure a unique negation, since it simplifies matters. To demonstrate the concept of orthogonality, observe that the interpolant $\gi$ given below is orthogonal to the interpolant $\gi'$ as the empty sequent $\seqempstr \sar \seqempstr$ can be derived with applications of $\cut$ and $\wk$ between the labelled sequents (which have been placed in between parentheses to improve readability) of the interpolants:

\begin{center}
\begin{tabular}{c @{\hskip 2em} c}
$\gi := \{(\seqempstr \sar w : p), (\seqempstr \sar u : \negnnf{q})\}$

&

$\gi' := \{(\seqempstr \sar w : \negnnf{p}, u : q)\}$
\end{tabular}
\end{center}

Before moving on to prove the main results of the section, we remark on the relationship between the syntactic method of interpolation from~\cite{LyoTiuGorClo20} presented here and the closely related semantic (yet, still proof-theoretic) method of interpolation by Kuznets \textit{et al.}~\cite{FitKuz15,Kuz16,Kuz16b,Kuz18,KuzLel18} applied to modal and intermediate logics equipped with a Kripke semantics.\footnote{This method may be qualified as both proof-theoretic and semantic since the method inductively constructs interpolants of formulae based on the structure of a given derivation, but proves the correctness of the inductive construction via \emph{semantic arguments}.} One notable difference is that the semantic method introduces meta-connectives $\owedge$ and $\ovee$ expressing meta-level conjunction and disjunction, respectively, and which are used to construct interpolants. This contrasts with the syntactic approach that uses sets of labelled (or, nested) sequents as explained above. Moreover, in the semantic method, the meta-connectives $\owedge$ and $\ovee$ are interpreted in a semantics extending the semantics of the logic under consideration. In this regard, the syntactic method can be seen as possessing a higher degree of parsimony, as the semantics of the considered logic need not be generalized to a meta-language. Additionally, the interpolants of the syntactic method are justified on purely syntactic and proof-theoretic grounds, whereas the interpolants of the semantic method are justified via semantic arguments. As pointed out in~\cite{LyoTiuGorClo20}, defining and justifying interpolants and their negations in a purely syntactic fashion independent of the underlying logic's semantics, appears to have the advantageous consequence that the method is applicable to logics for which no Kripke semantics is known (e.g. bi-intuitionistic linear logic~\cite{CloDawGorTiu13}), something which the semantic method does not appear immediately capable.






\subsection{Lyndon Interpolation via Proof-Theoretic Methods}

We leverage the syntactic method of interpolation from~\cite{LyoTiuGorClo20} to prove (the new result) that all context-free grammar logics with converse possess the effective Lyndon interpolation property, meaning that our proof not only shows the existence of a Lyndon interpolant for each valid implication, but also shows how to construct it (cf.~\cite{GorNgu05}). We first define a set of concepts necessary to apply the syntactic method, and then prove the main results.

As mentioned previously, in our more general setting, an interpolant is a set of labelled sequents. In defining our interpolants however, the multiset of relational atoms $\rel$ is left empty since $\rel$ can be recovered from the sequents with which the interpolants are used. This gives rise to the following definitions:

\begin{definition}[Flat Labelled Sequent, Interpolant] A \emph{flat labelled sequent}\index{Flat labelled sequent} is a labelled sequent of the form $\seqempstr \sar \Gamma$. An \emph{interpolant}\index{Interpolant} $\gi$ is defined to be a set of flat sequents, i.e.
$$
\gi := \{(\seqempstr \sar \Gamma_{1}), \ldots, (\seqempstr \sar \Gamma_{n}) \}
$$
We use $\gi$, $\gi'$, $\ldots$ (occasionally annotated) to denote interpolants.
\end{definition}

The relation of being \emph{orthogonal} was informally defined above, and intuitively holds between interpolants $\gi_{1}$ and $\gi_{2}$ \ifandonlyif the empty sequent is derivable from the labelled sequents in $\gi_{1}$ and $\gi_{2}$ using $\cut$ and possibly $\wk$ and/or $\ctrr$. This relationship is used to establish a generalized notion of negation, that is to say, if $\gi_{1}$ is orthogonal to $\gi_{2}$, then the two interpolants are negations of one another. For example, suppose that we are given the interpolant $\gi_{1} := \{(\seqempstr \sar w : \phi, u : \psi), (\seqempstr \sar v : \chi)\}$. Then, the following interpolants are orthogonal to $\gi_{1}$ and can be considered negations:
\begin{eqnarray*}
\gi_{2} & := & \{(\seqempstr \sar w : \negnnf{\phi}),(\seqempstr \sar u : \negnnf{\psi}, v : \negnnf{\chi})\}\\
\gi_{3} & := & \{(\seqempstr \sar w : \negnnf{\phi}, v : \negnnf{\chi}),(\seqempstr \sar u : \negnnf{\psi}, u : \negnnf{\psi})\}\\
\gi_{4} & := & \{(\seqempstr \sar w : \negnnf{\phi}, v : \negnnf{\chi}),(\seqempstr \sar u : \negnnf{\psi}, v : \negnnf{\chi})\}
\end{eqnarray*}
In each case, one can derive the empty sequent from $\gi_{1} \cup \gi_{i}$ using $\cut$, $\wk$, and $\ctrr$ for each $i \in \{2,3,4\}$.  
 Despite this fact, to make the negation of an interpolant deterministic, our definition below will choose $\gi_{4}$ to be the negation, as it straightforward to calculate such an interpolant as a function of $\gi_{1}$. Instead of using the general notion of negation presented above, we opt for a more specific notion, which we refer to as the \emph{orthogonal}:

\begin{definition}[Orthogonal]\label{def:orthogonal-operation-kms} For an interpolant $\gi := \{(\seqempstr \sar \Gamma_{1}), \ldots, (\seqempstr \sar \Gamma_{n})\}$, its \emph{orthogonal}\index{Orthogonal} $\orth{\gi}$ is defined as:
$$
\orth{\gi} ::= \{(\seqempstr \sar w_{1} : \negnnf{\phi_{1}}, \ldots, w_{n} : \negnnf{\phi_{1}}) \ | \ \text{for all } i \in \{1, \ldots, n\}, w_{i} : \phi_{i} \in \Gamma_{i}\}
$$
\end{definition}

When constructing an interpolant by induction on the height of the input derivation, the processing of $\charaboxr$ rules requires a special interpolant to be produced. The necessity of such interpolants will become apparent in the proof of \lem~\ref{lem:fundamental-interpolation-lemma} below. Such interpolants are defined as follows:

\begin{definition}\label{def:boxed-interpolant-kms} Suppose we have an interpolant of the following form:
$$
\gi := \{(\seqempstr \sar \Gamma_{1}, u : \phi_{1,1}, \ldots, u : \phi_{1,k_{1}}), \ldots, (\seqempstr \sar \Gamma_{n}, u : \phi_{n,1}, \ldots, u : \phi_{n,k_{n}})\}
$$
where $u$ does not occur in $\Gamma_{1}, \ldots, \Gamma_{n}$. Let $\chara \in \albet$, that is, let $\chara$ be a character in our alphabet $\albet$, and define
$$
\charaboxgi := \{(\seqempstr \sar \Gamma_{1}, w : \charabox \bigvee_{i = 1}^{k_{1}} \phi_{1,i}), \ldots, (\seqempstr \sar \Gamma_{n}, w : \charabox \bigvee_{i = 1}^{k_{n}} \phi_{n,i})\}
$$
\end{definition}

Let us now define the types of sequents that will be used in confirming the effective Lyndon interpolation property:

\begin{definition}[Interpolation Sequent] An \emph{interpolation sequent}\index{Interpolation sequent} is defined to be a syntactic object of the form $\rel \sar \Gamma \mid \Delta \sepgi \gi$, where $\rel$ is a set of relational atoms, $\Gamma$ and $\Delta$ are labelled formulae, and $\gi$ is an interpolant. 
\end{definition}

The vertical bar `$\sep$' occurring in an interpolation sequent partitions the sequent into a left part serving as the antecedent in the interpolation statement, and a right part serving as the consequent in the interpolation statement. To illustrate this point, the interpolation sequent shown below top partitions into the two labelled sequents shown below bottom:
\begin{center}
$\rel \sar \Gamma_{1} \sep w : p, w : \negnnf{p}, \Gamma_{2} \sepgi \{(\seqempstr \sar w : \top)\}$
\end{center}
\begin{center}
\begin{tabular}{c c}
$\Lambda_{1} := (\rel \sar \Gamma_{1}, w : \top)$

&

$\Lambda_{2} := (\rel \sar w : \negnnf{\top}, w : p, w : \negnnf{p}, \Gamma_{2})$
\end{tabular}
\end{center}
In the above example, the left component $\Gamma_{1}$ of the partition is placed in $\Lambda_{1}$, and the right component $w : p, w : \negnnf{p}, \Gamma_{2}$ is placed in $\Lambda_{2}$. We think of the interpolant $w : \top$ as being \emph{implied by} the left component of the partition, and so, we place it in $\Lambda_{1}$, and we think of the interpolant as \emph{implying} the right component of the partition, so we place the negation of the interpolant 
 in $\Lambda_{2}$. Observe that we can derive the labelled sequent $\rel \sar \Gamma_{1}, w : p, w : \negnnf{p}, \Gamma_{2}$ by applying $\cut$ (possibly with $\wk$) to $\Lambda_{1}$ and $\Lambda_{2}$, which \emph{syntactically establishes} that the interpolant is indeed an interpolant (given that the interpolant satisfies certain other properties; see \lem~\ref{lem:kmsli-preserves-prop-and-labels} below).

Each interpolation calculus $\kmsli$, for each grammar logic $\kms$, is given in \fig~\ref{fig:Km(s)LI}. The derivability relation is defined as follows:

\begin{definition} We write $\vdash_{\kmsli} \rel \sar \Gamma \mid \Delta \sepgi \gi$ to indicate that an interpolation sequent $\rel \sar \Gamma \mid \Delta \sepgi \gi$ is derivable in the interpolation calculus $\kmsli$.
\end{definition}

Since each interpolation calculus is obtained from a corresponding refined labelled calculus $\kmsl$, each propagation rule $\prcharadiar$ relies on the notion of a propagation graph, which we define for interpolation sequents below. The notion of a propagation path, the string of a propagation path, and their converses are defined as in \dfn~\ref{def:propagation-path-kms}, so we do not repeat them here.

\begin{definition}[Propagation Graph of Interpolation Sequent] Let $\Lambda := \rel \sar \Gamma \mid \Delta \sepgi \gi$ and $\Lambda' := \rel \sar \Gamma, \Delta$. We define the \emph{propagation graph}\index{Propagation graph!for an interpolation sequent} of an interpolation sequent $\Lambda$ as follows: $\prgr{\Lambda} := \prgr{\Lambda'}$. 
\end{definition}

\begin{figure}[t]
\noindent\hrule

\begin{center}
\begin{tabular}{c c}
\AxiomC{}
\RightLabel{$\idi$}
\UnaryInfC{$\iseq{\rel \sar \Gamma, w : \negnnf{p} \sep w : p, \Delta}{(\sar w : p)}$}
\DisplayProof

&

\AxiomC{$\iiseq{\rel \sar \Gamma \sep \Delta}{\gi}$}
\RightLabel{$\orthru$}
\UnaryInfC{$\iiseq{\rel \sar \Delta \sep \Gamma}{\orth{\gi}}$}
\DisplayProof
\end{tabular}
\end{center}

\begin{center}
\begin{tabular}{c c}
\AxiomC{}
\RightLabel{$\idii$}
\UnaryInfC{$\iseq{\rel \sar \Gamma  \sep w : \negnnf{p}, w : p, \Delta}{(\sar w : \top)}$}
\DisplayProof

&

\AxiomC{$\iiseq{\rel \sar \Gamma \sep  w : \phi, w : \psi, \Delta}{\gi}$}
\RightLabel{$\disr$}
\UnaryInfC{$\iiseq{\rel \sar \Gamma \sep w : \phi \lor \psi, \Delta}{\gi}$}
\DisplayProof
\end{tabular}
\end{center}

\begin{center}
\begin{tabular}{c}
\AxiomC{$\iiseq{\rel \sar \Gamma \sep w : \phi, \Delta}{\gi_{1}}$}
\AxiomC{$\iiseq{\rel \sar \Gamma \sep w : \psi, \Delta}{\gi_{2}}$}
\RightLabel{$\conr$}
\BinaryInfC{$\iiseq{\rel \sar \Gamma \sep w : \phi \land \psi, \Delta}{\gi_{1} \cup \gi_{2}}$}
\DisplayProof
\end{tabular}
\end{center}

\begin{center}
\begin{tabular}{c c}
\AxiomC{$\iiseq{\rel, R_{\chara}wu \sar \Gamma \sep  u : \phi, \Delta}{\gi}$}
\RightLabel{$\charaboxr^{\dag_{1}}$}
\UnaryInfC{$\iiseq{\rel \sar \Gamma \sep  w : \charabox \phi, \Delta}{\charaboxgi}$}
\DisplayProof

&

\AxiomC{$\iiseq{\rel \sar \Gamma \sep  w : \charadia \phi, u : \phi, \Delta}{\gi}$}
\RightLabel{$\prcharadiar^{\dag_{2}}$}
\UnaryInfC{$\iiseq{\rel \sar \Gamma \sep  w : \charadia \phi, \Delta}{\gi}$}
\DisplayProof
\end{tabular}
\end{center}

\hrulefill
\caption{The calculus $\kmsli$\index{$\kmsli$} for constructing interpolants for the grammar logic $\kms$, for a \cfcst system $\thuesys$. The calculus contains a $\charaboxr$ and $\prcharadiar$ rule for each $\chara \in \albet$, i.e. for each character in our alphabet. The side condition $\dag_{1}$ states that the rule is applicable only if $u$ is an eigenvariable, and $\dag_{2}$ states that the rule is applicable only if $\stra_{\ppath}(w,u) \in \thuesyslang{\chara}$.}
\label{fig:Km(s)LI}
\end{figure}

Perhaps the most unique feature of each interpolation calculus $\kmsli$ is the $\orthru$ rule. The rule lets us cut the number of rules needed in our calculus in half by permitting the components of the partition in an interpolation sequent to be `flipped'. The key to the correctness of the rule, is given in the lemma below, which shows that applying the orthogonal operation twice to an interpolant always retains the sequents of the original interpolant.


\begin{lemma}\label{lm:double-orthogonal}
If $(\seqempstr \sar \Gamma) \in \orth{\orth{\gi}}$, then there exists a $(\seqempstr \sar \Delta) \in \gi$ such that $\Delta \subseteq \Gamma$.
\end{lemma}

\begin{proof} We prove the lemma by contradiction, and assume that there exists a $(\seqempstr \sar \Gamma) \in \orth{\orth{\gi}}$ such that for all $(\seqempstr \sar \Delta) \in \gi$, $\Delta \not\subseteq \Gamma$. Let $\gi := \{(\seqempstr \sar \Gamma_{1}), \ldots, (\seqempstr \sar \Gamma_{n}) \}$. By our assumption, we know that for each $i \in \{1, \ldots, n\}$, there exists a formula $w_{i} : \phi_{i} \in \Gamma_{i}$ such that $w_{i} : \phi_{i} \not\in \Gamma$. Let $\Gamma' := w_{1} : \phi_{1}, \ldots, w_{n} : \phi_{n}$ and observe that by construction $\Gamma \cap \Gamma' = \emptyset$. However, by \dfn~\ref{def:orthogonal-operation-kms}, we know that $(\seqempstr \sar \Gamma') \in \orth{\gi}$, and since $(\seqempstr \sar \Gamma) \in \orth{\orth{\gi}}$, it follows that $\Gamma \cap \Gamma' \neq \emptyset$, giving a contradiction.
\end{proof}

Enough groundwork has been laid for us to prove our main results, though before we do so, we formally define the literal function $\lit(\cdot)$ and the Lyndon interpolation property:

\begin{definition}[Literal Function] For a formula $\phi \in \langkm{\thuesys}$, we define $\lit(\phi)$\index{Literal function} to be the set of all literals occurring in $\phi$. If the multiset $\Gamma := w_{1} : \phi_{1}, \ldots, w_{n} : \phi_{n}$ and the interpolant $\gi := \{(\seqempstr \sar \Gamma_{1}), \ldots, (\seqempstr \sar \Gamma_{n})\}$, then we respectively let
$$
\displaystyle{\lit(\Gamma) := \bigcup_{1 \leq i \leq n} \lit(\phi_{i})} \qquad \displaystyle{\lit(\gi) := \bigcup_{1 \leq i \leq n} \lit(\Gamma_{i})}
$$
\end{definition}

\begin{definition}[Lyndon Interpolation Property]\label{def:lyndon-interpolation-property-kms} A grammar logic $\kms$ has the \emph{Lyndon interpolation property}\index{Lyndon interpolation property} \ifandonlyif for every implication $\phi \cimp \psi \in \langkm{\thuesys}$ such that $\vdash_{\kms} \phi \cimp \psi$, there exists a formula $\chi$ such that (i) $\lit(\chi) \subseteq \lit(\phi) \cap \lit(\psi)$ and (ii) $\vdash_{\kms} \phi \cimp \chi$ and $\vdash_{\kms} \chi \cimp \psi$.
\end{definition}

\begin{lemma}\label{lem:kmsl-to-kmsli}
If $\vdash_{\kmsl} \rel \sar \Gamma, \Delta$, then $\vdash_{\kmsli} \iiseq{\rel \sar \Gamma \sep \Delta}{\gi}$, for some interpolant $\gi$.
\end{lemma}

\begin{proof} Straightforward, by induction on the height of the proof of $\rel \sar \Gamma, \Delta$.
\end{proof}

\begin{lemma}\label{lem:kmsli-preserves-prop-and-labels}
If $\vdash_{\kmsli} \iiseq{\rel \sar \Gamma \sep \Delta}{\gi}$, then $\lit(\gi) \subseteq \lit(\negnnf{\Gamma}) \cap \lit(\Delta)$, and all labels occurring in $\gi$ also occur in $\rel,\Gamma$ or $\Delta$.
\end{lemma}

\begin{proof} By induction on the height of the given derivation.

\textit{Base case.} The base case can be confirmed by observing the $\idi$ and $\idii$ rules.

\textit{Inductive step.} We consider the $\orthru$ and $\conr$ rules, as the $\disr$, $\charaboxr$, and $\prcharadiar$ cases follow directly from the inductive hypothesis. In all cases, it is straightforward to verify that all labels occurring in $\gi$ also occur in $\rel,\Gamma$ or $\Delta$.

$\orthru$. Let $l \in \lit(\gi)$. By IH, we know that $l \in \lit(\negnnf{\Gamma}) \cap \lit(\Delta)$. By \dfn~\ref{def:orthogonal-operation-kms}, it follows that $\negnnf{l} \in \lit(\orth{\gi})$, which implies that $\negnnf{l} \in \lit(\negnnf{\Delta}) \cap \lit(\Gamma)$ by the fact expressed in the previous sentence.

$\conr$. Let $l \in \lit(\gi_{1} \cup\gi_{2})$. Then, either $l \in \lit(\gi_{1})$ or $l \in \lit(\gi_{2})$. Without loss of generality, we assume the former case holds. This entails that $l \in \lit(\negnnf{\Gamma}) \cap \lit(w : \phi, \Delta)$, and since $\lit(w : \phi, \Delta) \subseteq \lit(w : \phi \land \psi, \Delta)$, we have that $l \in \lit(\negnnf{\Gamma}) \cap \lit(w : \phi \land \psi, \Delta)$.
\end{proof}

The following lemma establishes that each interpolation calculus $\kmsli$ correctly constructs interpolants, and is vital in proving our main theorem below (\thm~\ref{thm:interpolation-theorem}).

\begin{lemma}\label{lem:fundamental-interpolation-lemma}
For all $\rel$, $\Gamma$, $\Delta$, and $\gi$, if $\vdash_{\kmsli} \iiseq{\rel \sar \Gamma \sep \Delta}{\gi}$, then

(i) for all $(\seqempstr \sar \Xi) \in \gi$, we have $\vdash_{\kmsl} \rel \sar \Gamma, \Xi$ and

(ii) for all $(\seqempstr \sar \Theta) \in \orth{\gi}$, we have $\vdash_{\kmsl} \rel \sar \Theta, \Delta$.
\end{lemma}

\begin{proof} By  induction on the height of the given derivation of $\iiseq{\rel \sar \Gamma \sep \Delta}{\gi}$.

\textit{Base case.} If the interpolation sequent was derived by an instance of $\idi$, then both (i) and (ii) follow by using the $\id$ rule of $\kmsl$. If the interpolation sequent was derived by an instance of $\idii$, then (i) and (ii) are resolved as shown below, and make use of the definitions $\top := q \lor \negnnf{q}$ and $\bot := q \land \negnnf{q}$.
\begin{center}
\begin{tabular}{c c}
\AxiomC{}
\RightLabel{$\id$}
\UnaryInfC{$\rel \sar \Gamma, w : q, w : \negnnf{q}$}
\RightLabel{$\disr$}
\UnaryInfC{$\rel \sar \Gamma, w : q \lor \negnnf{q}$}
\RightLabel{=}
\dottedLine
\UnaryInfC{$\rel \sar \Gamma, w : \top$}
\DisplayProof

&

\AxiomC{}
\RightLabel{$\id$}
\UnaryInfC{$\rel \sar w : q \land \negnnf{q}, w : p, w : \negnnf{p}, \Delta$}
\RightLabel{=}
\dottedLine
\UnaryInfC{$\rel \sar w : \bot, w : p, w : \negnnf{p}, \Delta$}
\DisplayProof
\end{tabular}
\end{center}

\textit{Inductive step.}  We consider the (non-trivial) $\orthru$, $\conr$, and $\charaboxr$ cases; the $\disr$ and $\prcharadiar$ cases are straightforward.

$\orthru$. Suppose our derivation ends with and $\orthru$ inference of the following form:
\begin{center}
\AxiomC{$\iiseq{\rel \sar \Gamma \sep \Delta}{\gi}$}
\RightLabel{$\orthru$}
\UnaryInfC{$\iiseq{\rel \sar \Delta \sep \Gamma}{\orth{\gi}}$}
\DisplayProof
\end{center}

(i) Let $(\seqempstr \sar \Xi) \in \orth{\gi}$. To resolve the case, we need to show that $\vdash_{\kmsl} \rel \sar \Xi, \Delta$; however, this follows immediately from IH.

(ii) Let $(\seqempstr \sar \Theta) \in \orth{\orth{\gi}}$. We need to show that $\vdash_{\kmsl} \rel \sar \Gamma, \Theta$. By \lem~\ref{lm:double-orthogonal}, we know there exists a $(\seqempstr \sar \Theta') \in \gi$ such that $\Theta' \subseteq \Theta$. By IH, $\vdash_{\kmsl} \rel \sar \Gamma, \Theta'$, and by the admissibility of $\wk$ (\cor~\ref{cor:struc-admiss-kmsl}), it follows that $\vdash_{\kmsl} \rel \sar \Gamma, \Theta$.

$\conr$. Suppose our derivation ends with the following $\conr$ inference:
\begin{center}
\AxiomC{$\iiseq{\rel \sar \Gamma \sep w : \phi, \Delta}{\gi_{1}}$}
\AxiomC{$\iiseq{\rel \sar \Gamma \sep w : \psi, \Delta}{\gi_{2}}$}
\RightLabel{$\conr$}
\BinaryInfC{$\iiseq{\rel \sar \Gamma \sep w : \phi \land \psi, \Delta}{\gi_{1} \cup \gi_{2}}$}
\DisplayProof
\end{center}

(i) Let $(\seqempstr \sar \Xi) \in \gi_{1} \cup \gi_{2}$. By IH, we have that for any $(\seqempstr \sar \Xi_{1}) \in \gi_{1}$ and $(\seqempstr \sar \Xi_{2}) \in \gi_{2}$, $\vdash_{\kmsl} \rel \sar \Gamma, \Xi_{1}$ and $\vdash_{\kmsl} \rel \sar \Gamma, \Xi_{2}$. Therefore, the conclusion follows regardless of if $(\seqempstr \sar \Xi_{1}) \in \gi_{1}$ or $(\seqempstr \sar \Xi_{2}) \in \gi_{2}$.

(ii) Let $(\seqempstr \sar \Theta) \in \orth{\gi_{1} \cup \gi_{2}}$.
By IH, for all $(\seqempstr \sar \Theta_{1}) \in \orth{\gi_{1}}$ and $(\seqempstr \sar \Theta_{2}) \in \orth{\gi_{2}}$, $\vdash_{\kmsl} \rel \sar \Theta_{1}, w : \phi, \Delta$ and $\vdash_{\kmsl} \rel \sar \Theta_{2}, w : \psi, \Delta$. By \dfn~\ref{def:orthogonal-operation-kms}, there exists a $(\seqempstr \sar \Theta_{1}) \in \orth{\gi_{1}}$ and a $(\seqempstr \sar \Theta_{2}) \in \orth{\gi_{2}}$ such that $(\seqempstr \sar \Theta) = (\seqempstr \sar \Theta_{1}, \Theta_{2})$. Therefore, by the admissibility of $\wk$ (\cor~\ref{cor:struc-admiss-kmsl}), we can derive $\vdash_{\kmsl} \rel \sar \Theta_{1}, \Theta_{2}, w : \phi, \Delta$ and $\vdash_{\kmsl} \rel \sar \Theta_{1}, \Theta_{2}, w : \psi, \Delta$. A single application of $\conr$ gives the desired conclusion.

$\charaboxr$. Suppose our derivation ends with an inference of the following form:
\begin{center}
\AxiomC{$\iiseq{\rel, R_{\chara}wu \sar \Gamma \sep  u : \phi, \Delta}{\gi}$}
\RightLabel{$\charaboxr$}
\UnaryInfC{$\iiseq{\rel \sar \Gamma \sep  w : \charabox \phi, \Delta}{\charaboxgi}$}
\DisplayProof
\end{center}

(i) Let $(\seqempstr \sar \Xi) \in \charaboxgi$. By \dfn~\ref{def:boxed-interpolant-kms}, we know that $\Xi = \Xi', w : \charabox (\phi_{1} \lor \cdots \lor \phi_{n})$ with $(\seqempstr \sar, \Xi', u : \phi_{1}, \ldots, u : \phi_{n}) \in \gi$. We derive the desired conclusion as follows:
\begin{center}
\AxiomC{}
\RightLabel{\ih}
\dashedLine
\UnaryInfC{$\rel, R_{\chara}wu \sar \Gamma, \Xi', u : \phi_{1}, \ldots, u : \phi_{n}$}
\RightLabel{$\disr \times (n-1)$}
\UnaryInfC{$\rel, R_{\chara}wu \sar \Gamma, \Xi', u : \phi_{1} \lor \cdots \lor \phi_{n}$}
\RightLabel{$\charaboxr$}
\UnaryInfC{$\rel \sar \Gamma, \Xi', u : \charabox (\phi_{1} \lor \cdots \lor \phi_{n})$}
\DisplayProof
\end{center}

(ii)  Let $(\seqempstr \sar \Theta) \in \orth{\charaboxgi}$. We need to establish that $\vdash_{\kmsl} \rel \sar \Theta, w : \charabox \phi, \Delta$. By \dfn~\ref{def:orthogonal-operation-kms}, $\Theta$ may contain zero or more formulae of the form $w : \negnnf{\charabox (\psi_{1} \lor \cdots \lor \psi_{n})}$ such that there exists a $(\seqempstr \sar \Theta') \in \gi$ with $\{u : \psi_{1}, \ldots, u : \psi_{n}\} \subseteq \Theta'$. We refer to such formulae as \emph{modal-interpolant} formluae, and for the sake of simplicity we assume that one modal-interpolant formula exists in $\Theta$ with $n = 2$, that is, $w : \negnnf{\charabox (\psi_{1} \lor \psi_{2})}$; we prove the result for this simplifies case as the general case is analogous. We let $\Theta := \Theta', w : \negnnf{\charabox (\psi_{1} \lor \psi_{2})}$.

By assumption, we know that there exists a $(\seqempstr \sar \Xi) \in \gi$ such that $\{u : \psi_{1}, u : \psi_{2}\} \subseteq \Xi$, which implies that there exist $(\seqempstr \sar \Xi_{1}), (\seqempstr \sar \Xi_{2}) \in \orth{\gi}$ of the form $\Xi_{1} = u : \negnnf{\psi_{1}}, \Theta'$ and $\Xi_{2} = u : \negnnf{\psi_{2}}, \Theta'$. By \ih, we have:
$$
\vdash_{\kmsl} \rel, R_{\chara}wu \sar u : \negnnf{\psi_{1}}, \Theta', u : \phi, \Delta
$$
$$
\vdash_{\kmsl} \rel, R_{\chara}wu \sar u : \negnnf{\psi_{2}}, \Theta', u : \phi, \Delta
$$
By invoking the admissibility of $\wk$ (\cor~\ref{cor:struc-admiss-kmsl}), we obtain:
$$
\vdash_{\kmsl} \rel, R_{\chara}wu \sar w : \charadia (\negnnf{\psi_{1}} \land \negnnf{\psi_{2}}), u : \negnnf{\psi_{1}}, \Theta', u : \phi, \Delta
$$
$$
\vdash_{\kmsl} \rel, R_{\chara}wu \sar w : \charadia (\negnnf{\psi_{1}} \land \negnnf{\psi_{2}}), u : \negnnf{\psi_{2}}, \Theta', u : \phi, \Delta
$$
Applying the $\conr$ rule between the above to labelled sequents gives:
$$
\vdash_{\kmsl} \rel, R_{\chara}wu \sar w : \charadia (\negnnf{\psi_{1}} \land \negnnf{\psi_{2}}), u : \negnnf{\psi_{1}} \land \negnnf{\psi_{2}}, \Theta', u : \phi, \Delta
$$
Last, the desired conclusion is derived as follows:
\begin{center}
\AxiomC{$\rel, R_{\chara}wu \sar w : \charadia (\negnnf{\psi_{1}} \land \negnnf{\psi_{2}}), u : \negnnf{\psi_{1}} \land \negnnf{\psi_{2}}, \Theta', u : \phi, \Delta$}
\RightLabel{$\prcharadiar$}
\dashedLine
\UnaryInfC{$\rel, R_{\chara}wu \sar w : \charadia (\negnnf{\psi_{1}} \land \negnnf{\psi_{2}}), \Theta', u : \phi, \Delta$}
\RightLabel{$\charaboxr$}
\dashedLine
\UnaryInfC{$\rel \sar w : \charadia (\negnnf{\psi_{1}} \land \negnnf{\psi_{2}}), \Theta', w : \charabox \phi, \Delta$}
\RightLabel{=}
\dottedLine
\UnaryInfC{$\rel \sar w : \negnnf{\charabox (\psi_{1} \lor \psi_{2})}, \Theta', w : \charabox \phi, \Delta$}
\DisplayProof
\end{center}
\end{proof}

To prove (effective) Lyndon interpolation, we need to be able to construct interpolants that are \emph{formulae} from our more general notion of an interpolant. This construction is possible if all labelled formulae in the interpolant share the same label. In such a case, the interpolant is of the form $\gi := \{(\seqempstr \sar \Xi_{1}), \ldots , (\seqempstr \sar \Xi_{n})\}$ with $\Xi_{i} := \{w : \phi_{i,1}, \ldots , w : \phi_{i,k_{i}}\}$ for each $1 \leq i \leq n$, and its corresponding formula is $ \bigwedge_{i=1}^{n} \bigvee_{j=1}^{k_{i}} \phi_{i,j}$. Given such an interpolant $\gi$, we write $\bigwedge \bigvee \gi$ to mean its corresponding formula. The following two lemmata show how to construct the corresponding formula of an interpolant (in the form just described) proof-theoretically, which will inevitably let us construct Lyndon interpolants effectively:

\begin{lemma}\label{lem:interpolant-implied-by-A}
Let $\gi := \{(\seqempstr \sar \Xi_{1}), \ldots , (\seqempstr \sar \Xi_{n})\}$ with $\Xi_{i} := \{w : \phi_{i,1}, \ldots , w : \phi_{i,k_{i}}\}$ for each $1 \leq i \leq n$. For any multiset of relational atoms $\rel$ and multiset of labelled formulae $\Gamma$, if $\vdash_{\kmsl} \rel \sar \Gamma, \Xi$ for all $(\seqempstr \sar \Xi) \in \gi$, then $\vdash_{\kmsl} \rel \sar \Gamma, w : \bigwedge \bigvee \gi$.
\end{lemma}

\begin{proof} Assume that $\vdash_{\kmsl} \rel \sar \Gamma, \Xi_{i}$, for all $\Xi_{i} \in \gi$. By repeated application of the $\disr$ rule, we obtain
$$
\vdash_{\kmsl} \rel \sar \Gamma, w : \bigvee_{j=1}^{k_{i}} \phi_{i,j}
$$
for each $i \in  \{1, \ldots, n\}$. Repeated application of the $\conr$ rule gives us 
$$
\vdash_{\kmsl} \rel \sar \Gamma, w : \bigwedge_{i=1}^{n} \bigvee_{j=1}^{k_{i}} \phi_{i,j}
$$
which is the desired conclusion since $\bigwedge \bigvee \gi = \bigwedge_{i=1}^{n} \bigvee_{j=1}^{k_{i}} \phi_{i,j}$.
\end{proof}

\begin{lemma}\label{lem:interpolant-implies-B}
Let $\gi := \{(\seqempstr \sar \Xi_{1}), \ldots, (\seqempstr \sar \Xi_{n})\}$ be an interpolant with $\Xi_{i} := \{w : \phi_{i,1}, \ldots, w : \phi_{i, k_{i}}\}$ for each $1 \leq i \leq n$. For any multiset of relational atoms $\rel$ and multiset of labelled formulae $\Delta$, if $\vdash_{\kmsl} \rel \sar \Theta, \Delta$ for all $(\seqempstr \sar \Theta) \in \orth{\gi}$, then $\vdash_{\kmsl} \rel \sar w : \negnnf{\bigwedge \bigvee \gi}, \Delta$.
\end{lemma}

\begin{proof} We prove that $\vdash_{\kmsli} \rel w : \negnnf{\bigwedge \bigvee \gi}, \Delta$ by induction on the cardinality of $\gi$. 

\textit{Base case.} If $\gi$ is a singleton, then by assumption $\vdash_{\kmsl} \rel \sar w : \negnnf{\phi_{1,j}}, \Delta$ is derivable for all $j \in \{1, \ldots, k_{1}\}$. The conclusion follows by $k_{1} - 1$ applications of the $\conr$ rule.

\textit{Inductive step.} Suppose that $\gi := \{(\seqempstr \sar \Theta_{1}), \ldots, (\seqempstr \sar \Theta_{n+1})\}$ contains $n+1$ elements and assume that $\vdash_{\kmsl} \rel \sar \Theta, \Delta$ for all $(\seqempstr \sar \Theta) \in \orth{\gi}$. It follows that for each $j \in \{1, \ldots, k_{n+1}\}$, $\vdash_{\kmsl} \rel \sar \Theta', w : \negnnf{\phi_{n+1,j}}, \Delta$ for all $(\seqempstr \sar \Theta') \in \orth{\gi - \{ (\seqempstr \sar \Theta_{n+1}) \}}$, which implies that 
$$
\vdash_{\kmsl} \rel \sar w : \bigvee \bigwedge \negnnf{(\gi - \{ (\seqempstr \sar \Theta_{n+1}) \})}, w : \negnnf{\phi_{n+1,j}}, \Delta
$$
for each $j \in \{1, \ldots, k_{n+1}\}$. By applying the $\conr$ rule between each of the $k_{n+1} - 1$ labelled sequents above, followed by a single application of the $\disr$ rule, we obtain
$$
\vdash_{\kmsl} \rel \sar w : \bigvee \bigwedge \negnnf{(\gi - \{ (\seqempstr \sar \Theta_{n+1}) \})} \lor \bigwedge_{1 \leq j \leq k_{n+1}} \negnnf{\phi_{n+1,j}}, \Delta
$$
which gives our desired conclusion since
$$
\negnnf{\bigwedge \bigvee \gi} = \bigvee \bigwedge \negnnf{(\gi - \{ (\seqempstr \sar \Theta_{n+1}) \})} \lor \bigwedge_{1 \leq j \leq k_{n+1}} \negnnf{\phi_{n+1,j}}.
$$
\end{proof}

Let us now state and prove our main theorem, which entails that each grammar logic in our class possesses the effective Lyndon interpolation property.

\begin{theorem}\label{thm:interpolation-theorem}
If $\vdash_{\kmsl} w : \phi \cimp \psi$, then there exists a $\chi$ such that (i) $\lit(\chi) \subseteq \lit(\phi) \cap \lit(\psi)$, and (ii) $\vdash_{\kmsl} \seqempstr \sar w : \phi \cimp \chi$ and $\vdash_{\kmsl} \seqempstr \sar w : \chi \cimp \psi$.
\end{theorem}

\begin{proof} Suppose that $\vdash_{\kmsl} \seqempstr \sar w : \phi \cimp \psi$, that is, $\vdash_{\kmsl} \seqempstr \sar w : \negnnf{\phi} \lor \psi$. By the invertibility of $\disr$ (\cor~\ref{cor:invert-kmsl}), we have $\vdash_{\kmsl} \seqempstr \sar w : \negnnf{\phi}, w : \psi$. Therefore, by \lem~\ref{lem:kmsl-to-kmsli} and~\ref{lem:kmsli-preserves-prop-and-labels}, there exists an interpolant $\gi := \{(\seqempstr \sar \Xi_{1}), \ldots, (\seqempstr \sar \Xi_{n})\}$ with $\Xi_{i} = \{w:\chi_{i,1}, \ldots, w:\chi_{i,k_{i}}\}$ for $i \in \{1, \ldots, n\}$ such that $\vdash_{\kmsli} \seqempstr \sar w : \negnnf{\phi} \sep w : \psi \sepgi \gi$ and $\lit(\gi) \subseteq \lit(w : \negnnf{\negnnf{\phi}}) \cap \lit(w : \psi) = \lit(\phi) \cap \lit(\psi)$. By \lem~\ref{lem:fundamental-interpolation-lemma}, we know that:

(a) For all $(\seqempstr \sar \Xi) \in \gi$, $\vdash_{\kmsl} \seqempstr \sar w : \negnnf{\phi}, \Xi$, and

(b) for all $(\seqempstr \sar \Theta) \in \orth{\gi}$, $\vdash_{\kmsl} \seqempstr \sar \Theta, w : \psi$.

By \lem~\ref{lem:interpolant-implied-by-A}, claim (a) implies that $\vdash_{kmsl} \seqempstr \sar w : \negnnf{\phi}, \bigwedge \bigvee \gi$, and by \lem~\ref{lem:interpolant-implies-B}, claim (b) implies that $\vdash_{kmsl} \seqempstr \sar w : \negnnf{\bigwedge \bigvee \gi}, w : \psi$. The truth of claim (ii) is confirmed by the following derivations in $\kmsl$:
\begin{center}
\begin{tabular}{c c}
\AxiomC{$\seqempstr \sar w : \negnnf{\phi}, \bigwedge \bigvee \gi$}
\RightLabel{$\disr$}
\UnaryInfC{$\seqempstr \sar w : \negnnf{\phi} \lor \bigwedge \bigvee \gi$}
\RightLabel{=}
\dottedLine
\UnaryInfC{$\seqempstr \sar w : \phi \cimp \bigwedge \bigvee \gi$}
\DisplayProof

&

\AxiomC{$\seqempstr \sar w : \negnnf{\bigwedge \bigvee \gi}, w : \psi$}
\RightLabel{$\disr$}
\UnaryInfC{$\seqempstr \sar w :\negnnf{\bigwedge \bigvee \gi} \lor \psi$}
\RightLabel{=}
\dottedLine
\UnaryInfC{$\seqempstr \sar w : \bigwedge \bigvee \gi \cimp \psi$}
\DisplayProof
\end{tabular}
\end{center}

The truth of claim (i) follows from the fact that $\lit(\bigwedge \bigvee \gi) = \lit(\gi)$.
\end{proof}

\begin{corollary}
Every context-free grammar logic with converse has the effective Lyndon interpolation property.
\end{corollary}

\begin{proof} Follows from the soundness and completeness of $\kmsl$ (\cor ~\ref{cor:sound-comp-kmsl}) and \thm~\ref{thm:interpolation-theorem} above. Also, the property is effective because \thm~\ref{thm:interpolation-theorem} gives a procedure showing how to construct the interpolant of any valid implication.
\end{proof}


\chapter{Conclusion and Future Work}
\label{CPTR:Conclusion}

This thesis began by introducing labelled sequent calculi for a diverse class of logics, and then presented the method of refinement as a means of `simplifying' the systems. Since labelled calculi are straightforwardly obtainable from a logic's semantics, the method of refinement can be seen as a strategy for transforming the relational (i.e. Kripke) semantics of a modal or constructive logic into a proof system with an economical amount of sequential structure and which encodes semantic information within the functionality of logical rules. The method leverages results from the labelled paradigm to obtain labelled sequent calculi for modal and constructive logics via their semantics, and then performs structural rule elimination to reduce the syntactic structures utilized in the proof systems. As was seen, nested calculi regularly result as the output of this method suggesting that the method serves as a theoretical basis underpinning a variety of nested sequent systems. Nevertheless, the work in \sect~\ref{SECT:Refine-STIT} on refining labelled systems for deontic \stit logics showed that refinement can still be performed in the presence of certain structural rules which do not appear immediately eliminable via the method. The last technical chapter of the thesis (\cptr~\ref{CPTR:Applications}) harnessed refined labelled calculi for deontic \stit logics to provide proof-search algorithms and made use of the refined labelled calculi for context-free grammar logics with converse (along with the syntactic method of interpolation from~\cite{LyoTiuGorClo20}) to show that all such grammar logics possess the effective Lyndon interpolation property.


A promising avenue of future research concerns the development of a general theory of nested sequent systems for large classes of logics characterized by relational semantics. Not only would such a theory be useful in generating nested systems on demand for modal, constructive, and related logics---which have found significant applications in philosophy~\cite{BelPer90,BerLyo21}, legal theory~\cite{Bro11b,LorSar15}, artificial intelligence~\cite{McCHay69}, verification~\cite{ClaEmeSis86,SheAbb20}, and distributed computing~\cite{HalMos90}---but would allow for automated reasoning techniques to be applied to these logics by harnessing their nested systems (e.g.~\cite{FitKuz15,LyoBer19,LyoTiuGorClo20,TiuIanGor12}). Also, although refined calculi are often notational variants of nested systems, it seems plausible that through the implementation of additional methods, one can potentially simplify the refined systems even further to obtain calculi within the paradigm of hypersequents~\cite{Avr96}, linear nested sequents~\cite{Lel15}, or Gentzen-style sequents~\cite{PimLelRam19}.

Moreover, we would like to apply the refinement method to modal and constructive logics outside the classes discussed. As was hinted at and briefly discussed in \cptr~\ref{CPTR:Refinment-Constructive}, the use of grammar theoretic machinery in the refined labelled systems for first-order intuitionistic logics should allow for the systems to be straightforwardly transformed into systems for a diverse class of propositional and first-order (sub-)(bi-)intuitionistic logics---given that logical rules for the exclusion connective $\exc$ (the dual of intuitionistic implication) are added in the bi-intuitionistic cases. (NB. See~\cite{Cor87,Res94} for information on sub-intuitionistic logics and~\cite{GorPosTiu08,PinUus18,Rau80} for information on bi-intuitionistic logic.) Other fruitful candidates for applying refinement concern modal intuitionistic logics and bi-intuitionistic tense logics. Logics within both classes possess known nested sequent calculi (e.g. see~\cite{Str13} for modal intuitionistic nested systems and~\cite{GorPosTiu08} for a bi-intuitionistic tense nested system), and since such calculi are typically the output of refinement, it is reasonable to conjecture that labelled sequent systems for the logics are refinable. Moreover, Simpson~\cite{Sim94} defined labelled sequent systems for modal intuitionistic logics and defined propagation rules for such systems, which further strengthens the possibility that such systems can be refined due to the central role played by propagation rules. We are also interested in applying refinement to hybrid logics (e.g.~\cite[\sect~7.3]{BlaRijVen01}) and relevance logics (e.g.~\cite[\cptr~3]{Vig00}).
 
 


Another promising direction for future research regards refinement in the presence of structural rules that do not appear eliminable.  
 In \sect~\ref{SECT:Refine-STIT}, where the refinement of labelled systems for deontic \stit logics was discussed, the $\ioa$ and $\choicer$ rules were not eliminated, but allowed for the elimination of the other structural rules. Both rules had the property that the conclusion was free of active relational atoms. This suggests that rules of a similar shape ought to allow for other structural rules to be eliminated in a labelled sequent system, allowing for the calculus to be refined to a degree. We aim to investigate what sets of rules allow for refinement to be performed and the degree to which structural rules can be eliminated and replaced by propagation or reachability rules. Furthermore, as was seen in \sect~\ref{SECT:Refine-STIT}, the class $\{\mathsf{DS}^{0}_{n}\mathsf{L} \ | \ n \in \mathbb{N}\}$ of refined labelled calculi for deontic \stit logics only required labelled DAG derivations, showing that such calculi are close relatives of indexed-nested sequent systems (cf.~\cite{Fit15,MarStr17}). Therefore, the question arises: can refinement serve as a foundation for producing large classes of indexed-nested sequent calculi as well? As explained in~\cite{MarStr17}, it seems that some logics cannot be given a cut-free treatment in the nested sequent formalism (e.g. modal logics extended with Scott-Lemmon axioms~\cite{Lem77}), and so, indexed-nested sequents were introduced as a slight extension of nested sequents to capture such logics in a cut-free manner~\cite{Fit15}. Such systems employ sequents whose underlying data structure is a directed acyclic graph, demonstrating that such systems exhibit a higher degree of parsimony than their labelled counterparts (which use general graphs). As with nested sequent systems, this reduction in structure could compress the size of proofs (relative to labelled proofs of the same theorems), and lead to a savings in space as well as ease the confirmation of termination for associated proof-search algorithms. 
 
 
 
 
Last, it seems worthwhile to investigate if labelled sequent systems can be bypassed altogether and if refined labelled systems using propagation and reachability rules can be straightforwardly obtained from the semantics of a logic. If so, then perhaps general results could be obtained for modal, constructive, and related logics demonstrating that (indexed-)nested sequent calculi are immediately obtainable from a logic's relational semantics. 
 It is conceivable that such results could yield algorithms 
 which transform frame conditions or axioms into propagation and reachability rules, thus allowing for refined labelled, nested, or indexed-nested systems to be obtained in an automated fashion. Such work is reminiscent of the algorithms in~\cite{CiaGalTer08,CiaMafSpe13,CiaStr09,Lah13}, which transform axioms into structural rules, rather than propagation and reachability rules, in order to obtain classes of cut-free calculi for sizable classes of logics.










\backmatter

\chapter{Appendix}

\section{A \quad Fitting's Nested Calculi for First-Order Intuitionistic Logics}\label{app:fittings-nested-calculi}

In this section of the appendix we introduce Fitting's nested calculi for the first-order intuitionistic logics $\intfond$ and $\intfocd$ (see \dfn~\ref{def:axiomatization-IntFO}). We first define Fitting's notion of an \emph{available parameter}, which serves as a side condition on the $\existsr$ and $\alll$ rule in the nested calculus for first-order intuitionistic logic with non-constant domains; afterward, we define his nested calculi.

\begin{definition}[Available Parameter~\cite{Fit14}] Let $X\nbl Y \sar Z, \nbbl X_{1} \nbbr, \ldots, \nbbl X_{n}\nbbr \nbr$ be a nested sequent. If there exists a formula $A(\unda) \in Y,Z$, then the parameter $\unda$ is available in $Y \sar Z$ and in all boxed subsequents $X_{i}$ (with $i \in \{1, \ldots, n\}$).
\end{definition}


\begin{definition}[Fitting's Nested Calculi for $\intfond$ and $\intfocd$~\cite{Fit14}] The rules for Fitting's nested calculi are shown below. We define the nested calculus for $\intfond$ and $\intfocd$, and note that the side condition $\dag_{1}$ states that the parameter $\unda$ is either available or is an eigenvariable, and $\dag_{2}$ states that $\unda$ is an eigenvariable.

\begin{enumerate}


\item The nested calculus for first-order intuitionistic logic with non-constant domains ($\intfond$) consists of $\idfo$, $\conl$, $\conr$, $\disr$, $\disl$, $\negl$, $\negr$, $\impr$, $\impl$, $\lift$, $\existsl$, $\existsr$, $\alll$, and $\allrn$. 

\item The nested calculus first-order intuitionistic logic with constant domains ($\intfocd$) consists of $\idfo$, $\conl$, $\conr$, $\disr$, $\disl$, $\negl$, $\negr$, $\impr$, $\impl$, $\lift$, $\existsl$, $\existsr$, $\alll$, and $\allrc$ and omits the side condition $\dag_{1}$ on the $\existsr$ and $\alll$ rules.

\end{enumerate}

\begin{center}
\begin{tabular}{c c c}
\AxiomC{}
\RightLabel{$\idfo$}
\UnaryInfC{$X \nbl Y, p(\vec{\unda}) \sar p(\vec{\unda}), Z \nbr$}
\DisplayProof

&

\AxiomC{$X \nbl Y, A,B \sar Z \nbr $}
\RightLabel{$\conl$}
\UnaryInfC{$X \nbl Y, A \land B \sar Z \nbr$}
\DisplayProof

&

\AxiomC{$X \nbl Y, A(\unda/x) \sar Z  \nbr$}
\RightLabel{$\alll^{\dag_{1}}$}
\UnaryInfC{$X \nbl Y, \forall x A \sar Z \nbr$}
\DisplayProof
\end{tabular}
\end{center}

\begin{center}
\begin{tabular}{c c}
\AxiomC{$X \nbl Y, A \sar Z \nbr$}
\AxiomC{$X \nbl Y, B \sar Z \nbr$}
\RightLabel{$\disl$}
\BinaryInfC{$X \nbl Y, A \lor B \sar Z \nbr$}
\DisplayProof

&

\AxiomC{$X \nbl Y \sar A, Z \nbr$}
\AxiomC{$X \nbl Y \sar B, Z \nbr$}
\RightLabel{$\conr$}
\BinaryInfC{$X \nbl Y \sar A\land B, Z \nbr$}
\DisplayProof

\end{tabular}
\end{center}

\begin{center}
\begin{tabular}{c c c}
\AxiomC{$X \nbl Y \sar Z, \nbbl A \sar \seqempstr \nbbr  \nbr$}
\RightLabel{$\negr$}
\UnaryInfC{$X \nbl Y \sar Z, \neg A \nbr$}
\DisplayProof

&

\AxiomC{$X \nbl Y \sar A, Z \nbr$}
\RightLabel{$\negl$}
\UnaryInfC{$X \nbl Y, \neg A \sar Z \nbr$}
\DisplayProof

&

\AxiomC{$X \nbl Y \sar A,B, Z \nbr $}
\RightLabel{$\disr$}
\UnaryInfC{$X \nbl Y \sar A\lor B, Z \nbr$}
\DisplayProof
\end{tabular}
\end{center}

\begin{center}
\begin{tabular}{c c}
\AxiomC{$X \nbl Y \sar Z, \nbbl A \sar B\nbbr  \nbr$}
\RightLabel{$\impr$}
\UnaryInfC{$X \nbl Y \sar A \imp B, Z \nbr$}
\DisplayProof

&

\AxiomC{$X \nbl Y \sar A, Z \nbr$}
\AxiomC{$X \nbl Y, B \sar Z \nbr$}
\RightLabel{$\impl$}
\BinaryInfC{$X \nbl Y, A \imp B \sar Z \nbr$}
\DisplayProof
\end{tabular}
\end{center}

\begin{center}
\begin{tabular}{c c c}
\AxiomC{$X \nbl Y \sar A(\unda/x), Z  \nbr$}
\RightLabel{$\existsr^{\dag_{1}}$}
\UnaryInfC{$X \nbl Y \sar \exists x A, Z \nbr$}
\DisplayProof

&

\AxiomC{$X \nbl Y, A(\unda/x) \sar Z  \nbr$}
\RightLabel{$\existsl^{\dag_{2}}$}
\UnaryInfC{$X \nbl Y, \exists x A \sar Z \nbr$}
\DisplayProof

&

\AxiomC{$X \nbl Y \sar A(\unda/x), Z  \nbr$}
\RightLabel{$\allrc^{\dag_{2}}$}
\UnaryInfC{$X \nbl Y \sar \forall x A, Z \nbr$}
\DisplayProof
\end{tabular}
\end{center}

\begin{center}
\begin{tabular}{c c}
\AxiomC{$X \nbl Y \sar Z, \nbbl  \seqempstr \sar A(\unda/x)\nbbr   \nbr$}
\RightLabel{$\allrn^{\dag_{2}}$}
\UnaryInfC{$X \nbl Y \sar \forall x A, Z \nbr$}
\DisplayProof

&

\AxiomC{$X\nbl Y \sar Z, \nbbl Y', A \sar Z'\nbbr \nbr$}
\RightLabel{$\lift$}
\UnaryInfC{$X\nbl Y, A \sar Z, \nbbl Y' \sar Z'\nbbr \nbr$}
\DisplayProof
\end{tabular}
\end{center}
\end{definition}


\section{B \quad Classical Completeness of $\gtkms$ and $\gtdsn$}\label{app:classical-completeness}

In this appendix, we show that all instances of classical propositional tautologies in $\langkm{\albet}$ and $\langdsn$ are derivable in $\gtkms$ and $\gtdsn$, respectively. To prove this, we first show that the rules $\{\id, \disr, \conr\}$---all of which are contained in $\gtkms$ and $\gtdsn$---are complete relative to classical propositional logic. Second, we show how a proof of a classical propositional tautology can be proof-theoretically transformed into a proof of any \emph{instance} of the tautology in $\langkm{\albet}$ and $\langdsn$.

\begin{definition}[The Language $\lang$] We define the \emph{classical propositional language} $\lang$ as follows:
$$
\phi ::= p \ | \ \negnnf{p} \ | \ (\phi \lor \phi) \ | \ (\phi \land \phi)
$$
where $p \in \prop$. 
\end{definition}

\begin{definition}[Valuation, Tautology] We say that a function $\val : \lang \mapsto \{0,1\}$ mapping formulae of $\lang$ to the truth values in $\{0,1\}$ is a \emph{valuation} \ifandonlyif the following hold:
\begin{itemize}

\item[$\li$] $\val(p) \in \{0,1\}$, for each $p \in \prop$

\item[$\li$] $\val(\negnnf{p}) = 1 - \val(p)$, for each $p \in \prop$

\item[$\li$] $\val(\phi \lor \psi) = max\{\val(\phi),\val(\psi)\}$

\item[$\li$] $\val(\phi \land \psi) = min\{\val(\phi),\val(\psi)\}$

\end{itemize}
We say that a formula $\phi \in \lang$ is a \emph{classical propositional tautology} \ifandonlyif $\val(\phi) = 1$ for all valuations $\val$.
\end{definition}

\begin{definition}[Saturation] Let $\Lambda := \seqempstr \sar \Gamma$ be a labelled sequent. We say that $\Lambda$ is \emph{saturated} \ifandonlyif
\begin{itemize}

\item[$\li$] If $w : \phi \in \Gamma$, then $w : \negnnf{\phi} \not\in \Gamma$;

\item[$\li$] If $w : \phi \lor \psi \in \Gamma$, then $w : \phi, w : \psi \in \Gamma$;

\item[$\li$] If $w : \phi \land \psi \in \Gamma$, then either $w : \phi \in \Gamma$ or $w : \psi \in \Gamma$.

\end{itemize}
\end{definition}

\begin{algorithm}
\KwIn{A Labelled Sequent: $\seqempstr \sar \Upgamma$}
\KwOut{A Boolean: \texttt{True}, \texttt{False}}

\If{$w : p, w :\negnnf{p} \in \Upgamma$}
     {\Return \texttt{True};}

\If{$\seqempstr \sar \Upgamma$ is saturated}
     {\Return \texttt{False};}
     
\If{$w : \phi \lor \psi \in \Upgamma$, but either $w : \phi \not\in \Upgamma$ or $w : \psi \not\in \Upgamma$}
     {Let $\Upgamma' := w :\phi, w :\psi, \Upgamma$;\\
     \Return ProveCPL($\seqempstr \sar \Upgamma'$);}
     
\If{$w : \phi \land \psi \in \Upgamma$, but $w : \phi, w : \psi \not\in \Upgamma$}
{
    Let $\Upgamma_{1} := w :\phi, \Upgamma$;\\
    Let $\Upgamma_{2} := w :\psi, \Upgamma$;\\
    \If{$\provecpl$($\R \sar \Upgamma_{i}$) = \texttt{False} for some $i \in \{1,2\}$}
    {
    \Return \texttt{False};
    }\Else{
    \Return \texttt{True};
    }
}

\caption{$\provecpl$}\label{alg:ProveCL}
\end{algorithm}

\begin{lemma}[Termination of $\provecpl$]\label{lem:ProveCPL-termination}
Let $\phi \in \lang$. Then, $\provecpl$($w : \phi$) terminates.
\end{lemma}

\begin{proof} Let $\Lambda := \seqempstr \sar \Gamma$. We first define what it means for a labelled formula to be \emph{active} in $\seqempstr \sar \Gamma$ below:
\begin{itemize}

\item[$\li$] For all $p \in \prop$, neither $w : p$ nor $w : \negnnf{p}$ are \emph{active} in $\seqempstr \sar \Gamma$;

\item[$\li$] Consider a formula $\psi$ of the form $\chi \lor \xi$. Then, $w : \psi$ is \emph{active} in $\seqempstr \sar \Gamma$ \ifandonlyif $w : \psi \in \Gamma$, and either $w : \chi \not\in \Gamma$ or $w : \xi \not\in \Gamma$;

\item[$\li$] Consider a formula $\psi$ of the form $\chi \land \xi$. Then, $w : \psi$ is \emph{active} in $\seqempstr \sar \Gamma$ \ifandonlyif $w : \psi \in \Gamma$, and $w : \chi, w : \xi \not\in \Gamma$.
\end{itemize}
Next, we define the \emph{complexity} of a labelled sequent $\Lambda = \seqempstr \sar \Gamma$ (written $| \Lambda | = |\seqempstr \sar \Gamma|$) as follows:
$$
|\Lambda| := \sum_{\psi \in \mathrm{X}} \fcomp{\psi}
$$
with
$$
\mathrm{X} := \{\psi \ | \ w : \psi \in \Gamma \text{ and } w : \psi \text{ is active in $\seqempstr \sar \Gamma$.}\}
$$
The recursive call of line 7 ``deactivates'' a formula of the form $\psi \lor \chi$, and introduces $\psi$ and $\chi$ into the sequent, while the recursive call of line 11 ``deactivates'' a formula of the form $\psi \land \chi$ and either introduces $\psi$ or $\chi$ to a sequent. In the former case, $\fcomp{\psi} + \fcomp{\chi} < \fcomp{\psi \lor \chi}$, and in the latter case $\fcomp{\psi}, \fcomp{\chi} < \fcomp{\psi \land \chi}$, meaning that the complexity continually decreases---implying termination.
\end{proof}

\begin{lemma}[Correctness of $\provecpl$]\label{lem:ProveCPL-correctness}
Let $\thuesys$ be a \cfcst system with alphabet $\albet$, and $n, k \in \mathbb{N}$

(i) If $\provecpl$($w : \phi$) = \texttt{True}, then $w : \phi$ is provable in $\gtkms$ and $\gtdsn$.

(ii) If $\provecpl$($w : \phi$) = \texttt{False}, then there exists a valuation $\val$ such that $\val(\phi) = 0$.

\end{lemma}

\begin{proof} Claim (i) follows from the fact that each recursive call (lines 7 and 11) to $\provecpl$ corresponds to a bottom-up application of $\ctrr$, followed by an instance of $\disr$ or $\conr$. Since $\ctrr$ is hp-admissible in both $\gtkms$ and $\gtdsn$ (\lem~\ref{lem:struc-rules-admiss-kms} and \lem~\ref{lem:ctrr-admiss-dsn}, \resp), we know that if $\provecpl$($w : \phi$) = \texttt{True}, then a proof of $w : \phi$ was constructed in both $\gtkms$ and $\gtdsn$.

Let us now prove claim (ii). Since $\provecpl$($\seqempstr \sar w : \phi$) = \texttt{False}, we know that a saturated sequent $\seqempstr \sar \Gamma$ was generated. We will make use of $\seqempstr \sar \Gamma$ to construct a valuation $\val$ such that $\val(\phi) = 0$. We define $\val$ as follows: $\val(p) = 1$ \ifandonlyif $w : \negnnf{p} \in \Gamma$. Let us now prove the following claim by induction on the complexity of $\psi$: if $w : \psi \in \Gamma$, then $\val(\psi) = 0$.

\textit{Base case.} Let $p \in \prop$. Suppose that $\psi$ is of the form $p$, and assume that $w : p \in \Gamma$. Since $\Gamma$ is saturated, it follows that $w : \negnnf{p} \not\in \Gamma$. Hence, by the definition of $\val$, we know that $\val(p) = 0$. If $\psi$ is of the form $\negnnf{p}$, then the claim follows directly from the assumption that $w : \negnnf{p} \in \Gamma$ and the definition of $\val$.

\textit{Inductive step.} We consider the case where $\psi$ is of the form $\chi \lor \xi$ and the case were $\psi$ is of the form $\chi \land \xi$.

\emph{$\lor$}. Suppose that $w : \psi = w : \chi \lor \xi \in \Gamma$. Since $\seqempstr \sar \Gamma$ is saturated, we know that $w : \chi, w : \xi \in \Gamma$ by the recursive call on line 7. By \ih it follows that $\val(\chi) = \val(\xi) = 0$, implying that $\val(\psi) = 0$.

\emph{$\land$}. Suppose that $w : \psi = w : \chi \land \xi \in \Gamma$. Since $\seqempstr \sar \Gamma$ is saturated, we know that either $w : \chi \in \Gamma$ or $w : \xi \in \Gamma$ by the recursive call on line 11. By \ih it follows that either $\val(\chi) = 0$ or $\val(\xi) = 0$, implying that $\val(\psi) = 0$.
\end{proof}

\begin{customlem}{\ref{lem:classical-completeness-kms}}
All instances of classical propositional tautologies in $\langkm{\albet}$ are derivable in $\gtkms$.
\end{customlem}

\begin{proof} Let $\phi$ be a classical propositional tautology. Then, $\val(\phi) = 1$ for all valuations $\val$, implying that $\provecpl$($w : \phi$) $\neq$ \texttt{False} by \lem~\ref{lem:ProveCPL-correctness} above. By \lem~\ref{lem:ProveCPL-termination}, this implies that $\provecpl$($w : \phi$) = \texttt{True}, which implies that $\seqempstr \sar w : \phi$ has a proof in $\gtkms$ by \lem~\ref{lem:ProveCPL-correctness}. Let us call this proof $\Pi$. Observe that $\Pi$ is of the following form:
\begin{center}
\begin{tabular}{c c c}
$\Pi$

&

$= \Bigg \{$

&

\AxiomC{}
\RightLabel{$\id$}
\UnaryInfC{$\sar w : p_{0}, w : \negnnf{p_{0}}, \Gamma_{0}$}

\AxiomC{$\cdots$}

\AxiomC{}
\RightLabel{$\id$}
\UnaryInfC{$\sar w : p_{n}, w : \negnnf{p_{n}}, \Gamma_{n}$}

\noLine

\TrinaryInfC{$\vdots$}
\noLine
\UnaryInfC{$\seqempstr \sar w : \phi$}
\DisplayProof
\end{tabular}
\end{center}
In order to show that all \emph{instances} of classical propositional tautologies in $\langkm{\albet}$ are derivable, we have to show that for any substitution $\sub = (\psi_{0}/p_{0}, \ldots, \psi_{n}/p_{n})$ of formulae $\psi_{i} \in \langkm{\albet}$ for the propositional variables $p_{i}$ (with $0 \leq i \leq n$), $w : \phi\sub$ is derivable in $\gtkms$. To show this, we first apply the substitution $\sub$ to the entire derivation $\Pi$, i.e. to every labelled formula of every sequent in $\Pi$, yielding the following derivation $\Pi\sub$:
\begin{center}
\begin{tabular}{c c c}
$\Pi\sub$

&

$= \Bigg \{$

&

\AxiomC{}
\RightLabel{$\id$}
\UnaryInfC{$\sar w : \psi_{0}, w : \negnnf{\psi_{0}}, \Gamma_{0}\sub$}

\AxiomC{$\cdots$}

\AxiomC{}
\RightLabel{$\id$}
\UnaryInfC{$\sar w : \psi_{n}, w : \negnnf{\psi_{n}}, \Gamma_{n}\sub$}

\noLine

\TrinaryInfC{$\vdots$}
\noLine
\UnaryInfC{$\sar w : \phi\sub$}
\DisplayProof
\end{tabular}
\end{center}
Next, we think of each instance of $\id$ as an \emph{interface}, and employ \lem~\ref{lem:general-id-kms}, which implies that every sequent of the form  $\sar w : \psi_{i}, w : \negnnf{\psi_{i}}, \Gamma_{i}\sub$ (for $0 \leq i \leq n$) is derivable in $\gtkms$. Let us call the derivation of each such sequent $\Pi_{i}$. By attaching these derivations to the leaves, as shown below, we obtain a proof $\Pi'$ of the desired conclusion:
\begin{center}
\begin{tabular}{c c c}
$\Pi'$

&

$= \Bigg \{$

&

\AxiomC{$\Pi_{1}$}
\noLine
\UnaryInfC{$\sar w : \psi_{0}, w : \negnnf{\psi_{0}}, \Gamma_{0}\sub$}

\AxiomC{$\cdots$}

\AxiomC{$\Pi_{n}$}
\noLine
\UnaryInfC{$\sar w : \psi_{n}, w : \negnnf{\psi_{n}}, \Gamma_{n}\sub$}

\noLine

\TrinaryInfC{$\vdots$}
\noLine
\UnaryInfC{$\sar w : \phi\sub$}
\DisplayProof
\end{tabular}
\end{center}
\end{proof}

\begin{customlem}{\ref{lem:classical-completeness-dsn}}
All instances of classical propositional tautologies in $\langdsn$ are derivable in $\gtdsn$.
\end{customlem}

\begin{proof} Similar to the proof of \lem~\ref{lem:classical-completeness-kms}.
\end{proof}

\section{C \quad Additional Proofs}\label{app:proofs}

\begin{customlem}{\ref{lem:instance-of}} \ 

(i) The rule $\idfo$ is an instance of $\idfona$ and $\idfoca$.

(ii) The $\impl$ rule is an instance of the $\primp$ rule.

(iii) The $\existsr$ rule is an instance of the rules $\existsrna$ and $\existsrca$.

(iv) The $\alll$ rule is an instance of the rules $\alllna$ and $\alllca$.

\end{customlem}

\begin{proof} We prove claims (ii)--(iv) below:

(ii) In the $\primp$ inference below, observe that due to the $w \leq u$ relational atom occurring in the premises, we know that the propagation path $\ppath(w,u) := w, a, u$ occurs in the propagation graphs of both premises. Since $a \in \thuesysilang{a}$, and because $\stra_{\ppath}(w,u) = a$, it follows that $\stra_{\ppath}(w,u) \in \thuesysilang{a}$ showing that the side condition is satisfied, and therefore, the $\primp$ rule may be applied to the premises to derive the desired conclusion.
\begin{center}
\AxiomC{$\rel,w \leq u, \Gamma, w :\phi \imp \psi \sar \Delta, u :\phi$}
\AxiomC{$\rel,w \leq u, \Gamma, w :\phi \imp \psi, u :\psi \sar \Delta$}
\RightLabel{$\primp$}
\BinaryInfC{$\rel,w \leq u, \Gamma, w :\phi \imp \psi \sar \Delta$}
\DisplayProof 
\end{center}

(iii) We prove that $\existsr$ is an instance of $\existsrca$ since showing that $\existsr$ is an instance of $\existsrna$ is similar. By \dfn~\ref{def:S4-S5-FOInt}, we know that $\empstr \in \thuesysiilang{a}$. We also know that $\stra_{\emppath}(w,w) = \empstr$ by \dfn~\ref{def:propagation-path-kms}, implying that $\stra_{\emppath}(w,w) \in \thuesysiilang{a}$. Consequently, the side condition of $\existsrca$ is satisfied, and the following inference may be performed:
\begin{center}
\AxiomC{$\rel, \unda \in D_{w}, \Gamma \sar w: \phi(\unda/x), w: \exists x \phi, \Delta$}
\RightLabel{$\existsrca$}
\UnaryInfC{$\rel, \unda \in D_{w}, \Gamma \sar w: \exists x \phi, \Delta$}
\DisplayProof
\end{center}

(iv) We show that the $\alll$ rule is an instance of $\alllca$ since the proof for $\alllna$ is similar. By \dfn~\ref{def:S4-S5-FOInt}, we know that $\empstr \in \thuesysiilang{a}$, and since there is a propagation path $\emppath(u,u) = u$ from $u$ to itself with $\stra_{\emppath}(u,u) = \empstr$, it follows that $\stra_{\emppath}(u,u) \in \thuesysiilang{a}$. Furthermore, by \dfn~\ref{def:derivation-relation-language-kms}, we know that $a \dto^{*}_{\thuesysii} a$ holds, implying that $a \in \thuesysilang{a}$. Since there is a propagation path $\ppath(w,u) = w,a,u$ and $\stra_{\ppath}(w,u) = a$, we have that  $\stra_{\ppath}(w,u) \in \thuesysilang{a}$. Therefore, the side condition of $\alllca$ holds and the rule can simulate $\alll$ as shown below. 

\begin{center}
\AxiomC{$\rel, w \leq u, \unda \in D_{u}, \Gamma, u : \phi(\unda/x), w : \forall x \phi \sar \Delta$}
\RightLabel{$\alllca$}
\UnaryInfC{$\rel, w \leq u, \unda \in D_{u}, \Gamma, w : \forall x \phi \sar \Delta$}
\DisplayProof
\end{center}
\end{proof}

\begin{customlem}{\ref{lem:trans-elim-FOInt}} \ 

(i) The rule $\trans$ is eliminable in $\gtintfond + \rulesna - \{\refl, \nd, \ned\}$.

(ii) The rule $\trans$ is eliminable in $\gtintfocd + \rulesca - \{\refl, \nd, \cd, \ned\}$. 
\end{customlem}

\begin{proof} We prove the result by induction on the height of the given derivation and only show claim (ii) as it subsumes claim (i).

\textit{Base case.} The $\botl$ case is simple to verify and due to \lem~\ref{lem:instance-of}, we do not need to consider the $\idfo$ case, so we only consider the $\idfoca$ case below: 
\begin{flushleft}
\begin{tabular}{c c}
\AxiomC{}
\RightLabel{$\idfoca$}
\UnaryInfC{$\rel, w \leq u, u \leq v, w \leq v, \unda_{1} \in D_{u_{1}}, \ldots, \unda_{n} \in D_{u_{n}}, \Gamma, w : p(\vec{\unda}) \Rightarrow u : p(\vec{\unda}), \Delta$}
\RightLabel{$\trans$}
\UnaryInfC{$\rel, w \leq u, u \leq v, \unda_{1} \in D_{u_{1}}, \ldots, \unda_{n} \in D_{u_{n}}, \Gamma, w : p(\vec{\unda}) \Rightarrow u : p(\vec{\unda}), \Delta$}
\DisplayProof

&

$\leadsto$
\end{tabular}
\end{flushleft}
\begin{flushright}
\AxiomC{}
\RightLabel{$\idfoca$}
\UnaryInfC{$\rel, w \leq u, u \leq v, \unda_{1} \in D_{u_{1}}, \ldots, \unda_{n} \in D_{u_{n}}, \Gamma, w : p(\vec{\unda}) \Rightarrow u : p(\vec{\unda}), \Delta$}
\DisplayProof
\end{flushright}
By the side condition on $\idfoca$, we know that there exist propagation paths $\pi_{i}(u_{i},w)$ for each $i \in \{1, \ldots, n\}$ and a propagation path $\pi(w,u)$ in the premise of the top derivation such that $\stra_{\ppath_{i}}(u_{i},w)  \in \thuesysiilang{a}$ and $\stra_{\ppath}(w,u)  \in \thuesysilang{a}$. If the relational atom $w \leq v$ active in the $\trans$ inference occurs along a propagation path $\ppath_{i}(u_{i},w)$ or along the propagation path $\ppath(w,u)$, then by replacing each occurrence of $w, a, v$ and $v, \conv{a}, w$ in $\ppath_{i}(u_{i},w)$ and $\ppath(w,u)$ with $w, a, u, a, v$ and $v, \conv{a}, u, \conv{a}, w$ (respectively), we obtain new propagation paths $\ppath_{i}'(u_{i},w)$ and $\ppath'(w,u)$ that do not go rely on the relational atom $w \leq v$. We now argue that $\stra_{\ppath_{i}'}(u_{i},w)  \in \thuesysiilang{a}$ and $\stra_{\ppath'}(w,u)  \in \thuesysilang{a}$. The derivations below show that $a \dto^{*}_{\thuesysii} a \cate a$ and $\conv{a} \dto^{*}_{\thuesysii} \conv{a} \cate \conv{a}$, and we note that $a \pto a \cate a, \conv{a} \pto \conv{a} \cate \conv{a} \in \thuesysi$ (by \dfn~\ref{def:S4-S5-FOInt}).
$$
a \pto \conv{a} \cate a \pto a \cate \conv{a} \cate a \pto a \cate a \qquad \conv{a} \pto a \cate \conv{a} \pto \conv{a} \cate a \cate \conv{a} \pto \conv{a} \cate \conv{a}
$$
Since $\stra_{\ppath_{i}}(u_{i},w)  \in \thuesysiilang{a}$, $\stra_{\ppath}(w,u)  \in \thuesysilang{a}$, $a \dto^{*}_{\thuesysii} a \cate a$, $\conv{a} \dto^{*}_{\thuesysii} \conv{a} \cate \conv{a}$, and $a \pto a \cate a, \conv{a} \pto \conv{a} \cate \conv{a} \in \thuesysi$, by applying $a \dto^{*}_{\thuesysii} a \cate a$ and $\conv{a} \dto^{*}_{\thuesysii} \conv{a} \cate \conv{a}$ to $\stra_{\ppath_{i}}(u_{i},w)$, and $a \pto a \cate a$ and  $\conv{a} \pto \conv{a} \cate \conv{a}$ to $\stra_{\ppath}(w,u)$ for each occurrence of $a$ and $\conv{a}$ obtained from the path $w, a, v$ and $v, \conv{a}, w$ (respectively), we obtain $\stra_{\ppath_{i}'}(u_{i},w)$ and $\stra_{\ppath'}(w,u)$, respectively; the former of which is in $\thuesysiilang{a}$ and the latter of which is in $\thuesysilang{a}$. The confirms that the side condition of $\idfoca$ holds, showing that the second inference above is allowed.

\textit{Inductive step.} By \lem~\ref{lem:instance-of} we need not consider the $\impl$, $\existsr$, or $\alll$ cases. With the exception of $\primp$, $\existsrca$, and $\alllca$ cases, all remaining cases are resolved by invoking \ih and then applying the corresponding rule. We consider the $\primp$ and $\alllca$ cases; the $\existsrca$ case is similar.

$\primp$. Due to the application of $\primp$, we know that a propagation path $\ppath(w_{1},w_{2}) $ exists in the premises $\Lambda_{1}$ and $\Lambda_{2}$ (shown below) such that $\stra_{\ppath}(w_{1},w_{2}) \in \thuesysilang{a}$. If the active relational atom $w \leq v$ occurs along the propagation path $\ppath(w_{1},w_{2})$, then by replacing each occurrence of $w, a, v$ and $v, \conv{a}, w$ (obtained from the relational atom $w \leq v$) in $\ppath(w_{1},w_{2})$ with $w, a, u, a, v$ and $v, \conv{a}, u, \conv{a}, w$ (respectively), we obtain a new propagation path $\ppath'(w_{1},w_{2})$ that does not rely on $w \leq v$ (but instead, on $w \leq u, u \leq v$). Recall that $a \pto a \cate a, \conv{a} \pto \conv{a} \cate \conv{a} \in \thuesysi$, and since $\stra_{\ppath}(w_{1},w_{2}) \in \thuesysilang{a}$, we have that $\stra_{\ppath'}(w_{1},w_{2}) \in \thuesysilang{a}$ as we can replace each occurrence of $a$ that corresponds to the path $w, a, v$ and $\conv{a}$ that corresponds to the path $v, \conv{a}, w$ in $\stra_{\ppath}(w_{1},w_{2})$ with $a \cate a$ and $\conv{a} \cate \conv{a}$ by means of the production rules $a \pto a \cate a$ and $\conv{a} \pto \conv{a} \cate \conv{a}$, respectively. This shows that the side condition continues to hold if $\trans$ is applied first, and so, the rules may be permuted.
\begin{flushleft}
$\Lambda_{1} := \rel,w \leq u, u \leq v, w \leq v, \Gamma, w_{1} :\phi \imp \psi \sar \Delta, w_{2} :\phi$
\end{flushleft}
\begin{flushleft}
$\Lambda_{2} := \rel,w \leq u, u \leq v, w \leq v, \Gamma, w_{1} :\phi \imp \psi, w_{2} :\psi \sar \Delta$
\end{flushleft}
\begin{flushleft}
\begin{tabular}{c c}
\AxiomC{$\Lambda_{1}$}
\AxiomC{$\Lambda_{2}$}
\RightLabel{$\primp$}
\BinaryInfC{$\rel,w \leq u, u \leq v, w \leq v, \Gamma, w_{1} :\phi \imp \psi \sar \Delta$}
\RightLabel{$\trans$}
\UnaryInfC{$\rel, w \leq u, u \leq v, \Gamma, w_{1} :\phi \imp \psi \sar \Delta$}
\DisplayProof

&

$\leadsto$
\end{tabular}
\end{flushleft}
\begin{flushright}
\AxiomC{$\Lambda_{1}$}
\RightLabel{$\trans$}
\UnaryInfC{$\Lambda_{1}'$}

\AxiomC{$\Lambda_{2}$}
\RightLabel{$\trans$}
\UnaryInfC{$\Lambda_{2}'$}

\RightLabel{$\primp$}
\BinaryInfC{$\rel,w \leq u, u \leq v,\Gamma, w_{1} :\phi \imp \psi \sar \Delta$}
\DisplayProof
\end{flushright}
\begin{flushright}
$\Lambda_{1}' := \rel,w \leq u, u \leq v, \Gamma, w_{1} :\phi \imp \psi \sar \Delta, w_{2} :\phi$
\end{flushright}
\begin{flushright}
$\Lambda_{2}' := \rel,w \leq u, u \leq v, \Gamma, w_{1} :\phi \imp \psi, w_{2} :\psi \sar \Delta$
\end{flushright}

$\alllca$. By the application of $\alllca$, we know that propagation paths $\ppath_{1}(w_{1},w_{3})$ and $\ppath_{1}(w_{2},w_{3})$ exist in the premise of the top left derivation below such that $\stra_{\ppath_{1}}(w_{1},w_{3}) \in \thuesysiilang{a}$ and $\stra_{\ppath_{2}}(w_{2},w_{3}) \in \thuesysilang{a}$. If the relational atom $w \leq v$ occurs along the propagation path $\ppath_{1}(w_{1},w_{3})$ or $\ppath_{2}(w_{2},w_{3})$, then by replacing each path $w, a, v$ and $v,\conv{a}, w$ (obtained from the relational atom $w \leq v$) with $w, a, u, a, v$ and $v, \conv{a}, u, \conv{a}, w$ (respectively) in $\ppath_{1}(w_{1},w_{3})$ and $\ppath_{2}(w_{2},w_{3})$, we obtain new propagation paths $\ppath_{1}'(w_{1},w_{3})$ and $\ppath_{2}'(w_{2},w_{3})$ that do not rely on the relational atom $w \leq v$. Since $\stra_{\ppath_{1}}(w_{1},w_{3}) \in \thuesysiilang{a}$, $\stra_{\ppath_{2}}(w_{2},w_{3}) \in \thuesysilang{a}$, $a \dto^{*}_{\thuesysii} a \cate a$ and $\conv{a} \dto^{*}_{\thuesysii} \conv{a} \cate \conv{a}$ (as explained in the base case), and $a \pto a \cate a, \conv{a} \pto \conv{a} \cate \conv{a} \in \thuesysi$, we know that $\stra_{\ppath_{1}'}(w_{1},w_{3}) \in \thuesysiilang{a}$ and $\stra_{\ppath_{2}'}(w_{2},w_{3}) \in \thuesysilang{a}$, as we can replace each occurrence of $a$ arising from the path $w, a, u$ with $a \cate a$ and each occurrence of $\conv{a}$ arising from the path $u, \conv{a}, w$ with $\conv{a} \cate \conv{a}$ in $\ppath_{1}(w_{1},w_{3})$ and $\ppath_{2}(w_{2},w_{3})$. Therefore, if we apply the $\trans$ rule first, the side condition of $\alllca$ still holds, and so, we may permute the two rules as shown below.
\begin{flushleft}
\begin{tabular}{c c}
\AxiomC{$\rel, w \leq u, u \leq v, w \leq v, \unda \in D_{w_{1}}, w_{3} : \phi(\unda/x), w_{2} : \forall x \phi, \Gamma \Rightarrow \Delta$}
\RightLabel{$\alllca$}
\UnaryInfC{$\rel, w \leq u, u \leq v, w \leq v, \unda \in D_{w_{1}}, w_{2} : \forall x \phi, \Gamma \Rightarrow \Delta$}
\RightLabel{$\trans$}
\UnaryInfC{$\rel, w \leq u, u \leq v, \unda \in D_{w_{1}}, w_{2} : \forall x \phi, \Gamma \Rightarrow \Delta$}
\DisplayProof

&

$\leadsto$
\end{tabular}
\end{flushleft}
\begin{flushright}
\AxiomC{$\rel, w \leq u, u \leq v, w \leq v, \unda \in D_{w_{1}}, w_{3} : \phi(\unda/x), w_{2} : \forall x \phi, \Gamma \Rightarrow \Delta$}
\RightLabel{$\trans$}
\UnaryInfC{$\rel, w \leq u, u \leq v, \unda \in D_{w_{1}}, w_{3} : \phi(\unda/x), w_{2} : \forall x \phi, \Gamma \Rightarrow \Delta$}
\RightLabel{$\alllca$}
\UnaryInfC{$\rel, w \leq u, u \leq v, \unda \in D_{w_{1}}, w_{2} : \forall x \phi, \Gamma \Rightarrow \Delta$}
\DisplayProof
\end{flushright}
\end{proof}

\begin{customlem}{\ref{lem:refined-to-quasi-refined-FOInt}} \ 

(i) Every proof of a labelled sequent $\Lambda$ in $\intfondl$ can be algorithmically transformed into a proof of $\domcl(\Lambda)$ in $\intfondlq$.

(ii) Every proof of a labelled sequent $\Lambda$ in $\intfocdl$ can be algorithmically transformed into a proof of $\domcl(\Lambda)$ in $\intfocdlq$.
\end{customlem}

\begin{proof} We prove the result by induction on the height of the given derivation and consider (ii) as (i) is similar.

\textit{Base case.} The $\botl$ case is easy to verify. For the $\idfonc$ inference, due to the application of $\domcl$, we know that domain atoms $\vec{\unda} \in D_{w}$ exist in the output labelled sequent, ensuring that it is in fact an instance of $\idfoca$. 
\begin{center}
\begin{tabular}{c c c}
\AxiomC{}
\RightLabel{$\idfonc$}
\UnaryInfC{$\rel, \Gamma, w : p(\vec{\unda}) \Rightarrow u : p(\vec{\unda}), \Delta$}
\DisplayProof
&

$\leadsto$

&

\AxiomC{}
\RightLabel{$\idfoca$}
\UnaryInfC{$\domcl(\rel, \Gamma, w : p(\vec{\unda}) \Rightarrow u : p(\vec{\unda}), \Delta)$}
\DisplayProof
\end{tabular}
\end{center}

\textit{Inductive step.} We show the $\conr$, $\impr$, $\existsrc$, $\existslc$, $\alllc$, and $\allrc$ cases as all other cases are simple or similar. We consider each of the aforementioned cases below. 

$\conr$. Let our given $\conr$ inference be as shown below top-left. Also, let $\vec{\unda} := \unda_{1}, \ldots, \unda_{n}$ be all free parameters in $\phi, \Gamma \restriction w, \Delta \restriction w$ not occurring in $\psi, \Gamma \restriction w, \Delta \restriction w$, and let $\vec{\undb} := \undb_{1}, \ldots, \undb_{k}$ be all free parameters in $\psi, \Gamma \restriction w, \Delta \restriction w$ not occurring in $\phi, \Gamma \restriction w, \Delta \restriction w$. To obtain the desired conclusion, we first apply the admissibility of $\wk$ (Thm.~\ref{thm:properties-quasi-refined-FOInt}) in order to ensure that the contexts of the premises match, followed by an application of the $\conr$ rule.

\begin{flushleft}
\begin{tabular}{c c}
\AxiomC{$\rel, \Gamma \sar w :\phi, \Delta$}
\AxiomC{$\rel, \Gamma \sar w :\psi, \Delta$}
\RightLabel{$\conr$}
\BinaryInfC{$\rel, \Gamma \sar w :\phi \wedge \psi, \Delta$}
\DisplayProof

&

$\leadsto$
\end{tabular}
\end{flushleft}
\begin{flushright}
\AxiomC{}
\RightLabel{IH}
\dashedLine
\UnaryInfC{$\domcl(\rel, \Gamma \sar w :\phi, \Delta)$}
\RightLabel{=}
\dottedLine
\UnaryInfC{$\rel', \vec{\unda} \in D_{w}, \Gamma \sar w : \phi, \Delta$}
\RightLabel{$\wk$}
\dashedLine
\UnaryInfC{$\rel', \vec{\unda} \in D_{w}, \vec{\undb} \in D_{w}, \Gamma \sar w : \phi, \Delta$}

\AxiomC{}
\RightLabel{IH}
\dashedLine
\UnaryInfC{$\domcl(\rel, \Gamma \sar w :\psi, \Delta)$}
\RightLabel{=}
\dottedLine
\UnaryInfC{$\rel', \vec{\undb} \in D_{w}, \Gamma \sar w : \psi, \Delta$}
\RightLabel{$\wk$}
\dashedLine
\UnaryInfC{$\rel', \vec{\unda} \in D_{w}, \vec{\undb} \in D_{w}, \Gamma \sar w : \psi, \Delta$}

\RightLabel{$\conr$}
\BinaryInfC{$\rel', \vec{\unda} \in D_{w}, \vec{\undb} \in D_{w}, \Gamma \sar w : \phi \land \psi, \Delta$}
\RightLabel{=}
\dottedLine
\UnaryInfC{$\domcl(\rel, \Gamma \sar w :\phi \wedge \psi, \Delta)$}
\DisplayProof
\end{flushright}


$\impr$. There are two cases to consider: either (i) the auxiliary formulae $u : \phi$ and $u : \psi$ contain free occurrences of parameters $\vec{\unda} = \unda_{1}, \ldots, \unda_{n}$, or (ii) they do not. Case (ii) is easily resolved by applying \ih and then the $\impr$ rule, so we focus on case (i) which is shown below. Let $\rel', \vec{\unda} \in D_{u}$ be the multiset of relational atoms occurring within the premise of the bottom right inference below (that is, the multiset of relational atoms occurring in the labelled sequent obtained by applying $\domcl$ to the premise of the top left inference). To derive the desired conclusion, we first apply admissibility of $\wk$ (\lem~\ref{thm:properties-quasi-refined-FOInt}) to add in the domain atoms $\vec{\unda}' \in D_{w} := \unda_{i_{1}} \in D_{w}, \ldots, \unda_{i_{k}} \in D_{w}$ such that $\unda_{i_{j}} \in D_{w} \not\in \rel'$ for $i_{j} \in \{1, \ldots, n\}$ and $j \in \{1, \ldots, k\}$, i.e. for each parameter $\unda_{i_{j}}$ in $\vec{\unda}$ we add a new domain atom $\unda_{i_{j}} \in D_{w}$ given that it does not already occur in $\rel'$. After $\wk$ is applied, we apply the admissible $\nd$ rule (\thm~\ref{thm:admissible-rules-G3-Calc}) $n$ times to delete all domain atoms $\vec{\unda} \in D_{u}$, thus making $u$ an eigenvarible, and allowing for $\impr$ to be applied, giving the desired result.

\begin{flushleft}
\begin{tabular}{c c}
\AxiomC{$\rel, w \leq u, \Gamma, u :\phi \sar u :\psi, \Delta$}
\RightLabel{$\impr$}
\UnaryInfC{$\rel, \Gamma \sar w :\phi \imp \psi, \Delta$}
\DisplayProof

&

$\leadsto$
\end{tabular}
\end{flushleft}
\begin{flushright}
\AxiomC{}
\RightLabel{\ih}
\dashedLine
\UnaryInfC{$\domcl(\rel, w \leq u, \Gamma, u :\phi \sar u :\psi, \Delta)$}
\RightLabel{=}
\dottedLine
\UnaryInfC{$\rel', \vec{\unda} \in D_{u}, w \leq u, \Gamma, u :\phi \sar u :\psi, \Delta$}
\RightLabel{$\wk$}
\dashedLine
\UnaryInfC{$\rel', \vec{\unda}' \in D_{w}, \vec{\unda} \in D_{u}, w \leq u, \Gamma, u :\phi \sar u :\psi, \Delta$}
\RightLabel{$\nd \times n$}
\dashedLine
\UnaryInfC{$\rel', \vec{\unda}' \in D_{w}, w \leq u, \Gamma, u :\phi \sar u :\psi, \Delta$}
\RightLabel{$\impr$}
\UnaryInfC{$\rel', \vec{\unda}' \in D_{w}, \Gamma \sar w :\phi \imp \psi, \Delta$}
\RightLabel{=}
\dottedLine
\UnaryInfC{$\domcl(\rel, \Gamma \sar w :\phi \imp \psi, \Delta)$}
\DisplayProof

\end{flushright}

$\existsrc$-(i). In case (i), we assume that $\unda$ is not an eigenvariable. By the side condition imposed on $\existsrc$ then, we know there exists a labelled formula $u : \psi(\unda) \in \Gamma, \Delta$ and propagation path $\ppath(u,w)$ such that $\stra_{\ppath}(u,w) \in \thuesysiilang{a}$. Therefore, after applying \ih which applies $\domcl$, there will be a domain atom of the form $\unda \in D_{u}$ occurring in the output labelled sequent. We may then apply the $\existsrca$ rule since the side condition is satisfied, giving the desired conclusion.
\begin{flushleft}
\begin{tabular}{c c}
\AxiomC{$\rel, \Gamma \Rightarrow \Delta, w: \phi(\unda/x), w: \exists x \phi$}
\RightLabel{$\existsrc$}
\UnaryInfC{$\rel, \Gamma \Rightarrow \Delta, w: \exists x \phi$}
\DisplayProof

&

$\leadsto$
\end{tabular}
\end{flushleft}
\begin{flushright}
\AxiomC{}
\RightLabel{\ih}
\dashedLine
\UnaryInfC{$\domcl(\rel, \Gamma \Rightarrow \Delta, w: \phi(\unda/x), w: \exists x \phi)$}
\RightLabel{$\existsrca$}
\dashedLine
\UnaryInfC{$\domcl(\rel, \Gamma \Rightarrow \Delta, w: \exists x \phi)$}
\DisplayProof
\end{flushright}

$\existsrc$-(ii). In case (ii), we assume that $\unda$ is an eigenvariable. Then, after applying \ih there will be a \emph{single} domain atom of the form $\unda \in D_{w}$ (and no other domain atoms will contain the parameter $\unda$). This follows from the definition of $\domcl$ and the fact that $ w: \phi(\unda/x)$ is the only labelled formula containing $\unda$ (by the eigenvariable condition). We then apply the $\existsrcia$ rule to obtain the desired conclusion. 
\begin{flushleft}
\begin{tabular}{c c}
\AxiomC{$\rel, \Gamma \Rightarrow \Delta, w: \phi(\unda/x), w: \exists x \phi$}
\RightLabel{$\existsrc$}
\UnaryInfC{$\rel, \Gamma \Rightarrow \Delta, w: \exists x \phi$}
\DisplayProof

&

$\leadsto$
\end{tabular}
\end{flushleft}
\begin{flushright}
\AxiomC{}
\RightLabel{\ih}
\dashedLine
\UnaryInfC{$\domcl(\rel, \Gamma \Rightarrow \Delta, w: \phi(\unda/x), w: \exists x \phi)$}
\RightLabel{$\existsrcia$}
\dashedLine
\UnaryInfC{$\domcl(\rel, \Gamma \Rightarrow \Delta, w: \exists x \phi)$}
\DisplayProof
\end{flushright}

$\existslc$. After invoking \ih we know that a \emph{single} domain atom of the form $\unda \in D_{w}$ occurs within the output labelled sequent (and no other domain atoms contain the parameter $\unda$). This fact follows from the fact that $\unda$ is an eigenvariable, and so, by the definition of $\domcl$ only a single domain atom uniquely containing the parameter $\unda$ will be introduced on the basis of the labelled formula $w: \phi(\unda/x)$. Hence, we may directly apply $\existsl$ after applying the inductive hypothesis.
\begin{center}
\begin{tabular}{c c c}
\AxiomC{$\rel, \Gamma, w: \phi(\unda/x) \sar \Delta$}
\RightLabel{$\existslc$}
\UnaryInfC{$\rel, \Gamma, w: \exists x \phi \sar \Delta$}
\DisplayProof

&

$\leadsto$

&

\AxiomC{}
\RightLabel{\ih}
\dashedLine
\UnaryInfC{$\domcl(\rel, \Gamma, w: \phi(\unda/x) \sar \Delta)$}
\RightLabel{$\existsl$}
\UnaryInfC{$\domcl(\rel, \Gamma, w: \exists x \phi \sar \Delta)$}
\DisplayProof
\end{tabular}
\end{center}

$\alllc$-(i). In case (i), we assume that $\unda$ is not an eigenvariable. By the side condition imposed on $\alllc$, we know that $\unda$ is $\thuesysii$-available for $v$ and that there exists a propagation path $\ppath(w,v)$ such that $\stra_{\ppath}(w,v) \in \thuesysilang{a}$. The former implies that there exists a labelled formula $u : \psi(\unda) \in \Gamma, \Delta$ and propagation path $\ppath'(u,v)$ such that $\stra_{\ppath'}(u,v) \in \thuesysiilang{a}$. After invoking IH, a domain atom of the form $\unda \in D_{u}$ will occur in the output labelled sequent (since applying \ih applies $\domcl$), showing that the side condition of $\alllca$ is satisfied, and hence, the rule $\alllca$ may be applied.

\begin{flushleft}
\begin{tabular}{c c}
\AxiomC{$\rel, w : \forall x \phi, v : \phi(\unda/x), \Gamma \Rightarrow \Delta$}
\RightLabel{$\alllc$}
\UnaryInfC{$\rel, w : \forall x \phi, \Gamma \Rightarrow \Delta$}
\DisplayProof

&

$\leadsto$
\end{tabular}
\end{flushleft}
\begin{flushright}
\AxiomC{}
\RightLabel{\ih}
\dashedLine
\UnaryInfC{$\domcl(\rel, w : \forall x \phi, v : \phi(\unda/x), \Gamma \Rightarrow \Delta)$}
\RightLabel{$\alllca$}
\UnaryInfC{$\domcl(\rel, w : \forall x \phi, \Gamma \Rightarrow \Delta)$}
\DisplayProof
\end{flushright}

$\alllc$-(ii). In case (ii), we assume that $\unda$ is an eigenvariable and let $\Lambda := \domcl(\rel, w : \forall x \phi, v : \phi(\unda/x), \Gamma \Rightarrow \Delta)$.  By the side condition imposed on $\alllc$, we know that there exists a propagation path $\ppath(w,v)$ in the premise of the top left derivation such that $\stra_{\ppath}(w,v) \in \thuesysilang{a}$. Since $\unda$ is an eigenvariable, we know that after invoking \ih (which applies $\domcl$), there will be a domain atom of the form $\unda \in D_{v}$ occurring in $\Lambda$. Since $\empstr \in \thuesysiilang{a}$ and the empty path $\emppath(v,v) = v$ occurs in $\Lambda$, we know that $\stra_{\emppath}(v,v) = \empstr \in \thuesysiilang{a}$. Therefore, we have that $\unda$ is an eigenvarible, $\stra_{\emppath}(v,v) \in \thuesysiilang{a}$, and $\stra_{\ppath}(w,v) \in \thuesysilang{a}$ holds for $\Lambda$, meaning that $\alllcia$ may be applied, giving the desired result.

\begin{flushleft}
\begin{tabular}{c c}
\AxiomC{$\rel, w : \forall x \phi, v : \phi(\unda/x), \Gamma \Rightarrow \Delta$}
\RightLabel{$\alllc$}
\UnaryInfC{$\rel, w : \forall x \phi, \Gamma \Rightarrow \Delta$}
\DisplayProof

&

$\leadsto$
\end{tabular}
\end{flushleft}
\begin{flushright}
\AxiomC{}
\RightLabel{\ih}
\dashedLine
\UnaryInfC{$\domcl(\rel, w : \forall x \phi, v : \phi(\unda/x), \Gamma \Rightarrow \Delta)$}
\RightLabel{$\alllcia$}
\UnaryInfC{$\domcl(\rel, w : \forall x \phi, \Gamma \Rightarrow \Delta)$}
\DisplayProof
\end{flushright}

$\allrc$. Let $\Lambda := \domcl(\rel, w \leq u, \Gamma \sar  u : \phi(\unda/x), \Delta)$. We have two cases to consider: either (i) $u : \phi(\unda/x)$ contains no other parameters other than possibly $\unda$, or (ii) $u : \phi(\unda/x)$ is of the form $u : \phi(\vec{\unda})(\unda/x)$ and contains the parameters $\vec{\unda} := \unda_{1}, \ldots, \unda_{n}$. Case (i) is simple to resolve, so we show how to resolve case (ii). By the assumption of case (ii), $\Lambda$ is of the form $\rel', w \leq u, \vec{\unda} \in D_{u}, \unda \in D_{u}, \Gamma \sar u : \phi(\vec{\unda})(\unda/x), \Delta$. To resolve the case, as shown below, we first apply admissibility of $\wk$ (\lem~\ref{thm:properties-quasi-refined-FOInt}) to add in the domain atoms $\vec{\unda}' \in D_{w} := \unda_{i_{1}} \in D_{w}, \ldots, \unda_{i_{k}} \in D_{w}$ such that $\unda_{i_{j}} \in D_{w} \not\in \rel'$ for $i_{j} \in \{1, \ldots, n\}$ and $j \in \{1, \ldots, k\}$, i.e. for each parameter $\unda_{i_{j}}$ in $\vec{\unda}$ we add a new domain atom $\unda_{i_{j}} \in D_{w}$ given that it does not already occur in $\rel'$. We then apply the admissible $\nd$ rule (\thm~\ref{thm:admissible-rules-G3-Calc}) $n$ times to delete all domain atoms $\vec{\unda} \in D_{u}$, yielding a labelled sequent where $u$ and $\unda$ are eigenvariables. One application of $\allrnc$ gives the desired conclusion.

\begin{flushleft}
\begin{tabular}{c c}
\AxiomC{$\rel, w \leq u, \Gamma \sar  u : \phi(\vec{\unda})(\unda/x), \Delta$}
\RightLabel{$\allrc$}
\UnaryInfC{$\rel, \Gamma \sar w : \forall x \phi(\vec{\unda}), \Delta$}
\DisplayProof

&

$\leadsto$
\end{tabular}
\end{flushleft}
\begin{flushright}
\AxiomC{}
\RightLabel{\ih}
\dashedLine
\UnaryInfC{$\domcl(\rel, w \leq u, \Gamma \sar  u : \phi(\vec{\unda})(\unda/x), \Delta)$}
\RightLabel{=}
\dottedLine
\UnaryInfC{$\rel', w \leq u, \vec{\unda} \in D_{u}, \unda \in D_{u}, \Gamma \sar u : \phi(\vec{\unda})(\unda/x), \Delta$}
\RightLabel{$\wk$}
\dashedLine
\UnaryInfC{$\rel', w \leq u, \vec{\unda}' \in D_{w}, \vec{\unda} \in D_{u}, \unda \in D_{u}, \Gamma \sar u : \phi(\vec{\unda})(\unda/x), \Delta$}
\RightLabel{$\nd \times n$}
\dashedLine
\UnaryInfC{$\rel', w \leq u, \vec{\unda}' \in D_{w}, \unda \in D_{u}, \Gamma \sar u : \phi(\vec{\unda})(\unda/x), \Delta$}
\RightLabel{$\allr$}
\UnaryInfC{$\rel', \vec{\unda}' \in D_{w}, \Gamma \sar w : \forall x \phi(\vec{\unda}), \Delta$}
\RightLabel{=}
\dottedLine
\UnaryInfC{$\domcl(\rel, \Gamma \sar w : \forall x \phi(\vec{\unda}), \Delta)$}
\DisplayProof
\end{flushright}
\end{proof}

\begin{customlem}{\ref{lem:lsub-admiss-FOInt-refined}}
The rule $\lsub$ is hp-admissible in $\intfondl$ and $\intfocdl$.
\end{customlem}

\begin{proof} We prove the result by induction on the height of the given derivation.

\textit{Base case.} The base case follows from the fact that any application of $\lsub$ to $\idfonc$ or $\botl$ yields another instance of the rule.

\textit{Inductive step.} With the exception of the $\impr$ and $\allrnc$ rules, all cases follow by invoking \ih and then applying the corresponding rule. In the $\impr$ and $\allrnc$ cases it may be necessary to invoke \ih twice (before applying each respective rule) to ensure that the eigenvariable condition is met, similar to what was done in the proof of \lem~\ref{lem:lsub-admiss-FO-Int}.
\end{proof}

\begin{customlem}{\ref{lem:psub-admiss-FOInt-refined}}
The rule $\psub$ is hp-admissible in $\intfondl$ and $\intfocdl$. 
\end{customlem}

\begin{proof} We show the result for $\intfondl$ as the proof for $\intfocdl$ is similar, and prove the claim by induction on the height of the given derivation.

\textit{Base case.} Any application of $\psub$ to $\idfonc$ or $\botl$ yields another instance of the rule, resolving the base case.

\textit{Inductive step.} With the exception of the $\existsrn$, $\allln$, $\existsln$, and $\allrn$ rules, all cases follow by invoking \ih and then applying the corresponding rule. The non-trivial $\existsrn$ and $\allln$ cases occur when $\psub$ substitutes a parameter $\undb$ for the parameter $\unda$ (occurring in the auxiliary formula) or introduces the parameter $\unda$ of the auxiliary labelled formula when the parameter is an eigenvariable. The non-trivial $\existsln$ and $\allrn$ cases occur when $\psub$ introduces the eigenvariable $\unda$. Let us consider each non-trivial case in turn:

$\existsrn$. Either (i) $\unda$ is $\thuesysi$-available for $w$ or (ii) $\unda$ is an eigenvariable. In case (i), the non-trivial $\psub$ instance applies a substitution of the form $(\undb/\unda)$. By the assumption of the case, we know there exists a labelled formula $u :\psi(\unda) \in \Gamma, \Delta$ and a propagation path $\ppath(u,w)$ such that $\stra_{\ppath}(u,w) \in \thuesysilang{a}$. Observe that since $\psub$ applies $(\undb/\unda)$ to all labelled formulae of the sequent, the auxiliary formula will become $w : \phi(\undb/x)$, and we will know that a labelled formula $u :\psi(\undb) \in \Gamma, \Delta$ exists with a propagation path $\ppath(u,w)$ such that $\stra_{\ppath}(u,w) \in \thuesysilang{a}$. Hence, the side condition of the inference will continue to hold if \ih is applied first, allowing for the two rules to be permuted. In case (ii), where $\unda$ is an eigenvariable, the non-trivial case occurs when $\psub$ is of the form $(\unda/\undb)$. In such a case we invoke \ih twice to ensure that the eigenvariable condition is met (similar to the non-trivial case of $\existsln$ below), and then apply $\existsrn$.

$\allln$. By the side condition of the rule, either (i) $\unda$ is $\thuesysi$-available for $v$ or (ii) $\unda$ is an eigenvariable, and there exists a propagation path $\ppath(w,v)$ such that $\stra_{\ppath}(w,v) \in \thuesysilang{a}$. Let us consider case (i) first, where the non-trivial $\psub$ instance applies a substitution of the form $(\undb/\unda)$. Since $\unda$ is $\thuesysi$-available for $v$, we know that there exists a labelled formula $u :\psi(\unda) \in \Gamma, \Delta$ and a propagation path $\ppath(u,v)$ such that $\stra_{\ppath}(u,v) \in \thuesysilang{a}$. If we invoke \ih and apply $\psub$ first, then the auxiliary formula will be of the form $v : \phi(\undb/x)$, and by the side condition there will be a labelled formula $u :\psi(\undb) \in \Gamma, \Delta$ and a propagation path $\ppath(u,v)$ such that $\stra_{\ppath}(u,v) \in \thuesysilang{a}$. Hence, the side condition continues to hold after the invocation of IH, allowing for $\allln$ to be applied an the two rules to be permuted. In case (ii), the non-trivial case occurs when $\psub$ introduces the eigenvariable $\unda$. In such a case we invoke \ih twice to ensure that the eigenvariable condition is met (similar to the non-trivial case of $\allrn$ below), and then apply $\allln$.

$\existsln$. The non-trivial case arises when \ih (i.e. $\psub$) introduces the eigenvariable $\unda$, and is resolved as shown below:

\begin{flushleft}
\begin{tabular}{c c}
\AxiomC{$\rel, \Gamma, w: \phi(\unda/x) \sar \Delta$}
\RightLabel{$\existslnc$}
\UnaryInfC{$\rel, \Gamma, w : \exists x \phi \sar \Delta$}
\RightLabel{$\psub$}
\UnaryInfC{$\rel(\unda/\undb), \Gamma(\unda/\undb), (w : \exists x \phi)(\unda/\undb) \sar \Delta(\unda/\undb)$}
\DisplayProof

&

$\leadsto$
\end{tabular}
\end{flushleft}
\begin{flushright}
\AxiomC{$\rel, \Gamma, w: \phi(\unda/x) \sar \Delta$}
\RightLabel{\ih}
\dashedLine
\UnaryInfC{$\rel(\undc/\unda), \Gamma(\undc/\unda), (w: \phi(\unda/x))(\undc/\unda) \sar \Delta(\undc/\unda)$}
\RightLabel{=}
\dottedLine
\UnaryInfC{$\rel, \Gamma, w: \phi(\undc/x) \sar \Delta$}
\RightLabel{\ih}
\dashedLine
\UnaryInfC{$\rel(\unda/\undb), \Gamma(\unda/\undb), (w: \phi(\undc/x))(\unda/\undb) \sar \Delta(\unda/\undb)$}
\RightLabel{=}
\dottedLine
\UnaryInfC{$\R(\unda/\undb), \Gamma(\unda/\undb), w: \phi(\undc/x)(\unda/\undb) \sar \Delta(\unda/\undb)$}
\RightLabel{$\existsln$}
\UnaryInfC{$\R(\unda/\undb), \Gamma(\unda/\undb), (w : \exists x \phi)(\unda/\undb) \sar \Delta(\unda/\undb)$}
\DisplayProof
\end{flushright}

$\allrn$. The non-trivial case arises when \ih introduces the eigenvariable $\unda$, and is resolved as shown below:

\begin{flushleft}
\begin{tabular}{c c}
\AxiomC{$\rel, w \leq u, \Gamma \sar  u : \phi(\unda/x), \Delta$}
\RightLabel{$\allrn$}
\UnaryInfC{$\rel, \Gamma \sar w : \forall x \phi, \Delta$}
\RightLabel{$\psub$}
\UnaryInfC{$\R(\unda/\undb), \Gamma(\unda/\undb) \sar (w : \forall x \phi)(\unda/\undb), \Delta(\unda/\undb)$}
\DisplayProof

&

$\leadsto$
\end{tabular}
\end{flushleft}
\begin{flushright}
\AxiomC{$\rel, w \leq u, \Gamma \sar  u : \phi(\unda/x), \Delta$}
\RightLabel{\ih}
\dashedLine
\UnaryInfC{$\rel(\undc/\unda), (w \leq u)(\undc/\unda), \Gamma(\undc/\unda) \sar (u : \phi(\unda/x))(\undc/\unda), \Delta(\undc/\unda)$}
\RightLabel{=}
\dottedLine
\UnaryInfC{$\rel, w \leq u, \Gamma \sar u : \phi(\undc/x), \Delta$}
\RightLabel{\ih}
\dashedLine
\UnaryInfC{$\rel(\unda/\undb), (w \leq u)(\unda/\undb), \Gamma(\unda/\undb) \sar (u : \phi(\undc/x))(\unda/\undb), \Delta(\unda/\undb)$}
\RightLabel{=}
\dottedLine
\UnaryInfC{$\R(\unda/\undb), w \leq u, \Gamma(\unda/\undb) \sar u : \phi(\undc/x)(\unda/\undb), \Delta(\unda/\undb)$}
\RightLabel{$\allrn$}
\UnaryInfC{$\R(\unda/\undb), \Gamma(\unda/\undb) \sar (w : \forall x \phi)(\unda/\undb), \Delta(\unda/\undb)$}
\DisplayProof
\end{flushright}
\end{proof}

\begin{customlem}{\ref{lem:wk-admiss-FOInt-refined}}
The rule $\wk$ is hp-admissible in $\intfondl$ and $\intfocdl$. 
\end{customlem}

\begin{proof} We show the result for $\intfondl$ as the proof for $\intfocdl$ is similar. We prove the result by induction on the height of the given derivation.

\textit{Base case.} The base case is straightforward since any application of $\wk$ to $\idfonc$ or $\botl$ yields another instance of the rule.

\textit{Inductive step.} The only non-trivial cases concern $\impr$, $\existsrn$, $\allln$, $\existsln$, and $\allrn$, and arise when $\wk$ introduces a parameter or label identical to an eigenvariable of one of the aforementioned inferences. The cases involving the other rules are resolved by applying \ih followed by the corresponding rule. We show how to resolve the non-trivial $\impr$, $\existsln$, $\allrn$ cases below; all other cases are similar or simple. We let $\undc$ and $z$ be a fresh parameter and label, respectively.

\begin{flushleft}
\begin{tabular}{c c}
\AxiomC{$\rel, w \leq u, \Gamma, u :\phi \sar u :\psi, \Delta$}
\RightLabel{$\impr$}
\UnaryInfC{$\rel, \Gamma \sar w :\phi \imp \psi, \Delta$}
\RightLabel{$\wk$}
\UnaryInfC{$\rel, \rel', \Gamma, \Gamma' \sar w :\phi \imp \psi, \Delta, \Delta'$}
\DisplayProof

&

$\leadsto$
\end{tabular}
\end{flushleft}
\begin{flushright}
\AxiomC{$\rel, w \leq u, \Gamma, u :\phi \sar u :\psi, \Delta$}
\RightLabel{$\lsub$}
\dashedLine
\UnaryInfC{$\rel(z/u), (w \leq u)(z/u), \Gamma(z/u), (u :\phi)(z/u) \sar (u :\psi)(z/u), \Delta(z/u)$}
\RightLabel{=}
\dottedLine
\UnaryInfC{$\rel, w \leq z, \Gamma, z :\phi \sar z :\psi, \Delta$}
\RightLabel{\ih}
\dashedLine
\UnaryInfC{$\rel, \rel', w \leq z, \Gamma, \Gamma', z : \phi \sar z : \psi, \Delta, \Delta'$}
\RightLabel{$\impr$}
\UnaryInfC{$\rel, \rel', \Gamma, \Gamma' \sar w :\phi \imp \psi, \Delta, \Delta'$}
\DisplayProof
\end{flushright}

\begin{flushleft}
\begin{tabular}{c c}
\AxiomC{$\rel, \Gamma, w: \phi(\unda/x) \sar \Delta$}
\RightLabel{$\existsln$}
\UnaryInfC{$\rel, \Gamma, w : \exists x \phi \sar \Delta$}
\RightLabel{$\wk$}
\UnaryInfC{$\rel, \rel', \Gamma, \Gamma', w : \exists x \phi \sar \Delta, \Delta'$}
\DisplayProof

&

$\leadsto$
\end{tabular}
\end{flushleft}
\begin{flushright}
\AxiomC{$\rel, \Gamma, w: \phi(\unda/x) \sar \Delta$}
\RightLabel{$\psub$}
\dashedLine
\UnaryInfC{$\rel(\undc/\unda), \Gamma(\undc/\unda), (w: \phi(\unda/x))(\undc/\unda) \sar \Delta(\undc/\unda)$}
\RightLabel{=}
\dottedLine
\UnaryInfC{$\rel, \Gamma, w: \phi(\undc/x) \sar \Delta$}
\RightLabel{\ih}
\dashedLine
\UnaryInfC{$\rel, \rel', \Gamma, \Gamma', w: \phi(\undc/x) \sar \Delta, \Delta'$}
\RightLabel{$\existsln$}
\UnaryInfC{$\rel, \rel', \Gamma, \Gamma', w: \exists x \phi \sar \Delta, \Delta'$}
\DisplayProof
\end{flushright}

\begin{flushleft}
\begin{tabular}{c c}
\AxiomC{$\rel, w \leq u, \Gamma \sar  u : \phi(\unda/x), \Delta$}
\RightLabel{$\allrn$}
\UnaryInfC{$\rel, \Gamma \sar w : \forall x \phi, \Delta$}
\RightLabel{$\wk$}
\UnaryInfC{$\rel, \rel', \Gamma, \Gamma' \sar w : \forall x \phi, \Delta, \Delta'$}
\DisplayProof

&

$\leadsto$
\end{tabular}
\end{flushleft}
\begin{flushright}
\AxiomC{$\rel, w \leq u, \Gamma \sar  u : \phi(\unda/x), \Delta$}
\RightLabel{$\lsub$}
\dashedLine
\UnaryInfC{$\rel(z/u), (w \leq u)(z/u), \Gamma(z/u) \sar  (u : \phi(\unda/x))(z/u), \Delta(z/u)$}
\RightLabel{=}
\dottedLine
\UnaryInfC{$\rel, w \leq z, \Gamma \sar  z : \phi(\unda/x), \Delta$}
\RightLabel{$\psub$}
\dashedLine
\UnaryInfC{$\rel(\undc/\unda), (w \leq z)(\undc/\unda), \Gamma(\undc/\unda) \sar  (z : \phi(\unda/x))(\undc/\unda), \Delta(\undc/\unda)$}
\RightLabel{=}
\dottedLine
\UnaryInfC{$\rel, w \leq z, \Gamma \sar  z : \phi(\undc/x), \Delta$}
\RightLabel{\ih}
\dashedLine
\UnaryInfC{$\rel, \rel', w \leq z, \Gamma, \Gamma' \sar  z : \phi(\undc/x), \Delta, \Delta'$}
\RightLabel{$\allrn$}
\UnaryInfC{$\rel, \rel', \Gamma, \Gamma' \sar  w : \forall x \phi, \Delta, \Delta'$}
\DisplayProof
\end{flushright}
\end{proof}

\begin{customlem}{\ref{lem:dis-con-invert-FOInt-refined}}
The $\conl$, $\conr$, $\disl$, and $\disr$ rules are hp-invertible in $\intfondl$ and $\intfocdl$. 
\end{customlem}

\begin{proof} We prove the result by induction on the height of the given derivation, and consider only the $\conl$ case since the $\conr$, $\disl$, and $\disr$ cases are similar.

\textit{Base case.} The base cases are resolved as shown below:

\begin{flushleft}
\begin{tabular}{c c}
\AxiomC{}
\RightLabel{$\idfonc$}
\UnaryInfC{$\rel,\Gamma, v : \phi \land \psi, w :p(\vec{\unda}) \sar u :p(\vec{\unda}), \Delta$}
\DisplayProof

&

$\leadsto$
\end{tabular}
\end{flushleft}
\begin{flushright}
\AxiomC{}
\RightLabel{$\idfo$}
\UnaryInfC{$\rel,\Gamma, v : \phi, v : \psi, w :p(\vec{\unda}) \sar u :p(\vec{\unda}), \Delta$}
\DisplayProof
\end{flushright}

\begin{tabular}{c c c}
\AxiomC{}
\RightLabel{$\botl$}
\UnaryInfC{$\rel,\Gamma, w : \bot, u : \phi \land \psi \sar \Delta$}
\DisplayProof

&

$\leadsto$

&

\AxiomC{}
\RightLabel{$\botl$}
\UnaryInfC{$\rel,\Gamma, w : \bot, u : \phi, u : \psi \sar \Delta$}
\DisplayProof
\end{tabular}

\textit{Inductive step.} With the exception of the $\conl$ case, all cases are resolved by invoking \ih followed by the corresponding rule. If the last rule applied in the derivation is $\conl$, then there are two cases to consider: either (i) the conjunction we aim to invert is principal, or (ii) it is not. Case (i) is shown below top---the derivation of the premise gives the desired conclusion---and case (ii) is shown below bottom.

\begin{center}
\begin{tabular}{c c c}
\AxiomC{$\rel, \Gamma, w :\phi, w :\psi \sar \Delta$}
\RightLabel{$\conl$}
\UnaryInfC{$\rel, \Gamma, w :\phi \wedge \psi \sar \Delta$}
\DisplayProof

&

$\leadsto$

&

\AxiomC{$\rel, \Gamma, w :\phi, w :\psi \sar \Delta$}
\DisplayProof
\end{tabular}
\end{center}

\begin{center}
\begin{tabular}{c c c}
\AxiomC{$\rel, \Gamma, u : \chi, u : \theta, w :\phi \land \psi \sar \Delta$}
\RightLabel{$\conl$}
\UnaryInfC{$\rel, \Gamma, u : \chi \land \theta, w :\phi \wedge \psi \sar \Delta$}
\DisplayProof

&

$\leadsto$

&

\AxiomC{$\rel, \Gamma, u : \chi, u : \theta, w :\phi \land \psi \sar \Delta$}
\RightLabel{\ih}
\dashedLine
\UnaryInfC{$\rel, \Gamma, u : \chi, u : \theta, w :\phi, w : \psi \sar \Delta$}
\RightLabel{$\conl$}
\UnaryInfC{$\rel, \Gamma, u : \chi \land \theta, w :\phi, w : \psi \sar \Delta$}
\DisplayProof
\end{tabular}
\end{center}
\end{proof}

\begin{customlem}{\ref{lem:imp-invert-FOInt-refined}}
The $\primp$ and $\impr$ rules are hp-invertible in $\intfondl$ and $\intfocdl$. 
\end{customlem}

\begin{proof} The hp-invertibility of $\primp$ follows from \lem~\ref{lem:wk-admiss-FOInt-refined}. We argue that the rule $\impr$ is hp-invertible by induction on the height of the given derivation.

\textit{Base case.} The base cases are resolved as shown below:

\begin{flushleft}
\begin{tabular}{c c}
\AxiomC{}
\RightLabel{$\idfonc$}
\UnaryInfC{$\rel,\Gamma, w :p(\vec{\unda}) \sar v : \phi \imp \psi, u :p(\vec{\unda}), \Delta$}
\DisplayProof

&

$\leadsto$
\end{tabular}
\end{flushleft}
\begin{flushright}
\AxiomC{}
\RightLabel{$\idfonc$}
\UnaryInfC{$\rel,v\leq z,\Gamma, w :p(\vec{\unda}), z : \phi \sar z : \psi, u :p(\vec{\unda}), \Delta$}
\DisplayProof
\end{flushright}

\begin{center}
\begin{tabular}{c c c}
\AxiomC{}
\RightLabel{$\botl$}
\UnaryInfC{$\rel,\Gamma, w : \bot \sar v : \phi \imp \psi, \Delta$}
\DisplayProof

&

$\leadsto$

&

\AxiomC{}
\RightLabel{$\botl$}
\UnaryInfC{$\rel, v \leq z, \Gamma, w : \bot, z : \phi \sar z : \psi, \Delta$}
\DisplayProof
\end{tabular}
\end{center}

\textit{Inductive step.} With the exception of the $\impr$ rule, all cases are resolved by invoking \ih followed by the corresponding rule. If the last inference of the derivation is an instance of $\impr$, then there are two cases to consider: either (i) the implication we aim to invert is principal, or (ii) it is not. Case (i) is resolved below top---the derivation of the premise gives the desired conclusion---and case (ii) is resolved as shown below bottom.

\begin{center}
\begin{tabular}{c c c}
\AxiomC{$\rel, w \leq u, \Gamma, u :\phi \sar u :\psi, \Delta$}
\RightLabel{$\impr$}
\UnaryInfC{$\rel, \Gamma \sar w :\phi \imp \psi, \Delta$}
\DisplayProof

&

$\leadsto$

&

\AxiomC{$\rel, w \leq u, \Gamma, u :\phi \sar u :\psi, \Delta$}
\DisplayProof
\end{tabular}
\end{center}

\begin{flushleft}
\begin{tabular}{c c}
\AxiomC{$\rel, v \leq z, \Gamma, z :\chi \sar z :\theta, w :\phi \imp \psi, \Delta$}
\RightLabel{$\impr$}
\UnaryInfC{$\rel, \Gamma \sar v : \chi \imp \theta, w :\phi \imp \psi, \Delta$}
\DisplayProof

&

$\leadsto$
\end{tabular}
\end{flushleft}
\begin{flushright}
\AxiomC{$\rel, v \leq z, \Gamma, z :\chi \sar z :\theta, w :\phi \imp \psi, \Delta$}
\RightLabel{\ih}
\dashedLine
\UnaryInfC{$\rel, w \leq u, v \leq z, \Gamma, z :\chi, u : \phi \sar u : \psi, z :\theta, \Delta$}
\RightLabel{$\impr$}
\UnaryInfC{$\rel, w \leq u, \Gamma, u : \phi \sar u : \psi, v : \chi \imp \theta, \Delta$}
\DisplayProof
\end{flushright}
\end{proof}

\begin{customlem}{\ref{lem:exists-all-invert-FOInt-refined}} \ 

(i) The $\existsln$, $\existsrn$, $\allln$, and $\allrn$ rules are hp-invertible in $\intfondl$. 

(ii) The $\existslc$, $\existsrc$, $\alllc$, and $\allrc$ rules are hp-invertible in $\intfocdl$. 
\end{customlem}

\begin{proof} We prove both results simultaneously and let $\mathsf{X} \in \{\nnn,\ccc\}$. Hp-invertibility of $\existsrnc$ and $\alllnc$ follows from \lem~\ref{lem:wk-admiss-FOInt-refined}. Hp-invertibility of $\existslnc$ and $\allrnc$ is shown by induction on the height of the given derivation; we prove the $\allrnc$ case as the $\existslnc$ case is similar.

\textit{Base case.} The base cases are resolved as shown below:

\begin{flushleft}
\begin{tabular}{c c}
\AxiomC{}
\RightLabel{$\idfonc$}
\UnaryInfC{$\rel, \Gamma, w :p(\vec{\unda}) \sar u :p(\vec{\unda}), v : \forall x \phi, \Delta$}
\DisplayProof

&

$\leadsto$
\end{tabular}
\end{flushleft}
\begin{flushright}
\AxiomC{}
\RightLabel{$\idfonc$}
\UnaryInfC{$\rel, v\leq z,  \Gamma, w :p(\vec{\unda}) \sar u :p(\vec{\unda}), z : \phi(\undb/y), \Delta$}
\DisplayProof
\end{flushright}

\begin{center}
\begin{tabular}{c c c}
\AxiomC{}
\RightLabel{$\botl$}
\UnaryInfC{$\rel,\Gamma, v : \bot \sar w :\forall x \phi, \Delta$}
\DisplayProof

&

$\leadsto$

&

\AxiomC{}
\RightLabel{$\botl$}
\UnaryInfC{$\rel, w \leq u,  \Gamma,  v : \bot \sar u : \phi(\unda/x), \Delta$}
\DisplayProof
\end{tabular}
\end{center}

\textit{Inductive step.} With the exception of the $\allrn$ rule, all cases are resolved by invoking \ih followed by the corresponding rule. If the last inference of the derivation is an instance of $\allrn$, then there are two cases to consider: either (i) the labelled formula we aim to invert is principal, or (ii) it is not. Case (i) is resolved below top---the derivation of the premise gives the desired conclusion---and case (ii) is resolved as shown below bottom.

\begin{center}
\begin{tabular}{c c c}
\AxiomC{$\rel, w \leq u, \Gamma \sar  u : \phi(\unda/x), \Delta$}
\RightLabel{$\allrnc$}
\UnaryInfC{$\rel, \Gamma \sar w : \forall x \phi, \Delta$}
\DisplayProof

&

$\leadsto$

&

\AxiomC{$\rel, w \leq u, \Gamma \sar  u : \phi(\unda/x), \Delta$}
\DisplayProof
\end{tabular}
\end{center}

\begin{flushleft}
\begin{tabular}{c c}
\AxiomC{$\rel, v \leq z, \Gamma \sar  z : \psi(\undb/y), w : \forall x \phi, \Delta$}
\RightLabel{$\allrnc$}
\UnaryInfC{$\rel, \Gamma \sar v : \forall x \psi, w : \forall x \phi, \Delta$}
\DisplayProof

&

$\leadsto$
\end{tabular}
\end{flushleft}
\begin{flushright}
\AxiomC{$\rel, v \leq z, \Gamma \sar  z : \psi(\undb/y), w : \forall x \phi, \Delta$}
\RightLabel{\ih}
\dashedLine
\UnaryInfC{$\rel, v \leq z, w \leq u, \Gamma \sar  z : \psi(\undb/y), u : \phi(\unda/x), \Delta$}
\RightLabel{$\allrnc$}
\UnaryInfC{$\rel, w \leq u, \Gamma \sar w : \forall x \psi, u : \phi(\unda/x), \Delta$}
\DisplayProof
\end{flushright}
\end{proof}

\begin{customlem}{\ref{lem:ctrl-admiss-FOInt-refined}}
The rule $\ctrl$ is hp-admissible in $\intfondl$ and $\intfocdl$. 
\end{customlem}

\begin{proof} We prove both results simultaneously and let $\mathsf{X} \in \{\nnn,\ccc\}$. We prove the result by induction on the height of the given derivation.

\textit{Base case.} Any application of $\ctrl$ an instance of $\idfonc$ or $\botl$ yields another instance of the rule, which confirms the base case.

\textit{Inductive step.} For the inductive step, we assume that the derivation ends with an instance of a rule $\ru$ (from $\intfondl$ or $\intfocdl$) followed by an instance of $\ctrl$. If the principal formula of $\ru$ is not active in $\ctrl$, then the case is handled by invoking \ih followed by an instance of $\ru$. We therefore assume that the principal formula of $\ru$ is active in the $\ctrl$ inference. This assumption also implies that we need only consider the cases where $\ru$ is an instance of $\conl$, $\disl$, $\primp$, $\existslnc$, or $\alllnc$. We show how to resolve the $\conl$, $\existslnc$, and $\alllnc$ cases below; the $\disl$ and $\primp$ cases are similar or simple to verify.

\begin{flushleft}
\begin{tabular}{c c}
\AxiomC{$\rel, \Gamma, w :\phi \wedge \psi, w :\phi, w :\psi \sar \Delta$}
\RightLabel{$\conl$}
\UnaryInfC{$\rel, \Gamma, w :\phi \wedge \psi, w :\phi \wedge \psi \sar \Delta$}
\RightLabel{$\ctrl$}
\UnaryInfC{$\rel, \Gamma, w :\phi \wedge \psi \sar \Delta$}
\DisplayProof

&

$\leadsto$
\end{tabular}
\end{flushleft}
\begin{flushright}
\AxiomC{$\rel, \Gamma, w :\phi \wedge \psi, w :\phi, w :\psi \sar \Delta$}
\RightLabel{\lem~\ref{lem:dis-con-invert-FOInt-refined}}
\dashedLine
\UnaryInfC{$\rel, \Gamma, w :\phi, w : \psi, w :\phi, w :\psi \sar \Delta$}
\RightLabel{\ih $\times 2$}
\dashedLine
\UnaryInfC{$\rel, \Gamma, w :\phi, w : \psi \sar \Delta$}
\RightLabel{$\conl$}
\UnaryInfC{$\rel, \Gamma, w :\phi \wedge \psi \sar \Delta$}
\DisplayProof
\end{flushright}

\begin{flushleft}
\begin{tabular}{c c}
\AxiomC{$\rel, \Gamma, w : \exists x \phi, w: \phi(\unda/x) \sar \Delta$}
\RightLabel{$\existslnc$}
\UnaryInfC{$\rel, \Gamma, w : \exists x \phi, w : \exists x \phi \sar \Delta$}
\RightLabel{$\ctrl$}
\UnaryInfC{$\rel, \Gamma, w : \exists x \phi \sar \Delta$}
\DisplayProof

&

$\leadsto$
\end{tabular}
\end{flushleft}
\begin{flushright}
\AxiomC{$\rel, \Gamma, w : \exists x \phi, w: \phi(\unda/x) \sar \Delta$}
\RightLabel{\lem~\ref{lem:exists-all-invert-FOInt-refined}}
\dashedLine
\UnaryInfC{$\rel, \Gamma, w: \phi(\unda/x), w : \phi(\undb/x) \sar \Delta$}
\RightLabel{$\psub$}
\dashedLine
\UnaryInfC{$\rel(\unda/\undb), \Gamma(\unda/\undb), (w: \phi(\unda/x))(\unda/\undb), (w : \phi(\undb/x))(\unda/\undb) \sar \Delta(\unda/\undb)$}
\RightLabel{=}
\dottedLine
\UnaryInfC{$\rel, \Gamma, w: \phi(\unda/x), w : \phi(\unda/x) \sar \Delta$}
\RightLabel{\ih}
\dashedLine
\UnaryInfC{$\rel, \Gamma, w : \phi(\unda/x) \sar \Delta$}
\RightLabel{$\existslnc$}
\UnaryInfC{$\rel, \Gamma, w : \exists x \phi \sar \Delta$}
\DisplayProof
\end{flushright}

\begin{flushleft}
\begin{tabular}{c c}
\AxiomC{$\rel, \Gamma, u : \phi(\unda/x), w : \forall x \phi, w : \forall x \phi \sar \Delta$}
\RightLabel{$\alllnc$}
\UnaryInfC{$\rel, \Gamma, w : \forall x \phi, w : \forall x \phi \sar \Delta$}
\RightLabel{$\ctrl$}
\UnaryInfC{$\rel, \Gamma, w : \forall x \phi \sar \Delta$}
\DisplayProof

&

$\leadsto$
\end{tabular}
\end{flushleft}
\begin{flushright}
\AxiomC{$\rel, \Gamma, u : \phi(\unda/x), w : \forall x \phi, w : \forall x \phi \sar \Delta$}
\RightLabel{\ih}
\dashedLine
\UnaryInfC{$\rel, \Gamma, u : \phi(\unda/x), w : \forall x \phi \sar \Delta$}
\RightLabel{$\alllnc$}
\UnaryInfC{$\rel, \Gamma, w : \forall x \phi \sar \Delta$}
\DisplayProof
\end{flushright}
\end{proof}

\begin{customthm}{\ref{thm:cut-admiss-FOInt-refined}}
The rule $\cut$ is eliminable in $\intfondl$ and $\intfocdl$. 
\end{customthm}

\begin{proof} We show both results simultaneously and let $\mathsf{X} \in \{\nnn,\ccc\}$. We proceed by induction on the lexicographic ordering of pairs $(\fcomp{\phi},h_{1}+h_{2})$, where $\fcomp{\phi}$ is the complexity of the cut formula $\phi$, $h_{1}$ is the height of the derivation of the left premise of $\cut$, and $h_{2}$ is the height of the derivation of the right premise of $\cut$. We assume w.l.o.g. that $\cut$ is the last rule used in our given derivation, and that no other instances of $\cut$ appear in the given derivation. The general result follows by repeatedly applying the algorithm described below to successively eliminate topmost instances of $\cut$ until the derivation is free of $\cut$ instances.

As usual, we separate the proof into a large variety of cases, and additionally, we explicitly write out the assumption being made in each case.

\emph{1. The complexity of the cut formula is $0$, that is, $\fcomp{\phi} = 0$.}

\quad \emph{1.1 Both premises of $\cut$ are an instance of $\idfonc$ or $\botl$.}

\quad \quad \emph{1.1.1 Both premises of $\cut$ are instances of $\idfonc$ with the cut formula principal in both premises.} By the side conditions of the $\idfonc$ instances (in the top left derivation), we know that there exists a propagation path $\ppath(w,u)$ and a propagation path $\ppath'(u,v)$ such that $\stra_{\ppath}(w,u), \stra_{\ppath'}(u,v) \in \thuesysilang{a}$. This implies the existence of a propagation path $\ppath''(w,v)$ such that $\stra_{\ppath''}(w,v) \in \thuesysilang{a}$, showing that the output instance of $\idfonc$ is valid.

\begin{flushleft}
\begin{tabular}{c c}
\AxiomC{}
\RightLabel{$\idfonc$}
\UnaryInfC{$\rel, \Gamma, w : p(\vec{\unda}) \sar u : p(\vec{\unda}), \Delta$}

\AxiomC{}
\RightLabel{$\idfonc$}
\UnaryInfC{$\rel, \Gamma, u : p(\vec{\unda}) \sar v : p(\vec{\unda}), \Delta$}

\RightLabel{$\cut$}
\BinaryInfC{$\rel, \Gamma, w : p(\vec{\unda}) \sar v : p(\vec{\unda}), \Delta$}
\DisplayProof

&

$\leadsto$
\end{tabular}
\end{flushleft}

\begin{flushright}
\AxiomC{ }
\RightLabel{$\idfonc$}
\UnaryInfC{$\rel,\Gamma, w : p(\vec{\unda}) \sar v : p(\vec{\unda}), \Delta$}
\DisplayProof
\end{flushright}

\quad \quad \emph{1.1.2 Both premises of $\cut$ are instances of $\idfonc$ with the cut formula principal in the left premise and not the right.} Then, the conclusion of $\cut$ is an instance of $\idfonc$.

\quad \quad \emph{1.1.3 Both premises of $\cut$ are instances of $\idfonc$ with the cut formula principal in the right premise and not the left.} Then, the conclusion of $\cut$ is an instance of $\idfonc$.

\quad \quad \emph{1.1.4 Both premises of $\cut$ are instances of $\idfonc$ with the cut formula principal in neither premise.} Then, the conclusion of $\cut$ is an instance of $\idfonc$.

\quad \quad \emph{1.1.5 The left premise of $\cut$ is an instance of $\botl$.} Then, the conclusion of $\cut$ is an instance of $\botl$.

\quad \quad \emph{1.1.6 The right premise of $\cut$ is an instance of $\botl$.} Due to our assumption in case 1.1.5 above, we may assume that the left premise is an instance of $\idfonc$, which gives rise to two possibilities: either (i) the principal occurrence of $\bot$ in $\botl$ is the cut formula, or (ii) it is not. In case (i), the conclusion is an instance of $\idfonc$, and in case (ii), the conclusion is an instance of $\botl$.

\quad \emph{1.2 The left premise of $\cut$ is an instance of $\idfonc$ or $\botl$ and the right premise is not.} Note that by assumption 1, the principal formula of the right premise of $\cut$ is not the cut-formula. We split case 1.2 into two further cases, which are resolved as explained below:

\quad \quad \emph{1.2.1 The right premise of $\cut$ is derived with an instance of $\impr$, $\existsrnc$, $\alllnc$, $\existslnc$, or $\allrnc$.} The result follows by (i) potentially invoking the hp-admissibility of $\lsub$ (\lem~\ref{lem:lsub-admiss-FOInt-refined}) and $\psub$ (\lem~\ref{lem:psub-admiss-FOInt-refined}) on the premise of $\impr$, $\existsrnc$, $\alllnc$, $\existslnc$, or $\allrnc$ (which gives the right premise of $\cut$) to ensure the eigenvariable condition is met (if required), followed by the hp-invertibility of $\impr$, $\existsrnc$, $\alllnc$, $\existslnc$, or $\allrnc$ (\lem~\ref{lem:invert-FOInt-refined}) on the left premise of $\cut$ to ensure that its contexts match those of the sequent obtained from step (i). Then, (iii) we invoke \ih between the proofs of our two newly obtained sequents, followed by (iv) an application of the corresponding $\impr$, $\existsrnc$, $\alllnc$, $\existslnc$, or $\allrnc$ rule.

\quad \quad \emph{1.2.2 The right premise of $\cut$ is derived with a rule other than $\impr$, $\existsrnc$, $\alllnc$, $\existslnc$, or $\allrnc$ (and by assumption 1.2 above, cannot be an instance of $\idfonc$ or $\botl$).} Let $\ru$ be the rule used to derive the right premise of $\cut$. The result follows by (i) applying the hp-invertibility of $\ru$ (\lem~\ref{lem:invert-FOInt-refined}) to the left premise of $\cut$ to ensure its contexts match the contexts of the premise(s) of $\ru$, followed by (ii) an application of \ih between the proof(s) of the sequent(s) obtained from step (i) and the premise(s) of $\ru$, and last, (iii) an application of $\ru$ gives the desired conclusion.

\quad \emph{1.3 The right premise of $\cut$ is an instance of $\idfonc$ or $\botl$ and the left premise is not.} This case is argued similarly to case 1.2 above.

\quad \emph{1.4 Neither premise of $\cut$ is an instance of $\idfonc$ or $\botl$.} Let $\ruone$ be the rule used to derive the left premise of $\cut$ and $\rutwo$ be the rule used to derive the right premise of $\cut$. Since the cut formula is of the form $p(\vec{\unda})$ or $\bot$ by assumption 1 above, we know that the cut formula is not principal in $\ruone$ and $\rutwo$. To eliminate $\cut$, we perform the following: (i) We potentially apply hp-admissibility of $\lsub$ (\lem~\ref{lem:lsub-admiss-FOInt-refined}) and/or $\psub$ (\lem~\ref{lem:psub-admiss-FOInt-refined}) to the premise(s) of $\ruone$ and $\rutwo$ to ensure that the eigenvariable condition is met, given that $\ruone$ or $\rutwo$ is an instance of $\impr$, $\existsrnc$, $\alllnc$, $\existslnc$, or $\allrnc$. (ii) We apply hp-invertibility of $\rutwo$ to the proof(s) of the sequent(s) obtained from the premise(s) of $\ruone$ in step (i) and hp-invertibility of $\ruone$ to the proof(s) of the sequent(s) obtained from the premise(s) of $\rutwo$ in step (i) (thus invoking \lem~\ref{lem:invert-FOInt-refined}). (iii) We invoke \ih between the proofs of the sequents obtained from step (ii), followed by (iv) applications of $\ruone$ and $\rutwo$ to obtain a cut-free proof of the desired conclusion.




\emph{2. The complexity of the cut formula is greater than $0$, that is, $\fcomp{\phi} > 0$.}

\quad \emph{2.1 The cut formula is not principal in either premise of $\cut$.} Similar to the proof of case 1.4 above. Note that in case 1.4, our assumptions implied that the cut formula was not principal in the left or right premise of $\cut$. Here, in case 2.1, this fact holds by assumption, and so, we may argue similarly.

\quad \emph{2.2 The cut formula is principal in the left, but not the right, premise of $\cut$.} Let $\ru$ be the rule used to derive the right premise of $\cut$. Below top, we show how to resolve the case were $\ru$ is a unary rule, and below bottom, we show how to resolve the case where $\ru$ is a binary rule. In the unary case below top, $\ru$ may be an instance of $\impr$, $\existsrnc$, $\alllnc$, $\existslnc$, or $\allrnc$, and so, in such a case it may be necessary to apply the hp-admissibility of $\lsub$ (\lem~\ref{lem:lsub-admiss-FOInt-refined}) and/or $\psub$ (\lem~\ref{lem:psub-admiss-FOInt-refined}) to ensure the eigenvariable condition is met after the application of IH. We note that in the cases below, \ih may be applied since the sum of the heights $h_{1} + h_{2}$ has decreased by $1$.

\begin{flushleft}
\begin{tabular}{c c}
\AxiomC{$\rel, \Gamma \sar w : \phi, \Delta$}

\AxiomC{$\rel', \Gamma', w :\phi \sar \Delta'$}
\RightLabel{$\ru$}
\UnaryInfC{$\rel, \Gamma, w : \phi \sar \Delta$}

\RightLabel{$\cut$}
\BinaryInfC{$\rel, \Gamma \sar \Delta$}
\DisplayProof

&

$\leadsto$
\end{tabular}
\end{flushleft}
\begin{flushright}
\AxiomC{$\rel, \Gamma \sar w : \phi, \Delta$}
\dashedLine
\RightLabel{\lem~\ref{lem:invert-FOInt-refined}}
\UnaryInfC{$\rel', \Gamma' \sar w : \phi, \Delta'$}

\AxiomC{$\rel', \Gamma', w :\phi \sar \Delta'$}

\RightLabel{\ih}
\dashedLine
\BinaryInfC{$\rel', \Gamma' \sar \Delta'$}

\RightLabel{$\ru$}
\UnaryInfC{$\rel, \Gamma \sar \Delta$}
\DisplayProof
\end{flushright}

\resizebox{\columnwidth}{!}{
\begin{tabular}{c c c}
\AxiomC{$\rel, \Gamma \sar w : \phi, \Delta$}

\AxiomC{$\rel', \Gamma', w :\phi \sar \Delta'$}
\AxiomC{$\rel'', \Gamma'', w :\phi \sar \Delta''$}
\RightLabel{$\ru$}
\BinaryInfC{$\rel, \Gamma, w : \phi \sar \Delta$}

\RightLabel{$\cut$}
\BinaryInfC{$\rel, \Gamma \sar \Delta$}
\DisplayProof

&

$\leadsto$

&

\AxiomC{$\Pi_{1}$}

\AxiomC{$\Pi_{2}$}

\RightLabel{$\ru$}
\BinaryInfC{$\rel, \Gamma \sar \Delta$}
\DisplayProof
\end{tabular}
}

\begin{center}
\begin{tabular}{c c c}
$\Pi_{1}$

&

$= \bigg \{$

&

\AxiomC{$\rel, \Gamma \sar w : \phi, \Delta$}
\dashedLine
\RightLabel{\lem~\ref{lem:invert-FOInt-refined}}
\UnaryInfC{$\rel', \Gamma' \sar w : \phi, \Delta'$}
\AxiomC{$\rel', \Gamma', w :\phi \sar \Delta'$}
\RightLabel{\ih}
\dashedLine
\BinaryInfC{$\rel', \Gamma' \sar \Delta'$}
\DisplayProof
\end{tabular}
\end{center}

\begin{center}
\begin{tabular}{c c c}
$\Pi_{2}$

&

$= \bigg \{$

&

\AxiomC{$\rel, \Gamma \sar w : \phi, \Delta$}
\dashedLine
\RightLabel{\lem~\ref{lem:invert-FOInt-refined}}
\UnaryInfC{$\rel'', \Gamma'' \sar w : \phi, \Delta''$}
\AxiomC{$\rel'', \Gamma'', w :\phi \sar \Delta''$}
\RightLabel{\ih}
\dashedLine
\BinaryInfC{$\rel'', \Gamma'' \sar \Delta''$}
\DisplayProof
\end{tabular}
\end{center}

\quad \emph{2.3 The cut formula is principal in the right, but not the left, premise of $\cut$.} Similar to the previous case 2.2.

\quad \emph{2.4 The cut formula is principal in both premises of $\cut$.} By the cut-elimination theorem for $\gtintfond$ and $\gtintfocd$ (\thm~\ref{thm:cut-admiss-FO-Int}), we need only consider the cases where the cut formula is of the form $\forall x \psi$ or $\exists x \psi$, as the propositional cases (where the cut formula is of the form $\psi \lor \chi$, $\psi \land \chi$, or $\phi \imp \chi$) are resolved in a similar manner.

\quad \quad \emph{2.4.1 The cut formula is of the form $\exists x \psi$.} The case is resolved as shown below. Observe that we may the first use of \ih as the sum of the heights $h_{1} + h_{2}$ is one less than the original $\cut$, and we may invoke the second use of \ih since the cut formula (namely, $w : \psi(\unda/x)$) is of a smaller complexity, i.e. $\fcomp{\psi(\unda/x)} < \fcomp{\exists x \psi} = \fcomp{\phi}$.

\begin{flushleft}
\begin{tabular}{c c}
\AxiomC{$\R, \Gamma \Rightarrow \Delta, w: \psi(\unda/x), w: \exists x \psi$}
\RightLabel{$\existsrnc$}
\UnaryInfC{$\R, \Gamma \Rightarrow \Delta, w: \exists x \psi$}

\AxiomC{$\R, w: \psi(\undb/x), \Gamma \Rightarrow \Delta$}
\RightLabel{$\existslnc$}
\UnaryInfC{$\R, w : \exists x \psi, \Gamma \Rightarrow \Delta$}

\RightLabel{$\cut$}

\BinaryInfC{$\R, \Gamma \Rightarrow \Delta$}
\DisplayProof

&

$\leadsto$
\end{tabular}
\end{flushleft}

\begin{flushright}
\begin{tabular}{c}
\AxiomC{$\Pi_{1}$}
\AxiomC{$\R, w: \psi(\undb/x), \Gamma \Rightarrow \Delta$}
\RightLabel{$\wk$}
\dashedLine
\UnaryInfC{$\R, w: \psi(\undb/x), \Gamma \Rightarrow w : \psi(\unda/x), \Delta$}
\RightLabel{$\existslnc$}
\UnaryInfC{$\R, w : \exists x \psi, \Gamma \Rightarrow w : \psi(\unda/x), \Delta$}
\RightLabel{\ih}
\dashedLine
\BinaryInfC{$\R, \Gamma \Rightarrow w : \psi(\unda/x), \Delta$}

\AxiomC{$\R, w: \psi(\undb/x), \Gamma \Rightarrow \Delta$}
\RightLabel{$\psub$}
\dashedLine
\UnaryInfC{$\R, w: \psi(\unda/x), \Gamma \Rightarrow \Delta$}

\RightLabel{\ih}
\dashedLine
\BinaryInfC{$\R, \Gamma \Rightarrow \Delta$}
\DisplayProof
\end{tabular}
\end{flushright}

\begin{center}
$\Pi_{1} = \Bigg \{ \quad \R, \Gamma \sar \Delta, w: \psi(\unda/x), w: \exists x \psi$
\end{center}

\quad \quad \emph{2.4.2 The cut formula is of the form $\forall x \psi$.} The case is resolved as shown below. We my invoke the first use of \ih since the sum of the heights $h_{1} + h_{2}$ is one less than the original $\cut$, and we may invoke the second use of \ih since the cut formula $v : \psi(\unda / x)$ is of a smaller complexity, that is $\fcomp{\psi(\unda/x)} < \fcomp{\forall x \psi} = \fcomp{\phi}$. We note that the use of \lem~\ref{lem:deleting-relational-atoms-refined} is justified since there exists a propagation path $\ppath(w,v)$ in the propagation graph of $\R, \Gamma \Rightarrow \Delta$ such that  $\stra_{\ppath}(w,v) \in \thuesysilang{a}$ due to the application of $\alllnc$.

\begin{flushleft}
\begin{tabular}{c c c}
\AxiomC{$\R, w \leq u, \Gamma \Rightarrow \Delta, u : \psi(\undb/x)$}
\RightLabel{$\allrnc$}
\UnaryInfC{$\R, \Gamma \Rightarrow \Delta, w : \forall x \psi$}

\AxiomC{$\R, v : \psi(\unda / x), w : \forall x \psi, \Gamma \Rightarrow \Delta$}
\RightLabel{$\alllnc$}
\UnaryInfC{$\R, w : \forall x \psi, \Gamma \Rightarrow \Delta$}

\RightLabel{$\cut$}
\BinaryInfC{$\R, \Gamma \Rightarrow \Delta$}

\DisplayProof

&

$\leadsto$
\end{tabular}
\end{flushleft}

\begin{flushright}
\AxiomC{$\R, w \leq u, \Gamma \Rightarrow \Delta, u : \psi(\undb / x)$}
\RightLabel{$\lsub$}
\dashedLine
\UnaryInfC{$\R, w \leq v, \Gamma \Rightarrow \Delta, v : \psi(\undb / x)$}
\RightLabel{$\psub$}
\dashedLine
\UnaryInfC{$\R, w \leq v, \Gamma \Rightarrow \Delta, v : \psi(\unda / x)$}

\AxiomC{$\Pi_{1}$}

\AxiomC{$\Pi_{2}$}
\RightLabel{\ih}
\dashedLine
\BinaryInfC{$\R, w \leq v, v : \psi(\unda / x), \Gamma \Rightarrow \Delta$}

\RightLabel{\ih}
\dashedLine
\BinaryInfC{$\R, w \leq v, \Gamma \Rightarrow \Delta$}
\RightLabel{\lem~\ref{lem:deleting-relational-atoms-refined}}
\dashedLine
\UnaryInfC{$\R, \Gamma \Rightarrow \Delta$}
\DisplayProof
\end{flushright}

\begin{center}
\begin{tabular}{c c c}
$\Pi_{1}$ & 

$= \Bigg \{$

& 

\AxiomC{$\R, w \leq u, \Gamma \Rightarrow \Delta, u : \psi(\undb / x)$}
\RightLabel{$\allrnc$}
\UnaryInfC{$\R, \Gamma \Rightarrow \Delta, w : \forall x \psi$}
\RightLabel{$\wk$}
\dashedLine
\UnaryInfC{$\R, w \leq v, v : \psi(\unda / x), \Gamma \Rightarrow \Delta, w : \forall x \psi$}
\DisplayProof
\end{tabular}
\end{center}

\begin{center}
\begin{tabular}{c c c}
$\Pi_{2}$ & 

$= \Bigg \{$

& 

\AxiomC{$\R, v : \psi(\unda / x), w : \forall x \psi, \Gamma \Rightarrow \Delta$}
\RightLabel{$\wk$}
\dashedLine
\UnaryInfC{$\R, w \leq v, v : \psi(\unda / x), w : \forall x \psi, \Gamma \Rightarrow \Delta$}
\DisplayProof
\end{tabular}
\end{center}

\end{proof}




\printindex

\printglossaries

\bibliographystyle{alpha}
\bibliography{bibliography}

\end{document}